\documentclass[3p,times,11pt]{elsarticle}

\usepackage[english]{babel}
\usepackage[T1]{fontenc}
\usepackage{graphics}
\usepackage{graphicx}
\usepackage{tabularx}
\usepackage{fancyheadings}
\usepackage{amsmath}
\usepackage{amsbsy}
\usepackage{pifont}
\usepackage{float}
\usepackage{amssymb}
\usepackage{delarray}
\usepackage{rotating}
\usepackage{makeidx}
\usepackage{color}
\usepackage[multiple]{footmisc}
\usepackage{xspace}
\usepackage{subfig}
\usepackage{bm}
\usepackage{longtable}
\DeclareGraphicsExtensions{.png,.pdf,.eps}

\DeclareMathAlphabet{\mathbfit}{T1}{cmr}{bx}{it}
\DeclareMathAlphabet{\bbox}{T1}{cmr}{bx}{it}
\DeclareMathAlphabet{\bbbox}{T1}{cmr}{bx}{n}

\graphicspath{{Figures/}{./Figures/}}

\usepackage{bm}
\usepackage[hypertex,colorlinks=true,linkcolor=red,citecolor=blue]{hyperref}

\newcommand{\nn}{\nonumber}

\usepackage{comment}
\usepackage{array}
\usepackage{longtable}
\usepackage{multirow}
\usepackage{calc}
\usepackage{multirow}
\usepackage{hhline}
\usepackage{ifthen}
\usepackage{sectsty}

\newif\ifshowcitations\showcitationsfalse%
\newif\ifshowlinks\showlinksfalse%

\ifshowlinks%
  \usepackage[
        colorlinks=true,
         urlcolor=blue,       
         ]{hyperref}
  \newcommand*{\inspireurl}[1]{\\\href{#1}{INSPIRE-HEP entry}}
\else
  \makeatletter
  \newcommand*{\inspireurl}[1]{\@bsphack\@esphack}
  \makeatother
\fi
\ifshowcitations%
  \newcommand*{\citations}[1]{\\* #1}
\else
  \makeatletter
  \newcommand*{\citations}[1]{\@bsphack\@esphack}
  \makeatother
\fi

\newcommand{\be}{\begin{eqnarray}}
\newcommand{\ee}{\end{eqnarray}}
\newcommand{\ba}{\begin{array}}
\newcommand{\ea}{\end{array}}

\newcommand{\bea}{\begin{eqnarray}}
\newcommand{\eea}{\end{eqnarray}}

\newcommand{\bi}{\begin{itemize}}
\newcommand{\ei}{\end{itemize}}
\renewcommand{\nn}{{\nonumber}}

\newcommand{\im}{{\rm Im}}
\newcommand{\re}{{\rm Re}}

\restylefloat{figure}
\restylefloat{table}

\renewcommand{\theequation}{\arabic{section}.\arabic{equation}}%

\begin{document}

\begin{frontmatter}

\title{Transition distribution amplitudes and hard exclusive reactions with baryon number transfer}

\date{\today}
\author[CPHT]{B.Pire}
\author[SP,HE]{ K. Semenov-Tian-Shansky}
\author[NCBJ]{  L. Szymanowski}
\address[CPHT]{CPHT, CNRS, Ecole Polytechnique, IP Paris, 91128-Palaiseau, France}
\address[SP]{National Research Centre Kurchatov Institute: Petersburg Nuclear Physics Institute, 188300 Gatchina, Russia}
\address[HE]{Higher School of Economics,
National Research University, 194100 St. Petersburg, Russia}
\address[NCBJ]{National Centre for Nuclear Research, NCBJ, 02-093 Warsaw, Poland}

\begin{abstract}
\mbox

Baryon-to-meson and baryon-to-photon transition distribution amplitudes (TDAs) arise in the collinear factorized description of a class of hard exclusive reactions characterized by the exchange of a non-zero baryon number in the cross channel. These TDAs extend the concepts of generalized parton distributions (GPDs) and baryon distribution amplitudes (DAs). In this review we discuss
the
general properties and physical interpretation of baryon-to-meson and
baryon-to-photon TDAs. We argue that these non-perturbative objects are a convenient complementary tool to explore the structure of baryons at the partonic level. We present an overview of hard exclusive reactions admitting a description in terms of TDAs. We discuss the first signals from hard exclusive backward meson electroproduction at JLab with the 6 GeV electron beam and explore further experimental opportunities to access TDAs at JLab@12 GeV, \=PANDA, J-PARC and EIC.
\end{abstract}

\end{frontmatter}

\tableofcontents

 \section{Introduction}
 \label{Sec_Introduction}
 \mbox

The remarkable property of ``asymptotic freedom'' of Quantum Chromodynamics (QCD)~\cite{Politzer:1974fr,Marciano:1977su,Dokshitzer:1978hw,Mueller:1981sg} ensures the validity of  perturbative methods in the description
of strong interaction phenomena at short distances. It makes QCD a
self-consistent
relativistic quantum field theory, for which a perturbative analysis gives
a correct treatment of ultraviolet divergences. However, perturbative methods
fail to provide a description for the simplest QCD bound states (light hadrons) in terms of
fundamental quark and gluon degrees of freedom. The study of hadronic structure has become the goal of numerous experimental and theoretical studies aiming to the development of refined theoretical methods to address QCD in the strong coupling regime.

Among these methods, one of the most powerful and universal approaches is that of
hard-scattering factorization
(for a review see \textit{e.g.} the monograph~\cite{JCollins_pQCD}). It allows separating the interaction into a long-range (soft) part, described by means of universal non-perturbative light-cone dominated hadronic matrix
element of
non-local QCD operators, and a short-range (hard) part, for which perturbative QCD can be systematically applied. Establishing  factorization theorems leads to a rich phenomenological program for numerous observables measured in
$e^+ e^-$-annihilation, deep inelastic $e N$-, $\nu N$-, and $A N$-collisions, with $A$ being a photon, a meson or a nucleon.

The textbook example is provided by the factorized  description of Deep Inelastic Scattering (DIS). The corresponding factorization theorem allows writing the DIS cross sections
as a convolution of a perturbatively calculable coefficient function with
parton distribution functions (PDFs).
 The simplest quark PDF  is defined as the diagonal hadronic matrix element of a non-local quark--antiquark operator on the light-cone ($z^2=0$):
$
\hat{O} (-z/2, z/2)=\bar{\Psi}(-z/2) \gamma^+
\Psi (z/2)
$,
where $\gamma^+$ stands for the light-cone component of $\gamma^\mu$; and the use of the light-cone gauge $A^+=0$ is assumed.

While the early studies considered mostly inclusive or semi-inclusive cross sections, the advent of high luminosity electron beams and advanced detectors allowed to access the exclusive channels. An important step forward was the derivation of the factorization property  of the deeply virtual Compton scattering (DVCS) \textit{near-forward} amplitude in terms of perturbatively calculable coefficient functions and of generalized parton distributions (GPDs) defined as the
\textit{non-diagonal} matrix elements of the same operators as in DIS.
GPDs were found to be an extremely convenient tool to address the origin of the
nucleon's spin~\cite{Ji:1996ek}, to study the spatial distribution
of forces experienced by quarks and gluons inside hadrons~\cite{Polyakov:2002yz};
and to explore the three-dimensional structure
of hadrons at the partonic level~\cite{Burkardt:2000za,Ralston:2001xs,Diehl:2002he}. This opened a completely new chapter in the quest for a quark and gluon description of hadrons. There are many excellent reviews of these
advances, see \textit{e.g.} Refs.~\cite{Goeke:2001tz,Diehl:2003ny,Belitsky:2005qn,Boffi:2007yc,Mueller:2014hsa}.

A natural question, which emerges from these developments, is the following: since both forward and nearly-forward deeply virtual scattering amplitudes can be related in a fruitful way to hadronic matrix elements of quark and gluon operators, that capture the dynamics of quark and gluon confinement in hadrons, can similar ideas be applied to \textit{backward} reactions? This is the essence of the introduction of transition distribution amplitudes (TDAs)
\cite{Frankfurt:1999fp,Pire:2004ie,Pire:2005ax},
that are designed to play a role similar to GPDs in a complementary kinematical domain of DVCS (and similar reactions). The basic difference between GPDs and TDAs lies in the operator which defines them, the non-local three-quark operator on the light-cone:
$
 \hat{O}_{\rho \tau \chi}(z_1,z_2,z_3)= \varepsilon_{c_1 c_2 c_3}
 \Psi_\rho^{c_1} (z_1) \Psi_\tau^{c_2}(z_2) \Psi_\chi^{c_3} (z_3)  \,.
$
Here $z_i$ are light-cone distances ($z_i^2=0$); $\rho$, $\tau$, $\chi$
denote the Dirac indices; antisymmetrization is performed over the color group indices $c_i$.
Baryon-to-meson (respectively, baryon-to-photon) TDAs
are defined as matrix elements of this three quark light-cone operator
between a baryon state
$| B(p_B) \rangle$ and a meson state
$\langle{\mathcal{M}}(p_{\mathcal{M}})\rvert  $
(or a photon state $\langle\gamma(p_\gamma)\rvert  $).
Similarly to GPDs, TDAs are functions of the $3$ light-cone momentum fractions
$x_i$,
the skewness variable
$\xi$,
that, in contrast to the GPD case, is defined with respect to the longitudinal momentum transfer between the initial baryon and the final meson (or photon), and a momentum transfer squared $(p_B-p_{\mathcal{M} \,  (\gamma)})^2$,
as well as of the
factorization
scale $\mu$.

Baryon-to-meson (baryon-to-photon) TDAs occur within the collinear factorized description of a class of hard exclusive reactions with a non-zero baryon number exchange in the cross channel. Prominent examples of such reactions are the backward DVCS and backward hard electroproduction of mesons off nucleons
\cite{Lansberg:2011aa,Pire:2015kxa}
and nucleon--antinucleon annihilation into a lepton pair (or a heavy quarkonium) associated with production of a light meson~\cite{Lansberg:2012ha,Pire:2013jva}.

The non-local three-quark light-cone operator written above
has been used for a long time to define the baryon distribution amplitudes (DAs) through the vacuum to baryon matrix elements:
$
\langle B(p_B)\rvert
\hat{O}_{\rho \tau \chi}(z_1,z_2,z_3)
  | 0 \rangle  \,.
$
These non-perturbative objects
were extensively applied to provide
the factorized description of various reactions including the QCD description of the nucleon electromagnetic form factors
and heavy quarkonium decay~\cite{Lepage:1980fj,Chernyak:1983ej,Chernyak:1987nv}.
For a review see \textit{e.g.} Refs.~\cite{Stefanis:1999wy,Braun:1999te}.

Thus, baryon-to-meson (and baryon-to-photon) TDAs share common features both with baryon DAs and with GPDs and encode a conceptually close physical picture. They characterize partonic correlations inside a baryon and give access to the momentum distribution of the baryonic number inside a baryon. Similarly to GPDs, TDAs -- after the Fourier transform in the transverse plane -- represent  valuable information on the transverse location of hadron constituents.

Recent experimental studies
\cite{Park:2017irz,Li:2019xyp,Diehl:2020uja}
brought first evidences in favor of the validity of the reaction mechanism involving nucleon-to-meson TDAs for the description of backward pion or
$\omega$-electroproduction at JLab kinematical conditions.
The perspective
to access baryon-to-meson TDAs experimentally
\cite{Lutz:2009ff,Li:2017xcf,Li:2020nsk,WP:21}
rises a high demand for a consistent exposition of the considerable theoretical progress in the field achieved during the last two decades.

This review presents, in a broad context of applications of the collinear factorization approach in QCD, the physical content of TDAs as well as  their application
to the description of hadronic structure.
We  also
present an overview of existing phenomenological models of TDAs and their predictions
for the kinematical conditions of existing and planned experimental facilities. The
results of existing feasibility studies
\cite{Singh:2014pfv,Singh:2016qjg} are also reported.

The review is much inspired by  the original papers
\cite{Lansberg:2007ec,Lansberg:2007se,Pire:2010if,Pire:2011xv,Lansberg:2011aa,Lansberg:2012ha,Pire:2013jva,Pire:2015kxa,Pire:2016gut} published by the authors and their collaborators during these last 15~years.

\section{Collinear factorization framework for hard exclusive reactions}
\setcounter{equation}{0}
\label{Sec_CollinearFactFramework}
\mbox

This section is devoted to a brief overview of the main theoretical
concepts, which provide the basis for the description  of  hard exclusive processes in  the
near-forward kinematics within the collinear factorization
framework. These developments began with the analysis of deep inelastic scattering processes and the introduction of the partonic description of inclusive cross sections. Historically, it has contributed a lot to the confidence we now have in QCD as a consistent theory of strong interactions.
Following that, we recall the description of electromagnetic form factors at large momentum transfer
in terms of hadron distribution amplitudes. We finally sketch the main results of the collinear QCD description of near-forward scattering amplitudes in terms of GPDs and review their basic properties. These concepts
will be generalized in the following sections to provide a description of  hard exclusive  processes in the near-backward kinematics in terms of TDAs.

\subsection{Deep inelastic scattering and parton distribution functions }
\mbox

The QCD collinear factorization framework
\cite{Mueller:1981sg,JCollins_pQCD}
has been primarily developed for inclusive reactions such as deep inelastic scattering (DIS),
hadron production in
$e^+ e^-$-collisions,
lepton pair production in hadron--hadron collisions (Drell--Yan),
hadron (or jet) production at large transverse momentum in hadron--hadron collisions.

\begin{figure}[H]
\begin{center}
\includegraphics[width=0.4\textwidth]{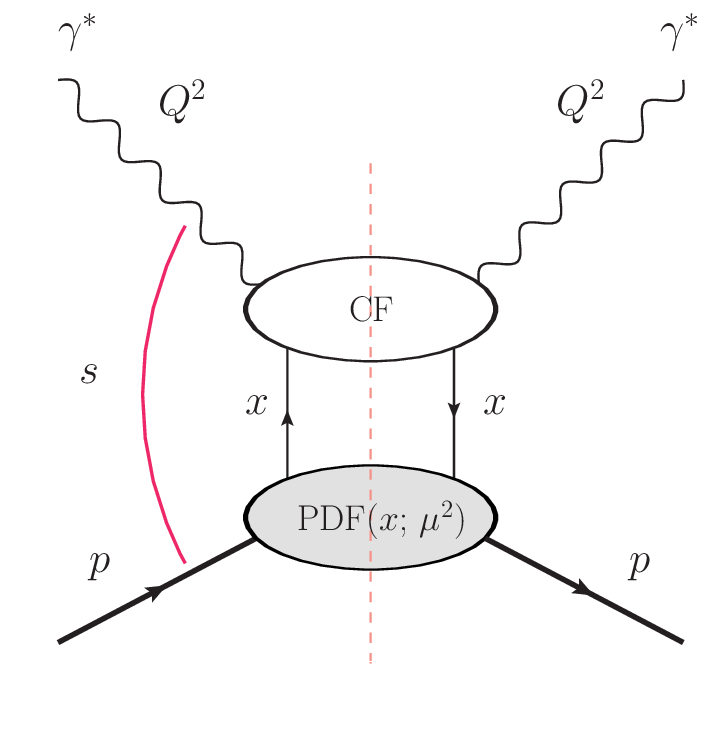}
\end{center}
     \caption{Factorization of DIS : the optical theorem relates the total
     cross section of DIS  ($\gamma^{*} \pi \to X$) to the imaginary part of the forward Compton amplitude
     ($\gamma^{*} \pi \to \gamma^{*} \pi$).}
\label{Fig_DIS}
\end{figure}

The collinear  factorization theorem for the DIS
(see  Fig.~\ref{Fig_DIS})
allows to present the corresponding
cross sections as a convolution of  perturbatively calculable coefficient functions
(CFs) with parton distribution functions (PDFs).
Below we consider the case of
a pseudoscalar meson target. Within the light-cone gauge ($A^+=0$) the corresponding quark PDF is defined as the diagonal
hadronic matrix element of the non-local quark--antiquark light-cone operator\footnote{Throughout this review we adopt the standard conventions for the
Sudakov decomposition of the relevant $4$-vectors. For the specification see
Eqs.~(\ref{Def_p_n}), (\ref{Def_Sudakov_dec}).}:
\begin{equation}
q_\pi(x,\mu^2) = \int \frac{dz^-}{4\pi} e^{ix p \cdot z } \langle \pi(p)|   \bar \Psi(-z/2) \gamma^+ \Psi (z/2) |  \pi(p) \rangle \big|_{z^+=0, \ z_T=0} \,.
\label{Def_pion_PDF}
\end{equation}
Here $x$ is the Bjorken variable; $z$ is the light cone distance ($z^2=0$);
$\gamma^+ \equiv 2 \gamma^\mu n_\mu$, with $n$ being a light cone vector ($n^2=0$);
and $\mu$ is the factorization scale.
Note that since for DIS the factorization occurs at the cross section level,
the matrix elements defining  PDFs are \textit{diagonal}. This comes from the fact that
an inclusive  DIS cross section is represented as the imaginary part of a forward Compton amplitude.

The collinear factorization framework for DIS was established through the experimental observation of the Bjorken scaling law, which states that PDFs are $Q^2$-independent. This scaling law is however violated by a logarithmic $Q^2$-dependence inferred from the
Dokshitzer--Gribov--Lipatov--Altarelli--Parisi
(DGLAP)
evolution equations
\cite{Gribov:1972ri,Altarelli:1977zs,Dokshitzer:1977sg}.

Similarly to the DIS case,
the QCD analysis of exclusive amplitudes will require the existence of a hard scale,
usually denoted as $Q^2$, which is large enough to prevent higher twist contributions to pollute the analysis of experimental data.
The onset of the collinear factorization (often named \textit{partonic} regime) is usually  probed through the observation of the adequacy of the scaling laws for cross-sections or  appropriate polarization observables.

\subsection{Electromagnetic form factors and distribution amplitudes}
\label{SubSec_EM_FFs_in_QCD_sec2}
\mbox

A subsequent application of the collinear factorization approach was proposed in
\cite{Efremov:1979qk,Lepage:1979zb,Lepage:1980fj}
for the description  of electromagnetic form factors (FFs) of hadrons at large invariant momentum transfer. In the mesonic case, the relevant hadronic matrix element of the light-cone
$\bar{\Psi} \Psi$
operator is
\begin{equation}
\langle \pi(p)|  \bar \Psi(-z/2) \gamma^+ \Psi (z/2) |  0 \rangle \,,
\label{DApi}
\end{equation}
while in the baryonic case, the  relevant matrix element of the three-quark  light-cone operator  $ \Psi \Psi \Psi $ is
\begin{equation}
\langle B(p)|  \varepsilon_{c_1 c_2 c_3}
\Psi_\rho^{c_1} (z_1) \Psi_\tau^{c_2}(z_2) \Psi_\chi^{c_3} (z_3)
  | 0 \rangle  \,.
  \label{DAbar}
\end{equation}
The corresponding hadronic quantities, the distribution amplitudes (DAs), are the Fourier transforms of the matrix elements
(\ref{DApi}),
(\ref{DAbar})
decomposed over an appropriate set of the Dirac structures
(for a review, see Refs.~\cite{Chernyak:1983ej,Stefanis:1999wy,Braun:1999te}).
The factorization scale dependence of hadron DAs is controlled by the
Efremov--Radyushkin--Brodsky--Lepage (ERBL)
\cite{Efremov:1978rn,Efremov:1979qk,Lepage:1979zb,Lepage:1979za,Lepage:1980fj} evolution equations.

\begin{figure}[H]
\begin{center}
\includegraphics[width=0.4\textwidth]{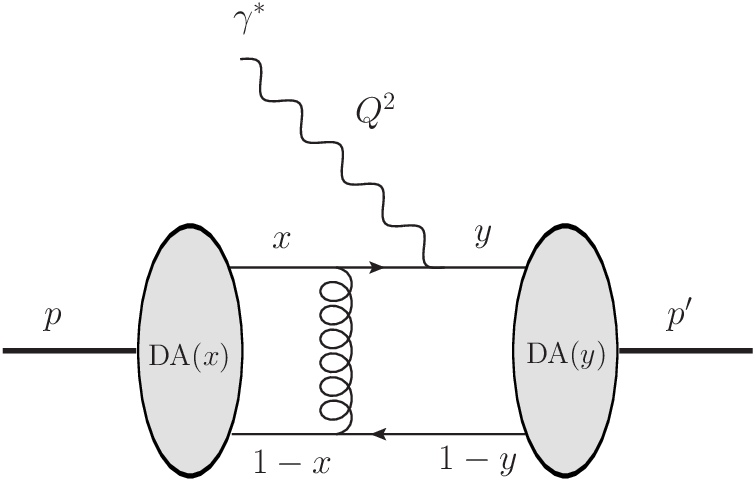} \ \ \ \ \ \ \ \ \
\includegraphics[width=0.4\textwidth]{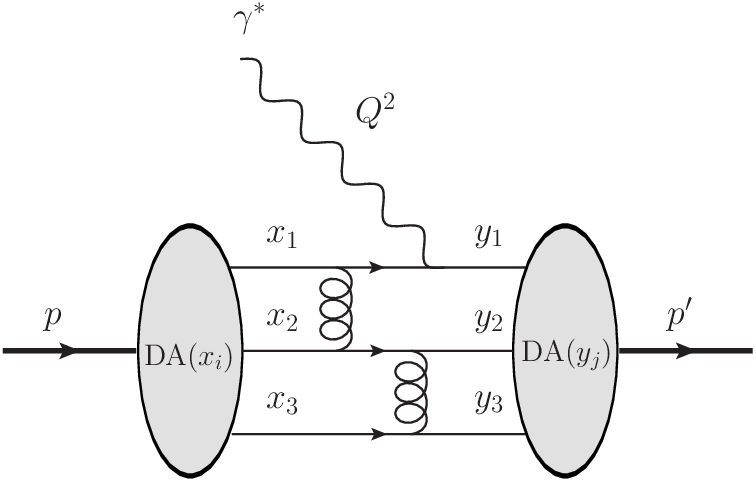}
\end{center}
     \caption{Examples of  Feynman graphs contributing to the
     meson (left panel) and
     baryon (right panel) electromagnetic form factor at the tree level.}
\label{Fig_FFs}
\end{figure}

 Fig.~\ref{Fig_FFs} provides examples of leading order diagrams contributing to the meson and baryon electromagnetic form factor.
The scaling law predicted by the QCD collinear factorization framework, namely
\begin{equation}
F_\pi(Q) \sim \frac{1}{Q^2}~;~~~ F_{1\, N}(Q) \sim \frac{1}{Q^4}\,,
\end{equation}
was a first indication of the success of this approach, as were the successes of the dimensional scaling laws~\cite{Matveev:1972gb,Brodsky:1973kr} for many fixed angle scattering processes~\cite{Amaryan:2021cnj}.

Further development of the collinear factorization approach,
however, revealed subtleties in the application
to fixed angle hadronic scattering (for a review, see \textit{e.g.} Ref.~\cite{Mueller:1981sg}).
The importance of the Sudakov suppression of some delicate integration regions was in particular discovered~\cite{Botts:1989nd},
which in turn may help to understand the suppression  of endpoint region contributions  for meson and baryon form factors~\cite{Li:1992nu,Li:1996gi}
(for an alternative point of view, see
\cite{Dagaonkar:2014yea,Dagaonkar:2015laa}).
Moreover, the absence of pinch singularities, which is a necessary element of the proof of factorization,  was shown in
\cite{Farrar:1989wb}
for the scattering amplitude for electroproduction processes at fixed angle,
putting on a firm ground the collinear factorized framework for various processes.

However, nowadays it is recognized that the hard scattering mechanism
alone does not provide a satisfactory description of the existing electromagnetic form factor data up to rather large values
of $Q^2$. A possible remedy was proposed within the framework based on the light-cone sum rules (LCSRs)
(see \textit{e.g.} Ref.~\cite{Colangelo:2000dp} for a review). Within this approach the ``soft contributions'' into form factors are systematically computed in terms of the
same DAs that  occur in the collinear factorization framework. The application of the
LCSR approach to nucleon form factors is presented in Refs.
\cite{Lenz:2003tq,Braun:2001tj,Lenz:2009ar}.

\subsection{Near-forward exclusive scattering and generalized parton distributions
}
\mbox

A significant breakthrough in QCD appeared when it was realized
\cite{Mueller:1998fv,Radyushkin:1996nd,Ji:1996nm,Collins:1996fb,Radyushkin:1997ki}
that some exclusive processes, such as the
deeply virtual Compton scattering (DVCS)
\begin{equation}
    \gamma^{*}(q) +N(p) \to \gamma (q')+ N(p') \,,
    \label{reacDVCS}
\end{equation}
and hard exclusive meson production (HMP)
\begin{equation}
    \gamma^{*}(q) +N(p) \to {\mathcal{M}} (p_{\mathcal{M}})+ N(p') \,,
\end{equation}
in the generalized Bjorken limit of large
$Q^2=-q^2$, $W^2=(p+q)^2$
with fixed
$x_B = Q^2/(2 p \cdot q)$,
and for a limited range of the invariant momentum transfer $t = (p-p')^2 \sim 0$
proceed via the short-distance scattering on a single parton and may be subject to a collinear factorized description. The condition $t \sim 0$ corresponds
to the final state photon (or meson) produced in the nearly forward direction in the
$\gamma^{*} N$ center-of-mass system (CMS). Therefore, the corresponding
kinematical regime is often referred to as \textit{near-forward} kinematics.

A much related subject is the discussion of the crossed reaction to the DVCS process
(\ref{reacDVCS}), deep exclusive small invariant mass ($s=(p_1+p_2)^2 \ll Q^2$) hadron pair production:
\begin{equation}
    \gamma^{*}(q) + \gamma (q')  \to H_1(p_1)+ H_2(p_2) \,,
    \label{reacGDA}
\end{equation}
where collinear factorized description is performed in terms of generalized distribution amplitudes (GDAs)~\cite{Mueller:1998fv,Diehl:1998dk} defined as the Fourier transforms of the matrix elements $\langle H_1(p_1)H_2(p_2)|  \bar\Psi(-z/2) \gamma^+ \Psi(z/2) |  0 \rangle $.

\begin{figure}[H]
\begin{center}
\includegraphics[width=0.4\textwidth]{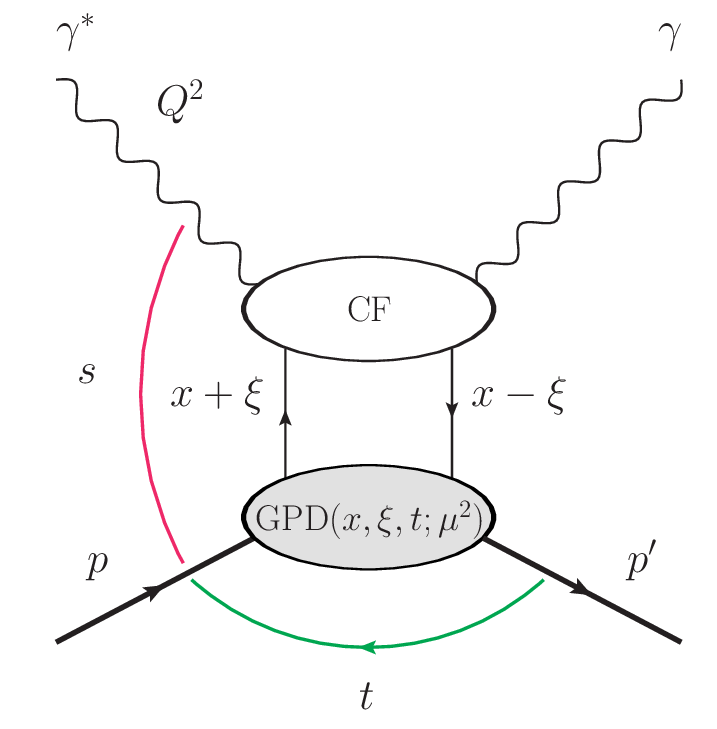}
\end{center}
     \caption{Collinear factorization mechanism for DVCS.}
\label{Fig_DVCS}
\end{figure}

For definiteness here we limit our considerations to the simplest case of
DVCS on a pseudoscalar meson target. The corresponding collinear
factorization mechanism is presented  in  Fig.~\ref{Fig_DVCS}.
The short-distance process amplitude (CF) involves only longitudinal momenta. It can be systematically computed in the perturbation theory. The DVCS amplitude is presented as a convolution of
the corresponding perturbative amplitude with the Fourier transforms of
\textit{off-diagonal} matrix elements of
quark (or gluon) light-cone operators
\cite{Dittes:1988xz,Mueller:1998fv}, named generalized parton distributions (GPDs).

Subsequent developments
\cite{Ji:1996ek,Ji:1996nm,Radyushkin:1997ki}
led to the detailed   understanding of this new tool for the study of the physics of confined quarks and gluons
and the description of hadron structure.
We refer the reader to a collection of excellent reviews
\cite{Goeke:2001tz,Diehl:2003ny,Belitsky:2005qn,Boffi:2007yc}
existing on this subject.
In the remaining part we will quote some crucial features
of the \textit{near-forward} GPD framework.
In  Sections \ref{Sec_StatusFAct},
\ref{Sec_DefAndProp}, with appropriate modification, these features
will be adapted to the
\textit{near-backward} TDA framework.

In particular, the leading chiral-even twist-$2$ quark GPD of a pseudoscalar meson is defined as the
hadronic matrix element of the
$\bar \Psi \Psi$ operator on the light cone\footnote{We imply the use of the light-cone gauge
$A^+=0$
for the gluon fields and omit the otherwise necessary gauge link.}:
\begin{equation}
H^q_\pi(x,\xi,t,\mu^2) =  \int \frac{dz^-}{4\pi} e^{ix P \cdot z} \langle \pi(p') |  \bar \Psi(-z/2)\gamma^+ \Psi (z/2) |  \pi(p) \rangle \big|_{z^+=0,\;{z_T}=0} \,,
\label{Def_pion_GPD}
\end{equation}
where $P \equiv \frac{p+p'}{2}$;  $\Delta \equiv p'-p$;
$x$ is the light cone momentum fraction variable,
$\xi=-\frac{\Delta^{+}}{2P^{+}}$,
the so-called skewness variable, characterizes the longitudinal momentum transfer between the initial and final hadron states,
$t=\Delta^2$
is the invariant squared momentum transfer and
$\mu$ is the factorization scale.
In the forward limit $\xi=0$, $t=0$ the GPD
(\ref{Def_pion_GPD})
reduces to the forward PDF (\ref{Def_pion_PDF}):
\begin{equation}
H^q_\pi(x,\xi=0,t=0,\mu^2)=q_\pi(x,\mu^2).
\end{equation}

The GPDs possess the restricted $[-1,\,1]$ support in the longitudinal
momentum-fraction variable $x$.
Their partonic interpretation in the momentum space leads to
the definition of three distinct regions corresponding to different directions of the longitudinal momentum flows:
\bi
\item
When $x>\xi$, both momentum fractions $x+\xi$ and $x-\xi$ are positive and  GPDs describe the emission and reabsorption of a quark.

\item
When $| x|  \le \xi$, the momentum fraction $x-\xi$ is interpreted as belonging to an antiquark with momentum fraction $\xi -x$, and  GPDs describe the emission of a quark--antiquark pair from the initial hadron.

\item
When $x< -\xi$, both momentum fractions $x+\xi$ and $x-\xi$ are negative and are interpreted as belonging to  antiquarks with momentum fraction $\xi -x$ and $-\xi -x$,  GPDs describing then the emission and reabsorption of an antiquark.
\ei

The scale evolution of hadronic matrix elements of light-cone operators turns to be firstly a property of the
operators in question.
Since the quark--antiquark
$\bar{\Psi} \gamma^+ \Psi$
operator occurring in the definition of quark GPDs is the same as that defining PDFs and mesonic DAs, the evolution equations for GPDs  are much related to the DGLAP evolution equations for PDFs and the ERBL evolution of
DAs~\cite{Mueller:1998fv}.
In fact, the GPD evolution in the outer  support regions  $x>\xi$ and $x<-\xi$
(often called the DGLAP regions)
is governed by the DGLAP-type equations, while in the inner support region
$| x|  \le \xi$ (referred to as the ERBL region), the evolution equations
turn to be of the ERBL-type.

Among the various GPD properties a special role is attributed
to the so-called polynomiality property of the
Mellin moments of GPDs in the momentum-fraction variable $x$.
This highly non-trivial constraint is a direct consequence of
the underlying Lorentz invariance.
Indeed, integrating over $x$ the matrix elements of bilocal operators
removes all reference to a light-cone axis defined by the vector $n$ and
provides matrix elements of \textit{local} operators.
The $N$th Mellin moment of a GPD turns to be a polynomial of order
(at most)
$N+1$ in the skewness variable $\xi$.
The coefficients at powers of $\xi$ can be expressed through the
form factors of the local twist-$2$ operators:
\begin{equation}
\widehat{O}^{\mu \mu_1...\mu_{N}}(0)= \bar{\Psi}(0)
\gamma^\mu
i \overleftrightarrow{D}^{\mu_1}...\,i\overleftrightarrow{D}^{\mu_{N}}  \Psi(0),
\label{local_op_derivatives-tw2}
\end{equation}
where
$
\overleftrightarrow{D}^\mu = \overleftrightarrow{\partial}^\mu -i g  A^{a \, \mu} t^a
$
is the covariant derivative. Here $t^a= \frac{\lambda^a}{2}$ with $\lambda^a$, $a=1,\ldots,8$ being the Gell-Mann matrices;
$
 \overleftrightarrow{\partial}^\mu
\equiv \frac{1}{2}\left(\overrightarrow{\partial}^{\mu}-\overleftarrow{\partial}^{\mu}\right)$.

The polynomiality property brings one of the most important practical application of the
GPD formalism. The study of the first Mellin moment of quark  (gluon) GPDs
provides access
to the  hadronic matrix elements of the quark (respectively gluon) part of the QCD energy momentum
tensor. This allows to address the origin of hadron's spin
\cite{Ji:1996ek,Cosyn:2019aio}
and build up a comprehensible
picture of the ``mechanical properties'' of hadrons
\cite{Polyakov:2018zvc,Lorce:2018egm}.

A convenient way to implement both the
support and the polynomiality properties of GPDs consists
in expressing them in terms of more fundamental quantities,
the double distributions (DDs)~\cite{Radyushkin:1996nd,Radyushkin:1996ru,Musatov:1999xp}
(also called spectral densities in~\cite{Mueller:1998fv}).
The basic idea is to present the relevant hadronic matrix elements
as the Fourier-transform with respect to two independent scalars
$P \cdot z$ and $\Delta \cdot z$.
The corresponding double distribution
$f(\alpha,\beta,t)$
has the restricted support in the spectral variables
$(\alpha,\beta)$, which is the rhombus
$\Omega: \; | \alpha | + | \beta | \le 1$.
The double distribution representation of GPDs
highlights the hybrid nature of GPDs that combine the
properties of forward parton densities in the
$\xi \to 0$
limit and
those of distribution densities in the
$\xi \to 1$
limit.

In the original formulation of DD representation, in order to
satisfy the polynomiality condition
in its complete form, the spectral part
must be supplemented with the so-called $D$-term
\cite{Polyakov:1999gs}:
\begin{equation}
H^q(x,\xi,t) =\int_\Omega d\alpha d\beta \, \delta(x-\beta - \xi \alpha) f(\beta,\alpha, t) +
\theta(\xi-| x| )
D \left( \frac{x}{\xi},t\right).
\end{equation}
The $D$-term produces the highest possible power of $\xi$ ($\xi^{N+1}$)
of the $N$th ($N$ -odd) Mellin moments of the
$C$-even GPD.
In fact, as pointed out in Ref.~\cite{Teryaev:2001qm}, the DD representation  turns to be defined
up to a ``gauge transformation''.
By altering the admittable analytic properties of the spectral densities
\cite{Radyushkin:2011dh}
one can also obtain representations
\cite{Belitsky:2000vk}
with a $D$-term implemented into the spectral density.
The double distribution of GPDs was  extensively employed to provide the
framework for various phenomenological models for GPDs.

The physical contents of GPDs  became more transparent
within the impact parameter space interpretation proposed in
\cite{Burkardt:2000za,Ralston:2001xs,Diehl:2002he}.
The basic idea is to employ the
mixed representation of GPDs
Fourier transformed from transverse momentum to transverse position, referred as
the ``impact parameter''.
This  allows to perform hadron tomography in the transverse plane
and highlight the new physical information encoded in GPDs with respect
to forward PDFs.

The transition to the impact parameter implies introducing hadron states
of definite light cone momentum and definite position ${\mathbf{b}}$ in the transverse plane
$| p^+, {\mathbf{b}} \rangle$
and consideration of transverse position dependent quark--antiquark operator.
The Fourier transform of GPDs is performed with respect to the transverse part ${\mathbf{D}}$ of  the
$4$-vector
$D = \frac{p'}{1-\xi} - \frac{p}{1+\xi}$.
It is  related to the Mandelstam invariant $t$ by
\begin{equation}
t=t_0 - (1-\xi^2)  {\mathbf{D}}^2,
\end{equation}
where
$t_0 = \frac{-4\xi^2 m^2}{1-\xi^2}$
is the largest possible value of $t$.
In the  DVCS  case, $t=t_0$ corresponds
to the final state photon  produced exactly in the forward  direction.

For the case of the spin-$0$ target the resulting impact parameter representation of GPDs reads~\cite{Diehl:2002he}:
\begin{equation}
\int \frac{d^2 {\mathbf{D}}}{4\pi^2}e^{-i({\bf D }\cdot{\bf b})} H(x,\xi,t) = {\mathcal{N}} \frac{1+\xi^2}{(1-\xi^2)^{5/2}}\langle p'^+,-\frac{\xi {\mathbf{b}}}{1-\xi} | \widehat{O}({\mathbf{b}}) | p^+,\frac{\xi {\mathbf{b}}}{1+\xi} \rangle \,,
\end{equation}
where ${\mathcal{N}}$ is a normalization factor and the transverse position dependent  non-local quark--antiquark operator is
\begin{equation}
\widehat{O}({\mathbf{b}}) = \int \frac{dz^-}{4\pi} e^{ixP^+z^-} \bar{\Psi}(0,-z^-/2,{\mathbf{b}}) \gamma^+ \Psi(0,z^-/2,{\mathbf{b}}) \,.
\label{Def_impact_GPD}
\end{equation}
A similar impact parameter representation was established in
\cite{Pire:2002ut} for the cross-channel counterparts of GPDs ---
GDAs of hadronic states with meson quantum numbers.

\begin{figure}[H]
\begin{center}
\includegraphics[width=0.48\textwidth]{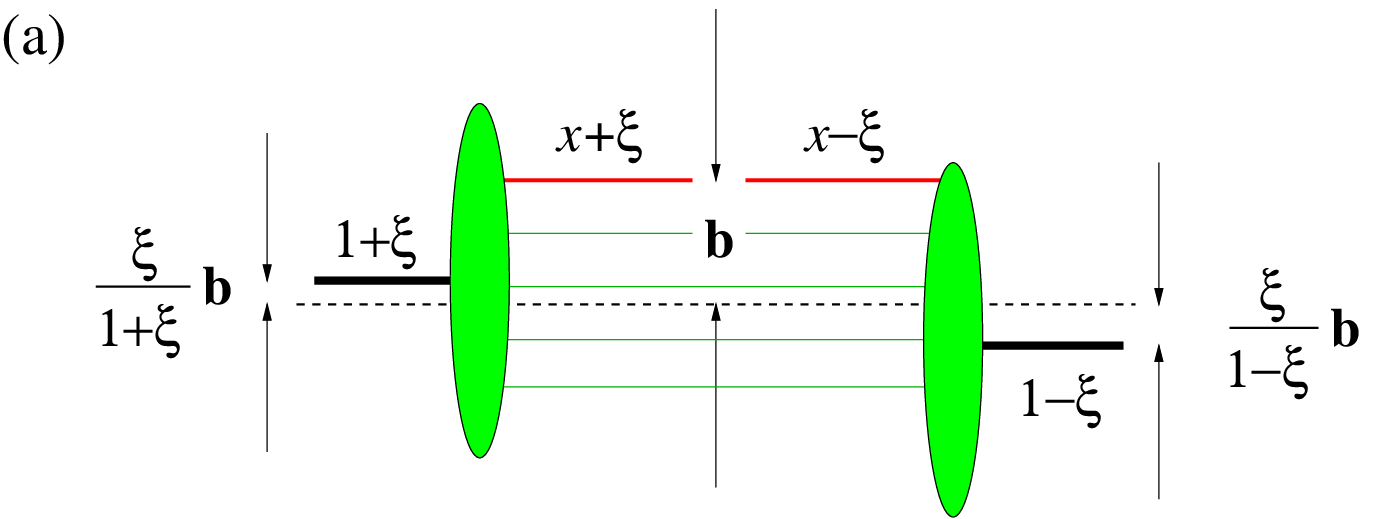} \ \ \ \
\includegraphics[width=0.48\textwidth]{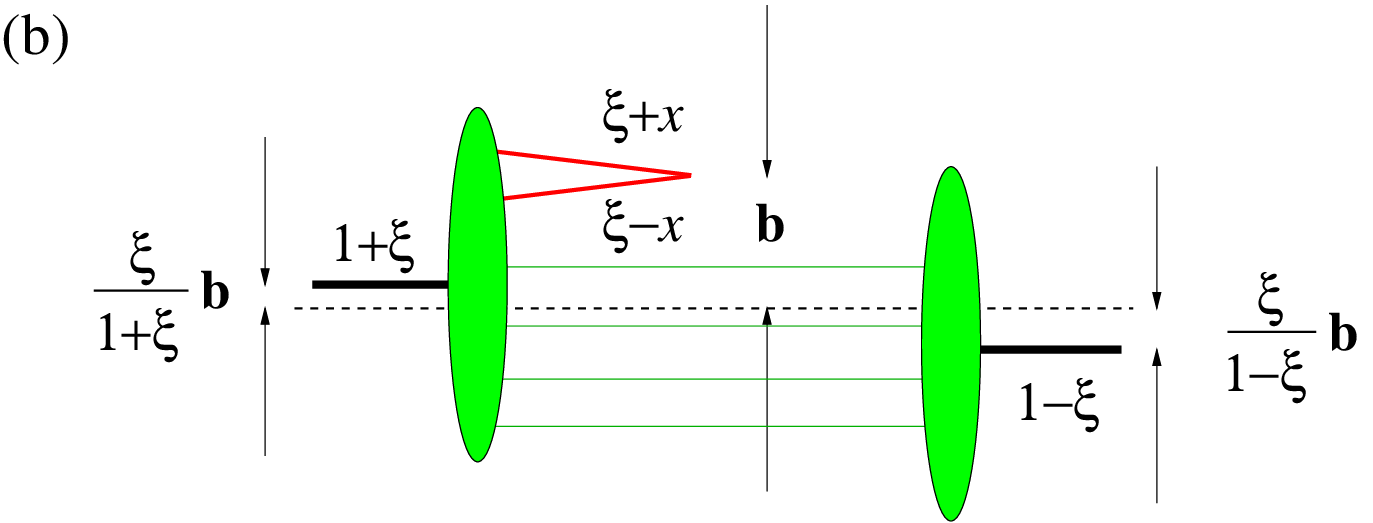}
\end{center}
     \caption{The impact parameter space representation of GPDs:  (a) DGLAP region of GPDs with $x \in  [\xi,1]$, (b) ERBL region of GPDs with $x\in [-\xi,\xi]$. [Reprinted Figure 12 from Ref.~\cite{Diehl:2003ny}. Copyright (2003) by Elsevier.]
 }
\label{DiehlImpactGPD}
\end{figure}

The interpretation of the GPD impact parameter representation
(\ref{Def_impact_GPD})
is illustrated in
 Fig.~\ref{DiehlImpactGPD} separately for the  DGLAP and the ERBL regions.
Depending on the $x$-range the relevant matrix element describes different processes.
\bi
\item In the DGLAP region $x> \xi$ (or $x<-\xi$) the
emission and subsequent reabsorption of a quark (or antiquark) at transverse position ${\mathbf{b}}$.
\item In the ERBL region $| x|  \le \xi$  the emission of a  quark--antiquark
pair at transverse position ${\mathbf{b}}$.
\ei
The non-zero skewness results in  $\xi$-dependent transverse shifts of centers of
incoming and outgoing particles. This reflects the fact that GPDs are more
complicated objects than PDFs which admit
a simple probability densities interpretation.

\section{Probing dynamics of the near-backward hard exclusive meson electroproduction}
\label{Sec_StatusFAct}
\setcounter{equation}{0}
\mbox

In this section we present the near-backward kinematical regime for
the reaction of hard exclusive meson electroproduction off a nucleon.
Following that, we discuss the present status of the collinear
factorization theorem providing a description of this reaction
in terms of nucleon-to-meson TDAs and nucleon DAs in the generalized Bjorken limit. We also shortly discuss the alternative description of hard processes within the Regge theory approach. We conclude this section with a summary of the color transparency effects to understand which further  experimental data are of crucial importance to test the onset of the TDA description.

\subsection{Light-cone kinematics for the near-backward regime}
\label{SubSec_Kinematics}
\mbox

Let us consider hard exclusive meson electroproduction
off a nucleon
\begin{equation}
e(k,s_e)+N(p_N,s_N) \to
\left(\gamma^{*}(q, \lambda_\gamma)+N(p_N,s_N) \right) + e(k',s'_e) \to e(k',s'_e)+N(p'_N,s'_N)+ {\mathcal{M}}(p_{\mathcal{M}})
\label{hard_meson_production}
\end{equation}
in the generalized Bjorken limit within the \emph{near-backward} kinematics regime.
For the moment we do not specify the nature of the final state meson
(this can be a light pseudoscalar $\pi$, $\eta$, $\eta'$ or  a light vector $\rho$, $\omega$)

\begin{figure}[H]
\begin{center}
\includegraphics[width=0.4\textwidth]{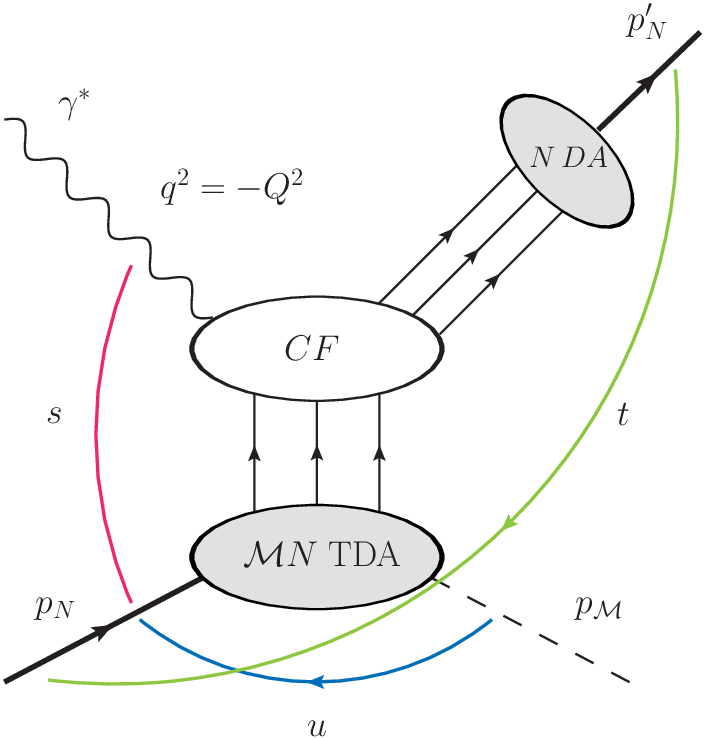}
\end{center}
     \caption{Kinematical quantities and the collinear factorization mechanism  for
      $\gamma^{*} N \to N {\mathcal{M}}$  in the  near-backward  kinematical regime (large $Q^2$, $W$; fixed $x_{B}$; $| u|  \sim 0$). The lower blob, denoted
  ${\mathcal{M}}N$ TDA, depicts the nucleon-to-meson~$\mathcal M$ transition
     distribution amplitude; $N$ DA blob depicts the nucleon distribution amplitude;
      $CF$  denotes the hard subprocess amplitude (coefficient function).}
\label{Fig_Kinematics_TDAs}
\end{figure}

The relevant kinematical quantities for the hard subprocess
\begin{equation}
\gamma^*(q,\lambda_\gamma)+  N(p_N,s_N) \to N(p'_N,s'_N)+ {\cal{M}}(p_{\cal{M}})
\label{Hard_subpr}
\end{equation}
 of the reaction
(\ref{hard_meson_production})
are presented in  Fig.~\ref{Fig_Kinematics_TDAs}.
We employ the usual notations for the photon virtuality, the
$\gamma^{*} N$
center-of-mass energy $W$, and the Bjorken variable $x_B$
\begin{equation}
Q^2=-q^2; \ \   W^2=(p_N+q)^2; \ \ x_B=\frac{Q^2}{2 p_N \cdot q},
\end{equation}
and introduce the standard Mandelstam variables
\begin{equation}
s \equiv W^2; \ \ t=(p'_N-p_N)^2; \ \ u=(p_{\mathcal{M}}-p_N)^2.
\label{Def_Mand_variables}
\end{equation}
The Mandelstam variables
(\ref{Def_Mand_variables})
satisfy
\begin{equation}
s+t+u=-Q^2+2m_N^2+m_{\mathcal{M}}^2 \,,
\label{Condition_for_Mand_var}
\end{equation}
where
$m_N$
and
$m_{\mathcal{M}}$
denote respectively the nucleon and the meson masses.

Within the generalized Bjorken limit,
the near-backward kinematics corresponds to large
$Q^2$
and
$W^2$
with fixed
$x_B$
and small invariant momentum transfer between the
final meson and the initial nucleon:
\begin{equation}
| u| = | (p_{\mathcal{M}}-p_N)^2|   \ll Q^2, \, W^2.
\label{Abs_u_backward}
\end{equation}
This corresponds to the final state meson produced
in a nearly backward direction in the $\gamma^{*}N$
center-of-mass-system (CMS). Note that the condition~(\ref{Condition_for_Mand_var}) together with
(\ref{Abs_u_backward})
implies that in the near-backward kinematics the invariant momentum transfer
$t$
between the initial and final state nucleons is large (by the absolute value)
and negative.

The near-backward kinematical regime can be seen as being complementary to
the more familiar \emph{near-forward} kinematics with large
$Q^2$
and
$W^2$,
fixed
$x_B$
and  small invariant momentum transfer $| t| $ between the
final and  initial nucleons:
\begin{equation}
| t| = | (p'_N-p_N)^2|  \ll Q^2, \, W^2\,,
\end{equation}
where the collinear factorization providing a description in terms of GPDs is at work. This kinematical regime corresponds to the final
state nucleon produced in the near-backward direction in the
$\gamma^{*} N$ CMS and hence the final
state meson produced in the near-forward direction.

A common procedure to address the reaction
(\ref{hard_meson_production})
in { both near-forward and} near-backward kinematics consists in the introduction of light-cone
coordinates.
The $z$-axis is naturally chosen along the colliding virtual photon and nucleon.
We introduce the light-cone vectors
$p$
and
$n$:
\begin{equation}
p^2=n^2=0; \ \ 2 p \cdot n=1.
\label{Def_p_n}
\end{equation}
Then for an arbitrary four-vector $l^\mu$  we introduce the Sudakov decomposition
\begin{equation}
l^\mu= l^+ p^\mu + l^-n^\mu+ l^\mu_T; \ \ \ l^2=l^+ l^-+ l^2_T,
\label{Def_Sudakov_dec}
\end{equation}
where $l^+=2 (n \cdot l)$;  $l^-=2 (p \cdot l)$;
and the $T$ subscript is adopted for the transverse part
of a four-vector $l^\mu$, which satisfies $(p \cdot l_T)=(n \cdot l_T)=0$.
In some cases we also use  boldface for the Euclidian two-dimensional
transverse components of vectors ${\mathbf{l}}=(l^1,l^2)$:
$
l^2=l^+ l^-- {\mathbf{l}}^2.
$

We employ the notation
$\Delta$
for the momentum transfer between the final state meson
and the initial state nucleon ($u$-channel momentum transfer):
\begin{equation}
\Delta \equiv p_{\mathcal{M}}-p_N; \ \ \ \Delta^2=u;
\label{Def_Delta}
\end{equation}
and define the average nucleon--meson momentum
$P$
and mass
$\bar{m}_{{\mathcal{M}}N}$:
\begin{equation}
P \equiv \frac{1}{2} \left( p_N +p_{\mathcal{M}} \right); \ \ \ P^2=\bar{m}^2_{{\mathcal{M}}N} =
\frac{1}{2}\left(m_{N}^{2}
+ {m_{\mathcal{M}}^{2}}\right)-\frac{\Delta^{2}}{4}.
\label{Def_P_TDA}
\end{equation}

Keeping the first-order corrections in squared masses and in the square of the transverse momentum transfer
$\Delta_T^2$,
we establish the following Sudakov decomposition for the momenta of the reaction
(\ref{hard_meson_production}) in the near-backward kinematical regime
(\textit{cf.} Ref.~\cite{Lansberg:2007ec}):
\begin{eqnarray}
 &&
\nonumber
p_N = (1+\xi) p + \frac{m_N^2}{1+\xi}n \,; \nonumber \\
 &&\nonumber
p_{\mathcal{M}} = (1-\xi) p +\frac{m^2_{\mathcal{M}}-\Delta_T^2}{1-\xi}n+ \Delta_T \, ;
\\  && \nn
\Delta =- 2 \xi p +\Bigl[\frac{m^2_{\mathcal{M}}-\Delta_T^2}{1-\xi}- \frac{m_N^2}{1+\xi}\Bigr]n
+ \Delta_T \,; \\
 && P=p+ \frac{1}{2}\Bigl[  \frac{m_N^2}{1+\xi}+ \frac{m^2_{\mathcal{M}}-\Delta_T^2}{1-\xi}\Bigr] n + \frac{1}{2} \Delta_T;
\\
 && \nn
p'_N \simeq -2\xi \frac{(\Delta_T^2-m_N^2)}{Q^2} p + \left[ \frac{Q^2}{2\xi \Bigl(1+   \frac{(\Delta_T^2-m_N^2)}{Q^2}\Bigr)} - \frac{m^2_{\mathcal{M}}-\Delta_T^2}{1-\xi}+ \frac{m_N^2}{1+\xi}\right]n-\Delta_T;
\nonumber \\
 && q \simeq- 2 \xi \Bigl(1+ \frac{(\Delta_T^2-m_N^2)}{Q^2}\Bigr)  p + \frac{Q^2}{2\xi\Bigl(1+   \frac{(\Delta_T^2-m_N^2)}{Q^2}\Bigr)} n\,.    \label{Sudakov_dec_momenta}
\end{eqnarray}
The square of the transverse momentum transfer
$\Delta_T^2$ is defined below in Eq.~(\ref{Delta_t2BMP})
and
$\xi$
stands for the $u$-channel skewness variable introduced with respect to the $u$-channel momentum transfer
\begin{equation}
\xi = -\frac{(p_{\mathcal{M}}-p_N) \cdot n}{(p_{\mathcal{M}}+p_N) \cdot n}.
\label{Def_xiBMP}
\end{equation}
For the leading twist accuracy calculations
we employ the approximate expression
for the skewness variable $\xi$
(\ref{Def_xiBMP})
neglecting order-of-mass and $\Delta_T^2$  corrections
\begin{equation}
\xi \simeq \frac{x_{B}}{2-x_{B}} = \frac{Q^2}{Q^2+2W^2} + {\mathcal{O}}(1/Q^2).
\label{Def_xi_approx}
\end{equation}

The $u$-channel transverse momentum transfer squared is expressed as
\begin{equation}
\Delta_T^2= \frac{1-\xi}{1+\xi}\left( \Delta^2-2 \xi \left[ \frac{m_N^2}{1+\xi} - \frac{m_{\mathcal{M}}^2}{1-\xi} \right] \right).
\label{Delta_t2BMP}
\end{equation}
We introduce $u_{0}$ corresponding to $\Delta_T^2=0$:
\begin{equation}
u_{0}=-\frac{2 \xi  \left(m_{\mathcal{M}}^2 (1+\xi)-m_N^2 (1-\xi)\right)}{1-\xi ^2}.
\label{Def_u0}
\end{equation}
It is the maximal possible value of $u$ for given $\xi$. For $\Delta_T^2=0$ ($u=u_0$) the  meson is produced exactly
in the backward direction in the $\gamma^{*} N$ CMS ($\theta_{\mathcal{M}}^{*}=\pi$). Note that, for small meson masses, $u_0$ is positive for most values of $\xi$.

In the initial state nucleon rest frame, which corresponds to the laboratory (LAB) frame of a fixed target experiment, the light-cone vectors
$p$
and
$n$
read
\begin{equation}
p\Big|_{\rm LAB}=\frac{m_N}{2(1+\xi)} \{1,0,0,-1\}; \ \ \ n \Big|_{\rm LAB}=\frac{1+\xi}{2m_N} \{1,0,0,1\}.
\end{equation}
With the help of the appropriate boost
we establish the expressions for the light-cone vectors
in the
$\gamma^{*} N$
CMS:
\begin{equation}
p\Big|_{\gamma^{*} N \,  {\rm CMS}}= \alpha(W^2,\,Q^2,\,m_N^2) \{1,0,0,-1\}; \ \ \
n\Big|_{\gamma^{*} N \, {\rm CMS}}=  \frac{1}{4 \alpha(W^2,\,Q^2,\,m_N^2) } \{1,0,0,1\},
\end{equation}
with
\begin{equation}
\alpha(W^2,\,Q^2,\,m_N^2) = \frac{ W^2+Q^2+m_N^2+\Lambda \left(W^2,-Q^2,m_N^2\right)}{4
   (1+\xi) W},
\label{Def_alpha_for_cos}
\end{equation}
where $\Lambda$
is the usual Mandelstam function
\begin{equation}
\Lambda(x,y,z)= \sqrt{x^2+y^2+z^2-2xy-2xz-2yz} \, .
\label{Def_lambda}
\end{equation}

The $\mathcal M$-meson  scattering angle in the $\gamma^{*} N$ CMS for the near-backward kinematical regime can be expressed as
\begin{equation}
\cos \theta_{\mathcal{M}}^{*}=\frac{-4(1-\xi )^2\alpha^2(W^2,\,Q^2,\,m_N^2) +{m^2_{\mathcal{M}}-\Delta_T^2} }
{\sqrt{\left(-4(1-\xi)^2\alpha^2(W^2,\,Q^2,\,m_N^2)+m^2_{\mathcal{M}}-\Delta_T^2 \right)^2-16 \alpha^2(W^2,\,Q^2,\,m_N^2) (1-\xi)^2\Delta_T^2}} \, \,,
\label{CosThetaV_CMS}
\end{equation}
where
$\alpha(W^2,\,Q^2,\,m_N^2)$
is given by (\ref{Def_alpha_for_cos}).
One may check that for
$\Delta_T^2=0$
indeed
$\cos \theta_{\mathcal{M}}^{*}=-1$,
corresponding to a meson
$\mathcal M$  produced exactly in the backward direction in the $\gamma^{*} N$ CMS.

\subsection{Status of the collinear factorization theorem}
\mbox

The research program aiming at the study of nucleon-to-meson
TDAs  in  hard exclusive backward
meson electroproduction reactions (and in the cross-channel counterpart reactions)
is generally analogous to the extraction of GPDs from  DVCS and hard exclusive near-forward meson electroproduction.

The possibility to access nucleon-to-meson TDAs in hard exclusive backward
meson electroproduction reactions
is based on the collinear factorization theorem that is similar to the familiar factorization theorems for the near-forward hard exclusive
electroproduction of mesons
\cite{Collins:1996fb}
and DVCS
\cite{Collins:1998be}.
This collinear factorization theorem was first conjectured in Refs.~\cite{Frankfurt:1999fp,Frankfurt:2002kz},
although never proven consistently.

There are several approaches for providing proofs for collinear
factorization theorems (see \textit{e.g.} Sec.5.1 of~\cite{Diehl:2003ny} and Ref.~\cite{Wallon:2011zx} for a review).
\bi
\item
The approach employed by J.~Collins et al.~\cite{Collins:1996fb,Collins:1998be} and by X.-D.~Ji and J. Osborne~\cite{Ji:1998xh}
relies on the general properties
of the Feynman diagrams
and makes use of the theoretical tools developed by
S.~Libby, G.~Sterman and J.~Collins~\cite{Sterman:1978bi,Sterman:1978bj,Libby:1978bx,Collins:1981ta}
(for a detailed description see  Refs.~\cite{CSS,JCollins_pQCD}).
The Coleman--Norton theorem
\cite{Coleman:1965xm}
allows to locate
the regions of the loop momentum space giving rise to the leading asymptotic
behavior of the relevant Feynman diagrams.
\item The alternative approach of A.~Radyushkin
\cite{Radyushkin:1997ki} addresses the spectral properties
of the Feynman diagrams using the $\alpha$-representation technique
\cite{Radyushkin:1983wh,Radyushkin:1983ea}.
\item A systematic framework to study QCD factorization  is provided by
the effective theory approach~\cite{Bauer:2002nz}, the so-called Soft Collinear Effective Theory (SCET).
This latter framework provides a consistent description of both hard and soft
spectator scattering mechanisms and was in particular applied to the description of
the nucleon form factor~\cite{Kivel:2010ns} and to the description of the wide-angle Compton scattering~\cite{Kivel:2013sya}.
\ei

In this section, mainly following Refs.
\cite{Collins:1996fb,Collins:1998be,Collins:1999yw},  we discuss some of the
essential steps in the construction of a factorization theorem for hard exclusive backward
meson electroproduction.
Let us thus consider the reaction (\ref{hard_meson_production}).
In the limit
\begin{equation}
-q^2 \equiv Q^2 \to \infty \ \ \text{with fixed} \ \ \frac{Q^2}{W^2} \ \ \text{and small} \ \ | u| =| (p_{\mathcal{M}}-p_N)^2| \,,
\end{equation}
the scattering amplitude of the hard
subprocess of (\ref{hard_meson_production})
\begin{equation}
\gamma^{*}(q) + N(p_N) \to {\mathcal{M}}(p_{\mathcal{M}})+ N(p'_N).
\label{React_for_fact}
\end{equation}
for a transversely polarized virtual photon, up to
$1/Q$ suppressed corrections,  reads
\cite{Lansberg:2007ec}:
\begin{eqnarray}
 &&
{\mathcal{A}}(\gamma^{*}_T N \to {\mathcal{M}}N) \simeq \frac{1}{Q^3}
\int_{-1+\xi}^{1+\xi} dx_1  dx_2  dx_3  \delta( \sum_{k=1}^3 x_k-2\xi)
\int_{0}^1 dy_1  dy_2  dy_3   \delta( \sum_{l=1}^3 y_l-1)  \nn \\ && \
T^{ij}(x_{1,2,3}, \, \xi, \, y_{1,2,3}; \, \mu^2) F^i(x_{1,2,3},\,\xi, \,u; \, \mu^2) \, \Phi^j(y_{1,2,3}; \, \mu^2).
\label{Fact_form}
\end{eqnarray}
Here $F^i$ stands for the matrix element of the light-cone $3$-quark operator of the relevant flavor contents between the initial nucleon and final meson states expressed through nucleon-to-meson TDAs;
$\Phi^j$ denotes the distribution amplitude of the final nucleon state; $T^{ij}$
is the hard part of the corresponding amplitude.
$x_{1,2,3}$ stand for the momentum fractions of quarks coming from the initial nucleon;
$y_{1,2,3}$ are the momentum fractions of quarks of the final state nucleon. The skewness
variable $\xi$
(\ref{Def_xiBMP})
is defined with respect to the longitudinal momentum
transfer between the initial nucleon and the final meson. The factorization scale $\mu$
is supposed to be of order $Q$. The factorization scale dependence of nucleon-to-meson TDAs
and nucleon DAs is given by the appropriate evolution equations.
We adopt a reference frame in which the external momenta of
(\ref{React_for_fact})
have small transverse components of the order $\sqrt{u_0-u} \sim {\mathcal{O}}(m_N)$ (see  Sec.~\ref{SubSec_Kinematics} for the details
of our kinematics conventions).

In order to employ  the power-counting technique  of Refs.~\cite{Collins:1996fb,Collins:1998be}
that allows to   identify the leading power
contributions to the amplitude of
(\ref{React_for_fact})
we establish the following counting rules for the relevant momenta denoted $(l^+,l^-,l_\bot)$:
\begin{eqnarray}
 &&
\text{hard part} \ \ H: \; \left( {\mathcal{O}}(Q), \,{\mathcal{O}}(Q), \, {\mathcal{O}}(m_N) \right);
\nn \\  &&
\text{incoming nucleon}: \ \ \left( {\mathcal{O}}(Q), \,{\mathcal{O}}(m_N^2/Q), \, {\mathcal{O}}(m_N) \right); \nn \\  &&
\text{outgoing meson}: \ \ \left( {\mathcal{O}}(Q), \,{\mathcal{O}}(m_{\mathcal{M}}^2/Q), \, {\mathcal{O}}(m_{\mathcal{M}}) \right); \nn \\  &&
\text{outgoing nucleon}: \ \ \left( \,{\mathcal{O}}(m_{N}^2/Q), {\mathcal{O}}(Q),  \, {\mathcal{O}}(m_N)\right); \nn \\  &&
\text{soft part} \ \ S: \text{all components of momenta small compared to} \; Q.
\end{eqnarray}

\begin{figure}[H]
\begin{center}
\includegraphics[width=0.35\textwidth]{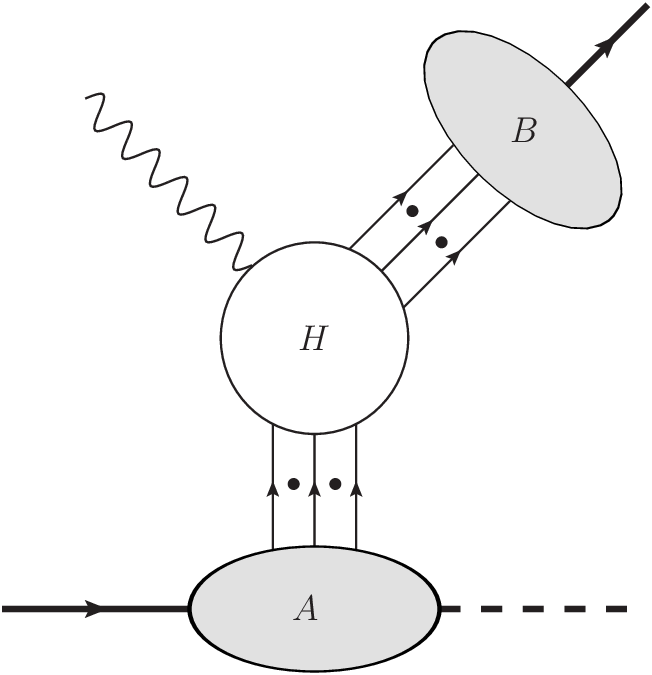}
\end{center}
     \caption{Reduced graph for the factorized description of hard exclusive backward
meson electroproduction. Black dots depict arbitrary number of additional collinear gluon
lines with suitable longitudinal polarization.}
\label{Fig_Lead_Fact}
\end{figure}

The factorization formula
(\ref{Fact_form})
corresponds to the contribution of the reduced graph
depicted  in  Fig.~\ref{Fig_Lead_Fact}. The lines of the subgraph $A$
are collinear to the incoming nucleon and outgoing meson;
the lines of the subgraph $B$ are collinear to the outgoing nucleon,
the lines of the hard subgraph $H$ have large components in both light-cone
directions. In addition to the quark lines depicted in  Fig.~\ref{Fig_Lead_Fact} there
might be an arbitrary number of collinear gluons with polarization
along the plus-direction connecting the subgraphs $A$ and $H$ and
of collinear gluons with polarization
along the minus-direction connecting the subgraphs $B$ and $H$.

Obviously,  this graph corresponds to the textbook
hard scattering mechanism
proposed a long time ago to provide
the leading power behavior
within the pQCD description of the
nucleon electromagnetic form factor
\cite{Lepage:1979zb,Chernyak:1983ej} (see  Fig.~\ref{Fig_NFF} {(a)}).

\begin{figure}[H]
\begin{center}
\includegraphics[width=0.35\textwidth]{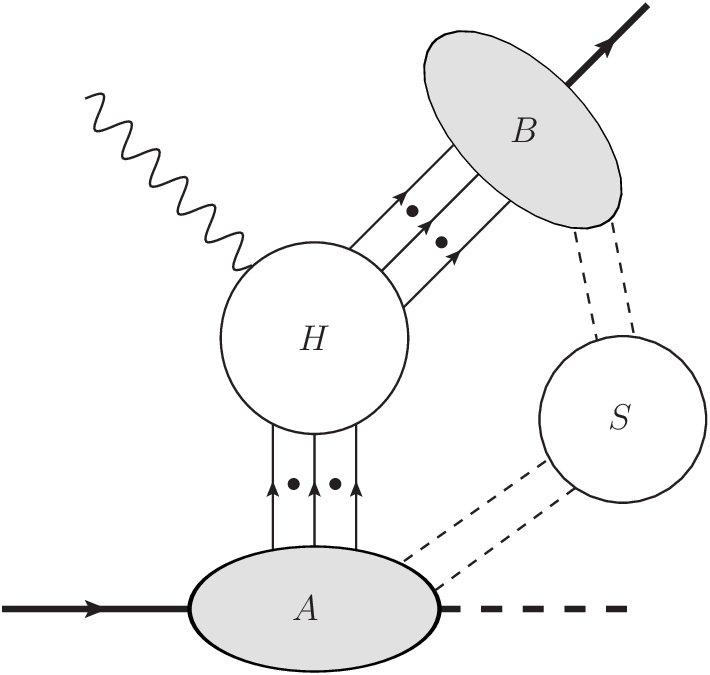}
\end{center}
     \caption{General reduced graph potentially contributing to the leading  asymptotic
     regime for hard exclusive backward
meson electroproduction. The soft subgraph $S$ is connected to the collinear subgraphs
$A$ and $B$ by an arbitrary number of only gluon lines. According to the power counting
formula (\ref{Power_Count_F}) this graph contributes to the leading asymptotic behavior (\ref{Power3}).
Graphs with extra lines connecting the $S$-subgraph to the hard part $H$  do not contribute to the leading asymptotic behavior due to
$1/Q$-suppression.}
\label{Fig_Gen_Fact}
\end{figure}

The  graph depicted in  Fig.~\ref{Fig_Lead_Fact} is
a particular example of a more general class of reduced graphs (see  Fig.~\ref{Fig_Gen_Fact})
for the backward reaction (\ref{React_for_fact}) that may include an additional soft subgraph
$S$ connected to $A$, $B$  (and $H$) by an arbitrary number of soft lines depicted
by dashed lines.

The power-counting formula that determines the
$Q^{p(\pi)}$
behavior of a particular reduced graph takes the same form as in~\cite{Collins:1996fb}:
\begin{eqnarray}
 &&
p(\pi)=3-n(H) - \text{No. (quark lines from} \;  S\; \text{to}\; A, \, B\text{)} \nn \\
&& - 3 \cdot \text{No. (quark lines from} \;  S\; \text{to}\; H\text{)}
- 2 \cdot  \text{No. (gluons lines from} \;  S\; \text{to}\; H\text{)},
\label{Power_Count_F}
\end{eqnarray}
where $n(H)$ is the number of collinear quarks and transversely polarized gluons
attached to the hard subgraph.

According to the power counting formula (\ref{Power_Count_F}), the
reduced graph depicted in  Fig.~\ref{Fig_Lead_Fact} corresponds to the leading
asymptotic behavior
\begin{equation}
Q^{p(\pi)}=Q^{-3}.
\label{Power3}
\end{equation}
More generally, the leading power behavior
is provided by the graphs depicted in  Fig.~\ref{Fig_Gen_Fact} with soft subgraphs $S$   connected to the collinear subgraphs $A$ and $B$ by an arbitrary number of only gluon lines.

\begin{figure}[H]
\begin{center}
\includegraphics[width=0.35\textwidth]{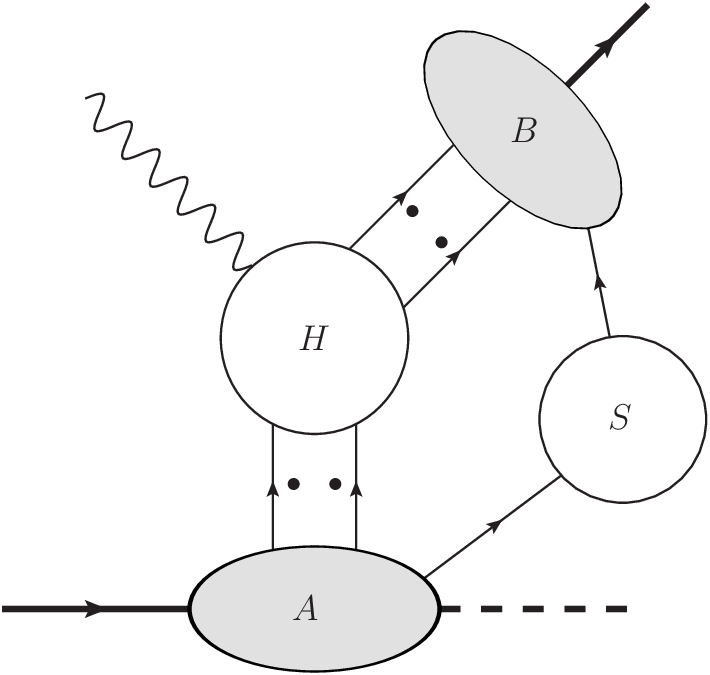} \ \ \ \
\includegraphics[width=0.35\textwidth]{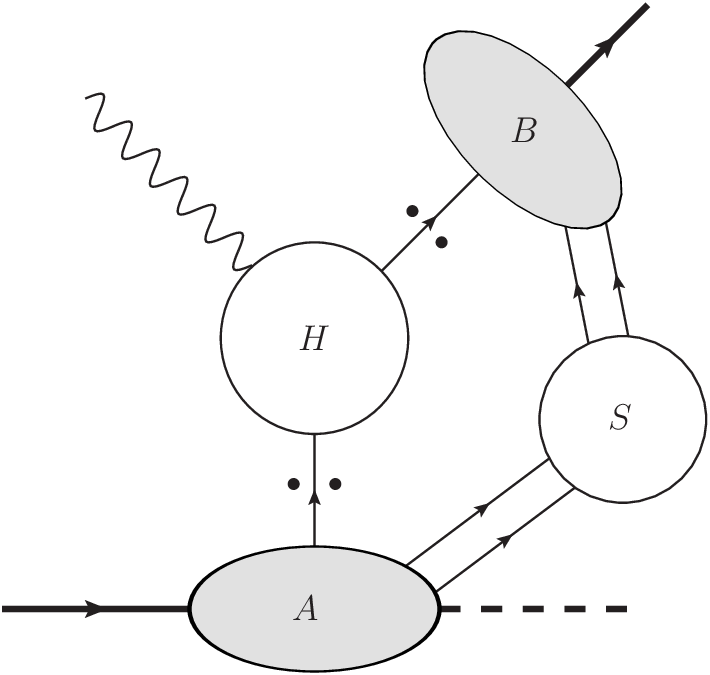}
\end{center}
     \caption{Examples of graphs corresponding to the soft rescattering mechanism
     for the backward reaction (\ref{React_for_fact}).}
\label{Fig_EP}
\end{figure}

\begin{figure}[H]
\begin{center}
\includegraphics[width=0.35\textwidth]{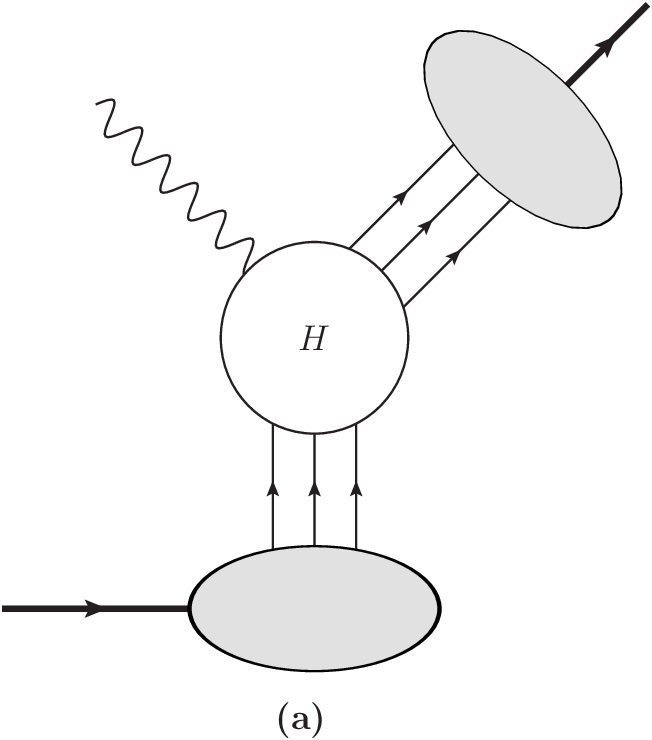} \ \ \ \
\includegraphics[width=0.35\textwidth]{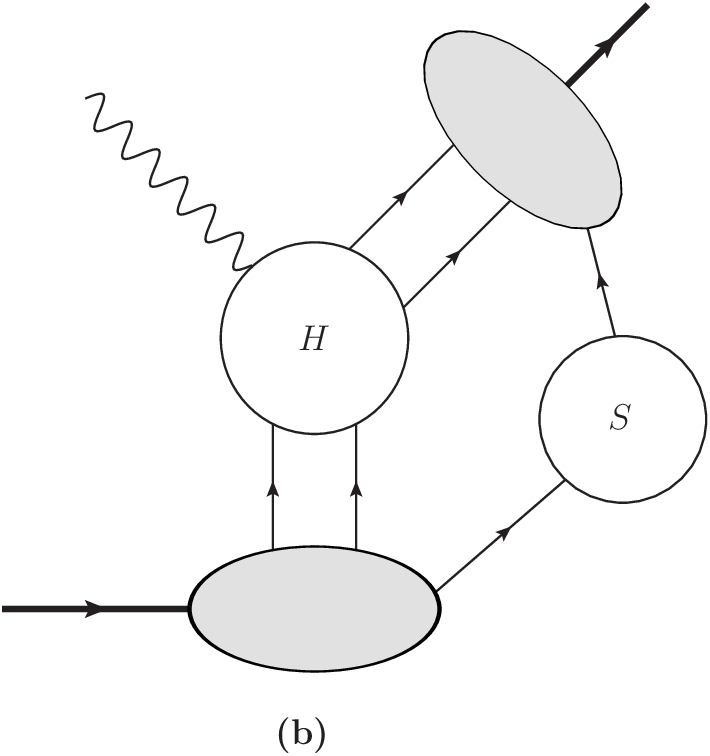}
\end{center}
     \caption{ {\bf (a)} Reduced diagram for hard scattering mechanism for nucleon e.m. form factor;
  {\bf (b)} reduced diagram for soft rescattering mechanism for nucleon e.m. form factor.}
\label{Fig_NFF}
\end{figure}

Another class of potentially leading regions corresponds to the soft rescattering
mechanism graphs with only one (or two) quark lines entering the hard subgraph
with the remaining quark lines connecting $A$ and $B$  through the soft subgraph $S$.
There also could be an arbitrary number of gluons connecting  $A$ and $B$  to $S$.
The power counting formula (\ref{Power_Count_F}) provides the same power-like
behavior (\ref{Power3})
An example of such reduced graphs is depicted in  Fig.~\ref{Fig_EP}.

Further steps in proving the collinear factorization theorem are:
\bi
\item Proving the suppression of the contributions of the soft rescattering
mechanism graphs in  Fig.~\ref{Fig_EP} and the dominance of the transverse polarization of the virtual photon.

\item Proving the effective suppression of the soft gluon subgraphs in the contributions
of generic reduced graphs in  Fig.~\ref{Fig_Gen_Fact}. This provides the ``color neutrality''
of the final state nucleon  and ensures actual dominance of the hard
reaction mechanism graph depicted in  Fig.~\ref{Fig_Lead_Fact}.
\item Implementing the gauge invariance that allows to sum the contributions of
collinear gluons polarized along the {$+$}-direction connecting $A$ and $H$
(respectively collinear gluons polarized along the $-$-direction connecting $B$ and $H$) into the Wilson lines
(\ref{Wilson_Line})
that appear in the gauge invariant three-local light-cone operators
entering into the definition of nucleon-to-meson TDAs (nucleon DAs), see  Sec.~\ref{SubSec_Definition}.
\item To obtain the final form of the factorization theorem
(\ref{Fact_form})
the hard subgraph $H$ has to be expanded in the small components of its
external momenta:
\begin{equation}
 \int d^4 l  \, d^4 k \ A(l) B(k) H(l,k) \simeq \int dl^+ dk^- H(l,k)
\Big|_{{l^-=k^+=0\atop  l_\bot=k_\bot=0}}
\int dl^- d^2l_\bot A(l) \int dk^+ d^2k_\bot B(k).
\end{equation}
The explicit expressions for the leading order hard scattering amplitudes is presented
in  Sec.~\ref{Sec_ExclProcLO}.

\item Presenting the collinear subgraphs $A(l)$ and $B(k)$ as the Fourier transforms
of hadronic matrix  elements of position space operators results in the
familiar light-cone operator definition of nucleon-to-meson TDAs and nucleon DAs.

\item The factorization scale $\mu$ dependence of nucleon-to-meson TDAs is governed by a generalization
of the DGLAP and ERBL evolution equations, see  Sec.~\ref{SubSec_Evolution}.
\ei

One of the most non-trivial steps in proving the factorization theorem turns to be
the demonstration of the suppression of the contributions of the soft rescattering
mechanism and the dominance of a particular polarization of the virtual photon.

For the case of near-forward deeply virtual meson production~\cite{Collins:1996fb} the corresponding proof was based
on rather non-trivial arguments~\cite{Mankiewicz:1997uy,Diehl:1998pd,Collins:1999un}.
The essential clause is the observation that the final state meson
is produced from a small size $q\bar{q}$ configuration  created in the hard
scattering. Therefore, the soft interactions happen only in the final state,
and not in the initial state.
The possibility of a proper generalization of these arguments for the
case of a three-quark intermediate configuration still remains an open question.

Another  perspective on the soft contributions corresponding to graphs in  Fig.~\ref{Fig_EP}
consists in treating them as the so-called endpoint singularities.
Indeed, within the momentum integrals in the factorization formula
(\ref{Fact_form})
such configurations correspond to the cross-over trajectories $x_k=0$,
separating the ERBL-like and the DGLAP-like support regions (see  Sec.~\ref{SubSec_Support}) of nucleon-to-meson
TDAs
$F^i(x_{1,2,3},\,\xi, \,u)$, and to the endpoint region $y_l=0$ of nucleon DAs
$ \Phi^j(y_{1,2,3})$.

The issue of the potential end point singularities has been rather controversial
in the literature. In particular, in context of the QCD description
of the nucleon electromagnetic form factor the importance of this type of contributions
was first highlighted in~\cite{Duncan:1979hi,Milshtein:1981cy,Milshtein:1982js}.
In Ref.~\cite{Lepage:1979zb} these arguments were parried
by the possible strong suppression of such contributions at large-$Q^2$
due to the Sudakov-type all-order resummation of corresponding non-renormalization group logarithms.
A detailed study employing the technique originally developed for the case of the pion form factor~\cite{Li:1992nu}
is presented in Ref.~\cite{Li:1992ce}, see also the discussion in~\cite{Bolz:1994hb}.

However, the present day experimental studies
\cite{Jones:1999rz,Puckett:2010ac,Riordan:2010id}
give evidences that in the region of
moderate $Q^2 \simeq 2.5\text{--}10~\mathrm{GeV}^{2}$ the factorization approach based
on the hard scattering mechanism (see  Fig.~\ref{Fig_NFF}(a)) cannot provide
a description of some part of existing experimental data. In particular, at this kinematical regime the form factor ratio
$F_2/F_1$ does not follow the asymptotic $Q^2 F_2/F_1 \sim {\rm const}$ behavior
suggested by the pQCD description based solely on the hard scattering mechanism.

This motivated the possible explanations relying on the soft rescattering mechanism,
 Fig.~\ref{Fig_NFF}(b).
A detailed study of the soft rescattering mechanism for the nucleon
form factor within the SCET approach is presented in Refs.~\cite{Kivel:2010ns,Kivel:2012mf}.
It suggests a large contribution from the soft rescattering mechanism in the region of
moderate $Q^2$ and brings essential end point singularities that may result in the
breakup of the collinear factorization for the FF $F_1$

The existing phenomenological approaches based on the QCD-motivated models of hadronic wave functions
\cite{Isgur:1984jm,Isgur:1988iw,Isgur:1989cy},
QCD sum rules
\cite{Ioffe:1982qb,Nesterenko:1982gc}
and light-cone sum rules
\cite{Braun:2001tj,Braun:2006hz}
provide additional evidences in favor of the relevance of the  soft-spectator mechanism
for a
realistic description of some scattering amplitudes at moderate $Q^2$.
The generalization of these arguments and the quantitative estimate of the
soft spectator mechanism for the reaction
(\ref{React_for_fact})
and  possible implications for the collinear factorization breakup
at the moment remain open issues.

The issue of collinear factorization breaking should not be confused with the issue of using asymptotic forms of distribution amplitudes.
In various phenomenological applications
it was noticed that the use of the simple asymptotic DAs
usually leads to a rather small contribution into the amplitudes of hard exclusive
reactions. A possible way out was suggested by V.~Chernyak and A.~Zhitnitsky
\cite{Chernyak:1983ej}.
It consists
in using DAs that differ considerably from the asymptotic form and are mostly
concentrated near the end points. Effectively, this can be seen as a way to partially
take into account the contribution of the soft spectator mechanism and greatly
improves the description of the data. The regularization of the potential end point
singularities then requires further theoretical efforts (see \textit{e.g.} the discussion
on the pQCD description of $\gamma \gamma^{*} \to \pi^0$ form factors in Ref.~\cite{Musatov:1997pu}).

Obviously, the rigorous proof of the collinear factorization theorem for hard exclusive backward meson
electroproduction (\ref{React_for_fact})
and a careful analysis of possible alternative reaction mechanisms
are important issues to put the TDA formalism on a firm ground.
However, given the considerable technical difficulties in proving the collinear
factorization theorems and the still controversial status of the validity of the collinear
factorized description even for much simpler hard exclusive reactions at intermediate
values of $Q^2$,  it turns
extremely important to simultaneously look for the experimental evidences of the
possible early onset of
factorization regime for the reaction (\ref{React_for_fact}). These include
\bi
\item the characteristic $1/Q^8$ scaling behavior of the transverse cross section $\frac{d^2 \sigma_T}{d \Omega_{\mathcal{M}}}$, see  Sec.~\ref{SubSec_Bkw_CS};
\item the dominance of the  transverse cross section $\sigma_T$;
\item the constant scaling behavior of the cross section ratio $Q^2 \sigma_L / \sigma_T$.
\ei

It is also worth mentioning that following  the
analogy with the collinear factorized description of the time-like Compton process
proposed in Refs.~\cite{Berger:2001zn,Berger:2001xd},
a version of the collinear factorization theorem
(\ref{Fact_form})
can be formulated in appropriate kinematical regime for the cross channel counterparts
of the reaction (\ref{React_for_fact}), see  Sec.~\ref{SubSec_Cross_Ch_Excl_R}.
These include nucleon--antinucleon annihilation into a highly virtual
lepton pair (or a heavy quarkonium) in association with a light meson $\mathcal M$:
\begin{eqnarray}
 &&\bar{N} (p_{\bar{N}}) + N (p_N) \rightarrow \gamma^{*}(q) + {\mathcal{M}}(p_{\mathcal{M}})  \rightarrow \ell^+(k_{\ell^+})
\ell^-(k_{\ell^-})+ {\mathcal{M}}(p_{\mathcal{M}}); \nonumber \\
 &&\bar{N} (p_{\bar{N}}) + N (p_N) \rightarrow J/\psi(p_\psi) + {\mathcal{M}}(p_{\mathcal{M}})  \rightarrow \ell^+(k_{\ell^+})
\ell^-(k_{\ell^-})+ {\mathcal{M}}(p_{\mathcal{M}});
\label{BarNNannihilation reaction0}
\end{eqnarray}
and
the exclusive meson-induced Drell--Yan process
\begin{equation}
{\mathcal{M}}(p_{\mathcal{M}}) \;+ N  (p_N)  \; \to  \ell^+(k_{\ell^+})
\ell^-(k_{\ell^-}) + N  (p'_N) \,,
\label{DY_meson_induced0}
\end{equation}
in the backward region.

The study of the cross-channel counterpart reactions
(\ref{BarNNannihilation reaction0}),
(\ref{DY_meson_induced0})
allows to challenge the collinear factorized description in terms
of nucleon-to-meson TDAs for a broader class of hard exclusive reaction
relying on the aforementioned criteria. In particular, the characteristic
$\sim (1 + \cos^2 \theta_\ell)$
angular distribution of the lepton pair allows a clear separation of the contribution corresponding to the
transverse cross section $\sigma_T$, see  Sec.~\ref{SubSec_CS_formuls_cross_ch}.
This also allows to test the universality of nucleon-to-meson TDAs
that is essential for the consistency of the approach.

\subsection{Regge-type models for forward and backward meson electroproduction versus partonic picture of hard processes}
\label{SubSec_Regge}
\mbox

Before the introduction of generalized parton distributions, forward photoproduction of mesons at high energy has been the subject of many phenomenological analysis based on the concepts of the Regge
exchanges~\cite{Yu:2018ydp}, for a recent review, see \textit{e.g.,} \cite{Laget:2019tou}. The Regge description is, thus, at the same time an alternative to the GPD case in the forward kinematics and an alternative to the
TDA case in the backward kinematics

The basic idea is to describe the amplitude as mainly due to a Regge trajectory exchange. In the backward kinematics, the Regge exchanges at work are baryonic exchanges. The dominant one is the nucleon trajectory
propagator, written as~\cite{Laget:2019tou}
\begin{equation}
    {\mathcal{P}}^R_N =(\frac{s}{s_0})^{\alpha_N-0.5} \alpha'_N \Gamma(0.5-\alpha_N) \frac{1-e^{-i\pi(\alpha_N+0.5)} }{2},
\end{equation}
where $\Gamma(\ldots)$ is Euler's gamma function.
The nucleon Regge trajectory is fitted as $\alpha_N = -0.37+ \alpha'_N u$, with $\alpha'_N = 0.98$; and
$s_0 = 1~{\rm GeV}^{2}$. In the case of $\pi$ meson production, the $\Delta$-Regge trajectory  exchange must as well be taken into account, but its contribution is often found to be quite small.

\begin{figure}[H]
\begin{center}
\includegraphics[width=0.4\textwidth]{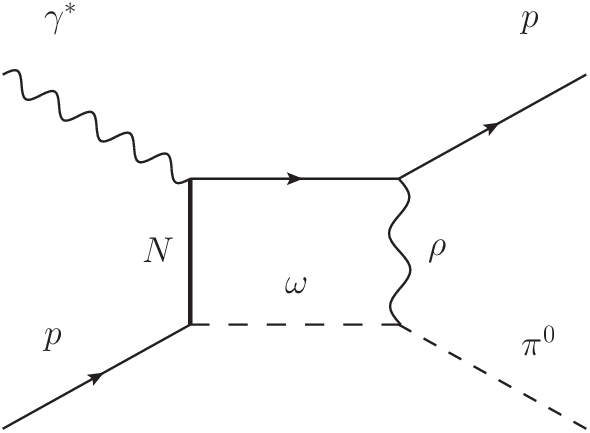}
\end{center}
     \caption{Example of a diagram contributing to
      $\gamma^{*} N \to N \pi^0$ in the Regge picture for the  near-backward  kinematical regime. The bold line denotes the nucleon Regge trajectory.}
\label{Lagetrescattering}
\end{figure}

To describe electroproduction, one needs to make further assumptions. Phenomenological damping factors are often used in the form of
$Q^2$-dependent form factors attached to the couplings present in the amplitude but some skill needs to be used to get a reasonable account of experimental data. The formalism  needs then to be complexified by adding the effects of  inelastic rescattering cuts (see  Fig.~\ref{Lagetrescattering}). The importance of these rescattering contributions at large $Q^2$ is at odds with the color transparency property of QCD, which implies that rescattering effects are less and less important when the characteristic momentum scale of the problem grows.

\subsection{Color transparency in forward and backward processes}
\label{SubSec_EM_FFs_in_QCD}
\mbox

A complementary argument to discover the  onset of the collinear factorization is based on the concept of color transparency introduced in~\cite{Mueller:1982bq,Brodsky:1982kg}. The basic idea is that leading twist
dominance of the scattering amplitude for an exclusive process is linked to a short distance shrinking of the hadronic wave functions probed by this process. Color transparency is very much related to the validity of the
factorization properties of exclusive amplitudes: if color transparency does not hold, one must consider final state interactions effects (such as additional phase-shifts) to be included in the scattering amplitude for
\textit{e.g.} electroproduction of a vector meson, thus modifying in a drastic way the simple leading twist factorized picture.

The fact that color transparency leads to a strong and $Q^2$-dependent suppression of final/initial state interactions of the  hadrons may be probed in a clean way by considering the hard exclusive reaction in a
nuclear environment. The nucleus then plays the role of a femto-detector of final state interactions. The size of the nucleus -- or more exactly its contacted size in the produced hadron reference system --  is then the main
parameter controlling how much of the expansion of the point-like configuration to a normal size hadron does  happen inside the nucleus. At very large energy, one expects the re-interaction cross section of the produced
hadron to be proportional to its transverse size, which is of order $1/Q^2$, where $Q$ is the large scale associated to the hard reaction. Color transparency is then probed by a nuclear transparency ratio for
which diverse definitions exist (for reviews, see~\cite{Frankfurt:1991rk,Kopeliovich:1991pu,Frankfurt:1992dx,Nikolaev:1992si,Jain:1995dd,Dutta:2012ii}).

Recent experimental data on $(e,e',p)$ reactions in a nucleus
\cite{Bhetuwal:2020jes} demonstrated that color transparency effects  leading to the above mentioned shrinking of the wave function were very small up to $Q^2= 11~{\rm GeV}^{2}$, thus demonstrating that the leading twist process does
not dominate  the nucleon form factor up to rather large values of $Q^2$.

Color transparency was also studied for other exclusive processes, and in particular for elastic scattering of nucleons at large angles~\cite{Carroll:1988rp}, where a spectacular rise of the nuclear transparency ratio lead
to many discussions and debates. Without entering the controversy, which certainly lacks more precise experimental data in a larger energy range to be closed, let us note that these results  were interpreted as the
evidence for a nuclear filtering mechanism~\cite{Ralston:1988rb} very related to the color transparency phenomenon.

These examples demonstrate how much the onset of the collinear factorization framework depends on the process one studies. In the absence of reliable ways to calculate non-leading twist contributions, there is no way, up to now, to predict the minimal scale, where a specific process is adequately described by the leading twist analysis. Experimental data are, therefore, crucial to decide this issue.

\section{Definition and properties of nucleon-to-meson and nucleon-to-photon {{TDA}}s}
\setcounter{equation}{0}
\label{Sec_DefAndProp}
\mbox

\subsection{Definition of nucleon-to-meson {{TDA}}s and nucleon-to-photon {{TDA}}s}
\label{SubSec_Definition}
\mbox

Similarly to nucleon DAs, nucleon-to-meson (and nucleon-to-photon) TDAs are defined through the hadronic matrix elements of a three-quark operator at light-like separations.
In this section we consider the case of the $uud$ operator on the
light-cone and present the parametrization for proton-to-neutral meson TDAs.
The implications of the SU$(2)$ flavor symmetry are presented in  Sec.~\ref{SubSec_Isospin}.

In an arbitrary gauge the $uud$ trilocal light-cone operator is defined as
(\textit{cf.}~\cite{Braun:1999te})
\begin{equation}
\widehat{O}_{\rho \tau \chi}^{\,uud}(\lambda_1n,\lambda_2n,\lambda_3n)=
\varepsilon_{c_1 c_2 c_3} u^{c'_1}_{\rho}(\lambda_1 n) W^{c'_1c_1}[\lambda_1,\lambda_0]
u^{c'_2}_{\tau}(\lambda_2 n) W^{c'_2c_2}[\lambda_2,\lambda_0] d^{c'_3}_{\chi}(\lambda_3 n) W^{c'_3c_3}[\lambda_3,\lambda_0].
\label{Def_O_uud_operator_arb_gauge}
\end{equation}
Here $\rho$, $\tau$ and $\chi$ stand for the Dirac indices of the
quark field operators,
$c_{1,\,2,\,3}$, $c'_{1,\,2,\,3}$
stand for the indices of the fundamental representation
of the SU$(3)_{\rm c}$ color group; the contraction with the totally antisymmetric tensor
$\varepsilon_{c_1 c_2 c_3}$
ensures that the operator
$\widehat{O}_{\rho \tau \chi}^{\,uud}$
is a color singlet.
To ensure the SU$(3)_{\rm c}$ gauge invariance the Wilson lines
$W[\lambda_i,\lambda_0]$
are included along the light-like paths:
\begin{equation}
W^{c'c}[\lambda_i, \,\lambda_0]= {\rm P}  \left\{ \exp \left(i g \int_{\lambda_0}^{\lambda_i}
d \lambda A^{+ \,a}( \lambda n) \left(t^a\right)^{c'c}
\right) \right\}.
\label{Wilson_Line}
\end{equation}
Here ${\rm P}$ denotes ordering along the light-like path from $\lambda_0 n$
to $\lambda_i n$; $g$ is the SU$(3)_{\rm c}$ coupling constant;
$t^a$ are the SU$(3)_{\rm c}$ generators in the fundamental representation.

In what follows we choose to use the light-cone gauge
$A^+=0$
for the gluon field. This allows us to omit the Wilson lines in
(\ref{Def_O_uud_operator_arb_gauge})
and to set
\begin{equation}
\widehat{O}_{\rho \tau \chi}^{\,uud}(\lambda_1n,\lambda_2n,\lambda_3n)=
\varepsilon_{c_1 c_2 c_3} u^{c_1}_{\rho}(\lambda_1 n)
u^{c_2}_{\tau}(\lambda_2 n)
d^{c_3}_{\chi}(\lambda_3 n).
\label{Def_O_uud_operator}
\end{equation}

Below we consider several of the most important examples of TDAs.

\subsubsection{Nucleon-to-pion {{TDA}}s}
\label{SubSubSec_Def_piN_TDAs}
\mbox

We begin with the simplest case of the leading twist-$3$ nucleon-to-pseudoscalar meson
TDAs. The parametrization for the leading twist $\pi N$ TDA involves $8$ independent Dirac structures.
Indeed, each of the three quarks and the nucleon have $2$ helicity states, while the pseudoscalar meson, obviously, has just $1$. This leads to $1 \cdot 2^4=16$ helicity amplitudes for the process $N \to qqq \, \pi$. However,  parity invariance relates helicity amplitudes
with all opposite helicities reducing the overall number of independent helicity amplitudes
by a factor of $2$.

The proton-to-$\pi^0$ $uud$ TDA is defined by the Fourier
transform of the $\pi^0 N^p$ matrix element of the
light-cone operator (\ref{Def_O_uud_operator}):
\begin{eqnarray}
 &&
4 (P \cdot n)^3 \int \left[ \prod_{j=1}^3 \frac{d \lambda_j}{2 \pi}\right]
e^{i \sum_{k=1}^3 x_k \lambda_k (P \cdot n)}
 \langle \pi^0(p_\pi)|  \,
\widehat{O}_{\rho \tau \chi}^{\,uud}(\lambda_1n,\lambda_2n,\lambda_3n)
\,| N^p(p_N,s_N) \rangle
\nonumber \\ &&
\equiv 4 {\mathcal{F}} \langle \pi^0(p_\pi)|  \,
\widehat{O}_{\rho \tau \chi}^{\,uud}(\lambda_1n,\lambda_2n,\lambda_3n)
\,| N^p(p_N,s_N) \rangle   \nonumber \\ &&
=
\delta( {x}_1+ {x}_2+ {x}_3-2  {\xi}) i \frac{f_N}{f_\pi}
\Bigl[
\sum_{\Upsilon= 1, 2} (v^{\pi N}_\Upsilon)_{\rho \tau, \, \chi} V_{\Upsilon}^{\pi N}(x_1,x_2,x_3, \xi, \Delta^2; \, \mu^2)
\nonumber \\ &&
+\sum_{\Upsilon= 1,\, 2 } (a^{\pi N}_\Upsilon)_{\rho \tau, \, \chi} A_{\Upsilon}^{\pi N}(x_1,x_2,x_3, \xi, \Delta^2; \, \mu^2)
+
\sum_{\Upsilon= 1,2,3,4} (t^{\pi N}_\Upsilon)_{\rho \tau, \, \chi} T_{\Upsilon}^{\pi N}(x_1,x_2,x_3, \xi, \Delta^2; \, \mu^2)
\Bigr],
 \label{Param_TDAs}
\end{eqnarray}
where the subscript $\Upsilon$  labels  the corresponding TDAs;  $f_\pi=93$~MeV
is the pion weak decay constant and
$f_N$
determines the value of the nucleon wave function at the origin.
Ref.~\cite{Chernyak:1987nv} provides an estimate
$f_{N}=5.0 \times 10^{-3} \mathrm{GeV}^{2}$.
The normalization factor
$\frac{f_N}{f_\pi}$
is introduced for
convenience for implementing the chiral
constraints for
$\pi N$
TDAs,  see  Sec.~\ref{SubSec_Chiral_Constr}.
In the second line of
(\ref{Param_TDAs}),
for further convenience, we introduce the compact notation
$\mathcal F$
for the conventional Fourier transform
\begin{equation}
{\mathcal{F}} \equiv {\mathcal{F}}(x_1,x_2,x_3)(\ldots)= (P \cdot n)^3 \int \left[ \prod_{j=1}^3 \frac{d \lambda_j}{2 \pi} \right] e^{i \sum_{k=1}^3 x_k \lambda_k (P \cdot n)} \,.
\label{Fourier_TDA}
\end{equation}

Each of the $8$ leading twist proton-to-$\pi^0$ TDAs
$V_{1,2}^{\pi N}$, $A_{1,2}^{\pi N}$, $T_{1,2,3,4}^{\pi N}$
is function
of $3$ longitudinal momentum fractions $x_i$, being the Fourier conjugates
of the corresponding light-cone distances; of the skewness variable $\xi$
defined in Eq.~(\ref{Def_xi}); of the invariant momentum transfer squared $\Delta^2$
(\ref{Def_Delta}); and of a factorization scale $\mu$.
The TDAs
$V_{1,2}^{\pi N}$
and
$T_{1,2,3,4}^{\pi N}
$
are defined symmetric under the interchange $x_1 \leftrightarrow x_2$,
while
$A_{1,2}^{\pi N}
$
are antisymmetric under the interchange $x_1 \leftrightarrow x_2$.

The intrinsic redundancy
of description, originating from the three-body nature of the
problem, results in multiple possible choices for the set of appropriate Dirac structures.
Obviously,
the corresponding $\pi N$ TDAs depend on a particular choice of
the set of Dirac structures. Here we employ the parametrization
first suggested in
Ref.~\cite{Lansberg:2007ec}:
\begin{align}
(v^{\pi N}_1)_{\rho \tau, \, \chi} &=  (\hat{p} C)_{\rho \tau}(U^+)_{\chi};   &
(v^{\pi N}_2)_{\rho \tau, \, \chi} & =
m_N^{-1} ( \hat{ p}  C)_{\rho \tau}( \hat{{\Delta}}_T U^+)_{\chi} \nn \\
(a^{\pi N}_1)_{\rho \tau, \, \chi} &= (  \hat{p} \gamma^5 C)_{\rho \tau}(\gamma^5 U^+ )_{\chi};   &
(a^{\pi N}_2)_{\rho \tau, \, \chi} & =
m_N^{-1}  ( \hat{ p}  \gamma^5 C)_{\rho\tau}(\gamma^5  \hat{{\Delta}}_T  U^+)_{\chi};
\nn \\
(t^{\pi N}_1)_{\rho \tau, \, \chi} & =
(\sigma_{p\mu} C)_{\rho \tau }(\gamma^\mu U^+ )_{\chi}; &
(t^{\pi N}_2)_{\rho \tau, \, \chi} & =  m_N^{-1}( \sigma_{p {\Delta}_T} C)_{\rho \tau} (U^+)_{\chi}; \nn \\
(t^{\pi N}_3)_{\rho \tau, \, \chi} & =m_N^{-1} ( \sigma_{p\mu} C)_{\rho \tau} (\sigma^{\mu {\Delta}_T}
 U^+)_{\chi};  &
(t^{\pi N}_4)_{\rho \tau, \, \chi} & = m_N^{-2}
 (\sigma_{p  {\Delta}_T} C)_{\rho \tau}
(\hat{  {\Delta}}_T U^+)_{\chi}.
   \label{Def_DirStr_piN_DeltaT}
\end{align}
We adopt Dirac's ``hat'' notation
$\hat{l} \equiv l_\mu \gamma^\mu$;
$\sigma^{\mu\nu}= \frac{1}{2} [\gamma^\mu, \gamma^\nu]$; $\sigma_{p \mu} \equiv p^\lambda \sigma_{\lambda \mu}$;
$\sigma_{p {\Delta}_T} \equiv   p^\lambda \Delta_T^\mu \sigma_{\lambda \mu}$;
$C$
is the charge conjugation matrix and
$U^+$
stands for the large component of the nucleon spinor.
The large and small components of  the nucleon Dirac spinor $U(p_N,s_N)$
are introduced as
\begin{equation}
U^+(p_N,s_N)=  \hat{p} \hat{n}  U(p_N,s_N); \ \ \ \
U^-(p_N,s_N)=  \hat{n} \hat{p}  U(p_N,s_N).
\end{equation}
From the Dirac equation
\begin{equation}
\hat{p}_N U(p_N,s_N)= m_N U(p_N,s_N)\,,
\label{Dirac_eq}
\end{equation}
we establish the following relations  for the large ($U^+=  \hat{p} \hat{n}   U $)
and small ($U^-=  \hat{n} \hat{p}  U $) components of the nucleon Dirac spinor:
\begin{eqnarray}
 &&\hat{p} \, U(p_N,s_N)= \frac{m_N}{1+\xi} U^+(p_N,s_N)  \ \ ; \  \ (p= \frac{1}{1+\xi} p_N  -\frac{m_N^2}{(1+\xi)^2} n)\,;\nonumber  \\
 && \hat{n}  \, U(p_N,s_N)=\frac{1+\xi}{m_N} U^-(p_N,s_N) \ \  ; \  \ (n= \frac{1+\xi}{m_N^2} p_N-\frac{(1+\xi)^2}{m_N^2} p )\,.
\end{eqnarray}

The Dirac structures $(v_{1,2}^{\pi N})_{\rho \tau, \, \chi}$ and
$(t_{1,2,3,4}^{\pi N})_{\rho \tau, \, \chi}$ are defined symmetric
under interchange of two first Dirac indices, and
$(a_{1,2}^{\pi N})_{\rho \tau, \, \chi}$ are defined antisymmetric:
\begin{equation}
(v_{1,2}^{\pi N})_{\rho \tau, \, \chi}= (v_{1,2}^{\pi N})_{\tau \rho, \, \chi};
\ \ \
(t_{1,2,3,4}^{\pi N})_{\rho \tau, \, \chi}=(t_{1,2,3,4}^{\pi N})_{\tau \rho, \, \chi};
\ \ \
(a_{1,2}^{\pi N})_{\rho \tau, \, \chi}= -(a_{1,2}^{\pi N})_{\tau \rho, \, \chi}.
\end{equation}

The parametrization of $\pi N$
TDAs employing the set of   Dirac structures
(\ref{Def_DirStr_piN_DeltaT})
is extremely convenient for practical application since
in the strictly backward limit
$\Delta_T=0$
only the contributions of $3$ invariant functions
$V_1^{\pi N}$, $A_1^{\pi N}$, $T_1^{\pi N}$
out of $8$ turn out to be relevant.
However, the drawback of the parametrization
(\ref{Def_DirStr_piN_DeltaT})
is that the corresponding set of $\pi N$ TDAs does not
satisfy the polynomiality property for the $x_i$-Mellin moments in its simple form,
see discussion in
 Sec.~\ref{SubSec_Polynomiality}.
For this issue it is convenient to employ the alternative
parametrization for the set of the Dirac structures introduced in
Ref.~\cite{Pire:2011xv}.
The correspondence between the
two definitions of
$\pi N$
TDAs is established in
Eq.~(\ref{Relation_DiracStr_DeltaT_to_covariant}).

We switch to the light-cone spinors in order for the Dirac
components of the nucleon and quark spinors to select particular configurations of the nucleon and quark spin.
This allows to express the $8$ leading twist-$3$
$\pi^0 p$ $uud$ TDAs
as linear combinations of the $8$  helicity  matrix elements
$T_{\lambda_1 \lambda_2, \lambda_3}^{\lambda_N}$
for the
$N^p (\lambda_N) \to u(x_1,\lambda_1) u(x_2,\lambda_2) d(x_3,\lambda_3)\pi^0$ transition
\cite{Pasquini:2009ki}:
\begin{equation}
\begin{aligned}
&V_{1}^{p \pi_{0}}=-i \frac{1}{2^{1/4} \sqrt{1+\xi}\left(P^{+}\right)^{3/2}} \frac{f_{\pi}}{f_{N}}\left(T_{\uparrow \downarrow,\uparrow}^{\uparrow}+T_{\downarrow\uparrow,\uparrow}^{\uparrow}\right);\\
&A_{1}^{p \pi_{0}}=i \frac{1}{2^{1/4} \sqrt{1+\xi}\left(P^{+}\right)^{3/2}} \frac{f_{\pi}}{f_{N}}\left(T_{\uparrow \downarrow,\uparrow}^{\uparrow}-T_{\downarrow\uparrow,\uparrow}^{\uparrow}\right);\\
&T_{1}^{p \pi_{0}}=i \frac{1}{2^{1/4} \sqrt{1+\xi}\left(P^{+}\right)^{3/2}} \frac{f_{\pi}}{f_{N}}\left[T_{\uparrow\uparrow,\downarrow}^{\uparrow}-\frac{\left(\Delta_{T}^{-}\right)^{2}}{\Delta_{T}^{2}} T_{\downarrow\downarrow,\downarrow}^{\uparrow}\right];\\
&V_{2}^{p \pi_{0}}=i \frac{m_N \Delta_{T}^{-}}{\Delta_{T}^{2}} \frac{1}{2^{1/4} \sqrt{1+\xi}\left(P^{+}\right)^{3/2}} \frac{f_{\pi}}{f_{N}}\left(T_{\uparrow \downarrow,\downarrow}^{\uparrow}+T_{\downarrow \uparrow,\downarrow}^{\uparrow}\right);\\
&A_{2}^{p \pi_{0}}=i \frac{m_N \Delta_{T}^{-}}{\Delta_{T}^{2}} \frac{1}{2^{1/4} \sqrt{1+\xi}\left(P^{+}\right)^{3/2}} \frac{f_{\pi}}{f_{N}}\left(T_{\uparrow \downarrow,\downarrow}^{\uparrow}-T_{\downarrow \uparrow,\downarrow}^{\uparrow}\right);\\
&T_{2}^{p \pi_{0}}=i \frac{m_N}{\Delta_{T}^{2}} \frac{1}{2^{1/4} \sqrt{1+\xi}\left(P^{+}\right)^{3/2}} \frac{f_{\pi}}{f_{N}}\left[\Delta_{T}^{+} T_{\uparrow \uparrow,\uparrow}^{\uparrow}-\Delta_{T}^{-} T_{\downarrow \downarrow,\uparrow}^{\uparrow}\right];\\
&T_{3}^{p \pi_{0}}=-i \frac{m_N}{\Delta_{T}^{2}} \frac{1}{2^{1/4} \sqrt{1+\xi}\left(P^{+}\right)^{3/2}} \frac{f_{\pi}}{f_{N}}\left[\Delta_{T}^{+} T_{\uparrow \uparrow,\uparrow}^{\uparrow}+\Delta_{T}^{-} T_{\downarrow \downarrow,\uparrow}^{\uparrow}\right];\\
&T_{4}^{p \pi_{0}}=i \frac{2 m_N^{2}\left(\Delta_{T}^{-}\right)^{2}}{\left(\Delta_{T}^{2}\right)^{2}} \frac{1}{2^{1/4} \sqrt{1+\xi}\left(P^{+}\right)^{3/2}} \frac{f_{\pi}}{f_{N}} T_{\downarrow \downarrow,\downarrow}^{\uparrow},
\end{aligned}\label{Def_TDA_through_T}
\end{equation}
where the
$\lambda_N, \lambda_{1,2,3}=\uparrow, \downarrow$ arrows denote the plus and minus helicities of the proton and quarks; and
$\Delta_{T}^{\pm} \equiv \Delta_{x} \pm i \Delta_{y}$.

It is also instructive to consider
the reciprocal relations, which express helicity amplitudes in terms of linear combinations of TDAs:
\begin{equation}
\begin{aligned}
&T_{\uparrow \downarrow,\uparrow}^{\uparrow} =i  2^{1/4} \sqrt{1+\xi}\left(P^{+}\right)^{3/2} \frac{f_{N}}{f_{\pi}}\left(\frac{V_{1}^{p \pi_{0}}-A_{1}^{p \pi_{0}}}{2}\right);\\
&T_{\downarrow\uparrow,\uparrow}^{\uparrow}=i  2^{1/4} \sqrt{1+\xi}\left(P^{+}\right)^{3/2} \frac{f_{N}}{f_{\pi}}\left(\frac{V_{1}^{p \pi_{0}}+A_{1}^{p \pi_{0}}}{2}\right);\\
&T_{\uparrow\uparrow,\downarrow}^{\uparrow} =-i 2^{1/4} \sqrt{1+\xi}\left(P^{+}\right)^{3/2} \frac{f_{N}}{f_{\pi}}\left(T_{1}^{p \pi_{0}}+\frac{\Delta_{T}^{2}}{2m_N^2} T_{4}^{p \pi_{0}}\right);\\
&T_{\uparrow \downarrow,\downarrow}^{\uparrow}=-i 2^{1/4} \sqrt{1+\xi}\left(P^{+}\right)^{3/2} \frac{\Delta_{T}^{2}}{m_N \Delta_{T}^{-}} \frac{f_{N}}{f_{\pi}}\left(\frac{V_{2}^{p \pi_{0}}+A_{2}^{p \pi_{0}}}{2}\right);\\
&T_{\downarrow \uparrow,\downarrow}^{\uparrow}=-i 2^{1/4} \sqrt{1+\xi}\left(P^{+}\right)^{3/2} \frac{\Delta_{T}^{2}}{m_N \Delta_{T}^{-}} \frac{f_{N}}{f_{\pi}}\left(\frac{V_{2}^{p \pi_{0}}-A_{2}^{p \pi_{0}}}{2}\right);\\
&T_{\uparrow \uparrow,\uparrow}^{\uparrow}=-i 2^{1/4} \sqrt{1+\xi}\left(P^{+}\right)^{3/2} \frac{\Delta_{T}^{2}}{m_N \Delta_{T}^{+}} \frac{f_{N}}{f_{\pi}}\left(\frac{T_{2}^{p \pi_{0}}-T_{3}^{p \pi_{0}}}{2}\right);\\
&T_{\downarrow \downarrow,\uparrow}^{\uparrow}=i 2^{1/4} \sqrt{1+\xi}\left(P^{+}\right)^{3/2} \frac{\Delta_{T}^{2}}{m_N \Delta_{T}^{-}} \frac{f_{N}}{f_{\pi}}\left(\frac{T_{2}^{p \pi_{0}}+T_{3}^{p \pi_{0}}}{2}\right);\\
&T_{\downarrow \downarrow,\downarrow}^{\uparrow} =-i \frac{\left(\Delta_{T}^{2}\right)^{2}}{2 m_N^{2}\left(\Delta_{T}^{-}\right)^{2}} 2^{1/4} \sqrt{1+\xi}\left(P^{+}\right)^{3/2} \frac{f_{N}}{f_{\pi}} T_{4}^{p \pi_{0}}.
\end{aligned}
\label{Def_TDA_through_Tinv}
\end{equation}

The relations
(\ref{Def_TDA_through_T}),
(\ref{Def_TDA_through_Tinv})
are very informative about the meaning of the different TDAs.
Each power of  $\Delta_{T}$  in the parametrization
(\ref{Param_TDAs})
with the set of  Dirac structures
(\ref{Def_DirStr_piN_DeltaT})
corresponds to one unit of orbital angular momentum (OAM).
Therefore, we conclude that
\bi
\item
 $T_{4}^{p \pi_{0}}$
contributes to the spin decomposition with the OAM
$L=2$;
\item
$V_{2}^{p \pi_{0}},A_{2}^{p \pi_{0}}, T_{2}^{p \pi_{0}} , T_{3}^{p \pi_{0}}$
contribute to the spin decomposition with the OAM
$L=1$;
\item
$V_{1}^{p \pi_{0}}, A_{1}^{p \pi_{0}}$ and $ T_{1}^{p \pi_{0}}$
correspond to the $L=0$ OAM, and thus survive in the $\Delta_{T} \to 0$ limit.
\ei

\subsubsection{Nucleon-to-vector-meson {{TDA}}s}
\label{SubSec_Def_NV_TDAs}
\mbox

Nucleon-to-vector-mesons TDA have been introduced in
\cite{Lansberg:2006uh,Pire:2015kxa}.
The parametrization  of the leading twist $V N$ TDA involves $24$ Dirac structures
corresponding to the number of independent helicity amplitudes of the
$N \to qqq V$ process.
The counting procedure follows the usual pattern:
each of the three quarks and the nucleon have $2$ helicity states, while the vector meson
has $3$. This leads to $3 \cdot 2^4=48$ amplitudes. Parity constraint relates helicity amplitudes
with all opposite helicities and reduces the number of independent helicity amplitudes
by a factor of $2$.

For the case of the proton-to-a-neutral-vector-meson (\textit{e.g.} $\rho^0$, $\omega$ or $\phi$) transition the corresponding TDAs are defined by the hadronic element
of the light-cone operator
(\ref{Def_O_uud_operator}):
\begin{eqnarray}
 &&
4 {\mathcal{F}} \langle V(p_V, \lambda_V)|   \widehat{O}_{\rho \tau \chi}^{uud}(\lambda_1n,\lambda_2n,\lambda_3n)| N^p(p_N,s_N) \rangle
\nonumber \\ &&
=
\delta(x_1+x_2+x_3-2\xi)  \times m_N \Bigl[
\sum_{ {\Upsilon= 1 {\mathcal{E}}, 1T, 1n, \atop  2 {\mathcal{E}}, 2T, 2n}}
(v^{V N}_\Upsilon)_{\rho \tau, \, \chi} V_{\Upsilon}^{VN}(x_1,x_2,x_3, \xi, \Delta^2; \, \mu^2)
\nonumber \\ &&
+
\sum_{ {\Upsilon= 1 {\mathcal{E}}, 1T, 1n, \atop  2 {\mathcal{E}}, 2T, 2n}}
(a^{V N}_\Upsilon)_{\rho \tau, \, \chi} A_{\Upsilon}^{VN}(x_1,x_2,x_3, \xi, \Delta^2; \, \mu^2)
+
\sum_{ {\Upsilon= 1 {\mathcal{E}}, 1T, 1n,   2{\mathcal{E}}, 2T, 2n, \atop
3 {\mathcal{E}}, 3T, 3n,  4 {\mathcal{E}}, 4T, 4n
}}
(t^{V N}_\Upsilon)_{\rho \tau, \, \chi} T_{\Upsilon}^{VN}(x_1,x_2,x_3, \xi, \Delta^2; \, \mu^2)
\Bigr], \nonumber \\ &&
\label{VN_TDAs_param}
\end{eqnarray}
where $\mathcal F$ stands for the Fourier transform operation (\ref{Fourier_TDA}).

The procedure for building the corresponding leading twist Dirac structures was described in Ref.~\cite{Lansberg:2006uh}.
The revised version\footnote{The original parametrization of Ref.~\cite{Lansberg:2006uh} erroneously lacked
$\gamma_5$
factors for the Dirac structures. See discussion in Ref.~\cite{Pire:2015kxa}.}
of the set of leading twist-$3$ Dirac structures for the $VN$ TDA parametrization reads
\begin{align}
(v_{1 {{\mathcal{E}} }}^{V N})_{\rho \tau, \, \chi}&= (\hat{p}C)_{\rho \tau} \bigl(\gamma^5 \hat{{\mathcal{E}}}^{*}U^+ \bigr)_\chi;          &  (v_{2 {\mathcal{E}}}^{V N})_{\rho \tau, \, \chi}&=  m_N^{-1}   (\hat{p}C)_{\rho \tau} \bigl(\gamma^5  \sigma^{\Delta_T {\mathcal{E}}^{*} } U^+ \bigr)_\chi;
\nn             \\
(v_{1 T}^{V N})_{\rho \tau, \, \chi}&=
m_N^{-1}
({{\mathcal{E}}}^{*}
\cdot \Delta_T)
(\hat{p}C)_{\rho \tau} \bigl(\gamma^5 U^+ \bigr)_\chi;        &  (v_{2 T}^{V N})_{\rho \tau, \, \chi} & =  m_N^{-2}
({{\mathcal{E}}}^{*} \cdot \Delta_T)
(\hat{p}C)_{\rho \tau}  \bigl(\gamma^5 \hat{\Delta}_T U^+ \bigr)_\chi;  \nn \\
(v_{1 n}^{V N})_{\rho \tau, \, \chi} & = m_N  ({{\mathcal{E}}^{*}}
\cdot n) (\hat{p}C)_{\rho \tau} \bigl(\gamma^5 U^+ \bigr)_\chi;   &  (v_{2 n}^{V N})_{\rho \tau, \, \chi} & =
({{\mathcal{E}}}^{*} \cdot n)
(\hat{p}C)_{\rho \tau}  \bigl(\gamma^5 \hat{\Delta}_T U^+ \bigr)_\chi;
\label{Dirac_v_NV}
\end{align}
\begin{align}
(a_{1 {\mathcal{E}}}^{V N})_{\rho \tau, \, \chi}& = (\hat{p} \gamma^5 C)_{\rho \tau} \bigl(  \hat{{\mathcal{E}}}^{*} U^+ \bigr)_\chi; &
(a_{2 {\mathcal{E}}}^{V N})_{\rho \tau, \, \chi} &=  m_N^{-1}   (\hat{p} \gamma^5C)_{\rho \tau} \bigl(   \sigma^{\Delta_T {\mathcal{E}}^{*} } U^+ \bigr)_\chi;
\nonumber \\
(a_{1 T}^{V N})_{\rho \tau, \, \chi} &=
m_N^{-1}
({{\mathcal{E}}}^{*}
\cdot \Delta_T)
(\hat{p} \gamma^5 C)_{\rho \tau} \bigl(  U^+ \bigr)_\chi; &
(a_{2 T}^{V N})_{\rho \tau, \, \chi} &=  m_N^{-2}
({{\mathcal{E}}}^{*} \cdot \Delta_T)
(   \hat{p}\gamma^5 C)_{\rho \tau}  \bigl(\hat{\Delta}_T U^+ \bigr)_\chi;
\nonumber \\
(a_{1 n}^{V N})_{\rho \tau, \, \chi}& = m_N  ({{\mathcal{E}}}^{*}
\cdot n) (\hat{p} \gamma^5 C)_{\rho \tau} \bigl(  U^+ \bigr)_\chi;  &
(a_{2 n}^{V N})_{\rho \tau, \, \chi} & =
({{\mathcal{E}}}^{*} \cdot n)
(   \hat{p} \gamma^5  C)_{\rho \tau}  \bigl(\hat{\Delta}_T U^+ \bigr)_\chi;
\label{Dirac_a_NV}
\end{align}
\begin{align}
(t_{1 {\mathcal{E}}}^{V N})_{\rho \tau, \, \chi} & =(\sigma_{p \lambda} C)_{\rho \tau} (\gamma_5 \sigma^{\lambda {\mathcal{E}}^{*}}  U^+)_\chi; &
(t_{2 {\mathcal{E}}}^{V N})_{\rho \tau, \, \chi}&=(\sigma_{p {\mathcal{E}}^{*}} C)_{\rho \tau} (\gamma_5   U^+)_\chi;
\nonumber \\
(t_{3 {\mathcal{E}}}^{V N})_{\rho \tau, \, \chi} & =m_N^{-1} (\sigma_{p \Delta_T} C)_{\rho \tau} (\gamma_5 \hat{{\mathcal{E}}}^{*} U^+)_\chi; &
(t_{4 {\mathcal{E}}}^{V N})_{\rho \tau, \, \chi}& = m_N^{-1} (\sigma_{p {\mathcal{E}}^{*}} C)_{\rho \tau} (\gamma_5 \hat{\Delta}_T U^+)_\chi;
\nonumber \\
(t_{1 T}^{V N})_{\rho \tau, \, \chi}& = m_N^{-1}  ({\mathcal{E}}^{*} \cdot \Delta_T) (\sigma_{p \lambda} C)_{\rho \tau} (\gamma_5 \gamma^\lambda  U^+)_\chi;
&(t_{2 T}^{V N})_{\rho \tau, \, \chi}&= m_N^{-2}  ({\mathcal{E}}^{*} \cdot \Delta_T) (\sigma_{p \lambda} C)_{\rho \tau} (\gamma_5 \sigma^{\lambda \Delta_T}  U^+)_\chi;
\nonumber \\
(t_{3 T}^{V N})_{\rho \tau, \, \chi}&= m_N^{-2}  ({\mathcal{E}}^{*} \cdot \Delta_T) (\sigma_{p \Delta_T} C)_{\rho \tau} (\gamma_5  U^+)_\chi; &
(t_{4 T}^{V N})_{\rho \tau, \, \chi} &=m_N^{-3} ({\mathcal{E}}^{*} \cdot \Delta_T) (\sigma_{p \Delta_T} C)_{\rho \tau} (\gamma_5 \hat{\Delta}_T U^+)_\chi;
\nonumber \\
(t_{1 n}^{V N})_{\rho \tau, \, \chi}&= m_N   ({\mathcal{E}}^{*} \cdot n) (\sigma_{p \lambda} C)_{\rho \tau} (\gamma_5 \gamma^\lambda  U^+)_\chi;&
(t_{2 n}^{V N})_{\rho \tau, \, \chi}&= ({\mathcal{E}}^{*} \cdot n) (\sigma_{p \lambda} C)_{\rho \tau} (\gamma_5 \sigma^{\lambda \Delta_T}  U^+)_\chi;
\nonumber \\
(t_{3 n}^{V N})_{\rho \tau, \, \chi}& =   ({\mathcal{E}}^{*} \cdot n) (\sigma_{p \Delta_T} C)_{\rho \tau} (\gamma_5  U^+)_\chi;&
(t_{4 n}^{V N})_{\rho \tau, \, \chi}&=m_N^{-1}  ({\mathcal{E}}^{*}\cdot n)   (\sigma_{p \Delta_T} C)_{\rho \tau} (\gamma_5 \hat{\Delta}_T U^+)_\chi.
\label{Dirac_t_NV}
\end{align}
The notations in
(\ref{Dirac_v_NV}),
(\ref{Dirac_a_NV}),
(\ref{Dirac_t_NV})
are the same as in
(\ref{Def_DirStr_piN_DeltaT}).
${\mathcal{E}}(p_V, \lambda_V)$ denotes
the polarization vector of the vector meson.
From the transversality of the polarization vector of the vector meson
\begin{equation}
{\mathcal{E}}^{*}(p_V, \lambda_V) \cdot p_V=0
\end{equation}
we establish the following condition for the ``$-$''-light-cone component of the polarization vector of the vector meson:
\begin{equation}
{\mathcal{E}}^{*}(p_V, \lambda_V) \cdot p= - \frac{m_V^2-\Delta_T^2}{(1-\xi)^2} ({\mathcal{E}}^{*}(p_V, \lambda_V) \cdot n)- \frac{1}{1-\xi} ({\mathcal{E}}^{*}(p_V, \lambda_V) \cdot \Delta_T).
\label{transvers_E}
\end{equation}
This relation is crucial for working out  the set of Dirac structures (\ref{Dirac_t_NV}).

Analogously as in the earlier  Sec.~\ref{SubSubSec_Def_piN_TDAs}, each  of the $24$ $VN$ TDAs defined in
(\ref{VN_TDAs_param})
is function of three longitudinal momentum fractions $x_1$, $x_2$, $x_3$, skewness parameter $\xi$, $u$-channel
momentum transfer squared $\Delta^2$ and of a factorization scale $\mu$.
TDAs
$V_{\Upsilon}^{VN}$
and
$T_{\Upsilon}^{VN}$
are defined symmetric under the interchange $x_1 \leftrightarrow x_2$,
while
$A_{\Upsilon}^{VN}$
are antisymmetric under the interchange $x_1 \leftrightarrow x_2$.

Similarly to the $\pi N$  TDA case,
the $VN$ TDA parametrization
(\ref{VN_TDAs_param})
with the set of  Dirac structures
(\ref{Dirac_v_NV}),
(\ref{Dirac_a_NV}),
(\ref{Dirac_t_NV})
is well suited to keep eye on the
$\Delta_T = 0$
limit.
Namely, in the limit $\Delta_T=0$
only
$7$
TDAs out of $24$  turn out to be relevant:
$V_{1 {\mathcal{E}}}^{VN}$,  $V_{1 n}^{VN}$,
$A_{1 {\mathcal{E}}}^{VN}$,  $A_{1 n}^{VN}$, $
T_{1 {\mathcal{E}}}^{VN}$, $T_{1 n}^{VN}$,
$T_{2 {\mathcal{E}}}^{VN}$.

\subsubsection{Nucleon-to-photon {{TDA}}s}
\label{SubSec_Def_Ngamma_TDAs}
\mbox

\begin{figure}[H]
\begin{center}
\includegraphics[width=0.4\textwidth]{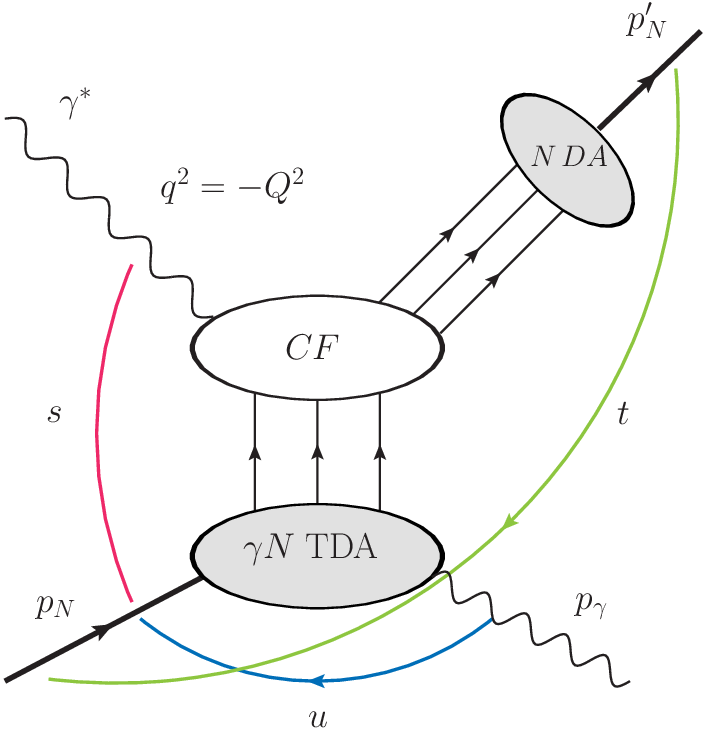}
\end{center}
     \caption{Kinematical quantities and the collinear factorization mechanism  for
      $\gamma^{*} N \to N \gamma$  in the  near-backward  kinematical regime (large $Q^2$, $W$; fixed $x_{B}$; $| u|  \sim 0$). The lower blob, denoted
  $\gamma N$ TDA, depicts the nucleon-to-photon transition
     distribution amplitude; $N$ DA blob depicts the nucleon distribution amplitude;
      $CF$  denotes hard subprocess amplitude (coefficient function).}
\label{Fig_Kinematics_TDAs_gammaN}
\end{figure}

The nucleon to photon TDAs, which enter the collinear factorized description of backward virtual Compton scattering
\begin{equation}
e(k)+N(p_N,s_N) \to
\left(\gamma^{*}(q, \lambda_\gamma)+N(p_N,s_N) \right) + e(k') \to e(k')+N(p'_N,s'_N)+ \gamma(p_\gamma,s_\gamma)
\label{bkw_DVCS}
\end{equation}
(see  Fig.~\ref{Fig_Kinematics_TDAs_gammaN}), may be obtained from the previous case; the absence of  helicity zero state of the real photon leads to a smaller number of TDAs.

The parametrization for nucleon-to-photon TDAs can be constructed
similarly to the nucleon-to-vector meson case (\ref{VN_TDAs_param}).
Counting the degrees of freedom fixes the number of independent $\gamma N$ TDAs to
$16$, since each quark, photon and proton have two helicity states (leading to $2^5=32$ helicity
amplitudes) and parity invariance relates amplitudes with opposite helicities for all particles.

We
can equally say that the photon has spin 1, which would normally lead to 24 TDAs, as in the case of nucleon-to-vector meson TDA,
  but QED gauge invariance provides further  relations between TDAs, which reduces the number of independent $\gamma N$ TDAs to 16.
Indeed, in the nucleon-to-photon case, QED gauge invariance  implies that the
matrix element vanishes
when the polarization vector
${\mathcal{E}}(p_\gamma, s_\gamma)$
is replaced by
$p_\gamma$.
At the leading-twist-$3$ accuracy, this provides $8$ relations:
\begin{eqnarray}
 && V_{1{\mathcal{E}}}^{\gamma N}(1-\xi) \frac{m_N}{2(1+\xi)}+V_{1T}^{\gamma N} \frac{\Delta_{T}^{2}}{m_N}+V_{1n}^{\gamma N} \frac{(1-\xi) m_N}{2} \nn
\\ && =V_{1{\mathcal{E}}}^{\gamma N}+\frac{V_{2 {\mathcal{E}}}^{\gamma N}}{2 m_N}(1-\xi) \frac{m_N}{2(1+\xi)}+V_{2T}^{\gamma N} \frac{\Delta_{T}^{2}}{m_N^{2}}+V_{2n}^{\gamma N} \frac{1-\xi}{2}=0; \nn \\ &&
A_{1 {\mathcal{E}}}^{\gamma N}(1-\xi) \frac{m_N}{2(1+\xi)}+A_{1T}^{\gamma N} \frac{\Delta_{T}^{2}}{m_N}+A_{1n}^{\gamma N} \frac{(1-\xi) m_N}{2} \nn \\ &&
=A_{1{\mathcal{E}}}^{\gamma N}+\frac{A_{2 {\mathcal{E}}}^{\gamma N}}{2 m_N}(1-\xi) \frac{m_N}{2(1+\xi)}+A_{2T}^{\gamma N} \frac{\Delta_{1}^{2}}{m_N^{2}}+A_{2n}^{\gamma N} \frac{1-\xi}{2}=0; \nn \\ &&
\frac{T_{1 {\mathcal{E}}}^{\gamma N}}{2}(1-\xi) \frac{m_N}{2(1+\xi)}+T_{1T}^{\gamma N} \frac{\Delta_{T}^{2}}{m_N}+T_{1n}^{\gamma N} \frac{(1-\xi) m_N}{2} \nn \\ &&
=T_{2 {\mathcal{E}}}^{\gamma N}+\frac{T_{3 {\mathcal{E}}}^{\gamma N}}{m_N}(1-\xi) \frac{m_N}{2(1+\xi)}+T_{3T}^{\gamma N} \frac{\Delta_{T}^{2}}{m_N^{2}}+T_{3n}^{\gamma N} \frac{1-\xi}{2}=0; \nn \\ &&
T_{1{\mathcal{E}}}^{\gamma N}+T_{2T}^{\gamma N} \frac{\Delta_{T}^{2}}{m_N^{2}}+T_{2n}^{\gamma N} \frac{1-\xi}{2}    =T_{3{\mathcal{E}}}^{\gamma N}+T_{4{\mathcal{E}}}^{\gamma N}+T_{4T}^{\gamma N} \frac{\Delta_{T}^{2}}{m_N^{2}}+T_{4n}^{\gamma N} \frac{1-\xi}{2}=0;
\end{eqnarray}

This effectively reduces the number of $\gamma N$ TDAs to $16$, as expected from the number
of helicity amplitudes for the process $N \to  qqq \gamma$:
\begin{eqnarray}
 &&
4 {\mathcal{F}} \langle \gamma(p_\gamma, s_\gamma) |  \widehat{O}_{\rho \tau \chi}^{uud}(\lambda_1n,\lambda_2n,\lambda_3n)| N^p(p_N,s_N) \rangle
\nonumber \\ &&
=
\delta(x_1+x_2+x_3-2\xi)  \, m_N \Bigl[
\sum_{ {\Upsilon= 1 {\mathcal{E}}, 1T \atop  2 {\mathcal{E}}, 2T}}      (v^{\gamma N}_\Upsilon)_{\rho \tau, \, \chi} V_{\Upsilon}^{\gamma N}(x_1,x_2,x_3, \xi, \Delta^2; \, \mu^2)
\nonumber \\ &&
+
\sum_{ {\Upsilon= 1 {\mathcal{E}}, 1T \atop  2 {\mathcal{E}}, 2T}}
(a^{\gamma N}_\Upsilon)_{\rho \tau, \, \chi} A_{\Upsilon}^{\gamma N}(x_1,x_2,x_3, \xi, \Delta^2; \, \mu^2)
+
\sum_{ {\Upsilon= 1 {\mathcal{E}}, 1T,   2{\mathcal{E}}, 2T \atop  3 {\mathcal{E}}, 3T,   4 {\mathcal{E}}, 4T }}
(t^{\gamma N}_\Upsilon)_{\rho \tau, \, \chi} T_{\Upsilon}^{\gamma N}(x_1,x_2,x_3, \xi, \Delta^2; \, \mu^2)
\Bigr].
\label{gammaN_TDAs_param}
\end{eqnarray}
The set of  QED gauge invariant Dirac structures\footnote{Note that the original set of the Dirac structures of
Eq.~(16) of~\cite{Lansberg:2006uh}
lacks $\gamma_5$.}
is given in
Eqs.~(\ref{v_QED_gauge_inv}), (\ref{a_QED_gauge_inv}), (\ref{t_QED_gauge_inv}):
\begin{eqnarray}
 &&
(v_{1 {{\mathcal{E}} }}^{\gamma N})_{\rho \tau, \, \chi}
= (\hat{p} C)_{\rho \tau}
\left[\left( \gamma_5 \hat{\mathcal{E}}^{*} U^{+}\right)_{\chi}-\frac{m_N}{1+\xi}
({\mathcal{E}}^{*} \cdot n)\left( \gamma_5 U^{+}\right)_{\chi}-\frac{2({\mathcal{E}}^{*} \cdot n)}{1-\xi}\left(\gamma_5 \hat{\Delta}_{T} U^{+}\right)_{\chi}\right]; \nn \\ &&
 (v_{1 T}^{\gamma N})_{\rho \tau, \, \chi}
=\frac{1}{m_N}
\left[\left( {\mathcal{E}}^{*} \cdot \Delta_{T}\right)-\frac{2 \Delta_{T}^{2}}{1-\xi}
({\mathcal{E}}^{*} \cdot n)\right]
(\hat{p} C)_{\rho \tau}\left(\gamma_5 U^{+}\right)_{\chi}; \nn \\ &&
 (v_{2 {\mathcal{E}}}^{\gamma N})_{\rho \tau, \, \chi}
=
\frac{1}{m_N}(\hat{p} C)_{\rho \tau}
\left[\left(\gamma_5 \sigma^{\Delta_{T} {\mathcal{E}}^{*}} U^{+}\right)_{\chi}-\frac{m_N
({\mathcal{E}}^{*} \cdot n)}{2(1+\xi)}\left(\gamma_5 \hat{\Delta}_{T} U^{+}\right)_{\chi}\right];
\nn \\ &&
 (v_{2 T}^{\gamma N})_{\rho \tau, \, \chi}
=\frac{1}{m_N^2}
\left[\left( {\mathcal{E}}^{*} \cdot \Delta_{T}\right)-\frac{2 \Delta_{T}^{2}}{1-\xi}
({\mathcal{E}}^{*} \cdot n)\right]
(\hat{p} C)_{\rho \tau}\left(\gamma_5 \hat{\Delta}_T U^{+}\right)_{\chi};
\label{v_QED_gauge_inv}
\end{eqnarray}
\begin{eqnarray}
 &&
(a_{1 {{\mathcal{E}} }}^{\gamma N})_{\rho \tau, \, \chi}
= \left(\hat{p} \gamma^{5} C\right)_{\rho \tau}\left[\left(\hat{\mathcal{E}}^{*} U^{+}\right)_{\chi}-\frac{m_N}{1+\xi}({\mathcal{E}}^{*} \cdot n)\left( U^{+}\right)_{\chi}-\frac{2({\mathcal{E}}^{*} \cdot n)}{1-\xi}\left(  \hat{\Delta}_{T} U^{+}\right)_{\chi}\right]; \nn \\  &&
(a_{1 T}^{\gamma N})_{\rho \tau, \, \chi}=
\frac{1}{m_N}
\left[\left( {\mathcal{E}}^{*} \cdot \Delta_{T}\right)-\frac{2 \Delta_{T}^{2}}{1-\xi}
({\mathcal{E}}^{*} \cdot n)\right]
(\hat{p} \gamma^5 C)_{\rho \tau}\left(  U^{+}\right)_{\chi}; \nn \\  &&
 (a_{2 {\mathcal{E}}}^{\gamma N})_{\rho \tau, \, \chi}
=
\frac{1}{m_N}(\hat{p}  \gamma^5 C)_{\rho \tau}
\left[\left( \sigma^{\Delta_{T} {\mathcal{E}}^{*}} U^{+}\right)_{\chi}-\frac{m_N
({\mathcal{E}}^{*} \cdot n)}{2(1+\xi)}\left(  \hat{\Delta}_{T} U^{+}\right)_{\chi}\right];
\nn \\  &&
 (a_{2 T}^{\gamma N})_{\rho \tau, \, \chi}
=\frac{1}{m_N^2}
\left[\left( {\mathcal{E}}^{*} \cdot \Delta_{T}\right)-\frac{2 \Delta_{T}^{2}}{1-\xi}
({\mathcal{E}}^{*} \cdot n)\right]
(\hat{p}  \gamma^5 C)_{\rho \tau}\left(  \hat{\Delta}_T U^{+}\right)_{\chi};  \label{a_QED_gauge_inv}
\end{eqnarray}
\begin{eqnarray}
 &&
(t_{1 {{\mathcal{E}} }}^{\gamma N})_{\rho \tau, \, \chi}
 =\left(\sigma_{p \mu} C\right)_{\rho \tau}
\left[\left(\gamma_5 \sigma^{\mu {\mathcal{E}}^{*}} U^{+}\right)_{\chi}-\frac{m_N
({\mathcal{E}}^{*} \cdot n)}{2(1+\xi)}\left(\gamma^5 \gamma^{\mu} U^{+}\right)_{\chi}-\frac{2({\mathcal{E}}^{*} \cdot n)}{(1-\xi)}\left(\gamma^5 \sigma^{\mu \Delta_{T}} U^{+}\right)_{\chi}\right]; \nn \\
 &&
 (t_{1 T}^{\gamma N})_{\rho \tau, \, \chi}
=\frac{1}{m_N}\left[\left({\mathcal{E}}^{*} \cdot \Delta_{T}\right)-\frac{2 \Delta_{T}^{2}}{1-\xi}({\mathcal{E}}^{*} \cdot n)\right]\left(\sigma_{p \mu} C\right)_{\rho \tau}\left(\gamma^5 \gamma^{\mu} U^{+}\right)_{\chi}; \nn \\
 &&
 (t_{2 {{\mathcal{E}} }}^{\gamma N})_{\rho \tau, \, \chi}
=\left[\left(\sigma_{p {\mathcal{E}}^{*}} C\right)_{\rho \tau}-\frac{2({\mathcal{E}}^{*} \cdot n)}{(1-\xi)}\left(\sigma_{p \Delta_{T}} C\right)_{\rho \tau}\right]\left(\gamma^5 U^{+}\right)_{\chi}; \nn \\
 &&
 (t_{2 T}^{\gamma N})_{\rho \tau, \, \chi}
=\frac{1}{m_N^{2}}
\left[\left({\mathcal{E}}^{*} \cdot \Delta_{T}\right)-\frac{2 \Delta_{T}^{2}}{1-\xi}
({\mathcal{E}}^{*} \cdot  n)\right]\left(\sigma_{p \mu} C\right)_{\rho \tau}\left(\gamma^5 \sigma^{\mu \Delta_{T}} U^{+}\right)_{\chi};
\nn \\
 &&
 (t_{3 {{\mathcal{E}} }}^{\gamma N})_{\rho \tau, \, \chi}
=
 \frac{1}{m_N}\left(\sigma_{p \Delta_{T}} C\right)_{\rho \tau}\left[\left( \gamma^5 \hat{\mathcal{E}}^{*} U^{+}\right)_{\chi}-\frac{m_N ({\mathcal{E}}^{*} \cdot n)}{(1+\xi)}\left(\gamma^5 U^{+}\right)_{\chi}-\frac{2({\mathcal{E}}^{*} \cdot n)}{(1-\xi)}\left(\gamma^5 \hat{\Delta}_{T} U^{+}\right)_{\chi}\right]; \nn \\
  &&
 (t_{3 T}^{\gamma N})_{\rho \tau, \, \chi}
=
 \frac{1}{m_N^{2}}\left[\left({\mathcal{E}}^{*} \cdot \Delta_{T}\right)-\frac{2 \Delta_{T}^{2}}{1-\xi}({\mathcal{E}}^{*} \cdot  n)\right]
\left(\sigma_{p \Delta_{T}} C\right)_{\rho \tau}\left(\gamma^5 U^{+}\right)_{\chi}; \nn \\
 &&
 (t_{4 {{\mathcal{E}} }}^{\gamma N})_{\rho \tau, \, \chi}
=\frac{1}{m_N}\left[\left(\sigma_{p {\mathcal{E}}^{*}} C\right)_{\rho \tau}-
\frac{2({\mathcal{E}}^{*} \cdot n)}{1-\xi}\left(\sigma_{p \Delta_{T}} C\right)_{\rho \tau}\right] \left(\gamma^5 \hat{\Delta}_{T} U^{+}\right)_{\chi}; \nn \\
 &&
 (t_{4 T}^{\gamma N})_{\rho \tau, \, \chi}
= \frac{1}{m_N^{3}}\left[\left({\mathcal{E}}^{*} \cdot \Delta_{T}\right)
-\frac{2 \Delta_{T}^{2}}{1-\xi}({\mathcal{E}}^{*} \cdot  n)\right]
\left(\sigma_{p \Delta_{T}} C\right)_{\rho \tau}\left(\gamma^5 \hat{\Delta}_{T} U^{+}\right)_{\chi}.
\label{t_QED_gauge_inv}
\end{eqnarray}
In the strictly backward
$\Delta_T= 0$
limit only
$4$ $\gamma N$
TDAs defined in (\ref{gammaN_TDAs_param})
$V_{1 {{\mathcal{E}} }}^{\gamma N}$, $A_{1 {{\mathcal{E}} }}^{\gamma N}$,
$T_{1 {{\mathcal{E}} }}^{\gamma N}$, $T_{2 {{\mathcal{E}} }}^{\gamma N}$
turn out to be relevant.
This is consistent with the helicity states counting,
since for $\Delta_T= 0$, there is no angular momentum exchanged, the helicity is
conserved. For definiteness, let us consider the proton-to-$uud \, \gamma$ transition.
We have three possible processes
$N^p(\uparrow) \to uud(\uparrow \downarrow \downarrow) + \gamma(\uparrow)$, where the quark
with helicity
$+1$ is either one of the
$u$'s or the $d$ and
also
$N^p(\uparrow) \to uud(\uparrow \uparrow \uparrow) + \gamma(\downarrow)$.
Therefore,
within the $\Delta_T=0$ limit the complete set of $16$ $\gamma N$ TDAs indeed reduces  to just $4$.

Now we establish the relation between
$\gamma N$ TDAs and the light-front helicity amplitudes
$T_{\lambda_1 \lambda_2, \lambda_3}^{\lambda_N , \, \lambda_\gamma}$,
where $\lambda_N, \lambda_{1,2,3} \equiv \uparrow \downarrow$ stand  for the light-front
helicities of, respectively, initial state nucleon and three final state
quarks; and $\lambda_\gamma \equiv \uparrow \downarrow$ denotes the
light-front
helicity of the final state photon.
For simplicity we present the result only for those  $\gamma N$ TDAs, which contribute within the $\Delta_T=0$ limit:
\begin{eqnarray}
 && V_{1 {\mathcal{E}}}^{ \gamma p}= \frac{1}{2^{1/4} \sqrt{1+\xi}\left(P^{+}\right)^{3/2}} \frac{1}{m_N}
\left[T_{\uparrow \downarrow,\downarrow}^{\uparrow , \, \uparrow}+T_{\downarrow \uparrow,\downarrow}^{\uparrow , \, \uparrow}\right]; \nonumber \\
 && A_{1 {\mathcal{E}}}^{ \gamma p}=- \frac{1}{2^{1/4} \sqrt{1+\xi}\left(P^{+}\right)^{3/2}}  \frac{1}{m_N}
\left[T_{\uparrow \downarrow,\downarrow}^{\uparrow, \, \uparrow}-T_{\downarrow \uparrow,\downarrow}^{\uparrow , \, \uparrow}\right]; \nonumber \\
 && T_{1 {\mathcal{E}}}^{ \gamma p}=-\frac{1}{2^{1/4} \sqrt{1+\xi}\left(P^{+}\right)^{3/2}} \frac{1}{m_{N}}
\left[T_{\downarrow \downarrow,\uparrow}^{\uparrow , \, \uparrow}+T_{\uparrow \uparrow,\uparrow}^{\uparrow, \, \downarrow}\right];\nonumber \\
 && T_{2 {\mathcal{E}}}^{ \gamma p}=-\frac{1}{2^{1/4} \sqrt{1+\xi}\left(P^{+}\right)^{3/2}} \frac{1}{m_{N}}
\left[T_{\downarrow \downarrow,\uparrow}^{\uparrow , \, \uparrow}-T_{\uparrow \uparrow,\uparrow}^{\uparrow , \, \downarrow}\right].
\label{Gamma_Hel}
\end{eqnarray}

Nucleon-to-photon TDAs  allow to access new physics information on the density probabilities for quark helicity configurations when a proton emits a photon. For instance the ratio
\begin{equation}
\frac {| V_{1  {\mathcal{E}}}^{ \gamma p}| ^2+ | A_{1  {\mathcal{E}}}^{ \gamma p}| ^2}{| T_{1  {\mathcal{E}}}^{ \gamma p}| ^2+ | T_{2  {\mathcal{E}}}^{\gamma p}| ^2}
\end{equation}
    gives access to the ratio $\frac{D_{h({u_1}) =-h({u_2})}(x_i)}
    {D_{h({u_1}) =+h({u_2})}(x_i)}$
where $D_{h({u_1}) =-h({u_2})}(x_i)$ (resp. $D_{h({u_1}) =+h({u_2})}(x_i)$) denotes the probability density that the helicities of the two $u$-quarks are opposite (resp. equal), which may be interpreted as the answer to the question: {``Is the nucleon brighter when $u$-quarks have equal helicities?''}.

Counting the $\Delta_T$ factors in the Dirac structures accompanying the TDAs allows to get access to the  orbital angular momentum contribution to nucleon spin. For instance, since  the spinor structure attached to $T_{4 {\mathcal{E}}}^{ \gamma p}$ contains $\Delta_T^{3}$, which implies  $L=3$, the
 $T_{4 {\mathcal{E}}}^{ \gamma p}$ TDA measures the helicity amplitude $T_{\downarrow \downarrow,\downarrow}^{\uparrow, \, \downarrow}$, and the ratio
\begin{equation}
\frac {| T_{4 {\mathcal{E}}}^{ \gamma p}| ^2}{| V_{1  {\mathcal{E}}}^{ \gamma p}| ^2+ | A_{1  {\mathcal{E}}}^{ \gamma p}| ^2+| T_{1  {\mathcal{E}}}^{ \gamma p}| ^2+ | T_{2  {\mathcal{E}}}^{ \gamma p}| ^2}
\end{equation}
 measures the ratio of density probabilities for three units \textit{versus} zero unit of orbital angular momentum between the three quarks when a proton emits a photon.

\subsubsection{Deuteron-to-nucleon and other nuclear {{TDA}}s}
\label{SubSec_Def_deutronN_TDAs}
\mbox

The TDA framework can be adapted as a tool for nuclear physics
to describe hard exclusive reactions on nuclei  dissociated by kicking out one unit of baryonic number.
From the theoretical viewpoint the simplest and best
known nuclear system is the deuteron. Therefore,
it has been usually considered as the most appropriate starting point to investigate
hard exclusive processes off nuclei~\cite{Berger:2001zb,Kirchner:2003wt,Guzey:2003jh,Cano:2003ju}.

\begin{figure}[H]
\begin{center}
\includegraphics[width=0.35\textwidth]{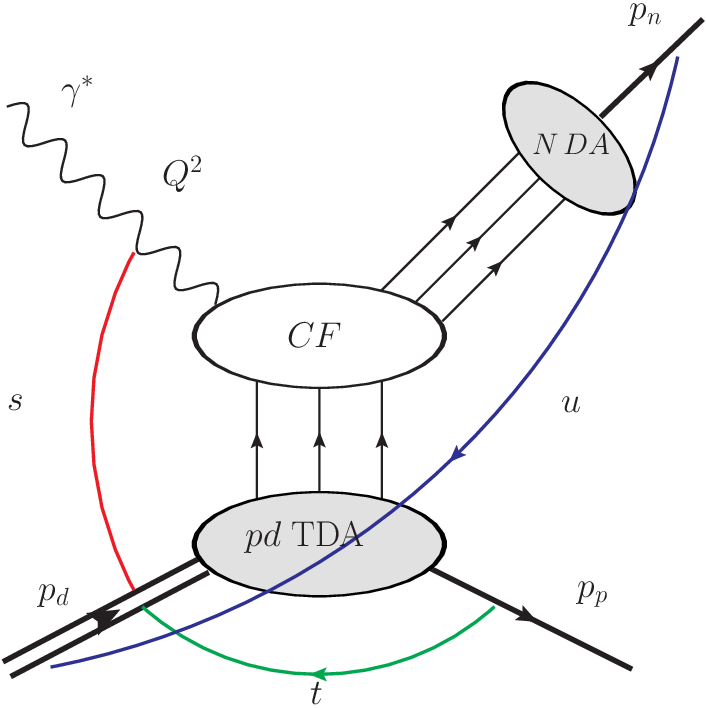} \ \ \ \ \
\includegraphics[width=0.35\textwidth]{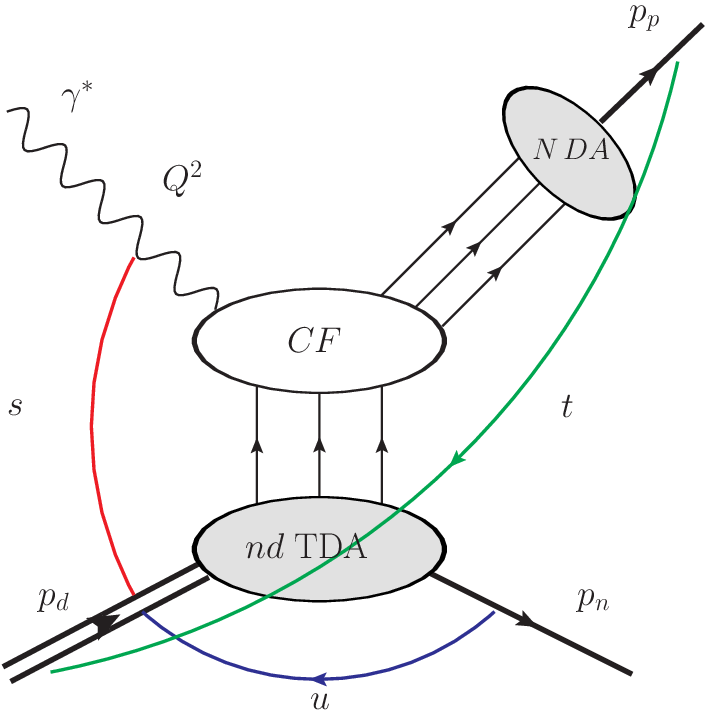}
\end{center}
     \caption{ Factorization mechanism for the deuteron deep electrodissociation with the
 baryon number $B=+1$ transfer in the cross channel.
 }
\label{Fig_Deutron_Diss}
\end{figure}

Deuteron-to-nucleon TDAs may be introduced to describe the hard
electrodissociation of a deuteron
\begin{equation}
e(k,\lambda_e)+d(p_d,s_d) \to
\left(\gamma^{*}(q, \lambda_\gamma)+d(p_d,s_d) \right) + e(k',\lambda'_e) \to e(k',\lambda'_e)+
p(p_p,s_p)+ n(p_n,s_n).
\label{Deut_Diss_reaction}
\end{equation}
in the generalized Bjorken limit
(large $Q^2=-q^2$ and $W^2=(p_d+q)^2$),
fixed
$x_B$
and small invariant momentum transfer between one of the
final nucleons and the target deuteron).
There are two regimes with either final state proton or neutron
produced in the near-backward direction in $\gamma^{*} d$ CMS
(see  Fig.~\ref{Fig_Deutron_Diss}):
\begin{equation}
| t| = | (p_p-p_d)^2|   \ll Q^2, \, W^2;
\end{equation}
and
\begin{equation}
| u| = | (p_n-p_d)^2|  \ll Q^2, \, W^2.
\label{Bkw_regime_d_diss}
\end{equation}

For definiteness below we consider the
kinematical
regime (\ref{Bkw_regime_d_diss}).
This requires the knowledge of deuteron-to-neutron $uud$ TDA.
Since deuteron is a spin-$1$ state,
for the leading twist-$3$
nucleon-to-deuteron  ($dN$) TDAs
it is natural to employ the same parametrization as for nucleon-to-vector meson TDAs
(see  Sec.~\ref{SubSec_Def_NV_TDAs}). This implies the use of the same set of
 leading twist Dirac structures
(\ref{Dirac_v_NV})--(\ref{Dirac_t_NV}) (with obvious modifications).
Each of the $24$ leading twist-$3$ $dN$ TDAs is
a function of three longitudinal momentum fractions $x_1$, $x_2$, $x_3$, skewness parameter $\xi$, and the momentum transfer squared $\Delta^2$, as well as the factorization scale $\mu$.

However, to describe the reaction
(\ref{Deut_Diss_reaction})
we have to employ deuteron-to-nucleon ($Nd$) TDAs defined through
the conjugated matrix element of the three-quark light-cone operator:
\begin{eqnarray}
 &&
4 {\mathcal{F}} \langle N^n(p_n, s_n)|
\bar{u}_\rho(\lambda_1n)
\bar{u}_\tau(\lambda_2n)
\bar{d}_\chi(\lambda_3n)
| d(p_d,s_d) \rangle
\nonumber \\ &&
=
\delta(x_1+x_2+x_3-2\xi)  \times m_N \Bigl[
\sum_{ {\Upsilon= 1 {\mathcal{E}}, 1T, 1n, \atop  2 {\mathcal{E}}, 2T, 2n  }} (v^{Nd}_\Upsilon)_{\rho \tau, \, \chi} V_{\Upsilon}^{Nd}(x_1,x_2,x_3, \xi, \Delta^2; \, \mu^2)
\nonumber \\ &&
+\sum_{{\Upsilon= 1 {\mathcal{E}}, 1T, 1n, \atop  2 {\mathcal{E}}, 2T, 2n  }} (a^{Nd}_\Upsilon)_{\rho \tau, \, \chi} A_{\Upsilon}^{Nd}(x_1,x_2,x_3, \xi, \Delta^2; \, \mu^2)
+
\sum_{{\Upsilon= 1 {\mathcal{E}}, 1T, 1n,   2{\mathcal{E}}, 2T, 2n, \atop  3 {\mathcal{E}}, 3T, 3n,  4 {\mathcal{E}}, 4T, 4n }} (t^{Nd}_\Upsilon)_{\rho \tau, \, \chi} T_{\Upsilon}^{Nd}(x_1,x_2,x_3, \xi, \Delta^2; \, \mu^2)
\Bigr].
\label{Param_ND_TDAs}
\end{eqnarray}

To express the Dirac structures
$s_{\rho \tau, \,\chi}^{(Nd)}= \{v_{\rho \tau, \,\chi}^{( Nd)}, \, a_{\rho \tau, \,\chi}^{(N d)}, \, t_{\rho \tau, \,\chi}^{(Nd)} \}$
occurring in the parametrization
$Nd$
TDAs through those of the parametrization of $dN$ TDAs
(\ref{Param_ND_TDAs})
we apply the Dirac conjugation  procedure described in  Appendix~A of Ref.~\cite{Pire:2016gut}
to the set of  Dirac structures
(\ref{Dirac_v_NV})--(\ref{Dirac_t_NV}).
This includes the complex conjugation, convolution with $\gamma_0$ matrices in the appropriate
spinor indices
\begin{equation}
s_{\rho \tau, \,\chi}^{( Nd)}=
\left(\gamma_{0}^{T}\right)_{\tau \tau'}
\left[s_{\rho' \tau', \,\chi'}^{dN} \right]^{\dagger}
\left(\gamma_{0}\right)_{\rho' \rho}
\left(\gamma_{0}\right)_{\chi' \chi};
\end{equation}
and subsequent substitution  $-\Delta \to \Delta \equiv p_n-p_d$.

It results in the following set of  Dirac structures:
\begin{align}
(v_{1 {{\mathcal{E}} }}^{N d})_{\rho \tau, \, \chi}
&= (C \hat{p})_{\rho \tau}
\bigl(
\bar{U}^+ \hat{{\mathcal{E}}} \gamma_5 \bigr)_\chi;
&(v_{2 {\mathcal{E}}}^{N d})_{\rho \tau, \, \chi}
&= -m_N^{-1}   (C\hat{p})_{\rho \tau}
\bigl(\bar{U}^+ \sigma^{\Delta_T {\mathcal{E}} }
\gamma^5
\bigr)_\chi;
\nn             \\
(v_{1 T}^{Nd})_{\rho \tau, \, \chi}
&=m_N^{-1}
({{\mathcal{E}}}
\cdot \Delta_T)
(C \hat{p})_{\rho \tau} \bigl( \bar{U}^+ \gamma^5 \bigr)_\chi;
&(v_{2 T}^{Nd})_{\rho \tau, \, \chi} & =  -m_N^{-2}
({{\mathcal{E}}} \cdot \Delta_T)
(C \hat{p})_{\rho \tau}
\bigl( \bar{U}^+ \hat{\Delta}_T \gamma^5  \bigr)_\chi;
\nn \\
(v_{1 n}^{Nd})_{\rho \tau, \, \chi}
&= -m_N  ({{\mathcal{E}}}
\cdot n) (C \hat{p})_{\rho \tau}
\bigl( \bar{U}^+ \gamma^5 \bigr)_\chi;
&(v_{2 n}^{Nd})_{\rho \tau, \, \chi} & =
({{\mathcal{E}}} \cdot n)
(C\hat{p})_{\rho \tau}
\bigl(
\bar{U}^+ \hat{\Delta}_T
\gamma^5
\bigr)_\chi;
\label{Dirac_v_dN}
\end{align}
\begin{align}
(a_{1 {\mathcal{E}}}^{Nd})_{\rho \tau, \, \chi}
&= ( C \gamma^5 \hat{p} )_{\rho \tau} \bigl(  \bar{U}^+ \hat{{\mathcal{E}}}  \bigr)_\chi;
&(a_{2 {\mathcal{E}}}^{ N d})_{\rho \tau, \, \chi}
&=  -m_N^{-1}   ( C \gamma^5 \hat{p})_{\rho \tau} \bigl( \bar{U}^+  \sigma^{\Delta_T {\mathcal{E}} } \bigr)_\chi;
\nonumber \\
(a_{1 T}^{N d})_{\rho \tau, \, \chi}
&=-m_N^{-1}
({{\mathcal{E}}}
\cdot \Delta_T)
( C \gamma^5 \hat{p})_{\rho \tau} \bigl(  \bar{U}^+ \bigr)_\chi;
&(a_{2 T}^{Nd})_{\rho \tau, \, \chi} &=
m_N^{-2}
({{\mathcal{E}}} \cdot \Delta_T)
( C \gamma^5 \hat{p})_{\rho \tau}  \bigl( \bar{U}^+ \hat{\Delta}_T  \bigr)_\chi;
\nonumber \\
(a_{1 n}^{Nd})_{\rho \tau, \, \chi}& = m_N  ({{\mathcal{E}}}
\cdot n) (  C \gamma^5 \hat{p})_{\rho \tau} \bigl(  \bar{U}^+ \bigr)_\chi;  &
(a_{2 n}^{Nd})_{\rho \tau, \, \chi} & =-
({{\mathcal{E}}} \cdot n)
(   C \gamma^5  \hat{p}  )_{\rho \tau}  \bigl( \bar{U}^+ \hat{\Delta}_T  \bigr)_\chi;
\label{Dirac_a_dN}
\end{align}
\begin{align}
(t_{1 {\mathcal{E}}}^{Nd})_{\rho \tau, \, \chi} & =-
(C \sigma_{p \lambda} )_{\rho \tau} (  \bar{U}^+ \sigma^{\lambda {\mathcal{E}}}  \gamma_5)_\chi;
&(t_{2 {\mathcal{E}}}^{Nd})_{\rho \tau, \, \chi}&=(C \sigma_{p {\mathcal{E}}} )_{\rho \tau}
( \bar{U}^+ \gamma_5 )_\chi;
\nonumber \\
(t_{3 {\mathcal{E}}}^{Nd})_{\rho \tau, \, \chi}
&=m_N^{-1} (C \sigma_{p \Delta_T} )_{\rho \tau} ( \bar{U}^+ \hat{{\mathcal{E}}} \gamma_5)_\chi;
&(t_{4 {\mathcal{E}}}^{ Nd})_{\rho \tau, \, \chi}& = -m_N^{-1} (C \sigma_{p {\mathcal{E}}} )_{\rho \tau} (\bar{U}^+ \hat{\Delta}_T \gamma_5)_\chi;
\nonumber \\
(t_{1 T}^{Nd})_{\rho \tau, \, \chi}& = -m_N^{-1}  ({\mathcal{E}} \cdot \Delta_T)
(C \sigma_{p \lambda} )_{\rho \tau} (\bar{U}^+ \gamma^\lambda  \gamma_5 )_\chi;
&(t_{2 T}^{Nd})_{\rho \tau, \, \chi}&= m_N^{-2}  ({\mathcal{E}} \cdot \Delta_T)
(C \sigma_{p \lambda} )_{\rho \tau} (\bar{U}^+ \sigma^{\lambda \Delta_T} \gamma_5)_\chi;
\nonumber \\
(t_{3 T}^{Nd})_{\rho \tau, \, \chi}&= m_N^{-2}  ({\mathcal{E}} \cdot \Delta_T)
(C \sigma_{p \Delta_T} )_{\rho \tau} (\bar{U}^+ \gamma_5 )_\chi; &
(t_{4 T}^{ Nd})_{\rho \tau, \, \chi} &=-m_N^{-3} ({\mathcal{E}} \cdot \Delta_T)
(C \sigma_{p \Delta_T} )_{\rho \tau} ( \bar{U}^+ \hat{\Delta}_T \gamma_5)_\chi;
\nonumber \\
(t_{1 n}^{ Nd})_{\rho \tau, \, \chi}&= m_N   ({\mathcal{E}} \cdot n)
(C \sigma_{p \lambda} )_{\rho \tau} (\bar{U}^+  \gamma^\lambda  \gamma_5)_\chi;&
(t_{2 n}^{ Nd})_{\rho \tau, \, \chi}&= ({\mathcal{E}} \cdot n) (C \sigma_{p \lambda} )_{\rho \tau} (\bar{U}^+ \sigma^{\lambda \Delta_T}  \gamma_5 )_\chi;
\nonumber \\
(t_{3 n}^{ Nd})_{\rho \tau, \, \chi}& =   -({\mathcal{E}} \cdot n) (C \sigma_{p \Delta_T} )_{\rho \tau} (  \bar{U}^+ \gamma_5)_\chi;&
(t_{4 n}^{ Nd})_{\rho \tau, \, \chi}&=m_N^{-1}  ({\mathcal{E}} \cdot n)   (C \sigma_{p \Delta_T} )_{\rho \tau} (\bar{U}^+ \hat{\Delta}_T \gamma_5)_\chi.
\label{Dirac_t_dN}
\end{align}
Here ${\mathcal{E}} \equiv {\mathcal{E}}(p_d,s_d)$ stands for the polarization vector of the
target deuteron and
$\bar{U}^{+} \equiv \bar{U}\left(p_{N}\right) \hat{n} \hat{p}$
stands for the large component of the Dirac spinor.

\subsection{Definition of $N {\mathcal{M}}$
and
$N  \gamma$
GDAs}
\label{SubSec_Definition_GDAs}
\mbox

Nucleon-to-meson (and nucleon-to-photon) TDAs are
related to nucleon--meson  (nucleon--photon) Generalized
Distribution Amplitudes (GDAs) defined by the matrix element
of the same three-quark light-cone operator between
${\mathcal{M}} N$ (respectively $\gamma N$) state and the vacuum.
A similar form of correspondence was established \textit{e.g.} between
pion GPD and $2 \pi$  GDAs~\cite{Diehl:1998dk,Polyakov:1998ze,Polyakov:1999gs,Diehl:2000uv}.

Nucleon--meson GDAs
occur in a factorized description of  deep inelastic meson electroproduction near threshold~\cite{Pobylitsa:2001cz,Braun:2006td}
\begin{equation}
 \gamma^{*}(\tilde{q},\lambda_\gamma) +N(p_N,s_N) \to {\mathcal{M}} (\tilde{p}_{\mathcal{M}})+ N(p'_N,s'_N),
\label{Cross_ch_react}
\end{equation}
in the kinematical regime where $Q^2 \gg W^2$.

For example, the proton-$\pi^0$ $uud$ GDA is defined by the Fourier
transform of the  matrix element of the
light-cone operator (\ref{Def_O_uud_operator}) between the
proton-$\pi^0$ state and the vacuum:
\begin{eqnarray}
 &&
4 (\tilde{P}  \cdot n)^3 \int \left[ \prod_{j=1}^3 \frac{d \lambda_j}{2 \pi}\right]
e^{i \sum_{k=1}^3 y_k \lambda_k (\tilde{P} \cdot n)}
 \langle     0| \,
\widehat{O}_{\rho \tau \chi}^{\,uud}(\lambda_1n,\lambda_2n,\lambda_3n)
\,|  \pi^0(\tilde{p}_\pi) N^p(p_N,s_N) \rangle
\nonumber \\ &&
\equiv 4 {\tilde{ F}} \langle     0 | \,
\widehat{O}_{\rho \tau \chi}^{\,uud}(\lambda_1n,\lambda_2n,\lambda_3n)
\,| \pi^0(\tilde{p}_\pi) N^p(p_N,s_N) \rangle  \nonumber \\ &&
=
\delta( {y}_1+ {y}_2+ {y}_3-1) i \frac{f_N}{f_\pi}
\Bigl[
\sum_{\Upsilon= 1, 2} (\tilde{v}^{\pi N}_\Upsilon)_{\rho \tau, \, \chi} \tilde{V}_{\Upsilon}^{\pi^0 N}(y_1,y_2,y_3, \zeta, \tilde{P}^2; \, \mu^2)
\nonumber \\ &&
+\sum_{\Upsilon= 1,\, 2 } (\tilde{a}^{\pi N}_\Upsilon)_{\rho \tau, \, \chi} \tilde{A}_{\Upsilon}^{\pi N} (y_1,y_2,y_3, \zeta, \tilde{P}^2; \, \mu^2)
+
\sum_{\Upsilon= 1,2,3,4} (\tilde{t}^{\pi N}_\Upsilon)_{\rho \tau, \, \chi} \tilde{T}_{\Upsilon}^{\pi N} (y_1,y_2,y_3, \zeta, \tilde{P}^2; \, \mu^2)
\Bigr]\,.
 \label{Param_GDAs}
\end{eqnarray}
Here we introduced the total momentum of the $\pi N$ state $\tilde{P}=p_N+\tilde{p}_\pi$;
$\tilde{\Delta}=p_N-\tilde{p}_\pi$; and $\mu$ is the factorization scale.
The variable $\zeta= \frac{p_N \cdot n}{\tilde{P} \cdot n}$
characterizes the
distribution of the plus momenta inside the
$\pi N$
system; $y_{1,2,3}$ are the light-cone momentum fractions. We choose the Dirac structures
$(\tilde{s}^{\pi N}_\Upsilon)_{\rho \tau,\, \chi}= \{ \tilde{v}^{\pi N}_{1,\,2}, \,
\tilde{a}^{\pi N}_{1,\,2}, \, \tilde{t}^{\,\pi N}_{1,\,2,\,3,\,4}
 \}_{\rho \tau,\, \chi}$
as being given by the crossing of the
$(s^{\pi N}_\Upsilon)_{\rho \tau,\, \chi}$ defined in
(\ref{Def_DirStr_piN_DeltaT}).

The explicit form of the crossing transformation which relates
$\pi N$ TDAs and GDAs reads
\begin{equation}
P' \leftrightarrow \Delta; \ \ \ \Delta' \leftrightarrow 2 P.
\label{Crossing_TDA_GDA1}
\end{equation}
It also implies an analytic continuation in the appropriate kinematical variables:
\begin{equation}
P^{\prime 2} \leftrightarrow \Delta^2\,; \ \ \ 2\zeta+1 \leftrightarrow \frac{1}{\xi}\,; \ \ \  y_i \leftrightarrow \frac{x_i}{2 \xi}\,, \ \ i=\{1,\,2,\,3\}\,.
\label{Crossing_TDA_GDA2}
\end{equation}

\subsection{Support properties of
$ {\mathcal{M}}N$
and
$\gamma N$
TDAs}
\label{SubSec_Support}
\mbox

Let us consider  the support properties of
$ {\mathcal{M}}N$
and
$ \gamma N$
TDAs in the longitudinal momentum fraction variables.
The support properties of TDAs naturally reflect the redundancy of
our description: we deal with $3$ momentum variables
$x_i$ $i=1,\,2,\,3$
that are subject to the longitudinal momentum conservation constraint
$\sum_i x_i= 2 \xi$.

For definiteness here
we consider the case of $\pi N$ TDAs, however our exposition obviously
does not depend on the nature of the final state meson/photon.
A consistent way to address the support properties of
$\pi N$ TDAs would be to adopt the approach of
Ref.~\cite{Radyushkin:1997ki}
based on the analysis of spectral properties of
amplitudes of a toy $\phi^3$-type model with the help of the $\alpha$-representation
technique
\cite{Radyushkin:1983ea,Radyushkin:1983wh}.
The results within a toy $\phi^3$-type model are then straightforwardly generalized to the QCD case. In
Ref.~\cite{Radyushkin:1997ki}
this method was successfully applied to work out the
support properties of forward parton distributions, GPDs and double distributions (DDs).
However, this method turns out to be tedious and lengthy. Therefore, following Ref.~\cite{Pire:2010if},
we work out the support domain of TDAs from the following symmetry considerations
and consistency requirements.
\begin{enumerate}
 \item The complete support domain of
 $\pi N$ TDAs must be symmetric in
 $x_i$s.
\item In the limiting case
 $\xi=1$
 this domain reduces to that of the nucleon DA. In the barycentric coordinates
 the support domain of the nucleon DA in the longitudinal momentum
 fractions
 $y_i$ ($i=1,\,2,\,3$)
 is an equilateral triangle defined by the requirements
 $0 \le y_i \le 1$
 and the longitudinal momentum conservation
 condition
 $\sum_i y_i=1$.
\item For any
 $x_i$
 set to zero we must recover the usual domain of definition of GPDs in the two remaining  variables.
\end{enumerate}
A cross check for our ``educated guess'' is provided by the spectral representation
for $\pi N$ TDAs in terms of quadruple distributions
(see  Sec.~\ref{SubSec_Spectral}) which naturally generalizes
the familiar spectral representation of GPDs in terms of double distributions.

\begin{figure}[H]
\begin{center}
\includegraphics[width=0.8\textwidth]{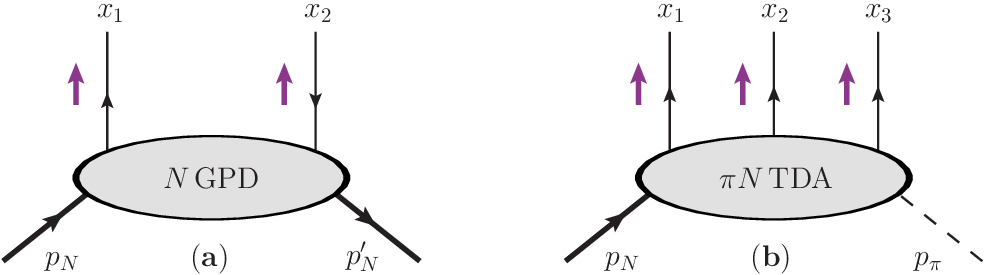}
\end{center}
     \caption{Longitudinal momentum flow in the ERBL regime for
GPDs (a) and $\pi N$ TDAs (b). Small arrows show the direction of flow of positive
momentum. Arrow on nucleon and quark/antiquark lines depict the flow of the baryon charge.}
\label{Fig_Momentum_flow_GPD_TDA}
\end{figure}

Let us first consider the GPD case (see
 Fig.~\ref{Fig_Momentum_flow_GPD_TDA}a
for the longitudinal momentum flow in the
Efremov--Radyushkin--Brodsky--Lepage regime) in terms suitable for further
generalization to TDAs.
Let
$x_1$
and
$x_2$
be the fractions
(defined with respect to average nucleon momentum
$P=\frac{p_1+p_2 }{2}$)
of the light-cone momentum carried by an active
quark and antiquark inside a nucleon. The longitudinal
momentum conservation implies\footnote{In our discussion of GPDs the variable $\xi$ refers to
the usual skewness variable defined with respect to the $t$-channel
longitudinal momentum transfer. }
$x_1+x_2=2 \xi$
with
$\xi \ge 0$.
The conventional GPD longitudinal momentum fraction  variable
$x$
is  defined as
\begin{equation}
x=\frac{x_1-x_2}{2}\,.
\label{Def_x_GPD}
\end{equation}

\bi
\item In the so-called Efremov--Radyushkin--Brodsky--Lepage (ERBL) region both
$x_1$
and
$x_2$
are positive:
$x_1, \,x_2 \in \,[0,\,2 \xi]$.
In terms of the $x$ variable
(\ref{Def_x_GPD})
it corresponds to the central region
$x \in \,[-\xi,\xi]$.
\item In the so-called
Dokshitzer--Gribov--Lipatov--Altarelli--Parisi
(DGLAP) region either
$x_1$
is positive
$x_1 \in \,[2\xi,\,1+\xi]$
and $x_2$ is negative
$x_2 \in \,[-1+\xi,0]$ or
vice versa
($x_1$ is negative
$x_1 \in \,[-1+\xi,0]$ and
$x_2$
positive
$x_2 \in \,[2\xi,\,1+\xi]$).
These two  DGLAP domains result in the
outer regions in terms of $x$
(\ref{Def_x_GPD})
$x \in [\xi,\,1]$
and
$x \in [-1,\,-\xi]$
respectively.
\ei
Thus, the support properties of GPDs in terms of the momentum fraction
variables $x_{1,2}$ can be summarized as
\begin{equation}
-1+\xi \le x_{1,2} \le 1+\xi; \ \ x_1+x_2=2 \xi.
\label{Support_GPD_x12}
\end{equation}

Now let us turn to the case of
$\pi N$
TDAs, see  Fig.~\ref{Fig_Momentum_flow_GPD_TDA}b.
The longitudinal momentum fraction variables
$x_1$, $x_2$
and
$x_3$
are defined with respect to the average hadron momentum
$P=\frac{p_1+p_\pi }{2}$
(\ref{Def_P_TDA})
and satisfy the momentum conservation constraint
$x_1+x_2+x_3=2 \xi$,
with the skewness variable
(\ref{Def_xi})
$\xi \ge 0$.
A natural
generalization of
(\ref{Support_GPD_x12}) for the three variables
satisfying the necessary symmetry and consistency conditions reads
\begin{equation}
-1+\xi \le x_{i} \le 1+\xi \ \ (i=1,\,2,\,3); \ \ x_1+x_2+x_3=2 \xi.
\label{TDA_support_xi}
\end{equation}
Contrary to the GPD case, the shape of the complete support of TDAs depends on
$\xi$.
Note that in the limit $\xi \to 0$, we recover Jaffe's results for the support property of $3$-particle parton distribution functions of higher twist
established in Ref.~\cite{Jaffe:1983hp}.

A convenient way to depict the support domain of
$\pi N$
TDAs
(\ref{TDA_support_xi})
is to employ the barycentric coordinates. The values of momentum fractions $x_i$
are specified by distances from a point on the plane to three sides of the equilateral triangle. The height of this
equilateral triangle is defined by
the momentum conservation constraint $x_1+x_2+x_3=2 \xi$.

\begin{figure}[H]
\begin{center}
\includegraphics[width=0.5\textwidth]{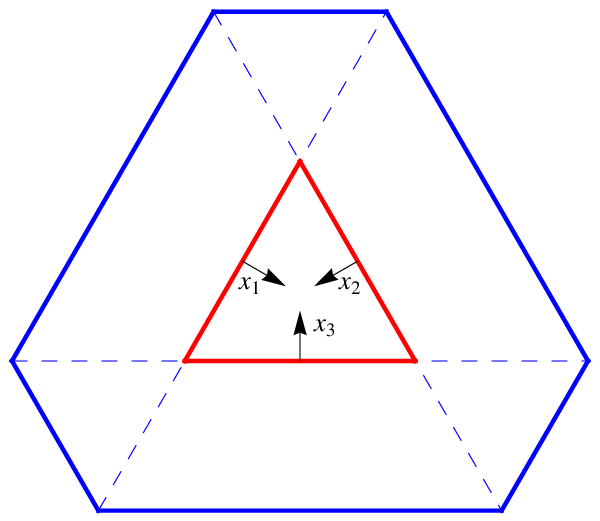}
\end{center}
     \caption{Support domain (\ref{TDA_support_xi}) of $\pi N$ TDAs in the longitudinal
     momentum fractions $x_i$ $i=1,\,2,\,3$
      in the
barycentric coordinates for $\xi=0.4$..}
\label{Fig_Support_TDAs_Barycentric}
\end{figure}

\bi
\item First of all we identify the ERBL-like domain,
in which the three longitudinal momentum fractions carried by the three quarks
are positive. In the barycentric coordinates, this ERBL-like region corresponds
to the interior of the equilateral triangle  with the height
$2 \xi$ bounded by the lines $x_i=0$ (see  Fig.~\ref{Fig_Support_TDAs_Barycentric}).
In  Sec.~\ref{SubSec_Evolution} we will show that the evolution properties of TDAs
within this domain are indeed governed by the ERBL-type evolution equations.

\item The DGLAP-like domains are bounded  by the lines
\begin{equation}
x_i=-1+ \xi\,; \ \  x_i=0 \,; \ \   x_i=1+\xi\,.
\label{Support_TDA_DGLAP}
\end{equation}
Three small equilateral triangles
in  Fig.~\ref{Fig_Support_TDAs_Barycentric}
correspond to DGLAP-like type I domains, where  only
one longitudinal momentum fraction is positive while the two other ones are negative.
Three trapezoid domains correspond to DGLAP-like type II, where two longitudinal momentum fractions are positive and one is negative.
\ei

\begin{figure}[H]
\begin{center}
\includegraphics[width=0.45\textwidth]{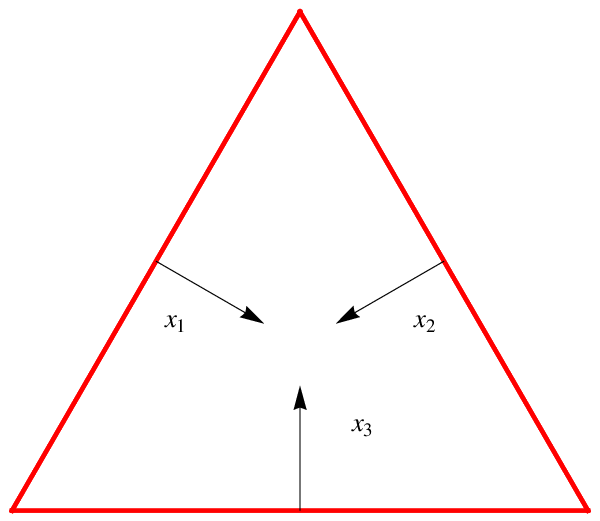} \ \ \
\includegraphics[width=0.45\textwidth]{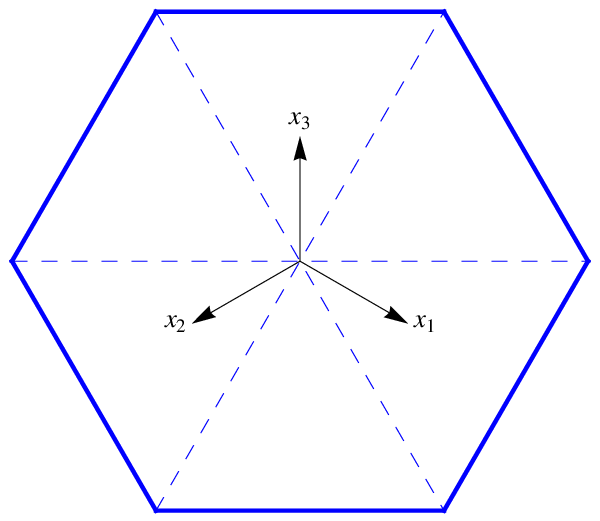}
\end{center}
     \caption{Physical domains for $\pi N$ TDAs in the
barycentric coordinates in the two limiting cases: $\xi=1$ (left) and
$\xi=0$ (right). Arrows show the positive directions for momentum fractions $x_i$ on the barycentric plane.
Note that in the case $\xi=0$ the equilateral triangle corresponding to the ERBL-like domain shrinks to a point. }
\label{Fig_Support_TDAs_Barycentric_special_cases}
\end{figure}

In the limiting case
$\xi=1$
the support domain of
$\pi N$ TDAs reduces
to a single ERBL-like domain, see the left panel of  Fig.~\ref{Fig_Support_TDAs_Barycentric_special_cases}.
The familiar support of nucleon DAs can be obtained by
a simple rescaling of the momentum fraction variables
\begin{equation}
x_i \to \frac{x_i}{2} \equiv y_i.
\end{equation}

In the second limiting case $\xi=0$ the ERBL-like domain shrinks to
a single point and we are left with six DGLAP-like type I and II domains
that form a regular hexagon in the barycentric coordinates, see the right panel of  Fig.~\ref{Fig_Support_TDAs_Barycentric_special_cases}.

In many cases it turns out to be convenient to switch to two independent momentum fraction variables instead
of $x_i$ that are subject to the
momentum conservation constraint
$\sum_i x_i=2\xi$. A natural choice of independent variables is given
by the so-called quark--diquark coordinates
$(w_i; \, v_i)$
(there exist $3$ equivalent choices
$i = 1,\, 2,\, 3$ of
quark--diquark coordinates, depending on which pair of quark momenta is selected to constitute the momentum of a diquark):
\begin{equation}
w_{i}=x_{i}-\xi; \ \ \  v_{i}=\frac{1}{2} \sum_{k, l=1}^{3}
\varepsilon_{i k l} x_{k},
\label{Def_qDq_coord}
\end{equation}
where $\epsilon_{ikl}$ is the antisymmetric tensor ($\epsilon_{123}=1$).

Within these coordinates, the support of
nucleon-to-meson TDAs can be parameterized as
\begin{equation}
-1 \leq w_{i} \leq 1 ; \quad-1+| \xi-\xi_{i}^{\prime}|   \leq v_{i} \leq 1-| \xi-\xi_{i}^{\prime}|  ,
\label{Support_TDA_wv}
\end{equation}
where
\begin{equation}
\xi_{i}^{\prime} \equiv \frac{\xi-w_{i}}{2}.
\label{Def_xip_i}
\end{equation}
The variables
$\xi'_i$
(\ref{Def_xip_i})
characterize the longitudinal momentum fraction carried by a corresponding diquark.
For example, if we choose the first and the second quarks to
form a diquark
\begin{equation}
w_3= x_3-\xi; \ \ \ x_1+x_2= 2 \xi_3'; \ \ \ x_1-x_2=2 v_3.
\label{qDq_3}
\end{equation}

\begin{figure}[H]
\begin{center}
\includegraphics[width=0.6\textwidth]{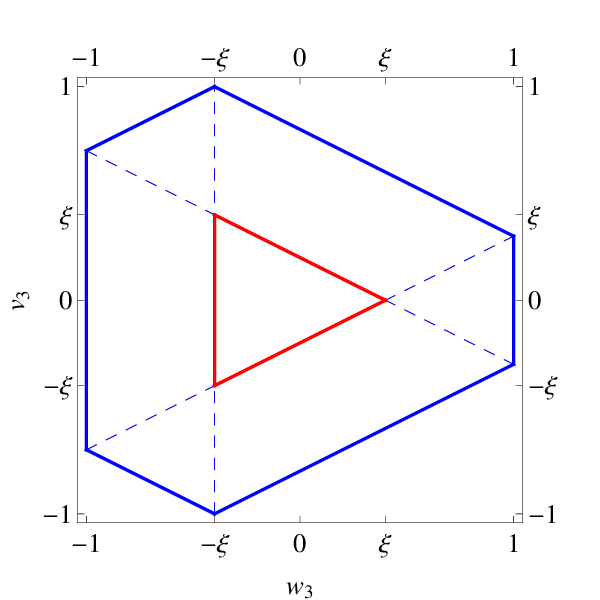}
\end{center}
     \caption{Support domain (\ref{Support_TDA_wv}) of $\pi N$ TDAs in the quark--diquark
     coordinates $w_3$, $v_3$ for $\xi=0.4$. Dashed lines show three cross-over trajectories
     $w_3=-\xi$; $v= \pm \xi'_3$ corresponding to $x_i=0$, $i=\{1,\,2,\,3\}$.}
\label{Fig_Support_TDAs_qDq}
\end{figure}

The support domain (\ref{Support_TDA_wv}) of $\pi N$ TDAs in the quark--diquark
coordinates $w_3$, $v_3$ for $\xi=0.4$ is presented in  Fig.~\ref{Fig_Support_TDAs_qDq}.
Note that the ERBL-like domain (the isosceles triangle in  Fig.~\ref{Fig_Support_TDAs_qDq}) turns to be separated from the DGLAP-like domains
by the cross-over trajectories
\begin{equation}
w_3=-\xi; \ \ \ v_3= \pm \xi_3'.
\label{Cross_over_trj_TDA}
\end{equation}

The quark--diquark coordinates can be seen as a natural three-body generalization of
the GPD longitudinal momentum fraction variable $x$ (\ref{Def_x_GPD}).
In  Sec.~\ref{SubSec_Bkw_meson_ampl} we show that the singularities of the  coefficient functions
of leading order backward meson electroproduction amplitudes lie on the cross-over
trajectories (\ref{Cross_over_trj_TDA}).
This generalizes the property of the elementary LO DVCS/HMP coefficient function that
is singular at $x= \pm \xi$.

In what follows, we will often present TDAs
$H_s^{{\mathcal{M}} N}=\{V_\Upsilon^{{\mathcal{M}} N}, \, A_\Upsilon^{{\mathcal{M}} N}, \,
T_\Upsilon^{{\mathcal{M}} N}  \}$
as functions of a selected pair
of quark--diquark coordinates instead of the longitudinal momentum fractions $x_{1,2,3}$:
\begin{equation}
H_s^{{\mathcal{M}} N}(x_1, \, x_2, \, x_3=2\xi-x_1-x_2, \, \xi, \, \Delta^2)  \to
H_s^{{\mathcal{M}} N}(w_i, \, v_i, \, \xi, \, \Delta^2).
\end{equation}
Note that three equivalent sets of variables can be employed for the same function
$H_s$.
This reflects the underlying symmetry of our description in terms of
the longitudinal momentum fractions
$x_{1,2,3}$.

\newpage

\subsection{Polynomiality property}
\label{SubSec_Polynomiality}
\mbox

The Mellin moments in the longitudinal momentum fraction $x$ of GPDs
are of particular importance since they are related
to the form factors of local derivative operators. These objects
often admit a clear physical interpretation and also can be studied within
the lattice approach to QCD. The remarkable consequence of the Lorentz
invariance of the underlying quantum field theory is the polynomiality property of GPDs:
the $x$- Mellin moments of  GPDs are polynomials
of a definite power in the skewness variable.

Analogously to GPDs, nucleon-to-meson (and nucleon-to-photon)
TDAs satisfy the polynomiality property.
We introduce the compact notations for the $x_i$-Mellin moments of TDAs:
\begin{eqnarray}
 &&
\langle x_1^{n_1} x_2^{n_2} x_3^{n_3} {\rm TDA} \rangle \nonumber \\
&& =
\int_{-1+\xi}^{1+\xi}  dx_1 \int_{-1+\xi}^{1+\xi}  dx_2 \int_{-1+\xi}^{1+\xi}  dx_3
\delta(x_1+x_2+x_3-2\xi)
 x_1^{n_1} x_2^{n_2} x_3^{n_3}
{\rm TDA}(x_1,x_2,x_3,\xi,\Delta^2)\,.
\label{Def_Mellin_moments_TDAs}
\end{eqnarray}
Following Ref.~\cite{Pire:2011xv}, we now argue that the Mellin moments
(\ref{Def_Mellin_moments_TDAs})
of the leading twist nucleon-to-meson (nucleon-to-photon) TDAs
are polynomials in the skewness variable
$\xi$
(\ref{Def_xi}).
It turns out that
in the case of TDAs the situation is somewhat tricky
due to the ambiguity of the choice of a set of  Dirac structures
in the parametrization of TDAs. The set of  Dirac structures
presented for $\pi N$, $VN$ and $\gamma N$ TDAs in
Secs.~\ref{SubSubSec_Def_piN_TDAs}--\ref{SubSec_Def_Ngamma_TDAs}
is convenient for the phenomenological applications since it
allows a clear distinction of the subset of TDAs relevant in the $\Delta_T \to 0$
limit. However, as pointed out in Ref.~\cite{Pire:2011xv}, the polynomiality
of the Mellin moments is spoiled by the factors
$\frac{1}{1+\xi}$
that have purely kinematical origin. The manifestation of these singularities
has much in common with the well known problem of construction of
a set of invariant amplitudes free of kinematical singularities
for a given scattering process (see \textit{e.g.} Appendix~II of Chapter~I of~\cite{Alfaro_red_book}).
In order to ensure the polynomiality property in an explicit form, one has
to switch to a parametrization of TDAs with a set of  Dirac structures
that do not bring kinematical singularities to the Mellin moments. The general
recipe consists in constructing the Dirac structures from the fully covariant
component, which have no reference to a particular kinematical setup. This
means that instead of the vectors $p$, $n$ and $\Delta_T$ one has to build
the Dirac structures from the vectors $P$ and $\Delta$.

For simplicity we are going to consider the
case of nucleon to pion TDAs. We omit the isospin labels
for both hadronic states, as they turn out to be irrelevant for our present purpose,
and consider the $3$-quark light-cone operator without reference to flavor:
\begin{equation}
\widehat{O}_{\rho \tau \chi} (\lambda_1n,\lambda_2n,\lambda_3n)
=  \Psi_{\rho}(\lambda_1n) \Psi_{\tau}(\lambda_2n) \Psi_{\chi}(\lambda_3n),
\label{3q_no_flavor}
\end{equation}
where
$\Psi$
stand for the quark field operators and the color indices are omitted.
We introduce an alternative parametrization for the leading twist-$3$ $\pi N$ TDAs through
\begin{eqnarray}
 &&
4 {\mathcal{F}} \langle \pi (p_\pi)|  \,
\widehat{O}_{\rho \tau \chi} (\lambda_1n,\lambda_2n,\lambda_3n)
\,| N (p_N,s_N) \rangle   \nonumber \\ &&
=
\delta( {x}_1+ {x}_2+ {x}_3-2  {\xi}) i \frac{f_N}{f_\pi m_N}
\sum_{s} (s^{\pi N} )_{\rho \tau, \, \chi} H_s^{\pi N}(x_1,x_2,x_3, \xi, \Delta^2; \, \mu^2),
 \label{Param_TDAs_Covariant_DS}
\end{eqnarray}
where we employ the shortened notations for the set of invariant $\pi N$ TDAs
\begin{equation}
H_s^{\pi N}=\bigl\{V_{1,2}^{\pi N},\, A_{1,2}^{\pi N}, \, T_{1,2,3,4}^{\pi N} \bigr\}\,.
\end{equation}
The sum in
(\ref{Param_TDAs_Covariant_DS})
stands over the set of the leading twist-$3$ Dirac structures
\begin{equation}
(s^{\pi N})_{\rho \tau, \, \chi}=
\bigl\{
(v_{1,2}^{\pi N})_{\rho \tau, \, \chi},\,
(a_{1,2}^{\pi N})_{\rho \tau, \, \chi},
(t_{1,2,3,4}^{\pi N})_{\rho \tau, \, \chi}
\bigr\}.
\label{s_piN_rho_tau_chi}
\end{equation}
The explicit expressions for the set of  Dirac structures in
(\ref{Param_TDAs_Covariant_DS})
read:
\begin{align}
(v_1^{\pi N})_{\rho \tau, \, \chi} &= (\hat{P}C)_{\rho \tau} (\hat{P} U)_\chi\,;
&(v_2^{\pi N})_{\rho \tau, \, \chi} &=(\hat{P}C)_{\rho \tau}  (\hat{\Delta}  U)_\chi\,;
\nn \\
(a_1^{\pi N})_{\rho \tau, \, \chi} &=(\hat{P} \gamma^5 C)_{\rho \tau} (\gamma^5 \hat{P} U)_\chi\,;
&(a_2^{\pi N})_{\rho \tau, \, \chi} &=(\hat{P} \gamma^5 C)_{\rho \tau} (\gamma^5 \hat{\Delta}  U )_\chi\,;
\nn \\
(t_1^{\pi N})_{\rho \tau, \, \chi} & =(\sigma_{P \mu} C)_{\rho \tau} (\gamma^\mu \hat{P} U)_\chi\,;
M&(t_2^{\pi N})_{\rho \tau, \, \chi} &=(\sigma_{P \mu} C)_{\rho \tau} (\gamma^\mu \hat{\Delta}  U)_\chi\,;
\nonumber \\
(t_3^{\pi N})_{\rho \tau, \, \chi} &=   m_N^{-1}(\sigma_{P \Delta } C)_{\rho \tau} (\hat{P} U)_\chi\,; &
(t_4^{\pi N})_{\rho \tau, \, \chi}& =  m_N^{-1}(\sigma_{P \Delta } C)_{\rho \tau} (\hat{\Delta}   U)_\chi\,.
\label{Dirac_structures_PiN_TDA_Cov}
\end{align}

As a consequence of the Dirac equation
(\ref{Dirac_eq})
we establish the following identities:
\begin{eqnarray}
 &&
(\hat{P} \, U(p_N,s_N))_{\chi} \nonumber \\
&& = \frac{m_N}{1+\xi} U^+(p_N,s_N)_{\chi} + \frac{1}{2} (\hat{\Delta}_T U(p_N,s_N))_{\chi}
+\frac{(1+\xi)}{2m_N} \left[  \frac{m^2_N}{1+\xi } + \frac{m_\pi^2-\Delta_T^2}{1-\xi} \right] U^-(p_N,s_N)_{\chi}\,;
\label{Hat_P_U}
\\
 &&(\hat{\Delta}\, U(p_N,s_N))_\chi \nonumber \\
&& = -2 \xi\frac{m_N}{1+\xi}  U^+(p_N,s_N)_{\chi}+ (\hat{\Delta}_T U(p_N,s_N))_{\chi}
-\frac{(1+\xi)}{m_N} \left[  \frac{m_N^2}{1+\xi } - \frac{m^2_\pi-\Delta_T^2}{1-\xi} \right] U^-(p_N,s_N)_{\chi}\,.
\label{Hat_Delta_U}
\end{eqnarray}
The last terms in Eqs.~(\ref{Hat_P_U}), (\ref{Hat_Delta_U})
are of sub-leading twist while the two first terms are of the leading twist.
Then, to the leading twist-$3$ accuracy, the relation of new $\pi N$ TDA definition (\ref{Param_TDAs_Covariant_DS}) with the set of  Dirac
structures
(\ref{Dirac_structures_PiN_TDA_Cov})
 of
$\pi N$ TDAs
to that of
Eq.~(\ref{Param_TDAs})
with the set of  Dirac
structures (\ref{Def_DirStr_piN_DeltaT})
is given by
\be
&&
\left.{ \{V_{1},A_{1},T_{1}\}^{\pi N}} \right|_{\text{set of Eq.~(\ref{Def_DirStr_piN_DeltaT})}}=
\left. \left(
\frac{1}{1+\xi} \{V_{1},A_{1},T_{1}\}^{\pi N}  -
\frac{2 \xi}{1+\xi} \{V_{2},A_{2},T_{2}\}^{\pi N}
\right)
\right|_{\text{set of Eq.~(\ref{Dirac_structures_PiN_TDA_Cov})}} \,;  \nonumber \\  &&
\left.{\{V_{2},A_{2}\}^{\pi N}} \right|_{\text{set of Eq.~(\ref{Def_DirStr_piN_DeltaT})}}=
\left. \left(  {\{V_{2},A_{2}\}^{\pi N}}  +
\frac{1}{2}   {\{V_{1},A_{1}\}^{\pi N}} \right) \right|_{\text{set of Eq.~(\ref{Dirac_structures_PiN_TDA_Cov})}}\,;
 \nonumber \\  &&
\left.{T_{3}^{\pi N}}
\right|_{\text{set of Eq.~(\ref{Def_DirStr_piN_DeltaT})}}=   \left. \left(
{T_{2}^{\pi N}}
+
\frac{1}{2}    {T_{1}^{\pi N}}  \right) \right|_{\text{set of Eq.~(\ref{Dirac_structures_PiN_TDA_Cov})}};
\nonumber \\ &&
\left.{T_{2}^{\pi N}} \right|_{\text{set of Eq.~(\ref{Def_DirStr_piN_DeltaT})}}=   \left.
\left(
\frac{1}{2} T_1^{\pi N}+T_2^{\pi N}+T_3^{\pi N}-2\xi T_4^{\pi N}
\right)
\right|_{\text{set of Eq.~(\ref{Dirac_structures_PiN_TDA_Cov})}}\,;
\nonumber \\ &&
\left.{T_{4}^{\pi N}} \right|_{\text{set of Eq.~(\ref{Def_DirStr_piN_DeltaT})}}=   \left. \left(  \frac{1+\xi}{2}  T_{3}^{\pi N}    +
 (1+\xi)  {T_{4}^{\pi N}} \right) \right|_{\text{set of Eq.~(\ref{Dirac_structures_PiN_TDA_Cov})}}\,.
 \label{Relation_DiracStr_DeltaT_to_covariant}
\ee

Now we proceed with the Mellin moments of $\pi N$ TDAs defined in
Eq.~(\ref{Param_TDAs_Covariant_DS}) with the fully covariant
set of  leading twist Dirac structures
(\ref{Dirac_structures_PiN_TDA_Cov}).
The
$(n_1,n_2,n_3)$-th
Mellin moments ($n_1+n_2+n_3=N$) of the TDAs in $x_1$, $x_2$, $x_3$
lead to derivative operations acting on the three quark fields:
\begin{eqnarray}
 &&
4 (P \cdot n)^{n_1+n_2+n_3+3} \int d^3x   \, x_1^{n_1}  x_2^{n_2}  x_3^{n_3}  \nonumber \\ &&
 \int   \left[ \prod_{k=1}^3 \frac{d \lambda_k}{2 \pi}   \right]
e^{i \sum_{k=1}^3 x_k \lambda_k (P \cdot n)}
\langle \pi(P + \frac{\Delta}{2}) |
\widehat{O}_{\rho \, \tau \, \chi}( \lambda_1 n, \,\lambda_2 n, \, \lambda_3 n )
|  N(P - \frac{\Delta}{2}) \rangle
\nonumber \\ &&
=(P \cdot n)^{n_1+n_2+n_3 }
 \frac{i f_N}{f_\pi m_N}
\sum_{s
}
(s^{\pi N})_{\rho \tau, \, \chi}
\int_{-1+\xi}^{1+\xi}  dx_1 \int_{-1+\xi}^{1+\xi}  dx_2 \int_{-1+\xi}^{1+\xi}  dx_3
\, x_1^{n_1}  x_2^{n_2}  x_3^{n_3}
\nonumber \\ && \times
\delta(x_1+x_2+x_3-2 \xi) H_s^{\pi N}(x_1, x_2, x_3, \xi,\Delta^2)
\nonumber \\ &&
=
4(-1)^{n_1+n_2+n_3}
\langle \pi(P + \frac{\Delta}{2}) |
 \left[ (i \vec{\partial}^+)^{n_1} \Psi_\rho( 0)  \right]
\left[ (i \vec{\partial}^+)^{n_2}  \Psi_\tau( 0)  \right]
\left[ (i \vec{\partial}^+)^{n_3}  \Psi_\chi( 0)  \right]
 |  N(P - \frac{\Delta}{2}) \rangle\,.
 \label{Moment_step_X}
\end{eqnarray}
Hence,  the Mellin moments of nucleon to meson TDAs  are expressed through the
form factors of the local twist-$3$ operators:
\begin{equation}
\widehat{O}_{\rho \tau \chi }^{\; \mu_1...\mu_{n_1}, \, \nu_1...\nu_{n_2}, \, \lambda_1...\lambda_{ n_3}}(0)=
\left[ i\vec{D}^{\mu_1}...\,i\vec{D}^{\mu_{n_1}}  \Psi_\rho \right]
\left[ i\vec{D}^{\nu_1}...\,i\vec{D}^{\nu_{n_2}}  \Psi_\tau \right]
\left[ i\vec{D}^{\lambda_1}...\,i\vec{D}^{\lambda_{n_3}}  \Psi_\chi \right]\,,
\label{local_op_derivatives}
\end{equation}
where
\begin{equation}
\vec{D}^\mu = \vec{\partial}^\mu -i g  A^{a \, \mu} t^a
\label{Def_cov_der}
\end{equation}
is the covariant derivative. Here $t^a= \frac{\lambda^a}{2}$ with $\lambda^a$, $a=1,\ldots,8$ being the Gell-Mann matrices.
Note that in (\ref{Moment_step_X}), (\ref{local_op_derivatives}) the color indices
are omitted.

Introducing the shortened notation
\begin{equation}
(\Delta^\mu)^i (P^\mu)^{n_1-i} \equiv \Delta^{\mu_1}...\Delta^{\mu_i} P^{\mu_{i+1}}... P^{\mu_{n_1}}
\end{equation}
we  write down the following parametrization for the $\pi N$ matrix element of the local
operator
(\ref{local_op_derivatives}):
\begin{eqnarray}
 &&
4 \langle \pi|
\widehat{O}_{\rho \tau \chi}^{ \;\mu_1...\mu_{n_1}, \, \nu_1...\nu_{n_2}, \, \lambda_1...\lambda_{ n_3}}(0)
| N \rangle
=i \frac{f_N}{f_\pi m_N}
\nonumber \\ &&
\times \Bigl[\sum_{s }
(s^{\pi N})_{\rho \tau, \, \chi}
\sum_{i=0}^{n_1} \sum_{j=0}^{n_2}  \sum_{k=0}^{n_3} A^{s\; (n_1, n_2, n_3)}_{ijk}(\Delta^2)
(\Delta^\mu)^i (P^\mu)^{n_1-i}
(\Delta^\nu)^j (P^\nu)^{n_2-j}
(\Delta^\lambda)^i (P^\lambda)^{n_3-i}
\nonumber \\ &&
+\left\{
(\hat{\Delta}C)_{\rho \tau} (\hat{P} U)_\chi
C^{V_1\; (n_1, n_2, n_3)}_{N+1}(\Delta^2)
+
(\hat{\Delta}C)_{\rho \tau} (\hat{\Delta} U)_\chi
C^{V_2\; (n_1, n_2, n_3)}_{N+1}(\Delta^2)
\right.
\nonumber \\ &&
\left.
+
(\hat{\Delta} \gamma^5 C)_{\rho \tau} (\gamma^5 \hat{P} U)_\chi
C^{A_1\; (n_1, n_2, n_3)}_{N+1}(\Delta^2)
+
(\hat{\Delta} \gamma^5 C)_{\rho \tau} (\gamma^5 \hat{\Delta} U)_\chi
C^{A_2\; (n_1, n_2, n_3)}_{N+1}(\Delta^2)
\right.
\nonumber \\ &&
\left.
+
(\sigma_{\Delta \mu} C)_{\rho \tau} (\gamma^\mu \hat{P} U)_\chi
C^{T_1\; (n_1, n_2, n_3)}_{N+1}(\Delta^2)
+
(\sigma_{\Delta \mu} C)_{\rho \tau} (\gamma^\mu \hat{\Delta} U)_\chi
C^{T_2\; (n_1, n_2, n_3)}_{N+1}(\Delta^2)
\right\}
\nonumber \\ &&
\times (\Delta^\mu)^{n_1} (\Delta^\nu)^{n_2} (\Delta^\lambda)^{n_3}
\Bigr]\,,
\label{FF_decomposition_operator}
\end{eqnarray}
where
the sum in the first term
is over the set of  Dirac structures
(\ref{Dirac_structures_PiN_TDA_Cov}); and
$A^{s\; (n_1, n_2, n_3)}_{ijk}(\Delta^2)$
and
$C^{V_{1,2}, \,A_{1,2}, \, T_{1,2}\; (n_1, n_2, n_3)}_{N+1}(\Delta^2)$
denote the appropriate invariant form factors.

Now from
(\ref{FF_decomposition_operator})
we establish the following relations for
$(n_1,\,n_2,\,n_3)$-th  Mellin moments
of the TDAs:
\begin{eqnarray}
 &&
\langle x_1^{n_1} x_2^{n_2} x_3^{n_3} \{V_{1,2}^{\pi N}, \, A_{1,2}^{\pi N}, \, T_{1,2}^{\pi N} \} \rangle
\nonumber \\ &&
 =
  \sum_{n=0}^{N} (-1)^{N-n}
(2\xi)^n
\sum_{i=0}^{n_1} \sum_{j=0}^{n_2}  \sum_{k=0}^{n_3}
\delta_{i+j+k,\,n} \;
A^{\{V_{1,2}, \, A_{1,2}, \, T_{1,2} \} \; (n_1, n_2, n_3)}_{ijk}(\Delta^2)
+  (-2 \xi)^{N+1} C^{\{V_{1,2}, \, A_{1,2}, \, T_{1,2} \} \; (n_1, n_2, n_3)}_{N+1}(\Delta^2)
\,;
\nonumber \\ &&
\nonumber \\ &&
\langle x_1^{n_1} x_2^{n_2} x_3^{n_3} \{  T_{3,4}^{\pi N} \} \rangle
 =
  \sum_{n=0}^{N} (-1)^{N-n}
(2\xi)^n
\sum_{i=0}^{n_1} \sum_{j=0}^{n_2}  \sum_{k=0}^{n_3}
\delta_{i+j+k,\,n} \;
A^{\{T_{3,4} \} \; (n_1, n_2, n_3)}_{ijk}(\Delta^2)\,,
\label{PolyProp_pN-TDA}
\end{eqnarray}
which demonstrates  that the $\pi N$ TDAs defined in
(\ref{Param_TDAs_Covariant_DS})
indeed satisfy the polynomiality property. Let us emphasize that,
contrary to the GPD case, the discrete symmetries do not impose any restrictions
for evenness/oddness of the Mellin moments of TDAs and $n_{1,2,3}$ are arbitrary integers.
\bi
\item For
$n_1+n_2+n_3=N$
the highest power of
$\xi$
occurring in
$(n_1,n_2,n_3)$-th Mellin moment
of
$\{V_{1,2}^{\pi N}, \, A_{1,2}^{\pi N}, \, T_{1,2}^{\pi N}\}$ is $N+1$.
\item For
$T_{3,4}^{\pi N}$
the highest power of
$\xi$
occurring in
$(n_1,n_2,n_3)$-th Mellin moment is
$N$.
\ei
Consequently, the TDAs
$\{V_{1,2}^{\pi N}, \, A_{1,2}^{\pi N}, \, T_{1,2}^{\pi N}\}$
include an analogue of the $D$-term contribution
\cite{Polyakov:1999gs}
that generates the highest possible power of $\xi$ for a given Mellin moment.

\subsection{A spectral representation}
\label{SubSec_Spectral}
\mbox

The double distribution representation of GPDs
\cite{Radyushkin:1997ki,Radyushkin:1998bz,Radyushkin:1998es,Musatov:1999xp}
incorporates both the polynomiality property of the Mellin moments
and the support properties of GPDs.
In
\cite{Teryaev:2001qm}
it was pointed out that the relation between a GPD
and the corresponding DD is a particular case of the Radon transform.
The polynomiality property is well known
in the framework of the Radon transform theory as the Cavalieri conditions
\cite{Gelfand_Graev}.

In this subsection, following Ref.~\cite{Pire:2010if}, we present
a generalization of a spectral representation for TDAs, that ensures
the support properties of  Sec.~\ref{SubSec_Support} and the polynomiality
property of the corresponding Mellin moments
(\ref{PolyProp_pN-TDA}).
Throughout this subsection we consider
$\pi N$ TDAs introduced within
the fully covariant parametrization
(\ref{Param_TDAs_Covariant_DS}) with the set of the Dirac structures
(\ref{Dirac_structures_PiN_TDA_Cov}),
that ensures the  polynomiality
property in its simple form.

\subsubsection{A symmetric form of the spectral representation for {{GPD}}s}

\label{SubSubSec_ToySpecGPD}

In the framework of the  DD representation a GPD\footnote{Throughout this subsubsection the skewness variable $\xi$ exceptionally refers to the
GPD kinematics. We also omit the $\Delta^2$ dependence of GPDs as it is irrelevant for the present
analysis.}
$H$
is given as a one dimensional section of the double distribution (DD)
$f(\alpha, \beta)$:
\begin{equation}
H(x,\,\xi)= \int_\Omega  d \beta
d \alpha \, \delta(x-\beta-\alpha \xi) f(\beta,\,\alpha)\,.
\label{DD_Rad}
\end{equation}
The spectral representation
(\ref{DD_Rad})
was originally recovered from the diagrammatic
analysis employing the $\alpha$-representation techniques
\cite{Radyushkin:1983ea,Radyushkin:1983wh}.
The restricted integration domain
\begin{equation}
\Omega=\{ | \beta|  \le 1; \ \ \ | \alpha|  \le 1-| \beta|  \};
\label{Spec_Cond_DD}
\end{equation}
ensures the support property of GPDs
$| x|  \le 1$ for any
$| \xi| \le 1$.
The polynomiality property of the
$x$-Mellin moments of
GPDs turns out to be an intrinsic feature of the
DD representation (\ref{DD_Rad}).

In order to generalize the spectral representation (\ref{DD_Rad})
for the three-parton case we
need to rewrite it
in a symmetric form as
a function of the momentum fraction variables
$x_{1,2}$
(\ref{Support_GPD_x12}).
The partonic momentum fractions
$x_{1,2}$
are supposed to have the following decomposition in terms of the spectral parameters:
\begin{equation}
x_1=\beta_1+(1+\alpha_1) \xi\,; \ \ \ x_2= \beta_2+(1+\alpha_2) \xi\,.
\end{equation}
This allows us  to write down the following spectral representation for GPD
$H(x_1,\,x_2=2 \xi-x_1,\,\xi)$:
\begin{eqnarray}
 &&
H(x_1,\,x_2=2 \xi-x_1,\,\xi) \nonumber \\ && = \int_{\Omega_1} d \beta_1 d \alpha_1
\int_{\Omega_2} d \beta_2 d \alpha_2
\delta(x_1-\xi-\beta_1-\alpha_1 \xi)
\delta(\beta_1+\beta_2) \delta(\alpha_1+\alpha_2)
F(\beta_1, \beta_2, \alpha_1, \alpha_2)\,.
\label{start_GPD0}
\end{eqnarray}
Here
$\Omega_{1,2}$
are the usual domains
(\ref{Spec_Cond_DD})
in the spectral parameter space.  The momentum conservation condition
$x_1+x_2=2 \xi$
is imposed by
two $\delta$-functions
$\delta(\beta_1+\beta_2) \delta(\alpha_1+\alpha_2)$.
The spectral density
$F(\beta_1, \beta_2, \alpha_1, \alpha_2)$
is thus a function of four variables that is subject to two constraints,
imposed by two
$\delta$-functions, hence effectively it is
a double distribution.

To show that the spectral representation
(\ref{start_GPD0})
is equivalent to the  usual DD representation
(\ref{DD_Rad})
the  two superfluous integrations must be lifted with the help of two $\delta$-functions.
As pointed out in
Ref.~\cite{Pire:2010if},
this problem can be solved by switching to the set of natural spectral variables
\begin{equation}
\alpha_{\pm}= \frac{\alpha_1 \pm \alpha_2}{2}\,;  \ \ \  \beta_{\pm}= \frac{\beta_1 \pm \beta_2}{2}\,.
\label{natural_variables_ab}
\end{equation}
and the appropriate combinations of the longitudinal momentum fraction variables
\begin{equation}
x_\pm = \frac{x_1 \pm x_2}{2}\,.
\end{equation}

Performing the integration in
$\beta_+$ and $\alpha_+$
is straightforward (see Sec.~IV.A of Ref.~\cite{Pire:2010if}).
It does not bring
additional restrictions on the remaining spectral parameters
$\alpha_-$, $\beta_-$. The result reads
\begin{equation}
H(x_1,\,x_2=2\xi-x_1,\,\xi)
=
\int_{-1}^1 d \beta_- \int_{-1+| \beta_-| }^{1-| \beta_-| } d \alpha_-
\delta(x_--\beta_--\alpha_-\xi)
2 F \left( \beta_-, -\beta_-, \alpha_-,-\alpha_- \right)
\,.
\end{equation}
Since
$x_-  \equiv x$,
by renaming the integration variables
$\alpha_- \to \alpha$, $\beta_- \to \beta$
and introducing the DD
\begin{equation}
f(\beta, \, \alpha) \equiv
2 F \left( \beta, -\beta, \alpha,-\alpha \right)
\label{f_through_F_GPD}
\end{equation}
one recovers the usual form of the DD representation
(\ref{DD_Rad}).

For completeness we also present the expressions for
a GPD in various domains in $x$, that results from
performing the $\alpha$-integration in Eq.~(\ref{DD_Rad})
with the help of the $\delta$-function. For arbitrary
$0 \le \xi \le 1$ we get the following:
\bi
\item $| x| >1$:
\begin{equation}
H(x,\xi)=0;
\label{GPD_0}
\end{equation}
\item DGLAP domain $-1 \le x \le -\xi$:
\begin{equation}
H \left(x, \xi \right)=\frac{1}{\xi} \int_{\frac{x-\xi}{1+\xi}}^{\frac{x+\xi}{1-\xi}} \mathrm{d} \beta F \left(\beta, \frac{x-\beta}{\xi} \right);
\label{GPD_DGLAP1}
\end{equation}
\item ERBL domain $-\xi \le x \le \xi$:
\begin{equation}
H \left(x, \xi \right)=\frac{1}{\xi} \int_{\frac{x-\xi}{1+\xi}}^{\frac{x+\xi}{1+\xi}} \mathrm{d} \beta F \left(\beta, \frac{x-\beta}{\xi} \right);
\label{GPD_ERBL}
\end{equation}
\item DGLAP domain $\xi \le x \le 1$:
\begin{equation}
H \left(x, \xi \right)=\frac{1}{\xi} \int_{\frac{x-\xi}{1-\xi}}^{\frac{x+\xi}{1+\xi}} \mathrm{d} \beta F\left(\beta, \frac{x-\beta}{\xi} \right).
\label{GPD_DGLAP2}
\end{equation}
\ei

\subsubsection{Quadruple distributions}
\label{SubSec_Quadr_distib}
\mbox

The spectral representation for a $\pi N$ TDA
$H(x_1,\,x_2, \, x_3=2 \xi-x_1-x_2, \,\xi)$
can be written as a straightforward generalization of
the spectral representation for GPDs
(\ref{start_GPD0}).

\begin{figure}[H]
\begin{center}
\includegraphics[width=4.8cm]{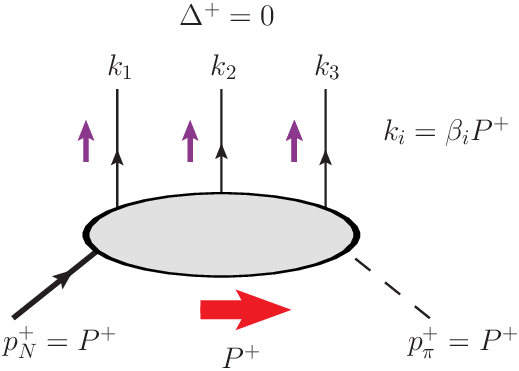} \ \
\includegraphics[width=4.9cm]{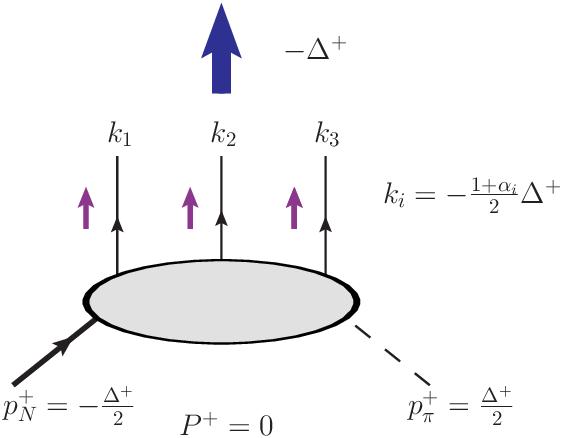} \ \
\includegraphics[width=5.5cm]{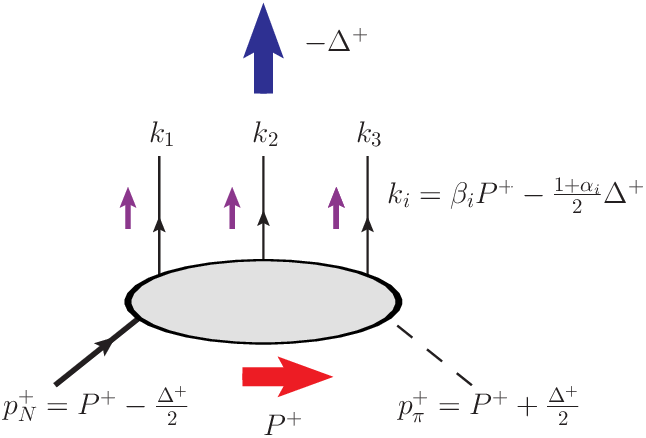}
\end{center}
     \caption{Momentum flow in a quadruple distribution (right panel) as a superposition
     of the momentum flow of a forward density (left panel) and of a distribution amplitude (central panel).}
\label{Fig_Mom_Flow}
\end{figure}

Following Section 3.8 of Ref.~\cite{Belitsky:2005qn},
we take a point of view on TDAs, as kinematic ``hybrids''
of forward 3-parton densities and of distribution amplitudes
and represent the corresponding momentum flow as a superposition
of the $s$-channel and $t$-channel momentum fluxes (see  Fig.~\ref{Fig_Mom_Flow}).

We introduce three sets of spectral parameters
$\beta_{1,2,3}$, $\alpha_{1,2,3}$.
The longitudinal momentum fractions of the three quarks
$x_{1,2,3}$
are supposed to have the following decomposition in terms of the spectral parameters:
\begin{equation}
x_1=\xi+\beta_1+\alpha_1 \xi\,; \ \ \ x_2=\xi+\beta_2+\alpha_2 \xi\,; \ \ \ x_3=\xi+\beta_3+\alpha_3 \xi\,.
\end{equation}
In order to satisfy the momentum conservation constraint $\sum_i x_i=2 \xi$
we  require that
\begin{equation}
\beta_1+ \beta_2+ \beta_3=0\, ; \ \ \ \alpha_1+\alpha_2+\alpha_3=-1\,.
\end{equation}

This allows us to write down the following spectral representation
for $\pi N$ TDAs:
\begin{eqnarray}
 &&
H(x_1,\,x_2,\,x_3=2 \xi -x_1-x_2,\,\xi) \nonumber \\ && =
\left[
\prod_{i=1}^3
\int_{\Omega_i} d \beta_i d \alpha_i
\right]
\delta(x_1-\xi-\beta_1-\alpha_1 \xi) \,
\delta(x_2-\xi-\beta_2-\alpha_2 \xi) \,
\nonumber \\ &&
\times
\delta(\beta_1+ \beta_2+ \beta_3)
\delta(\alpha_1+\alpha_2+\alpha_3+1)
 F(\beta_1, \, \beta_2, \, \beta_3, \, \alpha_1, \, \alpha_2, \alpha_3)\,.
\label{Spectral_for_GPDs_x123}
\end{eqnarray}
By
$\Omega_{i},\,i= \left\{ 1, \, 2, \,3 \right\}$
we denote the three copies of the usual domains
(\ref{Spec_Cond_DD})
in the spectral parameter space.
The spectral density
$F(\beta_1, \, \beta_2, \, \beta_3, \, \alpha_1, \, \alpha_2, \alpha_3)$
is a function of $6$ variables (sextuple  distribution).
However, these variables are subject to two constraints.
Therefore, effectively $F$ turns out to be a quadruple distribution.

To clarify the physical contents of the quadruple distribution
it is instructive to consider the following spectral representation in terms of quadruple distributions which can be established for $\pi N$
matrix element of the twist-$3$ non-local $3$-quark light-cone operator
(\ref{3q_no_flavor}):
\begin{eqnarray}
 &&
\langle \pi(p_\pi) |  \widehat{O}_{\rho \tau \chi}(\lambda_1 n,\, \,\lambda_2 n,\, \lambda_3n) |  p(p_N) \rangle
\nn
\\ &&
= \frac{1}{4 (P \cdot n)} i \frac{f_N}{f_\pi m_N}
\left[ \prod_{i=1}^3  \int_{-1+\xi}^{1+\xi} dx_i \right]
\delta(2 \xi -x_1-x_2-x_3) e^{-i \sum_k x_k \lambda_k (P \cdot n)}
\sum_s( s^{\pi N} )_{\rho \tau, \, \chi}
H_s(x_1,x_2,x_3,\xi) \nn \\ &&
=i \frac{f_N}{f_\pi m_N}
\left[
\prod_{i=1}^3
\int_{\Omega_i} d \beta_i d \alpha_i
\right]
e^{-\sum_k \beta_k \lambda_k (P \cdot n) - i \sum_l \alpha_l
\lambda_l (-\Delta \cdot n)}
\delta(\beta_1+ \beta_2+ \beta_3)
\delta(\alpha_1+\alpha_2+\alpha_3+1)
\nn \\ &&
\sum_s  ( s^{\pi N} )_{\rho \tau, \, \chi}
F_s(\beta_1, \, \beta_2, \, \beta_3, \, \alpha_1, \, \alpha_2, \alpha_3),
\label{Spec_rep_NL}
\end{eqnarray}
where the momentum flow in the quadruple distributions
is specified in  Fig.~\ref{Fig_Mom_Flow}. We denote here by
$F_s(\beta_1, \, \beta_2, \, \beta_3, \, \alpha_1, \,  \alpha_2, \alpha_3)$   the spectral density corresponding to a particular TDA $H_s$ occurring in the Dirac decomposition of the matrix element in the l.h.s. of
Eq.~(\ref{Spec_rep_NL}).
The formula (\ref{Spec_rep_NL}) can be seen as a straightforward generalization of the familiar spectral representation for the
matrix element of the composite operator $\phi \phi$  constructed out
of scalar fields (see \textit{e.g.} discussion around
Eq.~(3.206) in Section 3.8 of Ref.~\cite{Belitsky:2005qn}).

Now we need to verify that $\pi N$ TDAs within the spectral
representation
(\ref{Spectral_for_GPDs_x123})
satisfy the polynomiality property and
possess the correct support properties
(\ref{TDA_support_xi})
in the longitudinal momentum fraction
$x_i$.

Checking the polynomiality property is an easy task.
By a formal interchange of integration order we
show that the
$(n_1,\, n_2, \, n_3)$-th Mellin moment in
$(x_1,\, x_2, \, x_3)$
of
$\pi N$
TDA is, indeed, a polynomial of order
$N \equiv n_1+n_2+n_3$
of
$\xi$:
\begin{eqnarray}
 &&
\langle x_1^{n_1} x_2^{n_2} x_3^{n_3} H(x_1,\,x_2,\,x_3=2 \xi -x_1-x_2,\,\xi) \rangle \nn
\\ &&
=
\left[
\prod_{i=1}^3
\int_{\Omega_i} d \beta_i d \alpha_i
\right]
\left[
\prod_{j=1}^3
\int_{-1+\xi}^{1+\xi} d x_j
\right]
\delta(\beta_1+ \beta_2+ \beta_3)
\delta(\alpha_1+\alpha_2+\alpha_3+1)
 F(\beta_1, \, \beta_2, \, \beta_3, \, \alpha_1, \, \alpha_2, \alpha_3)
\nn \\ &&
\times x_1^{n_1} x_2^{n_2} x_3^{n_3}  \, \delta(x_1+x_2+x_3-2 \xi) \,
\delta(x_1-\xi-\beta_1-\alpha_1 \xi) \,
\delta(x_2-\xi-\beta_2-\alpha_2 \xi)
= P_{N}(\xi)\,.
\end{eqnarray}

Working out the support properties of (\ref{Spectral_for_GPDs_x123}) follows the
same stages as the derivation of  Sec.~\ref{SubSubSec_ToySpecGPD}.
The first step consists in switching to natural combinations of spectral parameters
and performing two out of the six integrals in (\ref{Spectral_for_GPDs_x123}).
At the second step we work out the support properties and derive the analogue
of Eqs.~(\ref{GPD_DGLAP1})--(\ref{GPD_DGLAP2}) for TDAs in the various domains of the support.

There exist there equivalent choices of convenient
combinations of spectral parameters
$\sigma_i$, $\rho_i$, $\omega_i$, $\nu_i$
that are adjusted with the
three sets of quark--diquark coordinates
$(w_i,\,v_i)$ (\ref{Def_qDq_coord}):
\begin{eqnarray}
 &&
\sigma_i= \beta_i; \ \ \ \rho_i=  \sum_{k, l=1}^{3} \varepsilon_{i k l} \beta_{k};
\nn \\
 &&
\omega_i=\alpha_i\,; \ \ \  \nu_i=\sum_{k, l=1}^{3} \varepsilon_{i k l} \alpha_{k}.
\label{Nat_spec_var_1}
\end{eqnarray}

Performing two integrals over spectral parameters
(see Sec.~IV and Appendix~A of Ref.~\cite{Pire:2010if})
results in three equivalent representation of $\pi N$ TDAs as
a function of quark--diquark coordinates $w_i,\,v_i$:
\begin{eqnarray}
 &&
H
(w_i,\,v_i,\,\xi)
\nonumber \\ &&
=
\int_{-1}^1 d \sigma_i
\int_{-1+\frac{| \sigma_i| }{2}}^{1-\frac{| \sigma_i| }{2}} d \rho_i
\int_{-1+| \sigma_i| }^{1-|  \rho_i- \frac{\sigma_i}{2}| -| \rho_i+ \frac{\sigma_i}{2}| } d \omega_i
\int_{-\frac{1}{2}+| \rho_i- \frac{\sigma_i}{2}| +\frac{\omega_i}{2}}^{\frac{1}{2}-| \rho_i+ \frac{\sigma_i}{2}| -\frac{ \omega_i}{2}} d \nu_i
\delta(w_i-\sigma_i-\omega_i \xi)
\nonumber \\ &&
\times  \delta(v_i-\rho_i-\nu_i \xi) \,
 F_i(\sigma_i,\, \rho_i,\, \omega_i,\, \nu_i)\,.
\label{Spectral_represent_Hi}
\end{eqnarray}
The quadruple distributions $F_i$ are expressed in three equivalent
forms in terms of the master sextuple  distribution defined in Eq.~(\ref{Spectral_for_GPDs_x123})
(\textit{cf.} Eq.~(\ref{f_through_F_GPD}) for the GPD case):
\begin{eqnarray}
 &&
F_1(\sigma_1,\, \rho_1,\, \omega_1,\, \nu_1) \equiv F(\sigma_1, \,\rho_1-\frac{\sigma_1}{2}\,,-\rho_1-\frac{\sigma_1}{2},\,
\omega_1,\, \nu_1-\frac{1+\omega_1}{2}, \, -\nu_1-\frac{1+\omega_1}{2})\,;
\nonumber \\
 &&
F_2(\sigma_2,\, \rho_2,\, \omega_2,\, \nu_2)
\equiv F(-\rho_2-\frac{\sigma_2}{2},\, \sigma_2,\,\rho_2-\frac{\sigma_2}{2},\,
 -\nu_2-\frac{1+\omega_2}{2},\,\omega_2,\,  \nu_2-\frac{1+\omega_2}{2})\,;
 \nonumber \\
  &&
F_3(\sigma_3,\, \rho_3,\, \omega_3,\, \nu_3) \equiv
F(\rho_3-\frac{\sigma_3}{2},\, -\rho_3-\frac{\sigma_3}{2},\, \sigma_3,\, \nu_3- \frac{1+\omega_3}{2},\, -\nu_3- \frac{1+\omega_3}{2}, \omega_3)\,.
\label{Quad_Distr_Fi}
\end{eqnarray}

The next step consists in establishing
a generalization of Eqs.~(\ref{GPD_0})--(\ref{GPD_DGLAP2})
expressing $\pi N$ TDAs in different parts of the
support domain
(\ref{Support_TDA_wv}).
It is done by performing two integrals over the spectral
parameters $\omega_i$, $\nu_i$ in
(\ref{Spectral_represent_Hi})
with the help of the two $\delta$-functions.
This analysis is rather technical and
requires careful treatment of the integration limits
in corresponding  multiple integrals. For the details of derivation we
refer the reader  to Sec.~V of Ref.~\cite{Pire:2010if}.

\begin{figure}[H]
\begin{center}
\includegraphics[width=0.6\textwidth]{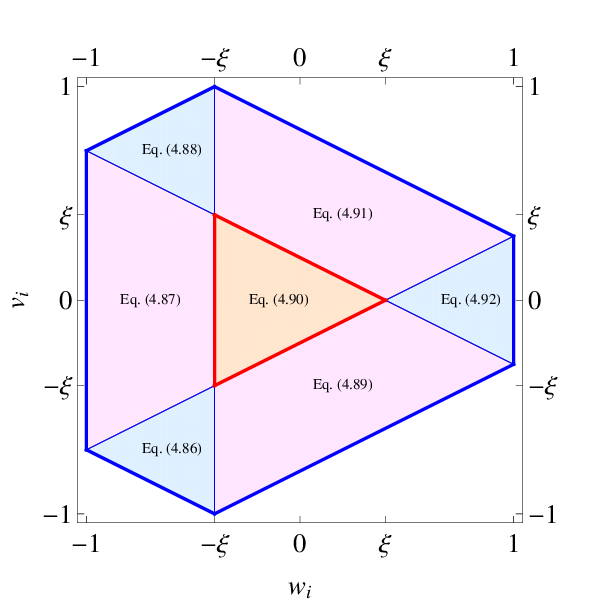}
\end{center}
     \caption{Spectral representation for $\pi N$ TDA (\ref{Spectral_represent_Hi})
     in the ERBL-like and DGLAP-like type I, II domains in the quark--diquark
     coordinates $w_i$, $v_i$ (\ref{Def_qDq_coord}) for $\xi=0.4$.}
\label{Fig_DomainsSpec_TDAs_qDq}
\end{figure}

For
$0 \le \xi \le 1$, depending on the part  of the
($w_i\,; \, v_i$)
support domain in question,
the quadruple integral
(\ref{Spectral_represent_Hi})
for
$H(w,v,\xi)$
reduces to the following expressions (see  Fig.~\ref{Fig_DomainsSpec_TDAs_qDq}):
\begin{itemize}

\item
DGLAP-like type I domain
$w_i  \in [-1;\,-\xi]$;
$v_i \in [\xi'_i ;\,  1-\xi'_i+\xi]$ ($\xi'_i \equiv \frac{\xi-w_i}{2}$):
\begin{equation}
H(w_i ,\,v_i ,\xi)= \frac{1}{\xi^2 }
\int_{\frac{2(v_i  \xi -\xi'_i)}{1-\xi^2}}^{\frac{w_i +\xi}{1-\xi}} d \sigma_i  \int_{-\frac{\sigma_i }{2}+ \frac{v_i -\xi'_i}{1-\xi}}^{ \frac{\sigma_i }{2}+ \frac{v_i +\xi'_i}{1+\xi}}
d \rho_i  \, F_i(\sigma_i ,\, \rho_i ,\, \frac{w_i -\sigma_i }{\xi}, \, \frac{v_i -\rho_i }{\xi})\,;
\label{DGLAP_type_I_1}
\end{equation}

\item DGLAP-like type II domain
$w_i  \in [-1;\,-\xi]$;
$v_i \in [-\xi'_i ;\,   \xi'_i]$:
\begin{equation}
H(w_i ,\,v_i ,\xi)= \frac{1}{\xi^2}
\int_{ \frac{w_i- \xi }{1+\xi }}^{\frac{w_i +\xi}{1-\xi}} d \sigma_i
\int_{-\frac{\sigma_i}{2}+ \frac{v_i -\xi'_i}{1+\xi}}^{ \frac{\sigma_i }{2}+ \frac{v_i +\xi'_i}{1+\xi}}
d \rho_i  \, F_i(\sigma_i ,\, \rho_i ,\, \frac{w_i -\sigma_i }{\xi}, \, \frac{v_i -\rho_i }{\xi})\,;
\label{DGLAP_type_II_1}
\end{equation}

\item DGLAP-like type I domain
$w_i  \in [-1;\,-\xi]$; $v_i \in [-1+\xi'_i-\xi  ;\,   -\xi'_i]$ :
\begin{equation}
H(w_i,\,v_i,\xi)= \frac{1}{\xi^2}
\int_{-\frac{2(v_i \xi +\xi'_i)}{1-\xi^2}}^{\frac{w_i+\xi}{1-\xi}} d \sigma_i \int_{-\frac{\sigma_i}{2}+ \frac{v_i-\xi'_i}{1+\xi}}^{ \frac{\sigma_i}{2}+ \frac{v_i+\xi'_i}{1-\xi}}
d \rho_i \, F_i(\sigma_i,\, \rho_i,\, \frac{w_i-\sigma_i}{\xi}, \, \frac{v_i-\rho_i}{\xi})\,;
\label{DGLAP_type_I_2}
\end{equation}

\item
DGLAP-like type II domain
$w_i \in [-\xi;\,1]$;
$v_i \in [-1+\xi-\xi'_i ;\, -| \xi'_i|  ]$:
\begin{equation}
H(w_i ,\,v_i ,\xi)= \frac{1}{\xi^2}
\int_{  -\frac{2(v_i \xi + \xi'_i)}{1-\xi^2}}^{\frac{w_i +\xi}{1+\xi}} d \sigma_i
\int_{-\frac{\sigma_i}{2}+ \frac{v_i -\xi'_i}{1+\xi}}^{ \frac{\sigma_i }{2}+ \frac{v_i +\xi'_i}{1-\xi}}
d \rho_i  \, F_i(\sigma_i ,\, \rho_i ,\, \frac{w_i -\sigma_i }{\xi}, \, \frac{v_i -\rho_i }{\xi})\,;
\label{DGLAP_type_II_3}
\end{equation}

\item ERBL-like domain
$w_i \in [-\xi;\,\xi]$;
$v_i \in [-\xi'_i ;\, \xi'_i]$:
\begin{equation}
H(w_i ,\,v_i ,\xi)= \frac{1}{\xi^2}
\int_{ \frac{w_i- \xi }{1+\xi }}^{\frac{w_i +\xi}{1+\xi}} d \sigma_i
\int_{-\frac{\sigma_i}{2}+ \frac{v_i -\xi'_i}{1+\xi}}^{ \frac{\sigma_i }{2}+ \frac{v_i +\xi'_i}{1+\xi}}
d \rho_i  \, F_i(\sigma_i ,\, \rho_i ,\, \frac{w_i -\sigma_i }{\xi}, \, \frac{v_i -\rho_i }{\xi})\,;
\end{equation}

\item DGLAP-like type II domain
$w_i \in [-\xi;\,1]$;
$v_i \in [| \xi'_i|  ;\, 1-\xi +\xi'_i]$:
\begin{equation}
H(w_i ,\,v_i ,\xi)= \frac{1}{\xi^2 }
\int_{\frac{2(v_i  \xi -\xi'_i)}{1-\xi^2}}^{\frac{w_i +\xi}{1+\xi}} d \sigma_i  \int_{-\frac{\sigma_i }{2}+ \frac{v_i -\xi'_i}{1-\xi}}^{ \frac{\sigma_i }{2}+ \frac{v_i +\xi'_i}{1+\xi}}
d \rho_i  \, F_i(\sigma_i ,\, \rho_i ,\, \frac{w_i -\sigma_i }{\xi}, \, \frac{v_i -\rho_i }{\xi})\,;
\label{DGLAP_type_II_2}
\end{equation}

\item $w_i \in [\xi;\,1]$ and $v_i \in [\xi'_i;-\xi'_i]$ (DGLAP-like type I domain):
\begin{equation}
H(w_i ,\,v_i ,\xi)= \frac{1}{\xi^2}
\int_{ \frac{w_i- \xi }{1-\xi }}^{\frac{w_i +\xi}{1+\xi}} d \sigma_i
\int_{-\frac{\sigma_i}{2}+ \frac{v_i -\xi'_i}{1-\xi}}^{ \frac{\sigma_i }{2}+ \frac{v_i +\xi'_i}{1-\xi}}
d \rho_i  \, F_i(\sigma_i ,\, \rho_i ,\, \frac{w_i -\sigma_i }{\xi}, \, \frac{v_i -\rho_i }{\xi})\,.
\label{DGLAP_type_I_3}
\end{equation}

\item For
$w_i$
and
$v_i$
outside the domain
$  w_i \in [-1;\,1]$ and $v_i \in [-1+| \xi_i-\xi'_i|  ;\,1-| \xi_i-\xi'_i| ]$
the spectral representation
(\ref{Spectral_represent_Hi})
provides vanishing result for $0 \le \xi \le 1$.

\item The corresponding expressions for $-1 \le \xi < 0$ are summarized in  Appendix~B of Ref.~\cite{Pire:2010if}.
\end{itemize}

From Eqs.~(\ref{DGLAP_type_I_1})--(\ref{DGLAP_type_I_3}) it is straightforward
to check that the TDAs $H(w_i,v_i,\xi)$ defined by the spectral representation (\ref{Spectral_represent_Hi})
are continuous on the cross-over lines $w_i=-\xi$; $v_i = \pm \xi'_i$.
This property turns out to be crucial for the calculation of convolutions of TDAs with
hard scattering amplitudes (see  Sec.~\ref{Sec_ExclProcLO}).

Thus the spectral representation (\ref{Spectral_represent_Hi})
indeed ensures both the polynomiality property of the Mellin moments of TDAs
and their support properties (\ref{Support_TDA_wv}).
The fact that there turns out to be three equivalent types of  spectral representations
depending on the particular choice of the quark--diquark coordinates mirrors
the redundancy of the description of the system of three partons. Nucleon-to-meson
TDAs turn out to be intricate functions, since the polynomiality property is
assured for a function with a $\xi$-dependent support. This requires the existence
of a highly non-trivial generalization of the duality property~\cite{Kumericki:2008di} interrelating
TDAs in different parts of their support domain.

\subsubsection{{$D$}-term-like contribution}
\mbox

Similarly to the GPD case,
$\pi N$
TDAs within the spectral
representation
(\ref{Spectral_represent_Hi})
 might not satisfy the polynomiality condition
in its complete form.
For example, as pointed out in  Sec.~\ref{SubSec_Polynomiality},
the maximal power of $\xi$ that occurs for the
$n_1+n_2+n_3=N$-th Mellin moments of TDAs
$V_{1,2}^{\pi N}$, $A_{1,2}^{\pi N}$, $T_{1,2}^{\pi N}$
defined in Eq.~(\ref{Param_TDAs_Covariant_DS})
is $N+1$ while the spectral representation
(\ref{Spectral_represent_Hi}) provides it to be $N$.
Thus for these TDAs the spectral representation (\ref{Spectral_represent_Hi})
has to be complemented by an analogue of the $D$-term.
This $D$-term-like contribution
has a pure ERBL-like support in $x_i$:
\begin{equation}
0 \le x_i \le 2 \xi.
\end{equation}
The simplest model for a $D$-term-like contribution into
$\pi N$ TDAs originates from the cross-channel
nucleon exchange contribution into the amplitude of backward
pion electroproduction amplitude (see  Sec.~\ref{SubSec_Cross_Ch_baryons}).
On the other hand, $\pi N$ TDAs
$T_{3,4}^{\pi N}$
do not require adding a $D$-term-like contribution.

\subsection{Implications of the
SU$(2)$
isotopic symmetry}
\label{SubSec_Isospin}
\mbox

In this section, following Ref.~\cite{Pire:2011xv},
we consider three-quark light-cone operator
with generic light flavors and
present the generalization of the
TDA description for transitions between
$I=\frac{1}{2}$ baryons to mesons belonging to $I=0$ and $I=1$ SU$(2)$-isospin
multiplets. We work out the isospin identities for nucleon-to-meson TDAs
and establish the set of independent TDAs providing description to all
isospin channels.

\subsubsection{SU$(2)$ isospin group properties of the three-quark operator}
\mbox

Let us first  specify the tensor properties  of the master three-quark light-cone operator
\begin{equation}
\widehat{O}^{\alpha \beta \gamma}_{\rho \tau \chi}(\lambda_1 n,\,\lambda_2 n,\lambda_3 n)
= 
\varepsilon_{c_1 c_2 c_3}
\Psi^{c_1\, \alpha}_\rho( \lambda_1 n) \Psi^{c_2 \, \beta}_\tau (\lambda_2 n)
\Psi^{c_3 \gamma}_\chi ( \lambda_3 n) \Big|_{n^2=0}
\label{Def_operator_O_isotopic}
\end{equation}
with respect to the SU$(2)$  isospin group.
In Eq.~(\ref{Def_operator_O_isotopic})
we assume the use of the light-cone gauge
$A^+=0$;
$c_{1,2,3}$
stand for the SU$(3)$ color indices;
$\rho$, $\tau$, $\chi$
are the Dirac indices and
$\alpha, \, \beta, \, \gamma=\{1,\,2\}$
stand for the SU$(2)$ isospin indices of the fundamental
representation.
We
distinguish between the upper (contravariant) and lower
(covariant) SU$(2)$ isospin indices.
The totally antisymmetric tensor
$\varepsilon_{\alpha \beta}$
is used to
 lower the SU$(2)$ indices
and $\varepsilon^{\alpha \beta}$
to rise the indices
($\varepsilon_{1 \,2}=\varepsilon^{1 \,2}=1$):
\[
\Psi ^\alpha \varepsilon_{\alpha \beta} = \Psi_\beta; \ \ \
\Psi_\alpha \varepsilon^{\alpha \beta} = \Psi^\beta.
\]
We also employ the Kronecker symbol
$\delta^\alpha_{\; \beta}= -\varepsilon^\alpha_{\; \beta}= \varepsilon_\beta^{\; \, \alpha}$.

In what follows we adopt the shortened notations for the spatial arguments of the operator (\ref{Def_operator_O_isotopic}):
$\widehat{O}^{\alpha \beta \gamma}_{\rho   \tau \chi}(\lambda_1 n,\,\lambda_2 n,\,   \lambda_3 n) \equiv \widehat{O}^{\alpha \beta \gamma}_{\rho   \tau \chi}(1,2,3)$.
For example,
the
$uud$
operator
(\ref{Def_O_uud_operator})
within these notations reads
\[
\widehat{O}_{\rho \tau \chi}^{\,uud}(1,2,3) \equiv
\varepsilon_{c_1 c_2 c_3} \Psi^{c_1 \, 1}_\rho(1) \Psi^{c_2\, 1}_\tau (2) \Psi^{c_3\, 2}_\chi (3).
\]
We will also omit the spatial arguments and the Dirac indices when they are irrelevant for the discussion.


The operator
(\ref{Def_operator_O_isotopic})
transforms according to the
\begin{equation}
2 \otimes 2 \otimes 2= 4\oplus 2 \oplus 2
\label{tensor_prod_decomp}
\end{equation}
representation of the isospin group
${\rm SU}(2)$.
To specify the operators transforming according to the isospin-$\frac{3}{2}$ and
isospin-$\frac{1}{2}$ representations we
single out the totally symmetric operator
$\widehat{O}^{\{\alpha \beta \gamma\}}$,
the totally antisymmetric part
$\widehat{O}^{[\alpha \beta \gamma]}$
and three operators
antisymmetric in pairs of indices
$[\alpha, \beta]$, $[\alpha, \gamma]$ and  $[\beta, \gamma]$:
\begin{equation}
\widehat{O}^{\alpha \beta \gamma}
=
\widehat{O}^{\{\alpha \beta \gamma\}}
+
\widehat{O}^{[\alpha \beta \gamma]}+
 \widehat{O}_1^{[\alpha \beta] \gamma}+ \widehat{O}_2^{ [\alpha \check{\beta}  \gamma] }+
\widehat{O}_3^{\alpha [\beta \gamma]}\,.
\label{Exp_sym_asym_SU(2)}
\end{equation}
The totally symmetric part
$\widehat{O}^{\{\alpha \beta \gamma\}}$
transforms according to the isospin-$\frac{3}{2}$ representation
while the totally antisymmetric part
$\widehat{O}^{[\alpha \beta \gamma]}$
turns to be zero for the SU$(2)$ isospin group.
Finally, the contraction of operators
$\widehat{O}_{1,\,2,\,3}$
with the appropriate $\varepsilon$-tensors provides spinors
transforming according to the fundamental  representation of the isospin group
SU$(2)$. Note, that due to the relation
\begin{equation}
\varepsilon_{\alpha \beta}  \widehat{O}_1^{[\alpha \beta] \delta }-
\varepsilon_{\alpha \gamma}  \widehat{O}_2^{ [\alpha \check{\delta}  \gamma] }+
\varepsilon_{\beta \gamma} \widehat{O}_3^{\delta [\beta \gamma]}=0,
\end{equation}
only two operators $\widehat{O}_{i}$ turn to be independent. Thus the tensor
decomposition of the operator
(\ref{Def_operator_O_isotopic})
involves
two operators transforming according to the isospin-$\frac{1}{2}$ representations
and one operator transforming according to the isospin-$\frac{3}{2}$ representation.
Therefore, the expansion
(\ref{Exp_sym_asym_SU(2)})
indeed matches
(\ref{tensor_prod_decomp}).


\subsubsection{Isospin and permutation symmetry identities for  the nucleon {{DA}}}
\label{SubSec_NucleonDA_isospin}
\mbox

For consistency below we consider the case of the leading twist-$3$ nucleon DA.
This allows to exemplify the implication of the
isospin invariance and to introduce  convenient unified
notations making explicit the isotopic and permutation symmetry
properties of the matrix element of  three-quark light-cone operators.

The invariance under the SU$(2)$ isospin group  provides
the following isospin decomposition for the matrix element
of the three-quark operator
$\widehat{O}^{\alpha \beta \gamma}_{\rho \tau \chi}(1,2,3)$
(\ref{Def_operator_O_isotopic})
between a nucleon doublet
$N=
\begin{pmatrix}
  p  \\   n
\end{pmatrix}
$
state and the vacuum:
\begin{eqnarray}
 &&
4 \langle 0 |  \widehat{O}^{\alpha \beta \gamma}_{\rho \tau \chi}
(1,2,3)
|  N_\iota(p_N) \rangle
\nonumber \\ &&
= \varepsilon^{\alpha \beta} \delta^\gamma_\iota {M_1^N}_{\rho \tau \chi}
(1,2,3)
+
\varepsilon^{\alpha \gamma} \delta^\beta_\iota {M_2^N}_{\rho \tau \chi}
(1,2,3)
+
\varepsilon^{\beta \gamma} \delta^\alpha_\iota {M_3^N}_{\rho \tau \chi}
(1,2,3)
\,.
\label{nucleon_DA_isospin_dec_1}
\end{eqnarray}
Due to the identity
\begin{equation}
\varepsilon^{\alpha \beta} \delta^\gamma_\iota+\varepsilon^{\beta \gamma} \delta^\alpha_\iota-\varepsilon^{\alpha \gamma} \delta^\beta_\iota=0\,,
\label{Id_tensor_str_NDA}
\end{equation}
the three  invariant isospin amplitudes
${M_{1,\,2,\,3}^N}$
in
(\ref{nucleon_DA_isospin_dec_1})
are not independent.
To work out the set of independent nucleon DAs
one has to establish the isospin identities  and to take into account
the properties of the matrix element
(\ref{nucleon_DA_isospin_dec_1})
under the group of permutations
of three quark field operators constituting the operator (\ref{Def_operator_O_isotopic}).

For this issue it is convenient to introduce the combinations of
the isotopic amplitudes
(\ref{nucleon_DA_isospin_dec_1})
which are symmetric under permutations of the appropriate quark fields in
the operator
$\widehat{O}^{\alpha \beta \gamma}_{\rho \tau \chi}(1,\,2,\,3)$:
\begin{eqnarray}
 &&
{M_2^N}_{\rho \tau \chi}(1,\,2,\,3)+{M_3^N}_{\rho \tau \chi}(1,\,2,\,3) \equiv M^{N\,\{12\}}_{\rho \tau \chi}(1,\,2,\,3)\,;
\nonumber \\
 &&
{M_1^N}_{\rho \tau \chi}(1,\,2,\,3)-{M_3^N}_{\rho \tau \chi}(1,\,2,\,3) \equiv M^{N\,\{13\}}_{\rho \tau \chi}(1,\,2,\,3)\,;
\nonumber \\
 &&
-{M_1^N}_{\rho \tau \chi}(1,\,2,\,3)-{M_2^N}_{\rho \tau \chi}(1,\,2,\,3) \equiv M^{N\,\{23\}}_{\rho \tau \chi}(1,\,2,\,3)\,.
\label{Def_combinations_NDA}
\end{eqnarray}
The combinations (\ref{Def_combinations_NDA}) satisfy the isospin identity
\begin{equation}
M^{N\,\{12\}}_{\rho \tau \chi}(z_1,\,z_2,\,z_3)+
M^{N\,\{13\}}_{\rho \tau \chi}(z_1,\,z_2,\,z_3)+
M^{N\,\{23\}}_{\rho \tau \chi}(z_1,\,z_2,\,z_3)=0\,.
\label{Isospin_Id_NDA}
\end{equation}
This allows to present the isospin decomposition (\ref{Def_combinations_NDA})
as
\begin{equation}
4 \langle 0 |  \widehat{O}^{\alpha \beta \gamma}_{\rho \tau \chi}
(1,2,3)
|  N_\iota(p_N) \rangle
=
\varepsilon^{\alpha \beta} \delta^\gamma_\iota
M^{N\,\{13\}}_{\rho \tau \chi }(1,2,3)
+
\varepsilon^{\alpha \gamma} \delta^\beta_\iota M^{N\,\{12\}}_{\rho  \tau \chi}(1,2,3)\,.
\label{Isospin_parmetrization_N_DA}
\end{equation}
The conventional $uud$ proton and $ddu$ neutron DAs are then expressed as
\begin{equation}
4\langle 0| \widehat{O}_{\rho \tau \chi}^{u u d}\left(1, \, 2, \,3 \right)
|  N_{p}\left(p_{N}\right)\rangle =-4\langle 0| \widehat{O}_{\rho \tau \chi}^{d d u}\left(1, \, 2, \,3 \right) |  N_{n}\left(p_{N}\right)\rangle=M_{\rho \tau \chi}^{N\{12\}}\left(1, \, 2, \,3 \right).
\end{equation}

The properties of the nucleon DA under concerted
permutations of spacial arguments and Dirac indices follow from the
anticommutation properties of quark field operators in
(\ref{Def_operator_O_isotopic}).
This allows to establish the following relations for the isospin amplitudes:
\begin{eqnarray}
 &&
M^{N\,\{12\}}_{\rho \tau \chi}(1,\,2,\,3)=M^{N\,\{12\}}_{ \tau \rho \chi }(2,\,1,\,3)\,; \ \ \
M^{N\,\{13\}}_{\rho \tau \chi}(1,\,2,\,3)=M^{N\,\{13\}}_{ \chi \tau \rho }(3,\,2,\,1)\,;
\nonumber \\
 &&
 M^{N\,\{23\}}_{\rho \tau \chi}(1,\,2,\,3)
=M^{N\,\{23\}}_{ \rho \chi \tau }(1,\,3,\,2)\,;  \nonumber \\
 &&
M^{N\,\{23\}}_{\rho \tau \chi}(1,2,3)=M^{N\,\{12\}}_{\tau  \chi \rho}(2,3,1)\,; \ \ \
M^{N\,\{13\}}_{\rho \tau \chi}(1,2,3)=M^{N\,\{12\}}_{\rho  \chi \tau}( 1, 3, 2)\,. \ \ \
\label{Relations_NDA_permutations}
\end{eqnarray}
The three first identities in
(\ref{Relations_NDA_permutations})
justify the symmetry properties of the combinations
of the invariant amplitudes defined in
(\ref{Isospin_parmetrization_N_DA}).
The two last identities express all invariant amplitudes in terms of a
single invariant amplitude
$M^{N\,\{12\}}$
with interchanged order of spatial arguments and the Dirac indices.

For the  leading twist-$3$ invariant amplitude
$M^{N\,\{12\}}$
symmetric under the exchange of the two first quark field operators, we
employ the standard parametrization
\begin{equation}
M^{N\,\{12\}}_{\rho \tau \chi}
(1,\,2,\,3)=
f_N
\Bigl[
V^p(1,\, 2, \, 3) v_{\rho \tau, \, \chi}^N+
A^p(1,\, 2, \, 3) a_{\rho \tau, \, \chi}^N+
T^p(1,\, 2, \, 3) t_{\rho \tau, \, \chi}^N
\Bigr]\,,
 \label{Decomposition_AVT_nucleon_DA}
\end{equation}
where $p$ is the  light-like vector ($p^2=0$) and  $\{v^N, a^N, t^N\}_{\rho \tau, \, \chi} $
are the conventional Dirac structures:
\begin{equation}
v_{\rho \tau, \, \chi}^N=(\hat{p} C)_{\rho \tau} (\gamma^5 U(p))_\chi\,;
\ \
a_{\rho \tau, \, \chi}^N=(\hat{p} \gamma^5 C)_{\rho \tau} ( U(p))_\chi
\,;
\ \
t_{\rho \tau, \, \chi}^N=(\sigma_{p \mu}  C)_{\rho \tau} ( \gamma^\mu \gamma^5 U(p))_\chi\,.
\label{DA structures}
\end{equation}

The symmetry relations
(\ref{symmetry_Dirac_Nucleon})
for the Dirac structures
(\ref{DA structures})
under the interchange of the two first Dirac indices together with
(\ref{Relations_NDA_permutations})
lead to the familiar symmetry properties:
\begin{equation}
V^p(1,2,3)=V^p(2,1,3)\,; \ \ \ T^p(1,2,3)=T^p(2,1,3)\,; \ \ \ A^p(1,2,3)=-A^p(2,1,3)\,.
\end{equation}
Finally, the symmetry relations
(\ref{Relations_NDA_permutations})
and isospin identity
(\ref{Isospin_Id_NDA})
together with the Fierz transformation properties
(\ref{Fierz_nucleon_structures})
of the Dirac structure set
(\ref{DA structures}),
result in the well known relation for twist-$3$ nucleon DAs~\cite{Avdeenko:1981twg,Chernyak:1984bm,Chernyak:1983ej}:
\begin{equation}
2T^p(1,2,3)= (V^p-A^p)(1,3,2)+(V^p-A^p)(2,3,1)\,.
\label{Isospi_ID_NDA}
\end{equation}
This reflects the fact that at leading twist-$3$ there is only one independent nucleon DA,
usually denoted as $\phi^N$:
\begin{equation}
\phi^N \equiv V^p-A^p\,.
\label{def_phi_N}
\end{equation}
 The DAs $V^p$, $A^p$ and $T^p$ are expressed through this latter function
according to
\begin{eqnarray}
 &&
2V^p(1,2,3)=   \phi^N(1,2,3)+ \phi^N(2,1,3)\,; \ \ \  2A^p(1,2,3)=   -\phi^N(1,2,3)+ \phi^N(2,1,3)\,;
\nonumber \\
 &&
2T^p(1,2,3)=\phi^N(1,3,2)+\phi^N(2,3,1)\,.
\end{eqnarray}

\subsubsection{Isospin and permutation symmetry identities for
$\Delta$-baryon {{DA}}}
\label{SubSec_DeltaDA_isospin}
\mbox

In this subsection we consider the application of the
isospin formalism to the case of $\Delta$-baryon.
With respect to the SU$(2)$ isospin group the $\Delta$-baryon
state is described by a spin-tensor
$| \Delta_{a\,\iota} \rangle$
with one covariant spinor
index $\iota=1,\,2$ and one vector (adjoint representation) index $a=1,\,2,\,3$.
It satisfies
\[
{P^{ I=\frac{3}{2}}}_{b a  \iota}^{\; \kappa} \, | \Delta_{b \kappa} \rangle=| \Delta_{a \iota} \rangle;
\ \ \
{P^{ I=\frac{1}{2}}}_{b  a  \iota}^{\; \kappa} \, | \Delta_{b \kappa}\rangle =0,\]
where
the explicit expressions for the isospin projecting operators read~\cite{SemenovTianShansky:2007hv}:
\begin{equation}
 {P^{ I=\frac{3}{2}}}_{b  a  \iota}^{\; \kappa}=
\frac{2}{3}
\bigl(
\delta_{ba} \delta^\kappa_{\; \iota}- \frac{i}{2} \varepsilon_{bac} (\sigma_c)^\kappa_{\; \iota}
\bigr)\,;
\ \ \
 {P^{ I= \frac{1}{2}}}_{b  a  \iota}^{\; \kappa}=
\frac{1}{3}
\bigl(
\delta_{ba} \delta^\kappa_{\; \iota} + i \varepsilon_{bac} (\sigma_c)^\kappa_{\; \iota}
\bigr)\,.
\label{isospin_projecting_oper}
\end{equation}

The isospin invariance suggests the following parametrization for the
matrix element of the three-quark operator
(\ref{Def_operator_O_isotopic})
between a $\Delta$-resonance state and the vacuum:
\begin{equation}
4 \langle 0 |  \widehat{O}^{\alpha \beta \gamma}_{\rho \tau \chi}(z_1,z_2,z_3)\,| \Delta_{a\,\iota}(p_\Delta) \rangle=
(f_a)^{\{\alpha \beta \gamma\}}_{\ \ \ \ \iota} \, M^\Delta_{\rho \tau \chi} ( 1, 2, 3)\,.
\label{isospin_parametrization_Delta_DA}
\end{equation}
Here
 $(f_a)^{\{\alpha \beta \gamma\}}_{\ \ \ \ \iota}$
stands for the only  tensor
one can construct out of the available tensor  structures
that is totally symmetric in the
three fundamental representation indices $\alpha$, $\beta$, $\gamma$:
\begin{equation}
(f_a)^{\{\alpha \beta \gamma\}}_{\ \ \ \ \iota}
 =
\frac{1}{3}
\left(
(\sigma_a)^{\alpha}_{\; \delta} \varepsilon^{\delta \beta} \delta^{\gamma}_{\; \iota}
+
(\sigma_a)^{\alpha}_{\; \delta} \varepsilon^{\delta \gamma} \delta^{\beta}_{\; \iota}+
(\sigma_a)^{\beta}_{\; \delta} \varepsilon^{\delta \gamma} \delta^{\alpha}_{\; \iota}
\right)\,.
\label{Def_ta_tensor}
\end{equation}

The convolutions of the symmetric isospin tensor
$(f_a)^{\{\alpha \beta \gamma\}}_{\ \ \ \ \iota}$
with the isospin projecting operators
(\ref{isospin_projecting_oper})
satisfy the following properties:
\begin{equation}
{P^{I=\frac{3}{2}}}^{\;\;  \kappa}_{b   a \iota}\,
(f_b)^{\{\alpha \beta \gamma\}}_{\ \ \ \ \kappa}=
(f_a)^{\{\alpha \beta \gamma\}}_{\ \ \ \ \iota}\,;
\ \ \
{P^{I=\frac{1}{2}}}^{\;\;  \kappa}_{b   a \iota}\,
(f_b)^{\{\alpha \beta \gamma\}}_{\ \ \ \ \kappa}=0\,.
\label{Projecting_ta_tensor}
\end{equation}

For the leading twist-$3$ invariant amplitude
$M^\Delta_{\rho \tau \chi} ( 1, 2, 3)$
we employ the parametrization\footnote{The  factor $-\frac{1}{\sqrt{2}}$
in (\ref{Parametrization_Delta_DA_FZ}) ensures matching with
the parametrization of~\cite{Farrar:1988vz} for the $uuu$ DA of $| \Delta^{++} \rangle$. Thus,
the DAs $V^\Delta$, $A^\Delta$, $T^\Delta$ and $\phi^{ \Delta_{3/2} }$
coincide with those
of Refs.~\cite{Farrar:1988vz,Braun:1999te}.}
\begin{eqnarray}
 &&
M^\Delta_{\rho \tau \chi} ( 1, 2, 3)
\nonumber \\ &&
=
-\frac{\lambda_\Delta^{\frac{1}{2}}}{\sqrt{2}}
\left\{
v^\Delta_{\rho \tau,\, \chi} V^\Delta(1,2,3)+ a^\Delta_{\rho \tau,\, \chi} A^\Delta(1,2,3)+
t^\Delta_{\rho \tau,\, \chi}T^\Delta(1,2,3)
\right\}-
\frac{f_\Delta^{\frac{3}{2}} }{\sqrt{2}} \varphi_{\rho \tau,\, \chi}^\Delta \, \phi^{ \Delta_{3/2} } (1,2,3)\,.
\label{Parametrization_Delta_DA_FZ}
\end{eqnarray}
Here
$(s^\Delta)_{\rho \tau,\, \chi} = \{v^\Delta, a^\Delta, t^\Delta, \varphi^\Delta \}_{\rho \tau,\, \chi} $
are the conventional  Dirac structures
\begin{eqnarray}
 &&
v^\Delta_{\rho \tau,\, \chi}= (\gamma_\mu C)_{\rho \tau} \, \mathcal{U}^\mu_\chi\,;
\ \ \
a^\Delta_{\rho \tau,\, \chi}= (\gamma_\mu \gamma_5 C)_{\rho \tau} \, (\gamma_5\mathcal{U}^\mu)_\chi\,;
\ \ \
t^\Delta_{\rho \tau,\, \chi}= \frac{1}{2} (\sigma_{\mu \nu} C)_{\rho \tau}(\gamma^\mu \mathcal{U}^\nu)_\chi\,;
\nonumber \\
 &&
\varphi_{\rho \tau,\, \chi}^\Delta= (\sigma_{\mu \nu} C)_{\rho \tau} (p^\mu \mathcal{U}^\nu- \frac{1}{2} m_\Delta \gamma^\mu \mathcal{U}^\nu)_\chi\,,
\label{Dirac_structures_Delta}
\end{eqnarray}
where $\mathcal{U}^\mu_\chi$ stands for the Rarita--Schwinger spin-tensor
describing the properties of a $\Delta$-baryon state with respect
to the Lorentz group.
The constants $\lambda_\Delta^{\frac{1}{2}}$,  $f_\Delta^{\frac{3}{2}}$ are defined in  Ref.~\cite{Farrar:1988vz}.

The familiar permutation properties of the invariant
amplitude $M^\Delta_{\rho \tau \chi} (1,2,3)$
can be established \textit{e.g.} from
the invariance of $\Delta^{++}$ $uuu$ DA under permutations of
the three $u$-quark fields.
This requires  to the complete symmetry of the invariant matrix element
under simultaneous permutations of
the arguments and of the Dirac indices:
\begin{equation}
 M^\Delta_{\rho \tau \chi} (1,2,3)
=
M^\Delta_{ \rho \chi \tau} (1,3,2)
=
M^\Delta_{\tau \rho \chi} (2,1,3)
  =
M^\Delta_{ \tau \chi \rho} (2,3,1)
=
M^\Delta_{  \chi \tau \rho} (3,2,1)
=
M^\Delta_{ \chi \rho  \tau} (3,1,2)
 \,.
 \label{symmetries_MI32}
\end{equation}

The implications of the relations
(\ref{symmetries_MI32})
for the invariant functions
$V^\Delta$, $A^\Delta$, $T^\Delta$
and
$\phi^{ \Delta_{3/2} }$
defined in
(\ref{Parametrization_Delta_DA_FZ})
can be established with the help of
the  symmetry properties
(\ref{symmetry_Dirac_structures_Delta})
and the twist-$3$ Fierz transformations (\ref{Fierz_Delta_structures})
for the Dirac structures
$(s^\Delta)_{\rho \tau,\, \chi}$.
To put these symmetry relations into a compact form it is
convenient to introduce
$\phi^{ \Delta_{1/2}} (1,2,3) \equiv V^\Delta(1,2,3)-A^\Delta(1,2,3)$
satisfying the  consistency condition
\begin{equation}
\phi^{\Delta_{1/2}}(1,2,3)=\phi^{\Delta_{1/2}}(3,2,1)\,.
\end{equation}
Then the relations can be written as
\begin{eqnarray}
 &&
2V^\Delta(1,2,3)=   \phi^{\Delta_{1/2}}(1,2,3)+\phi^{\Delta_{1/2}}(2,1,3)\,; \ \ \  2A^\Delta(1,2,3)=  -\phi^{\Delta_{1/2}}(1,2,3)+ \phi^{\Delta_{1/2}}(2,1,3)
\nonumber \\  &&
T^\Delta(1,2,3)=\phi^{\Delta_{1/2}}(2,3,1)\,.
\label{Isospin_and_Sym_Relations_DeltaDA}
\end{eqnarray}
Meanwhile, $\phi^{\Delta_{3/2}}(1,2,3)$ turns out to be totally symmetric
under permutation of its arguments.


\subsubsection{Isospin parametrization and permutation symmetry identities for nucleon-to-
$I=1$
 meson {{TDA}}s}
\label{SubSec_Isospin_I=1_meson_TDA}

In this subsection we present the isospin parametrization and permutation symmetry identities for ${\mathcal{M}} N$ TDAs, where ${\mathcal{M}}$ is a $I=1$ meson (\textit{e.g.} $\pi$, $\rho$). The corresponding meson state
$\langle {\mathcal{M}}_a|  $
transforms according to the adjoint representation
of the  SU$(2)$ isospin group.
 The isospin parametrization for
${\mathcal{M}} N$ TDAs then involves the isospin-$\frac{3}{2}$ and
isospin-$\frac{1}{2}$ invariant amplitudes and shares common features both with the
case of $\Delta$- and nucleon DAs.
It can be written as:
\begin{eqnarray}
 &&
4\langle  {\mathcal{M}}_a |  \widehat{O}^{\alpha \beta \gamma}_{\rho \tau \chi}(1,\,2,\,3) |  N_\iota \rangle
= (f_a)^{\{\alpha \beta \gamma \}}_{\ \ \ \ \iota}
M^{({\mathcal{M}} N)_{3/2}}_{\rho \tau \chi}
(1,2,3)
 +
\varepsilon^{\alpha \beta} (\sigma_a)^\gamma_{\ \iota} {M^{({\mathcal{M}} N)_{1/2}}_1}_{\rho \tau \chi}
(1,2,3) \nonumber \\ &&
+
\varepsilon^{\alpha \gamma} (\sigma_a)^\beta_{\ \iota} {M^{({\mathcal{M}} N)_{1/2}}_2}_{\rho \tau \chi}
(1,2,3)
+\varepsilon^{\beta \gamma} (\sigma_a)^\alpha_{\ \iota} {M^{({\mathcal{M}} N)_{1/2}}_3}_{\rho \tau \chi}
(1,2,3)
 \\
  &&
=(f_a)^{\{\alpha \beta \gamma\}}_{\ \ \ \ \iota} M^{({\mathcal{M}} N)_{3/2}}_{\rho \tau \chi}
(1,2,3)+
\varepsilon^{\alpha \beta} (\sigma_a)^\gamma_{\ \iota}  M^{({\mathcal{M}} N)_{1/2} \, \{1 3 \}}_{\rho \tau \chi}
(1,2,3)
+\varepsilon^{\alpha \gamma} (\sigma_a)^\beta_{\ \iota} M^{({\mathcal{M}} N)_{1/2} \, \{1 2 \}}_{\rho \tau \chi}
(1,2,3)\,,
\nonumber
\label{TDA_isospin_dec}
\end{eqnarray}
where
$(f_a)^{\{\alpha \beta \gamma\}}_{\ \ \ \ \iota}$
is the symmetric tensor defined in
(\ref{Def_ta_tensor}).
In the last line of
(\ref{TDA_isospin_dec}),
similarly to Eqs.~(\ref{Def_combinations_NDA}), we introduce the combinations
$M^{({\mathcal{M}} N)_{1/2}\,\{12\}}$,
$M^{({\mathcal{M}} N)_{1/2}\,\{13\}}$,
$M^{({\mathcal{M}} N)_{1/2}\,\{23\}}$
of the invariant isospin-$\frac{1}{2}$ amplitudes
$M^{({\mathcal{M}} N)_{1/2}}_{1,2,3}$
satisfying the isospin
identity
\begin{equation}
M^{({\mathcal{M}} N)_{1/2}\,\{12\}}_{\rho \tau \chi}(1,2,3)+
M^{({\mathcal{M}} N)_{1/2}\,\{13\}}_{\rho \tau \chi}(1,2,3)+
M^{({\mathcal{M}} N)_{1/2}\,\{23\}}_{\rho \tau \chi}(1,2,3)=0\,.
\label{Isospin_Id_piNTDA}
\end{equation}

The isospin-$\frac{1}{2}$ part is then treated according to the
pattern of the
nucleon DA providing the set of symmetry identities analogous to
(\ref{Relations_NDA_permutations}):
\begin{eqnarray}
 &&
M^{({\mathcal{M}} N)_{1/2}\,\{12\}}_{\rho \tau \chi}
(1,2,3)
=M^{({\mathcal{M}} N)_{1/2}\,\{12\}}_{ \tau \rho \chi }(2,1,3)\,; \ \ \
M^{({\mathcal{M}} N)_{1/2}\,\{13\}}_{\rho \tau \chi}(1,2,3)=M^{({\mathcal{M}} N)_{1/2}\,\{13\}}_{ \chi \tau \rho }(3,2,1)\,;
\nonumber \\
 &&
M^{({\mathcal{M}} N)_{1/2}\,\{23\}}_{\rho \tau \chi}(1,2,3)
=M^{({\mathcal{M}} N)_{1/2}\,\{23\}}_{ \rho \chi \tau }(1,3,2)\,;
\nonumber \\
 &&
M^{({\mathcal{M}} N)_{1/2}\,\{13\}}_{\rho \tau \chi}(1,2,3)=M^{({\mathcal{M}} N)_{1/2}\,\{12\}}_{\rho  \chi \tau}( 1, 3, 2)\,; \ \ \
M^{({\mathcal{M}} N)_{1/2}\,\{23\}}_{\rho \tau \chi}(1,2,3)=M^{({\mathcal{M}} N)_{1/2}\,\{12\}}_{\tau  \chi \rho}(2,3,1)\,.
\label{Relations_piNTDA_permutations}
\end{eqnarray}

Analogously to the case of $\Delta$-DA, Eq.~(\ref{symmetries_MI32}),
the isotopic and  permutation symmetries
for the isospin-$\frac{3}{2}$ invariant amplitude
$M^{({\mathcal{M}} N)_{3/2}}$
require it is symmetric under simultaneous permutations of arguments and the Dirac
indices:
\begin{eqnarray}
 &&
M^{({\mathcal{M}} N)_{3/2}}_{\rho \tau \chi} (1,2,3)
=
M^{({\mathcal{M}} N)_{3/2}}_{ \rho \chi \tau} (1,3,2)
=
M^{({\mathcal{M}} N)_{3/2}}_{\tau \rho \chi} (2,1,3)
 \nonumber \\ &&
=
M^{({\mathcal{M}} N)_{3/2}}_{ \tau \chi \rho} (2,3,1)
=
M^{({\mathcal{M}} N)_{3/2}}_{  \chi \tau \rho} (3,2,1)
=
M^{({\mathcal{M}} N)_{3/2}}_{ \chi \rho  \tau} (3,1,2)
 \,.
 \label{symmetries_MI32piNTDA}
\end{eqnarray}
Obviously the same isotopic parametrization applies to the case of
nucleon-$I=1$-meson GDAs (see  Sec.~\ref{SubSec_Definition_GDAs}).

To illustrate the consequences of the isospin and permutation symmetry relations
(\ref{Relations_piNTDA_permutations}),
(\ref{symmetries_MI32piNTDA}) we present in detail the case of $\pi N$ TDAs.
\bi
\item Proton-to-$\pi^0$ $uud$-, neutron-to-$\pi^0$ $ddu$-TDAs
of  Sec.~\ref{SubSubSec_Def_piN_TDAs}  are expressed through the invariant isospin
amplitudes
(\ref{TDA_isospin_dec})
as
\begin{eqnarray}
 &&
4\langle \pi^0 | \widehat{O}_{\rho \tau \chi}^{u u d}\left(1, \, 2, \,3 \right)
|  N_{p}\left(p_{N}\right)\rangle=
4\langle \pi^0 | \widehat{O}_{\rho \tau \chi}^{d d u}\left(1, \, 2, \,3 \right)
|  N_{n}\left(p_{N}\right)\rangle
\nn \\
&& = \frac{2}{3} M^{(\pi N)_{3/2}}_{\rho \tau \chi} (1,2,3)+
M^{(\pi N)_{1/2}}_{\rho \tau \chi} (1,2,3)\,.  \label{Isospin_uud_ppi0}
\end{eqnarray}
\item For proton-to-$\pi^+$ $ddu$-, and neutron-to-$\pi^-$ $uud$-TDAs
we get:
\begin{eqnarray}
 &&
4\langle \pi^+ | \widehat{O}_{\rho \tau \chi}^{d d u}\left(1, \, 2, \,3 \right)
|  N_{p}\left(p_{N}\right)\rangle=-
4\langle \pi^- | \widehat{O}_{\rho \tau \chi}^{u u d}\left(1, \, 2, \,3 \right)
|  N_{n}\left(p_{N}\right)\rangle \nn \\
&& = \frac{\sqrt{2}}{3} M^{(\pi N)_{3/2}}_{\rho \tau \chi} (1,2,3)- \sqrt{2} M^{(\pi N)_{1/2}}_{\rho \tau \chi} (1,2,3).
\label{Isospin_ddu_ppiplus}
\end{eqnarray}
\ei

Both for the isospin-$\frac{3}{2}$ part
$M^{(\pi N)_{3/2}}_{\rho \tau \chi}$
and for  the isospin-$\frac{1}{2}$ part
$ M^{(\pi N)_{1/2}}_{\rho \tau \chi}$
we employ the parametrization
(\ref{Param_TDAs_Covariant_DS})
with
the set of the fully covariant leading twist-$3$ Dirac structures
$(s^{\pi N})_{\rho \tau, \, \chi}$
(\ref{Dirac_structures_PiN_TDA_Cov}).

It is straightforward to check that the permutation symmetry relations
(\ref{Relations_piNTDA_permutations}),
(\ref{symmetries_MI32piNTDA})
result in the following symmetry
properties of the isospin-$\frac{1}{2}$ and isospin-$\frac{3}{2}$ $\pi N$ TDAs:
\begin{eqnarray}
 &&
V_{1,2}^{(\pi N)_{1/2,\,3/2}}(1,2,3)=V_{1,2}^{(\pi N)_{1/2,\,3/2}}(2,1,3)\,;  \ \ \
T_{1,2,3,4}^{(\pi N)_{1/2}}(1,2,3)=T_{1,2,3,4}^{(\pi N)_{1/2}}(2,1,3)\,;
\nonumber \\
 &&
A_{1,2}^{(\pi N)_{1/2,\,3/2}}(1,2,3)=-A_{1,2}^{(\pi N)_{1/2,\,3/2}}(2,1,3)\,; \ \ \
T_{1,2}^{(\pi N)_{3/2}}(1,2,3)=T_{1,2}^{(\pi N)_{3/2}}(2,1,3),
\label{sym_conditions_piNTDAs_isosp}
\end{eqnarray}
while
$T_{3,4}^{(\pi N)_{3/2}}$ are totally symmetric under the interchange of their
arguments.

\bi
\item
Let us first consider the isospin-$\frac{1}{2}$ part.  By analogy to the nucleon DA case
we introduce two independent isospin-$\frac{1}{2}$ $\pi N$ TDAs:
\begin{equation}
\phi^{(\pi N)_{1/2}}_{1,2}(1,2,3) \equiv V^{(\pi N)_{1/2}}_{1,2}(1,2,3)-A^{(\pi N)_{1/2}}_{1,2}(1,2,3)\,.
\label{independent1/2}
\end{equation}
Employing the Fierz transformations
(\ref{Fierz_for_piNTDA_structures}),  (\ref{Fierz for t34}),
one establishes the consequences of the isospin symmetry relation (\ref{Isospin_Id_piNTDA}) for
$T_{3,4}^{(\pi N)_{1/2}}$:
\begin{equation}
T_{3,4}^{(\pi N)_{1/2}}(1,2,3)+ T_{3,4}^{(\pi N)_{1/2}}(1,3,2)+ T_{3,4}^{(\pi N)_{1/2}}(2,3,1)=0
\label{symmetry_T34}
\end{equation}
and
\begin{eqnarray}
 &&
2 T_{1,2}^{(\pi N)_{1/2}}(1,2,3)
= \phi^{(\pi N)_{1/2}}_{1,2}(1,3,2) + \phi^{(\pi N)_{1/2}}_{1,2}(2,3,1) \nonumber \\ && +
2g_{1,2}(\xi,\Delta^2) T_3^{(\pi N)_{1/2}}(1,2,3) + 2h_{1,2}(\xi,\Delta^2) T_4^{(\pi N)_{1/2}}(1,2,3)\,;
\nonumber \\
 &&
2 V_{1,2}^{(\pi N)_{1/2}} (1,2,3)= \phi^{(\pi N)_{1/2}}_{1,2}(1,2,3)+\phi^{(\pi N)_{1/2}}_{1,2}(2,1,3)\,;
\nonumber \\
 &&
2 A_{1,2}^{(\pi N)_{1/2}}(1,2,3)= -\phi^{(\pi N)_{1/2}}_{1,2}(1,2,3)+\phi^{(\pi N)_{1/2}}_{1,2}(2,1,3)\,,
\label{VAT_piN_TDA_I12}
\end{eqnarray}
where
$g_{1,2}(\xi,\Delta^2)$,
$h_{1,2}(\xi,\Delta^2)$
are defined in (\ref{Fierz for t34}).

\item Now we treat the isospin-$\frac{3}{2}$ part.
We introduce two independent isospin-$\frac{3}{2}$ $\pi N$ TDAs:
\begin{equation}
\phi^{(\pi N)_{3/2}}_{1,2}(1,2,3) \equiv -V^{(\pi N)_{3/2}}_{1,2}(1,2,3)+A^{(\pi N)_{3/2}}_{1,2}(1,2,3)\,.
\label{independent3/2}
\end{equation}
These TDAs
satisfy the consistency conditions
\begin{equation}
\phi^{(\pi N)_{3/2}}_{1,2}(1,2,3) = \phi^{(\pi N)_{3/2}}_{1,2}(3,2,1)\,,
\label{consistency_for_isospin3/2TDAsym}
\end{equation}
following from (\ref{symmetries_MI32piNTDA}).
Now using
(\ref{sym_conditions_piNTDAs_isosp})
we express the isospin-$\frac{3}{2}$ $\pi N$ TDAs as
\begin{eqnarray}
 &&
2T_{1,2 }^{(\pi N)_{3/2}}(1,2,3)
\nonumber \\ && =\phi^{(\pi N)_{3/2}}_{1,2}(1,3,2) +
2g_{1,2}(\xi,\Delta^2) T_3^{(\pi N)_{3/2}}(1,2,3) + 2h_{1,2}(\xi,\Delta^2) T_4^{(\pi N)_{3/2}}(1,2,3)\,;
\nonumber \\
 &&
2 V^{(\pi N)_{3/2}}_{1,2}(1,2,3) = -\phi^{(\pi N)_{3/2}}_{1,2}(1,2,3)-\phi^{(\pi N)_{3/2}}_{1,2}(2,1,3)\,;
\nonumber \\
 &&
2 A^{(\pi N)_{3/2}}_{1,2}(1,2,3) =  \phi^{(\pi N)_{3/2}}_{1,2}(1,2,3)-\phi^{(\pi N)_{3/2}}_{1,2}(2,1,3)\,.
\label{Isospin_3/2_TDA_final_symmetry}
\end{eqnarray}
\ei

Thus the parametrization of  $\pi N$ TDAs  requires $8$ independent functions
($4$ for each, the isospin-$\frac{1}{2}$ and isospin-$\frac{3}{2}$, parts):
\begin{enumerate}
\item $\phi^{(\pi N)_{1/2}}_{1,2}$;
\item
$T_{3,4}^{(\pi N)_{1/2}}$
that satisfy the symmetry relations (\ref{symmetry_T34});
\item $\phi^{(\pi N)_{3/2}}_{1,2}$
that satisfy the symmetry relations (\ref{consistency_for_isospin3/2TDAsym});
\item $T_{3,4}^{(\pi N)_{3/2}}$ that are fully symmetric under interchanges
of their arguments.
\end{enumerate}


\subsection{Charge conjugation symmetry properties}
\label{SubSec_Charge_Conj}
\mbox

The application of the collinear factorization mechanism
involving nucleon-to-meson TDAs and nucleon DAs
to the cross-channel counterparts of the reaction (\ref{Hard_subpr}), see  Sec.~\ref{SubSec_Cross_Ch_Excl_R} requires to consider antinucleon DAs
and antinucleon-to-meson TDAs. They are defined through hadronic matrix elements
of the non-local three-antiquark operator on the light-cone
\begin{equation}
\widehat{\bar{O}}_{\alpha \beta \gamma \; \rho \tau \chi}(1,\,2,\,3)
 =
\varepsilon_{c_{1} c_{2} c_{3}}
\bar{\Psi}^{c_1 }_{\alpha \;\rho}(1)
\bar{\Psi}^{c_2 }_{\beta \; \tau}(2)
\bar{\Psi}^{c_3 }_{\gamma \;  \chi} (3).
\label{Def_OperatorObar}
\end{equation}
The notations in
(\ref{Def_OperatorObar}) are analogous to
(\ref{Def_operator_O_isotopic}):
$\alpha$, $\beta$, $\gamma$
stand for the antiquark SU$(2)$ flavor indices and
$\rho$, $\tau$, $\chi$
denote the Dirac spinor indices; $\varepsilon_{c_{1} c_{2} c_{3}}$ ensures
 antisymmetrization in  the color group indices $c_{1,2,3}$.
The gauge links in
(\ref{Def_OperatorObar}) are omitted in  the light-like gauge
$A^+=0$.

The charge conjugation symmetry allows to express the hadronic matrix
elements of the light-cone three-antiquark operator
(\ref{Def_OperatorObar})
through that of (\ref{Def_operator_O_isotopic}).

\subsubsection{Charge conjugation symmetry properties of antinucleon {{DA}}s}
\label{SubSec_AntiNucl_DA}
\mbox

In this subsection relying on the charge conjugation
symmetry properties we express the leading twist-$3$
antinucleon DAs in terms of the nucleon DAs.

The charge conjugation operator $\mathcal{C}$
acts on the antinucleon and nucleon states according to
\begin{equation}
\mathcal{C}| \bar{N}^\iota \rangle =  \eta_N | N_\iota \rangle ;
\ \ \
\mathcal{C}  | N_\iota \rangle  = \eta_N^{*} | \bar{N}^\iota \rangle ,
\label{CeffectstatesN}
\end{equation}
where
$\eta_N$
denotes the nucleon field charge parity ($| \eta_N| ^2=1$).

The effect of the charge conjugation operator on the quark field
$\Psi^\alpha$
is (see \textit{e.g.}~\cite{Itzykson}):
\begin{equation}
\mathcal{C} \Psi^\alpha \mathcal{C}^\dagger= \eta_q C \bar{\Psi}^T_\alpha,
\label{Cpar_quarks}
\end{equation}
where transposition refers to the Dirac index.
$C$ stands for the charge conjugation matrix and
$\eta_q$
is the charge parity of the quark field ($| \eta_q| ^2=1$).

Now employing (\ref{Cpar_quarks}) together with (\ref{CeffectstatesN})
we establish the link between the nucleon and antinucleon DAs:
\begin{eqnarray}
 &&
4\langle 0 |  \varepsilon_{c_1 c_2 c_3} \Psi^{c_1 \alpha}_\rho(1) \Psi^{c_2 \beta}_\tau(2) \Psi^{c_3 \gamma}_\chi(3)|  N_\iota(p) \rangle
 = 4\eta_N^{*} \langle 0|  \varepsilon_{c_1 c_2 c_3}
\mathcal{C}^\dagger \mathcal{C} \Psi^{c_1 \alpha}_\rho(1)
\mathcal{C}^\dagger \mathcal{C} \Psi^{c_2 \beta}_\tau(2)
\mathcal{C}^\dagger \mathcal{C} \Psi^{c_3 \gamma}_\chi(3) \mathcal{C}^\dagger | \bar{N}^\iota(p) \rangle
\nonumber \\ &&
= 4 \eta_N^{*} \eta_q^3
\langle 0|  \varepsilon_{c_1 c_2 c_3}
   \left( C\bar{\Psi}^{T \,c_1} _\alpha \right)_\rho
   \left( C\bar{\Psi}^{T \, c_2}_\beta \right)_\tau
   \left( C\bar{\Psi}^{T \, c_3}_\gamma \right)_\chi | \bar{N}^\iota(p) \rangle .
\label{Cconj_effect_NDA}
\end{eqnarray}

We define the antinucleon DA as the ${\rm SU}(2)$-isospin tensor:
\begin{equation}
4\langle 0 |  \varepsilon_{c_1 c_2 c_3}
  \left(  \bar{\Psi}^{  \,c_1} _\alpha \right)_\rho
   \left(  \bar{\Psi}^{  \, c_2}_\beta \right)_\tau
   \left(  \bar{\Psi}^{  \, c_3}_\gamma \right)_\chi | \bar{N}^\iota(p) \rangle
=
\varepsilon_{\alpha \beta} \delta_\gamma^\iota M^{ \bar{N} \{13\}}_{\rho \tau \chi}(1,2,3) +
\varepsilon_{\alpha \gamma} \delta_\beta^\iota M^{\bar{N} \{12\}}_{\rho \tau \chi}(1,2,3).
\end{equation}
The invariant amplitudes
$M^{\bar{N} \{12\}}$,
$M^{\bar{N} \{13\}}$
satisfy the set of  isospin and permutation symmetry
identities analogous to
(\ref{Relations_NDA_permutations}).
In particular,
\begin{equation}
M^{\bar{N} \{13\}}_{\rho \tau \chi}(1,2,3)=M^{\bar{N} \{12\}}_{\rho \chi \tau}(1,3,2).
\end{equation}
For $M^{\bar{N} \{12\}}$ we employ the parametrization
\begin{equation}
M^{\bar{N} \{12\}}_{\rho \tau \chi}(1,2,3)=
f_N
\left(
v_{\rho \tau, \chi}^{\bar{N}} V^{\bar{p}}(1,2,3)+ a_{\rho \tau, \chi}^{\bar{N}} A^{\bar{p}}(1,2,3) +
t_{\rho \tau, \chi}^{\bar{N}} T^{\bar{p}}(1,2,3)
\right)\,,
 \label{AntiPVAT}
\end{equation}
where the relevant Dirac structures are:
\begin{eqnarray}
 &&
(v_{\rho \tau, \chi}^{\bar{N}})^T \equiv (C^\dagger)_{\rho \rho'} (C^\dagger)_{\tau \tau'} (C^\dagger)_{\chi \chi'} v_{\rho' \tau', \chi'}^{N}=
(C \hat{p})^T_{\rho \tau} (\bar{V} \gamma^5)^T_\chi;
\nonumber \\
 &&
(a_{\rho \tau, \chi}^{\bar{N}})^T \equiv(C^\dagger)_{\rho \rho'} (C^\dagger)_{\tau \tau'} (C^\dagger)_{\chi \chi'} a_{\rho' \tau', \chi'}^{N}=
(C \gamma_5 \hat{p})^T_{\rho \tau}  (\bar{V} )^T_\chi
\nonumber \\
 &&
(t_{\rho \tau, \chi}^{\bar{N}})^T \equiv(C^\dagger)_{\rho \rho'} (C^\dagger)_{\tau \tau'} (C^\dagger)_{\chi \chi'} t_{\rho' \tau', \chi'}^{N}=
(C \sigma_{p \mu})^T_{\rho \tau}  (\bar{V} \gamma^\mu \gamma_5)^T_\chi\,.
\label{DiracStrBarNT}
\end{eqnarray}
We may lift the transposition with respect to the Dirac indices in
(\ref{DiracStrBarNT})
and obtain:
\begin{eqnarray}
 &&
v_{\rho \tau, \chi}^{\bar{N}}= (C \hat{p})_{\rho \tau} (\bar{V} \gamma^5)_\chi; \nonumber \\
 &&
a_{\rho \tau, \chi}^{\bar{N}}=(C \hat{p} \gamma_5 )_{\rho \tau}  (\bar{V} )_\chi; \nonumber \\
 &&  t_{\rho \tau, \chi}^{\bar{N}}=(C \sigma_{p \mu})_{\rho \tau}  (\bar{V} \gamma^\mu \gamma_5)_\chi.
\label{Def_DirStr_AntiN}
\end{eqnarray}
The symmetry properties of the Dirac structures
(\ref{Def_DirStr_AntiN}) with respect to the permutation
of the Dirac indices are analogous to (\ref{symmetry_Dirac_Nucleon}).
Therefore, from (\ref{Cconj_effect_NDA}) we conclude that, as a consequence of charge conjugation invariance, the leading twist-$3$ antiproton DAs defined in
(\ref{AntiPVAT}) are expressed  through the usual twist-$3$ proton DAs (\ref{Decomposition_AVT_nucleon_DA}) as:
\begin{equation}
\{V, \, A, \, T\}^{\bar{p}}(y_1,y_2,y_3)=\frac{1}{\eta_N^{*} \eta_q^3} \{V, \, A, \, T\}^{p}(y_1,y_2,y_3).
\end{equation}

\subsubsection{Charge conjugation symmetry properties of antinucleon-to-pion {{TDA}}s}
\mbox

Below we establish the relations between the antinucleon-to-pion and nucleon-to-pion TDAs.\footnote{This derivation also applies to arbitrary $I=1$ mesons, \textit{e.g.} $\rho^{\pm, 0}$.}
The derivation is analogous to  the case of the antinucleon DAs,  Sec.~\ref{SubSec_AntiNucl_DA}.
However, since the charged pions are not eigenstates of the charge conjugation operator
$\mathcal C$,
to work out the isotopic formalism for the antinucleon-to-pion TDAs
it turns out to be convenient
to employ the concept of  $G$-parity (see \textit{e.g.}~\cite{Ericson_Weise}).
The $G$-parity operation is defined as the combination of  charge conjugation and a
rotation around the second axis in the isospin space:
\begin{equation}
{\mathcal{G}}= {\mathcal{C}} e^{i \pi I_2}.
\end{equation}

First, we specify the effect of $\mathcal G$ on the quark and antiquark fields $\Psi^\alpha$:
\begin{equation}
{\mathcal{G}}  \Psi^\alpha {\mathcal{G}}^\dagger= {\mathcal{C}} e^{i \pi I_2} \Psi^\alpha e^{-i \pi I_2} {\mathcal{C}}^\dagger\,.
\end{equation}
According to our conventions of~\cite{Pire:2011xv},
we choose to transform the
$\bar{\Psi}_\alpha$
field  according to the covariant representation and to transform the
$\Psi^\alpha$
field  according to the contravariant representation of the isospin $  {\rm SU} (2)$.

The effect of the isospin rotation around the second axis then reads:
\begin{equation}
e^{i \pi I_2} (\Psi^\alpha)_\rho e^{-i \pi I_2}=  \left( \cos \frac{\pi}{2} \delta^\alpha_{\;  \beta} + \sin \frac{\pi}{2} (-i \sigma_2)^\alpha_{\; \beta}  \right) (\Psi^\beta)_\rho,
\end{equation}
where $\sigma_a$
are the Pauli matrices.
Thus, we conclude that
\begin{equation}
{\mathcal{G}} u_\rho {\mathcal{G}}^\dagger= \eta_q (-1) (C\bar{d})^T_\rho; \ \ \;{\mathcal{G}} d_\rho {\mathcal{G}}^\dagger= \eta_q   (C\bar{u})^T_\rho\,.
\label{transfud}
\end{equation}

We  also need to specify the $G$-parity properties  of the  pion ``out'' state and nucleon ``in'' state.
The triplet of pions has negative $G$ parity:
\begin{equation}
\langle \pi|   {\mathcal{G}}^\dagger= - \langle \pi|  .
\label{transpi}
\end{equation}
Employing our isospin conventions for the nucleon states
we conclude that:
\begin{equation}
{\mathcal{G}} | N_p \rangle = \eta_N^{*} (-1) | \bar{N}^{\bar{n}} \rangle ; \ \ \
{\mathcal{G}} | N_n \rangle = \eta_N^{*}  | \bar{N}^{\bar{p}} \rangle .
\label{transpn}
\end{equation}

Analogously to Eq.~(\ref{TDA_isospin_dec}),
for the antinucleon-to-pion TDAs we employ the
parametrization
\begin{eqnarray}
 &&
4\langle  \pi_a |  \hat{\bar{O}}_{\alpha \beta \gamma \; \rho \tau \chi}(1,\,2,\,3) |  \bar{N}^\iota \rangle
\nonumber \\
&&= (\bar{f}_a)_{\{\alpha \beta \gamma \}}^{\iota}
M^{(\pi \bar{N})_{3/2}}_{\rho \tau \chi}
(1,2,3)
+
\varepsilon_{\alpha \beta} (\sigma_a)^\iota_{\ \gamma}  M^{(\pi \bar{N})_{1/2} \, \{1 3 \}}_{\rho \tau \chi}
(1,2,3)
+\varepsilon_{\alpha \gamma} (\sigma_a)^\iota_{\ \beta} M^{(\pi \bar{N})_{1/2} \, \{1 2 \}}_{\rho \tau \chi}
(1,2,3)\,,
\label{BarNpi_isospin_dec}
\end{eqnarray}
where the totally symmetric in $\alpha$, $\beta$, $\gamma$ tensor $(\bar{f}_a)_{\{\alpha \beta \gamma\}}^{  \iota}$ reads:
\begin{eqnarray}
 &&
(\bar{f}_a)_{\{\alpha \beta \gamma\}}^{  \iota}
=
\frac{1}{3}
\left(
(\sigma_a^T)_{\alpha}^{\; \; \delta} \, \varepsilon_{\delta \beta} \, \delta^{\iota}_{\; \; \gamma}
+
(\sigma_a^T)_{\alpha}^{\; \delta} \varepsilon_{\delta \gamma} \delta^{\iota}_{\; \beta}+
(\sigma_a^T)_{\beta}^{\; \delta} \varepsilon_{\delta \gamma} \delta^{\iota}_{\; \alpha}
\right)\,,
\nonumber \\
 && {\rm  since} \ \ \
(\sigma_a^T)_{\alpha}^{\; \delta}=(\sigma_a)_\alpha^{\; \; \delta} \equiv \varepsilon_{\alpha \kappa}\, (\sigma_a)^\kappa_{\; \; \theta} \, \varepsilon^{\theta \delta}\,.
\label{Def_ta_tensor_bar}
\end{eqnarray}

The isospin and permutation symmetry identities for the  isospin-$\frac{1}{2}$
and isospin-$\frac{3}{2}$ invariant  $\pi \bar{N}$ TDAs
$ M^{(\pi \bar{N})_{1/2} \, \{1 2 \}}$,
$ M^{(\pi \bar{N})_{1/2} \, \{1 3 \}}$
and
$ M^{(\pi \bar{N})_{3/2}}$
are the same as those for the relevant $\pi N$ TDAs
(see  Sec.~\ref{SubSec_Isospin_I=1_meson_TDA}).
In particular,
\begin{equation}
M^{(\pi \bar{N})_{1/2} \, \{1 3 \}}_{\rho \tau \chi}(1,2,3)=
M^{(\pi \bar{N})_{1/2} \, \{1 2 \}}_{\rho  \chi \tau}(1,3,2).
\end{equation}

Now,
using $G$-parity operator $\mathcal G$ instead of $\mathcal{C}$,
we repeat the derivation of  Sec.~\ref{SubSec_AntiNucl_DA} and establish
the link between $\pi N$ and $\pi \bar{N}$ TDAs.
For example, let us consider neutron-to-$\pi^-$ $uud$ TDA:
\begin{eqnarray}
 &&
\langle \pi^-| u_\rho (1)u_\tau(2)d_\chi(3)|  N_n \rangle=
 -\frac{\sqrt{2}}{3} M^{(\pi N)_{3/2}}(1,2,3)+ \sqrt{2}  M^{(\pi N)_{1/2}}(1,2,3)
 \nonumber \\
 &&=
\langle \pi^-|  {\mathcal{G}}^\dagger {\mathcal{G}} u_\rho (1) {\mathcal{G}}^\dagger {\mathcal{G}} u_\tau(2)
{\mathcal{G}}^\dagger {\mathcal{G}} d_\chi(3) {\mathcal{G}}^\dagger {\mathcal{G}} |  N_n \rangle \nonumber \\ &&=(-1)     \eta_q^3 \eta_N^{*}
 (C)_{\rho \rho'} (C)_{\tau \tau'} (C)_{\chi \chi'}
 \langle \pi^-|  (\bar{d}_{\rho'})^T (1)  (\bar{d}_{\tau'})^T (2)  (\bar{u}_{\chi'})^T(3) |  \bar{N}^{\bar{p}} \rangle
  \nonumber \\
  &&=
  \eta_q^3 \eta_N^{*} (C)_{\rho \rho'} (C)_{\tau \tau'} (C)_{\chi \chi'}
  \left\{
  -\frac{\sqrt{2}}{3}  \left( M^{(\pi \bar{N})_{3/2}}_{\rho' \tau' \chi'}(1,2,3) \right)^T+ \sqrt{2}
   \left( M^{(\pi \bar{N})_{1/2}}_{\rho' \tau' \chi'}(1,2,3) \right)^T
  \right\}.
\end{eqnarray}

The  parametrization for $\pi \bar{N}$ TDAs (\ref{BarNpi_isospin_dec})
is consistent with that for $\pi N$ TDAs Eq.~(\ref{TDA_isospin_dec})
once
\begin{eqnarray}
  &&
  \left( M^{(\pi \bar{N})_{3/2}}_{\rho \tau \chi}(1,2,3) \right)^T = \frac{1}{\eta_N^{*} \eta_q^3}
  (C^\dagger)_{\rho \rho'} (C^\dagger)_{\tau \tau'} (C^\dagger)_{\chi \chi'}
  M^{(\pi N)_{3/2}}_{\rho' \tau' \chi'}(1,2,3);
\nonumber \\
 &&
 \left( M^{(\pi \bar{N})_{1/2}}_{\rho \tau \chi}(1,2,3) \right)^T = \frac{1}{\eta_N^{*} \eta_q^3}
  (C^\dagger)_{\rho \rho'} (C^\dagger)_{\tau \tau'} (C^\dagger)_{\chi \chi'}
  M^{(\pi N)_{1/2}}_{\rho' \tau' \chi'}(1,2,3),
\end{eqnarray}
where the transposition refers to the Dirac indices.

The Dirac structures
$s^{\pi \bar{N}} \equiv \{v^{\pi \bar{N}}_{1,2}, \, a^{\pi \bar{N}}_{1,2}, \, t^{\pi \bar{N}}_{1,2,3,4} \} $ occurring in the parametrization of
the isospin-$\frac{3}{2}$ and isospin-$\frac{1}{2}$ invariant amplitudes
$M^{(\pi \bar{N})_{3/2}}$ and $M^{(\pi \bar{N})_{1/2}}$
are defined by
\begin{equation}
\left(
s^{\pi \bar{N}}_{\rho \tau, \chi}
\right)^T= (C^\dagger)_{\rho \rho'} (C^\dagger)_{\tau \tau'} (C^\dagger)_{\chi \chi'} s_{\rho' \tau', \chi'}^{ \pi N},
\end{equation}
where
$s_{\rho' \tau', \chi'}^{ \pi N}$
are the Dirac structures occurring in the parametrization of $\pi N$ TDAs.
For the relevant $\pi \bar{N}$  TDAs
we get
\begin{equation}
\bigl\{V_{1,2}^{\pi N},\, A_{1,2}^{\pi N}, \, T_{1,2,3,4}^{\pi\bar{N}} \bigr\}
(x_1,x_2,x_3, \xi, \Delta^2)= \frac{1}{\eta_N^{*} \eta_q^3}
\bigl\{V_{1,2}^{\pi N},\, A_{1,2}^{\pi N}, \, T_{1,2,3,4}^{\pi N} \bigr\}
(x_1,x_2,x_3, \xi, \Delta^2).
\end{equation}

\subsection{{{QCD}} evolution equations for {{TDA}}s}
\label{SubSec_Evolution}
\mbox

QCD radiative corrections, as usual, lead  to logarithmic scaling violations
and must be treated consistently with the help of  corresponding evolution
equations.
The scale dependence of nucleon-to-meson TDAs is governed by
the evolution equations which turn to be an extension of the evolution equations for
baryon DAs~\cite{Efremov:1979qk,Lepage:1979zb,Brodsky:1981rp}
and GPDs
\cite{Mueller:1998fv,Ji:1996ek,Ji:1996nm,Radyushkin:1997ki,Blumlein:1997pi}.
In this section, following mainly Ref.~\cite{Braun:1999te}, we review
the evolution properties of the  non-local three-quark light-cone operators occurring
in the definition of baryon DAs and nucleon-to-meson TDAs.
We consider the implication of the conformal symmetry for solution of the evolution equations for nucleon DAs. We also provide an explicit form of the evolution equations for nucleon-to-pion TDAs derived in
Ref.~\cite{Pire:2005ax}.
Finally, we introduce the conformal partial wave (PW) expansion of nucleon-to-meson TDAs. We speculate on the perspectives of adopting the methods of handling the conformal PW expansion for GPDs to the case of TDAs.

\subsubsection{Evolution properties of the three-quark operator}
\label{SubSubSec_EvProp_3q}
\mbox

The  non-local three-quark operators relevant for TDAs and
their evolution involve quark fields with  definite
chirality
\begin{equation}
\label{arrows1}
\Psi^{\uparrow }= \frac{1}{2}\left(1 + \gamma^5  \right) \Psi; \ \ \
\Psi^{\downarrow}= \frac{1}{2}\left(1 - \gamma^5  \right) \Psi.
\end{equation}
The separation of the ``minus'' components of quark fields
leading to the dominant  twist-$3$ contribution is
 achieved by the substitution $q \to \hat n q$, with $\hat n=n^\mu \gamma_\mu$.
The  two relevant operators~\cite{Balitsky:1987bk}
correspond to the cases when the three quarks have
total chirality-$\frac{1}{2}$:
\begin{equation}
B^{\frac{1}{2}}_{\rho \tau \chi}(z_1,z_2,z_3)= \varepsilon^{c_1 c_2 c_3}(\hat n \Psi_{c_1}^{\uparrow})_\rho(z_1 n)
(\hat n \Psi_{c_2}^\downarrow)_\tau(z_2 n) (\hat n \Psi_{c_3}^\uparrow)_\chi(z_3 n);
\label{relop1}
\end{equation}
and  the total chirality-$\frac{3}{2}$:
\begin{equation}
B^{\frac{3}{2}}_{\rho \tau \chi}(z_1,z_2,z_3)= \varepsilon^{c_1 c_2 c_3}(\hat n \Psi_{c_1}^\uparrow)_\rho(z_1 n)
(\hat n \Psi_{c_2}^\uparrow)_\tau(z_2 n) (\hat n \Psi_{c_3}^\uparrow)_\chi(z_3 n)\;.
\label{relop3}
\end{equation}
Here $c_{1,2,3}$ stand for the color indices and $\rho$, $\tau$, $\chi$ are the Dirac spinor
indices.

Since the operators (\ref{relop1}), (\ref{relop3}) belong to different representations of the
Lorentz group they do not mix with each other under evolution.
The conditions imposed by the flavor symmetry  have no influence on the evolution equations since,
as explained in  Sec.~\ref{SubSec_Isospin}, they  result in
certain  symmetry properties of baryon DAs and nucleon-to-meson TDAs.
Therefore, for simplicity, we  assume  that all quarks in (\ref{relop1}), (\ref{relop3}) have different flavors.

\begin{figure}[H]
\begin{center}
\includegraphics[width=0.25\textwidth]{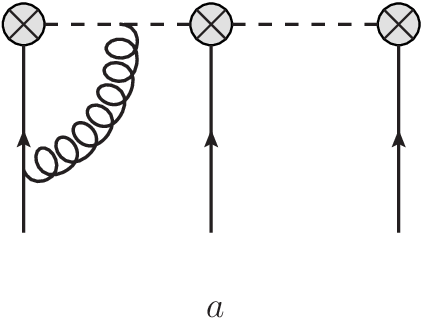} \ \ \ \ \ \ \ \ \
\includegraphics[width=0.25\textwidth]{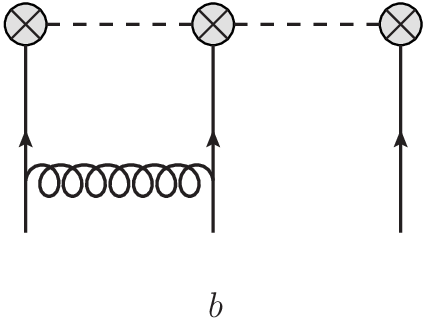} \ \ \ \ \ \ \ \ \
\includegraphics[width=0.25\textwidth]{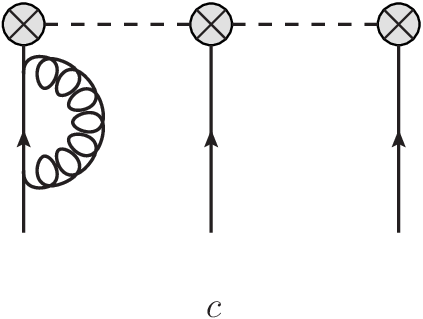}
\end{center}
     \caption{Examples of one-loop diagrams relevant for the evolution of the non-local three-quark operators (\ref{relop1}), (\ref{relop3}) in the Feynman gauge. {\bf a:} ``vertex correction'';
    {\bf b:}  ``exchange diagram''  {\bf c:} ``self energy correction''. Dashed lines represent path ordered gauge factors. The complete set of diagrams involves all possible permutations.}
\label{Fig_Evolution_graphs}
\end{figure}

The $B$ operators, in (\ref{relop1}), (\ref{relop3}), satisfy the renormalization group equation that can be put into the form~\cite{Bukhvostov:1985rn,Balitsky:1987bk}:
\begin{equation}
\label{rengr}
\mu \frac{d}{d \mu} B=
\left( \mu \frac{\partial }{\partial\,\mu} + \beta(g) \frac{\partial}{\partial g} \right)\;B = \mathbb{H} \cdot B \;,
\end{equation}
with $\mathbb{H}$ being the integral evolution
operator~\cite{Braun:1999te}:
\begin{equation}
\label{intop_Sec4}
\mathbb{H} = -\frac{\alpha_s}{2\pi}\left[ \left( 1+1/N_c \right) {\mathcal{H}} + 3C_F/2 \right];
\end{equation}
and $\beta(g)$ standing for the QCD $\beta$-function.
Here we employ the shortened notations of~\cite{Braun:1999te}
in which the color factors and the trivial contributions of the self
energy insertions are factored out. However,
in Eq.~(\ref{intop_Sec4}) we restored  the factor $-\frac{\alpha_s}{2\pi}$
that was not presented explicitly in Eq.~(2.23) of Ref.~\cite{Braun:1999te}.
The second term in (\ref{intop_Sec4}), proportional to  $C_F \equiv \frac{N_c^2-1}{2 N_c}$, corresponds to the self-energy corrections of each quark field (see  Fig.~\ref{Fig_Evolution_graphs}.c).

The operator ${\mathcal{H}}$  introduced in (\ref{intop_Sec4})
acts in a different way on the chirality-$\frac{3}{2}$ operator (\ref{relop3}) and on the
chirality-$\frac{1}{2}$ operator (\ref{relop1}).
\bi
\item In the case of the chirality-$\frac{3}{2}$ operator (\ref{relop3}),
to one loop order accuracy,
the operator ${\mathcal{H}}$
is determined by the  vertex correction diagrams (see  Fig.~\ref{Fig_Evolution_graphs}.a) corresponding to gluon exchanges between
quark lines and gluons forming Wilson lines. According to calculations of
Refs.~\cite{Lepage:1979za,Bukhvostov:1985rn,Peskin:1979mn} it reads:
\begin{equation}
\label{H3}
{\mathcal{H}}_{3/2} = {\mathcal{H}}^v_{1\,2} + {\mathcal{H}}^v_{2\,3}  +  {\mathcal{H}}^v_{1\,3} \;,
\end{equation}
where
\begin{eqnarray}\label{Hv12_Sec4}
&& {\mathcal{H}}^v_{1\,2} B(z_1,z_2,z_3) = - \int_0^1 \frac{d\alpha}{\alpha} \left\{
\bar \alpha\left[B(z^\alpha_{1\,2},z_2,z_3)- B(z_1,z_2,z_3)    \right] \right.
\nonumber \\
&& \left. + \bar \alpha \left[B(z_1,z^\alpha_{2\,1},z_3)- B(z_1,z_2,z_3) \right] \right\}\;,
\end{eqnarray}
with
\begin{equation}
\bar \alpha \equiv 1-\alpha; \ \ \  z^\alpha_{i\,k} \equiv z_i\bar \alpha +z_k \alpha.
\end{equation}

\item In the case of the chirality-$\frac{1}{2}$ operator (\ref{relop1})
the operator ${\mathcal{H}}$ is determined not only by the above
contributions but also by those  which correspond to Feynman diagrams with gluon
exchange between quark lines having opposite chiralities (see
 Fig.~\ref{Fig_Evolution_graphs}~b).
It can be written as
\begin{equation}
\label{H1_Sec4}
{\mathcal{H}}_{1/2} = {\mathcal{H}}_{3/2} - {\mathcal{H}}^e_{1\,2} - {\mathcal{H}}^e_{2\,3}\;,
\end{equation}
where
\begin{equation}
\label{He12_Sec4}
 {\mathcal{H}}^e_{1\,2} B(z_1,z_2,z_3) =
 \int_0^1
d_3 \alpha
 \;B(z^{\alpha_1}_{1\,2},z^{\alpha_2}_{2\,1},z_3)\;,
\end{equation}
with
\begin{equation}
\label{measure}
\int_0^1 d_3\alpha \equiv \int_0^1 d\alpha_1\, \int_0^1   d\alpha_2\,
\int_0^1
d\alpha_3\,
\delta(1- \alpha_1 - \alpha_2 - \alpha_3)\;.
\end{equation}
\ei

By performing the Taylor expansion of the generating functional at small distances
one switches to the local operators
\begin{equation}
B\left(z_{1}, z_{2}, z_{3}\right)=\sum_{N} \sum_{n_{1}+n_{2}+n_{3}=N} \frac{z_{1}^{n_{1}}}{n_{1} !} \frac{z_{2}^{n_{2}}}{n_{2} !} \frac{z_{3}^{n_{3}}}{n_{3} !}
(\overrightarrow{D}^+)^{n_{1}} \Psi(0)( \overrightarrow{D}^+)^{n_{2}} \Psi(0)(\overrightarrow{D}^+)^{n_{3}} \Psi(0),
\label{Loc_exp_B_operator}
\end{equation}
where $\overrightarrow{D}^+ \equiv \overrightarrow{D}^\mu \, n_\mu$  is the covariant derivative (\ref{Def_cov_der}).
Since the total number of derivatives $N$ is invariant under  the
evolution equation
(\ref{rengr}), (\ref{intop_Sec4})
takes a matrix form, with a square matrix of size $\frac{N(N+1)}{2}$
in a subspace corresponding to each given $N$.

The study of the evolution properties of TDAs
can then be performed by considering hadronic matrix elements of
the type
\begin{equation}
\langle A |  \mathcal{O}_{\Psi}|   B\rangle=\Psi\left(\partial_{1}, \partial_{2}, \partial_{3}\right)  \langle A |  B\left(z_{1}, z_{2}, z_{3}\right)| B\rangle \Big|_{z_{i}=0},
\end{equation}
where $\Psi\left(\partial_{1}, \partial_{2}, \partial_{3}\right)$
is a certain derivative operator.

First we review an approach for solving the ERBL-type evolution equations
of the baryon distribution amplitudes.
It turns out to be convenient to
reformulate this problem in the space of coefficient functions by representing
the sums of monomials in (\ref{Loc_exp_B_operator})
\begin{equation}
\mathcal{O}=\sum_{n_{1}+n_{2}+n_{3}=N} c_{n_{1}, n_{2}, n_{3}}
(\overrightarrow{D}^+)^{n_{1}} \Psi(0)( \overrightarrow{D}^+)^{n_{2}} \Psi(0)(\overrightarrow{D}^+)^{n_{3}} \Psi(0)
\end{equation}
by polynomials in three variables
\begin{equation}
\Psi\left(y_{1}, y_{2}, y_{3}\right)=\sum_{n_{1}+n_{2}+n_{3}=N} c_{n_{1} n_{2} n_{3}} y_{1}^{n_{1}} y_{2}^{n_{2}} y_{3}^{n_{3}}.
\end{equation}
The problem of solving the evolution equation can be reduced to the eigenvalue problem
\begin{equation}
\mathcal{H} \cdot \Psi_{N, q}=\mathcal{E}_{N, q} \Psi_{N, q},
\label{Eigenvalue_problem_H}
\end{equation}
where for given $N$ the index $q$ refers to the number of the ``energy level''.
The eigenvalues $\mathcal{E}_{N, q}$ are in a direct correspondence
with the anomalous dimensions:
\begin{equation}
\gamma_{N, q} \equiv\left(1+1/N_{c}\right) \mathcal{E}_{N, q}+3/2 C_{F}.
\end{equation}
In the next subsection we review the method  of solving
(\ref{Eigenvalue_problem_H}) based on the conformal partial wave expansion~\cite{Braun:1999te}.

\subsubsection{Conformal {{PW}} expansion for leading twist baryon {{DA}}s: basic definitions}
\label{SubSec_Conformal_PW}
\mbox

The methods based on exploiting the conformal symmetry of the Lagrangian of massless
QCD (see Ref.~\cite{Braun:2003rp} for a review) turned out to be extremely efficient in solving the evolution Eqs.~(\ref{Eigenvalue_problem_H}). The evolution operator $\mathcal H$ commutes with the generators of the
conformal group.
Therefore, the conformal symmetry strongly constrains the set of the eigenfunctions of
the evolution operator. In the case of the two-body operator the conformal symmetry
completely determines the solution of the evolution equations. However, in the case of
three-body operators  the conformal symmetry is not sufficient to uniquely determine the solution. It provides the solution up to an arbitrary function of one variable.

To take full advantage of the consequences of conformal symmetry
for the   general solution of Eq.~(\ref{Eigenvalue_problem_H})
one decomposes the corresponding eigenfunctions
over a suitable basis of conformal polynomials. There are several choices for conformal
basis. For example, in Refs.~\cite{Peskin:1979mn}
an expansion over the set of the Appell polynomials
\cite{Bateman:100233}
was employed.
However, a better suited basis was proposed in
\cite{Braun:1999te}.
In the following we adopt this latter form of the conformal basis.
The conformal expansion is performed with respect to the functions
$\Psi^{(12)3}_{N,n}$:
\begin{equation}
\Psi^{(12)3}_{N,n}(y_1,y_2,y_3)=(N+n+4)(y_1+y_2)^n P_{N-n}^{(2n+3,1)}(y_3-y_1-y_2) C_n^{\frac{3}{2}} \left( \frac{y_1-y_2}{y_1+y_2} \right).
\label{Conf_basis_Braun99}
\end{equation}
Here $ P_{N-n}^{(2n+3,1)}(\ldots)$ stand for the Jacobi polynomials and $C_n^{\frac{3}{2}}(\ldots)$ are the Gegenbauer polynomials
\cite{Bateman:100233}.

The superscript $(12)3$
refers to the order in which the conformal spins of the three quarks are
summed to form the total conformal spin $N + 3$.
First the conformal spins of the first and second quarks are summed
to form the total conformal spin $n + 2$, and then the conformal spin of the third quark is added.
This order is of
course arbitrary. The functions
$\Psi^{(12)3}_{N,n}$,
$\Psi^{1(23)}_{N,n}$
and
$\Psi^{(31)2}_{N,n}$
are related with each
other through the Racah-$6j$ symbols of the SL$(2)$ group:
\begin{eqnarray}
 &&
\Psi^{(32)1}_{N,n}(y_1,y_2,y_3)=(N+n+4)(y_3+y_2)^n P_{N-n}^{(2n-3,1)}(y_1-y_3-y_2) C_n^{\frac{3}{2}} \left( \frac{y_3-y_2}{y_3+y_2} \right)
\nonumber \\ &&
=\sum_{0 \le n' \le N} \Omega_{n n'}(N) \Psi^{(12)3}_{N,n'}(y_1,y_2,y_3).
\label{Conf_Basis_Reparam}
\end{eqnarray}
The explicit form of the matrices $\Omega_{n n'}(N)$ and additional details can be
found in  Appendix B of Ref.~\cite{Braun:2001qx}.

The conformal basis
(\ref{Conf_basis_Braun99})
is normalized according to
\begin{eqnarray}
&&
\langle \Psi^{(12)3}_{N,n}
\vert  \Psi^{(12)3}_{M,m}
\rangle
\nonumber \\ &&
\equiv 120 \int_0^1 dy_1 \int_0^1 dy_2 \int_0^1 dy_3 \delta(1-y_1-y_2-y_3) y_1 y_2 y_3 \Psi^{(12)3}_{N,n}(y_1,y_2,y_3) \Psi^{(12)3}_{M,m}(y_1,y_2,y_3)
\nonumber \\  &&
= \delta_{MN} \delta_{mn} \frac{60}{2N+5} f_{N,n},
\end{eqnarray}
where
$
f_{N,n}= \frac{(n+1)(n+2)}{2(2n+3)}(N-n+1)(N+n+4).
$

The conformal PW expansion of the leading twist-$3$ nucleon DA
$\phi_N$ within the $(12)3$ basis
(\ref{Conf_basis_Braun99})
reads
\begin{equation}
\phi_N(y_1,y_2,y_3)=120 y_1 y_2 y_3 \sum_{N=0}^\infty \sum_{n=0}^N \varphi_{N, n}^{(12)3}(\mu) \Psi^{(12)3}_{N,n}(y_1,y_2,y_3); \ \ \ \ (y_1+y_2+y_3=1)\,.
\label{Conf_PW_expansion_123}
\end{equation}
The conformal moments in the conformal PW expansion
(\ref{Conf_PW_expansion_123})
are defined as
\begin{eqnarray}
&&
\varphi_{N, n}^{(12)3}(\mu)
\nonumber \\ && = \frac{2N+5}{60 f_{N,n}}   120 \int_0^1 dy_1 \int_0^1 dy_2 \int_0^1 dy_3 \delta(1-y_1-y_2-y_3)
\Psi^{(12)3}_{N,n}(y_1,y_2,y_3)
\Phi(y_1,y_2,y_3).
\end{eqnarray}
To switch to a different basis (say $(31)2$) one
makes use of the $\Omega_{n n'}(N)$ matrix
(\ref{Conf_Basis_Reparam})
\begin{equation}
\varphi_{N, n}^{(31)2}(\mu)= \sum_{n'=0}^N \Omega_{n n'}(N) \varphi_{N, n}^{(12)3}(\mu).
\end{equation}

The conformal invariance is not sufficient to resolve mixing
of  the coefficient $\varphi_{N,n}(\mu)$ with the same $N$ and different
values of $n$. To diagonalize the mixing matrix and find the
corresponding anomalous dimensions the advanced approach
based on the complete integrability of the evolution equation
for the baryon DAs with helicity-$\frac{3}{2}$ was proposed in~\cite{Braun:1998id,Braun:1999te}.
The numerical algorithms based on this approach were found extremely effective.
This approach allows working out analytical expressions for corresponding
eigenvalues and eigenvectors in the large-$N$ limit by means of a
Wentzel--Kramers--Brillouin
(WKB)-type expansion.

The multiplicatively renormalizable contributions to the nucleon DA (\ref{Conf_PW_expansion_123})
are given by linear combination of the conformal basis polynomials
\begin{equation}
P_{N, q}\left(y_1,y_2,y_3\right)=\sum_{n=0}^{N} c_{N, n}^{(q)} \Psi_{N, n}^{(12) 3}\left(y_1,y_2,y_3\right),
\end{equation}
where the coefficient $c_{N, n}^{(q)}$ can be found by diagonalization
of the mixing matrix.
The scale dependence of the leading twist-$3$ nucleon DA can be formally written as
\begin{equation}
\phi_N\left(y_1,y_2,y_3, \mu^{2}\right)=y_{1} y_{2} y_{3} \sum_{N,\,q} \varphi_{N,q} P_{N,q}\left(y_1,y_2,y_3\right)\left(\frac{\alpha_{s}(\mu)}{\alpha_{s}\left(\mu_{0}\right)}\right)
^{\gamma_{N,q} / b_{0}},
\label{Nucl_DA_evolution}
\end{equation}
where $\gamma_{N,q}$ are the anomalous dimensions found from the solution of the
eigenvalue problem
(\ref{Eigenvalue_problem_H})
and
$b_0=11/3 N_c-2/3 n_f$
is the first coefficient of the
$\beta$-function of QCD. The coefficient $\varphi_{N,q}$ is scale-independent  and can be seen as non-perturbative inputs.

\subsubsection{Evolution equations for $\pi N$ {{TDA}}s}
\mbox

In this subsection we consider the derivation of the
evolution equation for nucleon-to-meson TDAs.
It repeats the main steps of the derivation of the evolution equations
for GPDs~\cite{Mueller:1998fv}.
The starting point is the renormalization group equation for the
three-quark operator
(\ref{rengr}) with the evolution operator
(\ref{intop_Sec4}).

As a definite example we present the evolution equation for the combination of
the leading twist-$2$ $\pi N$ TDAs (\ref{Def_TDA_through_Tinv})
\begin{equation}
T_{\uparrow \downarrow,\uparrow}^{\uparrow}(x_1,x_2, x_3,\xi,\Delta^2; \mu) \equiv
T_{\uparrow \downarrow,\uparrow}^{\uparrow}(x_i) \sim
\left( V_{1}^{p \pi_{0}}(x_i)-A_{1}^{p \pi_{0}}(x_i) \right)
\label{Combination_TDAs_for_ev}
\end{equation}
with $\pi N$ TDAs parameterized as in Eq.~(\ref{Param_TDAs})
with the set of  Dirac structures (\ref{Def_DirStr_piN_DeltaT}).

The combination (\ref{Combination_TDAs_for_ev}) is related to the matrix element of the three quark operator with the total
chirality $\frac{1}{2}$
between a nucleon with chirality $\frac{1}{2}$ and a pion:
\[
\langle \pi^0(p_\pi)|  \varepsilon^{c_1 c_2 c_3}(\hat n u_{c_1}^\uparrow)_\rho(z_1 n)
(\hat n u_{c_2}^\downarrow)_\tau(z_2 n) (\hat n d_{c_3}^\uparrow)_\chi (z_3 n)| N^\uparrow(p_N) \rangle.
\]

The evolution equation  takes the form~\cite{Pire:2005ax}:
\begin{eqnarray}\label{evoleq}
&&
\mu\frac{d}{d\,\mu}
\;
T_{\uparrow \downarrow,\uparrow}^{\uparrow}
(x_i)
= -\frac{\alpha_s}{2\pi}\left\{
\frac{3}{2}\,C_F\,
T_{\uparrow \downarrow,\uparrow}^{\uparrow}
(x_i)
- \left(1+\frac{1}{N_c} \right)
\right.
 \\
 &&\left[\left(\int_{-1+\xi}^{1+\xi}dx_1'
\left[\frac{x_1\rho(x_1',x_1) }{x_1'(x_1' -
x_1)}  \right]_+
+\int_{-1+\xi}^{1+\xi}dx_2'
\left[\frac{x_2\rho(x_2',x_2) }{x_2'(x_2' -
x_2)}  \right]_+ \right)
T_{\uparrow \downarrow,\uparrow}^{\uparrow}
(x_1',x_2',x_3)
\right.
\nonumber \\
&&+ 
\left(\int_{-1+\xi}^{1+\xi}dx_1'
\left[\frac{x_1\rho(x_1',x_1) }{x_1'(x_1' -
x_1)}  \right]_+
+\int_{-1+\xi}^{1+\xi}dx_3'
\left[\frac{x_3 \rho(x_3',x_3) }{x_3'(x_3' -
x_3)}  \right]_+ \right)
T_{\uparrow \downarrow,\uparrow}^{\uparrow}
(x_1',x_2,x_3')
\nonumber \\
&&+ 
\left(\int_{-1+\xi}^{1+\xi}dx_2'
\left[\frac{x_2\rho(x_2',x_2) }{x_2'(x_2' -
x_2)}  \right]_+
+ \int_{-1+\xi}^{1+\xi}dx_3'
\left[\frac{x_3 \rho(x_3',x_3) }{x_3'(x_3' -
x_3)}  \right]_+ \right)
T_{\uparrow \downarrow,\uparrow}^{\uparrow}
(x_1,x_2',x_3')
\nonumber \\
&&+
\frac{1}{2\xi-x_3}\left( \int_{-1+\xi}^{1+\xi}dx_1'
\frac{x_1}{x_1'}\rho(x_1',x_1)
+  \int_{-1+\xi}^{1+\xi}dx_2'
\frac{x_2}{x_2'}\rho(x_2',x_2)
 \right)
T_{\uparrow \downarrow,\uparrow}^{\uparrow}
(x_1',x_2',x_3)
\nonumber \\
&&\left. \left.
 +\frac{1}{2\xi - x_1} \left(
\int_{-1+\xi}^{1+\xi}dx_2'
\frac{x_2}{x_2'}\rho(x_2',x_2)
+\int_{-1+\xi}^{1+\xi}dx_3'
\frac{x_3}{x_3'} \rho(x_3',x_3)
\right)
T_{\uparrow \downarrow,\uparrow}^{\uparrow}
(x_1,x_2',x_3')
\right] \right\}\;, \nonumber
\end{eqnarray}
where $\left[ \ldots \right]_+$
denotes the standard ``{$+$}''-regularization prescription
for generalized functions
\cite{Gelfand_Graev}:
\begin{equation}
\left[f\left(x, x^{\prime}\right)\right]_{+}=f\left(x, x^{\prime}\right)-\delta\left(x-x^{\prime}\right) \int d x^{\prime} f \left(x^{\prime}, x^{\prime}\right).
\end{equation}

The integration region in each integral is restricted in two ways.
Firstly, the support of integrands is defined by functions $\rho(x,y)$
\begin{equation}
\rho(x,y)=\theta(x\ge y \ge 0)-\theta(x\le y \le 0),
\end{equation}
with
$\theta(x\ge y \ge 0) = \theta(x\ge y)\theta(y \ge 0)$.
This function  $\rho(x,y)$ is a generalization of the analogous one
which appears in equations describing the pure ERBL evolution
\cite{Mueller:1998fv}.
The second condition is the requirement that, although not
  denoted as a variable of integration,
the variables $x'_i$
must satisfy the condition
$x'_i \in [-1+\xi,1+\xi]$. For example, in the first integral
over $x_1'$ in the r.h.s. of (\ref{evoleq}) the variable $x_2'=2\xi - x_3 -x_1'$
must belong to the interval  $x_2' \in [-1+\xi,1+\xi]$.

The evolution equation for the combination of TDAs (\ref{Def_TDA_through_Tinv})
$T_{\uparrow \uparrow,\uparrow}^{\uparrow}(x_i) \sim \left( T_{2}^{p \pi_{0}}(x_i)-T_{3}^{p \pi_{0}}(x_i) \right)$,
that corresponds to the case where the
three quarks have total chirality  $3/2$, is obtained from
(\ref{evoleq}) by neglecting the two last lines.
The evolution equations for other TDA combinations listed in (\ref{Def_TDA_through_Tinv})
can be obtained according to the same pattern.

It is worth emphasizing, that, similarly to the case of GPDs, the form of the evolution
of nucleon-to-meson TDAs
turns out to be different in various domains of the $x_{i}$ support.
\bi
 \item In particular, when all  $x_{i} > 0$
 one is in the same kinematics as the usual ERBL equation for the
baryons, with the simple $x_{i} \to x_{i}/2\xi$ rescaling.
The solutions of Eq.~(\ref{evoleq}) in the ERBL-like domain are thus well known and are
expressed in terms of the conformal basis:
\begin{equation}
H(x_i, \xi, \mu^2)=x_1 x_2 x_3 \delta (x_1+ x_2 +x_3 - 2\xi)\sum_{N,q} h_{N,q}\, P_{N,q}(x_i /2\xi)
\left (\frac{
\alpha_s(\mu)}{\alpha_s(\mu_0)}
\right )^{\gamma_{n,q}/b_0},
\label{expansion1}
\end{equation}
where $b_0=11/3 N_c-2/3n_f$,
$\gamma_{n,q}$ are the corresponding anomalous dimensions.
The dimensionless  parameters
$h_{N,q}(\mu_0)$ represent nonperturbative input for the evolution. They are to be fixed
with the help of QCD sum rules technique or from
lattice calculations.

\item For the DGLAP-like domains, where one (or two)  $x_{i} < 0$, the solutions of the evolution Eqs.~(\ref{evoleq})   are for the moment unknown and deserve further study.
\ei

\subsubsection{Conformal partial wave expansion for nucleon-to-meson {{TDA}}s}
\label{SubSec_Conf_PW_exp_TDAs}
\mbox

GPD representations based on the conformal partial wave expansion
were found to be a convenient tool both to address the fundamental
properties of GPDs as well as for the phenomenological applications.
The key advantages of this approach consist in a possibility to
achieve the factorization of functional dependencies of GPDs on different
variables  and ensure diagonalization of the leading order (LO) evolution operator.
It also provides a consistent framework to study evolution effects at NLO.

There are several versions of this formalism in the present day literature.
In particular, the approach
\cite{Mueller:2005ed,Kirch:2005tt,Manashov:2005xp}
relies on the Mellin--Barnes integral technique
and the Sommerfeld--Watson transformation. This implies the analytic continuation
of conformal moments and of conformal partial waves to the
complex values of conformal spin.

An alternative approach is based on the so-called Shuvaev--Noritzsch transformation
\cite{Shuvaev:1999fm,Noritzsch:2000pr}.
A particular version of this formalism employing further  expansion
of conformal partial waves in terms of the cross-channel SO$(3)$ partial wave
is known as the dual parametrization of GPDs~\cite{Polyakov:2002wz,Polyakov:2008aa}. This latter picture makes a
direct contact with a cross channel representation of GPDs as infinite sums of
resonance exchanges with quantum numbers of mesons~\cite{Polyakov:1998ze}.
In Ref.~\cite{Muller:2014wxa} the dual parametrization of GPDs was found to be
completely equivalent to the formalism based on the Mellin--Barnes integral techniques.

In this section we sketch a possible generalization of conformal
PW expansion
approach to the case of nucleon-to-meson TDAs.
The conformal moments of nucleon-to-meson TDAs are formed with respect to the
conformal basis
(\ref{Conf_basis_Braun99}).
The corresponding polynomials take particularly simple form in terms
of the quark--diquark coordinates
(\ref{Def_qDq_coord}).
We employ the choice of quark--diquark coordinates (\ref{qDq_3}).
It is consistent with the order $(12)3$ in which the conformal spins of
quark components are added up to form the total conformal spin $N+3$ of the three-quark
operator. For simplicity in what follows we omit the subscript referring to the
particular choice  of quark--diquark coordinates:
$w_{3} \equiv w$; $v_{3} \equiv v$; $\xi'_{3}\equiv \xi'$.
The corresponding polynomials then read
\begin{equation}
\psi^{(12)3}_{N,n}(w,v,\xi)=\xi^N 2^{N-n} (N+n-4) \left(1- \frac{w }{\xi} \right)^n C_n^{\frac{3}{2}} \left( \frac{ v }{\xi'} \right)
P_{N-n}^{2n+3,1} \left( \frac{w }{\xi} \right).
\label{P_N_n_TDAs}
\end{equation}

The $(N,n)$-th conformal moments of a nucleon-to-meson TDA
$H(w_{},v_{},\xi,\Delta^2)$
$h_{N,n}^{(12)3}(\xi, \Delta^2)$
are defined as integrals  over the complete  support domain of TDAs in quark--diquark coordinates (\ref{Support_TDA_wv}):
\begin{equation}
h_{N,n}^{(12)3}(\xi, \Delta^2)= \int_{-1}^1 dw_{} \int_{-1+| \xi-\xi'_{}| }^{1-| \xi-\xi'_{}| } dv_{}
H(w_{},v_{},\xi,\Delta^2)
\psi^{(12)3}_{N,n}(w_{},v_{},\xi).
\label{Def_Conf_moment}
\end{equation}

We introduce the conformal PWs with the pure ERBL-like support (see  Fig.~\ref{Fig_Support_TDAs_qDq}):
\begin{eqnarray}
&&
p_{N,n}^{(12)3}(w ,v , \xi) =
\theta(-\xi\le w  \le \xi) \, \theta(-\xi'  \le v  \le \xi' ) \, \xi^{-N-2} \frac{1}{g_{N,n}}
\nonumber \\ &&
\times \left(1-   \frac{  v^2 }{{\xi'}^2}    \right)
C_n^{\frac{3}{2}} \left( -\frac{  v}{\xi'} \right)
\left(1-\frac{w}{\xi}\right)^{n+2}  \left(1+\frac{w}{\xi}\right)
P_{N-n}^{2n+3,1} \left( \frac{w}{\xi} \right),
\label{Conf_PW_qdq}
\end{eqnarray}
where $g_{N,n}$ is the convenient normalization factor
\begin{equation}
g_{N,n}=2^{N+ n+ 5} \frac{ (N-n+ 1)(n+1) (n+2)
  }{(2 n+3) (2 N+5)}.
\end{equation}
The conformal basis
(\ref{P_N_n_TDAs})
is normalized in a way that
\begin{equation}
\int_{-\xi}^\xi dw  \int_{-\xi' }^{\xi' } dv \, p_{N,n}^{(12)3}(w ,v , \xi) \, \psi^{(12)3}_{M,m}(w ,v ,\xi)=(-1)^n \delta_{MN} \delta_{mn}.
\label{OrthoTDA_conformal}
\end{equation}

The conformal PW expansion for a nucleon-to-meson TDA
can then be written as
\begin{equation}
H(w_{},v_{},\xi,\Delta^2)=
\sum_{N=0}^\infty \sum_{n=0}^N  p_{N,n}^{(12)3}(w_{},v_{}, \xi) h_{N,n}^{(12)3}(\xi, \Delta^2)\,.
\label{Conf_PW_exp_TDA_123}
\end{equation}
Due to the orthogonality relation
(\ref{OrthoTDA_conformal}),
the series
(\ref{Conf_PW_exp_TDA_123})
reproduces the conformal
moments of a TDA
(\ref{Def_Conf_moment}).

Since the conformal PWs
(\ref{Conf_PW_qdq})
have only the ERBL-like support, this series must be seen as a representation of a TDA in the space of
singular generalized functions.
Similarly to the case of GPDs, the series
(\ref{Conf_PW_exp_TDA_123}) requires proper resummation in order to
be defined rigorously in the mathematical sense. This turns to be a formidable problem
that still awaits its complete solution.

A possible approach could be a generalization of the Mellin--Barnes integral technique
that was efficiently employed in  the GPD case. This requires the analytic continuation
of both the conformal moments $h_{N,n}^{(12)3}$ and conformal PWs
(\ref{Conf_PW_qdq}) to the complex
values of the conformal spin of a $(12)$-quark pair $n+2$ and of the total conformal spin
of the three-quark operator $N+3$. The Sommerfeld--Watson transformation can be employed
to transform the formal series
(\ref{Conf_PW_exp_TDA_123}) into a double Mellin--Barnes type integral.
The analytic continuation of the conformal PWs
(\ref{Conf_PW_qdq})
can be performed with help of the familiar Schl\"afli integral representation for the
Jacobi polynomials~\cite{szeg1939orthogonal}:
\begin{equation}
P_{n}^{(\alpha, \beta)}(x)= \frac{2^{-n}}{2 \pi i} \oint_C dz (z^2-1)^n \left( \frac{1-z}{1-x} \right)^\alpha  \left( \frac{1+z}{1+x} \right)^\beta \frac{1}{(z-x)^{n+1}},
\end{equation}
where $C$ is the usual contour making one counter-clockwise turn around $z=x$.
Assuming the
proper asymptotic behavior of the conformal moments
$h_{N,n}^{(12)3}$
the uniqueness of the procedure can be ensured with help of the Carlson theorem~\cite{CarlsonTh}.

However, the mathematical aspects of handling the double
Mellin--Barnes-type integrals and symmetry issues related to switching between the
equivalent $(12)3$, $(23)1$ and $(31)2$ conformal basis
(\textit{c.f.} Eq.~(\ref{Conf_Basis_Reparam})) still require further study.
Also the intrinsic ambiguity of the analytic continuation of the anomalous dimensions
of three-quark operators reported in~\cite{Braun:1999te} may signal
possible severe complications for this approach.

Similarly to the GPD case, it is convenient to expand the conformal moments of TDAs
over the partial waves of the cross-channel SO$(3)$ rotation group.
For definiteness we consider the backward pion electroproduction reaction
\begin{equation}
\gamma^{*}(q)+N(p_N) \to N'(p'_N)+\pi(p_\pi);
\label{Direct_s_react}
\end{equation}
and its $u$-channel counterpart reaction
\begin{equation}
\gamma^{*}(q)+N'(\tilde{p}'_N) \to N(\tilde{p}_N)+\pi(p_\pi).
\label{Cross_u_react}
\end{equation}
The crossing transformation, relating (\ref{Cross_u_react}) to (\ref{Direct_s_react}), reads
\begin{eqnarray}
&&
\tilde{p}'_N \to -{p}'_N; \ \ \ \tilde{p}_N \to -{p}_N;  \nn \\
 &&
s\big|_{\text{reaction (\ref{Cross_u_react})}}=(q+\tilde{p}'_N)^2 \to u\big|_{\text{reaction (\ref{Direct_s_react})}}=(q-{p}'_N)^2.
\end{eqnarray}

Within the near-backward kinematics of  Sec.~\ref{SubSec_Kinematics}
the  cosine of the $u$-channel scattering angle
of the reaction (\ref{Direct_s_react})
can be expressed  to the leading order in $1/Q$-expansion as
\begin{equation}
\cos \theta_u
=
 -\frac{1- \xi \frac{m_N^2 -m_\pi^2}{u}}{\xi \sqrt{1+ \frac{(m^2_N-m_\pi^2)^2}{{u}^2}- \frac{2(m_N^2+m_\pi^2)}{u}}}+ {\mathcal{O}} \left(
\frac{1}{Q^2}
\right)\,,
\label{Cos_theta_u_dual}
\end{equation}
where $\xi$ is the $u$-channel skewness variable (\ref{Def_xiBMP}).
Setting  in (\ref{Cos_theta_u_dual}) hadronic masses to zero  corresponds to neglecting  the $\pi N$ threshold corrections. This results in
\begin{equation}
\cos \theta_u \simeq -\frac{1}{\xi}.
\end{equation}

To match with the polynomiality property of $\pi N$ TDAs
(\ref{PolyProp_pN-TDA}),
$h_{n,N}^{(12)3}(\xi, \Delta^2)$ turn to be polynomials of $\xi$ of order $N+1$ for
the TDAs $\{V_{1,2}^{\pi N}, \, A_{1,2}^{\pi N}, \, T_{1,2}^{\pi N}\}$  and
of order $N$ for the
$\{T_{3,4}^{\pi N}\}$ TDAs.
Calculations of $N$ and $\Delta(1232)$ $u$-channel exchange contributions into $\pi N$ TDAs
(see  Sec.~\ref{SubSec_Cross_Ch_baryons})
provide insight on the proper cross-channel SO$(3)$ PW basis.
\bi
\item
For
$H=\{V_{1,2}^{\pi N}, \, A_{1,2}^{\pi N}, \, T_{1,2}^{\pi N}\}$ TDAs
the expansion goes over the Gegenbauer polynomials
$
P'_{l+1}(1/\xi)= C_l^{\frac{3}{2}} (1/\xi):
$
\begin{equation}
h_{N,n}^{(12)3}(\xi, \Delta^2)= \xi^{N+1} \sum_{l=0}^{N+1} B_{N,n,l}^H (\Delta^2) C_l^{\frac{3}{2}} \left( \frac{1}{\xi} \right);
\label{SO3_PW_TDA_MOM}
\end{equation}

\item For
$H=\{ T_{3,4}^{\pi N}\}$ TDAs
the expansion goes over the Gegenbauer polynomials
$
P'_{l}(1/\xi)= C_{l-1}^{\frac{3}{2}} (1/\xi):
$
\begin{equation}
h_{N,n}^{(12)3}(\xi, \Delta^2)= \xi^{N} \sum_{l=0}^{N+1} B_{N,n,l}^H(\Delta^2) C_{l-1}^{\frac{3}{2}} \left( \frac{1}{\xi} \right);
\label{SO3_PW_TDA_MOMbis}
\end{equation}
\ei
Here the generalized form factors $B_{N,n,l}^H (\Delta^2)$ are functions of the
invariant $u$-channel momentum transfer of the reaction (\ref{Direct_s_react})
$\Delta^2=(p_\pi-p_N)^2 \equiv u$
and carry two conformal spin indices $N$, $n$ and the cross-channel SO$(3)$ rotation group index $l$.

Plugging
(\ref{SO3_PW_TDA_MOM}), (\ref{SO3_PW_TDA_MOMbis}) into
the conformal PW expansion (\ref{Conf_PW_exp_TDA_123})
results in an analogue of the dual representation of GPDs for the case of
nucleon-to-pion TDAs. Unfortunately, the summation of the
corresponding formal series employing a generalization of methods
of Ref.~\cite{Polyakov:2002wz} has been only achieved for the limiting
case $\xi=1$.

It is worth mentioning that  the double (conformal and the cross-channel SO$(3)$) PW expansion for nucleon--pion GDAs/nucleon-to-pion TDAs arises naturally
within the representation of corresponding matrix elements as infinite sums of the
$u$-channel $N^{*}$- and $\Delta$-family resonance exchanges
of arbitrary high half-integer spin and of arbitrary high mass (see  Fig.~\ref{Fig_Cross_Ch_Pole}).

Further development of the conformal partial wave expansion framework for
the case of nucleon-to-meson (and nucleon-to-photon) TDAs is highly demanded,
although mathematical difficulties considerably delay the progress.
In particular, the calculation of the LO elementary amplitudes of hard processes (see  Sec.~\ref{SubSec_Bkw_meson_ampl}) within the conformal partial wave expansion framework for nucleon-to-meson TDAs still remains an open issue. Such studies are particularly interesting
since, similarly to the case of GPD~\cite{Anikin:2007yh,Polyakov:2007rv,Kumericki:2008di,Muller:2015vha}, they can provide a new insight into the analytical properties
of the LO elementary amplitudes. Also an adaptation for the case of TDAs of an analogue of the Regge-theory inspired
Kumeri\v cki--M\"uller (KM)
Ansatz  for
the conformal moments of GPDs~\cite{Kumericki:2009uq} can be useful for phenomenological applications.

\subsection{Nucleon-to-meson {{TDA}}s in impact parameter space}
\label{SubSec_Impact}
\mbox

The use of the impact parameter representation to provide
a vivid physical interpretation to GPDs
and build up a comprehensible picture of hadrons in the transverse
plane has been pioneered by M.~Burkardt for GPDs in the zero
skewness limit in Refs.~\cite{Burkardt:2000za,Burkardt:2002hr}.
The extension of this framework for GPDs with non-zero skewness
was proposed in Refs.~\cite{Ralston:2001xs,Diehl:2002he}.

A possible novel physical aspect of nucleon-to-meson TDAs
is their sensitivity to the effect of diquark clustering in hadrons~\cite{Lichtenberg:1968zz,Carroll:1969ty,Cutkosky:1977kd,Anselmino:1987vk}.
An attempt to study diquark clustering in light-cone functions
(nucleon DA) and to provide dynamical explanation to the
asymmetry of nucleon DA was performed in
Ref.~\cite{Dziembowski:1990md}.
Below we consider a development of this idea. For this issue following
Ref.~\cite{Pire:2019nwa} we address
nucleon-to-meson TDAs integrated over
quark--diquark coordinate $v_i$
(\ref{Def_qDq_coord}). This provides access to
hadronic matrix elements of a bilocal
quark--diquark light-cone operator.
Moreover, the general structure of  integrated TDAs looks similar to
GPDs and it is, therefore, natural to adopt for them the impact parameter space representation.

\subsubsection{Integrated \texorpdfstring{$\pi N$}
{pi N} {{TDA}}s and quark--diquark picture of the nucleon}
\label{SubSubSec_IntegratedTDA}
\mbox

For definiteness,  we consider below the case of the
proton ($N^p$)-to-$\pi^0$ $uud$ TDA
(\ref{Param_TDAs}).
However, our results admit a straightforward
generalization for other isospin channels for
$\pi N$
TDAs as well as for more involved cases.
In order to make contact to a quark--diquark picture we
introduce $\pi N$ TDAs integrated over
the momentum fraction variable $v_i$ (\ref{Def_qDq_coord}).
For definiteness in what follows we choose $i=3$
and employ the set of quark--diquark coordinates
$w_3=x_3-\xi; \ \ v_3 \equiv \frac{x_1-x_2}{2}$.
The fraction of the longitudinal momentum
carried by the first and second quarks is given by
$x_1+x_2=2 \xi_3'$ with $\xi_3'\equiv \frac{\xi-w_3}{2}$.

We integrate the definition
(\ref{Param_TDAs})
in
$v_3$
over the interval
$(-\infty; \, \infty)$:
\begin{eqnarray}
&&
\int_{- \infty}^\infty
d v_3
4 {\mathcal{F}} \langle     \pi^0(p_\pi)| \,
\widehat{O}_{\rho \tau \chi}^{\,uud}(\lambda_1n,\lambda_2n,\lambda_3n)
\,| N^p(p_N,s_N) \rangle  \nn
\\  &&
=i \frac{f_N}{f_\pi}
\Bigl[
\sum_{\Upsilon= 1, 2} (v^{\pi N}_\Upsilon)_{\rho \tau, \, \chi}
\int_{-1+| \xi-\xi_{3}^{\prime}|  }^{1-| \xi-\xi_{3}^{\prime}|  } dv_3
V_{\Upsilon}^{\pi N}(w_3, v_3, \xi, \Delta^2)
 +\sum_{\Upsilon= 1,\, 2 } (a^{\pi N}_\Upsilon)_{\rho \tau, \, \chi}
\int_{-1+| \xi-\xi_{3}^{\prime}|  }^{1-| \xi-\xi_{3}^{\prime}|  } dv_3
A_{\Upsilon}^{\pi N}(w_3, v_3, \xi, \Delta^2)
\nonumber \\ &&
+
\sum_{\Upsilon= 1,2,3,4} (t^{\pi N}_\Upsilon)_{\rho \tau, \, \chi}
\int_{-1+| \xi-\xi_{3}^{\prime}|  }^{1-| \xi-\xi_{3}^{\prime}|  } dv_3
T_{\Upsilon}^{\pi N}(w_3, v_3, \xi, \Delta^2)
\Bigr].
\label{Integrated_Mtrx_el}
\end{eqnarray}
In the r.h.s. of
(\ref{Integrated_Mtrx_el}) the integral stands over the $v_3$ support
(\ref{Support_TDA_wv})
of $\pi N$ TDAs.

We employ
the following identity for the exponent in  the Fourier
transform (\ref{Fourier_TDA}):
\begin{eqnarray}
&&
{\mathcal{F}}(\ldots) =
(P \cdot n)^3 \int \left[ \prod_{j=1}^3 \frac{d \lambda_j}{2 \pi} \right]
e^{i( \frac{x_1-x_2}{2} (\lambda_1-\lambda_2) +
\frac{x_1+x_2}{2} (\lambda_1+\lambda_2)
+ x_3 \lambda_3) P \cdot n} \nn \\ &&
\equiv
(P \cdot n)^3 \int \left[ \prod_{j=1}^3 \frac{d \lambda_j}{2 \pi} \right]
e^{i( v_3 (\lambda_1-\lambda_2) +
\xi_3' (\lambda_1+\lambda_2)
+ x_3 \lambda_3) P \cdot n}.
\label{Exp_qdq}
\end{eqnarray}
In the l.h.s. of (\ref{Integrated_Mtrx_el})  the integration of the exponent
(\ref{Exp_qdq})
results in the delta function
$\delta(\lambda_1- \lambda_2)$,
which allows to
remove one of the
$\lambda_j$-integrals. Finally, this brings the arguments of the two
$u$-quark operators to the
same point on the light cone
$\lambda_1 n = \lambda_2 n \equiv \lambda_D n$
and gives rise to the hadronic matrix element of a bilocal light-cone operator:
\begin{equation}
\widehat{O}_{\rho \tau \chi}^{\{uu\}d}( \lambda_D n, \lambda_D n,\lambda_3 n)=
u_\rho(\lambda_D n) u_\tau(\lambda_D n) d_\chi(\lambda_3 n) \equiv
\widehat{D}_{\rho \tau }^{uu}( \lambda_D n) d_\chi(\lambda_3 n).
\label{diquark_quark_op_uud}
\end{equation}
It is natural to interpret
(\ref{diquark_quark_op_uud})
as the bilocal $uu$-diquark- $d$-quark
operator on the light-cone. Now we employ the translation invariance of the
matrix element to shift the arguments of the bilocal operator
(\ref{diquark_quark_op_uud})
to the symmetric points
$\pm \frac{\lambda}{2}$ on the light-cone introducing $\lambda \equiv \lambda_3-\lambda_D$ and $\mu \equiv \lambda_3+\lambda_D$. The integral
over $\mu$ can be performed producing the momentum conservation $\delta$-function:
\begin{equation}
\int \frac{d \mu}{2 \pi}
 e^{i  \left( x_D+x_3 -2\xi  \right) (P \cdot n) 2 \mu}= \frac{1}{2(P \cdot n)}
 \delta(x_D+x_3-2\xi); \ \ \ x_D \equiv x_1+x_2=2 \xi'_3.
\end{equation}
Finally, we get the equality relating the Fourier transform of the nucleon--pion matrix element
of the light-cone diquark--quark operator
(\ref{diquark_quark_op_uud})
to the $v_3$-integrated nucleon-to-pion TDAs:
\begin{eqnarray}
&&
2 (P \cdot n)
\int  \frac{d \lambda}{4 \pi}
 e^{i  \left( w_3 \lambda  \right) (P \cdot n) }
 \langle \pi(p_\pi)|
    \widehat{D}_{\rho \tau}^{\,uu}(- \frac{\lambda}{2} n)
 {d}_\chi ( \frac{\lambda}{2} n)|  N(p_N) \rangle \nn
\\ &&
=i \frac{f_N}{f_\pi}
\Bigl[
\sum_{\Upsilon= 1, 2} (v^{\pi N}_\Upsilon)_{\rho \tau, \, \chi}
\int_{-1+| \xi-\xi_{3}^{\prime}|  }^{1-| \xi-\xi_{3}^{\prime}|  } dv_3
V_{\Upsilon}^{\pi N}(w_3, v_3, \xi, \Delta^2)
+\sum_{\Upsilon= 1,\, 2 } (a^{\pi N}_\Upsilon)_{\rho \tau, \, \chi}
\int_{-1+| \xi-\xi_{3}^{\prime}|  }^{1-| \xi-\xi_{3}^{\prime}|  } dv_3
A_{\Upsilon}^{\pi N}(w_3, v_3, \xi, \Delta^2)
\nonumber \\ &&
+
\sum_{\Upsilon= 1,2,3,4} (t^{\pi N}_\Upsilon)_{\rho \tau, \, \chi}
\int_{-1+| \xi-\xi_{3}^{\prime}|  }^{1-| \xi-\xi_{3}^{\prime}|  } dv_3
T_{\Upsilon}^{\pi N}(w_3, v_3, \xi, \Delta^2)
\Bigr].
\label{TDA_result_integrated}
\end{eqnarray}
The $v_3$-integrated TDAs occurring in the r.h.s. of
Eq.~(\ref{TDA_result_integrated})
share many common features with GPDs:
\bi
\item They are functions of
one longitudinal momentum fraction
$w_3= \frac{x_3-x_D}{2} \in [-1;\, 1]$,
of skewness
$\xi$,
of the invariant momentum transfer
$\Delta^2$ (and of the factorization scale).
\item As a consequence of the Lorentz invariance, the Mellin moments
of $v_3$-integrated TDAs possess the usual polynomiality property in $\xi$.
\item For the $v_3$-integrated TDAs it is natural to specify the ERBL-region with
$w_3 \in [-\xi;\xi]$,
DGLAP-I region $w_3 \in [-1;\,-\xi]$ and DGLAP-II region $w_3 \in [\xi;\, 1]$.
\ei

A complementary picture can be obtained from the
$\pi N$ TDAs integrated over
$v_1=\frac{x_2-x_3}{2}$.
This results in a $\pi N$ matrix element of a quark--diquark operator
\begin{equation}
\widehat{O}_{\rho \tau \chi}^{ u \{ u d \}}( \lambda_1 n, \lambda_D n,\lambda_D n)=
u_\rho(\lambda_1 n) u_\tau(\lambda_D n) d_\chi(\lambda_D n) \equiv
\widehat{D}_{ \tau \chi }^{ud}( \lambda_D n) u_\rho(\lambda_1 n).
\label{diquark_quark_op_duu}
\end{equation}
with a diquark constructed out of the third and second quarks ($du$).
In this case, the Dirac structures in the r.h.s. of the analogue of Eq.~(\ref{TDA_result_integrated})
must be Fierz transformed (see  App.~\ref{App_Fierz}) to
get the appropriate order of the Dirac indices (first two indices correspond to a diquark):
\[
\{
v^{\pi N}_\Upsilon, \,
a^{\pi N}_\Upsilon, \,
t^{\pi N}_\Upsilon
\}_{\rho \tau, \, \chi}  \to \{
v^{\pi N}_\Upsilon, \,
a^{\pi N}_\Upsilon, \,
t^{\pi N}_\Upsilon
\}_{\tau \chi , \, \rho}.
\]
Such transformation produces certain combinations
of $v_1$-integrated $\pi N$ TDAs at the Dirac structures
$\{
v^{\pi N}_\Upsilon, \,
a^{\pi N}_\Upsilon, \,
t^{\pi N}_\Upsilon
\}_{\tau \chi , \, \rho}$.

Finally, the third complementary picture results from $v_2$-integrated
TDAs defined by the
$\pi N$ matrix element of a quark--diquark operator
\begin{equation}
\widehat{O}_{\rho \tau \chi}^{ \{u  \check{u} d \}}( \lambda_D n, \lambda_2 n,\lambda_D n)=
u_\rho(\lambda_D n) u_\tau(\lambda_2 n) d_\chi(\lambda_D n) \equiv
\widehat{D}_{ \rho \chi }^{ud}( \lambda_D n) u_\tau(\lambda_2 n)
\label{diquark_quark_op_udu}
\end{equation}
with a $ud$-diquark constituted of the first ($u$) and third ($d$) quarks.

\subsubsection{Integrated {{TDA}}s in impact parameter space}
\label{SubSubSec_IntTDA_ImpactPspace}
\mbox

We now propose an interpretation of
the $v$-integrated TDAs in the impact
parameter space. It allows us to use  these objects as a tool to study
the quark--diquark structure of hadrons in the transverse plane.
It is worth emphasizing that, contrary to GPDs, the $v$-integrated TDAs do not
possess a comprehensible forward limit in which a probabilistic interpretation
\cite{Burkardt:2000za,Burkardt:2002hr}
applies for GPDs. Thus, similarly to the case of GPDs
with non-zero skewness, we seek for an interpretation in terms of
 probability amplitudes.

The first step consists in introducing the initial nucleon and final meson states
with specified longitudinal momenta localized around a definite position ${\mathbf{b}}$
in the transverse plane:
\begin{equation}
| p_N^+,\, {\mathbf{b}}; \, s_N \rangle=  \int \frac{d^2 {\mathbf{p}}_N}{16 \pi^3} e^{-i {\mathbf{p}}_N \cdot {\mathbf{b}}} | p_N^+,\, {\mathbf{p}}_N; \, s_N \rangle; \ \ \  \langle p_\pi^+,\, {\mathbf{b}}|  = \int \frac{d^2 {\mathbf{p}}_\pi}{16 \pi^3}
e^{i {\mathbf{p}}_\pi \cdot {\mathbf{b}}}
\langle p_\pi^+,\, {\mathbf{p}}_\pi|  .
\label{Loc_states}
\end{equation}
Rigorous treatment requires forming wave packets with precisely localized states
(\ref{Loc_states})
using a smooth weight falling sufficiently fast at infinity with $| {\mathbf{p}}| $
in order to avoid infinities due to normalization. A possible choice is to
employ Gaussian wave packets with the same standard deviation
parameter for the initial nucleon and the final meson.

Switching to the impact parameter space representation is performed by
Fourier transforming the corresponding operator hadronic matrix element with respect
to the transverse component
${\mathbf{D}}$
of the vector $D$:
\begin{equation}
D=\frac{p_\pi}{1-\xi}- \frac{p_N}{1+\xi}.
\end{equation}
By construction, the transverse component
${\mathbf{D}}$
is invariant under the transverse boosts
\[
k^+ \to k^+; \ \;{\mathbf{k}} \to {\mathbf{k}}-k^+ {\mathbf{v}}.
\]
This ensures
that the invariant momentum transfer  depends on
${\mathbf{p}}_N$
and
${\mathbf{p}}_\pi$
only through ${\mathbf{D}}$:
\begin{equation}
\Delta^2=-2\xi \left( \frac{m_\pi^2}{1-\xi} - \frac{m_N^2}{1+\xi} \right)- (1-\xi^2) {\mathbf{D}}^2.
\label{Delta2_D}
\end{equation}

We consider the matrix elements of the diquark--quark
operator with an explicit dependence on the transverse position ${\mathbf{z}}$:
\begin{equation}
{\widehat {\mathcal{O}}}_{\rho \tau \chi}^{\{uu\} d}({{\mathbf{z}}})= \int \frac{d \lambda}{4 \pi}
e^{i  \left( w_3 \lambda  \right) (P \cdot n) }
 {u}_{\rho}(0,\,- \frac{\lambda}{2} , \, {\mathbf{z}})
 {u}_{\tau}(0,\,- \frac{\lambda}{2 }, \, {\mathbf{z}})
 {d}_\chi (0,\, \frac{\lambda}{2} , \, {\mathbf{z}}),
 \label{Dq_operator_trans}
\end{equation}
where we adopt the following convention for the
position arguments of quark fields: $q(z)=q(z^+,\,z^-,\,{\mathbf{z}})$.

In order to single out a particular combination of $\pi N$ TDAs
we contract the matrix element of the operator
(\ref{Dq_operator_trans})
over the Dirac indices with a suitable projector.
Assuming that for $\pi N$ TDAs we employ the parametrization of Eq.~(10) of Ref.~\cite{Pire:2011xv}
we, as an example, have chosen to contract the matrix element (\ref{Dq_operator_trans}) with
$
v^{-1}_{\tau \rho, \, \chi}=\left( C^{-1} \hat{P} \right)_{\tau \rho} \bar{U}_\chi(p_N,s_N).
$
Since we deal with unpolarized nucleon, we sum and average over the nucleon spin $s_N$.
This convolution singles out the following combination of $v_3$-integrated $\pi N$ TDAs:
\begin{eqnarray}
&&
{\mathcal{H}}^{\pi N}(w_3, \,\xi,\, {\mathbf{D}})=
 (4 m_N^2-\Delta^2)
\int_{-1+| \xi-\xi'_3| }^{1-| \xi-\xi'_3| } dv_3
\frac{1}{2} \Bigl[ (3m_N^2+m_\pi^2-\Delta^2) V_1^{\pi N}(w_3,\,v_3,\,\xi,\, \Delta^2)
\nn \\ &&
+ (-2\Delta^2-2m_N^2+2m_\pi^2) V_2^{\pi N}(w_3,\,v_3,\,\xi,\, \Delta^2)
\Bigr],  \label{Comb_H}
\end{eqnarray}
where $\Delta^2$ is expressed through ${\mathbf{D}}^2$.
The contraction with different projectors\footnote{A proper design of the projecting operation establishing connection
with corresponding diquark--quark helicity amplitudes still has to be developed.}
 will allow to single out
other combinations of $8$ leading twist-$3$ proton-to-$\pi^0$ $uud$ TDAs.
The transition to the impact parameter space then gives
\begin{eqnarray}
 &&
\int \frac{d^2 { \bf D}}{(2\pi)^2}\; e^{-i\,  {\mathbf{D}}\cdot {\mathbf{b}}}\,
      {\mathcal{H}}^{\pi N}(w,\xi, {\mathbf{D}})
\nonumber \\ &&
= {\mathcal{N}}^{-1}\frac{1+\xi^2}{(1-\xi^2)^{2}}\:\; \sum_{s_N}
\left( C^{-1} \hat{P}  \right)_{\tau \rho}  \bar{U}_\chi(p_N; \, s_N)
      \langle p_\pi^+,
     -\frac{\xi  {\mathbf{b}}}{1-\xi} \,
        \big|\,
        {\widehat {\mathcal{O}}}_{\rho \tau \chi}^{\{uu\} d}({{\mathbf{b}}})
        \,\big| \,
         p^+_N, \frac{\xi  {\mathbf{b}}}{1+\xi}; \, s_N \rangle  \;,
  \label{matrix-final_our}
\end{eqnarray}
where $\mathcal N$ is a normalization factor originating from the normalization
of the localized states (\ref{Loc_states}). Without use of  smooth wave packets  it would be singular as $\delta^{(2)}({\bf 0})$.

\begin{figure}[H]
 \begin{center}
\includegraphics[width=.48\textwidth]{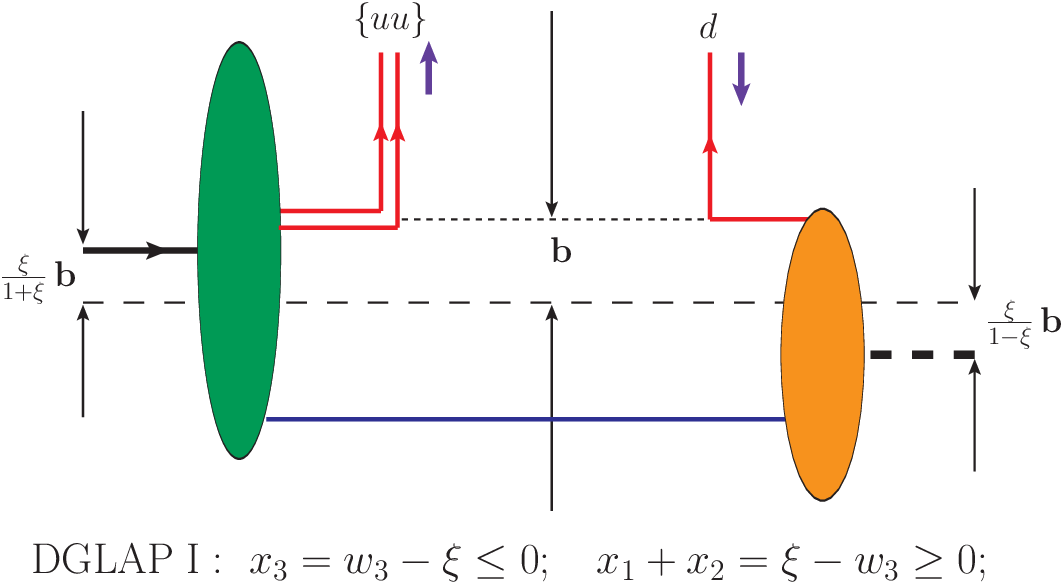}
\includegraphics[width=.48\textwidth]{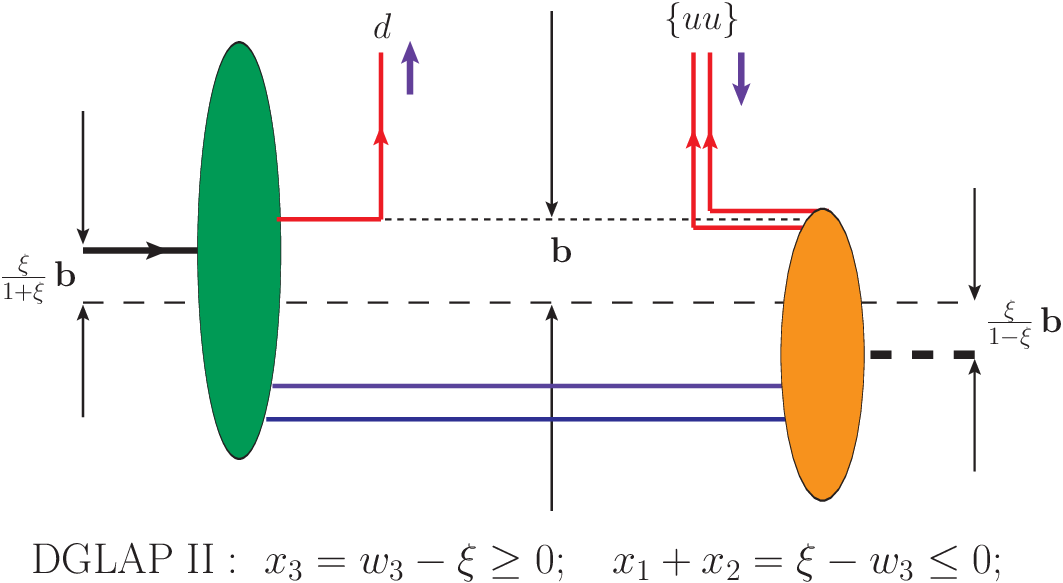} \\
  \vspace{1em}
\includegraphics[width=.48\textwidth]{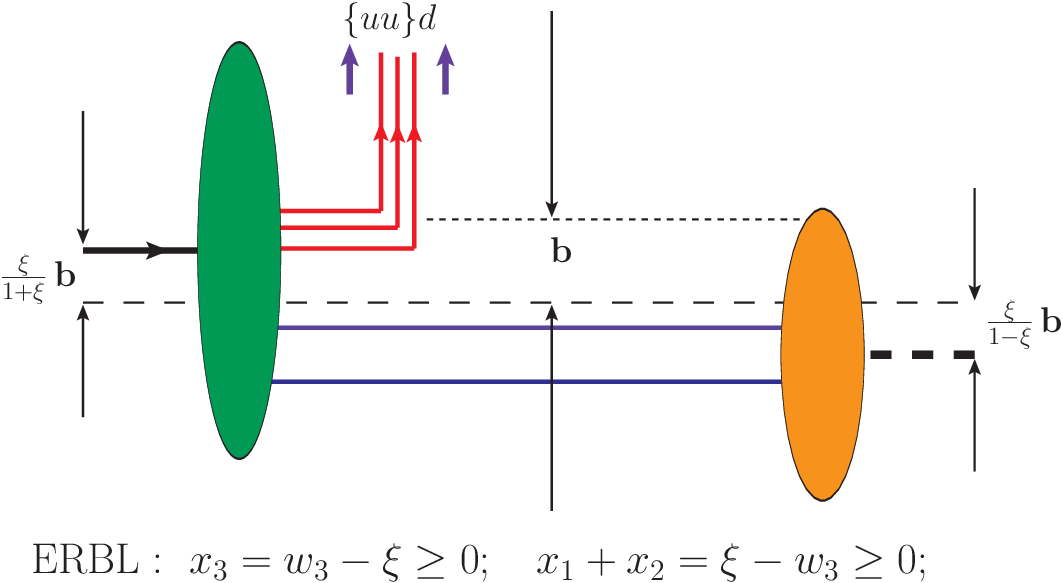}
\end{center}
\caption{Impact parameter space interpretation for the $v_3$-integrated $uud$ $\pi N$ TDA in the DGLAP-like~I, DGLAP-like~II  and in the ERBL-like domains. Solid arrows show the direction of the positive longitudinal momentum flow. }
\label{Fig_DGLAP_I_and_II}
\end{figure}

We end up with a picture that is completely analogous to the GPD case
(see  Fig.~\ref{DiehlImpactGPD}):
the hard probe interacts with a partonic configuration at the transverse
position ${\mathbf{b}}$. The initial state nucleon and the finite state meson are
localized around ${\bf 0}$, but they are shifted one from another by a transverse
separation of the order $\xi {\mathbf{b}}$. This interpretation is presented in
 Fig.~\ref{Fig_DGLAP_I_and_II}. It is also qualitatively
consistent with the low Fock component picture proposed in
\cite{Pasquini:2009ki}, see  Sec.~\ref{SubSec_LCQM_Barbara}.

\bi
\item In the DGLAP-like~I region $w_3 \le -\xi$ the impact parameter
specifies the location where a $uu$-diquark is pulled out of a proton and
then replaced by an antiquark $\bar{d}$ to form the final state meson.
\item In the DGLAP-like~II region $w_3 \ge \xi$ the impact parameter
specifies the location where a quark $d$ is pulled out of a proton and
then replaced by an antidiquark $\bar{u}\bar{u}$ to form the final state meson.
\item In the ERBL-like region $-\xi \le w_3 \le \xi$ the impact parameter
specifies the location where a three-quark cluster composed of a $uu$-diquark
and a $d$-quark is pulled out of the initial nucleon to form the final state meson.
\ei

\section{Dynamics and models}
\label{Sec_DynamMod}
\setcounter{equation}{0}
\mbox

In this section we review the existing approaches for modeling
nucleon-to-meson TDAs.

\subsection{Cross-channel baryon exchange contributions}
\label{SubSec_Cross_Ch_baryons}
\mbox

\begin{figure}[H]
\begin{center}
\includegraphics[width=0.3\textwidth]{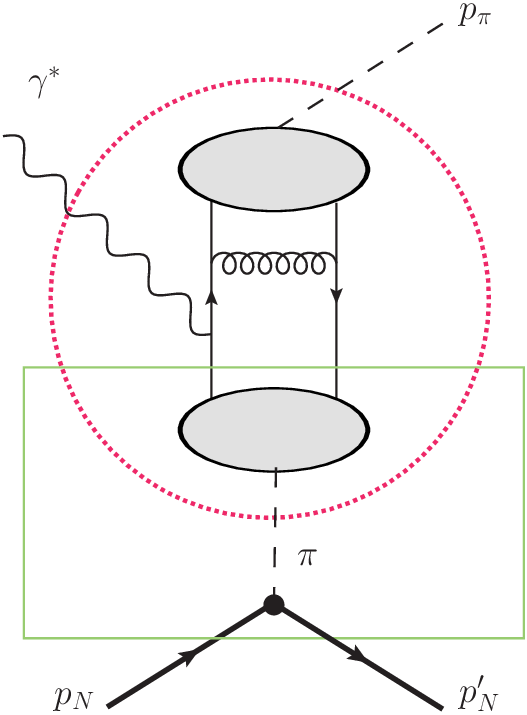} \ \ \ \ \ \
\includegraphics[width=0.3\textwidth]{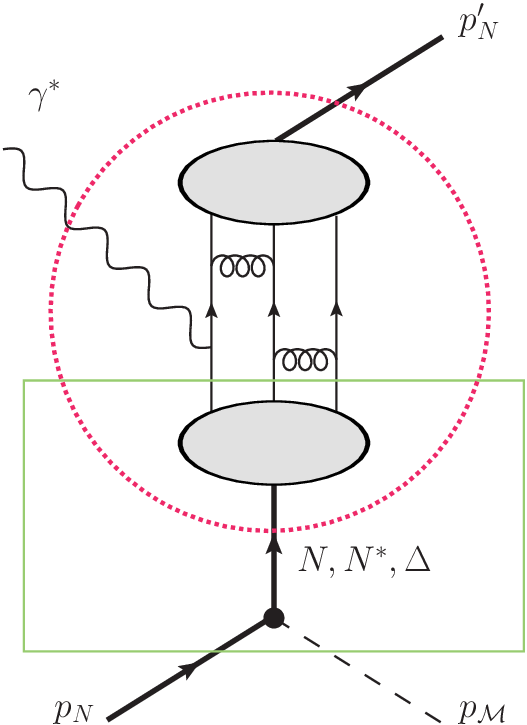}
\end{center}
     \caption{{\bf Left:} pion-exchange model for the nucleon polarized  GPD $\tilde{E}$; lower and upper blobs depict pion DAs;  the dashed circle contains
     a typical LO graph for the pion electromagnetic form factor in perturbative QCD; the rectangle contains the pion pole contribution into GPD $\tilde{E}$.
    {\bf Right:} cross-channel baryon exchange model for ${\mathcal{M}} N$ TDAs;  dashed circle contains a typical LO graph for the baryon electromagnetic form factor in perturbative QCD; the rectangle contains the cross channel baryon contribution into ${\mathcal{M}} N$ TDAs}
\label{Fig_Cross_Ch_Pole}
\end{figure}

Cross-channel meson  exchange contributions to
GPDs were extensively applied to GPD phenomenology
\cite{Mankiewicz:1997uy,Frankfurt:1998et}.
In the left panel of  Fig.~\ref{Fig_Cross_Ch_Pole} we show the pion exchange
contribution that was found substantial for the nucleon polarized GPD $\tilde{E}^{u-d}$,
see Ref.~\cite{Penttinen:1999th}.

Similarly to the case of GPDs, cross-channel hadron exchange contributions
provide a useful estimate for nucleon-to-meson TDAs. The corresponding hadrons possess quantum numbers of baryons. The relevant diagram is depicted in the right
panel of  Fig.~\ref{Fig_Cross_Ch_Pole}. For unflavored meson the lightest possible baryon occurring in the cross channel is obviously a nucleon. The problem of eventual non-negligible corrections due to an analytic continuation
from $\Delta^2 \sim m_N^2$ to the values of $\Delta^2$ typical for the direct channel reaction has the same status as for the case of GPDs.

In this subsection we review the cross-channel nucleon exchange contributions
into nucleon-to-pion and nucleon-to-vector-meson TDAs. We also consider a more involved
$\Delta(1232)$-exchange contribution into nucleon-to-pion TDAs.

\subsubsection{Cross-channel nucleon exchange contribution
for $\pi N$
{ {TDA}}s}
\label{SubSec_Nucle_ex_piN_TDA}
\mbox

Cross-channel nucleon exchange contribution to $\pi N$ TDAs corresponds to
the lower part of the graph depicted in the right panel of  Fig.~\ref{Fig_Cross_Ch_Pole}.
For the effective $\pi \bar{N} N$ interaction vertex we employ the
standard parametrization (see \textit{e.g.}~\cite{Ericson_Weise})
\begin{equation}
\mathcal{V}_{\rm eff}(\pi N N)=  i g_{\pi N N} \bar{N}_\alpha (\sigma_a)^\alpha_{\;\beta} \gamma_5 N^\beta \pi_a\,,
\end{equation}
where $g_{\pi NN} \simeq 13$ is the nucleon--pion dimensionless coupling constant
\cite{Matsinos:2019kqi}.

After reduction, the matrix element corresponding to the diagram in question reads:
\begin{eqnarray}
 &&
\langle \pi_a(p_\pi) |  \widehat{O}^{\alpha \, \beta \, \gamma}_{\rho \tau \chi}( \lambda_1 n, \,\lambda_2 n, \, \lambda_3 n )|  N_{\iota}(p_1,s_1) \rangle
\nonumber \\ &&
= \sum_{s_p}\langle 0|
\widehat{O}^{\alpha  \beta   \gamma}_{\rho \tau \chi}( \lambda_1 n, \,\lambda_2 n, \, \lambda_3 n )
|  N_\kappa(-\Delta, s_p) \rangle (\sigma_a)^\kappa_{\ \  \iota}
\frac{i g_{\pi NN} \, \bar{U}_\varrho (-\Delta, s_p) }{\Delta^2-m_N^2} \left( \gamma^5 U(p_1, s_1) \right)_\varrho\,.
\label{matrix_el}
\end{eqnarray}
The  operator matrix element in
the second line of (\ref{matrix_el})
is expressed through the leading twist-$3$ nucleon DAs
by means of the inverse Fourier transform
\begin{equation}
\mathcal{F}^{-1} (\lambda_k (-\Delta \cdot n) )(...)
=   \int d^3y \delta(1-y_1-y_2-y_3) e^{i (p \cdot n) 2 \xi \sum_{k=1}^3 y_k \lambda_k} (...)\,.
\end{equation}
Following that, $\pi N$ TDAs are computed from the matrix element
(\ref{matrix_el})
by the Fourier transform
(\ref{Fourier_TDA})
with  help of the master formula
\begin{eqnarray}
 &&
4 \mathcal{F}(x_1,x_2,x_3)
\frac{1}{4}
\bigl[ \mathcal{F}^{-1} (\lambda_k (-\Delta \cdot n) ) \bigl[
M^{N\,\{12\}}_{\rho   \tau \chi}(y_1, \, y_2,\, y_3) \bigr] \bigr] \nonumber \\ &&
= (p \cdot n)^3 \int_0^1 dy_1 dy_2 dy_3 \delta(1-y_1-y_2-y_3)
\bigl[ \prod_{k=1}^3 \frac{1}{2 \pi} \int d \lambda_k e^{i \lambda_k (x_k-2 \xi y_k) (p \cdot n)}  \bigr]
M^{N\,\{12\}}_{\rho   \tau \chi}(y_1, \, y_2,\, y_3)
 \nonumber \\ &&
 =\frac{1}{(2 \xi)^2} \delta(x_1+x_2+x_3-2\xi) \,
 \Theta_{\rm ERBL}(x_1,x_2,x_3) \,
 M^{N\,\{12\}}_{\rho   \tau \chi} \left( \frac{x_1}{2 \xi}, \frac{x_2}{2 \xi}, \frac{x_3}{2 \xi} \right)\,,
\end{eqnarray}
where we employ the notation
\begin{equation}
\Theta_{\rm ERBL}(x_1,x_2,x_3)  \equiv  \prod_{k=1}^3 \theta(0 \le x_k \le 2 \xi)\,.
\label{theta_ERBL}
\end{equation}

Obviously, the cross-channel nucleon exchange contributes only into
the isospin-$\frac{1}{2}$ $\pi N$ TDAs. One can establish the
following expressions for the set of invariant $\pi N$ TDAs defined in
 Sec.~\ref{SubSec_Polynomiality}:

\begin{eqnarray}
 &&
\bigl\{ V_1, \, A_1 , \, T_1  \bigr\}^{(\pi N)_{1/2}} (x_1,x_2,x_3,\xi,\Delta^2)\Big|_{N(940)}
\nonumber \\ &&
 =\Theta_{\rm ERBL}(x_1,x_2,x_3) \times  (g_{\pi NN}) \frac{m_N f_\pi}{\Delta^2-m_N^2} 2 \xi  \frac{1}{(2 \xi)^2}
 \bigl\{ V^p,\,A^p, \,T^p  \bigr\}\left( \frac{x_1}{2 \xi}, \frac{x_2}{2 \xi}, \frac{x_3}{2 \xi} \right);
  \nonumber \\  &&
\bigl\{ V_2, \, A_2 , \, T_2  \bigr\}^{(\pi N)_{1/2}} (x_1,x_2,x_3,\xi,\Delta^2)\Big|_{N(940)}
= \frac{1}{2} \bigl\{ V_1, \, A_1 , \, T_1  \bigr\}^{(\pi N)_{1/2}} (x_1,x_2,x_3,\xi,\Delta^2)\Big|_{N(940)};
  \nonumber \\  &&
\bigl\{ T_3,  \, T_4  \bigr\}^{(\pi N)_{1/2}} (x_1,x_2,x_3,\xi,\Delta^2)\Big|_{N(940)}=0,
\label{Nucleon_exchange_contr_VAT}
\end{eqnarray}
where $V^p(y_1,y_2,y_3)$, $A^p(y_1,y_2,y_3)$ and $T^p(y_1,y_2,y_3)$
stand for the usual leading twist-$3$ nucleon DAs~\cite{Chernyak:1984bm}.

The isotopic factors for the cross-channel nucleon exchange contributions for
$p \to \pi^0$ $uud$,
$n \to \pi^-$ $uud$ and
$p \to \pi^+$ $ddu$ TDAs
can be read off Eqs.~(\ref{Isospin_uud_ppi0}), (\ref{Isospin_ddu_ppiplus}).

Note that
(\ref{Nucleon_exchange_contr_VAT})
is a pure $D$- term like contribution. It is nonzero only in the ERBL-like region and its $(n_1,n_2,n_3)$-th ($n_1+n_2+n_3=N$)
Mellin moments
(\ref{PolyProp_pN-TDA})
give  rise to   monomials of
$\xi$  of the maximal possible power $N+1$.

The cross-channel nucleon exchange TDA model  can be easily generalized for the case
of different species of pseudoscalar mesons ($\eta$, $\eta'$, \textit{etc.}). The corresponding
coupling constants with nucleons can be found  \textit{e.g.} in Ref.~\cite{Dumbrajs:1983jd}.

\subsubsection{$\Delta(1232)$  exchange contribution for $\pi N$  { {TDA}}s}
\mbox

In this subsection we present the explicit expressions for
$\Delta(1232)$ exchange contribution to $\pi N$ TDAs.
The $\Delta$-exchange complements the nucleon exchange contribution presented in
 Sec.~\ref{SubSec_Nucle_ex_piN_TDA}
and contributes to the  isospin-$\frac{3}{2}$ $\pi N$ TDAs.
In particular, the $\Delta$-exchange gives a non-trivial contribution
to the least known $\pi N$ TDAs
$T_{3,\,4}^{\pi N}$.

The effective vertex for the $\Delta N \pi$ interaction reads (see \textit{e.g.}~\cite{SemenovTianShansky:2007hv}):
\begin{equation}
\mathcal{V}_{\rm eff}(\pi N  {\Delta}) =
g_{\pi N  {\Delta}}
\overline{N}_\kappa \,
{P^{ I=\frac{3}{2}}}_{b  a  \iota}^{\; \kappa}
 R_{\mu \, a}^{\,  \iota} \,
\partial^{\mu}   \pi_b
+ {\rm h.c.},
\label{Effective_Lag_Delta}
\end{equation}
where
$ R_{\mu \, a}^{\,  \iota}$ is the Rarita--Schwinger
spin-tensor describing spin-$\frac{3}{2}$ resonance field. It is
contracted with the
$P^{  I=\frac{3}{2}}$
isospin-$\frac{3}{2}$ projecting operator (\ref{isospin_projecting_oper}).
$g_{\pi N  {\Delta}}$ is a dimensional coupling constant.
Computation from  the PDG~\cite{PDG2020} average
${\Delta(1232)} \rightarrow \pi N$  decay width gives:
$
| g_{\pi N  {\Delta}}|  \approx 7.6 \ \;{\rm GeV}^{-1}\,.
$

After the reduction the matrix element corresponding to the
cross-channel $\Delta$-exchange contribution into $\pi N$ TDAs
(see  Fig.~\ref{Fig_Cross_Ch_Pole}) reads:
\begin{eqnarray}
 &&
\langle \pi_a(p_\pi) |  \widehat{O}^{\alpha  \beta   \gamma}_{\rho \tau \chi}( 1, \,2, \, 3 )|  N_{\iota}(p_1,s_1) \rangle
\nonumber \\ &&
= \sum_{s_\Delta}\langle 0 |
\widehat{O}^{\alpha  \beta   \gamma}_{\rho \tau \chi}
( 1, \,2, \, 3 )
|  \Delta_{b \kappa}(-\Delta, s_\Delta) \rangle
{P^{I= \frac{3}{2}}}^{\;\;  \kappa}_{b \ \  a \iota}
\frac{   g_{\pi N \Delta} \, \bar{\mathcal U}_\varrho^\nu  (-\Delta, s_\Delta) }{\Delta^2-m_\Delta^2} ( i P_\nu) \left(    U(p_1, s_1) \right)_\varrho\,,
\label{matrix_el_Delta}
\end{eqnarray}
where
$\mathcal{U}^\nu_{\varrho}$
is the Rarita--Schwinger spin-tensor
describing the  $\Delta$ resonance.
For the matrix element involving $\Delta$ in the second line of
(\ref{matrix_el_Delta}) we employ the parametrization
(\ref{isospin_parametrization_Delta_DA})
with
$M^\Delta_{\rho \tau \chi}$
given by (\ref{Parametrization_Delta_DA_FZ}).
As a consequence of the identity (\ref{Projecting_ta_tensor})
the cross-channel $\Delta$-exchange populates only the isospin-$\frac{3}{2}$ $\pi N$ TDAs.

Below, to the leading twist-$3$ accuracy,  we present the contribution of
(\ref{matrix_el_Delta})
into the invariant $\pi N$ TDAs using the
fully covariant parametrization of  Sec.~\ref{SubSec_Polynomiality}.

\begin{eqnarray}
 &&
 \left\{ V_{1,2}^{(  \pi N)_{3/2}}, \,  A_{1,2}^{(  \pi N)_{3/2}} \right\} (x_1,x_2,x_3,\xi,\Delta^2)\Big|_{\Delta(1232)}
  \nonumber \\ &&
=-\Theta_{\rm ERBL}(x_1,x_2,x_3)
 \frac{1}{(2 \xi)^2}\left\{ V^\Delta, \, A^\Delta \right\} \left( \frac{x_1}{2 \xi}, \frac{x_2}{2 \xi}, \frac{x_3}{2 \xi} \right)
 \frac{ g_{\pi N \Delta} \lambda_{\Delta}^{\frac{1}{2}} m_N f_\pi}{ \sqrt{2}(\Delta^2-m_\Delta^2) f_N} \;
R_{1,2}( \xi,  m_\Delta)\,;
\label{Contribution_of_Delta_exchange_finalVA}
\end{eqnarray}
\begin{eqnarray}
 &&
T_{1}^{(\pi N)_{3/2}}(x_1,x_2,x_3, \xi, \Delta^2) \Big|_{\Delta(1232)}
\nonumber \\ &&
=-\Theta_{\rm ERBL}(x_1,x_2,x_3)
\Bigm\{
\frac{1}{(2 \xi)^2} T^\Delta \left( \frac{x_1}{2 \xi}, \frac{x_2}{2 \xi}, \frac{x_3}{2 \xi} \right)
\frac{ g_{\pi N \Delta} \lambda_{\Delta}^{\frac{1}{2}} m_N f_\pi}{\sqrt{2} f_N (\Delta^2-m_\Delta^2) }
 R_{1}( \xi,  m_\Delta)
  \nonumber \\
  && +
 \frac{1}{(2 \xi)^2} \phi^\Delta \left( \frac{x_1}{2 \xi}, \frac{x_2}{2 \xi}, \frac{x_3}{2 \xi} \right)
 \frac{  g_{\pi N \Delta} f_{\Delta}^{\frac{3}{2}} m_N^2 f_\pi}{ \sqrt{2} f_N  }
 \left(
\frac{\widetilde{ R}_{1 }(\xi, m_\Delta )}{ \Delta^2-m_\Delta^2 } - \frac{1-\xi}{2 m_N^2} \right)\Bigm\}\,;
\end{eqnarray}
\begin{eqnarray}
 &&
 T_{2}^{(\pi N)_{3/2}}(x_1,x_2,x_3, \xi, \Delta^2) \Big|_{\Delta(1232)}
\nonumber \\ &&
=
-\Theta_{\rm ERBL}(x_1,x_2,x_3)
  \Bigl\{
\frac{1}{(2 \xi)^2} T^\Delta \left( \frac{x_1}{2 \xi}, \frac{x_2}{2 \xi}, \frac{x_3}{2 \xi} \right)
\frac{ g_{\pi N \Delta} \lambda_{\Delta}^{\frac{1}{2}} m_N f_\pi}{\sqrt{2}f_N (\Delta^2-m_\Delta^2) }
 R_{ 2}( \xi,  m_\Delta)
  \nonumber \\ &&
 +
 \frac{1}{(2 \xi)^2} \phi^\Delta \left( \frac{x_1}{2 \xi}, \frac{x_2}{2 \xi}, \frac{x_3}{2 \xi} \right)
 \frac{  g_{\pi N \Delta} f_{\Delta}^{\frac{3}{2}} m_N^2 f_\pi}{ \sqrt{2} f_N }
 \left(
\frac{\widetilde{ R}_{2 }(\xi, m_\Delta )}{ \Delta^2-m_\Delta^2 } + \frac{1-\xi}{4 m_N^2} \right)\Bigr\}\,;
\label{Contribution_of_Delta_exchange_finalT12}
\end{eqnarray}
\begin{eqnarray}
 &&
T_{3,4}^{(\pi N)_{3/2}}(x_1,x_2,x_3, \xi, \Delta^2)\Big|_{\Delta(1232)}
\nonumber \\ &&
=-\Theta_{\rm ERBL}(x_1,x_2,x_3)
\frac{1}{(2 \xi)^2} \phi^\Delta \left( \frac{x_1}{2 \xi}, \frac{x_2}{2 \xi}, \frac{x_3}{2 \xi} \right)
 \times
 \frac{  g_{\pi N \Delta} f_{\Delta}^{\frac{3}{2}} m_N^2 f_\pi}{\sqrt{2}(\Delta^2-m_\Delta^2) f_N}
 R_{3,4}(m_\Delta)\,.
   \label{Contribution_of_Delta_exchange_finalT34}
\end{eqnarray}
The functions $R_{1,2,3,4}$, $\widetilde{R}_{1,2}$ are determined by the residue at the
pole $\Delta^2=m_\Delta^2$.
They read as:

\begin{eqnarray}
 &&
R_1(\xi,   m_\Delta )=
\frac{(\xi -3) m_\Delta^2+2 m_N   m_\Delta \xi +4 \left(m_N^2-m_\pi^2 \right) \xi }{3 m_N m_\Delta}\,;
\nonumber \\  &&
R_2(\xi,  m_\Delta )=
\frac{-4 \xi  m_N^3+\left(4 \xi  m^2+6 m_\Delta^2\right) m_N-m_\Delta^3 (\xi -3)+4 m_\pi^2 m_\Delta \xi }{6 m_N m_\Delta^2}\,;
\nonumber \\  &&
R_3(m_\Delta )=- \frac{m_\Delta}{m_N}\,; \ \ \ R_4(m_\Delta )=1+ \frac{m_\Delta}{2m_N}\,;
\nonumber \\  &&
\widetilde{R_1}(\xi,   m_\Delta )= \frac{\left(m_N (m_N+m_\Delta)-m_\pi^2\right) \xi }{m_N^2}
-m_\Delta^2 \frac{(1-\xi)}{2 m_N^2}\,;
\nonumber \\  &&
\widetilde{R_2}(\xi,   m_\Delta )=
\frac{    m_N m_\Delta+(m_\pi^2+m_N^2) \xi }{2 m_N^2}+ m_\Delta^2 \frac{(1-\xi)}{4 m_N^2}\,.
\label{R_on_shell}
\end{eqnarray}
The numerical values of the
dimensional constants
$\lambda_{\Delta}^{\frac{1}{2}}$
and
$f_{\Delta}^{\frac{3}{2}}$
are quoted \textit{e.g.} in Ref.~\cite{Farrar:1988vz}:
\begin{eqnarray}
 &&
| \lambda_{\Delta}^{\frac{1}{2}}|  \equiv \sqrt{\frac{3}{2}} m_\Delta | f_{\Delta}^{\frac{1}{2}}| =(1.8 \pm 0.3) \times 10^{-2} \ \;{\rm GeV}^3\,;
\nonumber \\
 &&
| f_{\Delta}^{\frac{3}{2}}| =1.4 \times 10^{-2} \ \;{\rm GeV}^2\,.
\end{eqnarray}

The $\Delta$-exchange contribution into $\pi N$ TDAs
satisfies the isospin and permutation symmetry relations for the isospin-$\frac{3}{2}$ TDAs
established in  Sec.~\ref{SubSec_Isospin_I=1_meson_TDA}.
Employing the set of  the Fierz identities
(\ref{Fierz_for_piNTDA_structures}),
(\ref{Fierz for t34}), one may check that
the contributions involving
$V^\Delta$, $A^\Delta$, $T^\Delta$
decouple  and satisfy the symmetry relations
(\ref{symmetries_MI32piNTDA})
as a consequence of the symmetry relations
(\ref{Isospin_and_Sym_Relations_DeltaDA})
for
$\Delta$ DAs.
In order to satisfy the symmetry relation
(\ref{symmetries_MI32piNTDA})
for the contributions involving $\phi^\Delta$
a polynomial background regular at $\Delta^2=m_\Delta^2$
has been added to
$\pi N$ TDAs $T_{1,2}$ in
(\ref{Contribution_of_Delta_exchange_finalT12}).
The necessity for these background terms follows  from the fact that the Fierz identities
(\ref{Fierz for t34})
for the tensor structures
$t_3^{ \pi N }$,
$t_4^{ \pi N }$
involve coefficients
(\ref{Def_g12_h12})
with  explicit dependence on
$\xi$
and
$\Delta^2$.
The symmetry relation
(\ref{symmetries_MI32piNTDA})
fixes unambiguously the  background terms 
in
(\ref{Contribution_of_Delta_exchange_finalT12}).

It is also worth mentioning that the $\Delta(1232)$-exchange contribution
satisfies the polynomiality property of $\pi N$ TDAs.
Indeed, the kinematic prefactors $R_{1,2}$ and $\tilde{R}_{1,2}$ in Eqs.~(\ref{R_on_shell})
are
linear functions of $\xi$, while  $R_{3,4}$  are constants. Therefore, the $\Delta(1232)$-exchange contributes not only to the coefficients at
highest possible powers of $\xi$ for the $(n_1, \,n_2,\,n_3)$-th Mellin moments of $\pi N$ TDAs (as the nucleon exchange contribution)
but for some TDAs also to the coefficients at
powers of $\xi$ which are one power less than the highest possible (see discussion in  Sec.~\ref{SubSec_Polynomiality}).

\subsubsection{Cross-channel nucleon exchange contribution
for nucleon-to-vector-meson { {TDA}}s}
\label{SubSec_Nucle_ex_VN_TDA}
\mbox

In this subsection we present the cross-channel nucleon exchange contribution
to nucleon-to-vector-meson TDAs. The formulas can be employed both for the $I=0$ and $I=1$ vector-mesons. In the latter case the cross-channel
nucleon exchange contributes only  to the isospin-$\frac{1}{2}$
invariant amplitude
$M^{(V N)_{1/2} \, \{1 2 \}}_{\rho \tau \chi}$,
thus populating only isospin-$\frac{1}{2}$ $VN$ TDAs.

$VN$ TDAs computed from the cross-channel nucleon exchange
model satisfy the set of  isospin and permutation symmetry identities that,
for the $I=0$
meson case\footnote{For the case of $I=1$ mesons the same set of identities follows
from the relation analogous to  Eqs.~(\ref{Isospin_Id_piNTDA}), (\ref{Relations_piNTDA_permutations}) for the $I=\frac{1}{2}$ $VN$ TDAs.},
follow from the relation analogous to Eqs.~(\ref{Isospin_Id_NDA}), (\ref{Relations_NDA_permutations})
and the set of the Fierz identities summarized in   App.~\ref{identities}.
It is convenient to show the results for the groups of $VN$ TDAs interlinked through the
set of the isospin relations.
\bi
\item $V_{1 {\mathcal{E}}}^{VN}$, $A_{1 {\mathcal{E}}}^{VN}$, $T_{1 {\mathcal{E}}}^{VN}$, $T_{2 {\mathcal{E}}}^{VN}$
satisfy the isospin symmetry relations based on the Fierz transformation set (\ref{Fiers_1E_set}).
\begin{eqnarray}
 &&
\Bigl\{ V_{1 {\mathcal{E}}}^{VN}, \,  A_{1 {\mathcal{E}}}^{VN}, \,
T_{1 {\mathcal{E}}}^{VN},  T_{2 {\mathcal{E}}}^{VN} \Bigr\}
(x_1, x_2, x_3, \xi, \Delta^2) \Big|_{N(940)} \nn \\ && =
\Theta_{\rm ERBL}(x_1,x_2,x_3) \frac{1}{(2 \xi)^2}
K_{1 {\mathcal{E}}}^{VN}( \xi, \Delta^2)
\Bigl\{
V^p, \, A^p, \, -T^p,\,-T^p
\Bigr\} \left(  \frac{x_1}{2 \xi}, \frac{x_2}{2 \xi}, \frac{x_3}{2 \xi}  \right),
\end{eqnarray}
where
\begin{equation}
K_{1 {\mathcal{E}}}^{VN}( \xi, \Delta^2)= \frac{f_N}{\Delta^2-m_N^2} \left( G^V_{VNN} \frac{2\xi(1-\xi)}{1+\xi} + G^T_{VNN}\,  \xi  \left(\frac{2 \xi }{1+\xi }-\frac{ \Delta^2}{m_N^2}\right) \right).
\end{equation}

Below we present explicitly the isospin and permutation symmetry
relation linking
$V_{1 {\mathcal{E}}}^{VN}$, $A_{1 {\mathcal{E}}}^{VN}$, $T_{1 {\mathcal{E}}}^{VN}$ and $T_{2 {\mathcal{E}}}^{VN}$
TDAs:
\begin{eqnarray}
 &&
V_{1 {\mathcal{E}}}^{VN}(x_1,x_2,x_3,\xi, \Delta^2)+ \frac{1}{2}
\left(
V_{1 {\mathcal{E}}}^{VN}-A_{1 {\mathcal{E}}}^{VN}+T_{1 {\mathcal{E}}}^{VN}+T_{2 {\mathcal{E}}}^{VN}
\right)(x_3,x_1,x_2,\xi, \Delta^2)
\nonumber \\ &&
+ \frac{1}{2}
\left(
V_{1 {\mathcal{E}}}^{VN}-A_{1 {\mathcal{E}}}^{VN}+T_{1 {\mathcal{E}}}^{VN}+T_{2 {\mathcal{E}}}^{VN}
\right)(x_3,x_2,x_1,\xi, \Delta^2)=0.
\end{eqnarray}
The validity of this and the analogous
isospin and permutation  symmetry identities for different groups of $VN$ TDAs within the cross-channel nucleon exchange model
follows from the familiar isospin and permutation symmetry identities for
the leading twist nucleon DAs (\ref{Isospi_ID_NDA}).

\item $V_{1 T}^{VN}$, $A_{1 T}^{VN}$, $T_{1 T}^{VN}$ satisfy the isospin symmetry relations based on the Fierz transformation set (\ref{Fiers_1T_set})
\begin{eqnarray}
 &&
\Bigl\{ V_{1 { T}}^{VN}, \,  A_{1 { T}}^{VN}, \,
T_{1 { T}}^{VN} \Bigr\}
(x_1, x_2, x_3, \xi, \Delta^2) \Big|_{N(940)} \nn \\ && =
\Theta_{\rm ERBL}(x_1,x_2,x_3) \frac{1}{(2 \xi)^2}
K_{1 {T}}^{VN}( \xi, \Delta^2)
\Bigl\{
V^p, \, A^p, \, -T^p,
\Bigr\} \left(  \frac{x_1}{2 \xi}, \frac{x_2}{2 \xi}, \frac{x_3}{2 \xi}  \right),
\end{eqnarray}
where
\begin{equation}
K_{1 T}( \xi, \Delta^2)= \frac{f_N}{\Delta^2-m_N^2} \left( -G^V_{VNN} \frac{2 \xi  (1+3 \xi)}{1-\xi}   \right).
\end{equation}

\item $V_{1 n}^{VN}$, $A_{1 n}^{VN}$, $T_{1 n}^{VN}$ satisfy the isospin symmetry relations based on the Fierz transformation set (\ref{Fiers_1T_set})
\begin{eqnarray}
 &&
\Bigl\{ V_{1 { n}}^{VN}, \,  A_{1 { n}}^{VN}, \,
T_{1 { n}}^{VN} \Bigr\}
(x_1, x_2, x_3, \xi, \Delta^2) \Big|_{N(940)} \nn \\ && =
\Theta_{\rm ERBL}(x_1,x_2,x_3) \frac{1}{(2 \xi)^2}
K_{1 {n}}^{VN}( \xi, \Delta^2)
\Bigl\{
V^p, \, A^p, \, -T^p,
\Bigr\} \left(  \frac{x_1}{2 \xi}, \frac{x_2}{2 \xi}, \frac{x_3}{2 \xi}  \right),
\end{eqnarray}
where
\begin{eqnarray}
 &&
K_{1 n}( \xi, \Delta^2)
\nonumber \\ &&
= \frac{f_N}{\Delta^2-m_N^2}
\left(
\frac{(1+\xi) \left(m^2_V-\Delta_T^2\right)}{m_N^2 (1-\xi)^2}-\frac{1}{1+\xi} \right)
\left( G^V_{VNN}(-4 \xi) + G^T_{VNN} \frac{\xi (1-\xi)}{1+\xi}
\right).
\end{eqnarray}

\item $V_{2 {\mathcal{E}}}^{VN}$, $A_{2 {\mathcal{E}}}^{VN}$, $T_{3 {\mathcal{E}}}^{VN}$, $T_{4 {\mathcal{E}}}^{VN}$
satisfy the isospin symmetry relations based on the Fierz transformation set
(\ref{Fiers_2E_set}).
Within the cross channel nucleon exchange model this set decouples from the
$V_{1 T}^{VN}$, $A_{1 T}^{VN}$, $T_{1 T}^{VN}$ set.
\begin{eqnarray}
 &&
\Bigl\{ V_{2 {\mathcal{E}}}^{VN}, \,  A_{2 {\mathcal{E}}}^{VN}, \,
T_{3 {\mathcal{E}}}^{VN},  T_{4 {\mathcal{E}}}^{VN} \Bigr\}
(x_1, x_2, x_3, \xi, \Delta^2) \Big|_{N(940)} \nn \\ && =
\Theta_{\rm ERBL}(x_1,x_2,x_3) \frac{1}{(2 \xi)^2}
K_{2 {\mathcal{E}}}^{VN}( \xi, \Delta^2)
\Bigl\{
V^p, \, A^p, \, -T^p,\,-T^p
\Bigr\} \left(  \frac{x_1}{2 \xi}, \frac{x_2}{2 \xi}, \frac{x_3}{2 \xi}  \right),
\end{eqnarray}
where
\begin{equation}
K_{2 {\mathcal{E}}}^{VN}( \xi, \Delta^2)=\frac{f_N}{\Delta^2-m_N^2} \left( G^V_{VNN}(-2 \xi) + G^T_{VNN}\xi \right).
\end{equation}

\item $V_{2T}^{VN}$, $A_{2 T}^{VN}$, $T_{2T}^{VN}$, $T_{3 T}^{VN}$ satisfy the isospin symmetry relations based on the Fierz transformation set (\ref{Fiers_2T_set}).
\begin{eqnarray}
 &&
\Bigl\{ V_{2 {T}}^{VN}, \,  A_{2 {T}}^{VN}, \,
T_{2 {T}}^{VN},  T_{3 {T}}^{VN} \Bigr\}
(x_1, x_2, x_3, \xi, \Delta^2) \Big|_{N(940)} \nn \\ && =
\Theta_{\rm ERBL}(x_1,x_2,x_3) \frac{1}{(2 \xi)^2}
K_{2 {T}}^{VN}( \xi, \Delta^2)
\Bigl\{
V^p, \, A^p, \, -T^p,\,-T^p
\Bigr\} \left(  \frac{x_1}{2 \xi}, \frac{x_2}{2 \xi}, \frac{x_3}{2 \xi}  \right),
\end{eqnarray}
where
\begin{equation}
K_{2 T}( \xi, \Delta^2)= \frac{f_N}{\Delta^2-m_N^2} \left( G^T_{VNN}   \frac{\xi (1+\xi)}{1-\xi} \right).
\end{equation}

\item $V_{2n}^{VN}$, $A_{2 n}^{VN}$, $T_{2n}^{VN}$, $T_{3 n}^{VN}$ satisfy the isospin symmetry relations based on the Fierz transformation set (\ref{Fiers_2T_set}).
\begin{eqnarray}
 &&
\Bigl\{ V_{2 {n}}^{VN}, \,  A_{2 {n}}^{VN}, \,
T_{2 {n}}^{VN},  T_{3 {n}}^{VN} \Bigr\}
(x_1, x_2, x_3, \xi, \Delta^2) \Big|_{N(940)} \nn \\ && =
\Theta_{\rm ERBL}(x_1,x_2,x_3) \frac{1}{(2 \xi)^2}
K_{2 {n}}^{VN}( \xi, \Delta^2)
\Bigl\{
V^p, \, A^p, \, -T^p,\,-T^p
\Bigr\} \left(  \frac{x_1}{2 \xi}, \frac{x_2}{2 \xi}, \frac{x_3}{2 \xi}  \right),
\end{eqnarray}
where
\begin{equation}
K_{2n}^{VN}(\xi, \Delta^2)= \frac{f_N}{\Delta^2-m_N^2} \xi   \left(  \frac{1+\xi}{(1-\xi)^2} \frac{  \left(m^2_V-\Delta_T^2\right)}{m_N^2  }-\frac{1}{1+\xi}\right) G^T_{VNN}.
\end{equation}

\item Finally,
\begin{eqnarray}
 &&
 T_{4T}^{VN}(x_1, x_2, x_3, \xi, \Delta^2) \Big|_{N(940)}=0;
 \nonumber  \\
  &&
T_{4n}^{VN}(x_1, x_2, x_3, \xi, \Delta^2) \Big|_{N(940)}=0;
\end{eqnarray}
\ei

A useful summary of $G^{V,T}_{VNN}$ couplings can be found  \textit{e.g.} in the Table~9.2 of Ref.~\cite{Dumbrajs:1983jd}.
See also Ref.~\cite{Grein:1979nw} for the couplings of the $\omega$-meson
and  Refs.~\cite{Mergell:1995bf,Meissner:1997qt} for the case of $\phi(1020)$ meson coupling.

\subsection{Chiral constraints for nucleon-to-pion { {TDA}}s}
\label{SubSec_Chiral_Constr}
\mbox

In this subsection, following Ref.~\cite{Pire:2011xv}, we discuss the implication of the chiral constraints for $\pi N$
TDAs. We rely on the soft-pion theorem~\cite{Pobylitsa:2001cz}
for $\pi N$ GDAs (see  Sec.~\ref{SubSec_Definition_GDAs}) proposed in
\cite{Braun:2006td}
to be valid  at a scale $Q^2 \gg \Lambda_{\rm QCD}^3/m_\pi$.

\begin{figure}[H]
\begin{center}
\includegraphics[width=0.45\textwidth]{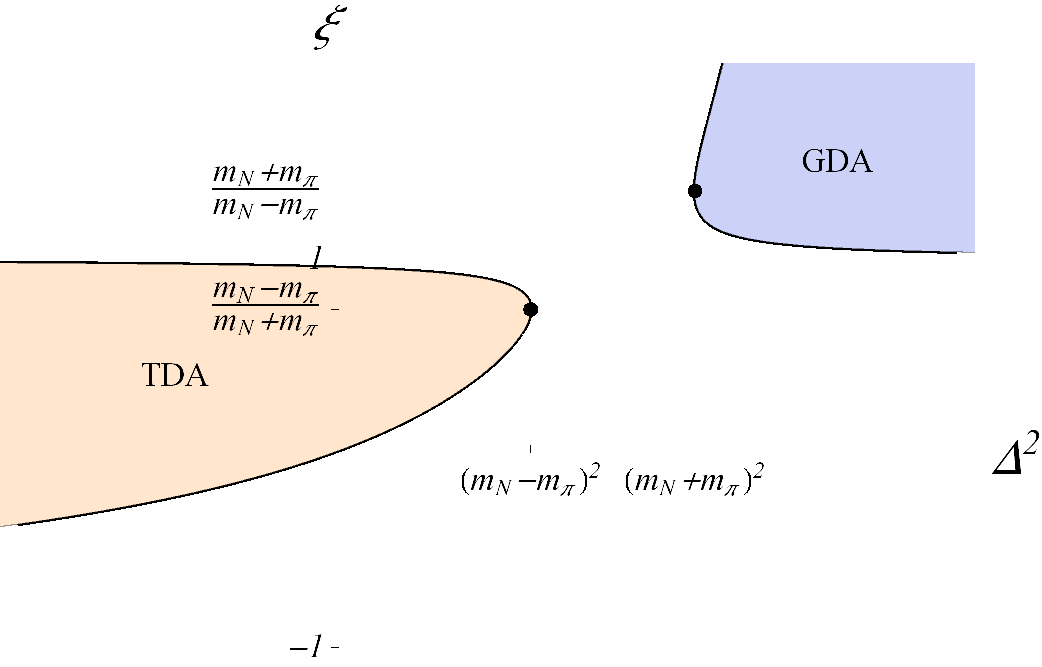} \ \ \
\includegraphics[width=0.45\textwidth]{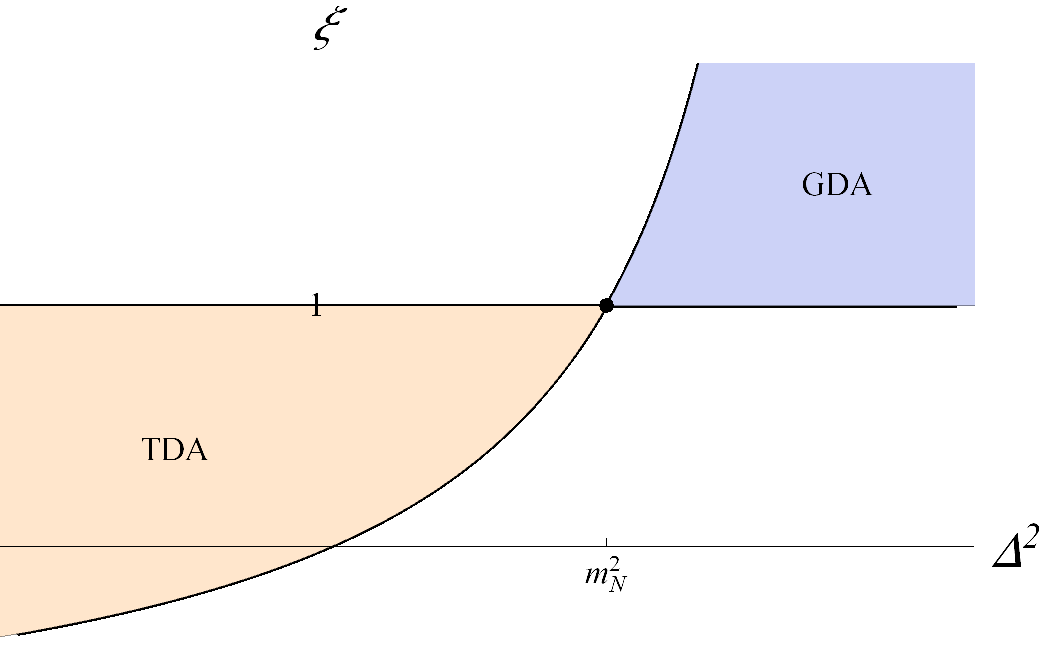}
\end{center}
     \caption{Physical domains for $\pi N$ TDAs/GDAs
          (bounded by the condition $\Delta_T^2 \le 0$) in  the
     $(\Delta^2,\xi)$-plane in the case $m_\pi \ne 0$ (left
panel) and in the chiral limit $m_\pi=0$ (right panel). [Reprinted Figure 1
from Ref.~\cite{Pire:2011xv}. Copyright (2011) by American Physical Society.]}
\label{Fig_Thresholds}
\end{figure}

The left panel of  Fig.~\ref{Fig_Thresholds}  depicts the physical domains for the direct channel reaction  (\ref{Hard_subpr})
and the cross channel reaction (\ref{Cross_ch_react})
for the physical value of the pion mass.
It is determined by the requirement that
the transverse momentum transfer $\Delta_T=-P'_T$ must be space-like:
\begin{equation}
\Delta_T^2= \frac{1-\xi}{1+\xi}\left( \Delta^2-2 \xi \left[ \frac{m_N^2}{1+\xi} - \frac{m_\pi^2}{1-\xi} \right] \right) \le 0\,.
\label{Delta_t2}
\end{equation}

One may distinguish  the direct-channel regime, with its threshold at
$\Delta^2 = (m_N-m_\pi)^2$ and $\xi=\frac{m_N-m_\pi}{m_N+m_\pi}$, and the cross-channel one, with its threshold at
$\Delta^2 = (m_N+m_\pi)^2$ and $\xi=\frac{m_N+m_\pi}{m_N-m_\pi}$.
The upper (lower) branch of the curve bordering the physical domain of the direct channel regime
tends to $\xi=1$ ($\xi=-1$) when $\Delta^2 \rightarrow -\infty$.
Note that the physical domain of the direct channel of
(\ref{Hard_subpr})
includes both negative and positive values of $\Delta^2$.

In the chiral limit ($m_\pi=0$) the two thresholds stick together (see the right panel of  Fig.~\ref{Fig_Thresholds}).
Relying on the crossing transformation
(\ref{Crossing_TDA_GDA1}), (\ref{Crossing_TDA_GDA2})
between $\pi N$ GDAs and $\pi N$ TDAs
we employ the soft-pion theorem for  $\pi N$ GDAs
to work out
constraints for threshold values of $\pi N$ TDAs
in the chiral limit ($m_\pi \rightarrow 0$).
The problem of validity of analytic continuation in $\Delta^2$ existing for $m_\pi \ne 0$ has the same status as that for the case of pion GPDs v.s. $2\pi$ GDAs~\cite{Polyakov:1998ze} (see also discussion
in~\cite{Kivel:2002ia}).
In order to rely on the soft-pion limit as
a reference point for  realistic modeling of $\pi N$ TDAs
we need to  assume  smallness of non-analytic corrections to the relevant matrix element
in the narrow domain in  $(\Delta^2,\,\xi)$-plane defined by
the inequalities
\begin{equation}
(m_N-m_\pi)^2 < \Delta^2 < (m_N+m_\pi)^2\,; \ \ \  \frac{m_N-m_\pi}{m_N+m_\pi} < \xi < \frac{m_N+m_\pi}{m_N-m_\pi}.
\end{equation}

The isospin technique presented in  Sec.~\ref{SubSec_Isospin} accounts for the isotopic symmetry and
permits an easy handling of all possible isospin channels.
It allows to fully take into account the consequences of the isotopic and permutation symmetries for
$\pi N$ GDAs.

\subsubsection{Soft-pion theorem for nucleon-to-pion { {GDA}}s}
\label{SubsubSec_Chiral_Constr}
\mbox

According to the partial conservation of the axial current (PCAC) hypothesis
(see \textit{e.g.}~\cite{Alfaro_red_book})
the following soft-pion theorem is valid for the matrix element of
the three-quark light-cone operator
$\widehat{O}_{\rho \tau \chi}^{\, \alpha \beta \gamma}$
(\ref{Def_operator_O_isotopic})
defining  $\pi N$ GDAs:
\begin{equation}
\langle 0 |  \widehat{O}^{\alpha \beta \gamma}_{\rho  \tau \chi}
 (1,\,2,\,3)
|  \pi_a N_\iota \rangle
\Big|_{ {\text{soft} \atop  \text{pion}}}=
-\frac{i}{f_\pi}
\langle 0 |
\left[
\widehat{Q}_5^a, \, \widehat{O}^{\alpha \beta \gamma}_{\rho  \tau \chi}
 (1,\,2,\,3)
\right]
 | N_\iota \rangle \,.
 \label{soft_pion_theorem_pobyl}
\end{equation}
The commutator of the chiral charge operator
$\widehat{Q}^5_a$
with quark field operators is given by
\begin{equation}
\left[\widehat{Q}^5_a,\ \Psi^\alpha_\eta \right]= - \frac{1}{2}
 (\sigma_a)^\alpha_{\;\delta}  \gamma^5_{\eta \tau} \Psi^\delta_\tau\,,
\end{equation}
where $\sigma_a$ are the Pauli matrices.
The commutator of the chiral charge operator
$\widehat{Q}^5_a$
with the operator
$\widehat{O}$
in
(\ref{soft_pion_theorem_pobyl})
is computed  with the help of the chain rule
$[A,BCD]=[A,B]CD+B[A,C]D+BC[A,D]$.
This allows to express $\pi N$ GDAs
in terms of the leading twist nucleon DAs
$M^{N \, \{12\}}$,
$M^{N \, \{13\}}$
defined in (\ref{nucleon_DA_isospin_dec_1}):
\begin{eqnarray}
  &&
4\langle 0 |  \widehat{O}^{\alpha \beta \gamma}_{\rho \tau \chi}
(1,\,2,\,3)
|  \pi_a N_\iota \rangle
\Big|_{{\text{soft} \atop  \text{pion}}} =
-\frac{i}{f_\pi}
\langle 0 |
\left[
\widehat{Q}_5^a, \, \widehat{O}^{\alpha \beta \gamma}_{\rho  \tau \chi}
(1,\,2,\,3)
\right]
 | N_\iota \rangle
\nonumber \\ &&
  =\frac{i}{2 f_\pi}
 \Bigl\{
 (\sigma_a)^\alpha_{\; \delta}  \varepsilon^{\delta \beta} \delta^\gamma_\iota \gamma^5_{\rho \eta} M^{N \, \{13\}}_{\eta \tau \chi}( 1, 2, 3)+
 (\sigma_a)^\alpha_{\; \delta } \varepsilon^{\delta \gamma} \delta^\beta_\iota  \gamma^5_{\rho \eta} M^{N \,\{12\}}_{\eta \tau \chi}( 1, 2, 3)
 \nonumber \\ &&
+(\sigma_a)^\beta_{\; \delta}  \varepsilon^{\alpha \delta} \delta^\gamma_\iota \gamma^5_{\tau \eta} M^{N \,\{13\}}_{\rho \eta \chi}( 1, 2, 3)+
 (\sigma_a)^\beta_{\; \delta}  \varepsilon^{\alpha \gamma} \delta^\delta_\iota \gamma^5_{\tau \eta} M^{N \, \{12\}}_{\rho \eta \chi}( 1, 2, 3)
 \nonumber \\ &&
 +(\sigma_a)^\gamma_{\; \delta}  \varepsilon^{\alpha \beta} \delta^\delta_\iota \gamma^5_{\chi \eta} M^{N \, \{13\}}_{\rho \tau \eta}( 1, 2, 3)+
 (\sigma_a)^\gamma_{\; \delta}  \varepsilon^{\alpha \delta} \delta^\beta_\iota \gamma^5_{\chi \eta} M^{N \, \{12\}}_{\rho \tau \eta}( 1, 2, 3)
   \Bigr\}\,.
  \label{soft_pion_full_glory}
\end{eqnarray}
The isospin parametrization for $\pi N$ GDAs
is analogous to that of $\pi N$ TDAs described in  Sec.~\ref{SubSec_Isospin}.
Employing the general isospin parametrization for the leading twist
nucleon DAs
(\ref{Isospin_parmetrization_N_DA})
and the symmetry relations
(\ref{Relations_NDA_permutations})
 together with the Fierz identities
(\ref{Fierz_nucleon_structures})
we establish the consequences of the soft-pion theorem
for the invariant isospin-$\frac{3}{2}$ and isospin-$\frac{1}{2}$ $\pi N$ GDAs.
The invariant amplitudes
$ M^{(\pi N)_{3/2}}_{\rho \tau \chi} (1,\,2,\,3)$
and
$M^{(\pi N)_{1/2}}_{\rho \tau \chi} (1,\,2,\,3)$
at the pion threshold are expressed in terms of
the leading twist nucleon DA $\phi^N$ (\ref{def_phi_N}):

\begin{eqnarray}
 &&
\nonumber M^{(\pi N)_{3/2}}_{\rho \tau \chi} (1,\,2,\,3) \Big|_{ {\text{soft} \atop  \text{pion}}}
 =\frac{i}{2 f_\pi}
 \Bigl\{
\gamma^5_{\rho \eta} M^{N \, \{12\}}_{\eta \tau \chi} (1,\,2,\,3)
-\gamma^5_{\chi \eta} M^{N \, \{12\}}_{\rho \tau \eta}(1,\,2,\,3)
\\ &&
+\gamma^5_{\rho \eta} M^{N \, \{13\}}_{\eta \tau \chi}(1,\,2,\,3)
-
 \gamma^5_{\tau \eta} M^{N \, \{13\}}_{\rho \eta \chi}(1,\,2,\,3)
 \Bigr\}
\nonumber \\ &&
=\frac{i f_N}{ f_\pi}
 \Bigl\{
- \left( \gamma^5_{\chi \eta} v^N_{\rho \tau, \, \eta} \right) \frac{1}{2} \left[
\phi^N(1,2,3)+\phi^N(2,1,3)+\phi^N(3,2,1)+\phi^N(3,1,2)
\right]
\nonumber \\ &&
- \left( \gamma^5_{\chi \eta} a^N_{\rho \tau, \, \eta} \right)
\frac{1}{2} \left[
-\phi^N(1,2,3)+\phi^N(2,1,3)-\phi^N(3,2,1)+\phi^N(3,1,2)
\right]
\nonumber \\ &&
- \left( \gamma^5_{\chi \eta} t^N_{\rho \tau, \, \eta} \right)
\frac{1}{2} \left[
\phi^N(1,3,2)+\phi^N(2,3,1)
\right]
  \Bigr\}\,;
\label{M_T_sym_soft_pion}
\end{eqnarray}
\begin{eqnarray}
 &&
\nonumber M^{(\pi N)_{1/2} \, \{ 12\}}_{\rho \tau \chi}(1,\,2,\,3)
\Big|_{{\text{soft} \atop  \text{pion}}}
 =\frac{i}{2 f_\pi} \frac{1}{3}
 \Bigl\{
 \gamma^5_{\rho \eta} M^{N \, \{12\}}_{\eta \tau \chi}(1,\,2,\,3)
 +
 3 \gamma^5_{\tau \eta} M^{N \, \{12\}}_{\rho \eta \chi}(1,\,2,\,3)
 \nonumber \\ &&
 -\gamma^5_{\chi \eta} M^{N \, \{12\}}_{\rho \tau \eta}(1,\,2,\,3)
 -2\gamma^5_{\rho \eta} M^{N \, \{13\}}_{\eta \tau \chi}(1,\,2,\,3)
 +
 2 \gamma^5_{\tau \eta} M^{N \, \{13\}}_{\rho \eta \chi}(1,\,2,\,3)
  \Bigr\}
   \nonumber \\ &&
   =\frac{i}{  f_\pi}  \Bigl\{
- \left( \gamma^5_{\chi \eta} v^N_{\rho \tau, \, \eta} \right) \frac{1}{12} \left[ -\phi^N(1,2,3)-\phi^N(2,1,3)-4(\phi^N(3,1,2)+\phi^N(3,2,1))  \right]
\nonumber \\ &&
- \left( \gamma^5_{\chi \eta} a^N_{\rho \tau, \, \eta} \right) \frac{1}{12}
\left[\phi^N(1,2,3)-\phi^N(2,1,3) - 4(\phi^N(3,1,2)-\phi^N(3,2,1)) \right]
\nonumber \\ &&
- \left( \gamma^5_{\chi \eta} t^N_{\rho \tau, \, \eta} \right)
\frac{5}{12}
\left[ \phi^N(1,3,2)+\phi^N(2,3,1) \right]
  \Bigr\}\,;
  \label{M12_12_piN}
\end{eqnarray}
\begin{equation}
 M^{(\pi N)_{1/2} \, \{ 13\}}_{\rho \tau \chi}(1,\,2,\,3)\Big|_{{\text{soft} \atop  \text{pion}}}=
  M^{(\pi N)_{1/2} \, \{ 12\}}_{\rho \chi \tau}(1,\,3,\,2)
  \Big|_{{\text{soft} \atop  \text{pion}}}.
    \label{M13_13_piN}
\end{equation}
 $\pi N$ GDAs in the soft-pion limit
(\ref{M_T_sym_soft_pion}), (\ref{M12_12_piN}), (\ref{M13_13_piN})
naturally satisfy the isospin-$\frac{3}{2}$ and isospin-$\frac{1}{2}$
symmetry and permutation symmetry relations (\ref{symmetries_MI32piNTDA}), (\ref{Isospin_Id_piNTDA}),
(\ref{Relations_piNTDA_permutations})
that provides an additional cross-check.

From (\ref{M_T_sym_soft_pion}), (\ref{M12_12_piN}), (\ref{M13_13_piN}), it is straightforward to verify that for particular
$\pi N$
GDAs we recover the same soft-pion theorems that were established in
Ref.~\cite{Braun:2006td}.
\bi
\item
For $p \pi^0$ GDA we get
\begin{eqnarray}
 &&
4 \langle 0 |  u_\rho(1)  u_\tau(2) d_\chi(3)|  p \pi^0 \rangle
\Big|_{{\text{soft} \atop  \text{pion}}}
=
 M^{(\pi N)_{1/2} \, \{ 12\}}_{\rho \tau \chi}(1,2,3)
\Big|_{{\text{soft} \atop  \text{pion}}}
    + \frac{2}{3}  M^{(\pi N)_{3/2}  }_{\rho \tau \chi}(1,2,3)\Big|_{{\text{soft} \atop  \text{pion}}}
\nonumber \\ &&
=
\frac{i f_N}{f_\pi} \Bigl\{
- \left( \gamma^5_{\chi \eta} v^N_{\rho \tau, \, \eta} \right)
\frac{1}{4} \left(  \phi^N(1,2,3) + \phi^N(2,1,3)\right)
- \left( \gamma^5_{\chi \eta} a^N_{\rho \tau, \, \eta} \right)
\frac{1}{4} \left(  -\phi^N(1,2,3) + \phi^N(2,1,3)\right)
\nn \\ &&
- \left( \gamma^5_{\chi \eta} t^N_{\rho \tau, \, \eta} \right)
\frac{3}{4} \left(  \phi^N(1,3,2) + \phi^N(2,3,1)\right)
\Bigr\}\,;
\end{eqnarray}
\item For $n \pi^+$ and $p \pi^-$ GDAs we get
\begin{eqnarray}
 &&
4\langle 0 |  u_\rho(1)  u_\tau(2) d_\chi(3)|  n \pi^+ \rangle
\Big|_{{\text{soft} \atop  \text{pion}}}
= -4\langle 0 |  d_\rho(1)  d_\tau(2) u_\chi(3)|  p \pi^- \rangle
\Big|_{{\text{soft}\atop  \text{pion}}}
\nonumber \\ &&
=
  \sqrt{2} M^{(\pi N)_{1/2} \, \{ 12\}}_{\rho \tau \chi}(1,2,3)
  \Big|_{{\text{soft} \atop  \text{pion}}}
  - \frac{ \sqrt{2}}{3}  M^{(\pi N)_{3/2}  }_{\rho \tau \chi}(1,2,3)
  \Big|_{{\text{soft} \atop  \text{pion}}}
  \nonumber \\ &&
  =\frac{i f_N}{f_\pi} \Bigm\{
  - \left( \gamma^5_{\chi \eta} v^N_{\rho \tau, \, \eta} \right)  \frac{1}{2 \sqrt{2}}
  \left[ -\phi^N(1,2,3)-\phi^N(2,1,3)-2((\phi^N(3,1,2)+\phi^N(3,2,1))\right]
    \nonumber \\ &&
 - \left( \gamma^5_{\chi \eta} a^N_{\rho \tau, \, \eta} \right)
 \frac{1}{2 \sqrt{2}}  \left[ \phi_N(1,2,3)-\phi^N(2,1,3)-2(\phi^N(3,1,2)-\phi^N(3,2,1))\right]
    \nonumber \\ &&
 - \left( \gamma^5_{\chi \eta} t^N_{\rho \tau, \, \eta} \right) \frac{1}{2 \sqrt{2}}
 \left[ \phi^N(1,3,2)+\phi^N(2,3,1) \right]
  \Bigm\}\,.
\end{eqnarray}
\ei

\subsubsection{Soft-pion theorem for $\pi N$ { {TDA}}s in the chiral limit }
\label{SubSec_Chiral_ConstrPiN}
\mbox

To establish the consequences of the soft-pion theorem
(\ref{soft_pion_theorem_pobyl})
for
$\pi N$ TDAs we apply the crossing transformation to the matrix element (\ref{soft_pion_full_glory}).
We assume smallness of possible
non-analytic corrections and rely on the soft-pion
theorem to work out the physical normalization of $\pi N$ TDAs. Below we
quote the results in the chiral limit $m_\pi=0$.

The relation between the Dirac structures
(\ref{s_piN_rho_tau_chi})
of the
$\pi N$
TDA parametrization
(\ref{Param_TDAs_Covariant_DS}) of  Sec.~\ref{SubSec_Polynomiality}
and those occurring in
(\ref{M_T_sym_soft_pion}), (\ref{M12_12_piN})
is given by:
\begin{eqnarray}
 &&
 \gamma^5_{\chi \eta} v^N_{\rho \tau, \, \eta}  = \frac{1}{m_N}
\left(
 {v_1}_{\rho \tau, \, \chi}^{(\pi N)}- \frac{1}{2}  {v_2 }_{\rho \tau, \, \chi}^{(\pi N)}
\right)\,;
\nonumber \\
 &&
 \gamma^5_{\chi \eta} a^N_{\rho \tau, \, \eta}  = \frac{1}{m_N}
\left(
 {a_1}_{\rho \tau, \, \chi}^{(\pi N)}- \frac{1}{2}  {a_2 }_{\rho \tau, \, \chi}^{(\pi N)}
\right)\,;
\nonumber \\
 &&
 \gamma^5_{\chi \eta} t^N_{\rho \tau, \, \eta}  = -\frac{1}{m_N}
\left(
 {t_1}_{\rho \tau, \, \chi}^{(\pi N)}- \frac{1}{2}  {t_2 }_{\rho \tau, \, \chi}^{(\pi N)}
\right)\,.
\end{eqnarray}

In the chiral limit $m_\pi=0$
 this results in the following contributions to
the independent isospin-$\frac{1}{2}$ and isospin-$\frac{3}{2}$ $\pi N$ TDAs
(\ref{independent1/2}), (\ref{independent3/2}) regular at $\Delta^2=m_N^2$:
\begin{eqnarray}
 &&
 \phi^{(\pi N)_{1/2}}_1(x_1,x_2,x_3, \xi=1, \Delta^2=m_N^2) \Big|_{{{\rm soft} \atop {\rm  pion}}}=
\frac{1}{24} \phi^N \left(\frac{x_1}{2},\frac{x_2}{2},\frac{x_3}{2} \right)+\frac{1}{6} \phi^N \left(\frac{x_3}{2},\frac{x_2}{2},\frac{x_1}{2} \right)\,;
\nonumber \\
 &&
 \phi^{(\pi N)_{1/2}}_2(x_1,x_2,x_3, \xi=1, \Delta^2=m_N^2) \Big|_{{{\rm soft} \atop {\rm  pion}}}=
- \frac{1}{2} \phi^{(\pi N)_{1/2}}_1(x_1,x_2,x_3, \xi=1, \Delta^2=m_N^2)\Big|_{{{\rm soft} \atop {\rm  pion}}}\,;
\nonumber \\  &&
\phi^{(\pi N)_{3/2}}_1(x_1,x_2,x_3, \xi=1, \Delta^2=m_N^2) \Big|_{{{\rm soft} \atop {\rm  pion}}}= \frac{1}{4}  \left( \phi^N\left(\frac{x_1}{2},\frac{x_2}{2},\frac{x_3}{2}\right)+ \phi^N\left(\frac{x_3}{2},\frac{x_2}{2},\frac{x_1}{2}\right) \right)\,;
\nonumber \\  &&
\phi^{(\pi N)_{3/2}}_2(x_1,x_2,x_3, \xi=1, \Delta^2=m_N^2) \Big|_{{{\rm soft} \atop {\rm  pion}}}=
- \frac{1}{2} \phi^{(\pi N)_{3/2}}_1(x_1,x_2,x_3, \xi=1, \Delta^2=m_N^2)\Big|_{{{\rm soft} \atop {\rm  pion}}}\,.
\end{eqnarray}

Now employing the isospin parametrization and the definitions
(\ref{VAT_piN_TDA_I12}),
(\ref{Isospin_3/2_TDA_final_symmetry})
we determine the physical normalization
of $\pi N$ TDAs from the soft-pion theorem
for all possible isospin channels within the parametrization\footnote{To switch to the parametrization of  Sec.~\ref{SubSubSec_Def_piN_TDAs}  one has to employ the formulas
(\ref{Relation_DiracStr_DeltaT_to_covariant}). }
(\ref{Param_TDAs_Covariant_DS}) of
 Sec.~\ref{SubSec_Polynomiality}.

\bi
\item For $p \to \pi^0$ $uud$ TDAs we get
\begin{eqnarray}
 &&
V_1^{\pi^0 p}(x_1,\,x_2,\, x_3, \xi=1, \Delta^2=m_N^2)=
-\frac{1}{16}   \left( \phi^N \left( \frac{x_1}{2},\frac{x_2}{2},\frac{x_3}{2} \right)+\phi^N \left( \frac{x_2}{2},\frac{x_1}{2},\frac{x_3}{2} \right) \right); \nn  \\ &&
V_2^{\pi^0 p}(x_1,\,x_2,\, x_3, \xi=1, \Delta^2=m_N^2)=- \frac{1}{2}V_1^{\pi^0 p}(x_1,\,x_2,\, x_3, \xi=1, \Delta^2=m_N^2); \nn \\
 &&
A_1^{\pi^0 p}(x_1,\,x_2,\, x_3, \xi=1, \Delta^2=m_N^2)=
-\frac{1}{16}   \left(-\phi^N \left( \frac{x_1}{2},\frac{x_2}{2},\frac{x_3}{2} \right)+\phi^N \left( \frac{x_2}{2},\frac{x_1}{2},\frac{x_3}{2} \right)\right); \nn  \\  &&
A_2^{\pi^0 p}(x_1,\,x_2,\, x_3, \xi=1, \Delta^2=m_N^2)=- \frac{1}{2}A_1^{\pi^0 p}(x_1,\,x_2,\, x_3, \xi=1, \Delta^2=m_N^2); \nn \\  &&
T_1^{\pi^0 p}(x_1,\,x_2,\, x_3, \xi=1, \Delta^2=m_N^2)=
\frac{3}{16}   \left(\phi^N \left( \frac{x_1}{2},\frac{x_3}{2},\frac{x_2}{2} \right)+\phi^N \left( \frac{x_2}{2},\frac{x_3}{2},\frac{x_1}{2} \right)\right); \nn \\  &&
T_2^{\pi^0 p}(x_1,\,x_2,\, x_3, \xi=1, \Delta^2=m_N^2)=- \frac{1}{2}T_1^{\pi^0 p}(x_1,\,x_2,\, x_3, \xi=1, \Delta^2=m_N^2); \nn \\  &&
T_3^{\pi^0 p}(x_1,\,x_2,\, x_3, \xi=1, \Delta^2=m_N^2)=T_4^{\pi^0 p}(x_1,\,x_2,\, x_3, \xi=1, \Delta^2=m_N^2)=0\,.
\label{SoftPion_p_pi0}
\end{eqnarray}

\item For
$n \to \pi^-$ $uud$
and
$p \to \pi^+$ $ddu$ TDAs we get
\begin{eqnarray}
 &&
V_1^{\pi^- n}(x_1,\,x_2,\, x_3, \xi=1, \Delta^2=m_N^2)=-V_1^{\pi^+ p}(x_1,\,x_2,\, x_3, \xi=1, \Delta^2=m_N^2) \nn \\ && =
\frac{1}{8 \sqrt{2}}    \left(\phi^N \left( \frac{x_1}{2},\frac{x_2}{2},\frac{x_3}{2} \right)+\phi^N \left( \frac{x_2}{2},\frac{x_1}{2},\frac{x_3}{2} \right) +
2\phi^N \left( \frac{x_3}{2},\frac{x_1}{2},\frac{x_2}{2} \right)+2 \phi^N \left( \frac{x_3}{2},\frac{x_2}{2},\frac{x_1}{2} \right)
\right); \nn  \\  &&
V_2^{\{\pi^+ p, \, \pi^-n\}}(x_1,\,x_2,\, x_3, \xi=1, \Delta^2=m_N^2)=- \frac{1}{2}V_1^{\{\pi^+ p, \, \pi^-n\}}(x_1,\,x_2,\, x_3, \xi=1, \Delta^2=m_N^2); \nn \\  &&
 A_1^{\pi^- n}(x_1,\,x_2,\, x_3, \xi=1, \Delta^2=m_N^2)=-A_1^{\pi^+ p}(x_1,\,x_2,\, x_3, \xi=1, \Delta^2=m_N^2) \nn \\ && =
\frac{1}{8 \sqrt{2}}    \left(-\phi^N \left( \frac{x_1}{2},\frac{x_2}{2},\frac{x_3}{2} \right)+\phi^N \left( \frac{x_2}{2},\frac{x_1}{2},\frac{x_3}{2} \right) -
2\phi^N \left( \frac{x_3}{2},\frac{x_1}{2},\frac{x_2}{2} \right)+2 \phi^N \left( \frac{x_3}{2},\frac{x_2}{2},\frac{x_1}{2} \right)
\right); \nn  \\  &&
A_2^{\{\pi^+ p, \, \pi^-n\}}(x_1,\,x_2,\, x_3, \xi=1, \Delta^2=m_N^2)=- \frac{1}{2}A_1^{\{\pi^+ p, \, \pi^-n\}}(x_1,\,x_2,\, x_3, \xi=1, \Delta^2=m_N^2); \nn \\  &&
 T_1^{\pi^- n}(x_1,\,x_2,\, x_3, \xi=1, \Delta^2=m_N^2)=-T_1^{\pi^+ p}(x_1,\,x_2,\, x_3, \xi=1, \Delta^2=m_N^2) \nn \\ && =
\frac{1}{8 \sqrt{2}}    \left(\phi^N \left( \frac{x_1}{2},\frac{x_3}{2},\frac{x_2}{2} \right)+\phi^N \left( \frac{x_2}{2},\frac{x_3}{2},\frac{x_1}{2}\right)
\right); \nn  \\  &&
T_2^{\{\pi^+ p, \, \pi^-n\}}(x_1,\,x_2,\, x_3, \xi=1, \Delta^2=m_N^2)=- \frac{1}{2}T_1^{\{\pi^+ p, \, \pi^-n\}}(x_1,\,x_2,\, x_3, \xi=1, \Delta^2=m_N^2); \nn \\  &&
T_{3,4}^{\{\pi^+ p, \, \pi^-n\}}(x_1,\,x_2,\, x_3, \xi=1, \Delta^2=m_N^2)=0.
\label{SoftPion_p_pi+}
\end{eqnarray}
\ei

The soft-pion theorem (\ref{SoftPion_p_pi0})
was employed in Ref.~\cite{Lansberg:2007ec}
to construct a simple phenomenological model for  $\pi N$ TDAs\footnote{Here it is presented using the parametrization of  Sec.~\ref{SubSubSec_Def_piN_TDAs}.
Also note the relative signs and the overall factor $\frac{1}{2}$ missed in the Erratum to~\cite{Lansberg:2007ec}.}:
\begin{eqnarray}
 &&
\left\{V_{1}^{p \pi^{0}}, A_{1}^{p \pi^{0}}\right\}\left(x_{i}, \xi\right)=-\frac{1}{2} \cdot \frac{1}{4 \xi}\left\{V^{p}, A^{p}\right\}\left(\frac{x_{i}}{2 \xi}\right) ; \quad T_{1}^{p \pi^{0}}\left(x_{i}, \xi\right)=\frac{1}{2} \cdot \frac{3}{4 \xi} T^{p}\left(\frac{x_{i}}{2 \xi}\right); \nn \\ &&
V_{1}^{p \pi^{+}}\left(x_{i}, \xi\right)=-\frac{\sqrt{2}}{2} \cdot \frac{1}{4 \xi}
\left(V^{p}\left( \frac{x_1}{2\xi},\frac{x_2}{2\xi},\frac{x_3}{2\xi} \right)+
V^{p}\left( \frac{x_3}{2\xi},\frac{x_1}{2\xi},\frac{x_2}{2\xi} \right)+
V^{p}\left( \frac{x_2}{2\xi},\frac{x_3}{2\xi},\frac{x_1}{2\xi} \right) \nn \right. \\ &&
\left.
-A^{p}\left( \frac{x_3}{2\xi},\frac{x_1}{2\xi},\frac{x_2}{2\xi} \right)-
A^{p}\left( \frac{x_3}{2\xi},\frac{x_2}{2\xi},\frac{x_1}{2\xi} \right)
 \right); \nn \\  &&
A_{1}^{p \pi^{+}}\left(x_{i}, \xi\right)=-\frac{\sqrt{2}}{2} \cdot \frac{1}{4 \xi}
\left(
A^{p}\left( \frac{x_1}{2\xi},\frac{x_2}{2\xi},\frac{x_3}{2\xi} \right)
-A^{p}\left( \frac{x_3}{2\xi},\frac{x_1}{2\xi},\frac{x_2}{2\xi} \right)
-A^{p}\left( \frac{x_2}{2\xi},\frac{x_3}{2\xi},\frac{x_1}{2\xi} \right)
\right. \nn \\ &&
\left.
+V^{p}\left( \frac{x_3}{2\xi},\frac{x_1}{2\xi},\frac{x_2}{2\xi} \right)-
V^{p}\left( \frac{x_3}{2\xi},\frac{x_2}{2\xi},\frac{x_1}{2\xi} \right)
\right); \nn \\  &&
T_{1}^{p \pi^{+}}\left(x_{i}, \xi\right)=-\frac{\sqrt{2}}{2} \cdot \frac{1}{4 \xi}
T^{p}\left(\frac{x_{i}}{2 \xi}\right),
\label{Model_TDA_JPhi}
\end{eqnarray}
where $V^p(y_1,y_2,y_3)$, $A^p(y_1,y_2,y_3)$ and $T^p(y_1,y_2,y_3)$
stand for the usual leading twist-$3$ nucleon DAs~\cite{Chernyak:1984bm}.

Despite its obvious drawbacks
(like the very narrow validity range limited to the close vicinity of the threshold
and lack of an intrinsic $\Delta^2$-dependence) the model (\ref{Model_TDA_JPhi}) for the first time provided a quantitative estimate
of the physical normalization for $\pi N$ TDAs.
In particular, the predictions of the   model
(\ref{Model_TDA_JPhi}) were
employed in the first feasibility study
\cite{Singh:2014pfv}
for accessing $\pi N$ TDAs with  \=PANDA through
$\bar{p} p \to \gamma^{*} \pi^0$.

\subsubsection{A model for quadruple distributions with input
from chiral dynamics}
\label{SubSec_SpectralPart_and_2comp_mod}
\mbox

A common strategy for modeling GPDs is to rely on extrapolation
of forward partonic densities with the help of an Ansatz for the corresponding
double distributions.
The latter are usually parameterized as forward partonic densities
times profile functions, which generate the skewness dependence.
The so-called Radyushkin's double distribution Ansatz
\cite{Musatov:1999xp}
saw an extensive application in the GPD phenomenology.

In this subsection, using the results of  Sec.~\ref{SubSec_Quadr_distib}, we consider a possible generalization~\cite{Lansberg:2012ha}
of this technique for the case of $\pi N$ TDAs. The key difference here is that nucleon-to-meson  TDAs lack a comprehensible forward limit $\xi=0$; however, due
to the soft-pion theorem presented
in  Sec.~\ref{SubsubSec_Chiral_Constr},
$\pi N$ TDAs
are constrained in a complementary
$\xi=1$
limit and reduce to particular combinations of nucleon DAs.

To rewrite the spectral representation
(\ref{Spectral_represent_Hi})
for $\pi N$ TDAs
in a suitable form we switch to the following
combinations of the spectral parameters:
\begin{eqnarray}
 &&
\kappa_i= \alpha_i+ \beta_i \,; \ \ \  \theta_i=\frac{1}{2} \sum_{k,l=1}^3 \varepsilon_{i k l}(\alpha_k+\beta_k); \nonumber \\
 &&
\mu_i= \alpha_i- \beta_i\,; \ \ \  \lambda_i= \frac{1}{2} \sum_{k,l=1}^3 \varepsilon_{i k l}(\alpha_k-\beta_k).
\end{eqnarray}
The index
$i=1,\,2,\,3$
here refers to a coordinated choice
of the quark--diquark coordinates
(\ref{Def_qDq_coord}).
After the change of integration variables the spectral representation
(\ref{Spectral_represent_Hi})
reads:
\begin{eqnarray}
 &&
H(w_i,\,v_i,\,\xi)=
\int_{-1}^1 d \kappa_i \int_{- \frac{1-\kappa_i}{2}}^{ \frac{1-\kappa_i}{2}} d\theta_i
\int_{-1}^1 d \mu_i \int_{- \frac{1-\mu_i}{2}}^{ \frac{1-\mu_i}{2}} d\lambda_i
\, \delta \left(w_i- \frac{\kappa_i-\mu_i}{2} (1-\xi) - \kappa_i \xi \right)  \nonumber \\ &&
\times
\delta\left(v_i- \frac{\theta_i-\lambda_i}{2} (1-\xi) - \theta_i \xi \right) \, \frac{1}{4} F_i(\kappa_i, \, \theta_i,\, \mu_i,\, \lambda_i ).
\label{Spectral_for_GPDs_kappa_theta}
\end{eqnarray}

Below we summarize the explicit expressions for
$\pi N$ TDAs from the spectral representation
(\ref{Spectral_for_GPDs_kappa_theta})
in the ERBL-like and DGLAP-like domains (see  Sec.~\ref{SubSec_Support}).
We employ the following shortened notation for
the spectral density:
\begin{equation}
F_i(....) \equiv F_i \left(\kappa_i, \, \theta_i,\,\frac{\kappa_i(1+\xi)-2w_i}{1-\xi}, \, \frac{\theta_i(1+\xi)-2v_i}{1-\xi} \right),
\end{equation}
where $F$ is the quadruple distribution (\ref{Quad_Distr_Fi}).

\begin{itemize}
\item DGLAP-like type I domain
$w_i \in [-1;\, -\xi]$; $v_i \in [\xi'_i; \, 1-\xi'_i+\xi]$:
\begin{equation}
 H(w_i,v_i,\xi)
 = \frac{1}{(1-\xi)^2}
\int_{-1}^{\frac{1-2v_i+w_i}{1+\xi}} d \kappa_i
\int_{\frac{\kappa_i}{2} - \frac{1}{1+\xi}(w_i-2v_i+ \frac{1-\xi}{2})}^{\frac{1-\kappa_i}{2}} d \theta_i
F_i
(....);
\label{TDA_skew1_DGLAP-like_type_I_1}
\end{equation}

\item DGLAP-like type II domain $w_i \in [-1;\, -\xi]$; $v \in [-\xi'_i; \,  \xi'_i ]$:
\begin{equation}
 H(w_i,v_i,\xi)
 =
\frac{1}{(1-\xi)^2}
\int_{-1}^{ \frac{1-\xi+2 w_i}{1+\xi} } d \kappa_i
\int_{\frac{\kappa_i}{2} - \frac{1}{1+\xi}(w_i-2v_i+ \frac{1-\xi}{2})}^{-\frac{\kappa_i}{2}+ \frac{1}{1+\xi}(w+2v+ \frac{1-\xi}{2})} d \theta_i
F_i(....);
\label{TDA_skew1_DGLAP-like_type_II_1}
\end{equation}

\item DGLAP-like type I domain $w_i \in [-1;\, -\xi]$; $v_i \in [-1+\xi'_i-\xi; \,  -\xi'_i ]$:
\begin{equation}
H(w_i,v_i,\xi)
=
\frac{1}{(1-\xi)^2}
\int_{-1}^{ \frac{1+2v_i+ w_i}{1+\xi} } d \kappa_i
\int_{-\frac{1-\kappa_i}{2} }^{-\frac{\kappa_i}{2}+ \frac{1}{1+\xi}(w_i+2v_i+ \frac{1-\xi}{2})} d \theta_i
F_i(....);
\label{TDA_skew1_DGLAP-like_type_I_2}
\end{equation}

\item DGLAP-like type II domain $w_i \in [-\xi;\, 1]$; $v_i \in [| \xi'_i| ;\,1-\xi+\xi'_i]$:
\begin{equation}
 H(w_i,v_i,\xi)   =
\frac{1}{(1-\xi)^2}
\int_{ \frac{-1+\xi+2w_i}{1+\xi}}^{ \frac{1-2v_i+ w_i}{1+\xi} } d \kappa_i
\int_{\frac{\kappa_i}{2} - \frac{1}{1+\xi}(w_i-2v_i+ \frac{1-\xi}{2})}^{\frac{1-\kappa_i}{2} } d \theta_i
F_i (....);
\label{TDA_skew1_DGLAP-like_type_II_2}
\end{equation}

\item ERBL-like   domain $w_i \in [-\xi;\, \xi]$; $v_i \in [-\xi'_i;\,  +\xi'_i]$:
\begin{equation}
 H(w_i,v_i,\xi)   =
\frac{1}{(1-\xi)^2}
\int_{ \frac{-1+\xi+2w_i}{1+\xi}}^{ \frac{1-\xi+ 2w_i}{1+\xi} } d \kappa_i
\int_{\frac{\kappa_i}{2} - \frac{1}{1+\xi}(w_i-2v_i+ \frac{1-\xi}{2})}^{-\frac{\kappa_i}{2}+ \frac{1}{1+\xi}(w_i+2v_i+ \frac{1-\xi}{2})} d \theta_i
F_i (....);
  \label{TDA_skew1_ERBL-like}
\end{equation}

\item DGLAP-like type II domain $w_i \in [-\xi;\, 1]$; $v_i \in [-1+\xi-\xi'_i;\, -| \xi'_i| ]$:
\begin{equation}
 H(w_i,v_i,\xi)  =
\frac{1}{(1-\xi)^2}
\int_{ \frac{-1+\xi+2w_i}{1+\xi}}^{ \frac{1+2v_i+ w_i}{1+\xi} } d \kappa_i
\int_{ - \frac{1-\kappa_i}{2}}^{-\frac{\kappa_i}{2}+ \frac{1}{1+\xi}(w_i+2v_i+ \frac{1-\xi}{2})} d \theta_i
F_i(....);
 \label{TDA_skew1_DGLAP-like_type_II_3}
\end{equation}

\item DGLAP-like type II domain $w_i \in [\xi;\, 1]$; $v_i \in [ \xi'_i;\, -\xi'_i]$:
\begin{equation}
 H(w_i,v_i,\xi)=  \frac{1}{(1-\xi)^2}
\int_{ \frac{-1+\xi+2w_i}{1+\xi}}^{ 1 } d \kappa_i
\int_{ - \frac{1-\kappa_i}{2}}^{\frac{1-\kappa_i}{2}} d \theta_i
F_i (....);
   \label{TDA_skew1_DGLAP-like_type_I_3}
\end{equation}

\item For
$w_i$
and
$v_i$
outside the domain
$w_i \in [-1;\,1]$ and $v_i \in [-1+| \xi_i-\xi'_i|  ;\,1-| \xi_i-\xi'_i| ]$
the spectral representation
(\ref{Spectral_for_GPDs_kappa_theta})
provides a vanishing result for $0 \le \xi \le 1$.

\end{itemize}

The suggested factorized Ansatz for quadruple distributions $F_i$
has the form
\begin{equation}
F_i(\kappa_i, \, \theta_i,\, \mu_i,\, \lambda_i )= 4 \Phi (\kappa_i, \, \theta_i) \, h(\mu_i,\, \lambda_i),
\label{Factorized_ansatz_xi=1}
\end{equation}
with the profile function
$h(\mu_i,\, \lambda_i)$
normalized as
\begin{equation}
\int_{-1}^1 d \mu_i \int_{- \frac{1-\mu_i}{2}}^{ \frac{1-\mu_i}{2}} d\lambda_i \, h(\mu_i,\, \lambda_i) =1.
\label{Norm_profile}
\end{equation}

With help of the spectral representation
(\ref{Spectral_for_GPDs_kappa_theta}),
one can check that in the limit $\xi \rightarrow 1$
$H_i$  indeed reduces to
\begin{equation}
H(w_i,\,v_i,\,\xi=1)= \Phi(w_i, \, v_i).
\end{equation}

As  the phenomenological input for
$\Phi(w_i, \, v_i)$ it is natural to use
the combinations of nucleon DAs
(\ref{SoftPion_p_pi0}),
(\ref{SoftPion_p_pi+})
to which
$\pi N$
TDAs reduce in the limit
$\xi \to 1$
due to the soft-pion theorem.

We denote the corresponding combinations of nucleon DAs as $\Phi(y_1,y_2,y_3)$.
These are functions of three momentum fractions $y_i$ ($0\le y_i \le 1$)
satisfying the condition $\sum_i y_i=1$.
According to the particular choice of quark--diquark coordinates in (\ref{Spectral_for_GPDs_x123}), one
has to employ in
(\ref{Factorized_ansatz_xi=1}):
\begin{eqnarray}
 &&
\Phi(\kappa_1, \theta_1) \equiv  \frac{1}{4} \Phi \left(\frac{\kappa_1+1}{2}, \,  \frac{1-\kappa_1+ 2 \theta_1}{4},\,\frac{1-\kappa_1- 2 \theta_1}{4}  \right);
\nonumber \\  &&
\Phi(\kappa_2, \theta_2) \equiv \frac{1}{4}  \Phi \left(  \frac{1-\kappa_2- 2 \theta_2}{4},\,\frac{\kappa_2+1}{2}, \,\frac{1-\kappa_2+ 2 \theta_2}{4}  \right);
\nonumber \\  &&
\Phi(\kappa_3, \theta_3) \equiv  \frac{1}{4}  \Phi \left( \frac{1-\kappa_3+ 2 \theta_3}{4},\,\frac{1-\kappa_3- 2 \theta_3}{4},\,  \frac{\kappa_3+1}{2} \right).
\end{eqnarray}

The profile function
$h(\mu_i,\, \lambda_i)$
also has the support of a baryon DA:
$-1 \le \mu_i \le 1$;
$-\frac{1-\mu_i}{2} \le \lambda_i \le \frac{1-\mu_i}{2}$.
Contrary to the GPD case, no symmetry constraint from hermiticity and time-reversal invariance
occurs for quadruple distributions. Therefore, we are free to employ an arbitrary shape for the profile function $h(\mu_i,\, \lambda_i)$.
The simplest possible choice is to assume that it is determined by the
asymptotic form of a baryon DA ($120 y_1 y_2 y_3$ with $\sum_i y_i=1$):
\begin{equation}
h(\mu_i,\, \lambda_i)=
\frac{15}{16} \, (1+\mu_i) ((1-\mu_i)^2-4 \lambda_i^2)\,.
\label{Profile_h}
\end{equation}
The normalization
(\ref{Norm_profile})
is obviously maintained and the  profile function vanishes at the borders of its domain
of definition.

For the $\Delta^2$-dependence the most straightforward solution is the factorized form of $\Delta^2$ dependence for quadruple distributions
(\ref{Factorized_ansatz_xi=1}):
\begin{equation}
F_i(\kappa_i, \, \theta_i,\, \mu_i,\, \lambda_i, \Delta^2 )= 4 \Phi(\kappa_i, \, \theta_i) \, h(\mu_i,\, \lambda_i) \times G(\Delta^2),
\label{Factorized_ansatz_xi=1_Delta2}
\end{equation}
where
$G(\Delta^2)$
is the
$\pi N$
transition form factor of the local three quark operator
$\widehat{O}^{\alpha \beta \gamma}_{\rho \tau \chi}(0,0,0)$
(\ref{Def_operator_O_isotopic}) for which \textit{e.g.} a dipole formula
$G(\Delta^2) \sim (M_D^2-\Delta^2)^{-2}$
can be adopted.
Let us however note that a factorized form of the
$\Delta^2$-dependence is known to be  oversimplified  and was much criticized in the GPD case
(see \textit{e.g.} discussion in~\cite{Kumericki:2008di}).

Let us briefly summarize the key features of the phenomenological model
for $\pi N$ TDAs based on the factorized Ansatz
(\ref{Spectral_for_GPDs_kappa_theta}).
\bi
\item The model can be employed for  $\pi N$ TDAs
(\ref{Param_TDAs_Covariant_DS})
corresponding to the fully covariant set of the Dirac structures (\ref{Dirac_structures_PiN_TDA_Cov}).
These TDAs satisfy the polynomiality property
in its simple form (see discussion in  Sec.~\ref{SubSec_Polynomiality}).

\item By construction, the threshold soft-pion limit of
$\pi N$
TDAs required by the soft-pion theorem (\ref{soft_pion_full_glory}) is assured.
\item As phenomenological input one can employ various phenomenological
solutions for leading twist nucleon DAs, for a review see~\cite{Stefanis:1999wy,Braun:1999te}. \item
The model provides  lively $x_i$ and $\xi$ dependencies for $\pi N$
TDAs.
\item The support properties of $\pi N$ TDAs are satisfied.
\item The polynomiality property
(\ref{PolyProp_pN-TDA})
of $x_i$ Mellin moments of $\pi N$
TDAs is generally ensured. However, to satisfy this property in its complete form
a $D$-term like contribution with a pure ERBL-like support is to be added to
$\pi N$
TDAs resulting from the spectral representation.
\ei

In conclusion, we present the phenomenological ``two component'' model
for $\pi N$ TDAs.  The model includes a spectral part contribution
built with the factorized Ansatz
(\ref{Factorized_ansatz_xi=1_Delta2}) for quadruple distributions
with the input from the soft-pion theorem and the cross-channel nucleon
exchange contribution of  Sec.~\ref{SubSec_Nucle_ex_piN_TDA}.

We quote the result within the  $\pi N$ parametrization
of
 Sec.~\ref{SubSubSec_Def_piN_TDAs}
corresponding to the set of the Dirac structures (\ref{Def_DirStr_piN_DeltaT}).
This parametrization turns to be better suited for the phenomenological applications\footnote{See the relation (\ref{Relation_DiracStr_DeltaT_to_covariant}) for the correspondence with $\pi N$ TDAs
of  Sec.~\ref{SubSec_Polynomiality}.} as it allows a clear separation of
contributions proportional to $\Delta_T^2$ that are presumably small
in the near-backward kinematics.
Note that within this parametrization the spectral part contribution with
input from the soft-pion theorem
occurs only for
$\pi N$ TDAs $V_1^{\pi N}$,  $A_1^{\pi N}$ and  $T_1^{\pi N}$.

\bi
\item
For the $uud$ proton-to-$\pi^0$ TDAs the suggested model reads:
\begin{eqnarray}
 &&
V_1^{\pi^0 p}(x_1,\,x_2,\,x_3, \xi) \Big|_{\text{set of Eq.~(\ref{Def_DirStr_piN_DeltaT})}}
\nn \\ && =
 \Bigl\{ \text{Spectral part with} \; \Phi \; \text{from Eq.~(\ref{SoftPion_p_pi0})} \Bigr\}+
\frac{1-\xi}{1+\xi}
\left\{
V_1^{(\pi N)_{1/2}}(x_1,\,x_2,\,x_3, \xi) \Big|_{N(940)}, \; \text{ Eq.~(\ref{Nucleon_exchange_contr_VAT})}\right\};
\nn \\  &&
 A_1^{\pi^0 p}(x_1,\,x_2,\,x_3, \xi) \Big|_{\text{set of Eq.~(\ref{Def_DirStr_piN_DeltaT})}}
\nn \\ && =
 \Bigl\{ \text{Spectral part with} \; \Phi \; \text{from Eq.~(\ref{SoftPion_p_pi0})} \Bigr\}+
\frac{1-\xi}{1+\xi}
\left\{
A_1^{(\pi N)_{1/2}}(x_1,\,x_2,\,x_3, \xi) \Big|_{N(940)}, \; \text{ Eq.~(\ref{Nucleon_exchange_contr_VAT})}\right\};
\nn \\  &&
 T_1^{\pi^0 p}(x_1,\,x_2,\,x_3, \xi) \Big|_{\text{set of Eq.~(\ref{Def_DirStr_piN_DeltaT})}}
\nn \\ && =
 \Bigl\{ \text{Spectral part with} \; \Phi \; \text{from Eq.~(\ref{SoftPion_p_pi0})} \Bigr\}+
\frac{1-\xi}{1+\xi}
\left\{
T_1^{(\pi N)_{1/2}}(x_1,\,x_2,\,x_3, \xi) \Big|_{N(940)}, \; \text{ Eq.~(\ref{Nucleon_exchange_contr_VAT})}\right\};
\nn \\  &&
 \{V_2, \, A_2, \, T_2\}^{\pi^0 p}(x_1,\,x_2,\,x_3, \xi) \Big|_{\text{set of Eq.~(\ref{Def_DirStr_piN_DeltaT})}} \nn \\ && =
\left\{
\{V_1, \, A_1, \, T_1\}^{(\pi N)_{1/2}}(x_1,\,x_2,\,x_3, \xi) \Big|_{N(940)}, \;
\text{Eq.~(\ref{Nucleon_exchange_contr_VAT})}
\right\}
\nn \\  &&
\{T_3, \, T_4\}^{\pi^0 p}(x_1,\,x_2,\,x_3, \xi) \Big|_{\text{set of Eq.~(\ref{Def_DirStr_piN_DeltaT})}}=0.
\label{2Component_model_pi0p}
\end{eqnarray}

\item For the $ddu$ proton-to-$\pi^+$ and $uud$ neutron-to-$\pi^-$ TDAs the suggested model reads:
\begin{eqnarray}
 &&
V_1^{\pi^+ p}(x_1,\,x_2,\,x_3, \xi) \Big|_{\text{set of Eq.~(\ref{Def_DirStr_piN_DeltaT})}}
=-V_1^{\pi^- n}(x_1,\,x_2,\,x_3, \xi) \Big|_{\text{set of Eq.~(\ref{Def_DirStr_piN_DeltaT})}}
\nn \\ && =
 \Bigl\{ \text{Spectral part with} \; \Phi \; \text{from Eq.~(\ref{SoftPion_p_pi+})} \Bigr\}-
 \sqrt{2}
\frac{1-\xi}{1+\xi}
\left\{
V_1^{(\pi N)_{1/2}}(x_1,\,x_2,\,x_3, \xi) \Big|_{N(940)}, \; \text{Eq.~(\ref{Nucleon_exchange_contr_VAT})}\right\};
\nn \\  &&
 A_1^{\pi^+ p}(x_1,\,x_2,\,x_3, \xi) \Big|_{\text{set of Eq.~(\ref{Def_DirStr_piN_DeltaT})}}
=-A_1^{\pi^- n}(x_1,\,x_2,\,x_3, \xi) \Big|_{\text{set of Eq.~(\ref{Def_DirStr_piN_DeltaT})}}
\nn \\ && =
 \Bigl\{ \text{Spectral part with} \; \Phi \; \text{from Eq.~(\ref{SoftPion_p_pi+})} \Bigr\}-
 \sqrt{2}
\frac{1-\xi}{1+\xi}
\left\{
A_1^{(\pi N)_{1/2}}(x_1,\,x_2,\,x_3, \xi) \Big|_{N(940)}, \; \text{Eq.~(\ref{Nucleon_exchange_contr_VAT})}\right\};
\nn \\  &&
 T_1^{\pi^+ p}(x_1,\,x_2,\,x_3, \xi) \Big|_{\text{set of Eq.~(\ref{Def_DirStr_piN_DeltaT})}}
=-T_1^{\pi^- n}(x_1,\,x_2,\,x_3, \xi) \Big|_{\text{set of Eq.~(\ref{Def_DirStr_piN_DeltaT})}}
\nn \\ && =
 \Bigl\{ \text{Spectral part with} \; \Phi \; \text{from eq.~(\ref{SoftPion_p_pi+})} \Bigr\}-
 \sqrt{2}
\frac{1-\xi}{1+\xi}
\left\{
T_1^{(\pi N)_{1/2}}(x_1,\,x_2,\,x_3, \xi) \Big|_{N(940)}, \; \text{Eq.~(\ref{Nucleon_exchange_contr_VAT})}\right\};
\nn \\  &&
 \{V_2, \, A_2, \, T_2\}^{\pi^+ p}(x_1,\,x_2,\,x_3, \xi) \Big|_{\text{set of Eq.~(\ref{Def_DirStr_piN_DeltaT})}}=
-\{V_2, \, A_2, \, T_2\}^{\pi^- n}(x_1,\,x_2,\,x_3, \xi) \Big|_{\text{set of Eq.~(\ref{Def_DirStr_piN_DeltaT})}}
\nn \\ && =- \sqrt{2} \left\{
\{V_1, \, A_1, \, T_1\}^{(\pi N)_{1/2}}(x_1,\,x_2,\,x_3, \xi) \Big|_{N(940)}, \, \text{Eq.~(\ref{Nucleon_exchange_contr_VAT}) } \right\};
\nn \\  &&
\{T_3, \, T_4\}^{\pi^+ p}(x_1,\,x_2,\,x_3, \xi) \Big|_{\text{set of Eq.~(\ref{Def_DirStr_piN_DeltaT})}}=-\{T_3, \, T_4\}^{\pi^- n}(x_1,\,x_2,\,x_3, \xi) \Big|_{\text{set of Eq.~(\ref{Def_DirStr_piN_DeltaT})}}=0.
\label{2Component_model_pi_plus_n}
\end{eqnarray}
\ei

Eqs.~(\ref{2Component_model_pi0p}),
(\ref{2Component_model_pi_plus_n}) clearly show that we avoid the double counting problem
between the spectral part and the cross-channel nucleon exchange parts for
$\pi N$ TDAs  $\{V_1, \, A_1, \, T_1\}^{\pi N}$
since the nucleon exchange parts turn into zero for
$\xi=1$; and therefore, do not affect the normalization in the
soft-pion limit $\xi=1$, $\Delta^2=m_N^2$.

\begin{figure}[H]
\begin{center}
\includegraphics[width=0.3\textwidth]{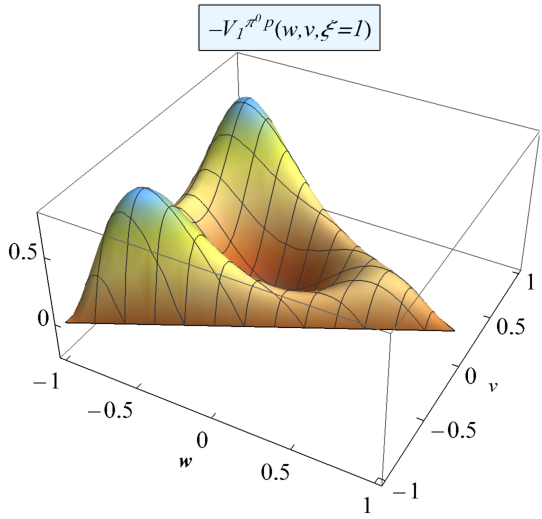}
\includegraphics[width=0.3\textwidth]{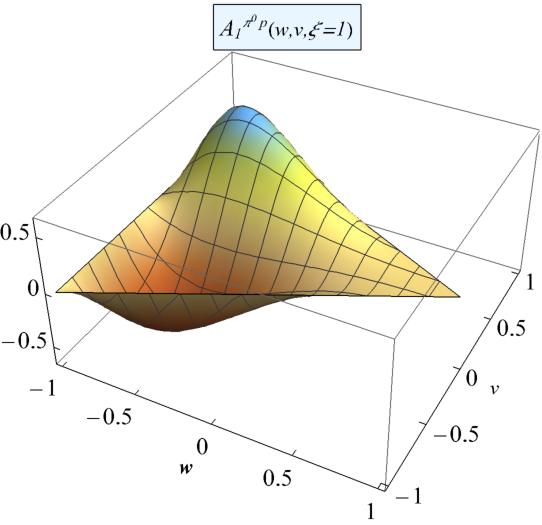}
\includegraphics[width=0.3\textwidth]{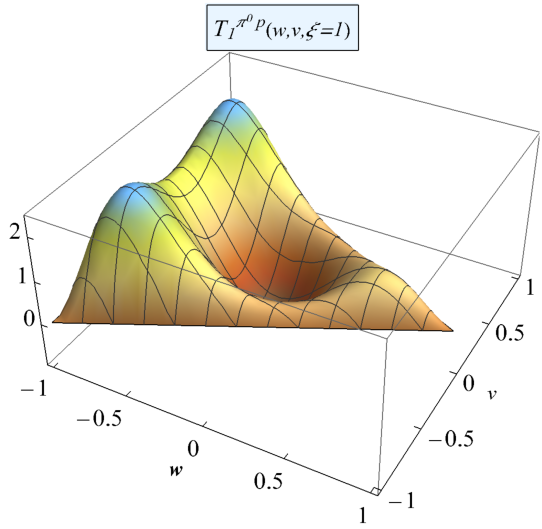}
\includegraphics[width=0.3\textwidth]{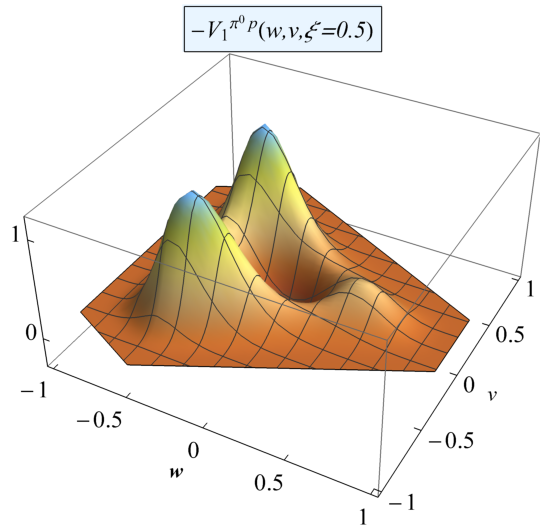}
\includegraphics[width=0.3\textwidth]{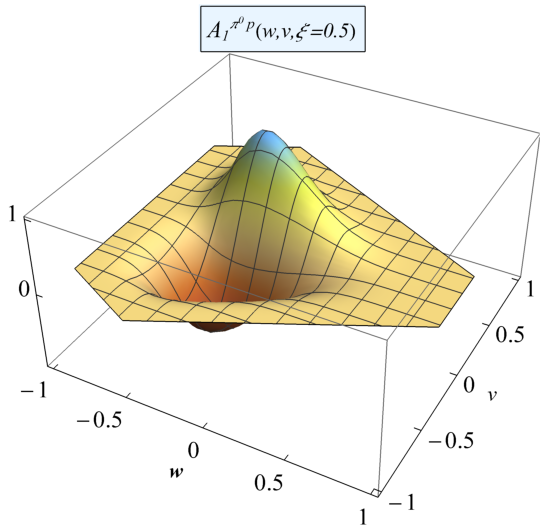}
\includegraphics[width=0.3\textwidth]{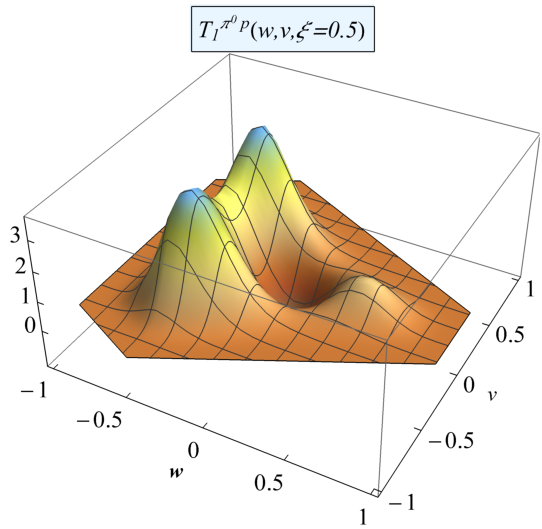}
\includegraphics[width=0.3\textwidth]{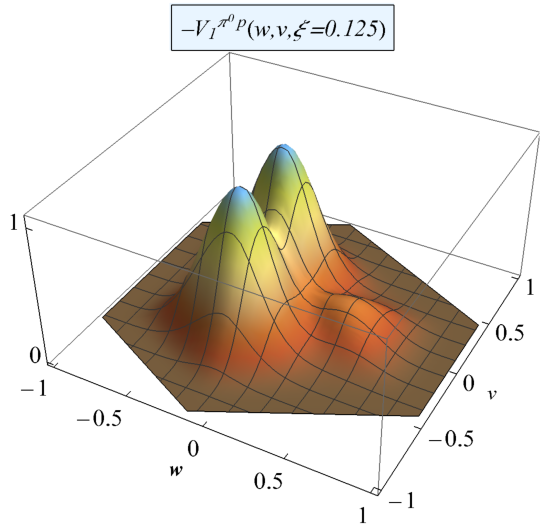}
\includegraphics[width=0.3\textwidth]{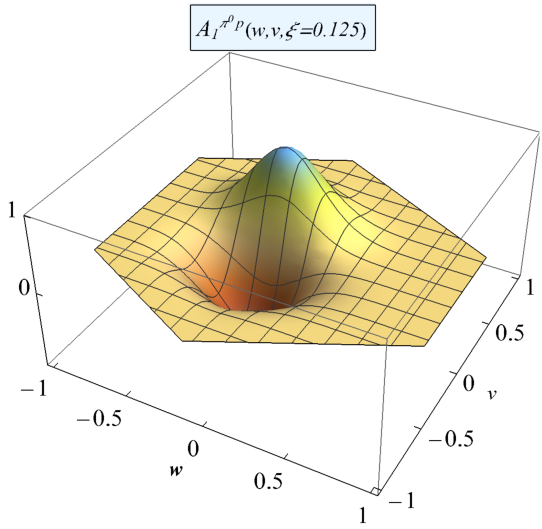}
\includegraphics[width=0.3\textwidth]{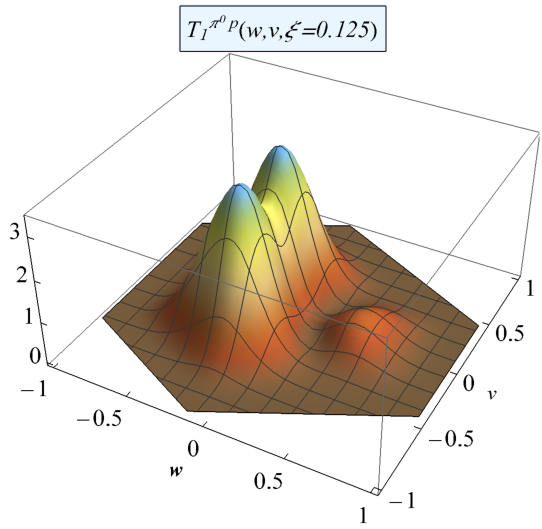}
     \caption{
    Spectral part of
    $\pi^0 p$ TDAs $V_1^{ \pi^0 p}$, $A_1^{  \pi^0 p}$ and $T_1^{ \pi^0 p}$
    (\ref{2Component_model_pi0p})
    computed in the model based on the factorized Ansatz (\ref{Factorized_ansatz_xi=1}) with the profile function (\ref{Profile_h})
    as   functions of quark--diquark coordinates $w \equiv w_3$, $v \equiv v_3$.
    Upper row shows the combination of nucleon DAs (\ref{SoftPion_p_pi0}) to which the corresponding TDAs are reduced in the limit $\xi=1$. In the two lower rows we show the effect of ``skewing'' for
    $\xi=\frac{1}{2}$ and $\xi=\frac{1}{8}$.
   COZ nucleon DAs~\cite{Chernyak:1987nv} are used as numerical input.}
     \label{Fig_TDA_plot_Spectral}
\end{center}
\end{figure}

In  Fig.~\ref{Fig_TDA_plot_Spectral}
we show the spectral part of
$\pi^0 p$ TDAs $V_1^{ \pi^0 p}$, $A_1^{  \pi^0 p}$ and $T_1^{ \pi^0 p}$
(\ref{2Component_model_pi0p})
computed in the model based on the factorized Ansatz (\ref{Factorized_ansatz_xi=1}) with the profile function (\ref{Profile_h})
as   functions of quark--diquark coordinates $w \equiv w_3$, $v \equiv v_3$ for $\xi=\frac{1}{2}$. We also
present the combinations of nucleon DAs (\ref{SoftPion_p_pi0}), to which the corresponding TDAs are reduced in the limit $\xi=1$.
Chernyak--Ogloblin--Zhitnitsky (COZ) phenomenological solution for nucleon DAs~\cite{Chernyak:1987nv} is used as numerical input.

\subsection{Nucleon-to-pion { {TDA}}s in the light-cone quark model}
\label{SubSec_LCQM_Barbara}
\mbox

\begin{figure}[H]
\begin{center}
\includegraphics[width=0.95\textwidth]{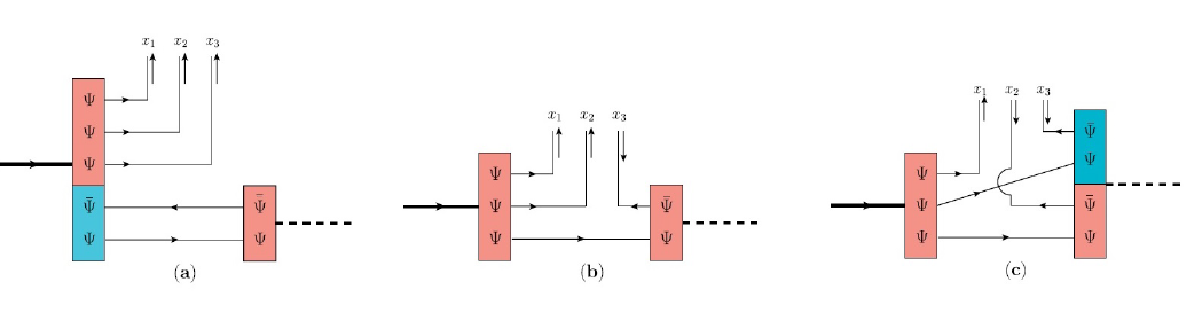}
  \caption{Interpretation of nucleon-to-meson TDAs within the
  light-front quantization framework.
 Small vertical arrows
show the direction of flow of longitudinal momenta. {\bf (a)}: Contribution in the ERBL-like region (all $x_i$ are positive); {\bf (b)}:
Contribution in the DGLAP-like I region (one of $x_i$ is negative). {\bf (c)}: Contribution in the DGLAP-like II
region (two $x_i$s are negative).
    }
     \label{Fig_LC_interpretation}
\end{center}
\end{figure}

A clear physical interpretation for nucleon-to-meson TDAs can be obtained
within the light-front quantization approach~\cite{Brodsky:1997de}.
The basic notions of this framework are the light-front wave functions
(LFWFs) of  hadrons. They are expressed as expansions over various Fock components of  hadronic states. For example, the LFWF of a nucleon or a meson state can be schematically represented by the  following decompositions:
\begin{eqnarray}
 &&
| N\rangle =\psi_{(3 q)}| q q q\rangle +\psi_{(3 q+1 g)}| q q q g\rangle +\psi_{(3 q+q \bar{q})}| q q q q \bar{q}\rangle +\cdots; \nn \\ &&
| {\mathcal{M}}\rangle =\psi_{(q \bar{q})}| q \bar{q}\rangle +\psi_{(q \bar{q}+ 1g)}| q \bar{q}g\rangle +
+\psi_{(q \bar{q}+q \bar{q})}| q \bar{q}  q \bar{q}\rangle +\cdots \,.
\end{eqnarray}
Depending on the particular support region in longitudinal momentum fractions, ${\mathcal{M}}N$ TDAs
give access to overlaps of the non-minimal Fock components of the nucleon
and meson LFVFs, see  Fig.~\ref{Fig_LC_interpretation}.
This provides additional information on the partonic correlations inside hadrons.
In particular, in the ERBL-like region ${\mathcal{M}}N$ TDAs probe the non-minimal
$(3 q+q \bar{q})$ Fock component of the nucleon LFWF.
Thus ${\mathcal{M}}N$
TDAs complement nucleon DAs that probe the lowest
$3 q$ Fock components of the nucleon LFWF.

In
Refs.~\cite{Pincetti:2008fh,PincettiPhD,Pasquini:2009ki}
both leading twist-$3$ nucleon DAs and $\pi N$ TDAs
were considered within a phenomenological model for LFWFs of the nucleon
based on the light-cone constituent quark model.
Previously, this model was successfully applied to the description of elastic form
factors
\cite{Dziembowski:1997vh},
parton distributions
\cite{Speth:1996pz},
GPDs~\cite{Boffi:2002yy} and transverse momentum dependent parton distribution functions (TMD PDFs)~\cite{Pasquini:2008ax}. In Ref.~\cite{Kofler:2017uzq} this model was applied to the
calculation of collinear PDFs.
In this subsection we present a short review
of the key results of Refs.~\cite{Pincetti:2008fh,Pasquini:2009ki} for $\pi N$ TDAs in the ERBL-like region.

The basic ingredients needed to compute the
matrix elements
$T^{s_N}_{\rho \tau, \,\chi}$
and express $\pi N$ TDAs in the ERBL-like region
are the non-minimal five-quark  ($qqqq\bar{q}$) component of a nucleon state and the valence
$(q \bar{q})$ component of the pion  state.

The five-quark component of a nucleon state was modeled within the
meson-cloud model of Refs.~\cite{Pasquini:2006dv,Pasquini:2007iz}.
This model assumes that a physical nucleon consists of a bare
nucleon dressed with a meson cloud. Thus the nucleon state is
represented as a meson--baryon Fock expansion formed as a superposition
of a three-valence-quark bare nucleon state and states containing virtual mesons
with recoiling baryons. The baryon--meson subsystems include configurations with
a nucleon or a $\Delta$-baryon accompanied with a pion as well as vector ($\rho$ and $\omega$) mesons. For the calculation of nucleon-to-pion TDAs it suffices to consider
meson--baryon subsystems containing nucleon or $\Delta$ and a pion.
This allows one to write the following dynamical representation:
\begin{equation}
\begin{aligned}
|  N(B \pi) ; p_{p}, \lambda\rangle=& \int \mathrm{d} y \mathrm{d}^{2} {\mathbf{k}}_{\perp} \int_{0}^{y} \left[ \prod_{i=1}^{3} \frac{\mathrm{d} \xi_{i}}{\sqrt{\xi_{i}}} \right] \int_{0}^{(1-y)}  \left[ \prod_{i=4}^{5} \frac{\mathrm{d} \xi_{i}}{\sqrt{\xi_{i}}} \right] \int \frac{1}{\left[2(2 \pi)^{3}\right]^{4}} \prod_{i=1}^{5} \mathrm{d}^{2}  {\mathbf{k}}_{i \perp}^{\prime} \\
&\times \delta\left(y-\sum_{i=1}^{3} \xi_{i}\right) \delta^{(2)}\left( {\mathbf{k}}_{\perp}-\sum_{i=1}^{3}  {\mathbf{k}}_{i \perp}^{\prime}\right) \delta\left(1-\sum_{i=1}^{5} \xi_{i}\right) \delta^{(2)}\left(\sum_{i=1}^{5}  {\mathbf{k}}_{i \perp}^{\prime}\right) \\
&\times \sum_{\lambda^{\prime} \lambda_{i}, \tau_{i}, c_{i}} \phi_{\lambda^{\prime} 0}^{\lambda(N, B \pi)}\left(y, {\mathbf{k}}_{\perp}\right) \sqrt{y(1-y)} \tilde{\Psi}_{\lambda^{\prime}}^{B,[f]}\left(y,  {\mathbf{k}}_{\perp} ;\left\{\xi_{i}, {\mathbf{k}}_{i \perp}^{\prime} ; \lambda_{i}, \tau_{i}, c_{i}\right\}_{i=1, \ldots, 3}\right) \\
&\times \tilde{\Psi}^{\pi,[f]}\left(1-y,- {\mathbf{k}}_{\perp} ;\left\{\xi_{i},  {\mathbf{k}}_{i \perp}^{\prime} ; \lambda_{i}, \tau_{i}\right\}_{i=4,5}\right) \prod_{i=1}^{5}|  \xi_{i} p_{p}^{+},  {\mathbf{k}}_{i \perp}^{\prime}+\xi_{i}  {\mathbf{p}}_{p \perp}, \lambda_{i}, \tau_{i}, c_{i} ; q\rangle
\end{aligned}
\label{N5q_state_Pasquini}
\end{equation}
The sum on the right-hand side of
(\ref{N5q_state_Pasquini})
stands over the baryon helicity $\lambda^{\prime}$,
quark helicities
$\lambda_i$,
quark color indices
$c_i$ and
isotopic indices
$\tau_i$;
$\tilde{\Psi}_{\lambda^{\prime}}^{B,[f]}$
and
$\tilde{\Psi}_{\lambda^{\prime}}^{\pi,[f]}$
denote the LFWFs of a baryon and a pion, respectively.
The splitting function $\phi_{\lambda^{\prime} 0}^{\lambda(N, B \pi)}(y,  {\mathbf{k}}_\perp)$
describes the probability amplitude to find a physical nucleon with
helicity $\lambda$
in a state consisting of a virtual baryon $B=N, \, \Delta$ with helicity $\lambda'$,
longitudinal momentum fraction $y$ and transverse momentum ${\mathbf{k}}_\bot$
and the pion with longitudinal momentum fraction $1-y$ and transverse momentum $-{\mathbf{k}}_\bot$. These functions have been tabulated in Ref.~\cite{Pasquini:2006dv}.

For the pion state the valence $q \bar{q}$ component is considered
\begin{equation}
|  \pi\left(p_{\pi}\right)\rangle=\sum_{\tau_{i}, \lambda_{i}, c_{i}} \int
\prod_{i=1}^2 \frac{\mathrm{d} \xi_i}{\sqrt{\xi_i}}
\left[ \prod_{i=1}^2 \mathrm{d}^{2}  {\mathbf{k}}_{i \perp} \right]
\Psi^{\pi,[f]}\left(\left\{\xi_{i}, \mathbf{k}_{i \perp} ; \lambda_{i}, \tau_{i}\right\}_{i=1,2}\right) \prod_{i=1}^{2}|  q^{\lambda_{i}} ; \xi_{i} p_{\pi}^{+}, \mathbf{p}_{i \perp}\rangle,
\label{Pion_state_Pasq}
\end{equation}
where $ {\mathbf{p}}_{\perp i}= {\mathbf{k}}_{i \perp}+\xi_{i} {\mathbf{p}}_{\pi \bot} $.

Using the expressions (\ref{N5q_state_Pasquini}), (\ref{Pion_state_Pasq}) together with the momentum representation for the
quark fields and the set of anticommutation relations for the quark creation/annihilation
operators one can establish the final expression for the matrix
elements
$T^{\lambda}_{\rho \tau, \,\chi}$ suitable for the calculation of
$\pi N$ TDAs in the ERBL-like domain:
\begin{equation}
\begin{aligned}
T
_{\rho \tau, \, \chi}^{\lambda}=&-24\left(\frac{1}{2 \xi}\right)^{3/2} \frac{1}{\sqrt{x_{1} x_{2} x_{3}}} u_{+\rho}\left(k_{1}^{+}, \lambda_{1}\right) u_{+\tau}\left(k_{2}^{+}, \lambda_{2}\right) u_{+\chi}\left(k_{3}^{+}, \lambda_{3}\right) \\
&\times \sum_{B=N, \Delta} \sum_{\lambda^{\prime}} \int \mathrm{d} y \mathrm{d}^{2} {\mathbf{k}}_{\perp} \phi_{\lambda^{\prime} 0}^{\lambda(N, B \pi)}\left(y,  {\mathbf{k}}_{\perp}\right) \sqrt{y(1-y)} \delta\left(1-y-\frac{p_{\pi}^{+}}{p_{p}^{+}}\right) \delta^{(2)}\left( {\mathbf{k}}_{\perp}+ {\mathbf{p}}_{\pi \perp}\right) \\
&\times \int
\left[\prod_{i=1}^3 {\mathbf{k}}_{i \perp} \right]
 \tilde{\Psi}_{\lambda^{\prime}}^{B,[f]}\left(\left\{\frac{x_{1}}{2 \xi}, {\mathbf{k}}_{1 \perp} ; \lambda_{1}, 1/2\right\}\left\{\frac{x_{2}}{2 \xi}, {\mathbf{k}}_{2 \perp} ; \lambda_{2}, 1/2\right\}\left\{\frac{x_{3}}{2 \xi}, {\mathbf{k}}_{3 \perp} ; \lambda_{3},-1/2\right\}\right).
\end{aligned}
\label{Tmatrix_el_Pasquini}
\end{equation}
The quark light-cone spinors of the longitudinal momenta $k_i^+= x_i P^+$ in (\ref{Tmatrix_el_Pasquini})
are given by
\begin{equation}
u_{+}\left(x_{i} P^{+}, \uparrow\right)=\sqrt{\frac{x_{i} P^{+}}{\sqrt{2}}}\left(
\begin{array}{l}1 \\
0 \\
1 \\
0
\end{array}
\right) \quad \text { and } \quad u_{+}\left(x_{i} P^{+}, \downarrow\right)=\sqrt{\frac{x_{i} P^{+}}{\sqrt{2}}}\left(
\begin{array}{c}0 \\
1 \\
0 \\
-1
\end{array}
\right), \quad i=1,2,3.
\end{equation}
Eq.~(\ref{Tmatrix_el_Pasquini}) relates $\pi N$ TDAs
to the baryon DAs in the $B \pi$ component of the nucleon
weighted by the probability amplitude that the nucleon fluctuates in the corresponding
subsystem with the pion momentum matching the pion momentum in the final state.
The momentum fraction
of the quarks in the baryon LFWF are defined with respect to the longitudinal momentum of the baryon. The integration over the transverse quark momenta corresponds to the projection of the baryon LFWF onto the zero orbital angular momentum component.
The sum over the helicity $\lambda'$ of the baryon permits baryon--pion fluctuations
which do not conserve the helicity of the parent nucleon.

The explicit expression for the $uud$ isospin component of the nucleon and $\Delta$
wave functions
$\Psi_{\lambda}^{p,[f]}\left(\left\{x_{i}, \mathbf{k}_{i \perp} ; \lambda_{i}\right\},\{u u d\}\right)$,
$\Psi_{\lambda}^{\Delta,[f]}\left(\left\{x_{i}, \mathbf{k}_{i \perp} ; \lambda_{i}\right\},\{u u d\}\right)$
can be found in Secs.~IV, V and  Appendix A of Ref.~\cite{Pasquini:2009ki}.
Similarly to the calculation of nucleon DA, the model allows the calculation
of $\pi N$ TDAs at a rather low normalization point $Q_0^2$.

\begin{figure}[H]
\begin{center}
\includegraphics[width=0.6\textwidth]{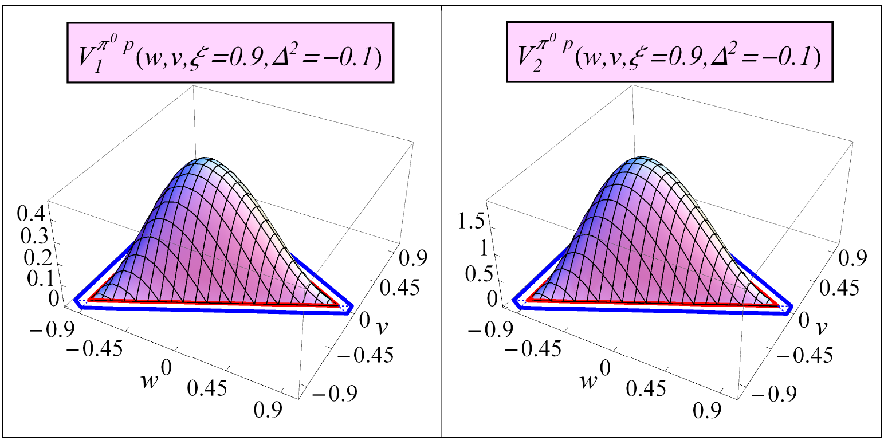}
\includegraphics[width=0.6\textwidth]{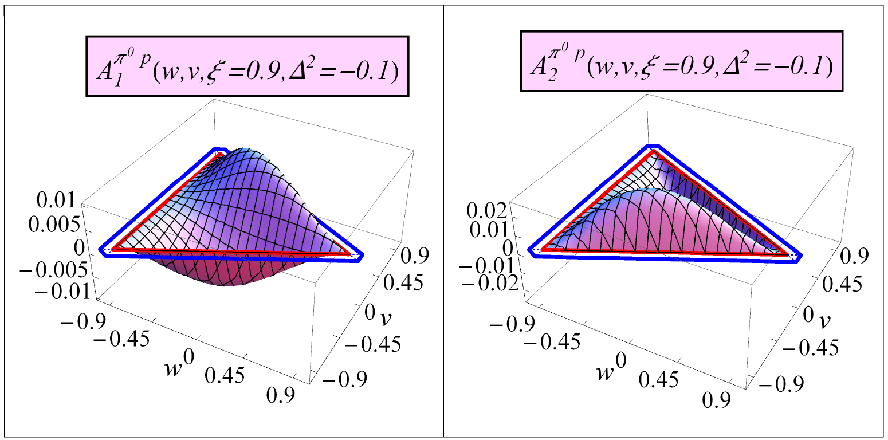}
  \caption{
  The $N^p \to \pi^0$ TDAs $V_{1,2}^{p \pi^0}$, $A_{1,2}^{p \pi^0}$ computed in the model of Ref.~\cite{Pasquini:2009ki} as functions of quark--diquark coordinates $w\equiv w_3$,  $v\equiv v_3$ (\ref{Def_qDq_coord})
  in the ERBL-like domain
  for $\xi=0.9$, $\Delta^2=-0.1~{\rm GeV}^{2}$. Red lines show the borders of the ERBL-like support domain for $\xi=0.9$. Blue lines delimit the DGLAP-like support domains for  $\xi=0.9$.}
     \label{Fig_Pasquini1}
\end{center}
\end{figure}

\begin{figure}[H]
\begin{center}
\includegraphics[width=0.6\textwidth]{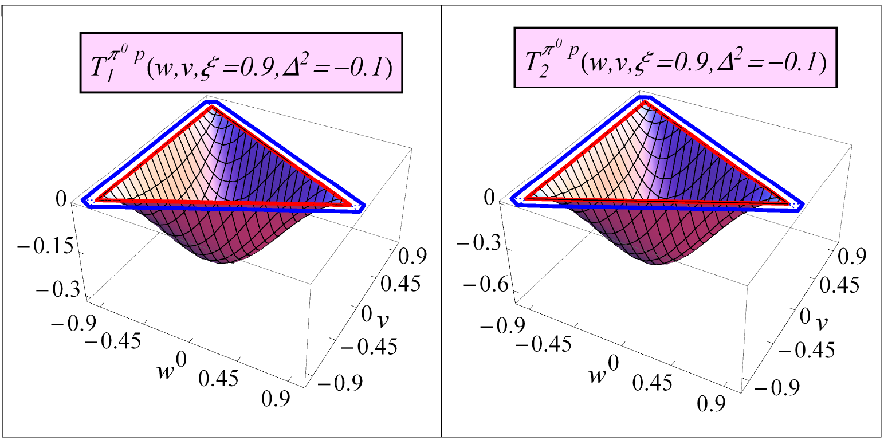}
\includegraphics[width=0.6\textwidth]{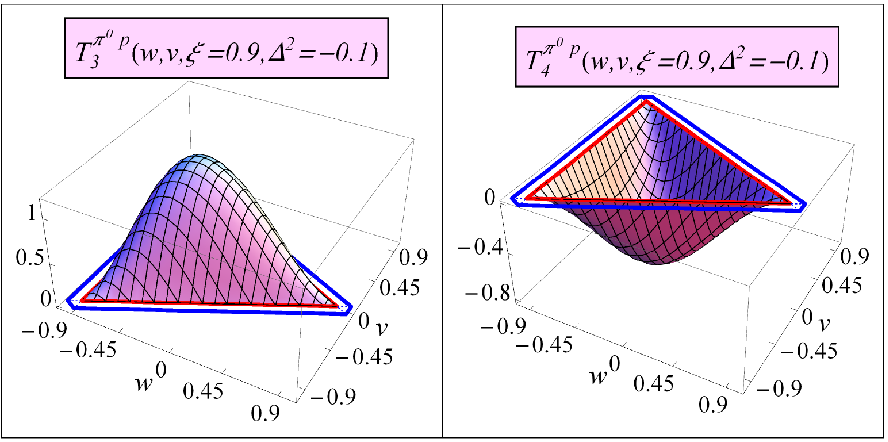}
  \caption{
  The $N^p \to \pi^0$ TDAs $T_{1,2,3,4}^{p \pi^0}$  computed in the model of Ref.~\cite{Pasquini:2009ki} as functions of quark--diquark coordinates $w\equiv w_3$,  $v\equiv v_3$ (\ref{Def_qDq_coord})
  in the ERBL-like domain
   for $\xi=0.9$, $\Delta^2=-0.1~{\rm GeV}^{2}$. Red lines show the borders of the ERBL-like support domain for $\xi=0.9$. Blue lines delimit the DGLAP-like support domains for  $\xi=0.9$.
   }
     \label{Fig_Pasquini2}
\end{center}
\end{figure}

The resulting $8$ proton-to-$\pi^0$ TDAs are presented in  Figs.~\ref{Fig_Pasquini1}, \ref{Fig_Pasquini2}. They
receive contribution from the fluctuations of the proton into
$p \pi^0$ and $\Delta^+ \pi^0$ subsystems. In particular, $\Delta$ plays a special role in the case of the tensor TDAs which involve configurations with helicity $\pm \frac{3}{2}$. The
interplay of the nucleon and $\Delta$ contributions with helicity $\pm \frac{1}{2}$ determines the different shapes of the vector and
axial-vector TDAs.

It is interesting to note that, similarly to the calculation of nucleon DAs in the same model, assuming the SU$(6)$ spin-flavor symmetry results in a
shape of the $\pi N$ TDAs that repeats the shape of the asymptotic nucleon DA
with a nearly symmetric contribution of the three quarks.

\section{Exclusive processes involving {{TDA}}s to lowest order accuracy}
\setcounter{equation}{0}
\label{Sec_ExclProcLO}
\mbox

The calculation of the lowest order in the strong coupling $\alpha_s$ and at the leading twist accuracy (LO) scattering amplitudes of hard exclusive reactions
admitting a description in terms of nucleon-to-meson (and nucleon-to-photon) TDAs shares many common features with the textbook LO pQCD calculation of the nucleon electromagnetic form factor in terms of the leading twist-$3$ nucleon DAs. This section contains a summary of results for the relevant LO scattering amplitudes.

\subsection{Nucleon form factor to the lowest order  accuracy: a compact formula}
\label{Sec_NFF}
\mbox

The isospin parametrization for the nucleon DAs
and nucleon-to-meson TDAs
presented in  Sec.~\ref{SubSec_Isospin}
makes the calculations of the LO scattering amplitudes
involving hadronic matrix element of the three-quark light-cone operator
particularly simple. In this
section we illustrate the application of this technique
and present a new compact LO formula for the nucleon electromagnetic form factor.

\begin{figure}[h]
\begin{center}
\includegraphics[width=0.5\textwidth]{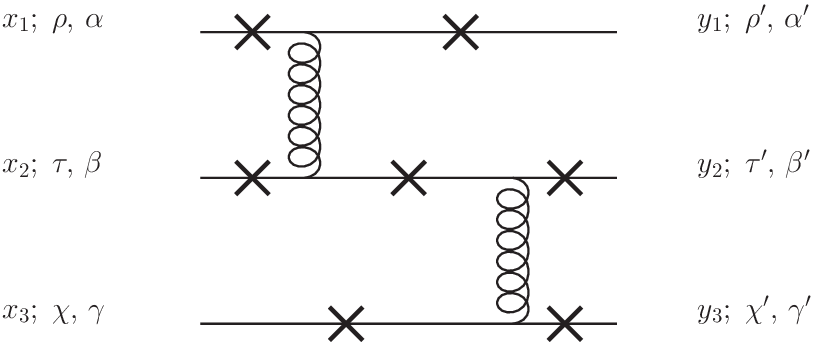}
  \caption{Basic diagram for the calculation of the hard scattering kernel for the LO nucleon form factor. By crosses we mark the
 possible locations of insertion of the electromagnetic current
  $(\sigma_{\rm e.m.})^f_{\; i}= \frac{Q^u+Q^d}{2} \delta^f_{\; i} + \frac{Q^u-Q^d}{2} \sigma^f_{\; i}$, where $i$ ($f$) stand for the SU$(2)$-isospin indices of the incoming (outgoing) nucleon; $\rho,\,\tau, \chi$ ($\rho',\,\tau', \chi'$)
  and $\alpha,\,\beta, \gamma$ ($\alpha',\,\beta', \gamma'$) stand for the Dirac and
  SU$(2)$-isospin indices of the incoming (outgoing) quarks.
  }
     \label{Fig_NFF_diags}
\end{center}
\end{figure}

To describe the outgoing nucleon
we use the following parametrization for the complex conjugated nucleon DA (\ref{Isospin_parmetrization_N_DA}):
\begin{eqnarray}
  &&
4 \langle N_f(p) |  \widehat{\bar{O}}_{\alpha \beta \gamma \; \rho \tau \chi}
(1,2,3)
| 0 \rangle  \equiv 4 \langle N_f(p) |   \varepsilon_{c_1 c_2 c_3} \bar{\Psi}_{\rho \alpha}^{c_1}(1)
\bar{\Psi}_{\tau \beta}^{c_2} (2)
\bar{\Psi}_{\chi \gamma}^{c_3} (3)
| 0 \rangle
\nonumber \\ &&
=- \bigl( \varepsilon_{\alpha \beta} \delta_\gamma^f  \overline{M}^{N\,\{13\}}_{\rho \tau \chi }(1,2,3)+
\varepsilon_{\alpha \gamma} \delta_\beta^f \overline{M}^{N\,\{12\}}_{\rho  \tau \chi}(1,2,3) \bigr)\,,
\label{Isospin_parmetrization_N_DA_conjugate}
\end{eqnarray}
with
\begin{equation}
\overline{M}^{N\,\{12\}}_{\rho  \tau \chi}(1,2,3)=
f_N
\Bigl[
V^p(1,2,3) \bar{v}_{\rho \tau, \, \chi}^N+
A^p(1,2,3) \bar{a}_{\rho \tau, \, \chi}^N+
T^p(1,2,3) \bar{t}_{\rho \tau, \, \chi}^N
\Bigr]
 \label{Decomposition_AVT_nucleon_DA_conjugate}
\end{equation}
and
\begin{equation}
\overline{M}^{N\,\{13\}}_{\rho  \tau \chi}(1,2,3)=\overline{M}^{N\,\{12\}}_{\rho  \chi  \tau}(1,3,2).
\end{equation}
In (\ref{Decomposition_AVT_nucleon_DA_conjugate}) we employ the set of the conjugated Dirac structures
$\bar{s}^N
\equiv (s^N_{\rho' \tau' \chi'})^{*} \gamma^0_{\rho' \rho} \gamma^0_{\tau' \tau} \gamma^0_{\chi' \chi}$:
\begin{eqnarray}
  &&
\bar{v}^N_{\rho \tau \chi} =(C^\dagger \hat{p})_{\tau \rho} ( \bar{U}(p) \gamma_5)_\chi\,; \nonumber \\
  &&
\bar{a}^N_{\rho \tau \chi} =(C^\dagger \gamma^5 \hat{p})_{\tau \rho} ( \bar{U}(p)  )_\chi\,; \nonumber \\
  &&
\bar{t}^N_{\rho \tau \chi} = (C^\dagger \sigma_{\mu p})_{\tau \rho} ( \bar{U}(p) \gamma_5 \gamma^\mu)_\chi \,.
\end{eqnarray}

The $7$ diagrams needed to compute the nucleon form factor (FF) with our system of isospin notations
are schematically presented in  Fig.~\ref{Fig_NFF_diags}.
The number of independent diagrams is further reduced to just $4$ (one of which
is identically zero) due to the $C$-conjugation symmetry between the initial and final
state nucleons. This represents a considerable simplification comparing
to the original approach, in which $14$ independent diagrams  have to be computed.
In this way we obtain the following simple formula the proton and neutron form factor
in terms of the leading twist-$3$ nucleon DA $\phi^N$:
\begin{equation}
Q^4 F_1^{p,n}(Q^2)
=\frac{(4 \pi \alpha_s)^2}{54} f_N^2
\int_0^1 d_3x  \;
\delta \left( \sum_{l=1}^3 x_j-1 \right)
\, \int_0^1 d_3y \; \delta \left( \sum_{l=1}^3 y_l-1 \right) \left\{ 2 (R_1^{p,n}+R_2^{p,n}+R_3^{p,n})+R_4^{p,n} \right\},
\label{N_FF_compact}
\end{equation}
where $\alpha_s$ is the strong coupling and
$f_N$
is the nucleon light-cone wave function normalization constant.
Ref.~\cite{Chernyak:1987nv} quotes the following value for $f_N$:
\begin{equation}
f_N=5.0 \cdot 10^{-3} \; {\rm GeV}^2.
\label{fN_CZvalue}
\end{equation}
The integration in (\ref{N_FF_compact}) stands over the usual support of
nucleon DAs
\begin{eqnarray}
  & & \int_0^1 d_3x  \;
\delta \left( \sum_{l=1}^3 x_j-1 \right)
\, \int_0^1 d_3y \; \delta \left( \sum_{l=1}^3 y_l-1 \right) \nn \\
&& \equiv
\int_0^1 dx_1 dx_2 dx_3 \delta(x_1+x_2+x_3-1)
\int_0^1 dy_1 dy_2 dy_3 \delta(y_1+y_2+y_3-1).
\end{eqnarray}
The integrands
$R_\alpha^{\{p,n\}}$
in
(\ref{N_FF_compact})
have the following structure
(no summation is implied over the repeating index $\alpha$):
\begin{equation}
R_\alpha^{\{p,n\}}= D_\alpha N_\alpha^{\{p,n\}},
\label{Str_R_NN}
\end{equation}
where $D_\alpha$ stand for the hard convolution kernels originating from the  partonic propagators
and  $N_\alpha^{\{p,n\}}$ are combinations of nucleon DAs. The explicit expressions
for the denominators $D_\alpha$ and numerators $N_\alpha^{\{p,n\}}$
are summarized in~Table~\ref{Table_Nucl_FF}.

\begin{longtable}{|c|p{2.5cm}|p{12.5cm}|}
%
\caption{Diagrams contributing to the LO nucleon FF. }
\label{Table_Nucl_FF}
 \\
\hline
$\alpha$ &  \qquad  Diagram  & Numerators \\
        &  \qquad  $D_\alpha$  &  \\
        \nopagebreak
\hline
1 & 
{\includegraphics[height=1.5cm,clip=true]{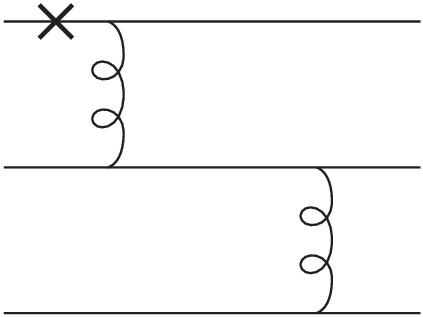}}
$\frac{1}{(1-x_1)^2 x_3 \, (1-y_1)^2 y_3}$ &
\raisebox{0.4cm}{\begin{tabular}{l| p{9.5cm}}
 $N_\alpha^{p}$ &
 $
 \frac{1}{3} \left(2 \phi^N(x_1,x_2,x_3)+ \phi^N(x_3,x_2,x_1)\right)\left(2 \phi^N(y_1,y_2,y_3)+ \phi^N(y_3,y_2,y_1)\right)$
\\
    &
\\
    &
$
+\frac{1}{3}\left(2 \phi^N(x_1,x_3,x_2)+ \phi^N(x_2,x_3,x_1)\right)\left(2 \phi^N(y_1,y_3,y_2)+ \phi^N(y_2,y_3,y_1)\right);$ \\ 
& \\
 $N_\alpha^{n}$
 & $
 \frac{1}{3}\left(\phi^N(x_1,x_3,x_2)- \phi^N(x_2,x_3,x_1)\right) \left( \phi^N(y_1,y_3,y_2)- \phi^N(y_2,y_3,y_1)\right) $ \\ &
 \\
    &
$
+\frac{1}{3}\left(\phi^N(x_3,x_2,x_1)- \phi^N(x_1,x_2,x_3)\right( \left(\phi^N(y_3,y_2,y_1)- \phi^N(y_1,y_2,y_3)\right)$
\\ &
\\ &
$
-\phi^N(x_1,x_2,x_3)\phi^N(y_1,y_2,y_3)
-\phi^N(x_1,x_3,x_2)\phi^N(y_1,y_3,y_2);$
\\
\\
\end{tabular}} \\
\hline
\hline
2 & 
{\includegraphics[height=1.5cm,clip=true]{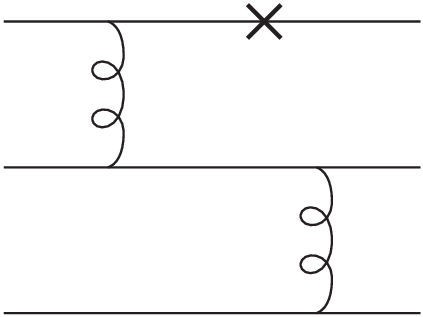}}
&
\raisebox{0.4cm}{\begin{tabular}{l| p{9.5cm}}
 $N_\alpha^{p}$ & $
0;
$ \\ \\ 
 $N_\alpha^{n}$
 & $0;
$
\\
\\
\end{tabular}} \\
\hline
\hline
3 & 
{\includegraphics[height=1.5cm,clip=true]{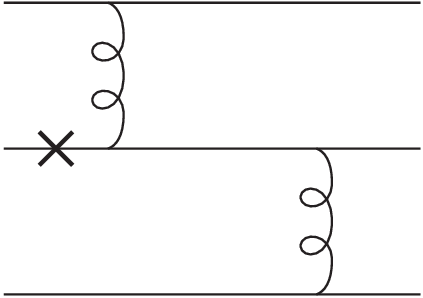}}
$\frac{1}{x_1 (1-x_2) x_3 \, y_1 (1-y_1) y_3}$ &
\raisebox{0.4cm}{\begin{tabular}{l| p{9.5cm}}
 $N_\alpha^{p}$ & $
   -\frac{1}{3} \left(\phi^N(x_1,x_3,x_2)+2 \phi^N(x_2,x_3,x_1)\right)\left(\phi^N(y_1,y_3,y_2)+2 \phi^N(y_2,y_3,y_1)\right);
$ \\ \\ 
 $N_\alpha^{n}$
 & $
   -\frac{1}{3}\left(\phi^N(x_2,x_3,x_1)-\phi^N(x_1,x_3,x_2)\right)
   \left(\phi^N(y_2,y_3,y_1)-\phi^N(y_1,y_3,y_2)\right)$ \\ &
   \\ &
$
+\phi^N(x_2,x_3,x_1)\phi^N(y_2,y_3,y_1);$
\\
\\
\end{tabular}} \\
\hline
\hline
4 & 
{\includegraphics[height=1.5cm,clip=true]{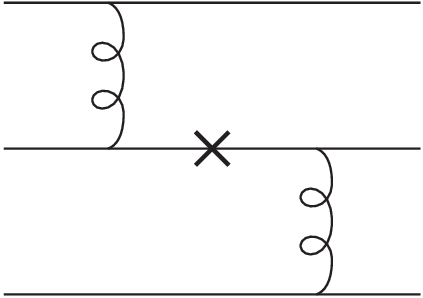}}
$\frac{1}{x_1 x_3 (1-x_3)\,y_1 (1-y_1) y_3}$ &
\raisebox{0.4cm}{\begin{tabular}{l| p{9.5cm}}
 $N_\alpha^{p}$ & $
\frac{1}{3} \left(\phi^N(x_1,x_2,x_3) -\phi^N(x_3,x_2,x_1) \right)\left(\phi^N(y_1,y_2,y_3) -\phi^N(y_3,y_2,y_1) \right);
   $ \\ \\ 
 $N_\alpha^{n}$
 & $
 \frac{1}{3}\left(\phi^N(x_1,x_2,x_3) +2 \phi^N(x_3,x_2,x_1)\right) \left(\phi^N(y_1,y_2,y_3) +2 \phi^N(y_3,y_2,y_1) \right)$ \\ &
 \\ &
$
-\phi^N(x_3,x_2,x_1)\phi^N(y_3,y_2,y_1);     $
\\
\\
\end{tabular}} \\
\hline
\end{longtable}

\bi
\item
With the use of the asymptotic form of the nucleon DA $\phi^N_{\rm as}(x_1,x_2,x_3)=120 x_1 x_2 x_3$
we recover the result of~\cite{Knodlseder:2015vmu}:
\begin{equation}
Q^4 F_1^p(Q^2)\Big| _{\phi^N_{\rm as}}=0;
\ \ \
Q^4 F_1^n(Q^2)\Big| _{\phi^N_{\rm as}}=\frac{(4 \pi \alpha_s)^2 f_N^2}{54}
1.8 \cdot 10^3.
\end{equation}

\item With the use of the COZ phenomenological solution of
~\cite{Chernyak:1987nv}
\begin{equation}
\phi^N_{\rm COZ}(x_1, \, x_2,\,x_3)=\phi^N_{\rm as}(x_1, \, x_2,\,x_3)
\left[
23.814 x_{1}^{2}+12.978 x_{2}^{2}+6.174 x_{3}^{2}+5.88 x_{3}-7.098
\right]
\end{equation}
we recover the result of Ref.~\cite{Chernyak:1987nv} both for the
proton and neutron FFs. In particular,
\[
Q^4 F_1^p(Q^2)
\Big| _{\phi^N_{\rm COZ}}
=\frac{(4 \pi \alpha_s)^2 f_N^2}{54} \,  1.45 \cdot 10^5; \ \ \
\frac{F_{1}^{n}\left(Q^{2}\right)}{F_{1}^{p}\left(Q^{2}\right)} \Big| _{\phi^N_{\rm COZ}} \simeq -0.47.
\]
\ei

We would like to attract the attention of the reader to a possible
controversy in the normalization of the proton helicity $\uparrow$
state (and hence of the nucleon DA $\phi^N$) first reported in Ref.~\cite{Ji:1986uh}.
It was considered by different authors~\cite{Brooks:2000nb,Thomson:2006ny}
(see particularly the discussion in  Sec.~III.C of Ref.~\cite{Thomson:2006ny}).
A dedicated study in Ref.~\cite{Knodlseder:2015vmu}
has confirmed the consistency of the conventions employed by V.~Chernyak~\cite{Chernyak:1987nv} (see also Ref.~\cite{Stefanis:1987vr}).
Throughout our analysis we rely on this latter set of conventions.

\subsection{Backward meson electroproduction: calculation of the hard amplitude}
\label{SubSec_Bkw_meson_ampl}
\mbox

In this section we present the calculation of the leading-twist leading order
in $\alpha_s$  scattering amplitudes of backward hard meson
electroproduction reactions within the collinear factorization approach
involving description in terms of nucleon DAs and nucleon-to-meson TDAs.

\subsubsection{Backward pseudoscalar meson electroproduction}
\label{SubSec_Ampl_Pkw_PS_meson}
\mbox

Let us first address the pseudoscalar meson case, and for definiteness consider
hard backward pion electroproduction:
\begin{equation}
\gamma^{*}(q, \lambda_\gamma) + N(p_N,s_N) \to N(p'_N,s'_N) + \pi( p_\pi)
\label{Reaction_ampl_p_pi0}
\end{equation}
in the near-backward kinematical regime (see  Sec.~\ref{SubSec_Kinematics}).
This requires the calculation of the same set of $21$ diagrams\footnote{With the use of the isotopic notations of  Sec.~\ref{SubSec_Isospin}
the number of the diagrams can be reduced to $7$, see  Sec.~\ref{Sec_NFF}. }
as the leading twist-$3$, leading order in $\alpha_s$ calculation
of the nucleon electromagnetic form factor.

For the phenomenological applications
it turns out to be convenient to employ the parametrization
of the leading twist-$3$ $\pi N$ TDAs introduced in
 Sec.~\ref{SubSubSec_Def_piN_TDAs}
with the set of the Dirac structures
(\ref{Def_DirStr_piN_DeltaT})
explicitly depending on
$\Delta_T$.
This allows a clear separation of the
$\Delta_T$-dependent and  $\Delta_T$-independent contributions
to the amplitude of the reaction
(\ref{Reaction_ampl_p_pi0}),
the latter controlling the only surviving contributions  in the
strictly backward kinematical regime
$\Delta_T=0$.

To the leading twist-$3$, leading order in $\alpha_s$ accuracy, the
helicity amplitudes
${\mathcal{M}}_{s_N s'_N}^{\lambda_\gamma }$
of the reaction
(\ref{Reaction_ampl_p_pi0})
admit the following parametrization
\begin{equation}
{\mathcal{M}}_{s_N s'_N}^{\lambda_\gamma }=
{\mathcal{C}}_\pi
\frac{1}{Q^4}
\sum_{k=1}^2
{\mathcal{S}}_{s_N s'_N}^{(k) \, \lambda_\gamma} {\mathcal{I}}^{(k)} (\xi, \, \Delta^2)\,,
\label{Hel_ampl_def_piN}
\end{equation}
with the overall normalization constant
\begin{equation}
{\mathcal{C}}_\pi\equiv-i \frac{(4 \pi \alpha_s)^2 \sqrt{4 \pi \alpha_{em}} f_N^2 }{54 f_\pi },
\label{Def_C_PiN}
\end{equation}
where
$\alpha_{em} \simeq \frac{1}{137}$  is the electromagnetic fine structure constant,
$\alpha_s$ is the strong coupling,
$f_N$
is the nucleon light-cone wave function normalization constant
(\ref{fN_CZvalue})
and
$f_\pi=93~{\rm MeV}$ is the pion weak decay constant.

The parametrization
(\ref{Hel_ampl_def_piN})  involves
two leading twist-$3$ spin structures
${\mathcal{S}}^{(1,2) \, \lambda_\gamma}_{s_N s'_N}$ defined as
\begin{eqnarray}
  &&{\mathcal{S}}^{(1) \, \lambda_\gamma}_{s_N s'_N} \equiv \bar U(p'_N,s'_N)
\hat{\mathcal{E}}(q,\lambda_\gamma)
 \gamma^5 U(p_N,s_N)\,; \nn \\
   &&    {\mathcal{S}}^{(2) \, \lambda_\gamma}_{s_N s'_N} \equiv \frac{1}{m_N}\bar U(p'_N,s'_N)
\hat{ \mathcal{E} }(q, \lambda_\gamma) \hat{\Delta}_T
 \gamma^5 U(p_N,s_N).
\end{eqnarray}
For the integral convolutions
$\mathcal{I}^{(k=1,2)}$ in (\ref{Hel_ampl_def_piN})
we employ the following notations
\begin{equation}
\mathcal{I}^{(k)} (\xi,\Delta^2) =
{\int^{1+\xi}_{-1+\xi} }\! \! \!d_3x  \; \delta \left( \sum_{j=1}^3 x_j-2\xi \right)
{\int^{1}_{0} } \! \! \! d_3y \; \delta \left( \sum_{l=1}^3 y_l-1 \right) \;
{\Biggl(2\sum_{\alpha=1}^{7}   T_{\alpha}^{(k)} +
\sum_{\alpha=8}^{14}   T_{\alpha}^{(k)} \Biggr)}.
\label{Def_I_k_convolutions}
\end{equation} 
The integration in
(\ref{Def_I_k_convolutions})
stands over the  support domain of $\pi N$ TDAs in the longitudinal momentum fraction
variables $x_i$ and over the support domain of nucleon DAs in the longitudinal momentum fraction
variables $y_i$:
\begin{eqnarray}
  &&
 {\int^{1+\xi}_{-1+\xi} }\! \! \!
d_3 x  \;  \delta \left( \sum_{j=1}^3 x_j-2\xi \right) \equiv  {\int^{1+\xi}_{-1+\xi} }\! \! \!
d x_1  {\int^{1+\xi}_{-1+\xi} }\! \! \!
d x_2  {\int^{1+\xi}_{-1+\xi} }\! \! \!
d x_3  \delta(x_1+x_2+x_3-2\xi);
\nonumber \\
  &&
{\int^{1}_{0} } \! \! \! d_3y
\; \delta \left( \sum_{l=1}^3 y_l-1 \right)
\equiv {\int^{1}_{0} } dy_1 {\int^{1}_{0} } dy_2 {\int^{1}_{0} } dy_3 \delta(y_1+y_2+y_3-1).
\label{Int_TDA_DA_support}
\end{eqnarray}
The index
$\alpha$
refers to the number of a diagram  (see Table~\ref{Table_Bkw_pion}) and the index
$k=1,\, 2$
runs for the contributions into the two invariant amplitudes of Eq.~(\ref{Hel_ampl_def_piN}).

The general structure of the convolution integrals
(\ref{Def_I_k_convolutions})
resembles much the familiar expressions for the leading twist
hard amplitudes of hard exclusive meson electroproduction off nucleons
within the collinear factorized description in terms of nucleon GPDs ($H(x,\xi)$)
and meson DAs ($\phi_{\mathcal{M}}$)~\cite{Diehl:2003ny}:
\begin{equation}
{\mathcal{A}}\Big| _{\text{HMP}} \sim \int_{-1}^{1} d x \frac{H(x, \xi)}{x \pm \xi \mp i 0} \int_{0}^{1} d y \frac{\phi_{\mathcal{M}}(y)}{y}.
\end{equation}
The coefficients
$T_{\alpha}^{(k)}$
can be represented as products of singular
hard kernels, originating from the partonic propagators,
times the combinations
$N_\alpha^{(k)} \equiv N_\alpha^{(k)}(x_1,x_2,\xi, \Delta^2; \, y_1, y_2,y_3)$
of
$\pi N$ TDAs and nucleon DAs
arising in the numerators:
\begin{equation}
T_{\alpha}^{(k)}\equiv  D_\alpha \times N_\alpha^{(k)}.
\label{Def_T_alpha}
\end{equation}
Note that no summation over the repeating index $\alpha$ is assumed in
(\ref{Def_T_alpha}).

\bi
\item
The explicit expressions for
$T_{\alpha}^{(k)}$
$\alpha=1,\ldots,14$
for $\gamma^{*} N^p \to \pi^0 N^p$ reaction
are presented in Table~\ref{Table_Bkw_pion}.
We adopt the shortened notations for the arguments
of $\pi N$ TDAs and nucleon DAs:
\begin{eqnarray}
  &&
\left\{V_{1,2}^{\pi^0 p}, \,A_{1,2}^{\pi^0 p}, \, T_{1,2,3,4}^{\pi^0 p} \right\} \equiv \left\{V_{1,2}^{\pi^0 p}, \,A_{1,2}^{\pi^0 p}, \, T_{1,2,3,4}^{\pi^0 p} \right\}(x_1,x_2,x_3,\xi,\Delta^2); \nn \\
  &&
\left\{V^{p}, \,A^{p}, \, T^{p} \right\} \equiv \left\{V^{p}, \,A^{p}, \, T^{p} \right\}
(y_1,y_2,y_3).
\label{Not_arg_short}
\end{eqnarray}

The diagrams with $\alpha=15,\ldots,21$ differ from
those with $\alpha=1,\ldots,7$ by a permutation of the $u$-quark lines $1$ and $2$.
They give the same contribution due to the symmetry  of the integration
domain and of the TDAs and DAs with respect to the change of variables $x_1 \leftrightarrow x_2$, $y_1 \leftrightarrow y_2$.
Therefore, the diagrams with $\alpha=15,\ldots,21$ are  accounted for by the factor $2$ in the last line of Eq.~(\ref{Def_I_k_convolutions}).

\item The result for
$\gamma^{*} p \rightarrow \pi^+ n$ channel can be read from the same
Table~\ref{Table_Bkw_pion}
with the obvious changes:
\begin{eqnarray}
  &&
\nonumber
Q_u \rightarrow Q_d; \ \ \ Q_d \rightarrow Q_u\,; \\
  &&
V^p, \, A^p, \, T^p \, \rightarrow \, V^n, \, A^n, \, T^n \nonumber  \;  \equiv - V^p, \, -A^p, \, -T^p\,; \\
  &&
V^{ \pi^0 p}_{1,2}, \, A^{ \pi^0 p}_{1,2}, \, T^{ \pi^0 p}_{1,2,3,4} \,
 \rightarrow
 V^{ \pi^+ p}_{1,2}, \, A^{ \pi^+ p}_{1,2}, \, T^{ \pi^+ p}_{1,2,3,4}\,.
\end{eqnarray}

\item The result presented in Table~\ref{Table_Bkw_pion} can be also
adopted for other pseudoscalar mesons. \textit{E.g.} for $\eta$-meson
one just has to replace the corresponding TDAs
\begin{equation}
V^{ \pi^0 p}_{1,2}, \, A^{ \pi^0 p}_{1,2}, \, T^{ \pi^0 p}_{1,2,3,4} \,
 \rightarrow
 V^{ \eta p}_{1,2}, \, A^{ \eta p}_{1,2}, \, T^{ \eta p}_{1,2,3,4}\,.
\end{equation}
\ei

\begin{longtable}{|c|p{4.5cm}|p{10cm}|}
\caption{$14$ out of $21$ diagrams contributing to the hard scattering amplitude of
$\gamma^* N^p \to N^p \pi^0$ reaction in the near backward kinematics. The crosses denote the virtual photon vertices.}
\label{Table_Bkw_pion}
\\
\hline
$\alpha$ &  \qquad \quad Diagram  & Numerators \\
        &  \qquad \qquad  $D_\alpha$  &  \\
        \nopagebreak
\hline
1 & 
{\includegraphics[height=1.5cm,clip=true]{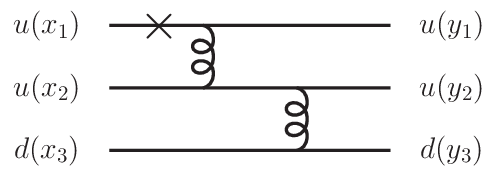}}
$\frac{Q_u (2\xi)^2}{(2\xi-x_{1}-i0)^2(x_{3}-i0)(1-y_{1})^2y_{3}}$ &
\raisebox{0.4cm}{\begin{tabular}{l| p{8.71cm}}
 $N_\alpha^{(1)}$ & $
-\left(V^p-A^p\right)
   \left(V_{1}^{p \pi^0}-A_{1}^{p \pi^0}\right)
   -
   4 T^{p} \left(T_{1} ^{p \pi^0}+ \frac{\Delta_T^2}{2m_N^2}T_{4 }^{p \pi^0}\right);
$ \\ \\ 
 $N_\alpha^{(2)}$
 & $ -\left(V^p-A^p\right)
  \left(V_{2}^{p \pi^0}-A_{2}^{p \pi^0}\right)
   -
   2 T^{p}  \left( T_{2 }^{p \pi^0} + T_{3 }^{p \pi^0} \right);
$
\\
\\
\end{tabular}} \\
\hline
\hline
2 & 
{\includegraphics[height=1.5cm,clip=true]{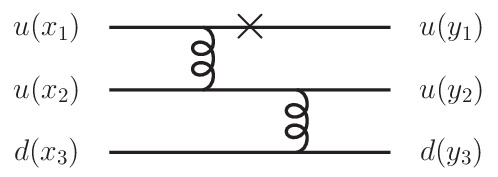}}
&
\raisebox{0.4cm}{\begin{tabular}{l| p{8.71cm}}
 $N_\alpha^{(1)}$ & $
0;
$ \\ \\ 
 $N_\alpha^{(2)}$
 & $0;
$
\\
\\
\end{tabular}} \\
\hline
\hline
3 & 
{\includegraphics[height=1.5cm,clip=true]{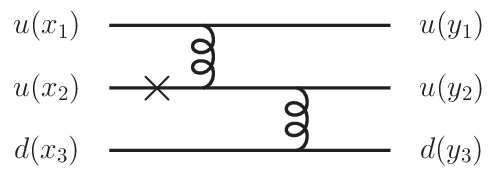}}
$\frac{Q_u (2 \xi)^2)}{(x_{1}-i0) (2\xi-x_{2}-i0)(x_{3}-i0)y_{1}(1-y_{1})y_{3}}$ &
\raisebox{0.4cm}{\begin{tabular}{l| p{8.71cm}}
 $N_\alpha^{(1)}$ & $
   4 T^{p} \left(T_{1} ^{p \pi^0}+ \frac{\Delta_T^2}{2m_N^2}T_{4 }^{p \pi^0}\right);
$ \\ \\ 
 $N_\alpha^{(2)}$
 & $
   2 T^{p}  \left( T_{2 }^{p \pi^0} + T_{3 }^{p \pi^0} \right);
$
\\
\\
\end{tabular}} \\
\hline
\hline
4 & 
{\includegraphics[height=1.5cm,clip=true]{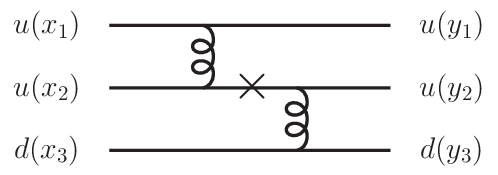}}
$\frac{Q_u (2\xi)^2}{(x_{1}-i0)
(2\xi-x_{3}-i0)(x_{3}-i0)y_{1}(1-y_{1})y_{3}}$ &
\raisebox{0.4cm}{\begin{tabular}{l| p{8.71cm}}
 $N_\alpha^{(1)}$ & $
-\left(V^p-A^p\right)
   \left(V_{1}^{p \pi^0}-A_{1}^{p \pi^0}\right);
   $ \\ \\ 
 $N_\alpha^{(2)}$
 & $ -\left(V^p-A^p\right)
  \left(V_{2}^{p \pi^0}-A_{2}^{p \pi^0}\right);
   $
\\
\\
\end{tabular}} \\
\hline
\hline
5 & 
{\includegraphics[height=1.5cm,clip=true]{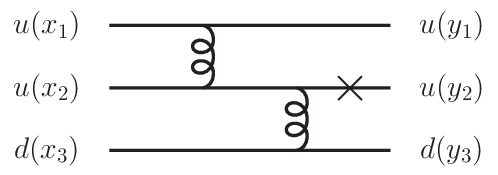}}
$\frac{Q_u (2\xi)^2}{(x_{1}-i0)(2\xi-x_{3}-i0)(x_{3}-i0)y_{1}(1-y_{2})y_{3}}$ &
\raisebox{0.4cm}{\begin{tabular}{l| p{8.71cm}}
 $N_\alpha^{(1)}$ & $
\left(V^p+A^p\right)
   \left(V_{1}^{p \pi^0}+A_{1}^{p \pi^0}\right);
   $ \\ \\ 
 $N_\alpha^{(2)}$
 & $ \left(V^p+A^p\right)
  \left(V_{2}^{p \pi^0}+A_{2}^{p \pi^0}\right);
   $
\\
\\
\end{tabular}} \\
\hline
\hline
6 & 
{\includegraphics[height=1.5cm,clip=true]{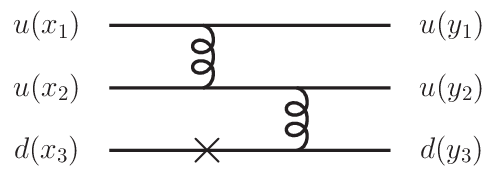}}
&
\raisebox{0.4cm}{\begin{tabular}{l| p{8.71cm}}
 $N_\alpha^{(1)}$ & $0;$ \\ \\ 
 $N_\alpha^{(2)}$ & $0;$
\\
\\
\end{tabular}} \\
\hline
\hline
7 & 
{\includegraphics[height=1.5cm,clip=true]{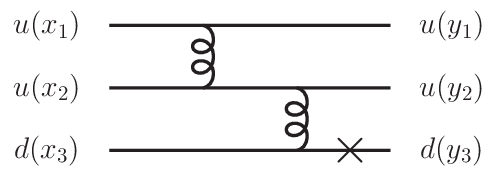}}
$\frac{Q_d(2\xi)^2}{(x_{1}-i0)(2\xi-x_{3}-i0)^2y_{1}(1-y_3)^2}$ &
\raisebox{0.4cm}{\begin{tabular}{l| p{8.71cm}}
 $N_\alpha^{(1)}$ & $
-2 \left(V^p V_{1}^{p \pi^0} +A^p A_{1}^{p \pi^0}\right);
    $ \\ \\ 
 $N_\alpha^{(2)}$
 & $ -2 \left(V^p V_{2}^{p \pi^0} +A^p A_{2}^{p \pi^0}\right);
   $
\\
\\
\end{tabular}} \\
\hline
\hline
8 & 
{\includegraphics[height=1.5cm,clip=true]{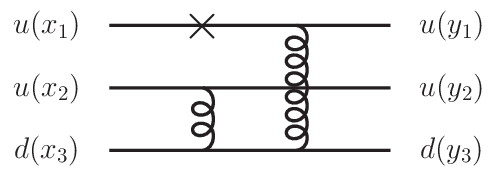}}
&
\raisebox{0.4cm}{\begin{tabular}{l| p{8.71cm}}
 $N_\alpha^{(1)}$ & $
0;
$ \\ \\ 
 $N_\alpha^{(2)}$
 & $0;
$
\\
\\
\end{tabular}} \\
\hline
\hline
9 & 
{\includegraphics[height=1.5cm,clip=true]{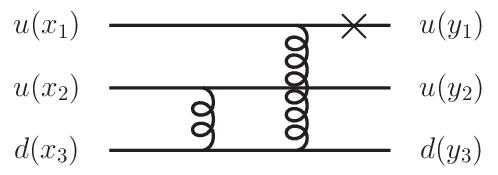}}
$\frac{Q_u (2\xi)^2}{(2\xi-x_{1}-i0)^2(x_{2}-i0)(1-y_{1})^2y_{2}}$ &
\raisebox{0.4cm}{\begin{tabular}{l| p{8.71cm}}
 $N_\alpha^{(1)}$ & $
-\left(V^p-A^p\right)
   \left(V_{1}^{p \pi^0}-A_{1}^{p \pi^0}\right)
   -
   4 T^{p} \left(T_{1} ^{p \pi^0}+ \frac{\Delta_T^2}{2m_N^2}T_{4 }^{p \pi^0}\right);
$ \\ \\ 
 $N_\alpha^{(2)}$
 & $ -\left(V^p-A^p\right)
  \left(V_{2}^{p \pi^0}-A_{2}^{p \pi^0}\right)
   -
   2 T^{p}  \left( T_{2 }^{p \pi^0} + T_{3 }^{p \pi^0} \right);
$
\\
\\
\end{tabular}} \\
\hline
\hline
10 & 
{\includegraphics[height=1.5cm,clip=true]{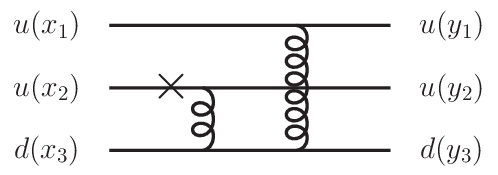}}
$\frac{Q_u (2\xi)^2}{(x_{1}-i0)(2\xi-x_{2}-i0)^2y_{1}(1-y_{2})^2}$ &
\raisebox{0.4cm}{\begin{tabular}{l| p{8.71cm}}
 $N_\alpha^{(1)}$ & $
-\left(V^p+A^p\right)
   \left(V_{1}^{p \pi^0}+A_{1}^{p \pi^0}\right)
   -
   4 T^{p} \left(T_{1} ^{p \pi^0}+ \frac{\Delta_T^2}{2m_N^2}T_{4 }^{p \pi^0}\right);
$ \\ \\ 
 $N_\alpha^{(2)}$
 & $ -\left(V^p+A^p\right)
  \left(V_{2}^{p \pi^0}+A_{2}^{p \pi^0}\right)
   -
   2 T^{p}  \left( T_{2 }^{p \pi^0} + T_{3 }^{p \pi^0} \right);
$
\\
\\
\end{tabular}} \\
\hline
\hline
11 & 
{\includegraphics[height=1.5cm,clip=true]{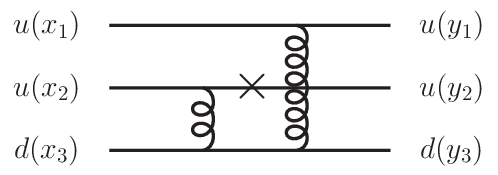}}
&
\raisebox{0.4cm}{\begin{tabular}{l| p{8.71cm}}
 $N_\alpha^{(1)}$ & $
0;
$ \\ \\ 
 $N_\alpha^{(2)}$
 & $0;
$
\\
\\
\end{tabular}} \\
\hline
\hline
12 & 
{\includegraphics[height=1.5cm,clip=true]{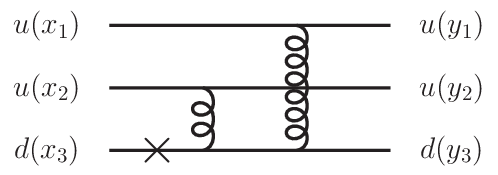}}
$\frac{Q_d (2 \xi)^2}{(x_{1}-i0)(x_{2}-i0)(2\xi-x_{3}-i0)y_{1}(1-y_{2})y_{2}}$ &
\raisebox{0.4cm}{\begin{tabular}{l| p{8.71cm}}
 $N_\alpha^{(1)}$ & $
\left(V^p+A^p\right)
   \left(V_{1}^{p \pi^0}+A_{1}^{p \pi^0}\right);
   $ \\ \\ 
 $N_\alpha^{(2)}$
 & $ \left(V^p+A^p\right)
  \left(V_{2}^{p \pi^0}+A_{2}^{p \pi^0}\right);
   $
\\
\\
\end{tabular}} \\
\hline
\hline
13 & 
{\includegraphics[height=1.5cm,clip=true]{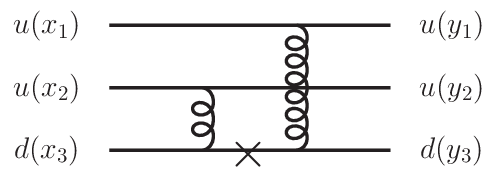}}
$\frac{Q_d (2 \xi)^2)}{(x_{1}-i0)(2\xi-x_{1}-i0)(x_{2}-i0)y_{1}(1-y_{2})y_{2}}$ &
\raisebox{0.4cm}{\begin{tabular}{l| p{8.71cm}}
 $N_\alpha^{(1)}$ & $
  - 4 T^{p} \left(T_{1} ^{p \pi^0}+ \frac{\Delta_T^2}{2m_N^2}T_{4 }^{p \pi^0}\right);
$ \\ \\ 
 $N_\alpha^{(2)}$
 & $
   -2 T^{p}  \left( T_{2 }^{p \pi^0} + T_{3 }^{p \pi^0} \right);
$
\\
\\
\end{tabular}} \\
\hline
\hline
14 & 
{\includegraphics[height=1.5cm,clip=true]{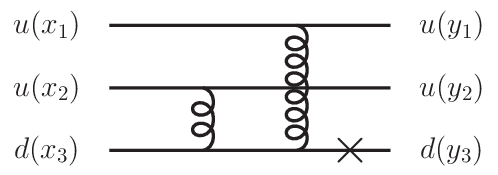}}
$\frac{Q_d (2\xi)^2}{(x_{1}-i0)(2\xi-x_{1}-i0)(x_{2}-i0)y_{1}y_{2}(1-y_{3})}$ &
\raisebox{0.4cm}{\begin{tabular}{l| p{8.71cm}}
 $N_\alpha^{(1)}$ & $
\left(V^p-A^p\right)
   \left(V_{1}^{p \pi^0}-A_{1}^{p \pi^0}\right);
   $ \\ \\ 
 $N_\alpha^{(2)}$
 & $ \left(V^p-A^p\right)
  \left(V_{2}^{p \pi^0}-A_{2}^{p \pi^0}\right);
   $
\\
\\
\end{tabular}} \\
\hline
\end{longtable}

We note that the $x_i$- and $y_i$- dependencies of the hard kernels $D_\alpha$
(\ref{Def_T_alpha}) are factorized:
\begin{equation}
D_\alpha \sim K_\alpha(x_1,x_2,x_3, \xi) \, {\mathcal{K}}_\alpha(y_1,y_2,y_3).
\label{D_factorized}
\end{equation}
\bi
\item The hard kernels ${\mathcal{K}}_\alpha(y_1,y_2,y_3)$
are similar to those occurring within the perturbative description of the nucleon
electromagnetic form factor. The convolutions of nucleon DAs with
${\mathcal{K}}_\alpha(y_1,y_2,y_3)$
do not produce any imaginary part for the amplitude since the nucleon DAs have pure ERBL support and are supposed to vanish at the borders
of their domain of definition.

\item To deal with the hard kernels $K_\alpha(x_1,x_2,x_3, \xi)$
it turns out to be convenient to switch  to the quark--diquark coordinates (see  Sec.~\ref{SubSec_Support}).
For each diagram  there exists
a preferred choice
$i=1,\,2,\,3$ of the quark--diquark coordinates
$(w_i, \, v_i)$
(\ref{Def_qDq_coord})
in which the corresponding convolution kernel
$K_\alpha$
reduces to one of the following types
\begin{eqnarray}
  &&
K_I^{(\pm, \pm)} (w_i,v_i)= \frac{1}{(w_i\pm \xi \mp i 0)} \frac{1}{(v_i \pm \xi'_i \mp i 0)}, \nonumber
\\
  &&
K_{II}^{(-, \pm)}(w_i,v_i)= \frac{1}{(w_i-\xi+ i 0)^2} \frac{1}{(v_i \pm \xi'_i \mp i 0)}. \label{Kernels_K}
\end{eqnarray}
The singularities of the hard kernels
(\ref{Kernels_K})
are located on the cross-over trajectories $w_i=-\xi, \, v_i=\pm \xi'$
separating the ERBL-like and the DGLAP-like support domains of $\pi N$ TDAs and
on the lines  $w_i=\xi$ that belong to the DGLAP-like support domains.
Thus the convolution integrals in
$x_i$
with
$\pi N$
TDAs in
(\ref{Def_I_k_convolutions})
may generate a  nonzero imaginary part for the amplitude.
$\pi N$ TDAs, indeed,
do not necessarily  vanish on the cross over trajectories
$x_i=0$,
separating ERBL-like and DGLAP-like regimes,
as well as on the lines
$x_i=2 \xi$.
\ei

We conclude that the calculation of the real and imaginary parts of
$\mathcal{I}^{(k)} (\xi,\Delta^2)$
requires  the consideration of the two types of integral convolutions of TDAs:
\begin{equation}
I_I^{(\pm, \pm)}(\xi)= \int_{-1}^1 dw  \int_{-1+| \xi-\xi'| }^{1-| \xi-\xi'| } dv
\frac{1}{(w\pm \xi \mp i 0)} \frac{1}{(v \pm \xi' \mp i 0)} H(w,v,\xi),
\label{I_I}
\end{equation}
and
\begin{equation}
I_{II}^{(-, \pm)}(\xi)= \int_{-1}^1 dw  \int_{-1+| \xi-\xi'| }^{1-| \xi-\xi'| } dv
\frac{1}{(w-\xi+ i 0)^2} \frac{1}{(v \pm \xi' \mp i 0)} H(w,v,\xi).
\label{I_II}
\end{equation}
In Eqs.~(\ref{I_I}), (\ref{I_II}) we adopt the convention that the first sign in the
$(\pm, \pm)$
index of a quantity
corresponds to the sign in the
$w \pm \xi$
denominator while the second sign in the index corresponds to the sign in the
$v \pm \xi'$
denominator. The real and imaginary parts of the integral convolutions
(\ref{I_I}), (\ref{I_II}) presented in Ref.~\cite{Lansberg:2011aa} contain
some misprints and sign errors. The revised expressions are summarized
in   App.~\ref{App_Conv_Re_Im}.

\subsubsection{Backward vector meson electroproduction}
\label{SubSec_Bkw_meson_formal}
\mbox

In this subsection, using generally the same system of notations as in  Sec.~\ref{SubSec_Ampl_Pkw_PS_meson},
we present the leading twist leading order in $\alpha_s$
amplitude of backward vector meson hard electroproduction  reaction
\begin{equation}
\gamma^{*}(q, \lambda_\gamma) + N(p_N,s_B) \to N(p'_N,s'_N) + V( p_V, \lambda_{V}  ).
\label{reaction_amp}
\end{equation}
within the collinear factorized description in terms of nucleon DAs and
nucleon-to-vector meson TDAs (\ref{VN_TDAs_param}).

The corresponding helicity amplitudes involve $6$ independent tensor structures
\begin{equation}
{\mathcal{M}}_{s_N s'_N}^{\lambda_\gamma \lambda_V}=
{\mathcal{C}}_V
\frac{1}{Q^4}
\sum_{k=1}^6
{\mathcal{S}}_{s_N s'_N}^{(k) \, \lambda_\gamma \lambda_V} {\mathcal{I}}^{(k)} (\xi, \, \Delta^2),
\label{Hel_ampl_def}
\end{equation}
where the overall normalization constant ${\mathcal{C}}_V$ reads
\begin{equation}
{\mathcal{C}}_V=-i \frac{(4 \pi \alpha_s)^2 \sqrt{4 \pi \alpha_{em}} f_N m_N}{54}.
\label{Def_C_V}
\end{equation}
There turns out to be $2$ tensor structures independent of $\Delta_T$:
\begin{eqnarray}
  &&
{\mathcal{S}}_{s_N s'_N}^{(1) \, \lambda_\gamma \lambda_V}= \bar{U}(p'_N,s'_N) \hat{\mathcal{E}}(q,\lambda_\gamma) \hat{\mathcal{E}}^{*}(p_V, \lambda_V) U(p_N,s_N);
\nonumber \\
  &&
{\mathcal{S}}_{s_N s'_N}^{(2) \, \lambda_\gamma \lambda_V}= m_N ({\mathcal{E}}^{*}(p_V, \lambda_V) \cdot n) \bar{U}(p'_N,s'_N) \hat{\mathcal{E}}(q,\lambda_\gamma)  U(p_N,s_N),
\end{eqnarray}
and $4$
$\Delta_T$-dependent tensor structures:
\begin{eqnarray}
  &&
{\mathcal{S}}_{s_N s'_N}^{(3) \, \lambda_\gamma \lambda_V}= \frac{1}{m_N} ({\mathcal{E}}^{*}(p_V, \lambda_V) \cdot \Delta_T) \, \bar{U}(p'_N,s'_N) \hat{\mathcal{E}}(q,\lambda_\gamma)  U(p_N,s_N);
\nonumber \\   &&
{\mathcal{S}}_{s_N s'_N}^{(4) \, \lambda_\gamma \lambda_V}=\frac{1}{m_N^2} ({\mathcal{E}}^{*}(p_V, \lambda_V) \cdot \Delta_T) \, \bar{U}(p'_N,s'_N) \hat{\mathcal{E}}(q,\lambda_\gamma) \hat{\Delta}_T U(p_N,s_N);
\nonumber \\   &&
{\mathcal{S}}_{s_N s'_N}^{(5) \, \lambda_\gamma \lambda_V}=\frac{1}{m_N}   \, \bar{U}(p'_N,s'_N) \hat{\mathcal{E}}(q,\lambda_\gamma)
\hat{{\mathcal{E}}}^{*}(p_V, \lambda_V)
\hat{\Delta}_T U(p_N,s_N);
\nonumber \\   &&
{\mathcal{S}}_{s_N s'_N}^{(6) \, \lambda_\gamma \lambda_V}=   ({\mathcal{E}}^{*}(p_V, \lambda_V) \cdot n) \, \bar{U}(p'_N,s'_N) \hat{\mathcal{E}}(q,\lambda_\gamma)
\hat{{\mathcal{E}}}^{*}(p_V, \lambda_V)
\hat{\Delta}_T U(p_N,s_N).
\end{eqnarray}
Here
$ \mathcal{E}(q, \lambda_\gamma)$
stands for the polarization vector of the incoming virtual photon and
${\mathcal{E}}^{*}(p_V, \lambda_V)$
is the polarization vector of the outgoing vector meson.

To the leading order in
$\alpha_s$,
within the collinear factorized description in terms of nucleon DAs and $VN$ TDAs, the amplitude of the
reaction
(\ref{reaction_amp})
can be computed from the same $21$ diagrams as listed in Table~\ref{Table_Bkw_pion}.

For definiteness we take $V$ to be a neutral vector meson   $\omega$, $\rho^0$ or $\phi$
and present the results for $\gamma^{*} N^p \to N^p V^0$ channel.
We adopt the common notations for the integral convolutions (\ref{Def_I_k_convolutions})
$\mathcal{I}^{(k)}(\xi, \Delta^2)$, $k=1, \, \ldots,\, 6$.
The explicit expressions for the corresponding coefficients  $T_{\alpha}^{(k)} \equiv  D_\alpha \times N_\alpha^{(k)}$ (no summation over $\alpha$ assumed)
are presented in Table~\ref{Table_Bkw_vector}.
The singular kernels
$D_\alpha$ turn out to be the same as in the pseudoscalar meson case and
$N_\alpha^{(k)} \equiv N_\alpha^{(k)}(x_1,x_2,x_3,\xi, \Delta^2; \, y_1, y_2,y_3)$
stand for the appropriate combinations of $VN$ TDAs and nucleon DAs $V^p$, $A^p$ and $T^p$ arising in the numerators. For the arguments of $VN$ TDAs and nucleon DAs
we adopt the shortened notations analogous to Eq.~(\ref{Not_arg_short}).
The
$\alpha=1, \ldots, 14$
index refers to the diagram number and the index
$k=1, \ldots, 6$
runs for the contributions into the $6$ invariant amplitudes of Eq.~(\ref{Hel_ampl_def}).
The $7$ diagrams with $\alpha=15,\ldots, 21$
that differ from those with $\alpha=1,\ldots, 7$
by the permutation of the $u$-quark
lines $1$ and $2$ and give the same result are not drawn.

The result for
$\gamma^{*} N^p \rightarrow \rho^+ N^n$ channel can be read from the same
Table~\ref{Table_Bkw_vector}
changing
\begin{eqnarray}
  &&
Q_u \rightarrow Q_d; \ \ \ Q_d \rightarrow Q_u; \\
  &&
V^p, \, A^p, \, T^p \, \rightarrow \, V^n, \, A^n, \, T^n \nonumber  \;  \equiv - V^p, \, -A^p, \, -T^p\,;
\end{eqnarray}
and substituting $ddu$ proton-to-$\rho^+$ TDAs.

 \begin{longtable}{|c|p{4.1cm} |c|}
\caption{14 of the 21 diagrams contributing to the hard-scattering amplitude with
their associated coefficient $T_\alpha^{(k)} \equiv D_\alpha \times N_\alpha^{(k)}$ (no summation over $\alpha$ assumed).    The $7$ diagrams with $\alpha=15,\ldots, 21$
that differ from those with $\alpha=1,\ldots, 7$
by the permutation of the $u$-quark
lines $1$ and $2$ and give the same result are not drawn.
The crosses represent the virtual-photon vertex.}
\label{Table_Bkw_vector} \\
\hline
$\alpha$ &  \qquad \quad Diagram  & Numerators \\
        &  \qquad \qquad  $D_\alpha$  &  \\
        \nopagebreak
\hline
1 & \raisebox{-0.0cm}
{\includegraphics[height=1.5cm,clip=true]{diag-01.eps}}
$\frac{Q_u (2\xi)^2}{(2\xi-x_{1}-i0)^2(x_{3}-i0)(1-y_{1})^2y_{3}}$ &
\begin{tabular}{p{0.85cm}|p{8.0cm}}
 $N_\alpha^{(1)}$ & $
-\left(V^p-A^p\right)
   \left(V_{1 {\cal E}}^{V N}+A_{1 {\cal E}}^{V N}\right)
   +
   2 T^{p} \left(T_{1 {\cal E}}^{V N}+T_{2 {\cal E}}^{V N}\right);
$ \\ 
 $N_\alpha^{(2)}$
 & $ -\left(V^p-A^p\right)
   \left(V_{1 n}^{V N}+A_{1 n}^{V N}\right)
   +
   4 T^{p}  \left( T_{1 n}^{V N} + \frac{\Delta_T^2 }{2 m_N^2} T_{4 n}^{V N} \right);
$
\\ 
\raisebox{-0.2cm}{  $N_\alpha^{(3)}$ } &
$
-\left(V^p-A^p\right)
   \left(V_{1 T}^{V N}+A_{1 T}^{V N}+V_{2 {\cal E}}^{V N}+A_{2 {\cal E}}^{V N}\right)
  $
\\
    &  $ +
   4 T^{p}  \left( T_{1 T}^{V N}+T_{3 {\cal E}}^{V N} + \frac{\Delta_T^2  }{2 m_N^2} T_{4T}^{V N} \right);$  \\ 
  $N_\alpha^{(4)}$ &  $
-\left(V^p-A^p\right)
   \left(V_{2 T}^{V N}+A_{2 T}^{V N} \right)
   +
   2 T^{p}  \left( T_{2 T}^{V N}+T_{3 T}^{V N}   \right);
$ \\ 
 $N_\alpha^{(5)}$  &  $
\left(V^p-A^p\right)
   \left(V_{2 {\cal E}}^{V N}+A_{2 {\cal E}}^{V N} \right)
   -
   2 T^{p}  \left( T_{3 {\cal E}}^{V N}-T_{4 {\cal E}}^{V N}   \right);
$ \\ 
 $N_\alpha^{(6)}$  &  $
-\left(V^p-A^p\right)
   \left(V_{2 n}^{V N}+A_{2 n}^{V N} \right)
   +
   2 T^{p}  \left( T_{2 n}^{V N}+T_{3 n}^{V N}   \right);
$ \\
\end{tabular} \\
\hline
\hline
2 & \raisebox{-0.0cm}
{\includegraphics[height=1.5cm,clip=true]{diag-02.eps}}
 &
\begin{tabular}{p{0.85cm}|p{8.0cm}}
 $N_\alpha^{(1)}$ &  
  $0;$ \\ 
 $N_\alpha^{(2)}$ &  
 $0;$ \\ 
 $N_\alpha^{(3)}$ &  
 $0;$ \\ 
 $N_\alpha^{(4)}$ &  
 $0;$ \\ 
 $N_\alpha^{(5)}$ &  
 $0;$ \\ 
 $N_\alpha^{(6)}$ &  
 $0;$ \\ 
\end{tabular} \\
\hline
\hline
3 & \raisebox{-0.0cm}
{\includegraphics[height=1.5cm,clip=true]{diag-03.eps}}
$\frac{Q_u (2 \xi)^2)}{(x_{1}-i0) (2\xi-x_{2}-i0)(x_{3}-i0)y_{1}(1-y_{1})y_{3}}$
 &
\begin{tabular}{p{0.85cm}|p{8.0cm}}
 $N_\alpha^{(1)}$ &  $-2 T^{p} \left(T_{1 {\cal E}}^{V N}+T_{2 {\cal E}}^{V N}\right);$ \\ 
 $N_\alpha^{(2)}$ & $-4 T^{p} \left(T_{1 n}^{V N}+\frac{\Delta_T^2}{2m_N^2}T_{4 n}^{V N}\right);$ \\ 
 $N_\alpha^{(3)}$ & $-4 T^{p} \left(T_{1 T}^{V N}+T_{3 {\cal E}}^{V N}+\frac{\Delta_T^2}{2m_N^2}T_{4 T}^{V N}\right);$ \\ 
 $N_\alpha^{(4)}$ & $-2 T^{p} \left(T_{2 T}^{V N}+T_{3 T}^{V N}\right);$ \\ 
 $N_\alpha^{(5)}$ & $2 T^{p} \left(T_{3 {\cal E}}^{V N}-T_{4 {\cal E}}^{V N}\right);$ \\ 
 $N_\alpha^{(6)}$ & $2 T^{p} \left(T_{2 n}^{V N}+T_{3 n}^{V N}\right);$ \\ 
\end{tabular} \\
\hline
\hline
 4 &  \raisebox{-0.0cm}
{\includegraphics[height=1.5cm,clip=true]{diag-04.eps}}
$\frac{Q_u (2\xi)^2}{(x_{1}-i0)
(2\xi-x_{3}-i0)(x_{3}-i0)y_{1}(1-y_{1})y_{3}}$
 &
\begin{tabular}{p{0.85cm}|p{8.0cm}}
 $N_\alpha^{(1)}$ &  $-\left(V^{p}-A^{p}\right) \left(V_{1 {\cal E} }^{V N}+A_{1 {\cal E}}^{V N}\right);$ \\ 
 $N_\alpha^{(2)}$ & $-\left(V^{p}-A^{p}\right) \left(V_{1 n }^{V N}+A_{1 n}^{V N}\right);$ \\ 
 $N_\alpha^{(3)}$ & $-\left(V^{p}-A^{p}\right) \left(V_{1 T }^{V N}+A_{1 T}^{V N}+ V_{2 {\cal E} }^{V N}+A_{2 {\cal E}}^{V N}\right);$ \\ 
 $N_\alpha^{(4)}$ & $-\left(V^{p}-A^{p}\right) \left(V_{2 T }^{V N}+A_{2 T}^{V N}\right);$ \\ 
 $N_\alpha^{(5)}$ & $\left(V^{p}-A^{p}\right) \left(V_{2 {\cal E} }^{V N}+A_{2 {\cal E}}^{V N}\right);$ \\ 
 $N_\alpha^{(6)}$ & $\left(V^{p}-A^{p}\right) \left(V_{2 n }^{V N}+A_{2 n}^{V N}\right);$  \\ 
\end{tabular} \\
\hline
\hline
5 & \raisebox{-0.0cm}
{\includegraphics[height=1.5cm,clip=true]{diag-05.eps}}
$\frac{Q_u (2\xi)^2}{(x_{1}-i0)(2\xi-x_{3}-i0)(x_{3}-i0)y_{1}(1-y_{2})y_{3}}$
 &
\begin{tabular}{p{0.85cm}|p{8.0cm}}
 $N_\alpha^{(1)}$ &  $\left(V^{p}+A^{p}\right) \left(V_{1 {\cal E} }^{V N}-A_{1 {\cal E}}^{V N}\right);$ \\ 
 $N_\alpha^{(2)}$ & $\left(V^{p}+A^{p}\right) \left(V_{1 n }^{V N}-A_{1 n}^{V N}\right);$ \\ 
 $N_\alpha^{(3)}$ & $\left(V^{p}+A^{p}\right) \left(V_{1T }^{V N}-A_{1 T}+V_{2 {\cal E} }^{V N}-A_{2 {\cal E}}^{V N}\right);$  \\ 
 $N_\alpha^{(4)}$ & $\left(V^{p}+A^{p}\right) \left(V_{2T }^{V N}-A_{2 T} \right);$ \\ 
 $N_\alpha^{(5)}$ & $-\left(V^{p}+A^{p}\right) \left(V_{2 {\cal E} }^{V N}-A_{2 {\cal E}}^{V N}\right);$ \\ 
 $N_\alpha^{(6)}$ & $-\left(V^{p}+A^{p}\right) \left(V_{2 n }^{V N}-A_{2 n}^{V N}\right);$ \\ 
\end{tabular} \\
\hline
\hline
6 & \raisebox{-0.0cm}
{\includegraphics[height=1.5cm,clip=true]{diag-06.eps}}
 &
\begin{tabular}{p{0.85cm}|p{8.0cm}}
 $N_\alpha^{(1)}$ &  
 $0;$ \\ 
 $N_\alpha^{(2)}$ & 
 $0;$ \\ 
 $N_\alpha^{(3)}$ & 
 $0;$ \\ 
 $N_\alpha^{(4)}$ & 
 $0;$ \\ 
 $N_\alpha^{(5)}$ & 
 $0;$ \\ 
 $N_\alpha^{(6)}$ & 
 $0;$ \\ 
\end{tabular} \\
\hline
\hline
7 & \raisebox{-0.0cm}
{\includegraphics[height=1.5cm,clip=true]{diag-07.eps}}
$\frac{Q_d(2\xi)^2}{(x_{1}-i0)(2\xi-x_{3}-i0)^2y_{1}(1-y_3)^2}$
 &
\begin{tabular}{p{0.85cm}|p{8.0cm}}
 $N_\alpha^{(1)}$ &    $-2 \left(V^{p} V_{1 {\cal E}}^{V N}-A^{p}  A_{1 {\cal E}}^{V N} \right);$ \\ 
 $N_\alpha^{(2)}$ & $-2 \left(V^{p} V_{1 n}^{V N}-A^{p}  A_{1 n}^{V N} \right);$ \\ 
 $N_\alpha^{(3)}$ &  $-2 \left(V^{p} (V_{1T}^{VN}+V_{2 {\cal E}}^{V N})-A^{p}  (A_{1T}^{VN}+A_{2 {\cal E}}^{V N}) \right);$ \\ 
 $N_\alpha^{(4)}$ & $-2 \left(V^{p} V_{2 T}^{V N}-A^{p}  A_{2T}^{V N} \right);$ \\ 
 $N_\alpha^{(5)}$ & $2 \left(V^{p} V_{2 {\cal E}}^{V N}-A^{p}  A_{2 {\cal E}}^{V N} \right);$  \\ 
 $N_\alpha^{(6)}$ &  $2 \left(V^{p} V_{2 n}^{V N}-A^{p}  A_{2 n}^{V N} \right);$\\ 
\end{tabular} \\
\hline
\hline
8 & \raisebox{-0.0cm}
{\includegraphics[height=1.5cm,clip=true]{diag-08.eps}}
 &
\begin{tabular}{p{0.85cm}|p{8.0cm}}
 $N_\alpha^{(1)}$ &  
 $0;$ \\ 
 $N_\alpha^{(2)}$ & 
 $0;$ \\ 
 $N_\alpha^{(3)}$ & 
 $0;$ \\ 
 $N_\alpha^{(4)}$ & 
 $0;$ \\ 
 $N_\alpha^{(5)}$ & 
 $0;$ \\ 
 $N_\alpha^{(6)}$ & 
 $0;$ \\ 
\end{tabular} \\
\hline
\hline
9 & \raisebox{-0.0cm}
{\includegraphics[height=1.5cm,clip=true]{diag-09.eps}}
 \ \ \ \ \ \ \ \  \ \ \ \ \
$\frac{Q_u (2\xi)^2}{(2\xi-x_{1}-i0)^2(x_{2}-i0)(1-y_{1})^2y_{2}}$ &
\begin{tabular}{p{0.85cm}|p{8.0cm}}
 $N_\alpha^{(1)}$ & $
-\left(V^p-A^p\right)
   \left(V_{1 {\cal E}}^{V N}+A_{1 {\cal E}}^{V N}\right)
   +
   2 T^{p} \left(T_{1 {\cal E}}^{V N}+T_{2 {\cal E}}^{V N}\right);
$ \\ 
 $N_\alpha^{(2)}$
 & $ -\left(V^p-A^p\right)
   \left(V_{1 n}^{V N}+A_{1 n}^{V N}\right)
   +
   4 T^{p}  \left( T_{1 n}^{V N} + \frac{\Delta_T^2 }{2 m_N^2} T_{4 n}^{V N} \right);
$
\\ 
\raisebox{-0.2cm}{  $N_\alpha^{(3)}$ } &
$
-\left(V^p-A^p\right)
   \left(V_{1 T}^{V N}+A_{1 T}^{V N}+V_{2 {\cal E}}^{V N}+A_{2 {\cal E}}^{V N}\right)
  $
\\
    &  $ +
   4 T^{p}  \left( T_{1 T}^{V N}+T_{3 {\cal E}}^{V N} + \frac{\Delta_T^2  }{2 m_N^2} T_{4T}^{V N} \right);$  \\ 
  $N_\alpha^{(4)}$ &  $
-\left(V^p-A^p\right)
   \left(V_{2 T}^{V N}+A_{2 T}^{V N} \right)
   +
   2 T^{p}  \left( T_{2 T}^{V N}+T_{3 T}^{V N}   \right);
$ \\ 
 $N_\alpha^{(5)}$  &  $
\left(V^p-A^p\right)
   \left(V_{2 {\cal E}}^{V N}+A_{2 {\cal E}}^{V N} \right)
   -
   2 T^{p}  \left( T_{3 {\cal E}}^{V N}-T_{4 {\cal E}}^{V N}   \right);
$ \\ 
 $N_\alpha^{(6)}$  &  $
-\left(V^p-A^p\right)
   \left(V_{2 n}^{V N}+A_{2 n}^{V N} \right)
   +
   2 T^{p}  \left( T_{2 n}^{V N}+T_{3 n}^{V N}   \right);
$ \\
\end{tabular} \\
\hline
\hline
10 & \raisebox{-0.0cm}
{\includegraphics[height=1.5cm,clip=true]{diag-10.eps}}
 \ \ \ \ \ \ \ \  \ \ \ \ \
$\frac{Q_u (2\xi)^2}{(x_{1}-i0)(2\xi-x_{2}-i0)^2y_{1}(1-y_{2})^2}$ &
\begin{tabular}{p{0.85cm}|p{8.0cm}}
 $N_\alpha^{(1)}$ & $
-\left(V^p-A^p\right)
   \left(V_{1 {\cal E}}^{V N}+A_{1 {\cal E}}^{V N}\right)
   +
   2 T^{p} \left(T_{1 {\cal E}}^{V N}+T_{2 {\cal E}}^{V N}\right);
$ \\ 
 $N_\alpha^{(2)}$
 & $ -\left(V^p-A^p\right)
   \left(V_{1 n}^{V N}+A_{1 n}^{V N}\right)
   +
   4 T^{p}  \left( T_{1 n}^{V N} + \frac{\Delta_T^2 }{2 m_N^2} T_{4 n}^{V N} \right);
$
\\ 
\raisebox{-0.2cm}{  $N_\alpha^{(3)}$ } &
$
-\left(V^p-A^p\right)
   \left(V_{1 T}^{V N}+A_{1 T}^{V N}+V_{2 {\cal E}}^{V N}+A_{2 {\cal E}}^{V N}\right)
  $
\\
    &  $ +
   4 T^{p}  \left( T_{1 T}^{V N}+T_{3 {\cal E}}^{V N} + \frac{\Delta_T^2  }{2 m_N^2} T_{4T}^{V N} \right);$  \\ 
  $N_\alpha^{(4)}$ &  $
-\left(V^p-A^p\right)
   \left(V_{2 T}^{V N}+A_{2 T}^{V N} \right)
   +
   2 T^{p}  \left( T_{2 T}^{V N}+T_{3 T}^{V N}   \right);
$ \\ 
 $N_\alpha^{(5)}$  &  $
\left(V^p-A^p\right)
   \left(V_{2 {\cal E}}^{V N}+A_{2 {\cal E}}^{V N} \right)
   -
   2 T^{p}  \left( T_{3 {\cal E}}^{V N}-T_{4 {\cal E}}^{V N}   \right);
$ \\ 
 $N_\alpha^{(6)}$  &  $
-\left(V^p-A^p\right)
   \left(V_{2 n}^{V N}+A_{2 n}^{V N} \right)
   +
   2 T^{p}  \left( T_{2 n}^{V N}+T_{3 n}^{V N}   \right);
$ \\
\end{tabular} \\
\hline
\hline
11 & \raisebox{-0.0cm}
{\includegraphics[height=1.5cm,clip=true]{diag-11.eps}}
 &
\begin{tabular}{p{0.85cm}|p{8.0cm}}
 $N_\alpha^{(1)}$ &  $0;$ \\ 
 $N_\alpha^{(2)}$ &  $0;$ \\ 
 $N_\alpha^{(3)}$ &  $0;$ \\ 
 $N_\alpha^{(4)}$ &  $0;$ \\ 
 $N_\alpha^{(5)}$ &  $0;$ \\ 
 $N_\alpha^{(6)}$ &  $0;$ \\ 
\end{tabular} \\
\hline
\hline
12 & \raisebox{-0.0cm}
{\includegraphics[height=1.5cm,clip=true]{diag-12.eps}}
$\frac{Q_d (2 \xi)^2}{(x_{1}-i0)(x_{2}-i0)(2\xi-x_{3}-i0)y_{1}(1-y_{2})y_{2}}$
 &
\begin{tabular}{p{0.85cm}|p{8.0cm}}
 $N_\alpha^{(1)}$ &  $\left(V^{p}+A^{p}\right) \left( V_{1 {\cal E}}^{V N}-A_{1 {\cal E}}^{V N}\right);$
 \\ 
 $N_\alpha^{(2)}$ &  $\left(V^{p}+A^{p}\right) \left( V_{1 {n}}^{V N}-A_{1 {n}}^{V N}\right);$  \\ 
 $N_\alpha^{(3)}$ & $\left(V^{p}+A^{p}\right) \left( V_{1 {T}}^{V N}-A_{1 {T}}^{V N}+V_{2 {\cal E}}^{V N}-A_{2 {\cal E}}^{V N}\right);$  \\ 
 $N_\alpha^{(4)}$ & $\left(V^{p}+A^{p}\right) \left( V_{2 {T}}^{V N}-A_{2 {T}}^{V N}\right);$ \\ 
 $N_\alpha^{(5)}$ & $-\left(V^{p}+A^{p}\right) \left( V_{2 {\cal E}}^{V N}-A_{2 {\cal E}}^{V N}\right);$  \\ 
 $N_\alpha^{(6)}$ & $-\left(V^{p}+A^{p}\right) \left( V_{2 n}^{V N}-A_{2 n}^{V N}\right);$  \\ 
\end{tabular} \\
\hline
\hline
13 & \raisebox{-0.0cm}
{\includegraphics[height=1.5cm,clip=true]{diag-13.eps}}
$\frac{Q_d (2 \xi)^2)}{(x_{1}-i0)(2\xi-x_{1}-i0)(x_{2}-i0)y_{1}(1-y_{2})y_{2}}$
 &
\begin{tabular}{p{0.85cm}|p{8.0cm}}
 $N_\alpha^{(1)}$ &  $2 T^{p} \left(T_{1 {\cal E}}^{V N}+T_{2 {\cal E}}^{V N}\right);$ \\ 
 $N_\alpha^{(2)}$ & $4 T^{p} \left(T_{1 n}^{V N}+\frac{\Delta_T^2}{2 m_N^2}T_{4 n}^{V N}\right);$ \\ 
 $N_\alpha^{(3)}$ & $4 T^{p} \left(T_{1 T}^{V N}+T_{3 {\cal E}}^{V N}+\frac{\Delta_T^2}{2 m_N^2}T_{4 T}^{V N}\right);$ \\ 
 $N_\alpha^{(4)}$ & $2 T^{p} \left(T_{2 T}^{V N}+T_{3 T}^{V N}\right);$ \\ 
 $N_\alpha^{(5)}$ & $-2 T^{p} \left(T_{3 {\cal E}}^{V N}-T_{4 {\cal E}}^{V N}\right);$ \\ 
 $N_\alpha^{(6)}$ & $-2 T^{p} \left(T_{2 n}^{V N}+T_{3 n}^{V N}\right);$ \\ 
\end{tabular} \\
\hline
\hline
14 &  \raisebox{-0.0cm}
{\includegraphics[height=1.5cm,clip=true]{diag-14.eps}}
$\frac{Q_d (2\xi)^2}{(x_{1}-i0)(2\xi-x_{1}-i0)(x_{2}-i0)y_{1}y_{2}(1-y_{3})}$
 &
\begin{tabular}{p{0.85cm}|p{8.0cm}}
 $N_\alpha^{(1)}$ &  $\left(V^{p}-A^{p}\right) \left(V_{1 {\cal E} }^{V N}+A_{1 {\cal E}}^{V N}\right);$ \\ 
 $N_\alpha^{(2)}$ & $\left(V^{p}-A^{p}\right) \left(V_{1 n }^{V N}+A_{1 n}^{V N}\right);$ \\ 
 $N_\alpha^{(3)}$ & $\left(V^{p}-A^{p}\right) \left(V_{1 T }^{V N}+A_{1 T}^{V N}+ V_{2 {\cal E} }^{V N}+A_{2 {\cal E}}^{V N}\right);$ \\ 
 $N_\alpha^{(4)}$ & $\left(V^{p}-A^{p}\right) \left(V_{2 T }^{V N}+A_{2 T}^{V N}\right);$ \\ 
 $N_\alpha^{(5)}$ & $-\left(V^{p}-A^{p}\right) \left(V_{2 {\cal E} }^{V N}+A_{2 {\cal E}}^{V N}\right);$ \\ 
 $N_\alpha^{(6)}$ & $-\left(V^{p}-A^{p}\right) \left(V_{2 n }^{V N}+A_{2 n}^{V N}\right);$  \\ 
\end{tabular} \\
\hline
\end{longtable}

\subsection{Cross-channel hard exclusive processes involving {{TDA}}s}
\label{SubSec_Cross_Ch_Excl_R}
\mbox

Similarly to the case of inclusive reactions, hard exclusive reactions that can be studied at hadronic facilities
provide a complementary access to
the partonic contents
of hadrons.
For example, in Ref.~\cite{Berger:2001zn}
the collinear factorization theorem was formulated for the exclusive
Drell--Yan production in
$\pi N$
collisions
$\pi N \to \ell^+ \ell^- N$
for large invariant
mass of the lepton pair $\ell^+ \ell^-$
and small invariant momentum transfer to the nucleon.
This reaction can be seen as a cross-channel counterpart
of the usual near-forward hard exclusive pion electroproduction
reaction and allows to study  polarized nucleon GPDs
$\tilde{H}$ and $\tilde{E}$.

Following the same pattern, within specific kinematics the cross-channel counterparts
of the backward DVCS process (\ref{bkw_DVCS})
and of hard exclusive backward meson leptoproduction reactions
(\ref{hard_meson_production})
can be challenged for collinear factorized description in terms of nucleon-to-photon and nucleon-to-meson TDAs. In Refs.~\cite{Pire:2004ie,Pire:2005ax,Lansberg:2007se} the corresponding collinear factorization theorems
were formulated for nucleon--antinucleon annihilation into a lepton pair in association with a photon or a light meson
 $N \bar{N} \to \ell^+ \ell^- \gamma$,  $N \bar{N} \to \ell^+ \ell^- \mathcal{M}$.
This class of reactions can be studied at the future \=PANDA facility at GSI-FAIR.
A special interest represents the situation when the lepton pair is produced at a resonance
energy of a heavy quarkonium \textit{e.g.} $N \bar{N} \to J/\psi \mathcal{M}$~\cite{Pire:2013jva}.

Finally, one may try a collinear factorized description in terms of
 meson-to-nucleon TDAs  of a different
cross-channel counterpart of the reaction (\ref{hard_meson_production}):
${\mathcal{M}} N \to \ell^+ \ell^- N$. In Ref.~\cite{Pire:2016gut} the factorized description
in terms of pion-to-nucleon TDAs
was presented for the backward $\pi N \to J/\psi N$ reaction that can be studied experimentally with the intense pion beam available at the J-PARC facility.

 \subsubsection{Nucleon--antinucleon annihilation into a lepton pair and a pion: $u$-channel and $t$-channel factorization regimes}
 \label{SubSec_PANDA_pi_gamma_star}
 \mbox

In this subsection, following Refs.~\cite{Lansberg:2007se,Lansberg:2012ha},
we present the collinear factorization framework for
nucleon--antinucleon  annihilation into a high invariant mass lepton pair
in association with a light meson
${\mathcal{M}}=\{\pi,\,\eta,\, \rho, \omega, \phi \, \ldots  \}$:
\begin{equation}
\bar{N} (p_{\bar{N}},s_{\bar{N}}) + N (p_N,s_N) \rightarrow \gamma^{*}(q,\lambda_\gamma) + {\mathcal{M}}(p_{\mathcal{M}})  \rightarrow \ell^+(k_{\ell^+},s_{\ell^+})+
\ell^-(k_{\ell^-}, s_{\ell^-})+ {\mathcal{M}}(p_{\mathcal{M}}).
\label{BarNNannihilation reaction}
\end{equation}
This analysis
identifies two similar factorization regimes for the reaction
(\ref{BarNNannihilation reaction})
corresponding to the forward and backward peaks of the meson production cross
section.

\begin{figure}[H]
\begin{center}
\includegraphics[width=0.47\textwidth]{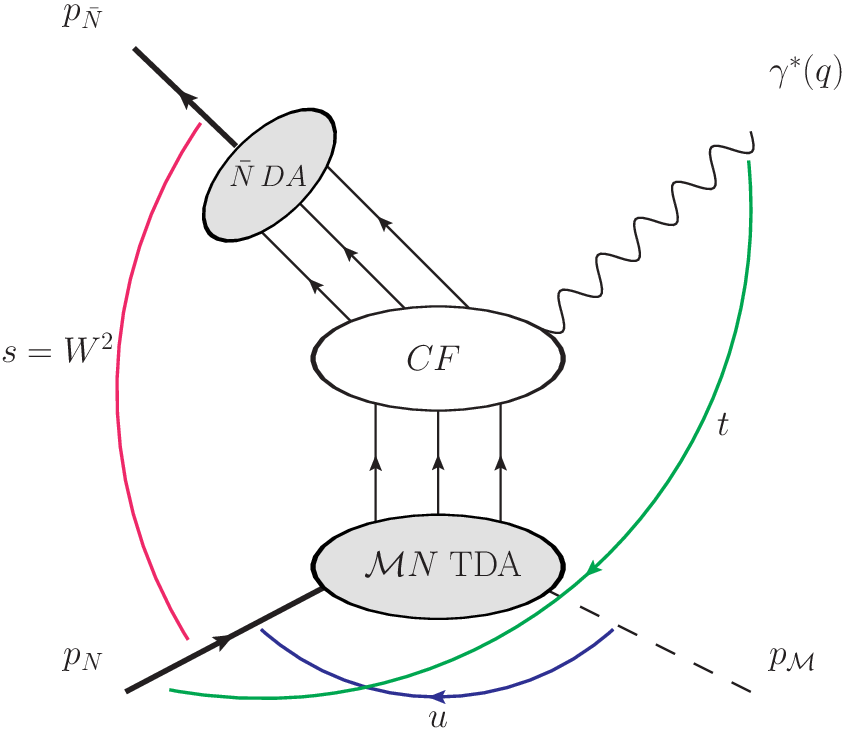} \ \ \
\includegraphics[width=0.47\textwidth]{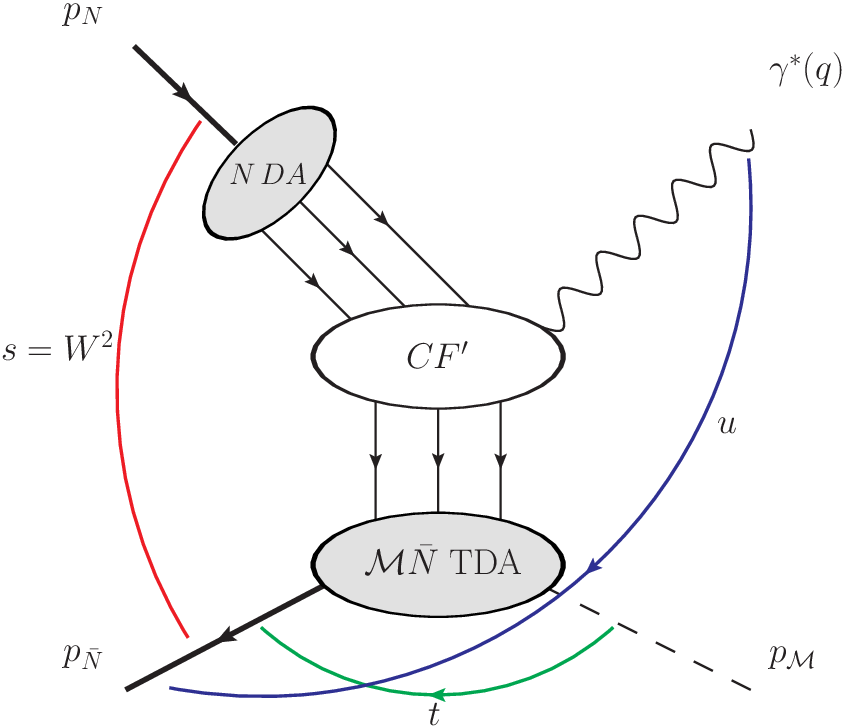}
\end{center}
     \caption{{\bf Left panel:} kinematical quantities and the collinear factorization mechanism  for
      $ \bar{N} N \to \gamma^{*} {\mathcal{M}}$  in the  near-backward  kinematical regime (large $q^2 \equiv Q^2$, large $W^2 \equiv (p_N+p_{\bar{N}})^2$ with $Q^2$ of order of $W^2$; $| u|  \sim 0$). The lower blob, denoted
  ${\mathcal{M}}N$ TDA, depicts the nucleon-to-meson~$\mathcal M$ transition
     distribution amplitude; $\bar{N}$ DA blob depicts the antinucleon distribution amplitude;
      $CF$ denotes the corresponding hard subprocess amplitudes (coefficient functions).
      {\bf Right panel:} kinematical quantities and the collinear factorization mechanism  for
      $\bar{N} N \to \gamma^{*} {\mathcal{M}}$  in the  near-forward  kinematical regime (large $q^2 \equiv Q^2$, large $W^2=(p_N+p_{\bar{N}})^2$  with $Q^2$ of order of $W^2$; $| t|  \sim 0$). The lower blob, denoted
  ${\mathcal{M}} \bar{N}$ TDA, depicts the antinucleon-to-meson~$\mathcal M$ transition
     distribution amplitude; $N$ DA blob depicts the nucleon distribution amplitude;
      $CF$ denotes the corresponding hard subprocess amplitudes (coefficient functions).}
\label{Fig_Kinematics_TDAs_PANDA}
\end{figure}

Keeping in mind the
\=PANDA layout it is natural to
choose the positive direction of the $z$-axis in  the direction
the antinucleon is moving in the
$\bar{N} N$ CMS frame. Within such convention we define the
near-backward and near-forward kinematical regimes.
\begin{itemize}
\item The kinematical regime in which the absolute value of the $u$-channel momentum transfer squared
$ u  \equiv  (p_N-p_{\mathcal{M}})^2 $
is   small   corresponds to a meson moving in the direction of the initial nucleon that is the
\emph{backward}
direction.
\item Analogously, the kinematical regime in which the absolute value of the $t$-channel
    momentum transfer squared
$ t  \equiv  (p_{\bar{N}}-p_{\mathcal{M}})^2 $
is small, corresponds to a meson
moving in the direction of the initial antinucleon that is the
\emph{forward}
direction.
\end{itemize}
The collinear factorization theorems formulated for
the near-backward  and near-forward factorization regimes of the reaction
(\ref{BarNNannihilation reaction})
are  schematically depicted on
 Fig.~\ref{Fig_Kinematics_TDAs_PANDA}.

For the near-backward kinematics the collinear factorization theorem
(see the left panel of
 Fig.~\ref{Fig_Kinematics_TDAs_PANDA})
is valid once
$s=(p_N+p_{\bar{N}})^2 \equiv W^2$
and the invariant mass of the lepton pair
$q^2 \equiv Q^2$ with $W^2 \sim Q^2$
are large; $| u| $
is small as compared to
$Q^2$
and
$W^2$.

Below we provide a detailed review of the corresponding kinematics that
is largely analogous to the analysis of  Sec.~\ref{SubSec_Kinematics} for
the leptoproduction reaction.
The $u$-channel light-cone vectors
$p^u$
and
$n^u$
 (${p^u}^2={n^u}^2=0$; $2 p^u \cdot n^u =1$) are defined with respect to the $z$-axis chosen along the colliding nucleon--antinucleon.
$P^u$ and $\Delta^u$
stand for the average momentum and $u$-channel momentum transfer
\begin{equation}
P^u=\frac{1}{2}(p_{\mathcal{M}}+p_N); \ \ \ \Delta^u=p_{\mathcal{M}}-p_N; \ \ \;{\Delta^u}^2 \equiv u;
\end{equation}
${\Delta^u}_T$ is the transverse component of
$\Delta^u$ ($ {\Delta^u_T}^2 \le 0$). We define the skewness
variable
\begin{equation}
\xi^u=- \frac{\Delta^u \cdot n^u}{2 P^u \cdot n^u},
\label{def_xiu}
\end{equation}
that characterizes the $u$-channel longitudinal momentum transfer. The
variable $\xi^u$ is expressed through the $W^2/Q^2$ ratio, see Eq.~(\ref{Expr_xi_u}).

The following Sudakov decomposition is established for the momenta of the reaction (\ref{BarNNannihilation reaction}):
\begin{eqnarray}
  &&
p_N= (1+\xi^u) p^u + \frac{m_N^2}{1+\xi^u} n^u; \nonumber \\
  &&
p_{\bar{N}}= \frac{2m_N^2(1+\xi^u)}{W^2-2 m_N^2+W \sqrt{W^2-4m_N^2}}p^u+ \frac{W^2-2 m_N^2+W \sqrt{W^2-4m_N^2}}{2(1+\xi^u)}n^u;
\nonumber \\   &&
p_{\mathcal{M}}=(1-\xi^u)p^u+ \frac{m_{\mathcal{M}}^2-{\Delta^u_T}^2}{1-\xi^u}n^u+{\Delta^u_T};
\nonumber \\   &&
\Delta^u=-2 \xi^u p^u+ \left[ \frac{m^2-{\Delta^u_T}^2}{1-\xi^u}- \frac{m_N^2}{1+\xi^u} \right] n^u+ {\Delta^u_T} ;
\nonumber \\   &&
q = \left[ 2 \xi^u + \frac{M^2}{W^2} (1+\xi^u) + O(1/W^4) \right] p^u+
\left[
\frac{W^2-M^2}{1+\xi^u} -\frac{m^2_{\mathcal{M}}-{\Delta^u_T}^2}{1-\xi^u} + O(1/W^2)
\right] n^u-\Delta^u_T.
\label{Sudakov_PANDA_u}
\end{eqnarray}

In the nucleon rest frame (\=PANDA laboratory frame)
the light-cone vectors $p^u$
and
$n^u$
read:
\begin{equation}
p^u \Big|_{N \; {\rm rest}}=\frac{m_N}{2(1+\xi^u)} \{1,0,0,-1\}; \ \ \ n^u  \Big|_{N \; {\rm rest}}=\frac{1+\xi^u}{2m_N} \{1,0,0,1\}.
\label{Def_pn_u}
\end{equation}
The expressions for the light-cone vectors
(\ref{Def_pn_u})
in the
$N \bar{N}$
CMS can be established with the help of the appropriate boost:
\begin{equation}
p^u \Big|_{\bar{N}N \; {\rm CMS}}=\{\alpha^u,0,0,-\alpha^u \}; \ \ \ n^u \Big|_{\bar{N}N \; {\rm CMS}}=\{\beta^u,0,0, \beta^u \},
\end{equation}
where
\begin{equation}
\alpha^u=\frac{W+\sqrt{W^2-4 m_N^2}}{4 (1+\xi^u)}; \ \ \ \beta^u= \frac{\left(W-\sqrt{W^2-4 m_N^2} \right) (1+\xi^u)}{4 m_N^2}.
\label{Def_alpha_beta_u}
\end{equation}

The meson $\mathcal{M}$ scattering angle in the $N \bar{N}$ CMS for the $u$-channel factorization regime is expressed as:
\begin{equation}
\cos \theta_{\mathcal{M}}^{*}=\frac{-(1-\xi^u)\alpha^u+\frac{m^2-{\Delta^u_T}^2}{1-\xi^u} \beta^u}
{\sqrt{(-(1-\xi^u)\alpha^u+\frac{m^2-{\Delta^u_T}^2}{1-\xi^u} \beta^u)^2-{\Delta^u_T}^2}}.
\label{CosThetaPi_uregime}
\end{equation}
Note that for ${\Delta^u_T}^2=0$ indeed $\cos \theta_{\mathcal{M}}^{*}=-1$, which means backward
scattering.

We also quote some useful relations for the kinematical quantities:
\begin{eqnarray*}
  &&
{\Delta^u_T}^2= \frac{1-\xi^u}{1+\xi^u} \left(u- 2\xi^u \left[\frac{m_N^2}{1+\xi^u} -\frac{m_{\mathcal{M}}^2}{1-\xi^u} \right] \right);
\nonumber  \\
  && Q^2 \equiv q^2 =\frac{2 \xi^u}{1+\xi^u} W^2+u-3m_N^2 +\frac{4m_N^2}{1+\xi^u}+{\mathcal{O}}(1/W^2)\,.
\nonumber
\label{Kin_quant_u}
\end{eqnarray*}
This allows us to express $\xi^u$ as
\begin{equation}
\xi^u \simeq \frac{Q^2-u-m_N^2}{2W^2-Q^2+u-3 m_N^2}.
\label{Expr_xi_u}
\end{equation}
We also introduce $u_{0}$ corresponding to ${\Delta_T^u}^2=0$:
\begin{equation}
u_{0}=-\frac{2 \xi  \left(m_{\mathcal{M}}^2 (1+\xi^u)-m_N^2 (1-\xi^2)\right)}{1-{\xi^u} ^2},
\end{equation}
that is the maximal possible value of $u$ for given $\xi$.

Analogously, the collinear factorization theorem  for the near-forward  kinematics presented in the right panel of
 Fig.~\ref{Fig_Kinematics_TDAs_PANDA} is valid once
$| t| $
is small as compared to
$Q^2$
and
$W^2$.

The $t$-channel light-cone vectors
$p^t$
and
$n^t$
(${p^t}^2={n^t}^2=0$; $2 p^t \cdot n^t =1$)
are defined with respect to the $z$-axis chosen along the colliding nucleon--antinucleon.
The average momentum and $t$-channel momentum transfer are defined as
\begin{equation}
P^t=\frac{1}{2}(p_{\mathcal{M}}+p_{\bar{N}}); \ \ \ \Delta^t=p_{\mathcal{M}}-p_{\bar{N}}; \ \ \
{\Delta^t}^2 \equiv t.
\end{equation}
${\Delta^t}_T$ is the transverse component of
$\Delta^t$ ($\Delta^t_T \cdot \Delta^t_T ={\Delta^t_T}^2 \le 0$). We define the skewness
variable
\begin{equation}
\xi^t=- \frac{\Delta^t \cdot n^t}{2 P^t \cdot n^t},
\label{defxit}
\end{equation}
which characterizes the $t$-channel longitudinal momentum transfer.

This allows to establish the Sudakov decomposition of the  particles
momenta:
\begin{eqnarray}
  &&
p_{\bar{N}}= (1+\xi^t) p^t + \frac{m_N^2}{1+\xi^t} n^t; \nonumber \\   &&
p_{N}= \frac{2M^2(1+\xi^t)}{W^2-2 m_N^2+W \sqrt{W^2-4m_N^2}}p^t+ \frac{W^2-2 m_N^2+W \sqrt{W^2-4m_N^2}}{2(1+\xi^t)}n^t;
\nonumber \\   &&
p_{\mathcal{M}}=(1-\xi^t)p^t+ \frac{m_{\mathcal{M}}^2-{\Delta^t_T}^2}{1-\xi^t}n^t+{\Delta^t_T};
 \\   &&
\Delta^t=-2 \xi^t p^t+ \left[ \frac{{\mathcal{M}}^2-{\Delta^t_T}^2}{1-\xi^t}- \frac{m_N^2}{1+\xi^t} \right] n^t+ {\Delta^t_T} ;
\nonumber \\   &&
q = \left[ 2 \xi^t + \frac{m_N^2}{W^2} (1+\xi^t) + O(1/W^4) \right] p^t+
\left[
\frac{W^2-m_N^2}{1+\xi^t} -\frac{m_{\mathcal{M}}^2-{\Delta^t_T}^2}{1-\xi^t} + {\mathcal{O}}(1/W^2)
\right] n^t-\Delta^t_T. \nonumber
\label{Sudakov_PANDA_t}
\end{eqnarray}

In the antinucleon rest frame
the light-cone vectors $p^t$
and
$n^t$
are:
\begin{equation}
p^t \Big|_{\bar{N} \; {\rm rest}}=\frac{m_N}{2(1+\xi^t)} \{1,0,0,1\}; \ \ \ n^t \Big|_{\bar{N} \; {\rm rest}}=\frac{1+\xi^t}{2m_N} \{1,0,0,-1\}.
\end{equation}
The explicit expressions for the light-cone vectors $p^t$ and $n^t$ in the $\bar{N}N$ CMS read:
\begin{equation}
p^t Big|_{\bar{N}N \; {\rm CMS}}=\{\alpha^t,0,0, \alpha^t \}; \ \ \ n^t \Big|_{\bar{N}N \; {\rm CMS}}=\{\beta^t,0,0, -\beta^t \},
\end{equation}
where $\alpha^t$ and $\beta^t$ are defined  as
\begin{equation}
\alpha^t=\frac{W+\sqrt{W^2-4 m_N^2}}{4 (1+\xi^t)}; \ \ \ \beta^t= \frac{\left(W-\sqrt{W^2-4 m_N^2} \right) (1+\xi^t)}{4 m_N^2}.
\label{Def_alpha_beta_t}
\end{equation}

The meson $\mathcal{M}$ scattering angle in the $N \bar{N}$ CMS for the $t$-channel factorization regime is  expressed as:
\begin{equation}
\cos \theta_{\mathcal{M}}^{*}=\frac{ (1-\xi^t)\alpha^t-\frac{m^2-{\Delta^t_T}^2}{1-\xi^t} \beta^t}
{\sqrt{((1-\xi^t)\alpha^t-\frac{m^2-{\Delta^t_T}^2}{1-\xi^t} \beta^t)^2-{\Delta^t_T}^2}}.
\label{Cos_theta_t}
\end{equation}
Note that for ${\Delta^t_T}^2=0$, indeed, $\cos \theta_{\mathcal{M}}^{*}=1$,  which means forward
scattering.

 We also work out the relations:
\begin{eqnarray}
  &&
{\Delta^t_T}^2= \frac{1-\xi^t}{1+\xi^t} \left(t- 2\xi^t \left[\frac{m_N^2}{1+\xi^t} -\frac{m_{\mathcal{M}}^2}{1-\xi^t} \right] \right);
\nonumber  \\   &&
Q^2 \equiv q^2 =\frac{2 \xi^t}{1+\xi^t} W^2+t-3m_N^2 +\frac{4m_N^2}{1+\xi^t}+{\mathcal{O}}(1/W^2);
\label{Kin_quant_t}
\end{eqnarray}
and express $\xi^t$ as:
\begin{equation}
\xi^t \simeq \frac{Q^2-t-m_N^2}{2W^2-Q^2+t-3 m_N^2}.
\end{equation}

The simplest example of the reaction
(\ref{BarNNannihilation reaction})
is the nucleon--antinucleon annihilation into a lepton pair
in association with a pion $N \bar{N} \rightarrow \gamma^{*} \pi \to \ell^+ \ell^- \pi$.
Within the $u$-channel factorized description in terms of $\pi N$ TDAs (and antinucleon DAs)  to the leading order in
$\alpha_s$,
the amplitude of  the hard $N \bar{N} \rightarrow \gamma^{*} \pi$ subprocess
$\mathcal{M}^{\lambda_\gamma}_{s_N s_{\bar{N}}}$
is expressed as
\begin{equation}
\mathcal{M}^{\lambda_\gamma}_{s_N s_{\bar{N}}}=
\mathcal{C}_\pi
\frac{1}{Q^4}
\Bigl[
\mathcal{S}_{s_N s_{\bar{N}}}^{(1)\, \lambda_\gamma  }
\mathcal{J}^{(1)}(\xi, \Delta^2)
-
\mathcal{S}_{s_N s_{\bar{N}}}^{(2) \, \lambda_\gamma}
\mathcal{J}^{(2)}(\xi, \Delta^2)
\Bigr],
\label{Def_ampl_MPANDA}
\end{equation}
where
$\mathcal{C}_\pi$ is the overall normalization constant
(\ref{Def_C_PiN}).
The spin structures
$\mathcal{S}^{(k) \, \lambda_\gamma}_{s_N s_{\bar{N}}}$, $k=1,\,2$
are defined as
\begin{eqnarray}
  &&
\mathcal{S}^{(1) \, \lambda_\gamma}_{s_N s_{\bar{N}}} \equiv
\bar{V}(p_{\bar{N}},s_{\bar{N}}) \hat{\mathcal{E}}^{*}_\gamma(q,\lambda_\gamma) \gamma_5 U(p_N,s_N);
\nonumber \\   &&
\mathcal{S}^{(2) \, \lambda_\gamma}_{s_N s_{\bar{N}}} \equiv
\frac{1}{m_N}
\bar{V}(p_{\bar{N}},s_{\bar{N}}) \hat{\mathcal{E}}^{*}_\gamma(q, \lambda_\gamma) \hat{\Delta}_T \gamma_5 U(p_N,s_N),
\label{Def_S_PANDA_process}
\end{eqnarray}
where
${\mathcal{E}}_\gamma(q,\lambda_\gamma)$
stands for the polarization vector of the virtual photon and $U$ and $V$ are the
nucleon Dirac spinors.
$\mathcal{J}^{(k)}$ $k=1,2$
denote the convolution integrals of
$\pi N$
TDAs and antinucleon DAs with the hard scattering kernels computed from the set of
$21$
relevant scattering diagrams (see~\cite{Lansberg:2007ec}):
\begin{equation}
{\mathcal{J}}^{(k)}
(\xi,\Delta^2) =
{\int^{1+\xi}_{-1+\xi} }\! \! \!d_3x  \; \delta \left( \sum_{j=1}^3 x_j-2\xi \right)
{\int^{1}_{0} } \! \! \! d_3y \; \delta \left( \sum_{l=1}^3 y_l-1 \right) \;
{\Biggl(2\sum_{\alpha=1}^{7}   R_{\alpha}^{(k)} +
\sum_{\alpha=8}^{14}   R_{\alpha}^{(k)} \Biggr)}.
\label{Def_JandJprime}
\end{equation}
The integrals  in
$x_i$'s ($y_i$'s) in
(\ref{Def_JandJprime})
stand  over the support of
$\pi N$
TDA (antinucleon DA)
(\ref{Int_TDA_DA_support}).
Within the $u$-channel factorization regime  of
$N^p \bar{N^p} \rightarrow \pi^0 \gamma^{*}$
the coefficients
$R_\alpha^{(k)}$
($\alpha=1,\ldots,14$)
correspond to the coefficients
$T_\alpha^{(k)}$
presented in Table~\ref{Table_Bkw_pion}
up to an overall irrelevant phase factor
${{\eta_N^{*}} }^{-1} \eta_q^{-3}$ originating
from the charge conjugation properties of antinucleon DAs (see  Sec.~\ref{SubSec_AntiNucl_DA})
and to the replacement
$-i 0 \rightarrow  i 0$
of the regulating prescriptions
in the denominators of hard scattering kernels $K_\alpha$. This latter change mirrors the fact that we consider
$\gamma^{*}$
in the final state rather than in the initial state as it is for backward pion electroproduction.

As argued in  Appendix C of Ref.~\cite{Lansberg:2012ha}, the amplitude of
$N \bar{N} \rightarrow \gamma^{*} \pi$
within the $t$-channel factorization mechanism can be obtained from that within the
$u$-channel factorization mechanism
(\ref{Def_ampl_MPANDA})
with the obvious change of the kinematical variables:
\begin{eqnarray}
  &&
p_N  \rightarrow p_{\bar{N}};  \ \ \   p_{\bar{N}}  \rightarrow p_{{N}}; \nonumber \\
  &&
\Delta^u \rightarrow \Delta^t \ \ \ (u  \rightarrow t); \nonumber \\
  &&
\xi^u  \rightarrow \xi^t.
\label{tu_change}
\end{eqnarray}
The proof relies on the $C$- or $\mathcal G$- parity properties of the amplitude. This ensures
the identical structure of the backward and the forward peaks of the unpolarized cross section
of the reaction (\ref{BarNNannihilation reaction}).
However,  the near-backward  and the near-forward regimes are treated somewhat unequally within the  \=PANDA experimental set-up
operating the antiproton beam. Indeed, once switching to the LAB system (which corresponds to the target nucleon at rest)
one may check that  the forward peak of the cross section as a function of
$\cos \theta_\pi^{\rm LAB}$
is narrowed, while  the backward peak is broadened  by the effect of the Lorentz boost from the
$N \bar{N}$
CMS to the LAB frame.

 \subsubsection{Nucleon--antinucleon annihilation into  $J/\psi$  and a pion}
 \label{SubSec_JiPsi}
 \mbox

In this subsection, following Ref.~\cite{Pire:2013jva}, we consider a variety of the  nucleon--antinucleon  annihilation into a high invariant mass lepton pair in association with a light meson reaction
(\ref{BarNNannihilation reaction})
in which the
invariant mass of lepton pair resonates at the mass of a heavy quarkonium.
For simplicity, below we address the case of
$\pi$
meson and consider the reaction
\begin{equation}
\label{Reac_JpsiPANDA}
  \bar N (p_{\bar N},s_{\bar{N}}) \; + \; N (p_N,s_N) \; \to  J/\psi(p_{\psi}, \lambda_\psi)\;+\; \pi(p_{\pi}).
\end{equation}
However, the suggested formalism admits a straightforward generalization
for production of a variety of light mesons (\textit{e.g.} $\rho, \eta, \omega, f_0, \varphi $) or  meson pairs (\textit{e.g.}
$\pi \pi, K \bar{K}$). Also, instead of $J/\psi$ one may consider another species of heavy quarkonium.

\begin{figure}[H]
\begin{center}
\includegraphics[width=0.47\textwidth]{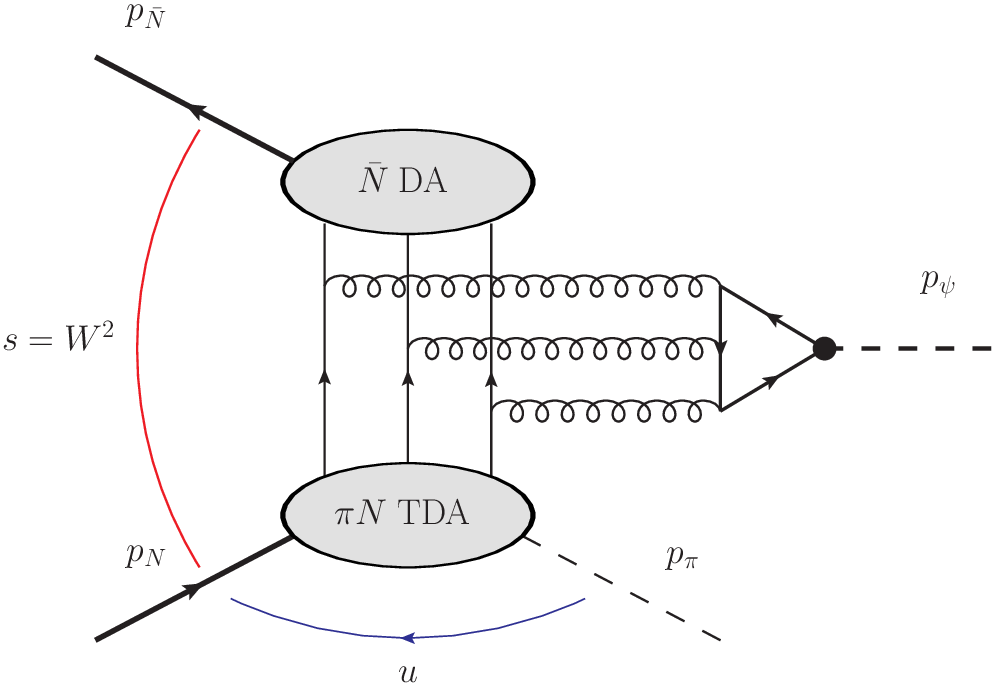} \ \ \
\includegraphics[width=0.47\textwidth]{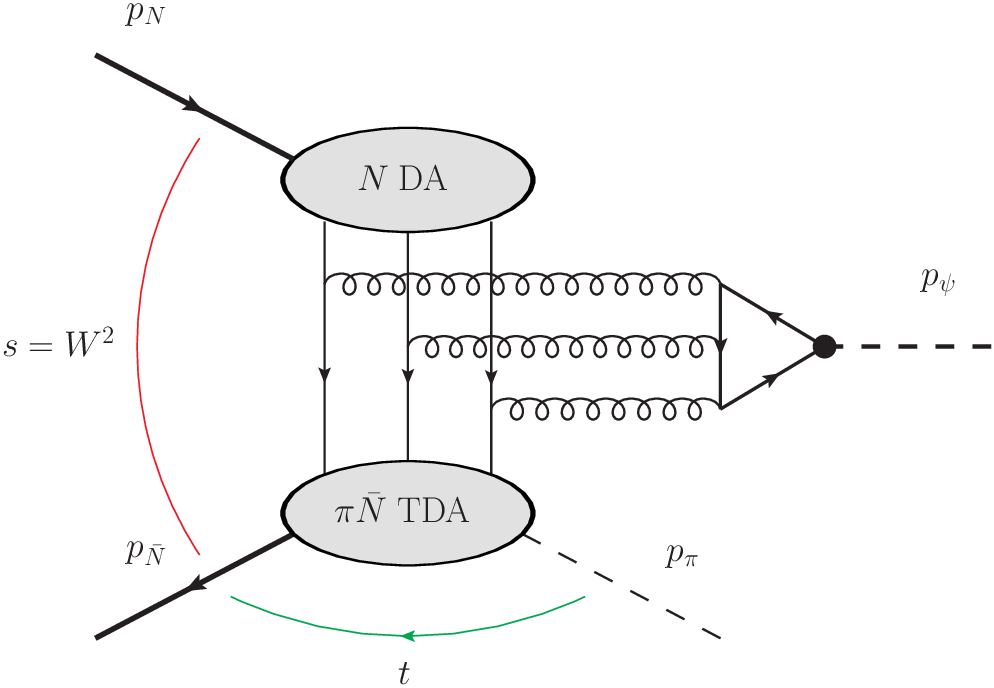}
\end{center}
     \caption{{\bf Left panel:} kinematical quantities and the collinear factorization mechanism  for
      $\bar{N} N \to J/\psi \pi$  in the  near-backward  kinematical regime (large $W^2 \equiv (p_N+p_{\bar{N}})^2$  of order of $M_\psi^2$; $| u|  \sim 0$). The lower blob, denoted
  $\pi N$ TDA, depicts the nucleon-to-pion transition
     distribution amplitude; $\bar{N}$ DA blob depicts the antinucleon distribution amplitude; black dot denotes the light-cone wave function of heavy quarkonium.
      {\bf Right panel:} kinematical quantities and the collinear factorization mechanism  for
         $\bar{N} N \to J/\psi \pi$ in the  near-forward  kinematical regime (large $W^2=(p_N+p_{\bar{N}})^2$  of order of $M_\psi^2$; $| t|  \sim 0$). The lower blob, denoted
  $\pi \bar{N}$ TDA, depicts the antinucleon-to-pion transition
     distribution amplitude; $N$ DA blob depicts the nucleon distribution amplitude;  black dot denotes the light-cone wave function of heavy quarkonium. [Reprinted Figure 1
from Ref.~\cite{Pire:2013jva}. Copyright (2013) by Elsevier.]
      }
\label{Fig_Kinematics_TDAs_JPsi_PANDA}
\end{figure}

As for the lepton pair production discussed earlier, the amplitude of this reaction
(\ref{Reac_JpsiPANDA})
admits a factorized description  within two distinct
kinematical regimes
(see  Fig.~\ref{Fig_Kinematics_TDAs_JPsi_PANDA}).
\begin{itemize}
\item The near-backward kinematics:
$s=(p_N+p_{\bar{N}})^2 \equiv W^2$ of the order of the hard scale introduced by $M_\psi^2$;
$u \equiv (p_\pi-p_{N})^2 \sim 0$.
This corresponds to a pion moving almost in the direction of the initial  nucleon in
$N \bar{N}$ center-of-mass system (CMS);
\item The  near-forward kinematics:
$s=(p_N+p_{\bar{N}})^2 \equiv W^2$ of the order of the hard scale introduced by $M_\psi^2$; $M_\psi^2$;
$t \equiv (p_\pi-p_{\bar{N}})^2 \sim 0$;
This corresponds to a pion moving almost in the direction of the initial antinucleon in
$N \bar{N}$
CMS.
\end{itemize}
This hard reaction mechanism can be contrasted to soft reaction mechanisms
suggested in Ref.~\cite{Gaillard:1982zm}.

The Sudakov decomposition of the relevant momenta for the near-backward
and near-forward regimes is analogous to Eqs.~(\ref{Sudakov_PANDA_u}), and (\ref{Sudakov_PANDA_t})
with the replacement
\begin{equation}
Q^2 \to M_\psi^2; \ \ \ \xi^u=\xi^t \simeq \frac{M_\psi^2}{2W^2-M_\psi^2}.
\label{Def_xi_tu_Jpsi}
\end{equation}
Due to the  $C$-invariance of strong interaction there exists a perfect symmetry between the forward and backward
kinematics regimes of the reaction
(\ref{Reac_JpsiPANDA}).
The reaction amplitude
within the $t$-channel factorization regime can be obtained from that  within the $u$-channel
factorization regime with the obvious change of the kinematical variables (\ref{tu_change}).
Therefore, in what follows we address only the $u$-channel factorization regime.

 The calculation of
$N + \bar{N} \to J/\psi+  \pi $
scattering amplitude follows the same main steps as the classical calculation~\cite{Brodsky:1981kj,Chernyak:1983ej,Chernyak:1987nv} of the
$J/\psi \to p + \bar{p}$ decay amplitude.
The large distance dynamics is encoded within the matrix elements of QCD light-cone operators between
the appropriate hadronic states: $\pi N$ TDAs, $\bar{N}$ DAs and heavy
quarkonium light-cone wave function.

\begin{figure}[H]
\begin{center}
\includegraphics[width=0.3\textwidth]{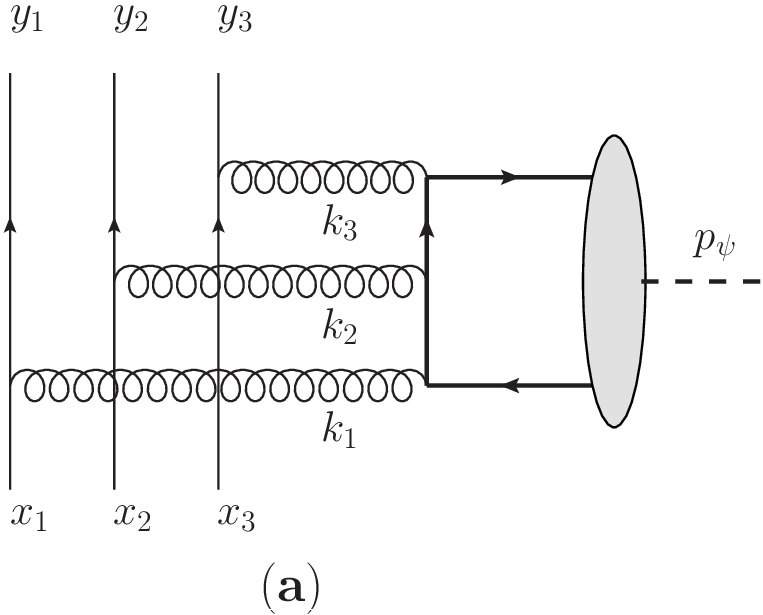}~~
\includegraphics[width=0.3\textwidth]{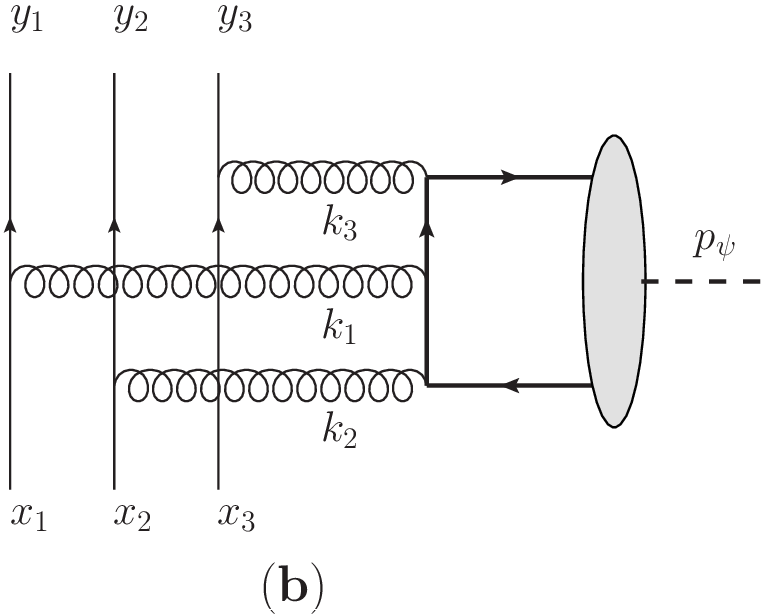}~~
\includegraphics[width=0.3\textwidth]{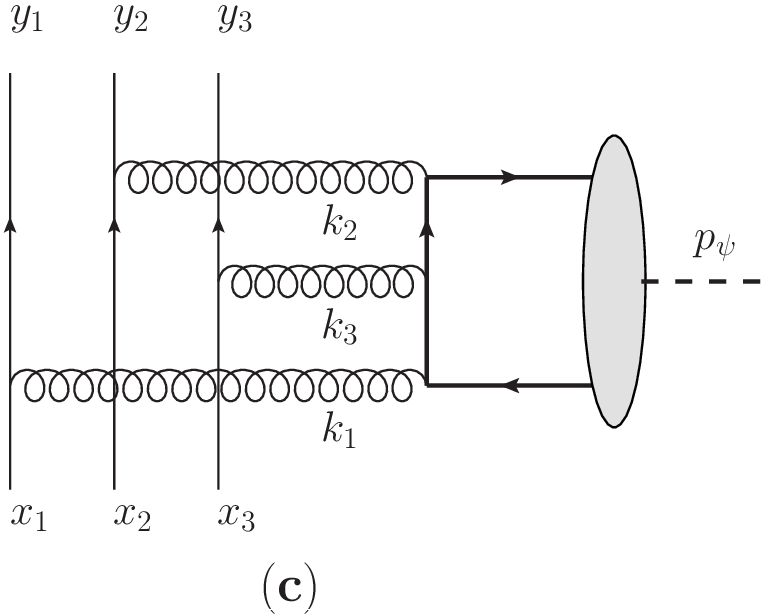}
\end{center}
     \caption{Feynman diagrams describing $J/\psi ~\pi$-production subprocess at the Born order. [Reprinted Figure 2 from Ref.~\cite{Pire:2013jva}. Copyright (2013) by Elsevier.] }
\label{Fig_diagrams}
\end{figure}

The hard part of the amplitude
given by the sum of the three diagrams presented in  Fig.~\ref{Fig_diagrams}
is computed within
perturbative QCD.

For $\pi N$ TDAs we employ the parametrization of  Sec.~\ref{SubSubSec_Def_piN_TDAs},
as it allows a clear separation of $\Delta_T$-independent part of the amplitude,
which dominates in the strictly backward $\Delta_T=0$ limit.
The $J/\psi$
heavy quarkonium is described by
the non-relativistic light-cone wave function
\cite{Chernyak:1983ej}:
\begin{equation}
\Phi_{\rho \tau}(z, p_\psi,\lambda_\psi)= \langle 0|  \bar{c}_\tau(z) c_\rho(-z) |  J/\psi (p_{\psi}, \lambda_\psi) \rangle
=
\frac{1}{4} f_\psi
\left[ 2 m_c \hat{\mathcal{E}}_\psi(p_{\psi}, \lambda_\psi) +\sigma_{ p_{\psi} \nu}   {\mathcal{E}}^\nu_\psi(p_{\psi}, \lambda_\psi)
\right]_{\rho \tau},
\label{WFNR}
\end{equation}
where $m_c$ is the $c$-quark mass and
${\mathcal{E}}_\psi$
stands for the charmonium polarization vector. With the use of the non-relativistic wave function
(\ref{WFNR})
we tacitly assume that each charm quark carries half of the momentum of the
$J/\psi$.
The normalization constant
$f_\psi$
is extracted from the charmonium leptonic decay width
$\Gamma(J/\psi \to e^+  e^- )$:
\begin{equation}
\Gamma(J/\psi \to e^+  e^- )= (4 \pi \alpha_{\rm e.m.})^2 \frac{e_c^2}{12 \pi } f_\psi^2 \frac{1}{M_\psi}, \ \ \ e_c=\frac{2}{3}.
\end{equation}
Using the values quoted in
\cite{PDG2020}
we get
\begin{equation}
| f_\psi| = 416 \pm 6 \; {\rm MeV}.
\label{fpsi_value}
\end{equation}
Following
\cite{Brodsky:1981kj},
in our calculation we set
\begin{equation}
M_\psi  \simeq  2m_c   \simeq \bar{M} ,
\label{mass approx}
\end{equation}
taking the average value
$\bar{M}= 3 \; {\rm GeV }$.

The leading order amplitude of
(\ref{Reac_JpsiPANDA})
is parameterized analogously to Eq.~(\ref{Def_ampl_MPANDA}):
\begin{equation}
{\mathcal{M}}_{s_N s_{\bar{N}}}^{\lambda_\psi}= {\mathcal{C}}_\psi \frac{1}{{\bar M}^5 } \Bigl[
\tilde{\mathcal{S}}^{(1) \,\lambda_\psi}_{ s_N s_{\bar{N}}}
\tilde{\mathcal{J}}^{(1)}(\xi,\Delta^2)
-
\tilde{\mathcal{S}}^{(2) \,\lambda_\psi}_{ s_N s_{\bar{N}}}
\tilde{\mathcal{J}}^{(2)}(\xi,\Delta^2)
\Bigr],
\label{Amplitude_master}
\end{equation}
where the tensor structures $\tilde{\mathcal{S}}^{(k)\, \lambda}_{ s_N s_{\bar{N}}}$
correspond to (\ref{Def_S_PANDA_process}) with
the virtual photon polarization vector replaced by the
polarization vector of the heavy quarkonium
${\mathcal{E}}_\psi(p_\psi, \lambda_\psi)$:
\begin{eqnarray}
  &&
\tilde{\mathcal{S}}^{(1) \,\lambda_\psi}_{ s_N s_{\bar{N}}}=\bar{V}(p_{\bar{N}},s_{\bar{N}} )\hat{\mathcal{E}}^{*}_\psi(p_\psi, \lambda_\psi) \gamma_5 U(p_N, s_N); \nn \\
  &&\tilde{\mathcal{S}}^{(2) \, \lambda_\psi}_{ s_N s_{\bar{N}}}=\frac{1}{m_N}  \bar{V}(p_{\bar{N}},s_{\bar{N}} )\hat{\mathcal{E}}^{*}_\psi(p_\psi, \lambda_\psi) \hat{\Delta}_T \gamma_5 U(p_N, s_N).
\end{eqnarray}

The calculation of $3$ diagrams presented on
 Fig.~\ref{Fig_diagrams}
yields the following result\footnote{We present the result for the $\bar{N}^{\bar{p}} N^p \to \pi^0 J/\psi$ channel.
For $\bar{N}^p N^n \to \pi^- J/\psi$ channel one has to replace proton-to-$\pi^0$ $uud$ TDAs
by neutron-to-$\pi^-$ $uud$ TDAs.}
 for
$\tilde{{\mathcal{J}}}^{(1,2)}(\xi,\Delta^2)$:
\begin{eqnarray}
 &&
\tilde{\mathcal{J}}^{(1)}(\xi,\Delta^2)
= { {\int^{1+\xi}_{-1+\xi} }\! \! \!
d_3 x
\, \delta \left(\sum_{j=1}^3 x_j-2\xi\right)
}
\;
{{\int^{1}_{0} }\! \! \! d_3y
\,
\delta\left(\sum_{l=1}^3 y_l-1\right) }
\nonumber
\\ &&
\left\{ \frac{\xi ^3 (x_1 y_3+x_3 y_1) (V_1^{\pi^0 p}-A_1^{\pi^0 p})
(V^{p}-A^{p}) }{  y_1 y_2 y_3 (x_1+i0) (x_2+i0) (x_3+i0)   (x_1 (2
   y_1-1)-2 \xi  y_1+i0) (x_3 (2 y_3-1)-2 \xi  y_3+i0)} \right.
\nonumber \\ &&
+\left. \frac{ \xi ^3 (x_1 y_2+x_2 y_1) (2 T_1^{\pi^0 p}
+
\frac{\Delta_T^2}{m_N^2}T_4^{\pi^0 p}
)
T^{p}}{  y_1 y_2 y_3 (x_1+i0) (x_2+i0) (x_3+i0)
(x_1 (2y_1-1)-2 \xi  y_1+i0) (x_2 (2 y_2-1)-2 \xi  y_2+i0)}  \right\};
   \label{Amplitude_result_I}
\end{eqnarray}
\begin{eqnarray}
 &&
\tilde{\mathcal{J}}^{(2)}(\xi,\Delta^2)
= { {\int^{1+\xi}_{-1+\xi} }\! \! \!
d_3 x
\, \delta \left(\sum_{j=1}^3 x_j-2\xi\right)
}
\;
{{\int^{1}_{0} }\! \! \! d_3y
\,
\delta\left(\sum_{l=1}^3 y_l-1\right) } \nonumber \\ &&
\left\{ \frac{\xi ^3 (x_1 y_3+x_3 y_1) (V_2^{\pi^0 p}-A_2^{\pi^0 p})
(V^{p}-A^{p}) }{  y_1 y_2 y_3 (x_1+i0) (x_2+i0) (x_3+i0)   (x_1 (2
   y_1-1)-2 \xi  y_1+i0) (x_3 (2 y_3-1)-2 \xi  y_3+i0)} \right.
\nonumber \\ &&
+\left. \frac{ \xi ^3 (x_1 y_2+x_2 y_1) (T_2^{\pi^0 p}+
T_3^{\pi^0 p}) T^{p} }{  y_1 y_2 y_3 (x_1+i0) (x_2+i0) (x_3+i0)
(x_1 (2   y_1-1)-2 \xi  y_1+i0) (x_2 (2 y_2-1)-2 \xi  y_2+i0)}  \right\},
   \label{Amplitude_result_Ip}
\end{eqnarray}
where we employ the shortened notations
(\ref{Not_arg_short})
for the arguments
of $\pi N$ TDAs and nucleon DAs.

The overall factor
${\mathcal{C}}_\psi$
in
(\ref{Amplitude_master})
is expressed as:
\begin{equation}
{\mathcal{C}}_\psi= (4 \pi \alpha_s)^3 \frac{f_N^2 f_\psi}{f_\pi}  \underbrace{ \frac{1}{2}}_{{J/\psi \; {\rm w.f.} \atop  { \rm  normalization} }}
\times \,  \underbrace{16}_{{\rm Dirac} \; {\rm trace}} \,  \times\underbrace{ \frac{5}{3} \cdot \frac{1}{3} \cdot \frac{1}{(3!)^2}}_{{\rm  color} \; {\rm factor}}=
(4 \pi \alpha_s)^3 \frac{f_N^2 f_\psi}{f_\pi}  \, \frac{10}{81},
\label{Def_C_Jpsi}
\end{equation}
where
$\alpha_s$
stands for the strong coupling,
$f_\psi$
is the normalization constant
of the wave function of heavy quarkonium
(\ref{fpsi_value}),
$f_N$
is the nucleon light-cone wave function normalization constant
(\ref{fN_CZvalue})
and
$f_\pi=93~{\rm MeV}$ is the pion weak decay constant.

Contrary to the case of hard exclusive meson production,
the hard scattering kernels in (\ref{Amplitude_result_I}), (\ref{Amplitude_result_Ip}) do not admit a factorized
form (\ref{D_factorized}). The dependencies on the $\pi N$  TDA longitudinal momentum fraction
variables $x_{1,2,3}$ and of the nucleon DA  longitudinal momentum fraction
variables $y_{1,2,3}$ turn out to be entangled.
However one may check that the poles in
$x_i$
of the singular convolution kernels
are indeed located either on the cross-over trajectories
$x_i=0$,
that separate the DGLAP-like and ERBL-like support domains of
$\pi N$ TDAs,
or within the DGLAP-like support domain (for the $y$-dependent poles in $x_i$), as it certainly should be.

The structure of the result
(\ref{Amplitude_master}), (\ref{Amplitude_result_I}), (\ref{Amplitude_result_Ip})
resembles much the well known expression for the
$J/\psi \to \bar{p} p$
decay amplitude
\cite{Chernyak:1987nv}:
\begin{equation}
{\mathcal{M}}_{J/\psi \to \bar{p} p}= (4 \pi \alpha_s)^3 \frac{f_N^2 f_\psi}{ {\bar{M}^5}} \,  \frac{10}{81}  \, \bar{U} \hat{\mathcal{E}}_\psi V \,  M_0,
\label{Decay_ampl_chernyak}
\end{equation}
where
\begin{eqnarray}
  &&
M_0=  {\int^{1 }_{0 } }\! \! \!
d_3 x  \delta \left(\sum_{j=1}^3 x_j-1\right)
 {\int^{1 }_{0 } }\! \! \!
 d_3 y   \delta \left(\sum_{k=1}^3 y_k-1\right)
 \\   &&
\left\{ \frac   {y_1 x_3  (V^{p}(x_{1,2,3})-A^{p}(x_{1,2,3}))  (V^{p}(y_{1,2,3})-A^{p}(y_{1,2,3})) }{  y_1 y_2 y_3 \,  x_1 x_2 x_3
(1-(2x_1-1)(2y_1-1))  (1-(2x_3-1)(2y_3-1))} \right.
\nonumber \\ &&
+\left. \frac{2 y_1 x_2 T^{p}(x_{1,2,3}) T^{p}(y_{1,2,3})}{  y_1 y_2 y_3 \, x_1   x_2   x_3   (1-(2x_1-1)(2y_1-1))  (1-(2x_2-1)(2y_2-1))}  \right\}.
   \label{Def_M0}
\end{eqnarray}
The
$J/\psi \to \bar{p} p$
decay amplitude
(\ref{Decay_ampl_chernyak})
results in the following expression for the decay width
~\cite{Chernyak:1987nv}:
\begin{equation}
\Gamma(J/\psi \to p \bar{p} )= (\pi \alpha_s)^6 \frac{1280 f_\psi^2 f_N^4 }{243 \pi {\bar{M}^9}} | M_0| ^2.
\label{Charm_dec_width}
\end{equation}

\subsubsection{Backward charmonium production in $\pi N$ collisions}
\label{SubSec_piN_JPARC}
\mbox

The collinear factorization mechanism
with the non-zero baryon number transfer in the cross channel
can also be applied for a different cross-channel conjugate
of the reaction (\ref{Reac_JpsiPANDA}):
the exclusive pion-induced Drell--Yan process
\begin{equation}
\pi^-(p_{\pi}) \;+ N^p (p_N,s_N)  \; \to  \ell^+ \ell^- + N^n (p'_N,s'_N) \,.
\label{DY_pion_induced}
\end{equation}
The intense pion beam available at the Japan Proton Accelerator Research Complex (J-PARC)
with pion beam momentum $P_\pi \sim 10 - 20$~GeV and center-of-mass energy squared
$W^2=m_N^2+m_\pi^2+2 E_\pi m_N \approx 2m_N P_\pi$
opens a possibility for the experimental study of the reaction (\ref{DY_pion_induced}).

A factorized description of (\ref{DY_pion_induced})
in the near-forward kinematical regime (large $W^2=(p_\pi+p_N)^2$, small
$| t| =| (p'_N-p_N)^2| $)
in terms of polarized nucleon GPDs $\tilde{H}$ and $\tilde{E}$ was proposed
in~\cite{Berger:2001zn} and discussed in~\cite{Goloskokov:2015zsa}.
The recent feasibility study
\cite{Sawada:2016mao}
of forward lepton pair production demonstrates that J-PARC is able to access
GPDs in this regime.

A natural generalization of this approach is  to try a factorized
description of the reaction (\ref{DY_pion_induced}) in the complementary
near-backward regime  in terms of pion-to-nucleon TDAs and nucleon DAs.

In this subsection, following Ref.~\cite{Pire:2016gut}, we  consider the near-backward regime
of the reaction (\ref{DY_pion_induced}) with the invariant mass of a lepton pair
resonating at the mass of charmonium:
\begin{equation}
\label{reacpsiJPARC}
\pi^-(p_{\pi}) \;+ N^p (p_N,s_N)  \; \to  J/\psi(p_{\psi}, \lambda_\psi) + N^n (p'_N,s'_N)
\to \ell^+ \ell^-+ N^n (p'_N,s'_N).
\end{equation}
This reaction might be preferable from the experimental perspective due to a larger
value of the resonance cross section.

\begin{figure}[H]
\begin{center}
\includegraphics[width=0.5\textwidth]{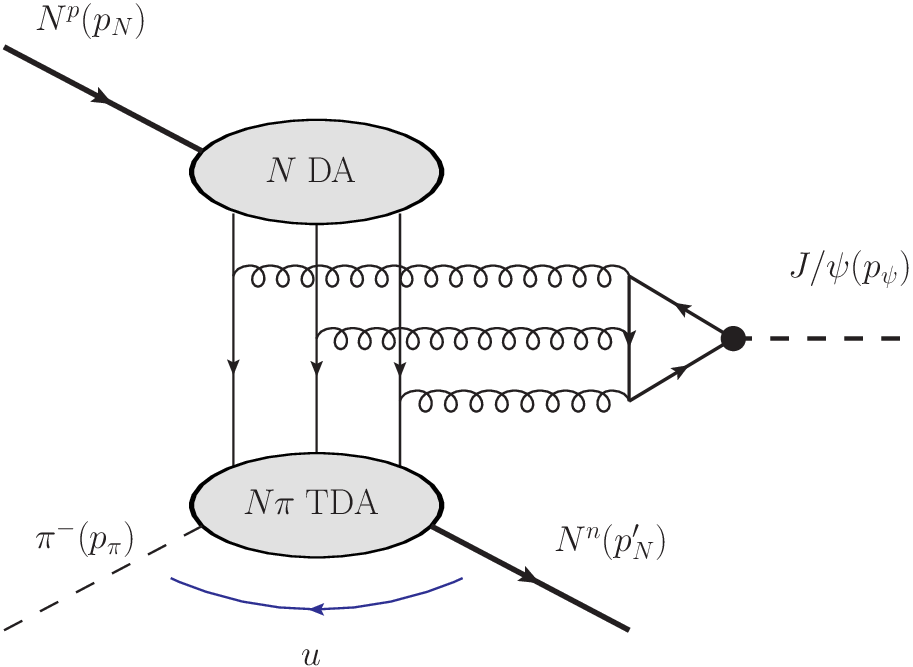}
\end{center}
\caption{Collinear factorization of the
$\pi^-(p_\pi)+ N^p(p_N) \to N^n(p'_N) + J/\psi(p_\psi)$
reaction in the $u$-channel regime.
$N$~DA stands for the distribution amplitude of the incoming nucleon;
$N \pi$
TDA stands for the transition distribution amplitude from a pion to a nucleon. [Reprinted Figure 1
from Ref.~\cite{Pire:2016gut}. Copyright (2017) by American Physical Society.]}
\label{Fig_factorization_JParc}
\end{figure}

The
$\pi N $
center-of-mass energy squared
$s=(p_\pi+p_N)^2 \equiv W^2$
and the charmonium mass squared
$M^2_\psi$
introduce the natural hard scale. In complete analogy with our
analysis of the nucleon--antinucleon annihilation process
\cite{Lansberg:2007se,Lansberg:2012ha}
we assume that this reaction admits a factorized description  in the
near-backward kinematical regime (see  Fig.~\ref{Fig_factorization_JParc}), where
$| u|  \equiv | \Delta^2| = | (p_2-p_\pi)^2|  \ll W^2, \, M_\psi^2$.
This corresponds to the final nucleon moving almost in the direction of the initial  pion in
$\pi N$
center-of-mass system (CMS).

The
$z$-axis is chosen along the direction of the pion beam in the
meson--nucleon CMS frame.
We introduce the
light-cone vectors
$p, n$
($p^2=n^2=0$;
$2p \cdot n =1$).
The Sudakov decomposition of the relevant momenta  reads
\begin{eqnarray}
  &&
p_\pi = (1+\xi) p +\frac{m_\pi^2}{1+\xi}n\, ;
\nonumber \\   &&
p_N =
\frac{2(1+\xi)m_N^2}{W^2+\Lambda(W^2,m_N^2,m_\pi^2)-m_N^2-m_\pi^2}\,
p +
\frac{W^2+\Lambda(W^2,m_N^2,m_\pi^2)-m_N^2-m_\pi^2}{2(1+\xi)} \,
n\,;   \nonumber \\   &&
\Delta \equiv (p_2-p_\pi)   =
-2\xi p + \left( \frac{ m_N^2-\Delta_T^2}{1-\xi } - \frac{m_\pi^2}{1+\xi}\right)n+\Delta_T;
\nonumber \\   &&
p_\psi =  p_1 - \Delta \, ; \ \ \ \
p'_N= p_\pi +\Delta,
\label{Sudakov_decomposition}
\end{eqnarray}
where
$\Lambda(x,y,z)$
is the Mandelstam function  (\ref{Def_lambda}) and
$m_N$
and
$m_\pi$
stand respectively for the nucleon and pion masses.
The transverse direction in
(\ref{Sudakov_decomposition})
is defined with respect to the
$z$
direction and
$\xi$
is the  skewness variable that characterizes the longitudinal
momentum transfer between the initial state pion and final state nucleon:
\begin{equation}
\xi \equiv  -\frac{(p'_N-p_\pi) \cdot n}{(p'_N+p_\pi) \cdot n}.
\label{Def_xi}
\end{equation}

Within the collinear factorization framework we neglect both the pion
and nucleon masses with respect to
$M_\psi$
and
$W$
and set
$\Delta_T=0$
within the coefficient function.
This results in the approximate expression for the  skewness variable
(\ref{Def_xi}):
\begin{equation}
\xi \simeq \frac{M_\psi^2}{2 W^2-M_\psi^2}.
\label{Xi_collinear}
\end{equation}

The leading twist-$3$
$uud$ $\pi^-$-to-neutron
($n \pi^-$)
TDAs occurring  in the factorized description
of (\ref{reacpsiJPARC}) in the near-backward regime
are defined from the Fourier transform
\begin{equation}
\bar{{\mathcal{F}}} \equiv
(p \cdot n)^3 \int
\left[
\prod_{j=1}^3 \frac{d \lambda_j}{2 \pi}
\right]
e^{i \sum_{k=1}^3 x_k \lambda_k (p \cdot n)}
\label{Fourier}
\end{equation}
of the $n\pi^-$ matrix element of the
trilinear antiquark operator on the light cone:
\begin{eqnarray}
  &&
4  \bar{\mathcal{F}}
\langle n(p_2)|  \varepsilon_{c_1 c_2 c_3}
\bar{u}_{\rho}^{c_1}(\lambda_1n)
\bar{u}_{\tau}^{c_2}(\lambda_2n)
\bar{d}_{\chi}^{c_3}(\lambda_3n)
| \pi^-(p_\pi) \rangle
\nonumber \\ &&
= \delta(x_1+x_2+x_3-2 \xi) i \frac{f_N}{f_\pi}
\sum_{s} s^{N\pi}_{\rho \tau, \, \chi} H_s^{
n \pi^- }(x_1,x_2,x_3, \xi, \Delta^2),
\label{FT_defining_npi_TDAs}
\end{eqnarray}
where the sum goes over the eight
leading twist-$3$ Dirac structures.

The connection between the pion-to-nucleon $N\pi$
TDAs defined in (\ref{FT_defining_npi_TDAs})
and $ \pi N$ TDAs within the parametrization of
 Sec.~\ref{SubSubSec_Def_piN_TDAs}
can be established with the help of
the Dirac conjugation
(complex conjugation and convolution with
$\gamma_0$ matrices
in the appropriate spinor indices)
of Eq.~(\ref{Param_TDAs}):
\begin{eqnarray}
  &&
-4 (p \cdot n)^3 \int \left[ \prod_{j=1}^3 \frac{d \lambda_j}{2 \pi}\right]
e^{-i \sum_{k=1}^3 \tilde{x}_k \lambda_k (p \cdot n)}
 \langle    n(p_N,s_N) | \,  \varepsilon_{c_1 c_2 c_3} \bar{u}^{c_1}_{\rho}(\lambda_1 n)
\bar{u}^{c_2}_{\tau}(\lambda_2 n) \bar{d}^{c_3}_{\chi}(\lambda_3 n)
\,| \pi^-(p_\pi) \rangle
\nonumber \\ &&
=
-\delta(\tilde{x}_1+\tilde{x}_2+\tilde{x}_3-2 \tilde{\xi}) i \frac{f_N}{f_\pi}
\sum_s   \underbrace{(\gamma_0^T)_{\tau \tau'} \left[ s_{\rho' \tau', \chi'}^{\pi N} \right]^\dagger
(\gamma_0)_{\rho'\rho }
(\gamma_0)_{\chi' \chi}}_{s_{\rho \tau, \chi}^{N \pi}} H_s^{ N \pi}
(\tilde{x}_{1},\tilde{x}_{2},\tilde{x}_{3}, \tilde{\xi} ,\Delta^2).
\label{DiracConjTDA}
\end{eqnarray}

For the relevant Dirac structures we get
\begin{eqnarray}
  &&
(v_1^{N \pi})_{\rho \tau, \chi}= (C \hat{p})_{\rho \tau} \bar{U}^+_\chi;
\ \ \ \
(v_2^{N \pi})_{\rho \tau, \chi}= (C \hat{p})_{\rho \tau}  \left( \hat{\tilde{\Delta}}_T\bar{U}^+ \right)_\chi=-
 (C \hat{p})_{\rho \tau}  \left( \hat{ \Delta}_T\bar{U}^+ \right)_\chi;
\nonumber \\   &&
(a_1^{N\pi})_{\rho \tau, \chi}= (C \hat{p} \gamma_5)_{\rho \tau} \left(\bar{U}^+ \gamma_5 \right)_\chi; \ \ \ \
(a_2^{N \pi})_{\rho \tau, \chi}= (C \hat{p} \gamma_5)_{\rho \tau} \left( \bar{U}^+ \hat{\tilde{\Delta}}_T \gamma_5 \right)_\chi=-
 (C \hat{p} \gamma_5)_{\rho \tau} \left( \bar{U}^+  \hat{\Delta}_T \gamma_5 \right)_\chi;
\nonumber \\   &&
(t_1^{N\pi })_{\rho \tau, \chi}= -(C \sigma_{p \mu})_{\rho \tau} \left(\bar{U}^+ \gamma_\mu \right)_\chi; \ \ \ \
(t_2^{N \pi})_{\rho \tau, \chi}= -(C \sigma_{p \tilde{\Delta}_T})_{\rho \tau} \left(\bar{U}^+ \right)_\chi=(C \sigma_{p  \Delta_T})_{\rho \tau} \left(\bar{U}^+ \right)_\chi;
 \nonumber \\   &&
(t_3^{N \pi})_{\rho \tau, \chi}= (C \sigma_{p \mu})_{\rho \tau}
\left(\bar{U}^+ \sigma_{  \mu \tilde{\Delta}_T} \right)_\chi=
- (C \sigma_{p \mu})_{\rho \tau}
\left(\bar{U}^+ \sigma_{  \mu  \Delta_T} \right)_\chi;
\nn \\   &&
(t_4^{N \pi})_{\rho \tau, \chi}= -(C \sigma_{p \tilde{\Delta}_T})_{\rho \tau}
\left(\bar{U}^+  \hat{\tilde{\Delta}}_T\right)_\chi=
-(C \sigma_{p \Delta_T})_{\rho \tau}
\left(\bar{U}^+  \hat{\Delta}_T \right)_\chi\,,
\end{eqnarray}
where
$\bar{U}^+\equiv \bar{U}(p_N,s_N) \hat{n} \hat{p}$
stands for the large component of the
$\bar{U}(p_N,s_N)$
Dirac spinor.

\begin{figure}[H]
 \begin{center}
 \includegraphics[width=0.7\textwidth]{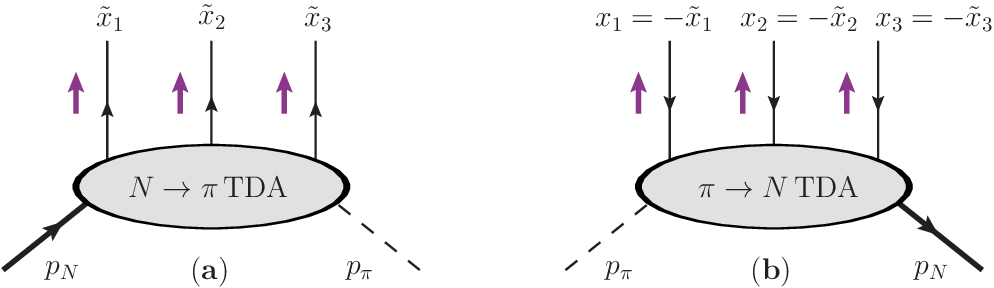}
   \end{center}
     \caption{Small arrows show the direction of the positive longitudinal momentum flow in the ERBL-like
      regime. Arrows on the nucleon and quark (antiquark) lines show the direction of flow of the baryonic charge.
{\bf (a)}: The longitudinal momentum flow for
$N \to \pi$ TDAs
defined in
(\ref{Param_TDAs}).
The longitudinal momentum transfer is $(p_\pi-p_N) \cdot n \equiv \tilde{\Delta} \cdot n$.
{\bf (b)}: The longitudinal momentum flow for
$\pi \to N$ TDAs
defined in
(\ref{FT_defining_npi_TDAs}).
The longitudinal momentum transfer is $(p_N-p_\pi) \cdot n \equiv  \Delta \cdot n$.
}
\label{Fig_Flow}
\end{figure}

The flow of the longitudinal momentum for
$N \to \pi$
   and
$\pi \to N$
TDAs  is presented in  Fig.~\ref{Fig_Flow}.
Therefore, we switch to the definition of momentum transfer
natural for the reaction (\ref{reacpsiJPARC}):
$\tilde{\Delta} \to -\Delta$.
This implies the change of sign of the skewness variable $\xi$
and the longitudinal momentum fractions:
\begin{equation}
\xi=-\tilde{\xi}; \ \ \ x_i=-\tilde{x}_i.
\end{equation}
The invariant momentum transfer squared remains unchanged:
$\tilde{\Delta}^2 \to \Delta^2$.

Now comparing
(\ref{DiracConjTDA})
to
(\ref{FT_defining_npi_TDAs})
we conclude that
\begin{equation}
 \left\{
V_{1,\, 2}^{n \pi^-}, \,
A_{1,\, 2}^{n \pi^-}, \,
T_{1,\, 2, \, 3, \,4}^{n\pi^-}
\right\}
(x_{1,2,3}, \xi, \Delta^2)
=
\left\{
V_{1,\, 2}^{ \pi^- n }, \,
A_{1,\, 2}^{  \pi^- n }, \,
T_{1,\, 2, \, 3, \,4}^{ \pi^- n  }
\right\}
(-x_{1,2,3}, -\xi, \Delta^2).
\label{NpiTDA_relation_to_piN}
\end{equation}

The leading order amplitude of the
$J/\psi ~N^n$
production subprocess of
(\ref{reacpsiJPARC})
is, up to the reverse of the direction of the fermion lines,
given by the sum of the same
three diagrams presented in  Fig.~\ref{Fig_diagrams}.
The amplitude of the reaction
(\ref{reacpsiJPARC})
may be written as:
\begin{eqnarray}
  &&
{\mathcal{M}}_{s_N s'_N}^{\lambda_\psi}
 = {\mathcal{C}}_\psi \frac{1}{{\bar M}^5 } \Bigl[
\bar{U}(p'_N,s'_N )\hat{\mathcal{E}}^{*}_{\psi}(p_\psi,\lambda_\psi) \gamma_5 U(p_N, s_N) \tilde{\mathcal{J}}^{(1)}(\xi,\Delta^2)
\nn  \\ &&
-\frac{1}{m_N}  \bar{U}(p'_N,s'_N )\hat{\mathcal{E}}^{*}_{\psi}(p_\psi,\lambda_\psi)
\hat{\Delta}_T \gamma_5 U(p_N, s_N) \tilde{\mathcal{J}}^{(2)}(\xi,\Delta^2)
\Bigr],
\label{Amplitude_masterJPARC}
\end{eqnarray}
where ${\mathcal{E}}_{\psi}$ is the charmonium polarization vector and
$\bar{U}$, $U$
stand  for the nucleon Dirac spinors of the final and initial state nucleons;
$\bar{M}$ stands for the mean mass (\ref{mass approx}). The normalization constant ${\mathcal{C}}_\psi$ is given by
(\ref{Def_C_Jpsi}).

The calculation  of the diagrams shown in  Fig.~\ref{Fig_diagrams} yields the same result for the integral
convolutions
$\tilde{\mathcal{J}}^{(1,2)} (\xi,\Delta^2)$ in Eq.~(\ref{Amplitude_masterJPARC})
as for
$J/\psi\, \pi$
production in
$\bar{p}p$
annihilation
(\ref{Amplitude_result_I}), (\ref{Amplitude_result_Ip})
up to the obvious replacement
of nucleon-to-pion ($\pi N $) TDAs
with pion-to-nucleon ($N \pi$) TDAs (\ref{NpiTDA_relation_to_piN}).
 The cross section estimates for the reaction
(\ref{reacpsiJPARC})
for the J-PARC kinematical conditions are presented in  Sec.~\ref{SubSec_Prosp_JPARC_EIC}.

\section{Phenomenology and experimental perspectives for { {TDA}}s}
\label{Sec_PhenomExp}
\setcounter{equation}{0}

\subsection{Cross section of backward meson electroproduction}
\label{SubSec_Bkw_CS}
\mbox

In this section we specify our conventions and present the results for the cross sections of backward meson electroproduction reactions.

\subsubsection{Backward pion electroproduction}
\label{SubSec_CS_formulas_Bkw_piN}
\mbox

\begin{figure}[h]
 \begin{center}
 \includegraphics[width=12cm]{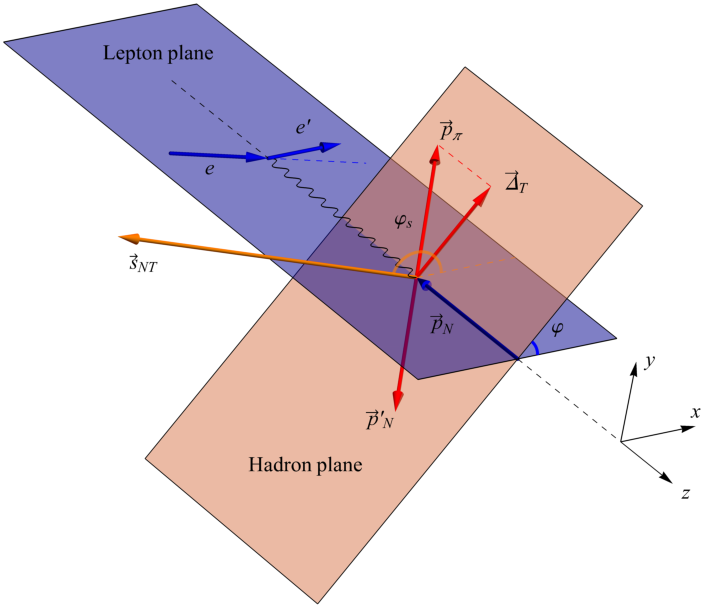}
   \end{center}
     \caption{The kinematics of the reaction (\ref{Reaction_piN_Sec7}) in the $\gamma^{*}N$ center-of-mass frame. $\varphi$ is the angle between the leptonic and hadronic reaction planes. $\varphi_s$ is the angle between the leptonic plane and the spin $\vec{s}_{N T}$ of the transversely polarized target nucleon.}
\label{Fig_Kinematics_piN_CMS}
\end{figure}

Let us consider the exclusive electroproduction of pions off nucleons within
the one photon exchange approximation
\begin{equation}
e(k,s_e)+ N(p_N,s_N) \rightarrow \bigl( \gamma^{*}(q, \lambda_\gamma) + N(p_N, s_N) \bigr) +e(k',s'_e)
  \rightarrow  e(k',s'_e)+ \pi(p_\pi) + N'(p'_N, s'_N),
\label{Reaction_piN_Sec7}
\end{equation}
where
$s_N$, $s'_N$ ($s_e$, $s'_e$)
stand for the polarization variables of the initial and final nucleon (lepton);
$\lambda_\gamma$
is the virtual photon polarization variable.
The kinematics of the reaction (\ref{Reaction_piN_Sec7}) in the
$\gamma^{*}N$ center-of-mass frame
is presented in  Fig.~\ref{Fig_Kinematics_piN_CMS}. We choose the direction
of the $z$-axis along the three momentum of the virtual photon.

The unpolarized cross section of the
reaction (\ref{Reaction_piN_Sec7}) can be decomposed into  the contributions
of the transverse ($T$) and longitudinal ($L$) polarizations of the virtual photon
and the transverse--transverse ($TT$) and longitudinal--transverse ($LT$)
interference contributions
\cite{Mulders:1990xw,Kroll:1995pv,Arens:1996xw,Huang:2000kd}:
\begin{eqnarray}
  &&
\frac{d^4 \sigma}{ds dQ^2 d \varphi dt}(e N \to e'N'\pi)= \frac{\alpha_{\rm em} (s-m_N^2)}{4 (2 \pi)^2 ( {k}_0^{\rm LAB})^2 m_N^2 Q^2 (1-\varepsilon)}
\nonumber \\ &&
\times
\Bigl(
{
\frac{d \sigma_T}{dt}
+ \varepsilon \frac{d \sigma_L}{dt}+
\varepsilon \cos 2\varphi \frac{d \sigma_{TT}}{dt}+
\sqrt{2 \varepsilon (1+\varepsilon)} \cos \varphi \frac{d \sigma_{LT}}{dt} }\
\Bigr).
\label{Def_CS_Kroll}
\end{eqnarray}
Here
$s=(p_N+q)^2 \equiv W^2$,
$t=(p'_N-p_N)^2$ and $Q^2=-q^2$
are the usual Lorentz invariants,
$\varphi$
is the angle between the leptonic and hadronic planes (see  Fig.~\ref{Fig_Kinematics_piN_CMS});
and
${k}_0^{LAB} $
is the initial state electron energy in the laboratory (LAB) frame  (electron beam energy).
$\varepsilon$
is the polarization parameter of the virtual photon that expresses the ratio
of longitudinal to transverse photon flux:
\begin{equation}
\varepsilon=
\frac{2(1-y)-2 x_B y \, m_{N}^{2}\left((p_N+k)^2-m_{N}^{2}\right)^{-1}}{1+(1-y)^{2}+2 x_B y\, m_{N}^{2}\left((p_N+k)^2-m_{N}^{2}\right)^{-1}}
=\left[1+2\frac{\bigl({k}_0^{  LAB}-{k'}_0^{  LAB} \bigr)^2+Q^2}{Q^2} \tan^2 \frac{\theta^{  LAB}_e }{2}\right]^{-1}.
\label{Def_polarization_epsilon}
\end{equation}
Here $x_B=\frac{Q^2}{2 p_N \cdot q}$ and $y=\frac{p_N \cdot q}{p_N \cdot k}$
are the usual dimensionless variables;
${k'}_0^{LAB} $
is the energy of the final state electron in the LAB frame and
$\theta^{LAB}_e $
is the electron scattering angle in the LAB frame.

Within the near-backward kinematics regime, in which the factorized description in terms
of $\pi N$ TDAs and nucleon DAs
applies for the $\gamma^{*} N \rightarrow N' \pi$ subprocess of the reaction
(\ref{Reaction_piN_Sec7}),
only the transverse cross section
$\frac{d \sigma_T}{d t}$
receives contribution to the leading twist-$3$ accuracy.

To express the transverse cross section
$\frac{d  \sigma_T}{d t}$
through the
helicity amplitudes
${\mathcal{M}}^{\lambda_\gamma}_{s_N \, s'_N}$
of $\gamma^{*} N \rightarrow N' \pi$ process
defined in
(\ref{Hel_ampl_def_piN}) we consider the four-fold differential cross section
\begin{equation}
\frac{d^4 \sigma}{ds dQ^2 d \varphi dt}(e N \to e'N'\pi)=
\frac{1}{(2\pi)^4} \frac{1}{128 m_N^2 \left(k_0^{LAB} \right)^2 \Lambda(s, m_N^2,-Q^2)}  \overline{ \big|    {\mathcal{M}}_{eN \to eN'\pi}\big|^2},
\label{CS_formal_d4}
\end{equation} 
where $\Lambda$ is
the Mandelstam function (\ref{Def_lambda}).
The squared amplitude averaged over polarizations of the initial lepton and nucleon reads
\begin{eqnarray}
  &&
\overline{ \big|    {\mathcal{M}}_{eN \to eN'\pi}\big|  ^2}= \frac{1}{4} \sum_{{s_e, \, s'_e \atop  s_N, \, s'_N}}
\big|    {\mathcal{M}}_{eN \to eN'\pi}\big|  ^2
\nn \\ && = \frac{4 \pi \alpha_{em}}{Q^2(1-\varepsilon)}
\sum_{s_N, \, s'_N}
\Bigl\{
\frac{1}{2}
\big(
| {\mathcal{M}}^1_{s_N s'_N}| ^2+
| {\mathcal{M}}^{-1}_{s_N s'_N}| ^2
\big)
+
\varepsilon | {\mathcal{M}}^0_{s_N s'_N}| ^2+\cdots
\Bigr\}.
\end{eqnarray}
Here by $\ldots$ we denote the $\cos{2\varphi}$-  and  $\cos{\varphi}$-modulated terms.
Comparing (\ref{CS_formal_d4}) to (\ref{Def_CS_Kroll}) we obtain the familiar expression
\begin{eqnarray}
  &&
\frac{d \sigma_T}{dt}
+ \varepsilon \frac{d \sigma_L}{dt}+\cdots=\frac{Q^2(1-\varepsilon)}{128 \pi^2 \alpha_{em}}
\frac{1}{(s-m_N^2) \Lambda(s,m_N^2,-Q^2)} \overline{ \big|    {\mathcal{M}}_{eN \to eN'\pi}\big|  ^2} \nn \\ &&
= \frac{1}{32 \pi (s-m_N^2) \Lambda(s,m_N^2,-Q^2)} \sum_{s_N, \, s'_N}
\Bigl\{
\frac{1}{2}
\bigl(
| {\mathcal{M}}^1_{s_N s'_N}| ^2+
| {\mathcal{M}}^{-1}_{s_N s'_N}| ^2
\bigr)
+
\varepsilon | {\mathcal{M}}^0_{s_N s'_N}| ^2+\cdots
\Bigr\}.
\end{eqnarray}

Employing Hand's convention
\cite{Hand:1963bb}
for the  virtual photon flux factor
$K_H= \frac{p_N \cdot q}{m_N}-\frac{Q^2}{2m_N}$
we define
\begin{equation}
\Gamma =
\frac{\alpha_{\rm em} K_H}{2 \pi^2 Q^2} \frac{{k'}_{0 }^L}{k_{0 }^L}
\frac{1}{1-\varepsilon}
=
\frac{\alpha_{\rm em}}{2 \pi^2} \frac{{k'}_{0 }^L}{k_{0 }^L} \frac{s-m_N^2}{2 m_N Q^2} \frac{1}{1-\varepsilon}.
\label{Def_Gamma_Hand}
\end{equation}
Then for the five-fold differential cross section of the reaction  (\ref{Reaction_piN_Sec7})
one gets~\cite{Burkert:2004sk}:
\begin{eqnarray}
  &&
 \frac{d^5 \sigma}{d E' d \Omega_{e'} d \Omega_\pi}(e N \to e'N'\pi)
\nonumber \\ &&
= \Gamma   \frac{\Lambda(s,m^2_N,m_\pi^2)}{128 \pi^2 s (s-m_N^2)}
\sum_{s_N, \, s'_N}
\Bigl\{
\frac{1}{2}
\bigl(
| {\mathcal{M}}^1_{s_N s'_N}| ^2+
| {\mathcal{M}}^{-1}_{s_N s'_N}| ^2
\bigr)
+
\varepsilon | {\mathcal{M}}^0_{s_N s'_N}| ^2+\cdots
\Bigr\}
= \Gamma  \Bigl(  \frac{d^2 \sigma_T}{d \Omega_\pi} +
\varepsilon  \frac{d^2 \sigma_L}{d \Omega_\pi}+\cdots
\Bigr).
\label{CS_working}
\end{eqnarray}
Here,
$d\Omega_{e'}$
is the differential solid angle for the scattered electron in the LAB frame;
$d\Omega_\pi$
is  the differential solid angle of the produced pion in the
$N' \pi$
CMS frame.
By dots we denote the $\cos{2\varphi}$-  and  $\cos{\varphi}$-modulated terms.
Within the reaction mechanism involving $\pi N$ TDAs and nucleon DAs in the near-backward kinematics  these terms
are of subleading twist and are suppressed by powers of $1/Q$.

The $\gamma^{*}N \to \pi N'$ helicity amplitudes
${\mathcal{M}}^{\lambda_\gamma}_{s_N \, s'_N}$
(\ref{Hel_ampl_def_piN})
can be parameterized as
\begin{equation}
\mathcal{M}^{\lambda_\gamma}_{s_N s'_N}=\mathcal{C}_\pi \frac{1}{Q^4} \bar{U}(p'_N,s'_N) \,  \Gamma_H \, U(p_N,s_N),
\label{Def_hel_GH}
\end{equation}
where
\begin{equation}
\Gamma_H= \hat{\mathcal{E}}(q, \lambda_\gamma) \gamma_5 \mathcal{I}^{(1)}-\hat{\mathcal{E}}(q,\lambda_\gamma) \frac{\hat{\Delta}_T}{m_N} \gamma_5 \mathcal{I}^{(2)}
\label{Def_Gamma_H_pi}
\end{equation}
and the overall normalization constant ${\mathcal{C}}_\pi$ is defined in (\ref{Def_C_PiN}).

To sum over the transverse polarizations of the virtual photon
we employ the relation
\begin{equation}
\sum_{\lambda_{\gamma \, T}}  {\mathcal{E}}^\nu(q,\lambda_\gamma) {\mathcal{E}}^{\mu *}(q, \lambda_\gamma)=-g^{\mu \nu}+\frac{1}{(p \cdot n)}(p^\mu n^\nu+p^\nu n^\mu).
\label{Sum_lambda_T}
\end{equation}
For the unpolarized target nucleon we get:
\begin{eqnarray}
  &&
| {\mathcal{M}}_T| ^2=
| {\mathcal{C}}_\pi| ^2 \frac{1}{Q^8}
\sum_{\lambda_{\gamma \,T}} {\rm Tr}
\Bigl\{
(\hat{p}'_N+m_N) \Gamma_H  (\hat{p}_N+m_N) \gamma_0 \Gamma_H^\dagger \gamma_0
\Bigr\}\nonumber \\ &&
=
| {\mathcal{C}_\pi}| ^2 \frac{1}{Q^8} \left(
\frac{2 Q^2 (1+\xi)}{\xi} \big|  \mathcal{I}^{(1)}\big|  ^2
-\frac{2 Q^2 (1+\xi)}{\xi} \frac{\Delta_T^2}{m_N^2} \big|  \mathcal{I}^{(2)}\big|  ^2 +{\mathcal{O}} \left (   Q^0    \right) \, \right),
\end{eqnarray}
where
$\xi$ is the skewness variable (\ref{Def_xiBMP}) defined with respect to the $u$-channel
longitudinal momentum transfer;
and
$\Delta_T^2$ is given by
(\ref{Delta_t2BMP}).

We establish the following formula for the LO unpolarized cross  (\ref{CS_working})
through the coefficients
$\mathcal{I}^{(1,2)}$
introduced in
(\ref{Def_I_k_convolutions}):
\begin{equation}
\frac{d^2 \sigma_T}{d \Omega_\pi}= | \mathcal{C}_\pi| ^2 \frac{1}{Q^6}
\frac{\Lambda(s,m^2_N,m_\pi^2)}{128 \pi^2 s (s-m_N^2)} \frac{1+\xi}{\xi}
\bigl(
| \mathcal{I}^{(1)}| ^2
-  \frac{\Delta_T^2}{m_N^2} | \mathcal{I}^{(2)}| ^2
\bigr).
\label{Work_fla_CS}
\end{equation}

Integrating over $\phi_\pi$ and performing the change of variable
\begin{equation}
dt= \frac{\Lambda(s,m_N^2,-Q^2) \Lambda(s,m_N^2,m_\pi^2) }{2s} d \cos \theta_\pi\,,
\end{equation}
we obtain the differential cross section of $\gamma^{*}N \to \pi N$
\begin{equation}
\frac{d  \sigma_T}{d t}= | \mathcal{C}_\pi| ^2 \frac{1}{Q^6}
\frac{1}{32 \pi  \Lambda(s,m_N^2,-Q^2)  (s-m_N^2)} \frac{1+\xi}{\xi}
\bigl(
| \mathcal{I}^{(1)}| ^2
-  \frac{\Delta_T^2}{m_N^2} | \mathcal{I}^{(2)}| ^2
\bigr).
\label{CS_bkw_dt_scaling}
\end{equation}

\subsubsection{Backward vector meson electroproduction}
\mbox

The set of  cross  section formulas for  vector meson electroproduction
within the one photon approximation
\begin{equation}
e(k,s_e)+ N(p_N,s_N) \rightarrow \bigl( \gamma^{*}(q,\lambda_\gamma) + N(p_N, s_N) \bigr) +e(k',s'_e)
  \rightarrow  e(k',s'_e)+ V(p_V, \lambda_V) + N'(p'_N, s'_N)
\label{Reaction_VN_Sec7}
\end{equation}
is  analogous to that for the pseudoscalar meson electroproduction summarized in
 Sec.~\ref{SubSec_CS_formulas_Bkw_piN}.
The unpolarized cross section is decomposed   into  the contributions
of the transverse ($T$) and longitudinal ($L$) polarizations of the virtual photon
and the transverse--transverse ($TT$) and longitudinal--transverse ($LT$)
interference contributions:
\begin{equation}
\frac{d^5 \sigma}{d E' d \Omega_{e'} d \Omega_V}(e N \to e'N'V)
= \Gamma
\Bigl(  \frac{d^2 \sigma_T}{d \Omega_V} +
\varepsilon  \frac{d^2 \sigma_L}{d \Omega_V}
+
\varepsilon \cos 2\varphi \frac{d^2 \sigma_{TT}}{d \Omega_V}+
\sqrt{2 \varepsilon (1+\varepsilon)} \cos \varphi \frac{d^2 \sigma_{LT}}{d\Omega_V}
\Bigr),
\label{CS_workingVM}
\end{equation}
where $\varphi$ is the angle between the leptonic and the hadronic reaction planes
(see  Fig.~\ref{Fig_Kinematics_piN_CMS}); $d\Omega_V=\sin \theta_V d\theta_V d\varphi_V$ is the solid angle of the produced
vector meson;
and $\Gamma$ is defined in Eq.~(\ref{Def_Gamma_Hand}).

Within the reaction mechanism involving $VN$ TDAs and nucleon DAs in the near-backward
kinematics  only the transverse
$\gamma^{*}N \to N' V$
cross section $\sigma_T$ receives contributions to the
leading twist-$3$ accuracy:
\begin{equation}
\frac{d^2 \sigma_T}{d \Omega_V}=
\frac{\Lambda(s,m^2_N,m_V^2)}{128 \pi^2 s (s-m_N^2)} \frac{1}{2} | {\mathcal{M}}_T| ^2.
\label{Work_fla_CSVM}
\end{equation}
By switching from $\cos \theta_V^{*}$ to the variable $t$ and
integrating over $\varphi_V$
we obtain
\begin{equation}
\frac{d  \sigma_T}{d t}=
\frac{1}{32 \pi  \Lambda(s,m_N^2,-Q^2)  (s-m_N^2)} \frac{1}{2} | {\mathcal{M}}_T| ^2.
\label{Work_fla_CSII}
\end{equation}

The square of the transverse hard
$\gamma^{*}N \to V N$
amplitude
$| {\mathcal{M}}_T| ^2$ in
(\ref{Work_fla_CS}), (\ref{Work_fla_CSII})
involves the summation over nucleon spins, over polarizations of the vector meson
and over the transverse polarization of the virtual photon
\begin{equation}
| {\mathcal{M}}_T| ^2=   \sum_{{\lambda_\gamma}_T, \, \lambda_V, \, s_N, \, s'_N } {\mathcal{M}}_{s_N s'_N}^{\lambda_\gamma \lambda_V} {{\mathcal{M}}_{s_N s'_N}^{\lambda_\gamma \lambda_V}}^{*},
\label{Hel_amp_squared_bkw_vector}
\end{equation}
where
${\mathcal{M}}_{s_N s'_N}^{\lambda_\gamma \lambda_V}$
stand for the helicity amplitudes (\ref{Hel_ampl_def})
of the hard reaction mechanism presented in  Sec.~\ref{SubSec_Bkw_meson_formal}.

The summation
over the transverse  polarizations of the virtual photon in
(\ref{Hel_amp_squared_bkw_vector})
is performed by means of
Eq.~(\ref{Sum_lambda_T}).
To the leading order in $1/Q$ one obtains~\cite{Pire:2015kxa}:
\begin{eqnarray}
  &&
| {\mathcal{M}}_T| ^2=   \sum_{{\lambda_\gamma}_T, \, \lambda_V, \, s_N, \, s'_N } {\mathcal{M}}_{s_N s'_N}^{\lambda_\gamma \lambda_V} {{\mathcal{M}}_{s_N s'_N}^{\lambda_\gamma \lambda_V}}^{*}
\nonumber \\
&&= | {\mathcal{C}}_V| ^2 \frac{1}{Q^6} \frac{2(1+\xi)}{\xi}  \left[  | {\mathcal{I}}^{(1)} | ^2
\left( 1+
\frac{ m_N^2(1-\xi)^2+(1+\xi)^2(m_V^2-\Delta_T^2)}{ m_V^2(1+\xi)^2}
\right)
   \right.
\nonumber \\
&&\left.
+ | {\mathcal{I}}^{(2)}  | ^2 \frac{m_N^2 (1-\xi)^2   }{4 m_V^2  } +
\left(
{\mathcal{I}}^{(1)}   {{\mathcal{I}}^{(2)*}}   +
{\mathcal{I}}^{(1)*}   {{\mathcal{I}}^{(2)}}
\right) \frac{m_N^2 (1-\xi)^2}{2 m_V^2 (1+\xi)}\right.
\nonumber \\ && \left. +\frac{\Delta_T^2}{m_N^2} \left\{
-| {\mathcal{I}}^{(3)}  | ^2\frac{ \left(m_V^2-\Delta_T^2\right)}{
   m_V^2  } +
\left(
{\mathcal{I}}^{(1)}   {{\mathcal{I}}^{(3)*}}   +
{\mathcal{I}}^{(1)*}   {{\mathcal{I}}^{(3)}}
\right)
\frac{m_N^2 (1-\xi)}{ m_V^2 (1+ \xi) }
\right.
\right.
\nonumber  \\ &&
\left.
\left.
 +
\left(
{\mathcal{I}}^{(2)}   {{\mathcal{I}}^{(3)*}}   +
{\mathcal{I}}^{(2)*}   {{\mathcal{I}}^{(3)}}
\right)
\frac{m_N^2(1-\xi)}{2 m_V^2 }+
| {\mathcal{I}}^{(4)}| ^2
\frac{\Delta_T^2(m_V^2-\Delta_T^2)}{m_N^2 m_V^2}
\right.
\right.
\nonumber  \\ &&
\left.
\left.
+\left(
{\mathcal{I}}^{(1)}   {{\mathcal{I}}^{(4)*}}   +
{\mathcal{I}}^{(1)*}   {{\mathcal{I}}^{(4)}}
\right)
\frac{ (m_V^2-\Delta_T^2)}{ m_V^2}-
| {\mathcal{I}}^{(5)}  | ^2
\left( 1+
\frac{ m_N^2(1-\xi)^2+(1+\xi)^2(m_V^2-\Delta_T^2)}{ m_V^2(1+\xi)^2}
\right)
\right.
\right.
\nonumber  \\ &&
\left.
\left.
+\left(
{\mathcal{I}}^{(1)}   {{\mathcal{I}}^{(5)*}}   +
{\mathcal{I}}^{(1)*}   {{\mathcal{I}}^{(5)}}
\right) \frac{2m_N^2(1-\xi)}{m_V^2(1+\xi)}
+\left(
{\mathcal{I}}^{(2)}   {{\mathcal{I}}^{(5)*}}   +
{\mathcal{I}}^{(2)*}   {{\mathcal{I}}^{(5)}}
\right)
\frac{m_N^2(1-\xi)}{2 m_V^2}
\right.
\right.
\nonumber  \\ &&
\left.
\left.
-\left(
{\mathcal{I}}^{(3)}   {{\mathcal{I}}^{(5)*}}   +
{\mathcal{I}}^{(3)*}   {{\mathcal{I}}^{(5)}}
\right)
\frac{m_V^2-\Delta_T^2}{m_V^2}+
\left(
{\mathcal{I}}^{(4)}   {{\mathcal{I}}^{(5)*}}   +
{\mathcal{I}}^{(4)*}   {{\mathcal{I}}^{(5)}}
\right)
\frac{\Delta_T^2(1-\xi)}{m_V^2(1+\xi)}
\right.
\right.
\nonumber  \\ &&
\left.
\left.
-| {\mathcal{I}}^{(6)}  | ^2 \frac{m_N^2(1-\xi)^2}{4m_V^2}-
\left(
{\mathcal{I}}^{(1)}   {{\mathcal{I}}^{(6)*}}   +
{\mathcal{I}}^{(1)*}   {{\mathcal{I}}^{(6)}}
\right)
\frac{m_N^2(1-\xi)}{2 m_V^2}
\right.
\right.
\nonumber  \\ &&
\left.
\left.
-\left(
{\mathcal{I}}^{(4)}   {{\mathcal{I}}^{(6)*}}   +
{\mathcal{I}}^{(4)*}   {{\mathcal{I}}^{(6)}}
\right)
\frac{\Delta_T^2(1-\xi)}{2 m_V^2}+
\left(
{\mathcal{I}}^{(5)}   {{\mathcal{I}}^{(6)*}}   +
{\mathcal{I}}^{(5)*}   {{\mathcal{I}}^{(6)}}
\right)
\frac{m_N^2(1-\xi)^2}{2 m_V^2(1+\xi)}
\right\} +{\mathcal{O}}(1/Q^2)
\right],
\label{M_T_Vect_squared}
\end{eqnarray}
where ${\mathcal{C}}_V$ is the overall normalization constant defined
in
(\ref{Def_C_V})
and ${\mathcal{I}}^{(k)}$, $k=1,\,\ldots,\,6$ stand for the
integral convolutions introduced in (\ref{Hel_ampl_def}).

\subsection{Polarization observables}
\mbox

Asymmetries, being ratios of  cross sections,
are often less sensitive to perturbative QCD corrections.
Therefore, they are usually considered to be  more reliable to test the
factorized description of hard reactions. Moreover, they are often less sensitive to  experimental uncertainties.

\subsubsection{Transverse target single spin asymmetry}
\label{SubSec_STSA}
\mbox

For  backward pion electroproduction  (\ref{Reaction_piN_Sec7}), as for forward exclusive reactions,
an interesting observable is the  transverse target single spin asymmetry (STSA)
\cite{Lansberg:2010mf}.

For this issue we introduce the transverse target polarization dependent amplitude.
To describe the nucleon target transverse polarization we
employ the familiar relation
\begin{equation}
U(p_N,s_N) \bar{U}(p_N,s_N)= \frac{1}{2} (1+ \gamma^5 \hat{s}_N)(p_N+m_N),
\end{equation}
where the axial vector  $s_N$ describing the target nucleon spin
has only spatial non-zero components in the target rest frame and
\begin{equation}
\bar{U}(p_N,s_N) \gamma^\mu \gamma_5 {U}(p_N,s_N)=m_N s^\mu_N; \ \  s_N^2=-1; \ \ s_N \cdot p_N=0.
\end{equation}
Since we limit ourselves to the leading twist-$3$ contribution we sum over the
transverse polarizations of the virtual photon.  Summing over the spin of the outgoing nucleon we obtain:
\begin{equation}
| \mathcal{M}_{T\,s_N}| ^2=| \mathcal{C}_\pi| ^2 \frac{1}{Q^8}
\sum_{\lambda_{\gamma \,T}}
{\rm Tr}
\Bigl\{
(\hat{p}'_N+m_N) \Gamma_H \frac{1+ \gamma_5 \hat{s}_N}{2} (\hat{p}_N+m_N) \gamma_0 \Gamma_H^\dagger \gamma_0
\Bigr\},
\label{Trace_to_compute}
\end{equation}
where $\Gamma_H$ is defined in
(\ref{Def_Gamma_H_pi}) and $\mathcal{C}_\pi$ is the overall normalization constant (\ref{Def_C_PiN}).
The nucleon spin dependent part of the trace
(\ref{Trace_to_compute}) reads
\begin{eqnarray}
  &&
\sum_{\lambda_{\gamma \,T}} {\rm Tr}
\Bigl\{
(\hat{p}'_N+m_N) \Gamma_H \frac{\gamma_5 \hat{s}_N}{2}  (\hat{p}_N+m_N) \gamma_0 \Gamma_H^\dagger \gamma_0
\Bigr\}  =2 \frac{Q^2(1+\xi)}{m_N \xi} \varepsilon(n, p, s_N, \Delta_T)
 \bigl( -i \mathcal{I}^{(1)} (\mathcal{I}^{(2)})^{*}+i \mathcal{I}^{(2)} (\mathcal{I}^{(1)})^{*} \bigr)
\nonumber \\ &&
=-2 \frac{Q^2(1+\xi)}{ \xi} \frac{| \Delta_T| }{m_N}| \vec{s}_N|  \sin(\varphi -\varphi_s) {\rm  Im}(\mathcal{I}^{(2)}(\mathcal{I}^{(1)})^{*}),
\label{Trace_2}
\end{eqnarray}
where $\varepsilon(n, p, s_N, \Delta_T)$ is the contraction of the corresponding
four-vectors with the Levi-Civita tensor.  We employ the conventions of~\cite{Itzykson}, in which
$  \varepsilon^{0123}=  1$ with $\gamma_5=- \frac{i}{4!} \varepsilon^{\mu \nu \rho \sigma} \gamma_\mu \gamma_\nu \gamma_\rho \gamma_\sigma $.
In the last line of (\ref{Trace_2}), we consider  $s_N$ as being purely transverse.
With the axis definition of  Fig.~\ref{Fig_Kinematics_piN_CMS} with unit vectors
$e_x=(0,1,0,0)$ and $e_y=(0,0,1,0)$ we define
\begin{equation}
{\Delta}_T=| \Delta_T| ( \cos \varphi e_x+ \sin \varphi e_y ); \ \ \
s_N \equiv {s_N}_T=\cos \varphi_s e_x+ \sin \varphi_s e_y\,.
\end{equation}
Using
\begin{equation}
\varepsilon(n, p, s_1, \Delta_T)= \frac{1}{2} | \Delta_T|  | \vec{s}_{N}|  \sin(\varphi -\varphi_s),
\end{equation}
where
$\varphi$
is the angle between the leptonic and hadronic planes; and
$\varphi_s$
is the angle between the leptonic plane and the transverse
target spin (see  Fig.~\ref{Fig_Kinematics_piN_CMS}),
we conclude:
\begin{equation}
| \mathcal{M}_{T\, s_N} | ^2=| \mathcal{C}_\pi| ^2 \frac{1}{Q^6} \frac{  (1+\xi)}{\xi}
\left(
 | \mathcal{I}^{(1)}| ^2
- \frac{\Delta_T^2}{m_N^2} | \mathcal{I}^{(2)}| ^2-2 \frac{| \Delta_T| }{m_N}| \vec{s}_N|   \, {\rm Im} (\mathcal{I}^{2}(\mathcal{I}^{(1)})^{*}) \sin (\varphi-\varphi_s)
\right)+ O(1/Q^8).
\end{equation}
The transverse target single spin asymmetry is defined as:
\begin{equation}
\mathcal{A}_{\rm STSA}= \frac{1}{| \vec{s}_N| }
\left(
\int_0^\pi d \tilde{\varphi} | \mathcal{M}_{T \, s_N}| ^2 - \int_\pi^{2\pi} d \tilde{\varphi} | \mathcal{M}_{T \, s_N}| ^2
\right) \left(
\int_0^{2\pi} d \tilde{\varphi} | \mathcal{M}_{T \, s_N} | ^2
\right)^{-1},
\label{Def_asymmetry}
\end{equation}
where
$\tilde{\varphi} \equiv \varphi-\varphi_s$.
The prediction for the STSA within the factorization framework involving
$\pi N$ TDAs and nucleon DAs reads
\begin{equation}
\mathcal{A}_{\rm STSA}= -\frac{4}{\pi} \frac{\frac{| \Delta_T| }{m_N}  \, {\rm Im} (\mathcal{I}^{(2)}(\mathcal{I}^{(1)})^{*})}{| \mathcal{I}^{(1)}| ^2
- \frac{\Delta_T^2}{m_N^2} | \mathcal{I}^{(2)}| ^2}.
\label{STSA_TDA}
\end{equation}

\bi
\item The factorized reaction mechanism predicts a $Q^2$-independent
STSA. A sizeable  $Q^2$-independent  STSA can be, therefore, seen as
a legitimate marking sign of the relevance of the factorized
reaction mechanism for hard exclusive meson electroproduction
within the near-backward kinematics regime.

\item
The STSA (\ref{STSA_TDA})
turns out to be directly sensitive to the imaginary parts of integral
convolutions $\mathcal{I}^{(k)}$. These imaginary parts originate
at leading order in $\alpha_s$
from the
integration over the cross-over trajectories $x_i=0$, separating
the ERBL-like and DGLAP-like support regions of $\pi N$ TDAs (see  Sec.~\ref{SubSec_Support}),
and over the $x_i=2\xi$ trajectories that lie entirely within the DGLAP-like support domain.
\ei

\subsubsection{Beam spin asymmetry}
\label{SubSec_Def_BSA}
\mbox

Experimental studies of hard exclusive meson
electroproduction
reactions employing polarized lepton beams can provide a variety of
observable quantities sensitive to the onset of the collinear
factorization regime both in the near-forward kinematics, where
a description in terms of   nucleon
GPDs is anticipated, and in the near-backward kinematics,
where we propose a description in terms of nucleon-to-meson TDAs.

This opportunity was recently considered
by
S.~Diehl and collaborators in Ref.~\cite{Diehl:2020uja}
presenting the extraction of Beam-Spin Asymmetry (BSA) of
hard exclusive pion electroproduction reaction
\begin{equation}
\vec{e}(k,h)+ p(p_N,s_N) \rightarrow \bigl( \gamma^{*}(q,\lambda_\gamma) + p(p_N, s_N) \bigr) +e(k')
  \rightarrow  e(k')+ \pi^+(p_\pi) + n(p'_N, s'_N)
\label{Def_react_polarized_beam_New}
\end{equation}
in a wide range of kinematics.

Taking into account the polarization of the electron beam a general
parametrization of the cross section of the reaction
(\ref{Def_react_polarized_beam_New})
within the one photon exchange approximation
can be written as~\cite{Arens:1996xw}:
\begin{equation}
\frac{d^5 \sigma}{d E' d \Omega_{e'}  d \Omega_{\pi}} =\Gamma \frac{d^2 \sigma}{d \Omega_{\pi}},
\end{equation} 
where $\Gamma$ is the virtual photon flux
(\ref{Def_Gamma_Hand})
and
\begin{eqnarray}
  &&
\frac{d^2 \sigma }{d \Omega_{\pi}}=\frac{d^2 \sigma_{{T}}}{d \Omega_{\pi}}+
\varepsilon
\frac{d^2 \sigma_{{L}}}{d \Omega_{\pi}}+
\sqrt{2\varepsilon(1+\varepsilon)}
\frac{d^2 \sigma_{{LT}}}{d \Omega_{\pi}} \cos \varphi+\varepsilon \frac{d^2 \sigma_{{TT}}}{d \Omega_{\pi}} \cos 2 \varphi  \nn \\ &&
+h\sqrt{2
\varepsilon
(1-\varepsilon)} \frac{d^2 \sigma_{{LT'}}}{d \Omega_{\pi}}
\sin \varphi
+h
\sqrt{1-\varepsilon^{2}}
\frac{d^2 \sigma_{{TT}^{\prime}}}{d \Omega_{\pi}}.
\label{CS_polarized_beam_detailed}
\end{eqnarray}
Here $\varphi$ is the angle between the leptonic and hadronic reaction planes
(see  Fig.~\ref{Fig_Kinematics_piN_CMS}),
$\varepsilon$ is the polarization parameter
(\ref{Def_polarization_epsilon})
and $h$ refers to the electron beam {$+$}  or $-$ helicity.
We refer the reader to Refs.~\cite{Arens:1996xw,Amaldi:1979vh,Drechsel:1992pn}
for the detailed account of the angular structure of the cross section and definition
of angular moments. We would like to stress that the conventions adopted in
Ref.~\cite{Arens:1996xw} differ from that of Ref.~\cite{Drechsel:1992pn}
by the definitions of the polarization vector of the longitudinally polarized
virtual photon.
As the result, the longitudinal and
longitudinal--transverse interference
cross sections defined in the two conventions turn to differ:
\begin{equation}
\sigma_L\Big|_{\text{\cite{Arens:1996xw}}}=
\frac{2 x_B m_N^2}{Q^2}\sigma_L\Big|_{\text{\cite{Drechsel:1992pn}}}; \ \ \
\sigma_{LT, \, LT'}\Big|_{\text{\cite{Arens:1996xw}}}=\frac{\sqrt{2 x_B} m_N}{Q} \sigma_{LT, \, LT'}\Big| _{\text{\cite{Drechsel:1992pn}}}.
\end{equation}

The key observable quantity of the analysis~\cite{Diehl:2020uja}
is the
$\sin \varphi$
moment of the cross  (\ref{CS_polarized_beam_detailed})
$A_{L U}^{\sin \varphi}$,
where the subscript $L$ refers to the longitudinal polarization of
the beam and the subscript $U$ refers to the unpolarized state of the target.
This quantity can be accessed by measuring the BSA defined as:
\begin{equation}
{\mathcal{A}}_{\rm BSA}\left(t, \varphi, x_{B}, Q^{2}\right) =\frac{d \sigma^{+}-d \sigma^{-}}{d \sigma^{+}+d \sigma^{-}}  =\frac{
\sqrt{2\varepsilon(1-\varepsilon)}
d \sigma_{LT'} \sin \varphi +\sqrt{1-\varepsilon^2} d\sigma_{TT'}}{d \sigma_T +\varepsilon d \sigma_L+
\sqrt{2\varepsilon(1+\varepsilon)} d \sigma_{LT} \cos \varphi+ \varepsilon  d \sigma_{TT'} \cos{2\varphi}
}\,,
\label{Def_BSA}
\end{equation}
where $\sigma^\pm$ stand for the differential cross section for
beam helicity parallel or antiparallel to the beam direction.
Experimentally it turns out to be easier to extract the
$\sin \varphi$-modulated harmonics $A_{L U}^{\sin \varphi}$ defined as\footnote{The subscripts $LU$ of the harmonics refer to the longitudinal
polarization of the beam $L$ and unpolarized
($U$) target.}:
\begin{equation}
A_{L U}^{\sin \varphi}=\frac{\sqrt{2 \varepsilon(1-\varepsilon)}
d\sigma_{L T^{\prime}}}{d\sigma_{T}+\varepsilon d\sigma_{L}},
\label{ALUsin_phi}
\end{equation}
where  $\sigma_{L}$ and $\sigma_{T}$ denote the unpolarized cross sections
with longitudinally and transversely polarized virtual photon
(\textit{cf.} Eq.~(\ref{Def_CS_Kroll})).

The BSA
(\ref{Def_BSA})
turns out to be an extremely convenient observable
to address the onset of the collinear factorization regime for the
reaction (\ref{Def_react_polarized_beam_New}).
For both the near-forward and the near-backward collinear factorization reaction mechanisms, the interference cross section
$\sigma_{LT'}$
occurring in
(\ref{ALUsin_phi})
turns to be a subleading twist effect.
Thus, for sufficiently large $Q^2$, both in the near-forward and near-backward kinematical regimes, once the factorization is achieved, the
$A_{L U}^{\sin \varphi}$
must decrease as $1/Q^2$. Therefore, a small value of
$A_{L U}^{\sin \varphi}$
decreasing with a growth of $Q^2$ when $Q^2 >Q^2_{min}$ is a distinctive marking sign
for the onset of the collinear factorization regime. However, this onset does not have to appear  at the same value of $Q^2_{min}$ in the near-forward and in the near-backward
kinematics.

A quantitative theoretical estimate of $A_{L U}^{\sin \varphi}$
(\ref{ALUsin_phi})
represents a formidable task as this requires a proper mastering
of higher twist effects and reliable theoretical models for
poorly known non-perturbative quantities.
In the near forward regime, it has been suggested in Ref.~\cite{Ahmad:2008hp,Goloskokov:2011rd}
that the interference cross section
$\sigma_{LT'}$ can be expressed
through the convolutions
of chiral-odd and chiral-even
nucleon GPDs with corresponding subprocess amplitudes
that turn to be of twist-$2$ for the longitudinal and of twist-$3$ for the
transverse amplitudes:
\begin{equation}
\sigma_{L T^{\prime}} \sim \operatorname{Im}\left[\left\langle\bar{E}_{T-\text { eff }}\right\rangle^{*}\left\langle\tilde{H}_{\text {eff }}\right\rangle+\left\langle H_{T-\text { eff }}\right\rangle^{*}\left\langle\tilde{E}_{\text {eff }}\right\rangle\right].
\end{equation}

To provide an estimate of $A_{L U}^{\sin \varphi}$
in the near-backward regime
it is necessary to consider the interference
between the leading twist transverse amplitude of
the convolution in terms of twist-$3$ $\pi N$ TDAs ($H^{\rm tw.3}_{\pi N}$) and
nucleon DAs ($\phi^{\rm tw.3}_N$) and the next-to-leading order subprocess
longitudinal amplitude of the convolution involving
twist-4 TDAs ($H^{\rm tw.4}_{\pi N}$) and DAs ($\phi^{\rm tw.4}_N$).
Symbolically this can be written as
\begin{equation}
\sigma_{L T^{\prime}} \sim \operatorname{Im}\left[\left\langle H_{i}^{\mathrm{tw}. 3} \phi_{j}^{\mathrm{tw} 3}\right\rangle\left(\left\langle H_{i}^{\mathrm{tw}. 4} \phi_{j}^{\mathrm{tw}. 3}\right\rangle+\left\langle H_{i}^{\mathrm{tw}. 3} \phi_{j}^{\mathrm{tw}. 4}\right\rangle\right)^{*}\right].
\end{equation}
Detailed theoretical studies of corresponding  twist-$4$ longitudinal
amplitude require non-perturbative input to build up phenomenological models for
twist-$4$ nucleon DAs
\cite{Braun:1999te,Braun:2000kw}
and twist-$4$  $\pi N$ TDAs. These latter quantities were never considered in the
literature and remain totally unconstrained theoretically.
A related problem has recently been considered~\cite{Kivel:2019wjh} for the analysis of the $J/\psi \to  \bar{N}N$ decay including twist-$4$ effects.

\subsection{Cross sections for antinucleon
initiated reactions}
\label{SubSec_CS_formuls_cross_ch}
\mbox

In this section we summarize the cross section formulas for the
antinucleon initiated reactions
that we address within the collinear factorization framework
in terms of nucleon-to-meson TDAs and nucleon DAs in  Sec.~\ref{SubSec_Cross_Ch_Excl_R}.

\subsubsection{Nucleon--antinucleon annihilation
into a lepton pair in association with a light meson }
\label{SubSec_NbarNlpair_meson}
\mbox

\begin{figure}[H]
 \begin{center}
 \includegraphics[width=12cm]{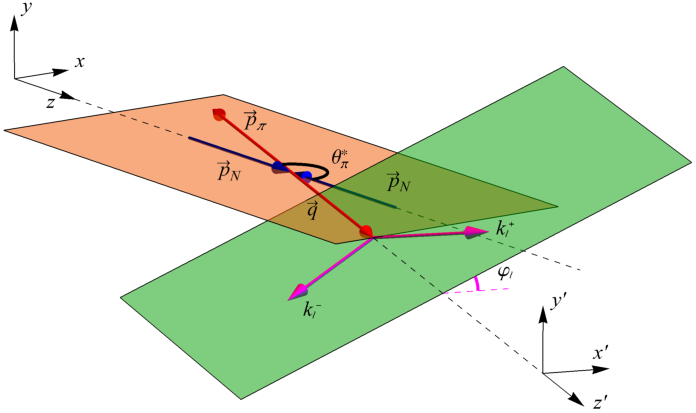}
   \end{center}
     \caption{The kinematics of the reaction (\ref{React_PANDA}) in the $\bar{N}N$ center-of-mass frame in the near-backward regime. We choose the coordinate system $(x,y,z)$ with the $z$-axis aligned with the antinucleon three-momentum. The coordinate system $(x',y',z')$ is defined with the $z'$-axis aligned with the three-momentum of the virtual photon $\vec{q}$. $\varphi_\ell$ is the lepton pair azimuthal angle (the angle between the hadronic and leptonic reaction planes). Lepton pair polar angle $\theta_\ell$  is defined in the rest frame of the virtual photon. The transition from the $\bar{N}N$  CMS to $\gamma^{*}(q)$ rest frame is performed by the appropriate boost in the $z'$ direction.}
\label{Fig_Kinematics_PANDA_CMS}
\end{figure}

Let us first review the formulas for the cross section of nucleon--antinucleon annihilation
into a lepton pair in association with a pion
\begin{equation}
N(p_N,s_N)+\bar{N}(p_{\bar{N}},s_{\bar{N}})
\rightarrow
\gamma^{*}(q,\lambda_\gamma)+\pi(p_\pi)
\rightarrow
\ell^+(k_{\ell^+})+\ell^-(k_{\ell^-})+\pi(p_\pi)
\label{React_PANDA}
\end{equation}
presented in Refs.~\cite{Lansberg:2007se,Lansberg:2012ha}.
The summary of kinematical formulas relevant for the near-backward and
near forward kinematical regimes in which a factorized description in terms
of $\pi N$ TDAs and nucleon DAs applies is given in  Sec.~\ref{SubSec_PANDA_pi_gamma_star}.
Below, for definiteness, we address the near-backward kinematical regime
(large time like photon virtuality $q^2 \equiv Q^2$, large $W^2 \equiv (p_N+p_{\bar{N}})^2$ with $Q^2$ of order of $W^2$; $| u|  \sim 0$)); therefore,
$\Delta=p_\pi-p_N$; $\Delta^2 \equiv u$ and
the skewness variable
$\xi$ refers to the $u$-channel scattering regime variable
(\ref{def_xiu}).
The kinematics of the reaction (\ref{React_PANDA}) in the $\bar{N} N$ center-of-mass frame
is presented in  Fig.~\ref{Fig_Kinematics_PANDA_CMS}. We choose the direction
of the $z$-axis along the three momentum of the antinucleon $\vec{p}_{\bar{N}}$.

A general formula for the unpolarized differential cross section of the reaction
(\ref{React_PANDA}) reads
(see {\it} e.g.~\cite{Borodulin:2017pwh}):
\begin{equation}
d \sigma = \frac{(2 \pi)^4}{2
{\Lambda(W^2,m_N^2,m_N^2)}} | \overline{\mathcal{M}_{ N \bar{N} \rightarrow \ell^+ \ell^- \pi}}| ^2
\delta^{(4)}(p_N+p_{\bar{N}}-p_\pi-k_{\ell^+}-k_{\ell^-})
 \frac{d^3p_\pi}{(2\pi)^3 2E_\pi} \frac{d^3k_{\ell^+}}{(2\pi)^3 2E_{\ell^+}} \frac{d^3k_{\ell^-}}{(2\pi)^3 2E_{\ell^-}},
\end{equation}
where
$\Lambda$
is the Mandelstam function (\ref{Def_lambda}).
The $3$-particle
differential Lorentz invariant phase space
is decomposed into two $2$-particle phase subspaces: those of
$\gamma^{*} \pi$
and
$\ell^+ \ell^-$
systems. The former can be easily computed in
the $\bar{N} N$
CMS while the latter is computed in
the $\ell^+ \ell^-$
CMS yielding the result:
\begin{equation}
d \sigma = \frac{1}{2 (2 \pi)^5
{\Lambda(W^2,m_N^2,m_N^2)}}
| \overline{\mathcal{M}_{ N \bar{N} \rightarrow \ell^+ \ell^- \pi }}| ^2
\frac{d \Omega^{*}_\pi}{8 W^2 }
{\Lambda(W^2,Q^2,m_\pi^2)} \frac{d \Omega_\ell}{8} dQ^2,
\end{equation}
where
$d\Omega^{*}_\pi \equiv  d \cos \theta_\pi^{*} d \varphi_\pi^{*}$
is the pion solid angle
in the $N \bar{N}$ CMS (corresponding to the rest frame of the virtual photon).
By $d \Omega_\ell
\equiv
d \cos \theta_\ell d \varphi_\ell$
we denote the produced lepton solid angle in $\ell^+ \ell^-$ CMS. By expressing
$ \cos \theta_\pi^{*}$
through
$u = (p_{N}-p_\pi)^2$
\begin{equation}
du = \frac{d \cos \theta_\pi^{*}}{2 W^2}
{\Lambda(W^2,m^2_N,m_N^2)}
{\Lambda(W^2,Q^2,m_\pi^2)}
\end{equation}
and  integrating over the azimuthal angle
$\varphi_\pi^{*}$
of the produced pion and over the azimuthal angle of the lepton
$\varphi_\ell$
the following formula for the unpolarized differential cross section of the reaction
(\ref{React_PANDA})
is established:
\begin{equation}
\frac{d \sigma}{du d Q^2 d \cos \theta_{\ell}}=
 \frac{\int d \varphi_\ell | \overline{\mathcal{M}_{N \bar{N} \rightarrow \ell^+ \ell^- \pi}}| ^2 }
 {64 W^2 (W^2-4m_N^2) (2 \pi)^4}.
 \label{Unpol_CS_PANDA}
\end{equation}

The average-squared amplitude
$| \overline{\mathcal{M}^{ N \bar{N} \rightarrow \ell^+ \ell^- \pi }}| ^2$
is expressed through the helicity amplitudes
$\mathcal{M}^{\lambda}_{s_N s_{\bar{N}}}$
(\ref{Def_ampl_MPANDA})
of the hard subprocess
$N \bar{N} \rightarrow \gamma^{*} \pi$:
\begin{equation}
| \overline{\mathcal{M}_{ N \bar{N} \rightarrow \ell^+ \ell^- \pi }}| ^2=
\frac{1}{4}
\sum_{s_N, \, s_{\bar{N}}, \, \lambda_\gamma\, \lambda'_\gamma}
\frac{1}{Q^4}
e^2
\mathcal{M}^{\lambda_\gamma}_{s_N s_{\bar{N}}}
 {\rm Tr}
\left\{
\hat{k}_{\ell^-} \hat{\mathcal{E}}(q,\lambda_\gamma) \hat{k}_{\ell^+} \hat{\mathcal{E}}^{*}(q,\lambda'_\gamma)
\right\}
\left( \mathcal{M}^{\lambda'_\gamma}_{s_N s_{\bar{N}}} \right)^{*},
\label{WF_Cross_sec}
\end{equation}
where the factor $\frac{1}{4}$ corresponds to averaging over polarizations
of the initial state nucleon and antinucleon.

Within the factorized description in terms of $\pi N$ TDAs (and nucleon DAs)
to the leading twist accuracy, only the contribution of
transverse polarization of the virtual photon
is relevant.
After computing the relevant trace in
(\ref{WF_Cross_sec})
in the
$\ell^+ \ell^-$
CMS,
and integrating over the lepton polar angle $\varphi_\ell$,
we obtain:
\begin{equation}
 \int d \varphi_{\ell} \, \vert \overline{\mathcal{M}^{ N \bar{N} \rightarrow \ell^+ \ell^- \pi }}\vert ^2
\Big|\mbox{}_{{\rm Leading }\,{\rm  twist}-3}=
\vert \overline{\mathcal{M}_{T}}\vert ^2 \frac{2 \pi e^2(1+\cos^2 \theta_\ell)}{Q^2},
\label{MT_squared_PANDA}
\end{equation}
where
\begin{eqnarray}
  &&
| \overline{\mathcal{M}_{T}}| ^2 = \frac{1}{4} \sum_{s_N, \, s_{\bar{N}}, \, \lambda_{\gamma \,T}}
\mathcal{M}^{\lambda}_{s_N s_{\bar{N}}}
\left(\mathcal{M}^{\lambda}_{s_N s_{\bar{N}}} \right)^{*} \nn \\ &&
=\frac{1}{4}
| {\mathcal{C}}_\pi| ^2 \frac{1}{Q^6} \left(
\frac{2  (1+\xi)}{\xi} \big|  \mathcal{J}^{(1)}\big|  ^2
-\frac{2 (1+\xi)}{\xi} \frac{\Delta_T^2}{m_N^2} \big|  \mathcal{J}^{(2)}\big|  ^2 +{\mathcal{O}} \left (   1/Q^2    \right) \, \right).
\label{M_T_piN_squared_PANDA}
\end{eqnarray}
Here $\mathcal{J}^{(1,2)}$ are the integral convolutions defined in
(\ref{Def_JandJprime})
and
$\mathcal{C}_\pi$
is the overall normalization constant (\ref{Def_C_PiN}).

The unpolarized cross section formulas (\ref{Unpol_CS_PANDA}), (\ref{MT_squared_PANDA}) can be easily generalized
for the reaction involving a vector meson:
\begin{equation}
N(p_N,s_N)+\bar{N}(p_{\bar{N}},s_{\bar{N}})
\rightarrow
\gamma^{*}(q)+V(p_V, \lambda_V)
\rightarrow
\ell^+(k_{\ell^+})+\ell^-(k_{\ell^-})+V(p_V, \lambda_V).
\label{React_PANDA_Vector_Meson}
\end{equation}
This implies replacing (\ref{M_T_piN_squared_PANDA}) with the
cross conjugated counterpart of Eq.~(\ref{M_T_Vect_squared}) averaged over polarizations
of the initial state nucleon and antinucleon. The relation between the
integral convolutions of the electroproduction channel $\mathcal{I}^{(k)}$
and those of the nucleon--antinucleon annihilation channel $\mathcal{J}^{(k)}$
is completely analogous to the pion case. Namely,
the regulating prescriptions
in the denominators of hard scattering kernels $D_\alpha$ listed in Table~\ref{Table_Bkw_vector}
are replaced according to
$-i 0 \rightarrow  i 0$. The overall phase factor ${{\eta_N^{*}} }^{-1} \eta_q^{-3}$ is obviously irrelevant for the squared amplitude.

\subsubsection{Nucleon--antinucleon annihilation
into a charmonium state in association with a pion}
\mbox

In this section we present the unpolarized cross section formulas for
the nucleon--antinucleon annihilation into a $J/\psi$ in association with a pion
\begin{equation}
\bar N (p_{\bar N},s_{\bar{N}}) \; + \; N (p_N,s_N) \; \to  J/\psi(p_{\psi}, \lambda_\psi)\;+\; \pi(p_{\pi})
\label{React_Jpsi_PANDA_repeat}
\end{equation}
within the collinear factorized description presented in  Sec.~\ref{SubSec_JiPsi}.
We present the result for the near-backward kinematical regimes, so
$\Delta=p_\pi-p_N$; $\Delta^2 \equiv u$ and
the skewness variable
$\xi$ refers to the $u$-channel scattering regime variable $\xi^u$
(\ref{Def_xi_tu_Jpsi}).

The squared amplitude
(\ref{Amplitude_master})
averaged over spins of the initial particles reads
\begin{equation}
| \overline{{\mathcal{M}}^{\lambda_\psi }}| ^2= \frac{1}{4} \sum_{s_N s_{\bar{N}}}
{\mathcal{M}}_{s_N s_{\bar{N}}}^{\lambda_\psi} ({\mathcal{M}}_{s_N s_{\bar{N}}}^{\lambda_\psi})^{*}.
\end{equation}
To the leading twist-$3$ accuracy only the transverse polarization of
$J/\psi$
is relevant.
Summing over the transverse polarization with help of
\begin{equation}
\sum_{\lambda_{\psi \,T}}  {\mathcal{E}}^\nu_\psi(\lambda_\psi) {\mathcal{E}^{*\mu}_\psi}(\lambda_\psi)
=-g^{\mu \nu}+\frac{1}{(p \cdot n)}(p^\mu n^\nu+p^\nu n^\mu),
\end{equation}
we get
\begin{equation}
| \overline{\mathcal{M}_{T}}| ^2 \equiv \sum_{\lambda_{\psi \, T}}
| \overline{{\mathcal{M}}^{\lambda }}| ^2
=
\frac{1}{4} | \mathcal{C}_\psi| ^2 \frac{2(1+\xi)}{\xi {\bar{M}}^8}  \left( | \tilde{\mathcal{J}}^{(1)}(\xi, \Delta^2)| ^2 - \frac{\Delta_T^2}{m_N^2} | \tilde{\mathcal{J}}^{(2)}(\xi, \Delta^2)| ^2 \right),
\label{M_T_Jpsi_squared}
\end{equation}
where the integral convolutions
$\tilde{\mathcal{J}}^{(1,2)}$
are defined in
(\ref{Amplitude_result_I}),
(\ref{Amplitude_result_Ip})
and
$\mathcal{C}_\psi$ is the overall normalization factor (\ref{Def_C_Jpsi}).

Therefore, to the leading twist-$3$ accuracy the differential cross section of
the reaction (\ref{React_Jpsi_PANDA_repeat})
reads
\begin{equation}
\frac{d \sigma}{d u}= \frac{1}{16 \pi \Lambda^2(W^2,m_N^2,m_N^2) } | \overline{\mathcal{M}_{T}}| ^2,
\label{CS_def_delta2}
\end{equation}
where $| \overline{\mathcal{M}_{T}}| ^2$ is given by (\ref{M_T_Jpsi_squared})
and $\Lambda$ is the Mandelstam function (\ref{Def_lambda}).

\subsection{Experimental evidences for the onset of factorized regime: scaling laws,
L-T separation, asymmetries}
\label{SubSec_Exp_evidences}
\mbox

The collinear factorized description of hard exclusive meson
electroproduction reaction in terms of nucleon GPDs and meson DAs
in the forward kinematical region  turns to be challenging both
for the case of production of pseudoscalar and vector mesons.
This issue has been extensively discussed in literature, see \textit{e.g.}
~\cite{Goloskokov:2011rd,Kroll:2012sm,Goldstein:2013gra,Favart:2015umi}.
In particular, the dominance of the leading twist longitudinal amplitude for
$\pi^0$
hard electroproduction contradicts the experimental data up to very high values of $Q^2$
\cite{Bedlinskiy:2012be,Dlamini:2020ulg}.

By no doubts, the experimental verification of the validity of the collinear factorized
description of hard exclusive meson
electroproduction reactions in the near-backward kinematical regime in terms
of nucleon-to-meson TDAs and nucleon DAs will turn to be
at the least equally challenging.
The rapid fall of the cross section with growth of $Q^2$  together with
the unknown magnitudes of subleading twist and NLO corrections that are
hard to estimate theoretically make the identification
of the signal from the leading twist mechanism even more endeavouring.
Therefore, it is essential to carefully specify the marking signs of reaching
the corresponding collinear factorization regime and point out the observable
quantities for which one may expect to check the early onset of the appropriate scaling behavior.

Below we list the specific distinguishing features
of the collinear factorized description of near-backward
hard exclusive meson electroproduction reactions
that may be helpful to provide reliable hints for the
validity of the TDA-based description.
\bi

\item The key prediction of the reaction mechanism involving
nucleon-to-meson TDAs and nucleon DAs for near-backward
meson hard electroproduction reaction is the manifestation
of a characteristic backward peak of the cross section.

\item Twist counting rules result in the
scaling behavior of the unpolarized cross  (\ref{Def_CS_Kroll})
in $Q^2$ for fixed $\xi$ (or, equivalently, $x_B$)
and specific counting rules for different
cross section contributions.
Particularly, from Eq.~(\ref{CS_bkw_dt_scaling}) it can be concluded\footnote{Note that $s$ scales like $Q^2$ for fixed $\xi$. }
 the transverse cross section
$\frac{d\sigma_T}{dt}$ scales like $ Q^{-10}$ for fixed $\xi$.
The longitudinal cross section
$\frac{d\sigma_L}{dt}$
and the interference cross sections are
suppressed (at least) by an extra  $Q^{-2}$ power.
This has to be compared with the specific behavior of the
longitudinal cross section
$\frac{d\sigma_L}{dt} \sim Q^{-6}$
in the near-forward kinematics within the description based on
nucleon GPDs and meson DAs.

\item Dominance of the transverse cross section $\sigma_T$
with respect to the longitudinal cross section $\sigma_L$
and the interference cross sections $\sigma_{LT}$, $\sigma_{LT'}$,
$\sigma_{TT}$. This makes essential the complete Rosenbluth-type
separation of the experimental cross section to control the predictions
of the twist counting rules. Considerable progress in that direction
achieved for the case of  near-backward  $\omega$-meson production is
reported in~\cite{Li:2019xyp}. A similar analysis is planned in the
framework of the dedicated experiment for the pion channel~\cite{Li:2020nsk}.

\item The polarization observables can be extremely helpful to scrutinize the onset of the leading twist regime.
   In some cases of forward electroproduction reactions, certain polarization asymmetries turn out to be less sensitive to NLO corrections than cross-sections, and have been shown to exhibit an early onset of scaling behavior already at the
rather low values of $Q^2$ available with present day fixed target experimental setups.
Although this argument is far from being a proof and should not prevent from carrying the detailed NLO studies,
such a behavior may be also expected in the backward regime.
In particular, the TDA-based reaction mechanism
predicts a non-vanishing and $Q^2$-independent Transverse Target Single Spin Asymmetry
(see  Sec.~\ref{SubSec_STSA}).
Preliminary estimates within the two-component $\pi N$ TDA model suggest
$10 \div 15$\% STSA for
$\gamma^{*} N_T \to \pi N$ for the $x_B$-range typical for JLab, see  Sec.~\ref{SubSec_Model_P_JLab}.
Another useful polarization observable is the beam spin asymmetry BSA
for hard exclusive meson electroproduction (see  Sec.~\ref{SubSec_Def_BSA}).
Despite the fact that the BSA is a subleading twist quantity within the TDA-based framework, the study
of its $Q^2$-dependence can help to uncover hints for the validity of the collinear factorized description
both in the near-forward and near backward regimes. Recently the first extraction of BSA
for hard exclusive pion electroproduction in a broad range of $-t$ was reported in
Ref.~\cite{Diehl:2020uja} (for an overview see  Sec.~\ref{SubSec_First_R_pions_JLab}).
\ei

Essential complementary evidences for the relevance of the collinear factorized
mechanism involving nucleon-to-meson TDAs and nucleon DAs can be provided
by the study of the cross-channel antinucleon beam induced reactions at \=PANDA
and pion beam induced reactions at J-PARC, see  Sections \ref{SubSec_Cross_Ch_Excl_R}, \ref{SubSec_CS_formuls_cross_ch}.
\bi
\item The suggested reaction mechanism predicts the appearance
of pronounced forward and backward cross section peaks for
$\bar{N}N \to \gamma^{*} {\mathcal{M}}$
and $\bar{N}N \to J/\psi {\mathcal{M}}$ reactions.
Charge conjugation symmetry results in the exact symmetry between
the forward and backward peaks. Twist counting rules
result in dominance of the transverse cross section in the vicinity
of the peaks and provide specific $Q^2$-scaling predictions for the
case of $\bar{N}N \to \gamma^{*} {\mathcal{M}}$ reactions.

\item In pion nucleon collisions, the TDA based factorization  framework ensures the presence of backward cross section peaks for both $\pi N \to J/\psi N$ and $\pi N \to \gamma^{*} N$ reactions, complementary to the forward peaks due to the GPD based factorization framework.

\item The special advantage of the cross channel reactions is the extremely easy experimental control
of the  dominance of the transverse polarization of the produced virtual photon (or $J/\psi$),
that is the essential manifestation of the validity of the factorization mechanism involving TDAs.
Indeed, this transverse polarization
results in a specific
$(1 + \cos^2 \theta_\ell)$
angular distribution of the  lepton  in the polar angle $\theta_\ell$ defined in the rest frame of the lepton pair.

\item Finally, the study of the cross channel reactions allows to address the
universality of the TDA-based description, which is an indispensable requirement
of the crossing symmetry and may provide a complementary test of the approach.
\ei

\subsection{Model predictions for JLab}
\mbox

In this Section we review the practical issues raised
when employing the existing phenomenological models of nucleon-to-meson
TDAs to get  cross section estimates for hard near-backward  electroproduction
reactions in the typical kinematical conditions of JLab@12 GeV experiments.
The corresponding results can be helpful for planning the next generation experiments at
JLab as well as to perform feasibility studies for the future EIC.

\subsubsection{Model predictions for near-backward pion electroproduction JLab}
\label{SubSec_Model_P_JLab}
\mbox

Below we present our estimates of the unpolarized cross section
for near-backward pion electroproduction for $ep \to ep \pi^0$ and
$ep \to e n \pi^+$
channels for the kinematical conditions of JLab experiments.
In our cross section studies we rely on the cross channel nucleon exchange  model for
$\pi N$ TDAs
described in   Sec.~\ref{SubSec_Nucle_ex_piN_TDA}.
We also present the estimates of the STSA (\ref{STSA_TDA}) within the  two-component model for
$\pi N$ TDAs (\ref{2Component_model_pi0p}), (\ref{2Component_model_pi_plus_n}),
see  Sec.~\ref{SubSec_SpectralPart_and_2comp_mod}.

As the phenomenological input both the cross channel nucleon exchange
and the two-component $\pi N$ TDA models employ phenomenological solutions for the
leading twist-$3$ nucleon DA $\phi^N$.
The choice of the phenomenological solution for the leading twist nucleon DA and of the corresponding
value of the strong coupling $\alpha_s$ represents a complicated problem.
A detailed discussion of this issue is presented \textit{e.g.} in Ref.~\cite{Stefanis:1999wy}.
The existing models of the leading twist-$3$ nucleon DA parametrizations
can be separated into two distinct classes.
\begin{enumerate} 
\item The models with the shape of nucleon DA close to the asymptotic form~\cite{Lepage:1980fj}
\begin{equation}
\phi_{ {\rm as}}^N(y_1,y_2,y_3)=120 y_1 y_2 y_3
\label{As_f_DA}
\end{equation}
already at a low normalization scale.
Prominent examples are:
\bi
\item the Bolz--Kroll (BK) model~\cite{Bolz:1996sw};
\item the Braun--Lenz--Wittmann (BLW) LO and NLO solutions
\cite{Braun:2006hz,Lenz:2009ar}.
\ei
It is worth mentioning that the advanced present day lattice calculations of the nucleon DAs
\cite{Gockeler:2008xv,Braun:2014wpa,Bali:2019ecy}
favor  solutions close to the asymptotic form.

\item The Chernyak--Zhitnitsky (CZ)-type models are
based on  QCD sum rule estimates and propose a considerably different shape of the nucleon DA, at a low normalization scale.
Examples of this type of nucleon DA models are
\bi
\item Chernyak--Zhitnitsky~\cite{Chernyak:1984bm};
\item King--Sachrajda (KS)~\cite{King:1986wi};
\item Chernyak--Ogloblin--Zhitnitsky (COZ)
\cite{Chernyak:1987nv};
\item  Gari--Stefanis (GS)
\cite{Gari:1986ue} and the so-called ``heterotic solution'' (HET) by N.~Stefanis~\cite{Stefanis:1999wy};
\item a more recent example of a DA solution of this type is the Braun--Lenz--Wittmann  NNLO (BLWNNLO) model
that complements the original BLW  solution with the contribution of the NNLO conformal partial waves
(see Appendix C of~\cite{Lenz:2009ar}).
\ei
\end{enumerate}

Both types of nucleon DA models were employed to provide a description of the nucleon electromagnetic form factors.
As it is well known, the asymptotic form of the nucleon DA
(\ref{As_f_DA})
results in a vanishing pQCD contribution for the proton form factor $F_1(Q^2)$. For that reason, using the nucleon DA with a shape close to the asymptotic form
implies that the standard pQCD contribution must be
complemented by the contribution of soft or end-point corrections (see \textit{e.g.} discussion in Refs.
\cite{Radyushkin:1990te,Bolz:1996sw,Braun:2006hz,Anikin:2013aka}).
The nucleon electromagnetic form factor appears as a building block of the backward amplitude within the
cross-channel nucleon exchange model for nucleon-to-pseudoscalar meson and nucleon-to-vector meson TDAs.
Therefore, we need to assure that the pQCD contribution into
the nucleon electromagnetic form factor is close to the experimental value.
This implies that we are forced to use the CZ-type solutions for nucleon DA.

In the following phenomenological estimates we have chosen to employ the CZ-type
solutions for the leading twist nucleon DAs. We set
compromise values of the strong coupling
$\alpha_s$ and nucleon light-cone wave function normalization constant $f_N$:
\begin{equation}
\alpha_s= 0.3; \ \ \ f_N=5 \cdot 10^{-2} \; {\rm GeV}^2.
\end{equation}

\begin{figure}[H]
 \begin{center}
 \includegraphics[width=8cm]{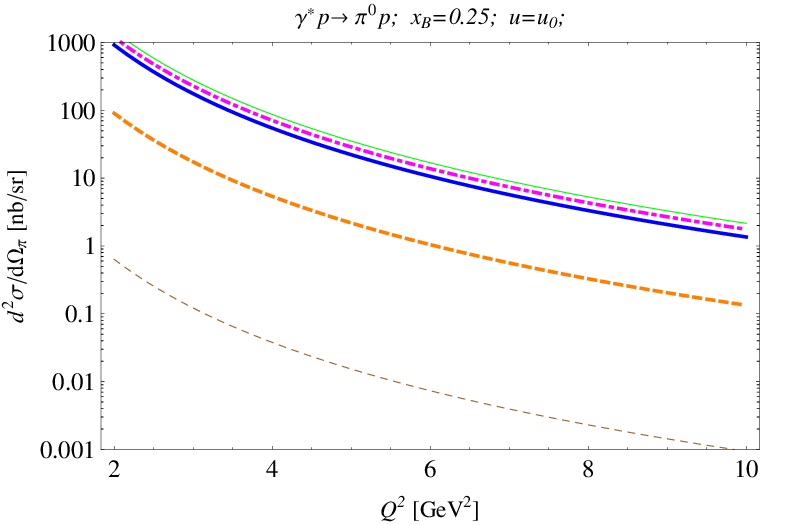} \ \
  \includegraphics[width=8cm]{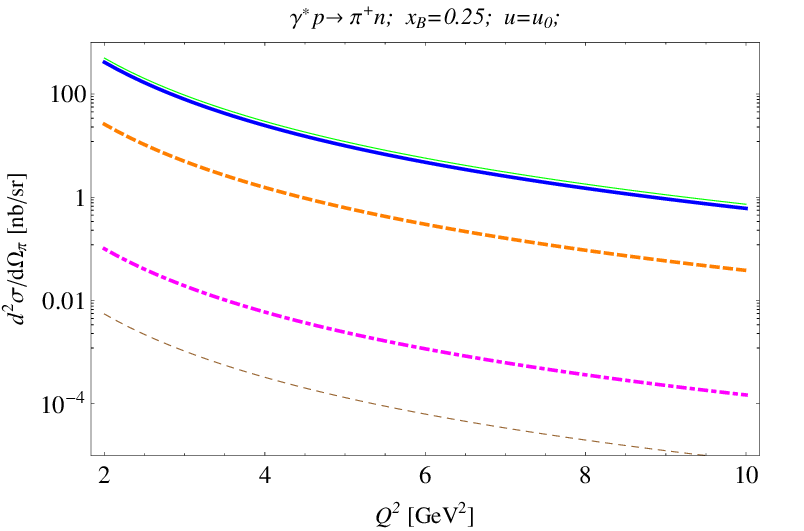}
   \end{center}
     \caption{Unpolarized $\gamma^{*} N \to \pi N$
cross section
$\frac{d^2 \sigma_T}{d \Omega_\pi}$
(\ref{Work_fla_CS})
for pion produced in exactly  backward direction ($u=u_0$) as a function of $Q^2$ for $x_B=0.25$ within the
cross channel nucleon exchange model for $\pi N$ TDAs. Different  solutions for nucleon DAs are used as phenomenological input:
COZ --- thick solid blue lines; KS --- thin solid green lines; BLW NNLO thick orange dashed lines;
HET solution --- thick dash-dotted magenta lines; BLW NLO thin solid brown lines.
}
\label{Fig_CS_BkwPi_Q2}
\end{figure}

In  Fig.~\ref{Fig_CS_BkwPi_Q2} we show the unpolarized $\gamma^{*} N \to \pi N$
cross section
$\frac{d^2 \sigma_T}{d \Omega_\pi}$
(\ref{Work_fla_CS})
for pion produced in exactly  backward direction
($u=u_0$ (\ref{Def_u0}), corresponding to $\Delta_T^2=0$) for
$\gamma^{*} p \to \pi^0 p$ and $\gamma^{*} p \to \pi^+ n$ reaction channels
as a function of $Q^2$ within the two component model for $\pi N$ TDAs.
As phenomenological input we employ the CZ-type solutions for the nucleon DA
(COZ --- thick solid blue lines, KS --- thin solid green lines, BLW NNLO thick orange dashed lines and
HET solution --- thick dash-dotted magenta lines). The magnitude of the corresponding cross section
turns to be large enough for detailed investigations in high luminosity experiments at JLab$@$12 GeV
as well as in a future high luminosity electron--ion collider (EIC).

 With  short brown dashes we also show for completeness  the cross sections obtained employing the BLW NLO solution with the shape close to the asymptotic form.
As clearly seen, the use of this DA as input results in negligibly small cross section.
A peculiar feature of the ``heterotic solution'' by N. Stefanis is that it results
in a considerable size of the cross section for the $\pi^0 p$ channel (of the same order magnitude as with  COZ and KS)
and in a very small cross section in the $\pi^+ n$ channel.

\begin{figure}[H]
 \begin{center}
 \includegraphics[width=8cm]{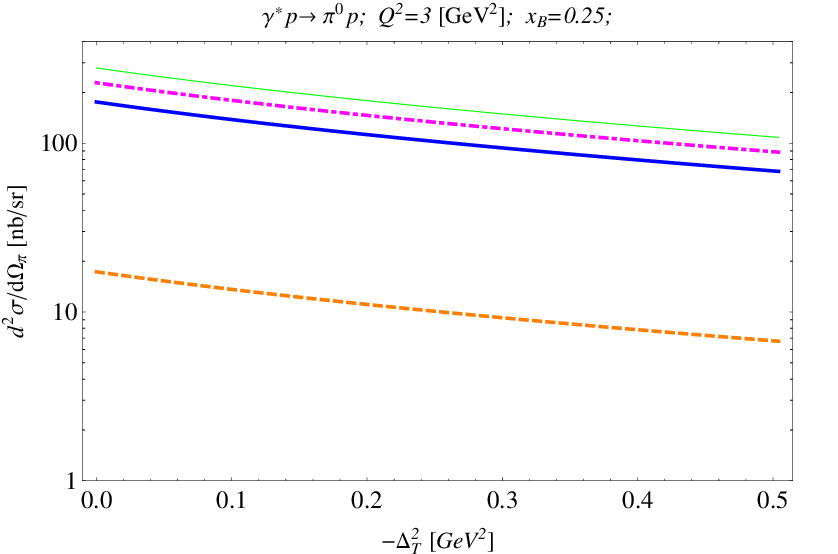} \ \
  \includegraphics[width=8cm]{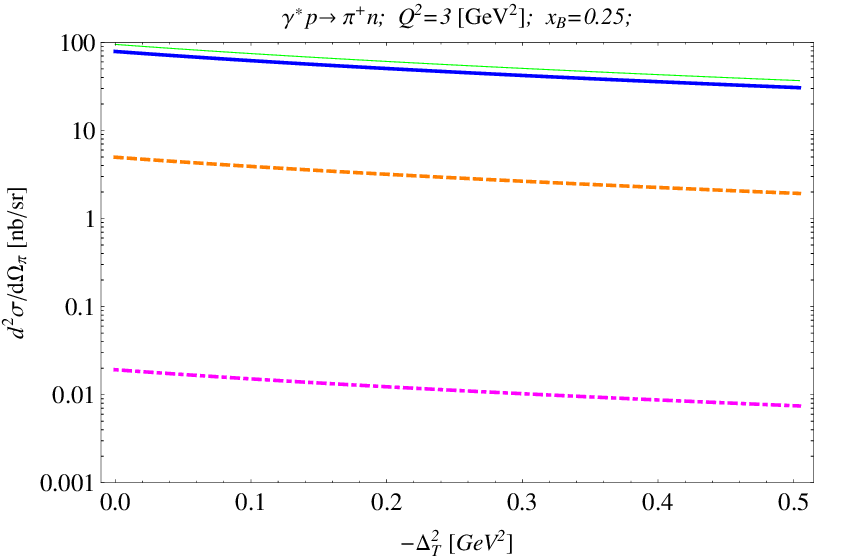}
   \end{center}
     \caption{Unpolarized $\gamma^{*} N \to \pi N$
cross section
$\frac{d^2 \sigma_T}{d \Omega_\pi}$
(\ref{Work_fla_CS})
as a function of $-\Delta_T^2$ for $Q^2=3~{\rm GeV}^{2}$ $x_B=0.25$ within the cross channel nucleon
 exchange model for $\pi N$ TDAs. Different  solutions for nucleon DAs are used as phenomenological input:
COZ --- thick solid blue lines; KS --- thin solid green lines; BLW NNLO thick orange dashed lines;
HET solution --- thick dash-dotted magenta lines.
}
\label{Fig_CS_BkwPi_Delta2}
\end{figure}

\begin{figure}[H]
 \begin{center}
 \includegraphics[width=8cm]{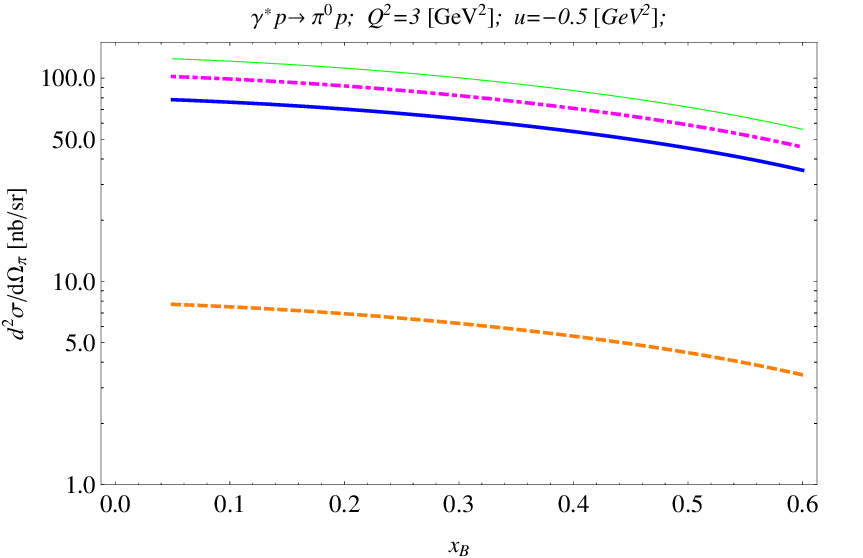} \ \
  \includegraphics[width=8cm]{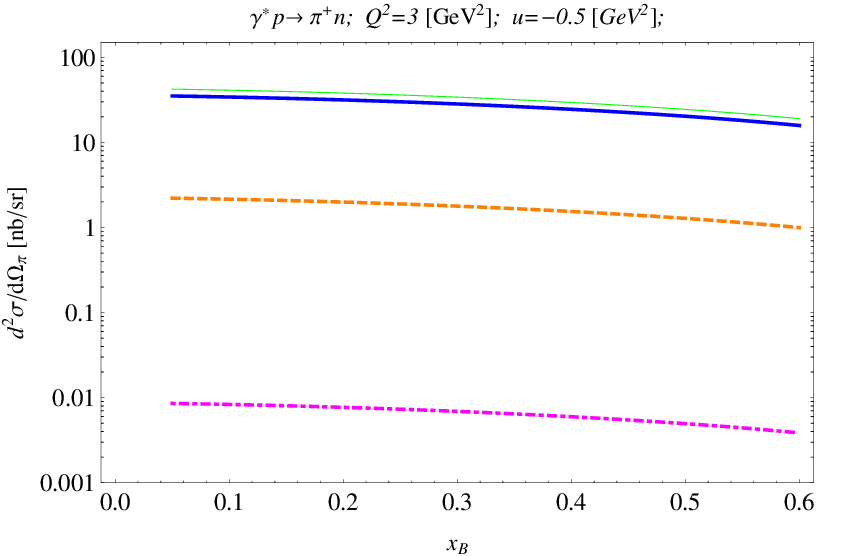}
   \end{center}
     \caption{Unpolarized $\gamma^{*} N \to \pi N$
cross section
$\frac{d^2 \sigma_T}{d \Omega_\pi}$
(\ref{Work_fla_CS})
as a function of $x_B$ for $Q^2=3~{\rm GeV}^{2}$, $u=-0.5~{\rm GeV}^{2}$ within the cross channel nucleon
 exchange model for $\pi N$ TDAs. Different  solutions for nucleon DAs are used as phenomenological input:
COZ --- thick solid blue lines; KS --- thin solid green lines; BLW NNLO thick orange dashed lines;
HET solution --- thick dash-dotted magenta lines.
}
\label{Fig_CS_BkwPi_xB}
\end{figure}

In  Fig.~\ref{Fig_CS_BkwPi_xB} we present the $\Delta_T^2$ dependence of the
unpolarized cross section
$\frac{d^2 \sigma_T}{d \Omega_\pi}$
in the vicinity of the backward peak
for the two possible reaction channels
for $Q^2=3~{\rm GeV}^{2}$, $x_B=0.25$ within the cross channel nucleon exchange
model for $\pi N$ TDAs. The results are presented for a selection
of CZ-type input nucleon DAs (COZ, KS, BLW NNLO and HET).
Finally,  Fig.~\ref{Fig_CS_BkwPi_xB} presents the unpolarized cross sections
$\frac{d^2 \sigma_T}{d \Omega_\pi}$
for the two channels as a function of $x_B$ for  $Q^2=3~{\rm GeV}^{2}$, $u=-0.5~{\rm GeV}^{2}$
within the same model.

Let us now turn to the estimates of the unpolarized $\gamma^{*} N \to \pi N$
cross section within the two component $\pi N$ TDAs model (\ref{2Component_model_pi0p}).
In addition to the cross channel nucleon exchange contribution this model includes the so-called
spectral part based on an Ansatz for quadruple distributions with input from chiral dynamics
(see  Sec.~\ref{SubSec_SpectralPart_and_2comp_mod}).
It must be stressed that in its present form this model is to be considered just as a
very first attempt. In particular, the choice of the profile function
(\ref{Profile_h})
as well as the very form of the factorized Ansatz
(\ref{Factorized_ansatz_xi=1})
were designed  just \textit{ad hoc} and its predictions
must be considered with  care and mostly for
illustrative purposes.
Also, for the moment the model for the spectral part
lacks  $\Delta^2$-dependence.
However, relying  on the two component model (\ref{2Component_model_pi0p}) it is possible to make
some crude estimates for new important observables (\textit{e.g.} transverse target single spin asymmetry,
 see  Sec.~\ref{SubSec_STSA}) that are sensitive
to the imaginary part of the leading order amplitude
(see  App.~\ref{App_Conv_Re_Im}).
The latter turns to be determined by the behavior of nucleon-to-meson TDAs on
the cross-over trajectories
(\ref{Cross_over_trj_TDA})
$w_i=-\xi$, $v_i = \pm \xi'_i$,
which separate the DGLAP-like and the ERBL-like support domains
and on the lines $w_i=\xi$ entirely belonging to the DGLAP-like
support domains of TDAs (see  Sec.~\ref{SubSec_Support} for the details of TDA support regions).
Scrutinizing the behavior of TDAs in  the DGLAP-like
support domains turns to be essential for
matching the predictions of the TDA-based description
with the Regge-like behavior of the cross section anticipated
for smaller values of $x_B$. The construction of a suitable Ansatz
for quadruple distributions that ensures such behavior is highly demanded and deserves dedicated studies.

\begin{figure}[H]
 \begin{center}
 \includegraphics[width=8cm]{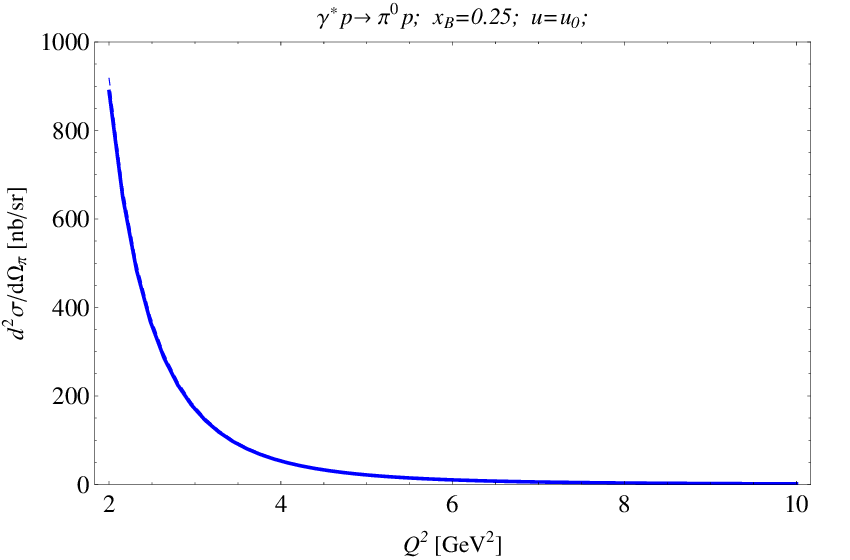} \ \
  \includegraphics[width=8cm]{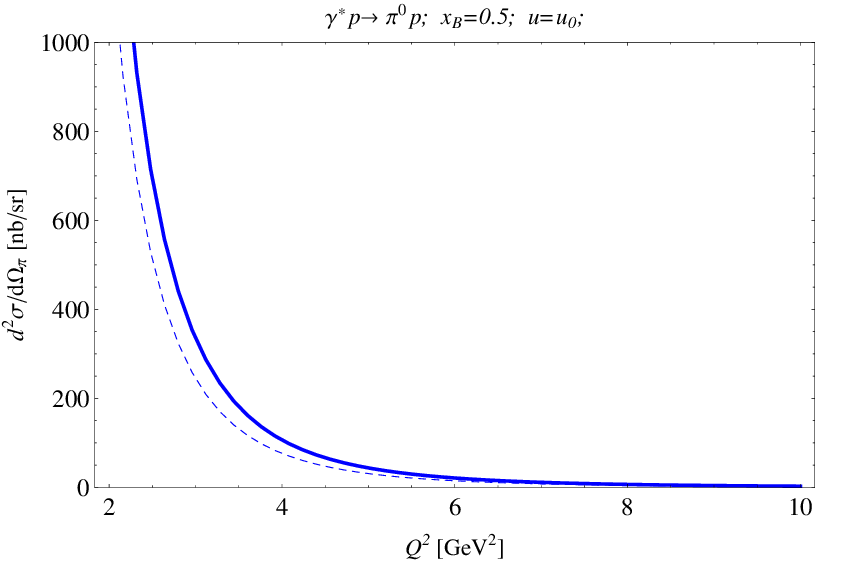}
   \end{center}
     \caption{
     Unpolarized $\gamma^{*} p \to \pi^0 p$
cross section
$\frac{d^2 \sigma_T}{d \Omega_\pi}$
(\ref{Work_fla_CS})
for pion produced in exactly  backward direction ($u=u_0$) as a function of $Q^2$ for $x_B=0.25$  and $x_B=0.5$
within the
cross channel nucleon exchange model for $\pi N$ TDAs (dashed curves) versus the two component  model for $\pi N$ TDAs (solid curves). COZ  solutions for nucleon DAs is used as phenomenological input.
     }
\label{Cs_pi0_2Component_model}
\end{figure}

In  Fig.~\ref{Cs_pi0_2Component_model} we present the unpolarized  cross section for
$\gamma^{*}p \to p \pi^0$
within the two component model for $\pi N$ TDAs
(\ref{2Component_model_pi0p})
as a function of $Q^2$
for exactly backward production of the pion ($u=u_0$)
for two values of $x_B$. As a phenomenological input we employ the
COZ solution for nucleon DA. The dashed line  shows the contribution
of the cross channel nucleon exchange part while  the solid line shows the sum of the cross channel nucleon exchange and spectral part contributions to the cross section.
For $x_B=0.25$ the contribution of the spectral part is negligibly small and
could not be distinguished on the plot while for  $x_B=0.5$
it already gives about 30\% of the whole cross section\footnote{Note that the relative sign between the two contributions  is quite arbitrary. Here we have specially chosen this sign to maximize the effect of including
the spectral part. With a different choice
of this sign the effect turns to be somewhat $5 \div 10 \%$ smaller.}.

It turns out that for smaller $x_B$
the spectral part gives just a negligible contribution to the cross section.
For larger values of $x_B \ge 0.3$ the contribution of the spectral part becomes relatively more important.
Quite expectedly, it entirely dominates in the large-$x_B$ region
since the cross channel nucleon exchange
contribution turns  exactly to zero for $\xi=1$.

\begin{figure}[H]
 \begin{center}
 \includegraphics[width=8cm]{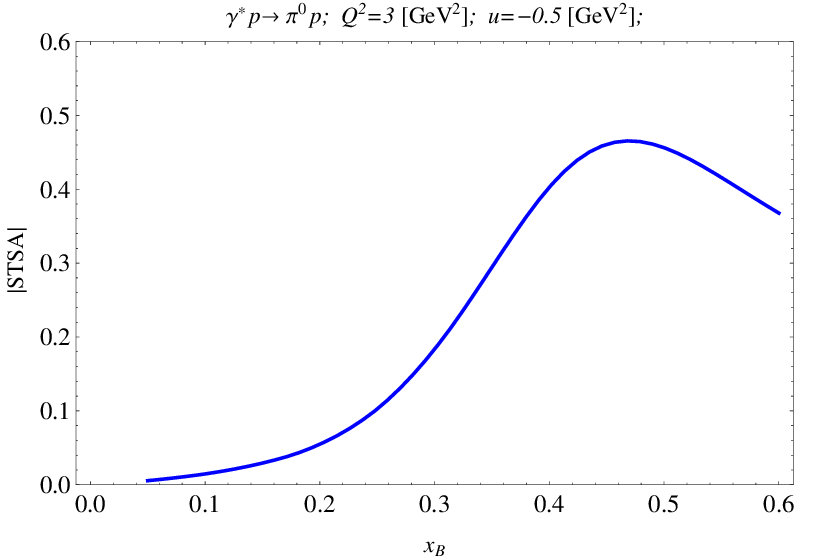} \ \
  \includegraphics[width=8cm]{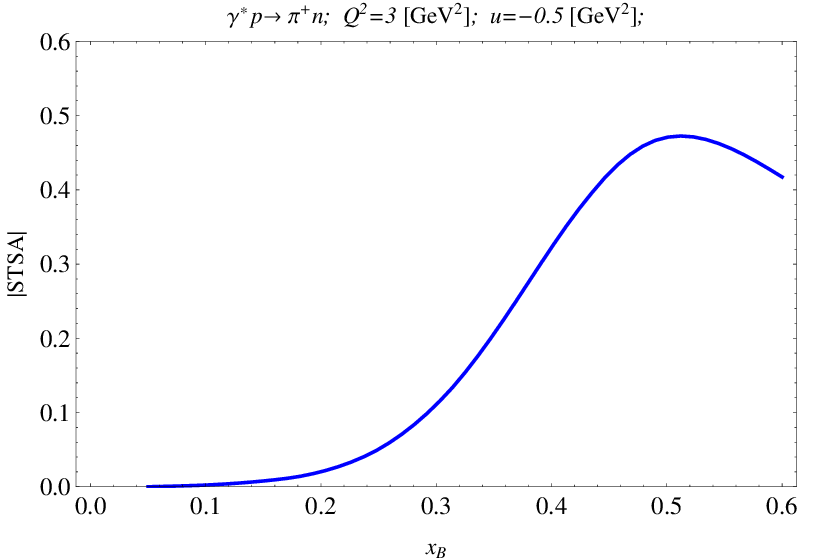}
   \end{center}
     \caption{Absolute value of the  Single Transverse target Spin Asymmetry (STSA) (\ref{STSA_TDA})
as a function of $x_B$ for $Q^2=3~{\rm GeV}^{2}$ $u=-0.5~{\rm GeV}^{2}$ within the two component model for $\pi N$ TDAs. COZ   solution for nucleon DA is used as phenomenological input.
}
\label{Fig_STSA_xB}
\end{figure}

In  Fig.~\ref{Fig_STSA_xB} we present the Single Transverse target Spin Asymmetry (STSA)
(\ref{STSA_TDA})
within the two component $\pi N$ TDA model
(\ref{2Component_model_pi0p}), (\ref{2Component_model_pi_plus_n})
as a function of $x_B$
for $Q^2=3~{\rm GeV}^{2}$ $u=-0.5~{\rm GeV}^{2}$ for the two reaction channels. The
COZ solution for nucleon DAs is used as the phenomenological input.
The STSA (\ref{STSA_TDA}) is sensitive to the
imaginary part of the amplitude
(\ref{Hel_ampl_def_piN}).
Since within the two component TDA model only the
invariant function
$\mathcal{I}^{(1)}$
obtains contribution from the spectral part,
the magnitude of the STSA is mostly determined by the
interference term between the cross channel nucleon exchange contribution
to the entirely real invariant function $\mathcal{I}^{(2)}$
and the spectral part contribution to the imaginary part of $\mathcal{I}^{(1)}$:
\[
{\rm STSA} \sim {\rm Im} \left(\mathcal{I}^{(2)} {\mathcal{I}^{(1)}}^{*} \right) \sim
(\mathcal{I}^{(2)}) \Big|_{N(940)} \times {\rm Im}(\mathcal{I}^{(1)})\Big|_{\rm Spectral}.
\]
Starting from $x_B \sim 0.25$ the
two component $\pi N$ TDA model
predicts a sizable value of the STSA $\ge 10\%$.
The refined theoretical predictions for STSA and its generalization
for the case of backward production of vector mesons require, as already mentioned, new Ans\"atze
for quadruple distributions allowing to implement the Regge behavior for small-$x_B$
and a proper interplay between the $x_B$- and $\Delta^2$-dependencies for resulting TDAs.

We would like to stress that a sizable value of the STSA together with its
$Q^2$-independence for near-backward exclusive meson electroproduction reaction can be interpreted as a
strong evidence for the validity of the collinear factorized
description involving nucleon-to-meson TDAs and nucleon DAs.
To the best of our knowledge, the alternative reaction mechanisms
(\textit{e.g.} the Regge framework)
generally fail to provide a large and $Q^2$-independent STSA for backward meson electroproduction reactions.
Therefore, we strongly suggest to consider the possibility for planning future experiments employing
 targets with transverse polarization.

\subsubsection{A summary on different backward meson electroproduction at JLab}
\mbox

For the convenience of the reader we would like now to quote the results for
different backward meson electroproduction  channels that can be studied at JLab.
\bi
\item Backward electroproduction of the $\eta$-meson $ep \to ep \eta$. The cross section
section can be estimated similarly to the $\pi^0$ production case employing the
cross channel nucleon exchange model with the obvious
change of phenomenological coupling
$g_{\pi NN} \rightarrow g_{\eta NN}$.
Estimates of
$g_{\pi NN}$
and
$g_{\eta NN}$
phenomenological couplings  taken from
the Table~9.1 of
Ref.~\cite{Dumbrajs:1983jd}
give
$\frac{g_{\eta NN}^2}{g_{\pi NN}^2} \sim 0.3$. Therefore, a rough
estimate of the backward $\eta$-electroproduction unpolarized cross section
is about 30\% of the expected magnitude for the near-backward $\pi^0$
production cross section (\textit{cf.} Figs.~\ref{Fig_CS_BkwPi_Q2}, \ref{Fig_CS_BkwPi_xB}).

\item In Ref.~\cite{Pire:2015kxa} we provided the cross section estimates for the
near-backward vector meson production including
$\rho(770)$, $\omega(782)$ and $\phi(1020)$ channels employing the cross channel nucleon exchange
model for $VN$ TDAs (see  Sec.~\ref{SubSec_Nucle_ex_VN_TDA}). For the predicted magnitudes of the cross sections we  refer to the collection
of plots presented in Ref.~\cite{Pire:2015kxa}. Also in  Sec.~\ref{SubSec_First_results_bkw_vect}
we review the first JLab Hall C results for the near-backward $\omega$-meson electroproduction~\cite{Li:2019xyp} that include a comparison with
the model predictions of Ref.~\cite{Pire:2015kxa}.
\ei

\subsection{First results from JLab for hard backward meson electroproduction}
\mbox

In this Section we review the results of the first  experimental studies
aiming  on testing the validity of the collinear factorized description in terms
of nucleon-to-meson TDAs (and nucleon DAs)
of hard exclusive meson electroproduction reactions in the near-backward kinematics.

\newpage

\subsubsection{Backward pion electroproduction at JLab}
\label{SubSec_First_R_pions_JLab}
\mbox

The first dedicated phenomenological analysis using the data from JLab HALL B
\cite{Mecking:2003zu}
(CLAS collaboration)
of near-backward pion electroproduction off nucleons
was performed by Alex Kubarovskiy
\cite{Kubarovskiy:2012yz}
and by Kijun Park and collaborators
\cite{Park:2017irz}.

The analysis of Ref.~\cite{Kubarovskiy:2012yz} addressed the $ep \to ep \pi^0$ channel using
the  JLab \@ 6 GeV ``e1-6'' data set collected in 2001.
It demonstrated the possibility to extract the signal cross section
for  large $-t$ kinematical regime.
The relevant values of $Q^2$ were $Q^2= 1.5 \div 4~{\rm GeV}^{2}$,
center-of-mass energy $W \ge 2$ GeV with
the corresponding range of
$x_B$ from 0.15 to 0.6.
Unfortunately, the statistics was rather poor and only few experimental points
approached the backward kinematical regime. The magnitude of
the cross sections seemed to be roughly consistent with the predictions
of the TDA framework obtained with the cross-channel nucleon exchange model employing
CZ and BLW NNLO input nucleon DAs.

In the pioneering analysis of K.~Park~\cite{Park:2017irz}
the near-backward kinematical regime of
$ep \to e'n \pi^+$ reaction was investigated
employing the same JLab \@ 6 GeV ``e1-6'' data set.
Earlier, in the methodologically similar analysis presented
in Ref.~\cite{Park:2012rn},
the same data were analyzed to extract GPDs, focusing on
the near-forward pions.

\begin{figure}[H]
 \begin{center}
 \includegraphics[width=7cm]{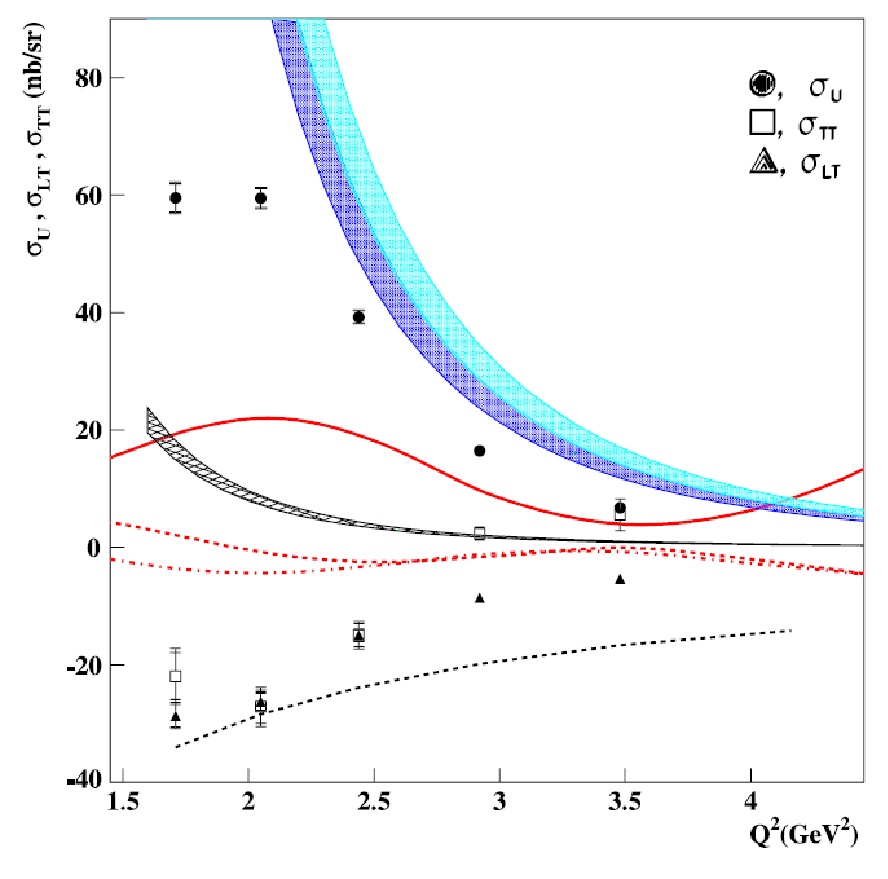}
   \end{center}
     \caption{ The structure functions
     $\sigma_U \equiv \sigma_T+ \varepsilon \sigma_L$ $(\bullet)$,
     $\sigma_{TT}$ $(\square)$ and $\sigma_{LT}$ $(\blacktriangle)$ as a function of $Q^2$.
   The bands refer to model calculations of $\sigma_T$ within the TDA description, black band: BLW NNLO~\cite{Lenz:2009ar}, dark blue band: COZ~\cite{Chernyak:1987nv}, and light blue band: KS~\cite{King:1986wi}.  The
lower black dashed curve represents an educated guess to a fit of the higher twist cross section $\sigma_{LT}$ and $\sigma_{TT}$ in the TDA picture.
   The red curves show the competing predictions of the Regge exchange model~\cite{Laget:2009hs} for: $\sigma_U$ (bold solid), $\sigma_{LT}$ (dashed), $\sigma_{TT}$ (dot-dashed). [Reprinted Figure 4
from Ref.~\cite{Park:2017irz}. Copyright (2018) by Elsevier.]}
\label{Fig_CS_Kijun}
\end{figure}

The key result of~\cite{Park:2017irz} is presented in  Fig.~\ref{Fig_CS_Kijun}.
It shows the $Q^2$-dependence of
$\sigma_U \equiv \sigma_T+ \varepsilon \sigma_L$,
$\sigma_{TT}$
and
$\sigma_{LT}$
cross sections (see Eq.~(\ref{Def_CS_Kroll}) for the definition)
obtained at the average kinematics $W =2.2$~GeV and
$\langle-u \rangle =0.5~{\rm GeV}^{2}$.
It was noted that all three cross sections have a similar strong $Q^2$-dependence.
Let us stress that  Fig.~\ref{Fig_CS_Kijun} cannot be used to claim the consistency with
the scaling behavior suggested by the reaction mechanism based on the collinear factorization, since this $Q^2$ dependence of the cross sections is shown at fixed   $W$  and not at fixed $x_B$. Instead it represents the very first extraction of the cross-sections in near backward kinematical regime and  can be used for a plausible
order-of-magnitude comparison with the predictions of different TDA models within the collinear framework.

The separation of $\sigma_L$ and $\sigma_T$ from $\sigma_U$
was unfortunately impracticable in that analysis due to the limitations of the available data set.
Therefore, the crucial test of the validity of the TDA approach which implies
$\sigma_T \gg \sigma_L$
could not be performed.
However, for larger values of $Q^2$  the order of magnitude of the cross section $\sigma_U$
was found to be consistent   with the predictions for $\sigma_T$
obtained within the TDA framework with the cross-channel nucleon exchange model
(see  Sec.~\ref{SubSec_Nucle_ex_piN_TDA})
employing
COZ, KS and BLW NNLO solutions for input nucleon DAs.
Moreover, it was observed that the interference cross sections
$\sigma_{TT}$ and $\sigma_{LT}$ are roughly equal in magnitude,
which is around 50\% of $\sigma_U$, thus signaling the possible relevance of
higher twist corrections for the kinematics in question.

An interesting experimental possibility to address the validity of the collinear factorized
description of hard exclusive near-backward pion electroproduction
$\vec{e} + p \rightarrow e^{\prime} + n + \pi^{+}$
through the study
of the corresponding beam spin asymmetry (BSA)
(\ref{Def_BSA})  (see
discussion in  Sec.~\ref{SubSec_Def_BSA})
was recently considered
by Stefan Diehl and collaborators in Ref.~\cite{Diehl:2020uja}.

It turns out to be convenient to introduce the following parametrization for the $\phi$-angle
harmonics of the cross  (\ref{CS_polarized_beam_detailed})
\begin{equation}
d \sigma=d \sigma_{0} \left(1+A_{U U}^{\cos \varphi} \cos \varphi+A_{U U}^{\cos 2 \varphi} \cos 2 \varphi+h_{e}
A_{L U}^{\sin \varphi} \sin \varphi\right).
\end{equation}
The subscripts $ij$ of harmonics $A_{ij}$ refer to the longitudinal ($L$) or unpolarized
($U$) state of the beam and the target, respectively.
The BSA
(\ref{Def_BSA})  then
reads
\begin{equation}
{\mathcal{A}}_{\rm BSA}\left(t, \varphi, x_{B}, Q^{2}\right)=
\frac{A_{L U}^{\sin \varphi} \sin \varphi}{1+A_{U U}^{\cos \varphi} \cos \varphi+A_{U U}^{\cos 2 \varphi} \cos 2 \varphi}.
\label{Def_BSA_Exp}
\end{equation}
The key observable quantity extracted from experiment is the $\sin \varphi$
moment
$A_{L U}^{\sin \varphi}$
(\ref{ALUsin_phi}).

As explained in  Sec.~\ref{SubSec_Def_BSA}, the BSA
turns to be an extremely convenient observable
to address the onset of the collinear factorization regime for the
hard exclusive pion electroproduction reaction since both
in the near-forward and near-backward kinematics regimes
the relevant collinear factorization mechanisms result
in a characteristic $1/Q^2$-scaling behavior of
$A_{L U}^{\sin \varphi}$ at fixed $x_B$.

\begin{figure}[H]
 \begin{center}
 \includegraphics[width=6cm]{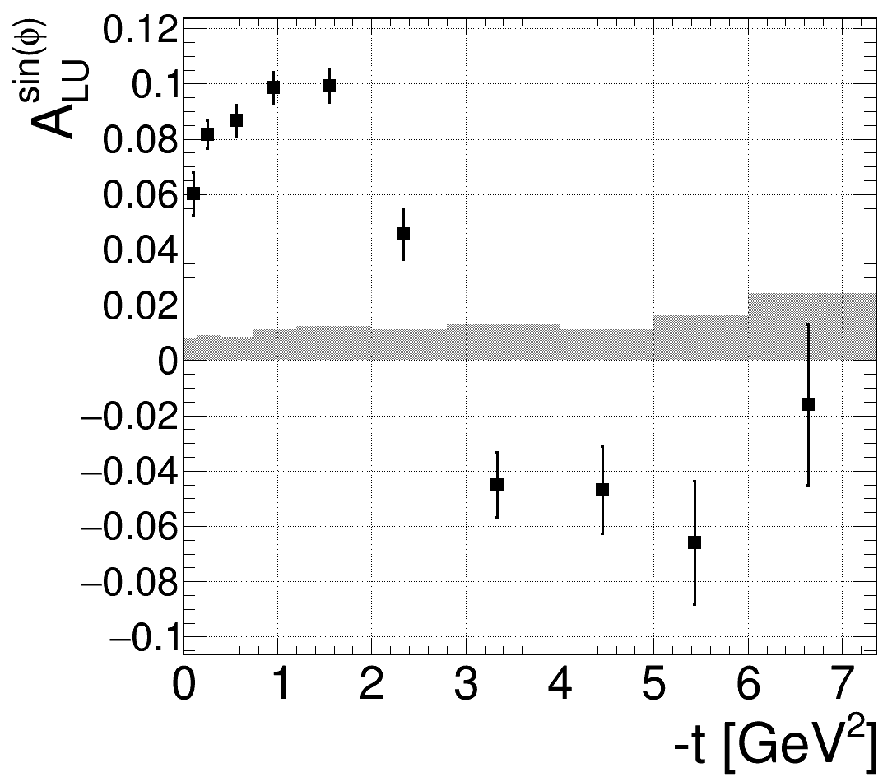}
   \end{center}
     \caption{ $A_{L U}^{\sin \varphi}$ as a function of $-t$. The shaded area represents the
systematic uncertainty, see detailed discussion in~\cite{Diehl:2020uja}. [Reprinted Figure 4 from Ref.~\cite{Diehl:2020uja}. Copyright (2020) by American Physical Society.]}
\label{Fig_A_LU_t}
\end{figure}

\begin{figure}[H]
 \begin{center}
  \includegraphics[width=10cm]{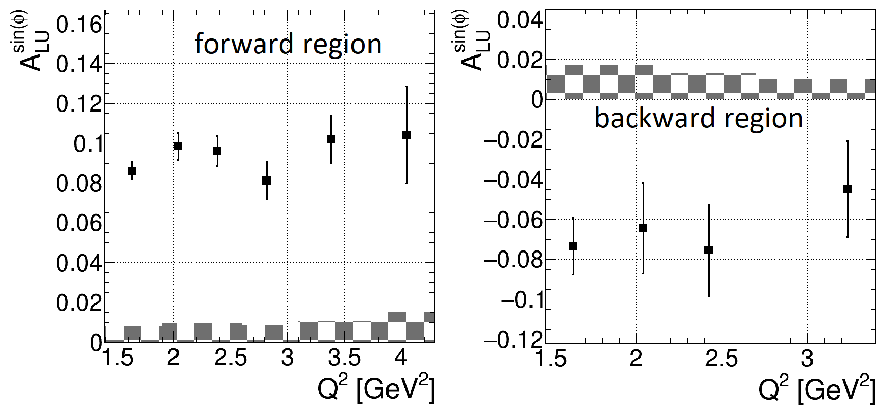}
  \includegraphics[width=10cm]{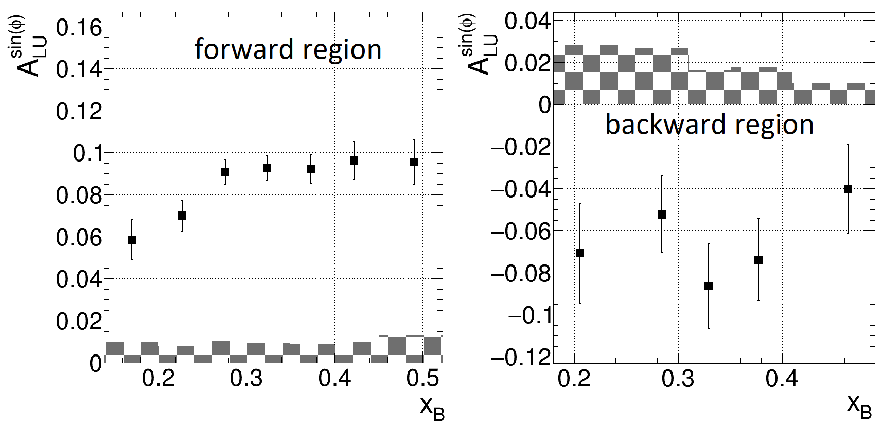}
    \end{center}
     \caption{ $A_{L U}^{\sin \varphi}$
as a function of $Q^2$ (top) and $x_B$ (bottom) for pions produced
in the near-forward (left)
and  near-backward (right) regions. The
shaded area represents the systematic uncertainty, see detailed discussion in~\cite{Diehl:2020uja}. [Reprinted Figure 5
from Ref.~\cite{Diehl:2020uja}. Copyright (2020) by American Physical Society.]}
\label{Fig_A_LU_Q2x}
\end{figure}

In the analysis of Ref.~\cite{Diehl:2020uja}  the BSA for the
reaction
$\vec{e}\, p \rightarrow e^{\prime} \,n \,\pi^{+}$
was extracted from the CLAS measurements with 5.498~GeV
longitudinally polarized electron beam.
The corresponding $Q^2$ range was
from $1 \div 4.6~{\rm GeV}^{2}$, and $x_B$ from $0.1 \div 0.6$
within an extremely broad range in  $-t$ up to 6.6~GeV$^2$.
The key results of Ref.~\cite{Diehl:2020uja}  are displayed in  Figs.~\ref{Fig_A_LU_t}, \ref{Fig_A_LU_Q2x}.

 Fig.~\ref{Fig_A_LU_t} presents the
kinematical region for the extraction of $A_{L U}^{\sin \varphi}$
(\ref{ALUsin_phi})
up to
$-t=6.6~{\rm GeV}^{2}$, which is close to the maximal accessible
$-t$ value. The data are binned in $t$ and integrated over the
complete $Q^2$ distribution ranging from $1$ to 4.5~GeV$^2$ and
$x_B$ ranging from 0.1 to 0.6.
Small $-t$ values correspond to the
near-forward kinematics, while large $-t$ (equivalent  to small
values of $| u| $) correspond to the
near-backward kinematics.
In the  near-forward region $A_{L U}^{\sin \varphi}$ has positive values
that are qualitatively  consistent with the predictions of present day GPD models.
In the region of intermediate $-t$,  $A_{L U}^{\sin \varphi}$
makes a distinctive transition
down to negative and rather small values
in the near-backward region.
This sign change may be interpreted as an indication for the change in the reaction mechanism
corresponding to the
transition between the GPD and TDA factorization regimes.

Fig.~\ref{Fig_A_LU_Q2x} shows $A_{L U}^{\sin \varphi}$ as a function of
$Q^2$, integrated over
$x_B$ in the top plots and as a function of $x_B$, integrated over
$Q^2$, in the bottom plots, for
near-forward (left)
and near-backward (right) kinematics regions.
These plots show that the effect of the sign change between the forward and
the backward regions is manifest in all $Q^2$ and $x_B$ bins.
Also the small absolute value of $A_{L U}^{\sin \varphi}$ for large $Q^2$ in the near backward regime
and some hints  that it is decreasing with growth of $Q^2$
(although for the moment the effect is not statistically significant)
favors the early onset of scaling behavior for near-backward pion electroproduction reaction.

A dedicated study of the exclusive backward electroproduction of a $\pi^0$  above the resonance region has recently been approved with JLab Hall C~\cite{Li:2020nsk}. It will apply the Rosenbluth separation technique
that provides the model-independent $L/T$ differential cross-sections in the backward kinematics region. While the outgoing nucleon and electron will be detected by the standard spectrometers of the JLab Hall C,
the $\pi^0$ exclusive production process will be selected by using the missing mass reconstruction technique. The goal of this E12-20-007 experiment is to perform measurements at several different $Q^2$
values, ranging from 2 to 6.25 GeV$^2$, at a common value of
$x_B = 0.36$
with the complete
$\sigma_L$, $\sigma_T$, $\sigma_{LT}$ and $\sigma_{TT}$
separation.

\subsubsection{First results for hard backward vector meson electroproduction at JLab}
\label{SubSec_First_results_bkw_vect}
\mbox

A considerable progress in investigating hard exclusive meson electroproduction
in the near-backward kinematical regime was recently achieved in
Ref.~\cite{Li:2019xyp} by Garth Huber, Bill Li and collaborators.
Their paper presents the pioneering study of backward-angle cross sections of the hard exclusive $\omega$-meson electroproduction:
$e p \rightarrow e^{\prime} p \omega$.

The analyzed data were part of experiment E01-004
($F_\pi$-2), which used $2.6 \div 5.2$ GeV electron beams on a liquid
hydrogen target and the high precision particle spectrometers
of JLab Hall C. The relevant data set contains two
central $Q^2$ values: $Q^2=1.60~{\rm GeV}^{2}$ and 2.45~GeV$^2$ at common
central $W=2.21$~GeV.
A detailed description of the analysis framework is presented in Ref.~\cite{Li:2017xcf}.

For the first time the hard exclusive $\omega$-meson electroproduction
cross  (\ref{CS_workingVM})
was extracted in the near-backward kinematical
regime and the complete Rosenbluth separation of
the transverse ($T$), longitudinal ($L$), and $LT$, $TT$ interference
terms was performed. In particular, this allowed  to
compare the individual $\sigma_L$
and $\sigma_T$ contributions to the predictions of the
collinear factorization framework involving $\omega N$ TDAs (and nucleon DAs).
In particular, this provides an opportunity to
test
the $\sigma_T$ dominance, that is an important prediction
of the TDA framework.

\begin{figure}[H]
 \begin{center}
 \includegraphics[width=10cm]{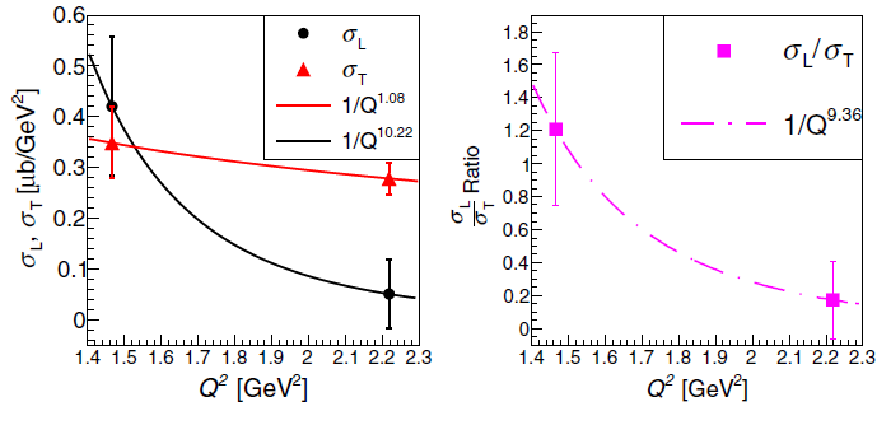}
   \end{center}
     \caption{
     Left: $\sigma_L$ and $\sigma_T$ at $u=u_0$ (\ref{Def_u0}) as function of $Q^2$
for the lowest  $u$ bin. Right:  $\frac{\sigma_L}{\sigma_T}$ ratio as
function of $Q^2$.
Fitted lines are for visualization purpose only. Copyright (2019) by American Physical Society.]}
\label{Fig_CS_Q2_VM}
\end{figure}

The left panel of  Fig.~\ref{Fig_CS_Q2_VM} presents the $Q^2$ dependence
of the cross sections
$\frac{d \sigma_L}{dt}$,
$\frac{d \sigma_T}{dt}$
for the lowest $u$ bin
as function of Q2
The right of  Fig.~\ref{Fig_CS_Q2_VM} presents the ratio of the longitudinal
and the transverse cross sections for the $u=u_0$ bin
as a function of $Q^2$.
$\sigma_T$ shows a flat $Q^2$ dependence
while $\sigma_L$  rapidly decreases with growth of  $Q^2$.
The falling off of the $\frac{\sigma_L}{\sigma_T}$ ratio with growth of  $Q^2$
is qualitatively consistent with the prediction of the TDA collinear factorization framework.

\begin{figure}[H]
 \begin{center}
 \includegraphics[width=10cm]{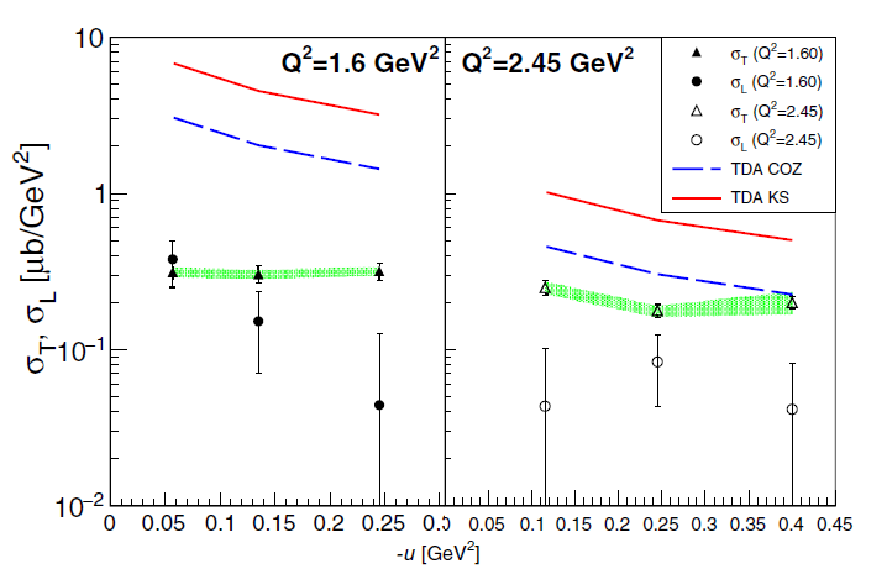}
   \end{center}
     \caption{
$\sigma_T$ ($\blacktriangle$, $\triangle$), $\sigma_L$ ($\bullet$, $\circ$) as function of $-u$, at
$Q^2=1.6~{\rm GeV}^{2}$ (left), 2.45~GeV$^2$ (right).
The predictions for $\sigma_T$ within the TDA framework employ the cross channel nucleon
exchange model for $\omega N$ TDAs (see  Sec.~\ref{SubSec_Nucle_ex_VN_TDA}).
Blue dashed lines show the result using Chernyak--Ogloblin--Zhitnitsky (COZ)
\cite{Chernyak:1987nv}
nucleon DA model as phenomenological input; red solid lines correspond to
using the  King--Sachrajda (KS)~\cite{King:1986wi} input nucleon DA.
The green bands
indicate correlated systematic uncertainties for $\sigma_T$, the uncertainties
for $\sigma_L$ have similar magnitude.
     [Reprinted Figure 5
from Ref.~\cite{Li:2019xyp}. Copyright (2019) by American Physical Society.]}
\label{Fig_CS_u_VM}
\end{figure}

 Fig.~\ref{Fig_CS_u_VM} presents the separated transverse and longitudinal cross sections
$\sigma_T$ and $\sigma_L$ as functions of $-u$ for
$Q^2=1.6~{\rm GeV}^{2}$ (left panel) and $Q^2=2.45~{\rm GeV}^{2}$ (right panel).
It clearly
shows the dominance of $\sigma_T$ for larger value of  $Q^2$.
The magnitude of the extracted $\sigma_T$ turns to be consistent with the
predictions of the TDA framework obtained within the cross-channel nucleon
exchange model of Ref.~\cite{Pire:2015kxa} (see  Sec.~\ref{SubSec_Nucle_ex_VN_TDA}).
Blue dashed lines show the result obtained employing the COZ
nucleon DA model~\cite{Chernyak:1987nv} as phenomenological input;
red solid lines correspond to
using the  KS~\cite{King:1986wi} input nucleon DA.

\begin{figure}[H]
 \begin{center}
 \includegraphics[width=10cm]{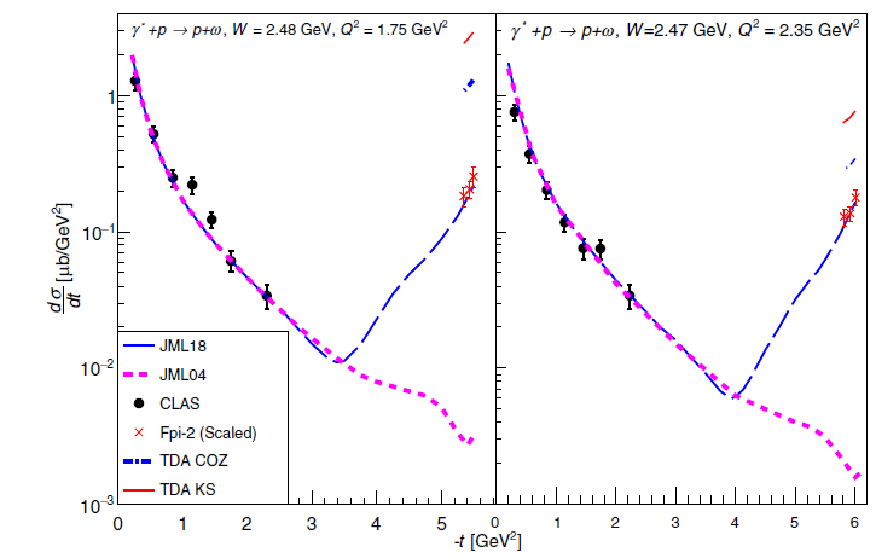}
   \end{center}
     \caption{
Exclusive $\omega N$ electroproduction cross section as a
function of $-t$ at $Q^2=1.75$ (left panel) and $Q^2 =2.35~{\rm GeV}^{2}$
(right panel). The CLAS Collaboration data are the black dots in
the near-forward kinematics region ($-t < 2.5~{\rm GeV}^{2}$), and the
$F_\pi-2$ are the red crosses in the backward region ($-t > 5~{\rm GeV}^{2}$),
scaled to the kinematics of the CLAS Collaboration data (see discussion
in Ref.~\cite{Li:2019xyp} for the details of scaling procedure).
The short curves above the $F_\pi-2$  data are TDA
predictions based on the cross channel nucleon exchange model with
COZ   (blue solid) and KS
 (red solid) DAs employed as the phenomenological input.
 The blue and magenta dashed thick lines
are the predictions within the Regge-based approach
JML04~\cite{Laget:2004qu} and JML18~\cite{Laget_TBP}
respectively.
     [Reprinted Figure 6
from Ref.~\cite{Li:2019xyp}. Copyright (2019) by American Physical Society.]}
\label{Fig_CS_omega_t}
\end{figure}

Finally,  Fig.~\ref{Fig_CS_omega_t} presents  the exclusive
$\omega N$-electroproduction cross section as a
function of $-t$ at $Q^2=1.75$ (left panel) and $Q^2 =2.35~{\rm GeV}^{2}$
(right panel) over the complete range in $-t$. It combines
together the CLAS collaboration data~\cite{Morand:2005ex}
and the properly scaled (see discussion
in Ref.~\cite{Li:2019xyp} for the details of scaling procedure)
${\rm F}_\pi$-2 data covering both the near-forward and near-backward kinematical
regimes.

The important finding presented in  Fig.~\ref{Fig_CS_omega_t}
is the first evidence of  the existence of a
backward-angle peak of the $\omega$-electroproduction cross section
at $-t > 5~{\rm GeV}^{2}$
for both values of
$ Q^2$.
The strength of the backward peak is about $1/10$ of the forward-angle cross section.
Previously, the ``forward--backward'' peak phenomenon was
only observed in  photoproduction data
\cite{Anderson:1969jw,Anderson:1969dv,Boyarski:1967sp}.
 Fig.~\ref{Fig_CS_omega_t} shows the predictions for the near-backward regime
cross section within the TDA framework employing the cross channel nucleon exchange
model of $\omega N$ TDAs of Ref.~\cite{Pire:2015kxa} (see  Sec.~\ref{SubSec_Nucle_ex_VN_TDA})
with COS (blue solid) and KS (red solid) nucleon DAs used as the phenomenological input.
It also shows the predictions within the Regge-based approach
JML04~\cite{Laget:2004qu} (blue dashed) and JML18~\cite{Laget_TBP} (magenta dashed thick line).

\subsection{Model predictions and feasibility studies for \=PANDA}
\label{SubSec_Exp_studies_PANDA}
\mbox

The long awaited completion and launch of the \=PANDA experiment at GSI-FAIR
by the end of this decade will offer unique
possibilities for new investigations of the hadron structure (see \textit{e.g.} Ref.~\cite{Fischer:2021kcr} for the recent status of the \=PANDA). In the proton--antiproton  annihilation mode the
invariant center-of-mass energy $W^2 \equiv (p_N+ p_{\bar{N}})^2$ of the \=PANDA is supposed to be in the range $W^2= 5 \div 25~{\rm GeV}^{2}$.
One of the primary goals of the \=PANDA experimental program
\cite{Lutz:2009ff,Wiedner:2011mf}
is the dedicated measurements of the nucleon electromagnetic
form factors in the time-like region through the proton--antiproton  annihilation
reactions into $e^+ e^-$ and $\mu^+ \mu^-$ in a broad kinematical range.
The detailed feasibility studies for these measurements have been performed in
\cite{Sudol:2009vc}.

The high intensity
of the antiproton beam, together with the outstanding performance
and particle identification capability
of the \=PANDA detector, will also provide access to exclusive channels
such as
\begin{equation}
\bar{p} (p_{\bar{N}})  +
p (p_N) \to \gamma^{*}(q)+ \gamma(q') \to \ell^+(k_{\ell^+})+\ell^-(k_{\ell^-})+ \gamma(q')
\end{equation}
and
\begin{equation}
\bar{p} (p_{\bar{N}})  +
p (p_N)
\to \gamma^{*}(q) + {\mathcal{M}}(p_{\mathcal{M}}) \to \ell^+(k_{\ell^+})+\ell^-(k_{\ell^-}) + {\mathcal{M}}(p_{\mathcal{M}}),
\label{p_barp_gamma_star_M_PANDA}
\end{equation}
where ${\mathcal{M}}$ is a light meson (or a light meson system). This will allow the systematic
studies of the cross channel counterparts of the lepton beam induced
DVCS and HMP reactions.
It worth mentioning that the reactions
(\ref{p_barp_gamma_star_M_PANDA})
represent the principle hadronic background for the
time like nucleon form factor studies. Therefore, gaining additional information on these
channels is important for the primary experimental goals of \=PANDA.

The theoretical framework to access nucleon-to-meson TDAs in these reactions was
elaborated in Refs.~\cite{Pire:2005ax,Lansberg:2007se}
(see  Sections \ref{SubSec_Cross_Ch_Excl_R}, \ref{SubSec_CS_formuls_cross_ch} for a review).
The detailed cross section estimates for the \=PANDA conditions
employing the cross channel nucleon exchange model for the relevant
nucleon-to-meson TDAs
have been performed in Ref.~\cite{Lansberg:2012ha}
for the near-froward and near-backward kinematical regimes of
the reactions
$\bar{p} p \to \gamma^{*} \pi^0 \to \ell^+ \ell^- \pi^0$,
$\bar{p} p \to \gamma^{*} \eta \to \ell^+ \ell^- \eta$
and
$\bar{p} n \to \gamma^{*} \pi^- \to \ell^+ \ell^- \pi^-$.

\begin{figure}[h]
 \begin{center}
 \includegraphics[width=8cm]{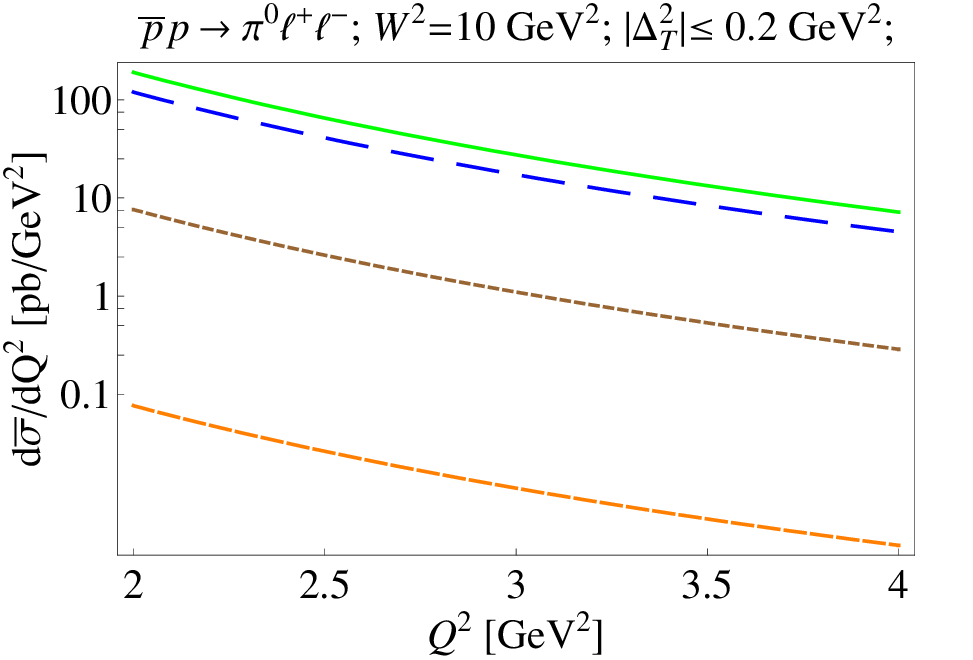} \ \ \
 \includegraphics[width=8cm]{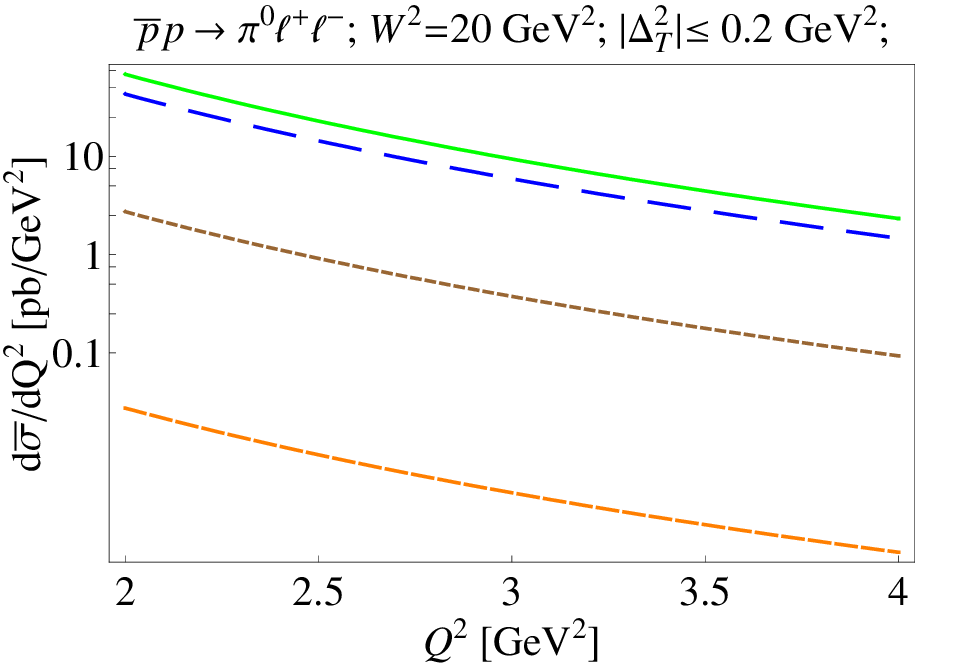}
     \caption{Integrated cross section $d \bar{\sigma}  /dQ^2$ for
     $\bar{p}p \rightarrow \ell^+\ell^- \pi^0$ as a function of $Q^2$ for different values of $W^2=10$ and $20~{\rm GeV}^{2}$
for various phenomenological nucleon DA solutions: COZ (long blue dashes);
   KS (solid green line);   BLW NLO~\cite{Braun:2006hz}
(medium orange dashes) and the BLW NNLO~\cite{Lenz:2009ar}  (short brown dashes)..
[Reprinted Figure 4
from Ref.~\cite{Lansberg:2012ha}. Copyright (2012) by American Physical Society.]}
\label{FigCSpi0PANDA}
\end{center}
\end{figure}

 Fig.~\ref{FigCSpi0PANDA} presents our model predictions for the
unpolarized cross  section (\ref{Unpol_CS_PANDA})
of the near-backward
$\bar{p} p \to \gamma^{*} \pi^0 \to \ell^+ \ell^- \pi^0$.
To quantify the contribution of the anticipated
backward peak the cross section is integrated over the lowest $u$-bin corresponding to a cut in
$\Delta_T^2$ (or, equivalently, $\theta_\pi^{*}$)
\begin{equation}
\frac{d \bar{\sigma} }{dQ^2}({\Delta_T^2}_{\rm cut}) \equiv \int_{u_{0}}^{u_{\rm cut}} du \int d \theta_{\ell} \frac{d \sigma}{du dQ^2 d \cos \theta_\ell }.
\label{Def_itegrated_CS}
\end{equation}
We plot the cross section for $W^2=10$ and $20~{\rm GeV}^{2}$
with different (COZ~\cite{Chernyak:1987nv}, KS~\cite{King:1986wi}, BLW NLO~\cite{Braun:2006hz} and BLW NNLO~\cite{Lenz:2009ar}) input phenomenological solutions for nucleon DA.

Ref.~\cite{Singh:2014pfv}
presented the first
feasibility study for measuring
$\bar{p}p \to \gamma^{*}\pi^0$
with the \=PANDA detector in the
near-forward and near-backward kinematical regimes
in which the factorized description in terms of
$\pi N$
TDAs and nucleon DAs could be challenged.
Additional technical details for this study can be found in the
Ph.D. thesis of M.C. Mora-Espi~\cite{Mora:2012}.

The studies were performed for the center-of-mass
energy squared $W^2 = 5~{\rm GeV}^{2}$ and $W^2 = 10~{\rm GeV}^{2}$
assuming the statistics expected for an integrated luminosity of $2$~fb$^{-1}$.
Definitely, $W^2 = 5~{\rm GeV}^{2}$ invariant center-of-mass energy turns to be insufficient to
justify the validity of the collinear factorized description. This case was considered
in~\cite{Singh:2014pfv} just for indicative purposes.
As the theoretical input for the event generator the cross section estimates
within a simple $\pi N$ TDA model
(\ref{Model_TDA_JPhi})
of Ref.~\cite{Lansberg:2007ec} were employed.

It was demonstrated that the future measurement of the differential
production cross section in bins of $Q^2$ is feasible with
\=PANDA, with averaged statistical uncertainty of 12\% at
$W^2 = 5~{\rm GeV}^{2}$, and with averaged statistical uncertainty of
24\% at $W^2 = 10~{\rm GeV}^{2}$.
The study included the analysis of the principal hadronic background
reaction $\bar{p}p \to \pi^0 \pi^+ \pi^-$.
It was shown that the  \=PANDA
particle identification capabilities will allow a reliable suppression
of the hadronic background. The pion pollution in the signal sample will stay
at the level of a few percent at lower $Q^2$, and under control
below 20\% for larger values of $Q^2$.

The feasibility studies also addressed the possibility to
test the early onset of the factorization regime admitting
the leading order description of the reaction in terms of $\pi N$ TDAs and nucleon DAs.
The cross sections obtained from the
simulations in $Q^2$ and $\cos \theta_\ell$
were fitted to test the $Q^2$-scaling behavior of the cross section
and the  specific
$(1+\cos \theta_\ell^2)$-shape
of the lepton angular distribution that can be considered
as ultimate marking signs for the onset of the factorization regime (see discussion
in Sec.~\ref{SubSec_Exp_evidences}).

The reported results were considered quite promising concerning the experimental
perspectives for addressing the issue of validity of the
factorized description of the
in terms of $\pi N$ TDAs and accessing $\pi N$ TDAs with
\=PANDA.

The investigation of the reaction
(\ref{p_barp_gamma_star_M_PANDA})
for the invariant mass of the lepton pair resonating
at the mass of heavy quarkonium
\begin{equation}
\bar{p} (p_{\bar{N}}, s_{\bar{N}})  +
p (p_N, s_N)
\to J/\psi(p_\psi,\lambda_\psi) + {\mathcal{M}}(p_{\mathcal{M}}) \to \ell^+(k_{\ell^+})+\ell^-(k_{\ell^-}) + {\mathcal{M}}(p_{\mathcal{M}}),
\label{p_barp_Jpsi_M_PANDA}
\end{equation}
constitutes a natural complement
to the aforementioned studies.
The noticeable experimental advantage of the resonance case is a larger expected
cross section and a cleaner signal selection due to the resonant lepton pair production.
On the other hand, the studies of heavy quarkonium decays
\cite{Lundborg:2005am},
to which the reactions
(\ref{p_barp_Jpsi_M_PANDA})
produce a considerable background,
constitute an important part of the \=PANDA research program~\cite{Wiedner:2011mf}.
Therefore, these reaction channels deserve special attention.

The cross section estimates of the
\begin{equation}
\bar{p} (p_{\bar{N}}, s_{\bar{N}})  +
p (p_N, s_N)
\to J/\psi(p_\psi,\lambda_\psi) + \pi^0(p_{\pi})
\label{p_barp_Jpsi_pi_PANDA}
\end{equation}
reaction
for the kinematics conditions of the \=PANDA
are presented in Ref.~\cite{Pire:2013jva}.
This study was performed employing the cross channel nucleon exchange
model for $\pi N$ TDAs, see  Sec.~\ref{SubSec_Nucle_ex_piN_TDA}.
Within  this model the integral convolutions
$\tilde{{\mathcal{J}}}^{(1,2)}(\xi,\Delta^2)$
defined in
(\ref{Amplitude_master})
are expressed in terms of the integral convolution
of nucleon DAs
$M_0$
(\ref{Def_M0})
occurring in the leading order pQCD expression for
the charmonium $\bar{N}N $ decay width (\ref{Charm_dec_width}).
The explicit expressions for
$\tilde{{\mathcal{J}}}^{(1,2)}(\xi,\Delta^2)$
are given by
\begin{eqnarray}
  &&
\tilde{{\mathcal{J}}}(\xi, \Delta^2)\Big|_{N(940)}=
\frac{  f_\pi \,   g_{\pi NN}  m_N (1-\xi) } {   (\Delta^2-m_N^2) (1+\xi )} M_0;
\nonumber \\
  &&
\tilde{{\mathcal{J}}}(\xi, \Delta^2)\Big|_{N(940)}=
\frac{  f_\pi \,   g_{\pi NN}  m_N   } {   (\Delta^2-m_N^2)  } M_0.
\end{eqnarray}
with $\xi$ specified in
(\ref{Def_xi_tu_Jpsi}).
The charmonium  $\bar{N}N $ decay width (\ref{Charm_dec_width})
depends strongly on the input phenomenological nucleon DA model and
is particularly sensitive to the value of $\alpha_s$: $\sim \alpha_s^6$
and $f_N$: $\sim f_N^4$
(see discussion in Sec.~4 of Ref.~\cite{Pire:2013jva}).
Therefore, in our cross section estimates we have chosen to fix the
value of $M_0$ (and the value of $\alpha_s$) from the requirement
that the experimental value of the charmonium $\bar{N}N $ decay width
is reproduced by (\ref{Charm_dec_width}).
This simplification allows to provide a rough order of magnitude estimate of the cross section
to perform feasibility studies. Obviously, once the experimental data will be available,
the reaction
(\ref{p_barp_Jpsi_pi_PANDA})
(and reactions (\ref{p_barp_Jpsi_M_PANDA}) involving other light mesons)
can be used to provide valuable information allowing to
discriminate between different phenomenological DA solutions and improve
models for nucleon-to-meson TDAs.

\begin{figure}[H]
 \begin{center}
\includegraphics[width=8cm]{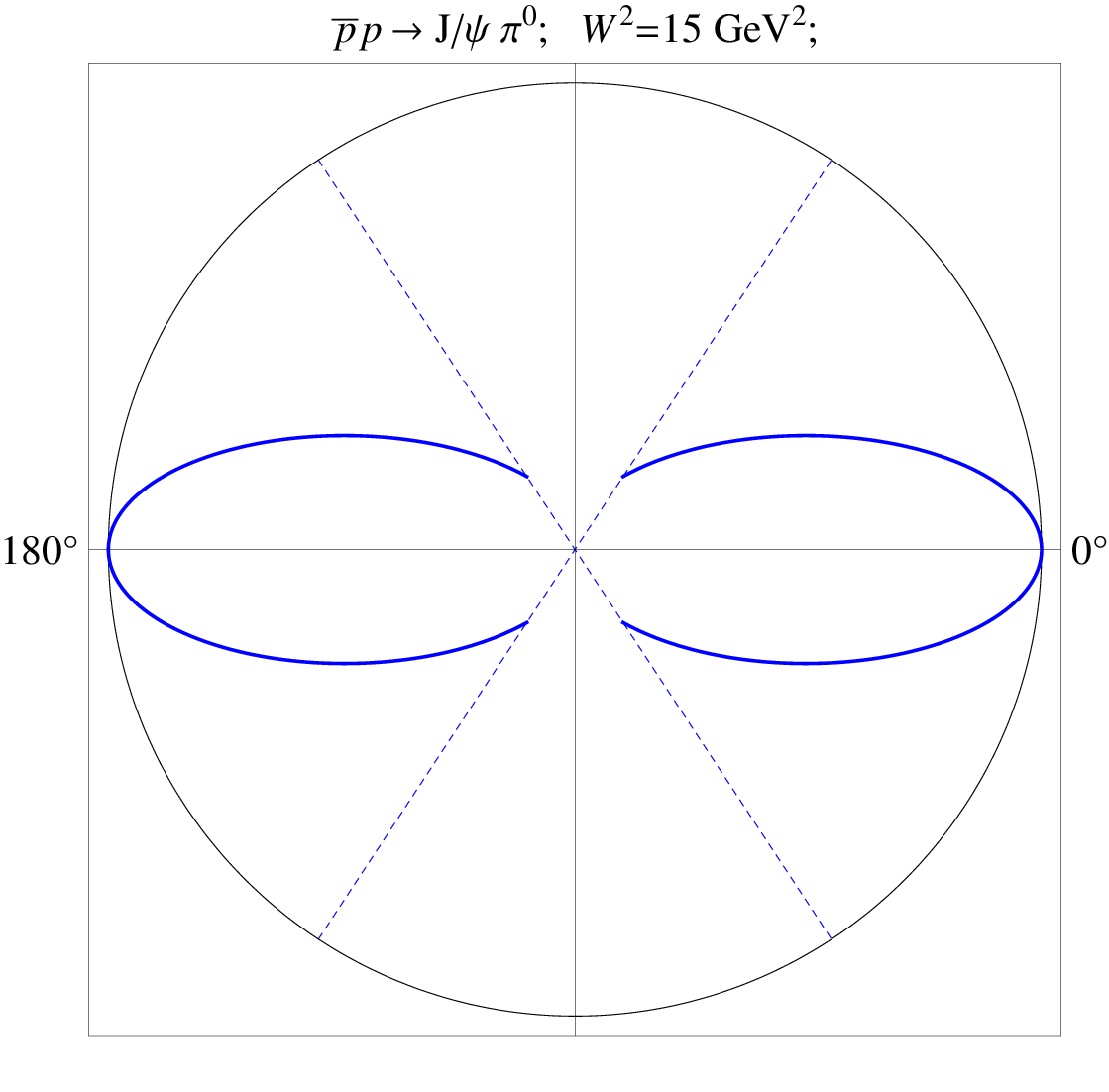}
 \end{center}
     \caption{Angular distribution for the
$d \sigma/d \Delta^2$
cross section for near forward
($\cos \theta_\pi^{*}\ge0$)
and near backward
($\cos \theta_\pi^{*}\le 0$) scattering regimes for   $-1 \, {\rm GeV}^2 \le  \Delta^2 \le \Delta^2_{0}$.
Dashed lines show the effect of the cutoff
$\Delta^2 \ge -1~{\rm GeV}^{2}$
for the values of the pion CMS scattering angle $\theta_\pi^{*}$.  [Reprinted Figure 5
from Ref.~\cite{Pire:2013jva}. Copyright (2013) by Elsevier.]}
\label{Fig_CS_angular_JPsi}
  \end{figure}

The important prediction of the reaction mechanism involving $\pi N$ TDAs and nucleon DAs
is the existence of the near-forward and near-backward cross section peaks for the reaction (\ref{p_barp_Jpsi_pi_PANDA}).
These two peaks are perfectly symmetric in $\bar{p}p$ CMS due to charge conjugation symmetry.
This is illustrated in  Fig.~\ref{Fig_CS_angular_JPsi}
by the polar plot presenting the ratio of the cross section to the maximal value of the exactly forward  (or exactly backward)
$ \theta_\pi^{*}=0^\circ$  ($ \theta_\pi^{*}=180^\circ$) cross-section
\begin{equation}
\frac{ \frac{d \sigma} {d \Delta^2} (W^2, \Delta_T^2)} {\frac{d \sigma} {d \Delta^2} (W^2, \Delta_T^2=0)},
\label{Ratio_CS}
\end{equation}
as a function of the CMS
scattering pion angle $\theta_\pi^{*}$
for $W^2=15~{\rm GeV}^{2}$.
The point on this polar plot is characterized by a distance to the origin (ratio of the cross-sections (\ref{Ratio_CS})) and the angle $\theta_\pi^{*}$ counted as the angle between the position vector and the abscissa. The form of the plot gives an idea of the shape of the near-forward and near-backward peaks of the cross-section.

To single the near-forward and the near-backward kinematical regimes of the
reaction (\ref{p_barp_Jpsi_pi_PANDA})
the
kinematical cut
$-1 \, {\rm GeV}^2 \le \Delta^2  \le \Delta_0^2$ is imposed,
where $\Delta_0^2$ stands for the minimal value of
invariant momentum transfer
$t_0$ or $u_0$.
The left half of the plot corresponds
to the near-backward factorization regime and right half of the graph corresponds
to the  near-forward factorization (see  Fig.~\ref{Fig_Kinematics_TDAs_JPsi_PANDA}).
With the dashed lines we show the effect of the kinematical cut
$\Delta^2= -1~{\rm GeV}^{2}$
for the values of the CMS scattering angle.

\begin{figure}[H]
 \begin{center}
 \includegraphics[width=8cm]{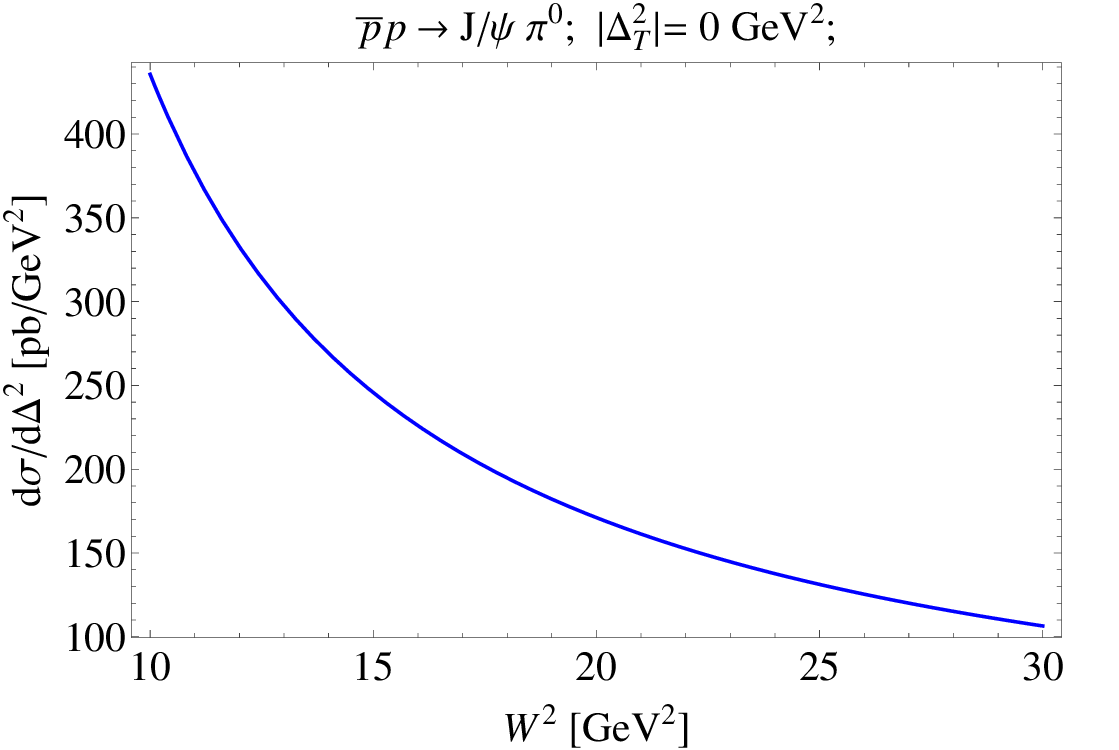}
    \end{center}
     \caption{Differential cross section
$\frac{d \sigma}{d \Delta^2}$ for $p \bar{p} \to J/\psi \,  \pi^0$
as a function of
$W^2$
for
$\Delta_T^2=0$. [Reprinted Figure 3
from Ref.~\cite{Pire:2013jva}. Copyright (2013) by Elsevier.]}
\label{Fig_CS_Jpsi_W2}
\end{figure}

In  Fig.~\ref{Fig_CS_Jpsi_W2}
we show our estimates of the differential cross section
$\frac{d \sigma}{d \Delta^2}$ for the $p \bar{p} \to J/\psi \,  \pi^0$ reaction
as a function
of
$W^2$
for
$\Delta_T^2=0$ corresponding to exactly forward (or backward) production of a pion.
Our cross section estimates
($\sim 100 \div 300$ pb/GeV$^2$ for $\Delta_T^2=0$)
turn to be consistent by the order of magnitude
with the corresponding cross sections obtained within
the phenomenological analysis of Ref.~\cite{Lin:2012ru}.

\begin{figure}[H]
 \begin{center}
  \includegraphics[width=8cm]{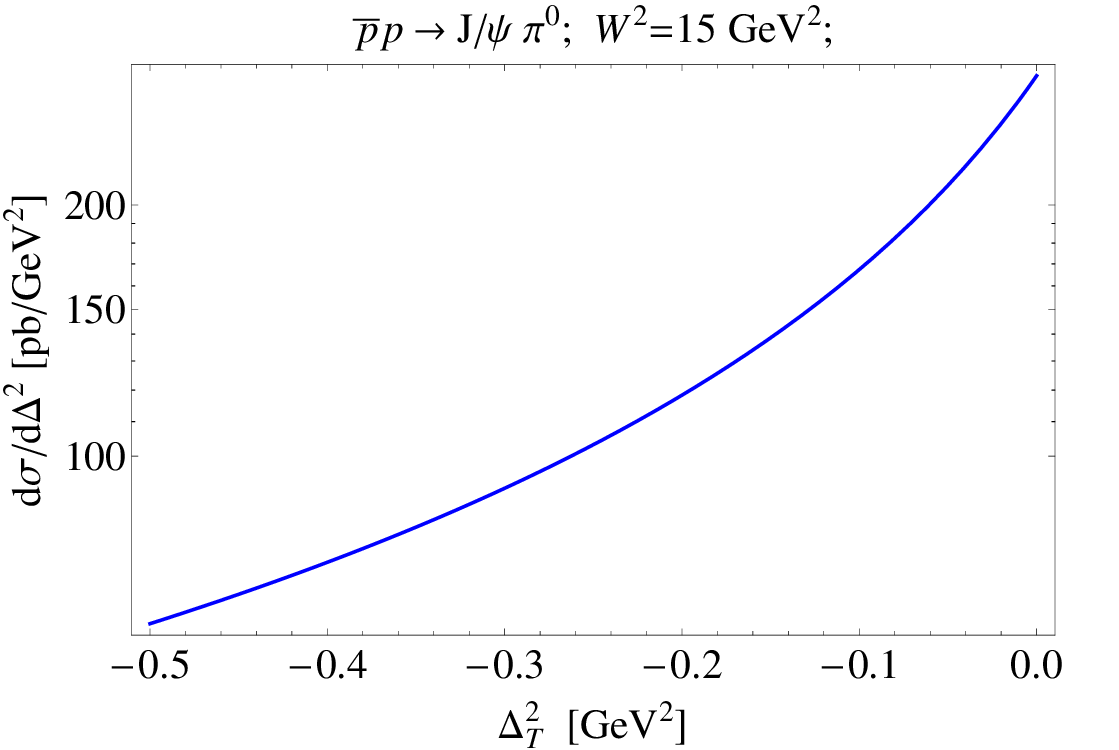}
    \includegraphics[width=8cm]{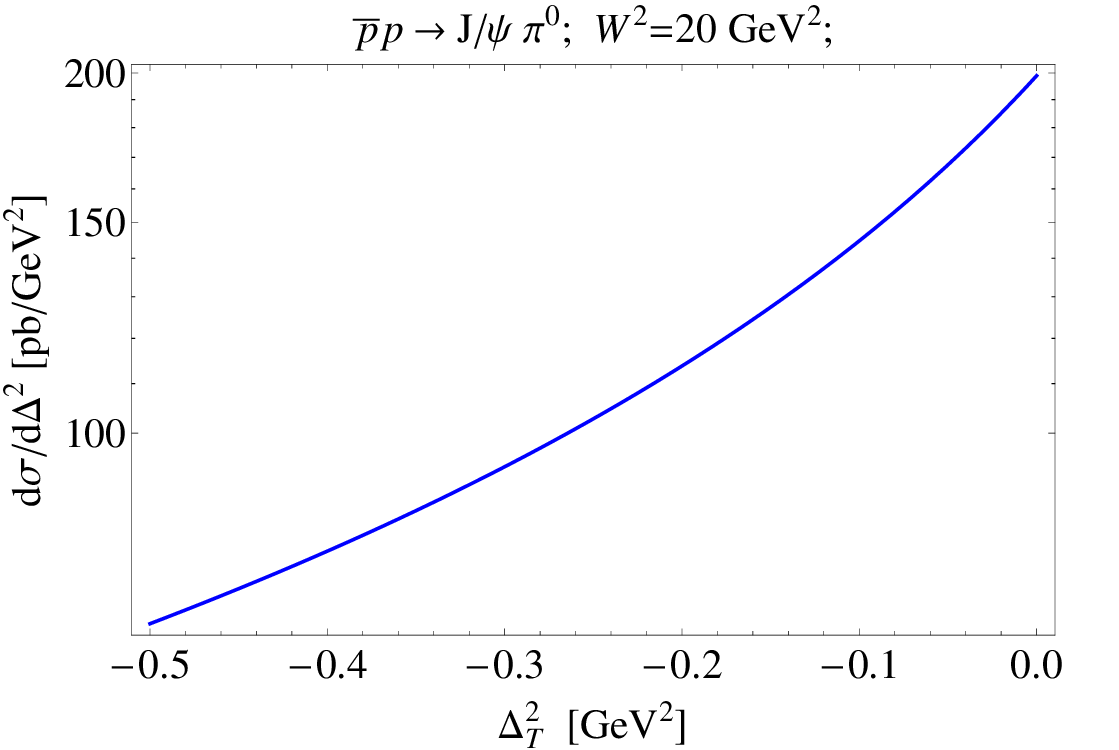}
   \end{center}
     \caption{ Differential cross section
$\frac{d \sigma}{d \Delta^2}$ for $p \bar{p} \to J/\psi \,  \pi^0$
as a function of
$\Delta_T^2$
for
$W^2=15~{\rm GeV}^{2}$
(left panel) and
$W^2=20~{\rm GeV}^{2}$
(right panel). [Reprinted Figure 4
from Ref.~\cite{Pire:2013jva}. Copyright (2013) by Elsevier.]}
\label{Fig_CS_Jpsi_DeltaT2}
\end{figure}

The $\Delta_T^2$ dependence of the cross section in the vicinity
of the forward (or backward)
$\frac{d \sigma}{d \Delta^2}$
cross section peaks is presented in
 Fig.~\ref{Fig_CS_Jpsi_DeltaT2}
for several values of
$W^2$.

Similarly to the case of non-resonant
lepton pair production in association with a light meson (see  Sec.~\ref{SubSec_NbarNlpair_meson})
the key distinguishing feature of the reaction mechanism
involving nucleon-to-meson TDAs (and nucleon DAs) for the near-forward and near-backward
kinematical regimes is the dominance of the
transverse polarization of charmonium to the leading twist-$3$ accuracy.
This results in the
specific $(1+\cos^2 \theta_\ell)$ angular distribution
of the decay lepton pair in the lepton polar angle $\theta_\ell$.

The results of dedicated feasibly studies of  the reaction
$\frac{d \sigma}{d \Delta^2}$ for the $p \bar{p} \to J/\psi \,  \pi^0 \to e^+e^-\,  \pi^0$
for the \=PANDA condition
based on the cross section estimates of~\cite{Pire:2013jva}
are presented in~\cite{Ma:2014pka,Singh:2016qjg}
Additional technical details for this analysis can be found in the PhD thesis of
B.Ma~\cite{MA:2014scq}.
The study includes full Monte Carlo (MC) simulations
of events for the \=PANDA kinematical conditions from both the signal and the principle
background  channels such as
$\bar{p}p \to \pi^+ \pi^-\, \pi^0$,
$\bar{p}p \to J/\psi \pi^0 \pi^0 \to e^+e^- \pi^0 \pi^0$
as well as non-resonant pion production
 $\bar{p}p \to \gamma^{*}\pi^0 \to e^+e^- \pi^0$.
A detailed discussion of the particle identification and selection
procedure and
signal to background ratios for each
background type included in the simulation study was presented.

\begin{figure}[H]
  \includegraphics[width=0.33\textwidth]{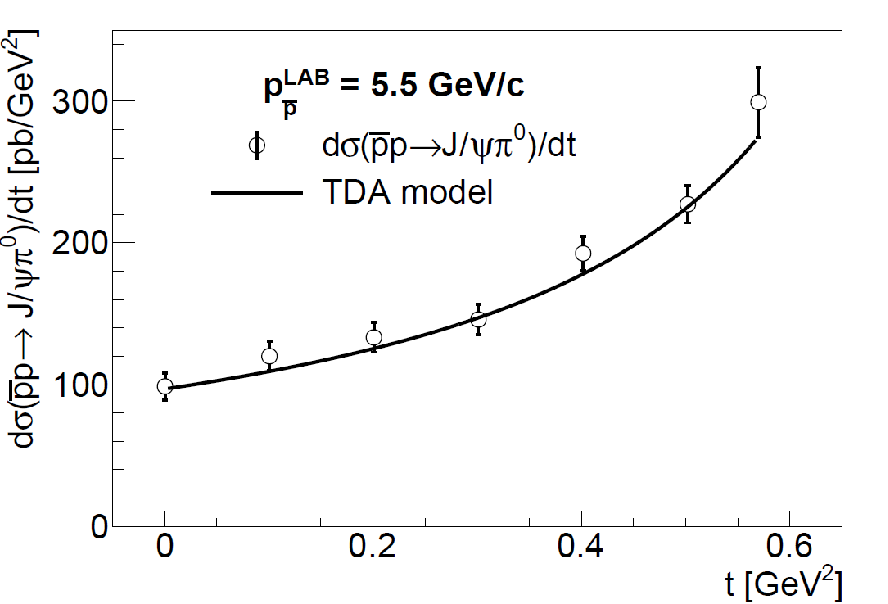}
  \includegraphics[width=0.33\textwidth]{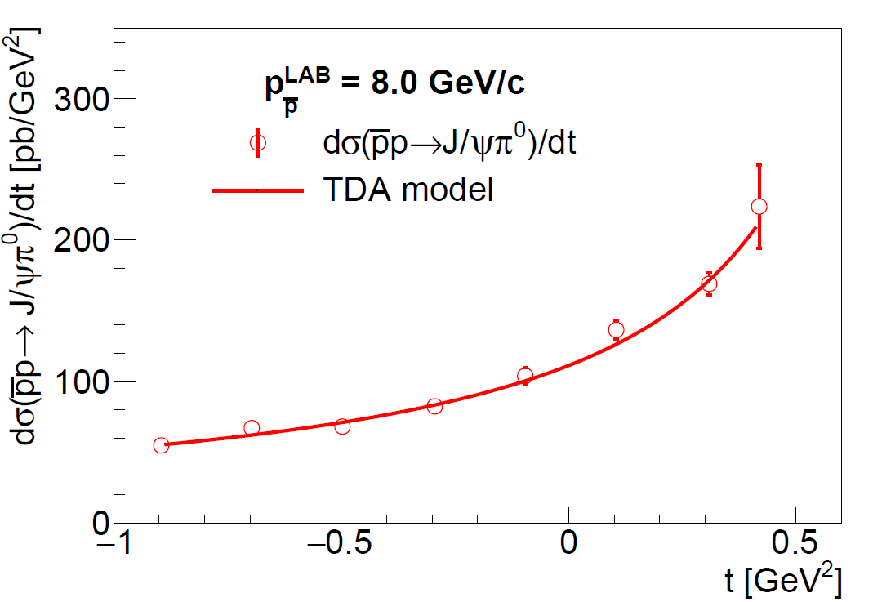}
  \includegraphics[width=0.33\textwidth]{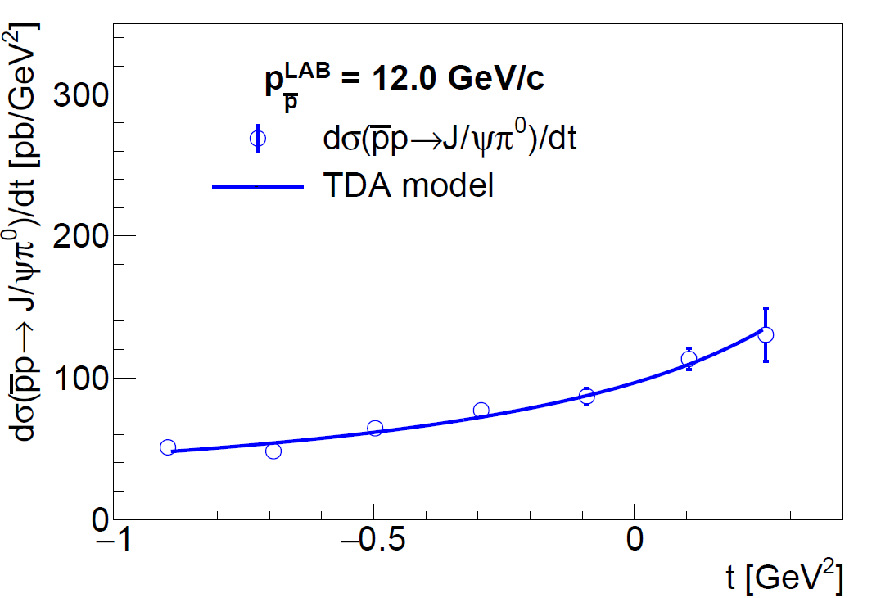}\\
  \includegraphics[width=0.33\textwidth]{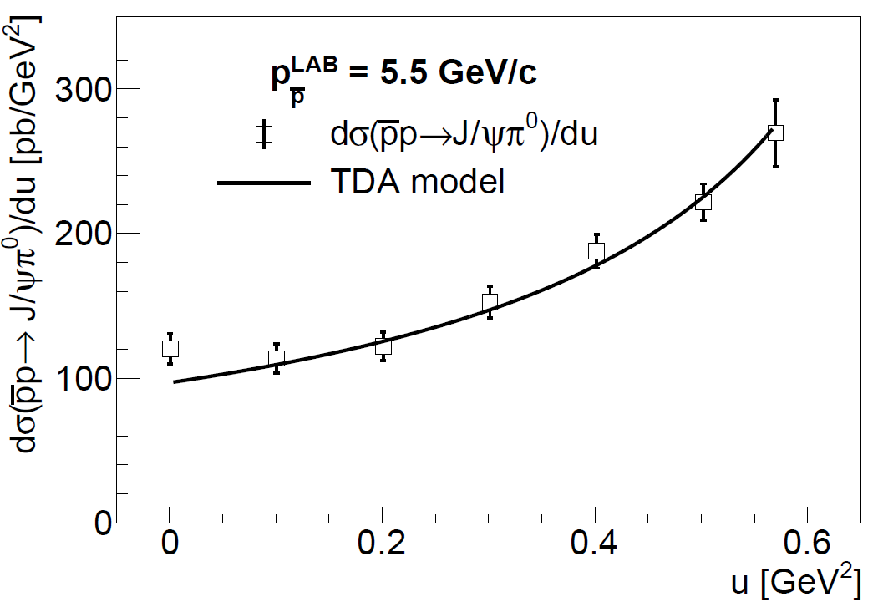}
  \includegraphics[width=0.33\textwidth]{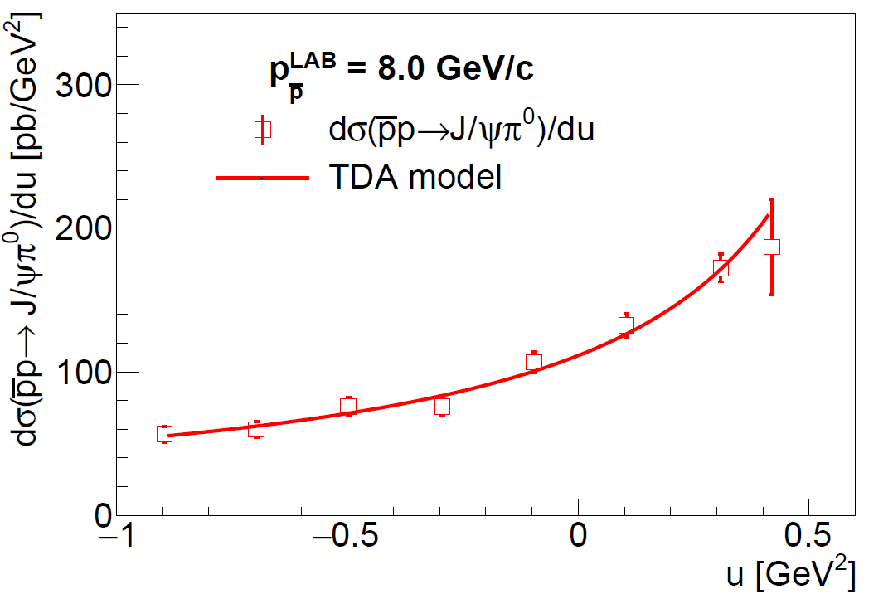}
  \includegraphics[width=0.33\textwidth]{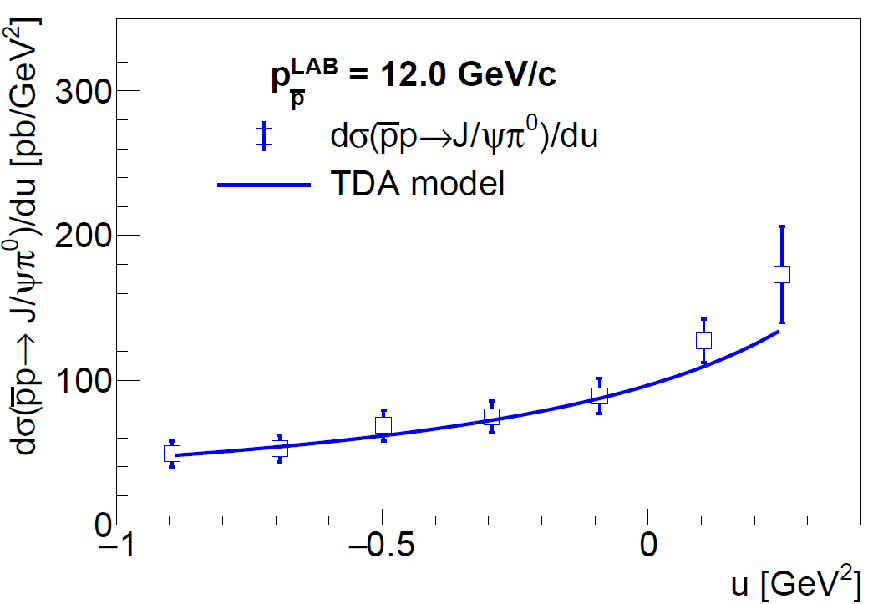}
  \caption{\label{Fig_model_comp_PANDAJpsi} Comparison between the cross sections
  of  $\bar{p} p \rightarrow J / \psi \pi^{0} \rightarrow e^{+} e^{-} \pi^{0}$
    extracted from the fully efficiency corrected yields expected from
    2~fb$^{-1}$ integrated luminosity (data points) and the TDA model
    prediction (full curves) at the three incident $\bar{p}$
    momenta in the LAB frame ($W^2=2m_N\left(m_N+\sqrt{m_N^2+(p_{\bar{p}}^{\rm LAB})^2}\right)$): 5.5~GeV$/c$ corresponding to $W^2=12.3~{\rm GeV}^{2}$ (left column), 8.0~GeV$/c$ corresponding to $W^2=16.9~{\rm GeV}^{2}$ (middle
      column) and 12.0~GeV$/c$ corresponding to $W^2=24.4~{\rm GeV}^{2}$ (right column). Top row:
    Near-forward kinematics regime as a function
    of $t$. Bottom row: Near-backward kinematics  regime as a function of $u$. [Reprinted Figure 21
from Ref.~\cite{Singh:2016qjg}. Copyright (2017) by American Physical Society.]}
\end{figure}

 Fig.~\ref{Fig_model_comp_PANDAJpsi} presents a
comparison between the
cross sections extracted from the fully efficiency corrected yields expected from
the integrated luminosity of 2~fb$^{-1}$
to the prediction of
Ref.~\cite{Pire:2013jva}
that was used as input for the signal event
generator. The measurements have a satisfactory precision
of about $8\div10\%$ relative uncertainty. This level of
precision will allow a quantitative test of the prediction
of TDA models for the reaction
$\bar{p} p \rightarrow J / \psi \pi^{0} \rightarrow e^{+} e^{-} \pi^{0}$
with \=PANDA.

\begin{figure}[H]
  \includegraphics[width=0.33\textwidth]{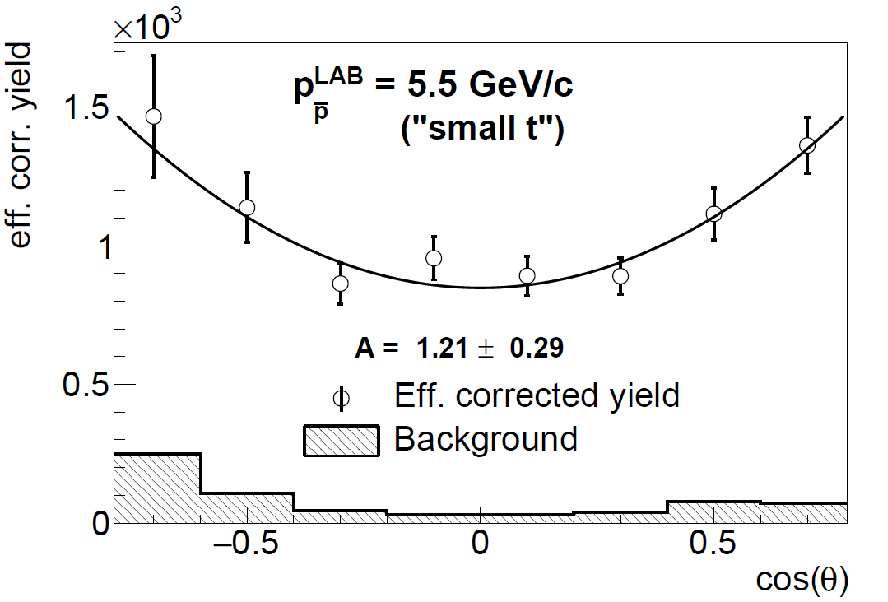}
  \includegraphics[width=0.33\textwidth]{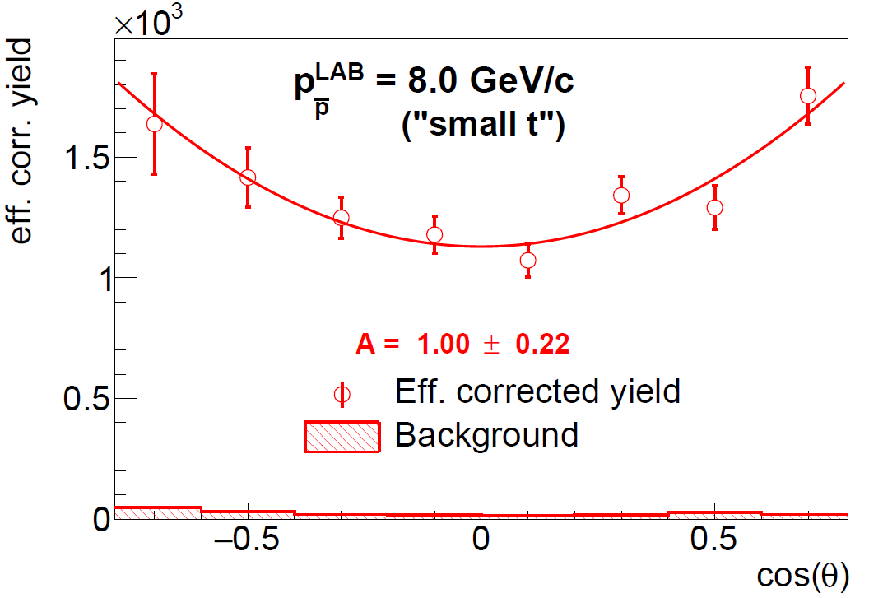}
  \includegraphics[width=0.33\textwidth]{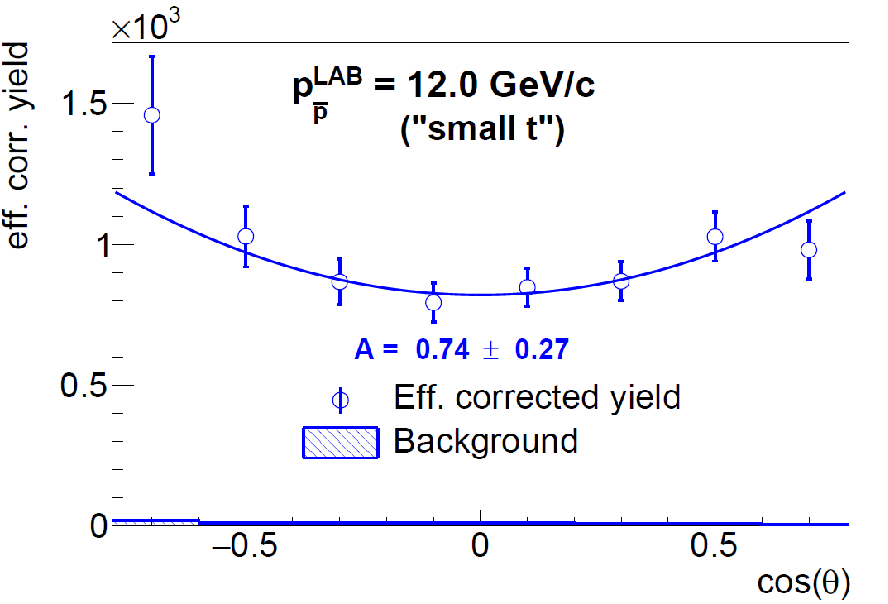}\\
  \includegraphics[width=0.33\textwidth]{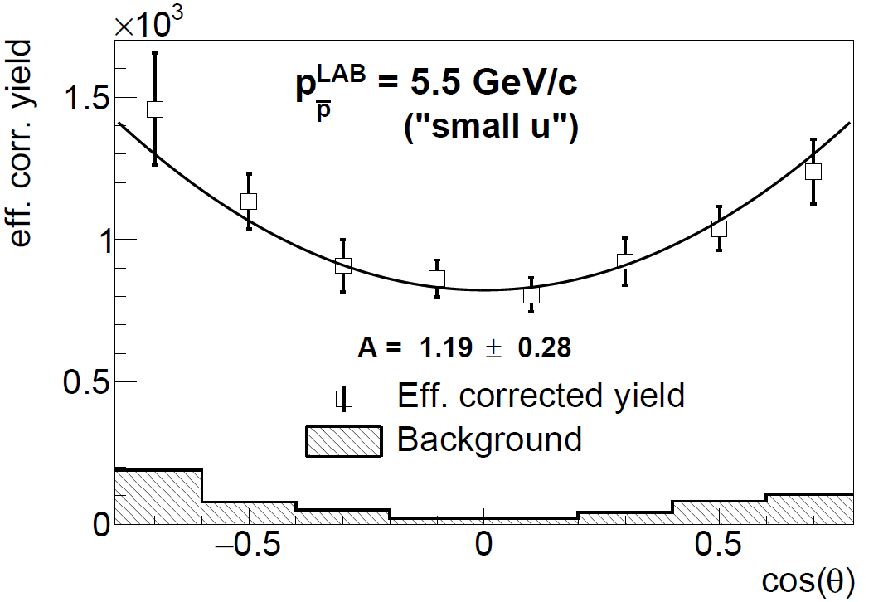}
  \includegraphics[width=0.33\textwidth]{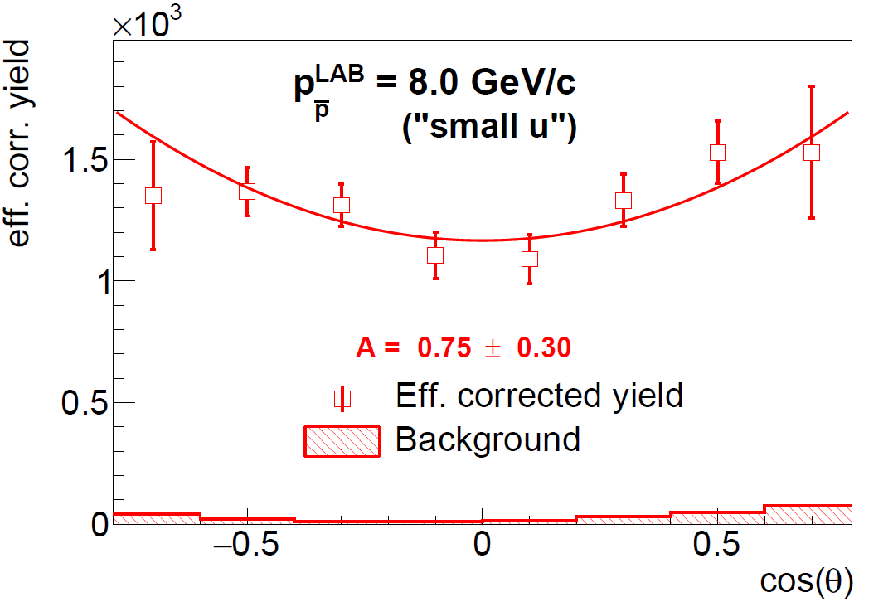}
  \includegraphics[width=0.33\textwidth]{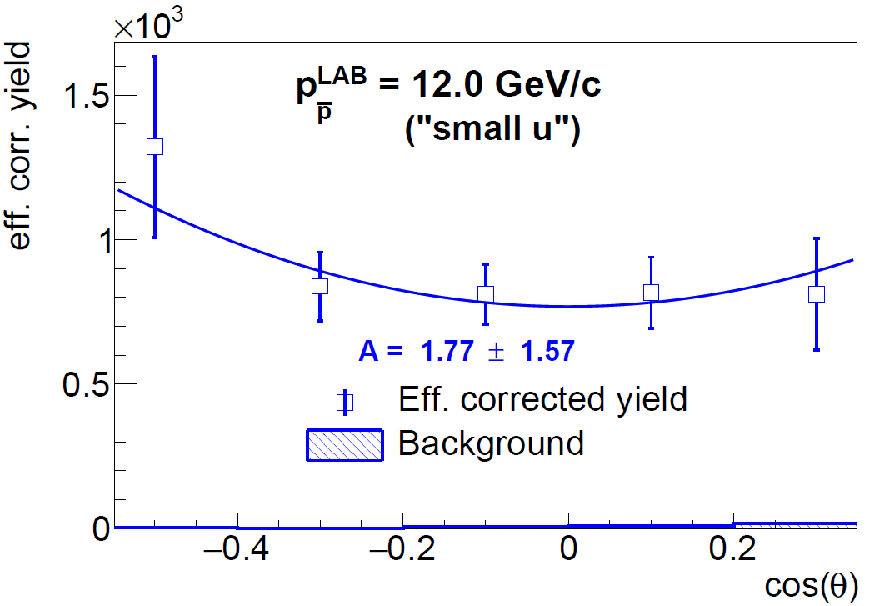}
  \caption{\label{Fig_epcth_fit} Efficiency corrected yield of
    $\bar{p} p \rightarrow J / \psi \pi^{0} \rightarrow e^{+} e^{-} \pi^{0}$ (open markers) and background yield (shaded
      histograms) as a function of $\cos\theta_\ell$ and the result of the fit
    with the function $B\times(1+A \cos^2 \theta_\ell)$ (solid lines) at the
    three incident $\bar{p}$ momenta in the LAB frame: 5.5~GeV$/c$ corresponding to $W^2=12.3~{\rm GeV}^{2}$ (left column), 8.0~GeV$/c$ corresponding to $W^2=16.9~{\rm GeV}^{2}$ (middle
      column) and 12.0~GeV$/c$ corresponding to $W^2=24.4~{\rm GeV}^{2}$ (right column). Top row: Near-forward kinematic approximation validity
    range as a function of $t$. Bottom row: Near-backward kinematic
    approximation validity range as a function of $u$ [Reprinted Figure 22
from Ref.~\cite{Singh:2016qjg}. Copyright (2017) by American Physical Society.]}
\end{figure}

The analysis also included a discussion of the sensitivity for
testing the validity of the TDA-based deception for the reaction.
For this issue an attempt to reconstruct the
specific
$\sim (1+\cos^2 \theta_\ell)$
form of the dependence of the differential cross section
with respect to the lepton polar emission angle $\theta_\ell$ in the
$J/\psi$
rest frame relative to the direction of motion of the
$J/\psi$. The generated pseudodata was produced in bins in
$\cos\theta_\ell$ and subsequently fitted with
$A \times (1+B \cos^2\theta_\ell)$ function. The results are presented in  Fig.~\ref{Fig_epcth_fit}.
It was concluded that the extraction of the angular distribution
turns to be feasible with  \=PANDA with an integrated luminosity of
$2$~fb$^{-1}$, except in some kinematic zones where the signal reconstruction efficiency
turns to be too low.

\subsection{Prospect for  {J}-{{PARC}} and { {EIC}}}
\label{SubSec_Prosp_JPARC_EIC}
\mbox

Let us now gather some remarks on less studied environments where the TDA framework may be applied provided more studies are devoted to it in the precise conditions of these facilities. We have in mind firstly the intense
medium energy meson beam facility J-PARC that is under construction in Japan and secondly the high luminosity high energy electron--ion colliders proposed in the USA~\cite{Boer:2011fh}, in
Europe~\cite{AbelleiraFernandez:2012cc} or in China~\cite{Anderle:2021wcy}.
\begin{itemize}
\item The invariant center-of-mass energy $W^2$ with the pion beam at JPARC ranges from $10$ to $40~{\rm GeV}^{2}$.
\item The planned invariant center-of-mass energy range at EIC is extremely broad. However, the expected rates at $W^2 > 100~{\rm GeV}^{2}$ are reported to be very small.
\end{itemize}

\subsubsection{Studying {{TDA}}s with pion beams}
\mbox

The collinear factorization framework for the description of
the pion beam induced charmonium production reaction
\begin{equation}
\pi^-(p_\pi) + p(p_1,s_N) \to J/\psi(p_\psi, s_\psi) + n(p_2,s'_N)
\label{reacJparc_once_more}
\end{equation}
in terms if meson-to-nucleon TDAs (and nucleon DAs)
 in the near-backward
kinematics  region, where $u =(p_2-p_\pi)^2 \ll W^2 =(p_1+p_\pi)^2$
has been presented in  Sec.~\ref{SubSec_piN_JPARC}.

The feasibility study
\cite{Sawada:2016mao}
challenged the reaction (\ref{reacJparc_once_more})
in the near-forward kinematics in order to access
polarized nucleon GPDs as studied in Ref.~\cite{Berger:2001zn,Goloskokov:2015zsa} for J-PARC conditions.
These results let us hope that the complementary near-backward
regime can be considered to get access to the  pion-to-nucleon $N\pi$ TDAs.

For this reason, following Ref.~\cite{Pire:2016gut}, we provide  the cross section estimates for the near-backward
kinematics regime of the reaction (\ref{reacJparc_once_more})  within the TDA framework.
The leading twist-$3$ unpolarized differential cross section of
the reaction (\ref{reacJparc_once_more})
in this case corresponds
to the production of transversely polarized charmonium.
This gives
rise to the characteristic angular distribution of the charmonium decay lepton pair.
Employing the  amplitude
(\ref{Amplitude_masterJPARC}) the cross section of
the reaction (\ref{reacJparc_once_more})
can be written as
\begin{eqnarray}
  &&
\frac{d \sigma}{d \Delta^2}= \frac{1}{16 \pi \Lambda^2(s,m_N^2,m_\pi^2) } | \overline{\mathcal{M}_{T}}| ^2 \nonumber  \\ &&
=\frac{1}{16 \pi \Lambda^2(s,m_N^2,m_\pi^2)}
\frac{1}{2} | \mathcal{C}_\psi| ^2 \frac{2(1+\xi)}{\xi {\bar{M}}^8}  \left( |
\tilde{\mathcal{J}}^{(1)}
(\xi, \Delta^2)| ^2 - \frac{\Delta_T^2}{m_N^2} |
\tilde{\mathcal{J}}^{(2)}
(\xi, \Delta^2)| ^2 \right),
\label{CS_def_delta2JPARC}
\end{eqnarray}
where the invariant functions
$\tilde{\mathcal{J}}^{(1,2)}$ are introduced in Eq.~(\ref{Amplitude_masterJPARC});
$C_\psi$ is the normalization constant (\ref{Def_C_Jpsi});
$\bar{M}$ is the mean mass (\ref{mass approx})
and $\Lambda(x,y,z)$
is the Mandelstam function (\ref{Def_lambda}).

\begin{figure}[H]
 \begin{center}
 \includegraphics[width=5cm]{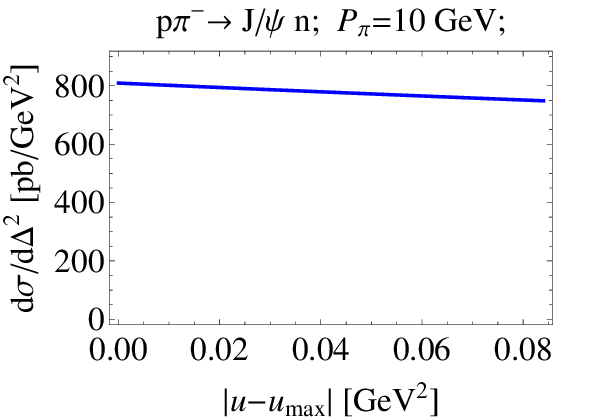}
  \includegraphics[width=5cm]{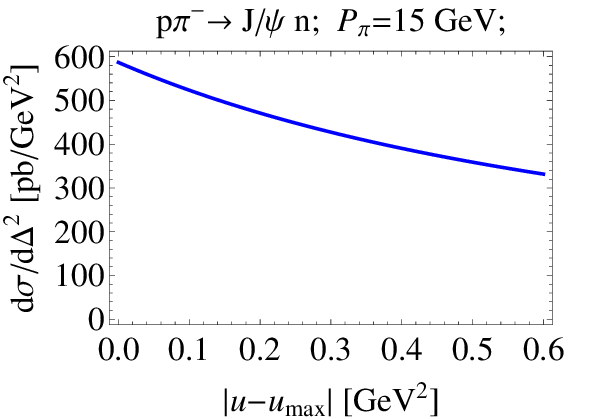}
  \includegraphics[width=5cm]{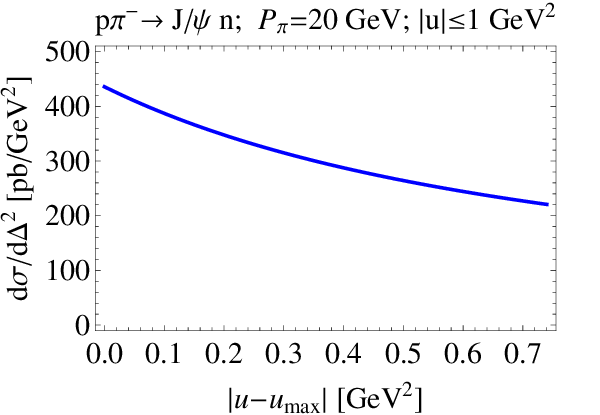}
   \end{center}
     \caption{
     Differential cross sections for $\pi^- p \to J/\psi n$  as a function of
     $| u-u_{\max}| $ for three values of the
     pion beam momentum specific for J-PARC. Left column: $P_\pi=10$~GeV corresponding to
     $W^2=19.7~{\rm GeV}^{2}$; central column: $P_\pi=15$~GeV corresponding to
     $W^2=29.1~{\rm GeV}^{2}$; right column: $P_\pi=20$~GeV corresponding to
     $W^2=38.5~{\rm GeV}^{2}$.
       [Reprinted Figure 3
from Ref.~\cite{Pire:2016gut}. [Copyright (2017) by American Physical Society.]
}
\label{JParc3}
\end{figure}

The invariant functions
$\tilde{\mathcal{J}}^{(1,2)}$
corresponding to the convolutions of the hard
subprocess amplitude with pion-to-nucleon TDAs
and nucleon DAs depend on the skewness variable (\ref{Xi_collinear}) and cross-channel invariant momentum transfer $\Delta^2 \equiv u$.
Within the simple cross channel nucleon exchange  model
(see  Sec.~\ref{SubSec_Nucle_ex_piN_TDA})
$\tilde{\mathcal{J}}^{(1,2)}$ are expressed as
\begin{eqnarray}
  &&
\tilde{\mathcal{J}}^{(1)}(\xi, \Delta^2)\Big|_{N(940)}=-\sqrt2
\frac{  f_\pi \,   g_{\pi NN}  m_N (1+\xi) } {   (\Delta^2-m_N^2) (1-\xi )} M_0;
\nonumber \\   &&
\tilde{\mathcal{J}}^{(2)}(\xi, \Delta^2)\Big|_{N(940)}=- \sqrt2
\frac{  f_\pi \,   g_{\pi NN}  m_N   } {    (\Delta^2-m_N^2)  } M_0,
\end{eqnarray}
where
$M_0$
(\ref{Def_M0})
occurs in the well-known expression for the
$J/\psi \to \bar{p} p$
decay width
(\ref{Charm_dec_width})
within the pQCD approach
\cite{Chernyak:1987nv}.
In  Fig.~\ref{JParc3}
we show the differential cross section
$\frac{d \sigma}{d \Delta^2}$
(\ref{CS_def_delta2JPARC})
for
 $\pi^- p \to J/\psi \,  n$
as a function of
$| u-u_{\max}| $ for $| u|  \le 1~{\rm GeV}^{2}$,
where
$u_{\max}$
is the threshold value
of the momentum transfer squared corresponding to the final state neutron
produced exactly in the backward direction.
We present our results for several values of the pion beam energy
$P_\pi$ typical for the J-PARC experimental set up. The cross-sections turn out to be large enough to expect a sizeable rate with a realistic luminosity; a careful feasibility study obviously needs to be carried in the near future.

\subsubsection{Studying {{TDA}}s at large energy with an electron--ion collider ({{EIC}})}
\mbox

\begin{figure}[ht]
 \begin{center}
     \includegraphics[width=8cm]{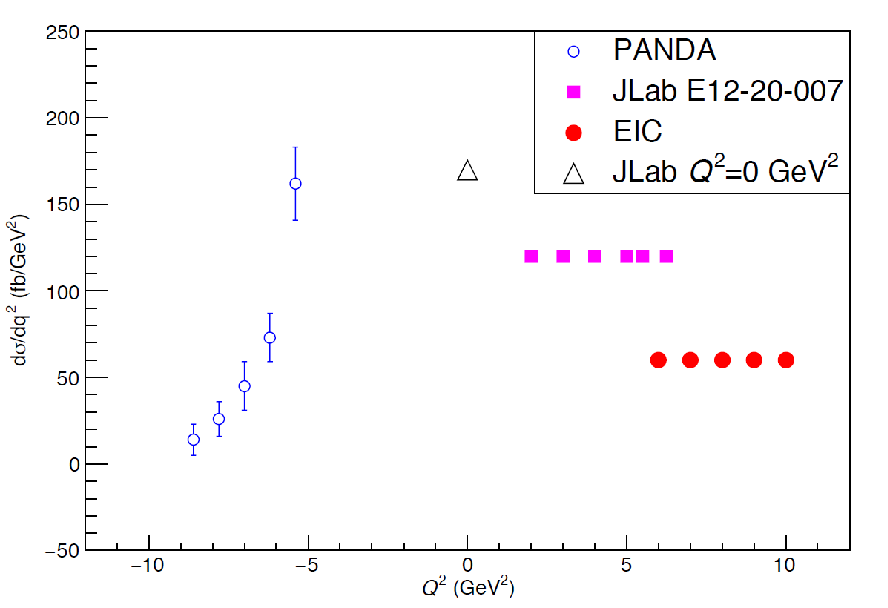}
   \end{center}
     \caption{Projected $Q^2$ coverage for \=PANDA, JLab 12 GeV (new PAC 48 proposal) and EIC $u$-channel exclusive $\pi^0$ electroproduction for
  $W=10$~GeV.
       [Reprinted Figure by W.B.~Li
from EIC Yellow Report \cite{AbdulKhalek:2021gbh}].
}
\label{EIC}
\end{figure}

Although detailed predictions have not yet been worked out for EIC energies, one can anticipate that these studies will allow this new domain of physics to be further explored. Higher $Q^2$ would be accessible in a domain of moderate $\gamma^{*} N$ energies, \textit{i.e.} rather small values of the usual $y$ variable and not too small values of $\xi$ (see Fig.~\ref{EIC}). The peculiar EIC kinematics, as compared to fixed target experiments, allows, in principle, a thorough analysis of the backward region pertinent to TDA studies. More phenomenological prospective studies are clearly needed.
The detection of $u$-channel exclusive electroproduction: $e+p\rightarrow e^\prime+ p^\prime + \pi^0$
 seems easily feasible
thanks to the 4$\pi$ coverage of EIC detector package.
A preliminary study documented in Ref.~\cite{AbdulKhalek:2021gbh}, shows the feasibility of
detecting exclusive $\pi^0$ production at $u \sim u_0$. The scattered electrons
are well within the standard detection specification. The two photons
(from decaying $\pi^0$) project a ring pattern at the zero degree
calorimeter (tagging detector along the incidence proton beam) close to
the effective acceptance, while recoiled proton enters forward EM
calorimeter at high pseudorapidity. The detector optimization and
efficiency for detecting these process is currently undergoing~\cite{EIC:RDHandbook}.

\section{Outlook and conclusions}
\setcounter{equation}{0}
\mbox

All facets of the TDA concept have not yet been developed in a satisfactory way. In this section, we address a few issues where progress may be anticipated.
We first (Section~\ref{SubSec_Non_pert_approach})
 emphasize the importance and usefulness to study
TDAs within the lattice QCD and the functional approach to QCD based on the Dyson--Schwinger equations.
We then outline some preliminary thoughts concerning further types of hard exclusive
reactions that may admit a description within the TDA framework and could be accessed experimentally. In particular
(Section~\ref{SubSec_Bkw_DVCS}) we review the status of a TDA-based description of
backward DVCS and backward time-like Compton scattering (TCS) and
we indicate the application of the TDA framework for  processes involving nuclei (Section~\ref{SubSec_Nuclear_Ph}).
Finally, we provide in  Section~\ref{SubSec_List_of_problems} a (most likely incomplete) list of key theoretical issues that
require further investigation in order to put the TDA framework on a firmer ground.

\subsection{Non-perturbative {{QCD}} approaches for { {TDA}}s}
\label{SubSec_Non_pert_approach}
\mbox

A useful complementary information for building phenomenological models of TDAs
can be obtained from calculation within the existing non-perturbative
numerical approach, namely lattice QCD, and the functional approach based
on the QCD Dyson--Schwinger equations.

\subsubsection{Lattice { {QCD}} studies for { {TDA}}s}
\mbox

Putting QCD on the lattice to study hadronic matrix elements has already produced many interesting results. Many quantities related to PDFs, GPDs and baryon DAs were calculated in various schemes, with encouraging results (for a review, see
\cite{Hagler:2009ni,Lin:2017snn}).

The common way to access hadronic matrix elements of non-local operators on the lattice goes through the calculation of their
first
Mellin moments. This approach gave interesting information on DAs, PDFs and GPDs, but was never tried for TDAs. The nearest problem which was addressed in such a way is the determination of the proton to pion form factor relevant to the calculation of the proton decay matrix element in grand-unified theories
\cite{Aoki:2017puj}.

New ideas to directly calculate PDFs and GPDs have recently been proposed
\cite{Ji:2013dva,Radyushkin:2017cyf}
to circumvent the use of the Mellin moments, and their first results
\cite{Lin:2014zya,Orginos:2017kos}
are quite encouraging. These interesting progresses  may also be used to study TDAs. Needless to say, these calculations will require enormous computer resources.

\subsubsection{Dyson--Schwinger approach for {{TDA}}s}
\mbox

The Dyson--Schwinger approach
\cite{Roberts:1994dr}
has been applied to many hadronic quantities
\cite{Cloet:2007pi},
including the $\pi$-meson wave function on the light front
\cite{Chang:2013pq}
leading to useful models for the
$\pi$-meson PDFs~\cite{Shi:2018mcb} and GPDs~\cite{Mezrag:2014jka}.
There is some recent progress on the nucleon case, where the quark PDFs have been calculated
\cite{Bednar:2018htv}
using nucleon bound state amplitudes derived from the Faddeev equations including scalar and axial-vector diquarks. These studies employ the same approximations that proved  to be successful for the description of the nucleon electromagnetic form-factors
\cite{Cloet:2008re}.
Since the Dyson--Schwinger equations capture  non-perturbative features of QCD, calculating TDAs within this approach would implement some of the physics of confinement in modeling TDAs. Working in the quasi two-body approximation for the baryon
(the quark--diquark picture) seems to be a realistic first step for these expected progresses.

\subsection{Backward DVCS, backward { {TCS}} and the nucleon to photon { {TDA}}s}
\label{SubSec_Bkw_DVCS}
\mbox

The study of deeply virtual Compton scattering (DVCS) in the near-forward region has been instrumental in the development of the collinear QCD description of deep exclusive reactions in terms of GPDs.
The peculiar feature of this reaction is that the Born term of the hard subprocess is a pure QED process. This property is shared with the exclusive photoproduction of a lepton pair, named time-like Compton scattering
(TCS)~\cite{Berger:2001xd}.
In this latter reaction the time-like nature of the highly virtual final state photon provides the large scale needed for a perturbative expansion of the coefficient function.

Both these processes mix with the pure QED Bethe--Heitler (BH) processes which are dominant in  most kinematical regions. Thanks to the specific properties of the interference cross-section the BH process serves as an amplifier for the DVCS/TCS signal.
For instance the charge-exchange property of the produced lepton pair in TCS and in the QED process allows one to easily separate the interference cross section through a charge-odd observable.

\begin{figure}[H]
\begin{center}
\includegraphics[width=0.34\textwidth]{Kin_Fact_TDA_gammaN.eps}
 \ \ \ \ \ \ \ \ \ \ \
 \includegraphics[width=0.4\textwidth]{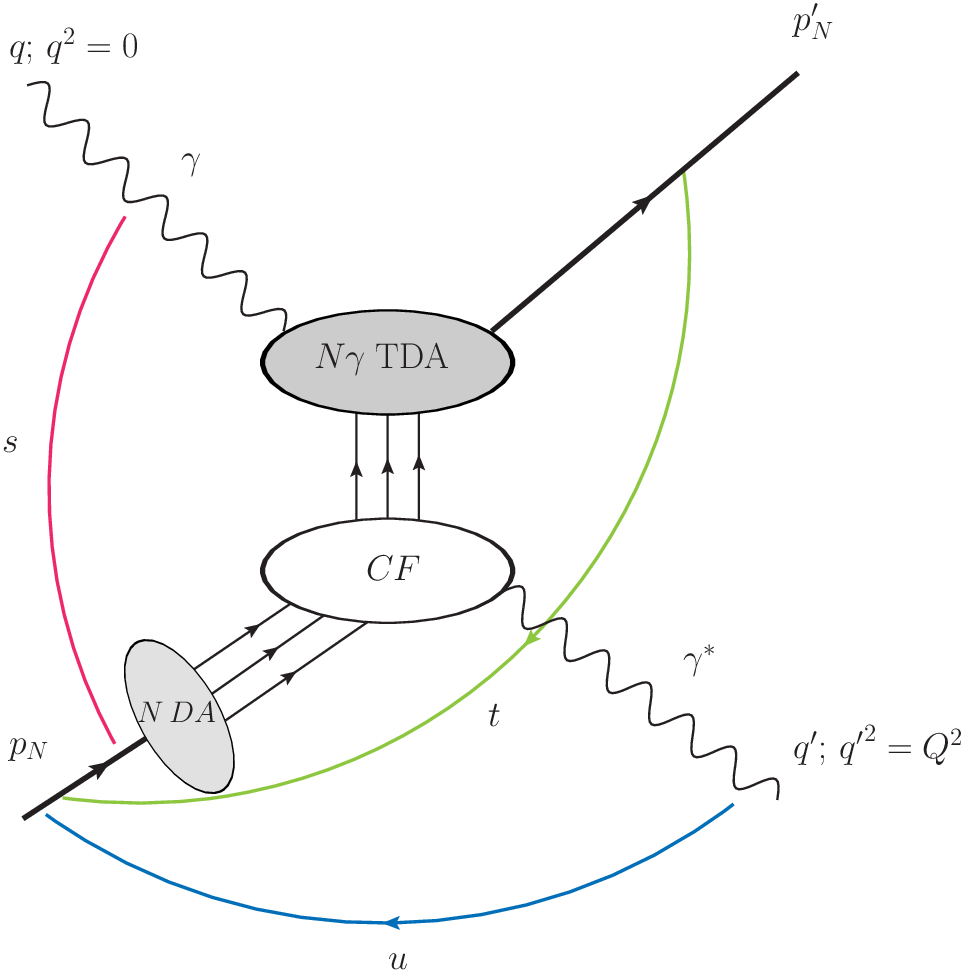}
\end{center}
\caption{
Left panel: collinear factorization mechanism  for the DVCS
($\gamma^{*} N \to \gamma N'$)  in the  near-backward  kinematics regime;
  $\gamma N$ TDA stands for the transition
     distribution amplitude from a nucleon to photon; $N$ DA stands for the nucleon distribution amplitude; $CF$  denotes hard subprocess amplitudes. Right panel: the analogous reaction mechanism for the  near-backward TCS ($\gamma N \to \gamma^{*} N'$).
 $N \gamma$ TDA stands for the transition
     distribution amplitude from a photon to nucleon.
}
\label{fig:bTCS}
\end{figure}

The case of near-backward DVCS was preliminary discussed in~\cite{Pire:2004ie,Lansberg:2006uh}; and the case of near-backward TCS is currently under investigation
(see Fig.~\ref{fig:bTCS} for the corresponding reaction mechanisms).
The photon-to-nucleon ($N  \gamma$) transition distribution amplitudes, which  enters the backward TCS amplitudes,
turn to be related by crossing to the
nucleon-to-photon ($\gamma N$) TDAs defined in  Section~\ref{SubSec_Def_Ngamma_TDAs}.
The corresponding relation is similar to the relation between
the nucleon-to-pion ($\pi N$)
and pion-to-nucleon ($N \pi$)
TDAs discussed in the context of backward production of a
$J/\psi$ with a pion beam~\cite{Pire:2016gut} (see
discussion in  Section~\ref{SubSec_piN_JPARC}):
\begin{equation}
    TDA^{\gamma  N} (x_i,\xi, u) = TDA^{N \gamma} (-x_i,-\xi, u)\,.
\end{equation}

The experimental difficulties  to access  backward DVCS and
backward TCS   processes with existing experimental facilities
turn out to be quite different.
However, the  theoretical and physical contents of the two processes are very similar. Indeed their hard amplitudes are equal (up to a complex conjugation) at the Born order, and turn to differ in a controllable way at the next to leading order in $\alpha_s$
\cite{Muller:2012yq}.

A peculiar feature of the near-backward kinematics (both for  DVCS and TCS)
is that the QED Bethe--Heitler process, which shows a pronounced peak in the forward region,
does not show any backward peak (see Fig.~23 of Ref.~\cite{Li:2020nsk}).
At the same time, within the VDM picture of nucleon-to-photon TDAs  the QCD process behaves at small $-u$ as in the backward vector meson electroproduction case.
 Therefore, a considerable backward cross section can be expected.

On the one hand, this considerably simplifies the analysis as all observed
signal in the near-backward regime can be attributed to the contribution of the hard QCD mechanism.
On the other hand, this deprives
the near-backward regime of natural amplification of the hard mechanism signal, which
is bad news if the corresponding cross section is very small.

Phenomenological models for the $\gamma N$ (and $N \gamma $) TDAs
still need to be constructed. A possible way is to use the
vector-meson-dominance (VMD) framework
(see \textit{e.g.}
\cite{Hakioglu:1991pn,Schildknecht:2005xr}).
This will allow to relate $\gamma N$ TDAs to nucleon-to-transversely-polarized-vector-meson TDAs. The detailed phenomenology of the reactions involving these new TDAs remains to be worked out.

A separate interesting open question is the dispersive analysis
of backward DVCS amplitude. This issue is closely connected to
the most debated manifestation of the $J=0$ fixed pole contribution
into the Compton amplitude (see Refs.~\cite{Brodsky:2008qu,Muller:2015vha}).

\subsection{{ {TDA}}s as a tool for  nuclear physics}
 \label{SubSec_Nuclear_Ph}
\mbox

Generalized parton distribution of nuclei has been recognized as a convenient
tool to obtain novel information on nuclear structure in terms of quarks and
gluons (for a review see \textit{e.g.}~\cite{Fucini:2020vpr}). Recent measurement
of the DVCS on the $^4$He target reported in Ref.~\cite{Hattawy:2017woc}
demonstrated the possibility to disentangle signals of the coherent and incoherent
channels of nuclear DVCS. This opens new exciting perspectives in the field.
Definitely, studies of nuclei structure through hard exclusive reactions
open a new chapter in the high energy nuclear physics and will receive
considerable attention in the experimental programs of existing  (JLab)
and future (\=PANDA, EIC) hadron physics experimental facilities.

It is, therefore, tempting to consider TDAs in a similar role.
TDAs defined as matrix element of
the three-quark light-cone operator between hadronic states
of unequal baryonic charge
 naturally appear in hard processes on nuclei.
An example is provided by a
near-forward transition occurring between two nuclei with baryon numbers differing by one unit.
The simplest process of this kind, the electrodissociation of a deuteron nucleus,
was considered in  Section~\ref{SubSec_Def_deutronN_TDAs}:
\begin{equation}
e(k)+d(p_d) \to
\left(\gamma^{*}(q)+d(p_d) \right) + e(k') \to e(k')+
p(p_p)+ n(p_n).
\label{Deut_Diss_reaction_sec8}
\end{equation}
The collinear factorized description involving the deuteron-to-nucleon TDAs
(\ref{Param_ND_TDAs})
can be applied in the generalized Bjorken limit with either
$| t| =| (p_d-p_p)^2| $
or
$| u| = | (p_d-p_n)^2| $
near their minimal values, and much smaller than the hard scale
$Q^2 =-q^2, \,W^2=(p_d+q)^2$ (see  Fig.~\ref{Fig_Deutron_Diss}).

If the deuteron were only an almost unbound system of a proton and a neutron sharing almost equally its energy--momentum, the deuteron-to-nucleon TDAs would be strongly peaked near
$\xi= \frac{1}{3}$,
with a width related to the Fermi motion effects inside deuteron.
Accessing these TDAs outside this peak will enable to probe the non-nucleonic Fock components
of the deuteron wave-function
\cite{Brodsky:1983vf}.

Such a study may be carried on a deuteron target in the present JLab set-up
\cite{Yero:2020cbq,Yero:2020urm},
or in a collider set-up at a future EIC experiment where the kinematic boost
may help to measure the outgoing nucleons momenta  with better precision and thus the $\xi$-dependence of the dissociation cross-section.

The time-like version of a  similar study may be carried with the FAIR antiproton beam and the \=PANDA facility in a deuteron target configuration with the reaction:
\begin{equation}
    \bar{p}(p_N) + d(p_d) \to \gamma^{*}(q) + n(p'_N)
\end{equation}
at large $q^2$ and with $ | t| = | (p_d-p'_N)^2| $  near its minimal value, as well as in the reaction
with the reaction:
\begin{equation}
    \bar{p}(p_N) + d(P_d) \to J/\psi(p_\psi) +  n(p'_N),
\end{equation}
in the same kinematics.

\subsection{A list of  key theoretical problems}
\label{SubSec_List_of_problems}
\mbox

In addition to the points we just somewhat elaborated before,
let us now provide a list of theoretical questions that require
further clarification.

\bi

\item Exploring the soft-hard transition in backward meson production
 as function of $Q^2$.
This study is naturally put inside a broader context of generic
studies of soft-hard transition for electroproduction processes
\cite{WeissTalk20,WP:21}.

\item The status of the collinear factorization theorem for the backward
reactions needs further elucidation.

\item For the moment the relevance of the evolution effects for nucleon-to-meson TDAs
remains completely unexplored. A first step could be a development of the TDA analogue
of the evolution code~\cite{Vinnikov:2006xw}. A more refined approach can be designed
relying on the conformal partial wave expansion framework sketched in  Section~\ref{SubSec_Conf_PW_exp_TDAs}.

\item It is essential to get a better understanding of the relevance of  higher twist
(twist-$4$) corrections to backward meson production. This analysis can also  be useful in
identifying observables that are less sensitive to higher twist corrections.

\item The  calculation of the NLO perturbative corrections
to hard meson production amplitudes has also to be performed. This is a
complicated task since, to the best of our knowledge, no NLO calculation
has been published even for the simpler case of the nucleon form factor. Also there exists neither any NLO calculations for fixed angle Compton scattering nor meson photoproduction at fixed angle.

\item Further efforts are required in building up flexible phenomenological models
for ${\mathcal{M}}N$ and $\gamma N$ TDAs. A possible option is the implementation of the
Regge-like behavior for small $\xi$ within the spectral representation
of TDAs in terms of quadruple distributions. Also a more realistic modeling of the
$\Delta^2$-dependence of TDAs is highly needed.

\item GPDs satisfy positivity constraints~\cite{Pire:1998nw,Pobylitsa:2002iu} which relate them (thorough inequalities) to parton distribution functions.
 Similar constraints should exist for TDAs, with nucleon matrix elements of twist four multiparton operators, somewhat related to but different from double parton distributions (for a review, see~\cite{Diehl:2017wew}),
entering the game.

\item The exploration of partonic distributions in the small-$x$ domain sensitive to transverse partonic momenta is one of the main motivations for the construction of the EIC. Therefore, an extension of the description of processes with collinear TDAs towards the small-$x$ regime constitutes a natural further step in the development of this concept, which requires the understanding of the relationship between the collinear TDAs  and their descendants depending on transverse partonic momenta. This type of correspondence was established in
Ref.~\cite{Lorce:2011dv} for the case of quark--antiquark operator in terms of the so-called Generalized Parton Correlation Functions (GPCFs).
\item TDAs may be a tool to understand other reactions within some phenomenological models, as was advocated for instance in Ref.~\cite{Goritschnig:2012vs} for the annihilation of proton--antiproton into a
charmed-anticharmed meson pair.

\item The QCD light cone sum rule technique has been applied with some success to the analysis of the nucleon form factors~\cite{Braun:2001tj}. Recently, some nucleon-to-meson matrix elements have been evaluated in this
framework~\cite{Haisch:2021hvj}  in order to quantify proton decay rates in grand unified theories. This technique could be applied to determine a few moments of various nucleon-to-meson TDAs, and thus constrain TDA
modeling.
\ei

Moreover, as already stressed, the future of the field to which this review is devoted depends very much on future experimental measurements, both with electron and antiproton beams. We are confident that new data will appear in the near future and help to clarify the physics content of hard exclusive reactions with baryonic exchange. If these experimental tests of the factorized QCD framework are indeed successful, our framework will allow us to relate the observables which will be accessed  to hadronic matrix elements of three quark operators which have a well-defined field-theoretic status. The fact that these objects are out of the reach of perturbative studies -- except for their renormalization group evolution equations -- is characteristic of many quantities related to the confining nature of QCD. The challenge at stake is the understanding of the very nature of the atomic nucleus, made of quarks and gluons in a yet mysterious but hopefully decidable way.


\renewcommand{\thesection}{}
\makeatletter
\def\@seccntformat#1{\csname #1ignore\expandafter\endcsname\csname the#1\endcsname\quad}

\section{Acknowledgements}
\mbox

We acknowledge first the early and very fruitful collaboration of Jean-Philippe Lansberg in this long study. We benefited a lot from numerous discussions with our experimentalist friends  Harut Avakian, Stefan Diehl, Michel Guidal, Garth Huber, Bill Lee, Binsong Ma, Frank Maas, Maria-Carmen Mora-Espi,  Kijun Park, Beatrice Ramstein, Justin Stevens, Carlo Yero, Manuel Zambrana. We are indebted for many enlightening discussions with Igor Anikin, Volodia Braun, Wim Cosyn, Markus Diehl, Alexander Goritschnig, Kolya Kivel, Jean-Marc Laget,  C\'edric Lorc\'e, C\'edric Mezrag, Herv\'e Moutarde, Dieter M\"uller,  Barbara Pasquini, Maxim Polyakov, Anatoly Radyushkin, Wolfgang Schweiger, Nico Stefanis,  Mark Strikman, Oleg Teryaev, Samuel Wallon, Christian Weiss.
We also would like to thank  Victor Chernyak, John Collins, Florian Hauenstein, Chueng Ji,  Lothar Tiator for correspondence.
We would like to specially thank Barbara Pasquini for providing the results of
calculations of $\pi N$ TDAs in the light-cone quark model and the corresponding computer code.

This work was supported in part by the  {European Union's Horizon 2020 research and innovation programme} under Grant Agreement No.{824093} and by
the P2IO LabEx ({ANR-10-LABX-0038})  managed by the {Agence Nationale de la Recherche}, France.
The work of K.S. is supported by the {Foundation for
the Advancement of Theoretical Physics and Mathematics} ``BASIS''. The work of L.S. is supported by the grant {2019/33/B/ST2/02588} of the {National Science Center in Poland}.

\renewcommand*{\thesection}{\Alph{section}}

\appendix

\setcounter{section}{0}
\setcounter{equation}{0}
\renewcommand{\thesection}{\Alph{section}}
\renewcommand{\theequation}{\thesection\arabic{equation}}

\section{Fierz identities}
\label{App_Fierz}
\mbox

Employing the Fierz identity for the
$\gamma$-matrices
(see \textit{e.g.}
\cite{Borodulin:2017pwh})
one may establish the following useful identity for arbitrary
Dirac structures $\Gamma$, $\Gamma'$:
\begin{eqnarray}
  &&
(\Gamma C)_{\rho \tau} (\Gamma' U)_\chi \nonumber \\ &&
=\frac{1}{4}
\bigl\{
C_{\chi \tau} (\Gamma \Gamma' U)_\rho+
(\gamma^\mu C)_{\chi \tau} (\Gamma \gamma^\mu \Gamma' U)_\rho+
(\gamma^5 C)_{\chi \tau} (\Gamma \gamma^5 \Gamma' U)_\rho-
(\gamma^5 \gamma^\mu C)_{\chi \tau} (\Gamma \gamma^5 \gamma^\mu \Gamma' U)_\rho
\nonumber \\ &&
-\frac{1}{2} (\sigma^{\mu \nu} C)_{\chi \tau}  (\Gamma \sigma_{\mu \nu} \Gamma' U)_\rho
\bigr\}\,.
\label{Master_Fierz}
\end{eqnarray}
Here $U$ stands for an arbitrary spin-tensor with one Dirac index and $C$ is the charge conjugation matrix.

\subsection{Nucleon { {DA}}}
\mbox

To the leading twist-$3$ the parametrization of the nucleon DA involves
the following Dirac structures
\begin{equation}
v_{\rho \tau, \, \chi}^N=(\hat{p} C)_{\rho \tau} (\gamma^5 U )_\chi\,;
\ \
a_{\rho \tau, \, \chi}^N=(\hat{p} \gamma^5 C)_{\rho \tau} ( U )_\chi
\,;
\ \
t_{\rho \tau, \, \chi}^N=(\sigma_{p \mu} C)_{\rho \tau} ( \gamma^\mu \gamma^5 U )_\chi\,.
\label{DA structures_App}
\end{equation}
The Dirac structures
(\ref{DA structures_App})
satisfy symmetry relations:
\begin{equation}
v_{\rho \tau, \, \chi}^N=v_{ \tau \rho, \, \chi}^N\,; \ \ \
a_{\rho \tau, \, \chi}^N=-a_{\tau \rho , \, \chi}^N\,; \ \ \
t_{\rho \tau, \, \chi}^N=t_{ \tau \rho, \, \chi}^N\,.
\label{symmetry_Dirac_Nucleon}
\end{equation}

With the help of
(\ref{Master_Fierz})
one may work out the following set of the Fierz identities valid to the leading twist-$3$ accuracy:
\begin{eqnarray}
  &&
v_{\rho \tau, \, \chi}^N= \frac{1}{2} \left(v^N-  a^N-  t^N \right)_{\chi \tau, \, \rho};
\;
a_{\rho \tau, \, \chi}^N= \frac{1}{2} \left(-v^N+  a^N-  t^N \right)_{\chi \tau, \, \rho};
\nonumber \\   &&
t_{\rho \tau, \, \chi}^N=   \left(-v^N-  a^N \right)_{\chi \tau, \, \rho}.
\label{Fierz_nucleon_structures}
\end{eqnarray}

\subsection{$\Delta(1232)$  { {DA}}}
 \mbox

Leading twist Dirac structures employed in the parametrization
(\ref{Parametrization_Delta_DA_FZ})
of
$\Delta(1232)$
resonance DA read:
\begin{eqnarray}
  &&
v^\Delta_{\rho \tau,\, \chi}= (\gamma_\mu C)_{\rho \tau} \, \mathcal{U}^\mu_\chi\,;
\ \ \
a^\Delta_{\rho \tau,\, \chi}= (\gamma_\mu \gamma_5 C)_{\rho \tau} \, (\gamma_5\mathcal{U}^\mu)_\chi\,;
\ \ \
t^\Delta_{\rho \tau,\, \chi}= \frac{1}{2} (\sigma_{\mu \nu} C)_{\rho \tau}(\gamma^\mu \mathcal{U}^\nu)_\chi\,;
\nonumber \\
  &&
\varphi_{\rho \tau,\, \chi}^\Delta= (\sigma_{\mu \nu} C)_{\rho \tau} (p^\mu \mathcal{U}^\nu- \frac{1}{2} m_\Delta \gamma^\mu \mathcal{U}^\nu)_\chi\,.
\label{Dirac_structures_Delta_App}
\end{eqnarray}
The Dirac structures
(\ref{Dirac_structures_Delta_App})
satisfy symmetry relations:
\begin{equation}
v_{\rho \tau, \, \chi}^\Delta=v_{ \tau \rho, \, \chi}^\Delta\,; \ \ \
a_{\rho \tau, \, \chi}^\Delta=-a_{\tau \rho , \, \chi}^\Delta\,; \ \ \
t_{\rho \tau, \, \chi}^\Delta=t_{ \tau \rho, \, \chi}^\Delta\,; \ \ \
\varphi_{\rho \tau, \, \chi}^\Delta=\varphi_{ \tau \rho, \, \chi}^\Delta\,;
\label{symmetry_Dirac_structures_Delta}
\end{equation}
The set of the corresponding Fierz identities valid to the leading twist-$3$ accuracy reads:
\begin{eqnarray}
  &&
v^\Delta_{\rho \tau, \, \chi}= \bigl( \frac{1}{2} v^\Delta -\frac{1}{2} a^\Delta +t^\Delta \bigr)_{\chi \tau, \rho}\,;
\ \ \
a^\Delta_{\rho \tau, \, \chi}=  \bigl( -\frac{1}{2} v^\Delta +\frac{1}{2} a^\Delta +t^\Delta \bigr)_{\chi \tau, \rho}\,;
\nonumber \\
  &&
t^\Delta_{\rho \tau, \, \chi}=  \bigl( \frac{1}{2} v^\Delta +\frac{1}{2} a^\Delta  \bigr)_{\chi \tau, \rho}\,;
\ \ \
\varphi_{\rho \tau, \, \chi}^\Delta=\varphi_{\chi \tau, \rho}^\Delta\,.
\label{Fierz_Delta_structures}
\end{eqnarray}

\subsection{$\pi N$ { {TDA}}}
\label{Subsec_Fierz_piN_TDA}
\mbox

In this subsection we review the properties of the Dirac structures (\ref{Dirac_structures_PiN_TDA_Cov})
occurring in the parametrization of the $\pi N$ TDA.
Firstly, one may check that the Dirac structures
(\ref{Dirac_structures_PiN_TDA_Cov})
satisfy symmetry relations:
\begin{equation}
(v_{1,2}^{\pi N})_{\rho \tau, \, \chi}=(v_{1,2}^{\pi N})_{ \tau \rho, \, \chi} \,; \ \ \
(a_{1,2}^{\pi N})_{\rho \tau, \, \chi}=-(a_{1,2}^{\pi N})_{ \tau \rho, \, \chi} \,; \ \ \
(t_{1,2,3,4}^{\pi N})_{\rho \tau, \, \chi}=(t_{1,2,3,4}^{\pi N})_{ \tau \rho, \, \chi} \,.
\label{symmetry_Dirac_structures_PiN_TDA}
\end{equation}

The set of the corresponding Fierz identities for the structures
$s_{1,2}^{\pi N}$
is similar to that for the case of the nucleon DA (\ref{Fierz_nucleon_structures}):
\begin{eqnarray}
  &&
(v_{1,2}^{\pi N})_{\rho \tau, \, \chi}=
\frac{1}{2}  \left( v_{1,2}^{\pi N}
-
a_{1,2}^{\pi N}
-
t_{1,2}^{\pi N}
\right)_{\chi \tau, \, \rho}; \ \ \
(a_{1,2}^{\pi N})_{\rho \tau, \, \chi}=
\frac{1}{2}\left(-  v_{1,2}^{\pi N}
+
a_{1,2}^{\pi N}-
t_{1,2}^{\pi N} \right)_{\chi \tau, \; \rho};
\nonumber \\
  &&
(t_{1,2}^{\pi N})_{\rho \tau, \, \chi}=
\left(-   v_{1,2}^{\pi N}
- a_{1,2}^{\pi N} \right)_{\chi \tau, \; \rho}.
\label{Fierz_for_piNTDA_structures}
\end{eqnarray}
The result for
$(t_{3,4}^{\pi N})$
is a  bit more involved:
\begin{eqnarray}
  &&
(t_{3}^{\pi N})_{\rho \tau, \, \chi}
 =
(t_{3}^{\pi N})_{\chi \tau, \, \rho}+
g_1(\xi, \Delta^2)
\left( v_{1}^{\pi N}+a_{1}^{\pi N}+t_{1}^{\pi N} \right)_{\chi \tau, \, \rho} +
g_2(\xi, \Delta^2)
\left(
v_{2}^{\pi N}+a_{2}^{\pi N}+t_{2}^{\pi N}
\right)_{\chi \tau, \, \rho}\,;
 \nonumber \\   &&
 (t_{4}^{\pi N})_{\rho \tau, \, \chi}
 =
(t_{4}^{\pi N})_{\chi \tau, \, \rho}+
h_1(\xi, \Delta^2)
\left( v_{1}^{\pi N}+a_{1}^{\pi N}+t_{1}^{\pi N} \right)_{\chi \tau, \, \rho} +
h_2(\xi, \Delta^2)
\left(
v_{2}^{\pi N}+a_{2}^{\pi N}+t_{2}^{\pi N}
\right)_{\chi \tau, \, \rho} \,,
 \label{Fierz for t34}
\end{eqnarray}
where
\begin{eqnarray}
  &&
g_1(\xi, \Delta^2)=\frac{-\Delta ^2 (1-\xi)-2 \left(m_\pi^2+m_N^2\right) \xi }{4 m_N^2} \,;
\nonumber \\   &&
g_2(\xi, \Delta^2)=
\frac{\Delta ^2 (1-\xi)+2 \left(m_\pi^2+m_N^2\right) \xi-4m_N^2}{8 m_N^2}\,;
\nonumber \\   &&
h_1(\xi, \Delta^2)=\frac{-\Delta ^2(1-\xi) -2 \left(m_\pi^2-m_N^2\right) \xi }{2 m_N^2}\,;
\nonumber \\   &&
h_2(\xi, \Delta^2)=\frac{\Delta ^2 (1-\xi )+2 \left(m_\pi^2+m_N^2\right) \xi }{4 m_N^2}\,.
 \label{Def_g12_h12}
\end{eqnarray}

\subsection{$VN$ { {TDA}}s}
\label{identities}
\mbox

In this subsection we consider the set of the Fierz identities for the Dirac structures
(\ref{Dirac_v_NV})--(\ref{Dirac_t_NV})
occurring in the parametrization of the leading-twist-$3$ nucleon-to-vector meson ($VN$) TDAs
(\ref{VN_TDAs_param}).
These Fierz identities are employed
to establish the consequences of the isotopic and permutation symmetry
for $VN$ TDAs.
\begin{eqnarray}
  &&
(v_{1 {\mathcal{E}}}^{V N})_{\rho \tau, \, \chi}= \frac{1}{2} \left(
 v_{1 {\mathcal{E}}}^{VN}-  a_{1 {\mathcal{E}}}^{VN}+
  t_{1 {\mathcal{E}}}^{VN}+  t_{2 {\mathcal{E}}}^{VN} \right)_{\chi \tau, \, \rho}; \ \ \
 (a_{1 {\mathcal{E}}}^{VN})_{\rho \tau, \, \chi}=\frac{1}{2} \left(
- v_{1 {\mathcal{E}}}^{VN} + a_{1 {\mathcal{E}}}^{VN} +
   t_{1 {\mathcal{E}}}^{VN} + t_{2 {\mathcal{E}}}^{VN}  \right)_{\chi \tau, \, \rho};
\nonumber \\   &&
(t_{1 {\mathcal{E}}}^{VN})_{\rho \tau, \, \chi}=
\frac{1}{2} \left(   v_{1 {\mathcal{E}}}^{VN} +
 a_{1 {\mathcal{E}}}^{VN} +
 t_{1 {\mathcal{E}}}^{VN} -
 t_{2 {\mathcal{E}}}^{VN}  \right)_{\chi \tau, \, \rho}; \ \ \
 (t_{2 {\mathcal{E}}}^{VN})_{\rho \tau, \, \chi}=\frac{1}{2} \left( v_{1 {\mathcal{E}}}^{VN} +
 a_{1 {\mathcal{E}}}^{VN} -
 t_{1 {\mathcal{E}}}^{VN} +
   t_{2 {\mathcal{E}}}^{VN} \right)_{\chi \tau, \, \rho};
\label{Fiers_1E_set}
\end{eqnarray}
\begin{eqnarray}
   &&
(v_{1T,1n}^{VN})_{\rho \tau,\,\chi}= \frac{1}{2} \left(v_{1T,1n}^{VN} - a_{1T,1n}^{VN} + t_{1T,1n}^{VN} \right)_{\chi \tau,\,\rho}; \ \ \
 (a_{1T,1n}^{VN})_{\rho \tau,\,\chi}=  \frac{1}{2} \left(- v_{1T,1n}^{VN}+ a_{1T,1n}^{VN} +   t_{1T,1n}^{VN} \right)_{\chi \tau,\,\rho};
\nonumber \\   &&
(t_{1T,1n}^{VN})_{\rho \tau,\,\chi}=   \left(v_{1T,1n}^{VN} +  a_{1T,1n}^{VN} \right)_{\chi \tau,\,\rho};
\label{Fiers_1T_set}
\end{eqnarray}
\begin{eqnarray}
  &&
(v^{VN}_{2 {\mathcal{E}}})_{\rho \tau,\,\chi}= \frac{1}{2}  \left( v^{VN}_{2 {\mathcal{E}}} -  a^{VN}_{2 {\mathcal{E}}} +
 t^{VN}_{3 {\mathcal{E}}}  -t^{VN}_{4 {\mathcal{E}}} \right)_{\chi \tau,\,\rho}; \ \ \
  (a^{VN}_{2 {\mathcal{E}}})_{\rho \tau,\,\chi}=  \frac{1}{2} \left(-v^{VN}_{2 {\mathcal{E}}} +  a^{VN}_{2 {\mathcal{E}}}
+ t^{VN}_{3 {\mathcal{E}}} - t^{VN}_{4 {\mathcal{E}}} \right)_{\chi \tau,\,\rho}; \nonumber \\   &&
(t^{VN}_{3 {\mathcal{E}} })_{\rho \tau,\,\chi}= \frac{1}{2 }  \left(v_{1T}^{VN}  +     v_{2 {\mathcal{E}}}^{VN} +
  a_{1T}^{VN}  +     a_{2 \mathcal{E}}^{VN}  +
 t_{3{\mathcal{E}}}^{VN} -  t_{1T}^{VN} +  t_{4 {\mathcal{E}}}^{VN} \right)_{\chi \tau,\,\rho};
\nonumber \\   &&
(t^{VN}_{4{\mathcal{E}}})_{\rho \tau,\,\chi}= \frac{1}{2} \left( - v^{VN}_{2 {\mathcal{E}}}+  v^{VN}_{1 T}
-  a^{VN}_{2 {\mathcal{E}}}+ a^{VN}_{1 T}
+   t^{VN}_{3{\mathcal{E}}}-  t^{VN}_{1T} +  t^{VN}_{4{\mathcal{E}}} \right)_{\chi \tau,\,\rho};
\label{Fiers_2E_set}
\end{eqnarray}
\begin{eqnarray}
  &&
(v_{2T,2n}^{VN})_{\rho \tau,\,\chi}=\frac{1}{2} \left(v_{2T,2n}^{VN} -   a_{2T,2n}^{VN} + t_{2T,2n}^{VN} + t_{3T,3n}^{VN}
\right)_{\chi \tau,\,\rho};
\nonumber \\   &&
(a_{2T,2n}^{VN})_{\rho \tau,\,\chi}=\frac{1}{2} \left(-v_{2T,2n}^{VN} + a_{2T,2n}^{VN} + t_{2T,2n}^{VN} + t_{3T,3n}^{VN}
\right)_{\chi \tau,\,\rho};
\nonumber \\   &&
(t_{2T,2n}^{VN})_{\rho \tau,\,\chi}=\frac{1}{2} \left(
v_{2T,2n}^{VN} + a_{2T,2n}^{VN} + t_{2T,2n}^{VN} - t_{3T,3n}^{VN}
\right)_{\chi \tau,\,\rho};
\nonumber \\   &&
(t_{3T,2n}^{VN})_{\rho \tau,\,\chi}=\frac{1}{2} \left(
v_{2T,2n}^{VN} + a_{2T,2n}^{VN} - t_{2T,2n}^{VN} + t_{3T,3n}^{VN}
\right)_{\chi \tau,\,\rho};
\label{Fiers_2T_set}
\end{eqnarray}
\begin{eqnarray}
  &&
(t_{4T}^{VN})_{\rho \tau,\,\chi}= \frac{1}{2}  \frac{\Delta_T^2}{m_N^2} \left(v_{1T}^{VN}  -   a_{1T}^{VN}
-   t_{1T}^{VN} \right)_{\chi \tau,\,\rho}+(t_{4T}^{VN})_{\chi \tau,\,\rho}; \nonumber \\   &&
(t_{4n}^{VN})_{\rho \tau,\,\chi}= \frac{1}{2}  \frac{\Delta_T^2}{m_N^2} \left(v_{1n}^{VN}  -   a_{1n}^{VN}
-   t_{1n}^{VN} \right)_{\chi \tau,\,\rho}+(t_{4n}^{VN})_{\chi \tau,\,\rho}.
   \label{Fiers_last_set}
\end{eqnarray}

\section{Calculation of the convolution integrals}
\label{App_Convolutions}
\mbox

In this Appendix we review the properties of the
relevant singular generalized functions and express the real and imaginary parts
of the typical convolution integrals
(\ref{I_I}), (\ref{I_II})
of nucleon-to-meson TDAs with the LO hard scattering kernels.

\subsection{The relevant generalized functions}
\mbox

Sohotsky's formula (see \textit{e.g.} Chapter~II of
\cite{Vladimirov})
reads:
\begin{equation}
\frac{1}{x \pm i 0}= \mp i \pi \delta (x)+ \mathcal{P} \frac{1}{x}\,,
\label{Soh_formula_Original}
\end{equation}
where
$\mathcal{P} $
stands for the Cauchy principal value prescription.
The generalized function
$\mathcal{P}\frac{1}{x^2}$
is then defined as
$
\frac{d}{dx} \mathcal{P}\frac{1}{x}=-\mathcal{P}\frac{1}{x^2}
$.
For arbitrary smooth test function $\varphi(x)$ defined on
$(-\infty; \infty)$
and decreasing fast enough at infinity,
\begin{equation}
\left(\mathcal{P} \frac{1}{x^2}, \, \varphi(x) \right)=
\mathcal{P} \! \! \int_{-\infty}^\infty dx \frac{\varphi(x)-\varphi(0)}{x^2}\,.
\label{Px2_def}
\end{equation}
Employing
(\ref{Soh_formula_Original}), (\ref{Px2_def})
one can establish the familiar relation:
\begin{equation}
\frac{1}{(x \pm i 0)^2}=-\frac{d}{dx} \frac{1}{x \pm i 0}=
\pm i \pi \delta'(x)+ \mathcal{P} \frac{1}{x^2}\,.
\label{Vlad_fla}
\end{equation}

The formula
(\ref{Vlad_fla})
applies for the conventional generalized function $\frac{1}{(x \pm i0)^2}$.
In our case, we have to consider a different class of
generalized functions dealing with the test functions $\varphi(x)$
defined on the interval
$[A; B]$. We assume $A<0$, $B>0$
so that the singularity point
$x=0$
belongs to the interval $[A; B]$.
The analogues of formulas
(\ref{Soh_formula_Original}),
(\ref{Vlad_fla})
for
the singular generalized functions
$\frac{1}{x \pm i0}$
and
$\frac{1}{(x \pm i0)^2}$
read (see Appendix C of Ref.~\cite{Lansberg:2011aa}):
\begin{equation}
\left(\frac{1}{x \pm i0}, \, \varphi(x) \right)_{[A;\,B]}= \mp i \pi \varphi(0)+
{\mathcal{P}} \int_A^B dx \frac{1}{x} \varphi(x);
\label{Soh_formula}
\end{equation}
and
\begin{equation}
\left(\frac{1}{(x \pm i 0)^2},\, \varphi(x) \right)_{[A,B]}
= \mp i \pi \varphi'(0)+ \varphi(0) \frac{(B-A)}{AB}+
\mathcal{P}\int_{A}^B dx \frac{1}{x} \frac{\varphi(x)-\varphi(0)}{x}\,,
\label{PV1/x2}
\end{equation}
where ${\mathcal{P}} $ denotes the principal value integral prescription.

\subsection{Real and imaginary parts of
TDA convolutions with hard scattering kernels}
\label{App_Conv_Re_Im}
\mbox

First we consider the integrals (\ref{I_I}). With help of
Sohotsky's formula (\ref{Soh_formula}) they can be rewritten as
\begin{eqnarray}
  &&
I_I^{(+, \pm)}(\xi)\nn \\ && = \int_{-1}^1 dw
\frac{1}{(w+\xi- i 0)}
\Bigl(
{\mathcal{P}} \int_{-1+| \xi-\xi'| }^{1-| \xi-\xi'| } dv
\frac{1}{(v \pm \xi')} H(w,v,\xi)
\pm
i \pi H(w,\mp \xi',\xi)
\Bigr)
\nn \\ &&
= \mathcal{P}\int_{-1}^1 dw
\frac{1}{(w+\xi)}
{\mathcal{P}} \int_{-1+| \xi-\xi'| }^{1-| \xi-\xi'| } dv
\frac{1}{(v \pm \xi')} H(w,v,\xi)
\mp \pi^2 H(-\xi, \mp \xi,\xi)
\nn \\ &&
\pm i \pi
\mathcal{P}\int_{-1}^1 dw
\frac{1}{(w+\xi)}H(w,\mp \xi',\xi)+i \pi  {\mathcal{P}}\int_{-1}^1 dv \frac{1}{(v \pm \xi)} H(-\xi,v,\xi);
\label{I_I11}
\end{eqnarray}
and
\begin{eqnarray}
  &&
I_I^{(-, \pm)}(\xi)\nn \\
&& = \int_{-1}^1 dw
\frac{1}{(w-\xi+ i 0)}
\Bigl(
{\mathcal{P}} \int_{-1+| \xi-\xi'| }^{1-| \xi-\xi'| } dv
\frac{1}{(v \pm \xi')} H(w,v,\xi)
\pm
i \pi H(w,\mp \xi',\xi)
\Bigr)
\nn \\ &&
= \mathcal{P}\int_{-1}^1 dw
\frac{1}{(w-\xi)}
{\mathcal{P}} \int_{-1+| \xi-\xi'| }^{1-| \xi-\xi'| } dv
\frac{1}{(v \pm \xi')} H(w,v,\xi)
\pm \pi^2 H(\xi, 0,\xi)
\nn \\ &&
\pm i \pi
\mathcal{P}\int_{-1}^1 dw
\frac{1}{(w-\xi)}H(w,\mp \xi',\xi)-i \pi  {\mathcal{P}}\int_{-1+\xi}^{1-\xi} dv \frac{1}{v} H(\xi,v,\xi).
\label{I_I12}
\end{eqnarray}
The final expressions for the real and imaginary parts
of
$I_I^{(\pm, \pm)}(\xi)$
read
\begin{eqnarray}
  &&
\re I^{(+,\pm)}_I(\xi)=  \mathcal{P} \!\! \int_{-1}^1 dw \,\frac{1}{(w+\xi)}  \, \mathcal{P} \!\! \int_{-1+| \xi-\xi'| }^{1-| \xi-\xi'| } dv\,
  \frac{1}{(v \pm \xi') }H(w,v,\xi)    {\boldsymbol \mp} \pi^2 H(-\xi,\mp \xi,\xi);
\nonumber \\   &&
\re I^{(-,\pm)}_I(\xi)= \mathcal{P} \!\! \int_{-1}^1 dw \, \frac{1}{(w-\xi)}  \mathcal{P} \!\! \int_{-1+| \xi-\xi'| }^{1-| \xi-\xi'| } dv\,
\frac{1}{(v \pm \xi') }H(w,v,\xi)  {\boldsymbol \pm} \pi^2 H(\xi,0,\xi);
\nonumber \\
  &&
\im  I^{(+,\pm)}_I(\xi)=  {\boldsymbol \pm} \pi \mathcal{P} \!\! \int_{-1}^1 dw  \frac{1}{(w+\xi) } H(w,\mp \xi', \xi)+
\pi \mathcal{P} \!\! \int_{-1 }^{1 }dv \frac{1}{(v \pm \xi)} H(-\xi,v,\xi);
\nonumber \\   &&
\im  I^{(-,\pm)}_I(\xi)=  {\boldsymbol \pm} \pi \mathcal{P} \!\! \int_{-1}^1 dw  \frac{1}{(w-\xi)} H(w,\mp \xi', \xi)-
\pi \mathcal{P}  \!\! \int_{-1+\xi }^{1 -\xi}dv \frac{1}{v  } H(\xi,v,\xi).
\label{FixedI_I}
\end{eqnarray}
The signs that differ from those of Eq.~(39) of~\cite{Lansberg:2011aa}
are marked with the { bold font}.

Now we turn to the integrals (\ref{I_II}).
They can be rewritten as
\begin{equation}
 I_{II}^{(-, \pm)}(\xi)
=\int_{-1}^1 dw \frac{1}{(w-\xi+ i 0)^2}
\Biggl\{{\boldsymbol \pm} i \pi H(w,\mp \xi',\xi)+
J^{(\pm)}(w,\xi)\Biggr\},
\label{I2_start}
\end{equation}
where
we employ the notation:
\begin{equation}
J^{(\pm)}(w,\xi)=\mathcal{P} \!\! \int_{-1+| \xi-\xi'| }^{1-| \xi-\xi'| } dv
\frac{1}{v \pm \xi'} H(w,v,\xi)\,.
\label{Def_J}
\end{equation}
To express the real
and imaginary parts of (\ref{I_II}), we apply
(\ref{PV1/x2})
to (\ref{I2_start}).
The factors which differ from that of Eq.~(42) of~\cite{Lansberg:2011aa}
are marked with the {bold font}.
The final expression read as
\begin{eqnarray}
  &&
\re I_{II}^{(-, \pm)}(\xi) \nonumber \\
&& = \pm \pi^2  \left( \frac{d H(w,\,\mp \xi',\xi)}{d w} \right)_{w=\xi}-\frac{2}{ {\boldsymbol 1 \boldsymbol -\boldsymbol \xi^{{\bf 2}}}} J^{(\pm)}(\xi,\,\xi)+
\mathcal{P} \!\! \int_{-1}^1 dw  \frac{1}{(w-\xi)} \frac{ \left( J^{(\pm)}(w,\xi)-J^{(\pm)}(\xi,\xi) \right)}{(w-\xi)};
\label{Re_I2_fixed}
\end{eqnarray}
\begin{eqnarray}
  &&\im I_{II}^{(-, \pm)}(\xi) \nonumber \\
&&=  {\boldsymbol \mp} \frac{2}{ {\boldsymbol 1 - \boldsymbol \xi^{{\bf 2}}}} \pi H(\xi,0,\xi)
 {\boldsymbol \pm}
 \pi \mathcal{P} \!\! \int_{-1}^1 dw \frac{1}{(w-\xi)}
\frac{(H(w,\,\mp \xi',\xi)- H(\xi,0,\xi))}{(w-\xi)}
- \pi \left(
\frac{d J^{(\pm)}(w,\xi)}{dw}
\right)_{w=\xi}.
\label{Im_I2_fixed}
\end{eqnarray}

\renewcommand{\thesection}{}
\makeatletter
\def\@seccntformat#1{\csname #1ignore\expandafter\endcsname\csname the#1\endcsname\quad}

\section{References}

\bibliographystyle{elsarticle-num}
\bibliography{Bib_TDAreview}

\begin{thebibliography}{100}
\expandafter\ifx\csname url\endcsname\relax
  \def\url#1{\texttt{#1}}\fi
\expandafter\ifx\csname urlprefix\endcsname\relax\def\urlprefix{URL }\fi
\expandafter\ifx\csname href\endcsname\relax
  \def\href#1#2{#2} \def\path#1{#1}\fi

\bibitem{Politzer:1974fr}
H.~Politzer, {Asymptotic Freedom: An Approach to Strong Interactions}, Phys.
  Rept. 14 (1974) 129--180.
\newblock \href {http://dx.doi.org/10.1016/0370-1573(74)90014-3}
  {\path{doi:10.1016/0370-1573(74)90014-3}}.

\bibitem{Marciano:1977su}
W.~J. Marciano, H.~Pagels, {Quantum Chromodynamics: A Review}, Phys. Rept. 36
  (1978) 137.
\newblock \href {http://dx.doi.org/10.1016/0370-1573(78)90208-9}
  {\path{doi:10.1016/0370-1573(78)90208-9}}.

\bibitem{Dokshitzer:1978hw}
Y.~L. Dokshitzer, D.~Diakonov, S.~Troian, {Hard Processes in Quantum
  Chromodynamics}, Phys. Rept. 58 (1980) 269--395.
\newblock \href {http://dx.doi.org/10.1016/0370-1573(80)90043-5}
  {\path{doi:10.1016/0370-1573(80)90043-5}}.

\bibitem{Mueller:1981sg}
A.~H. Mueller, {Perturbative QCD at High-Energies}, Phys. Rept. 73 (1981) 237.
\newblock \href {http://dx.doi.org/10.1016/0370-1573(81)90030-2}
  {\path{doi:10.1016/0370-1573(81)90030-2}}.

\bibitem{JCollins_pQCD}
J.~Collins, {Foundations of perturbative QCD}, Cambridge University Press,
  2013.

\bibitem{Ji:1996ek}
X.-D. Ji, {Gauge-Invariant Decomposition of Nucleon Spin}, Phys. Rev. Lett. 78
  (1997) 610--613.
\newblock \href {http://arxiv.org/abs/hep-ph/9603249}
  {\path{arXiv:hep-ph/9603249}}, \href
  {http://dx.doi.org/10.1103/PhysRevLett.78.610}
  {\path{doi:10.1103/PhysRevLett.78.610}}.

\bibitem{Polyakov:2002yz}
M.~V. Polyakov, {Generalized parton distributions and strong forces inside
  nucleons and nuclei}, Phys. Lett. B 555 (2003) 57--62.
\newblock \href {http://arxiv.org/abs/hep-ph/0210165}
  {\path{arXiv:hep-ph/0210165}}, \href
  {http://dx.doi.org/10.1016/S0370-2693(03)00036-4}
  {\path{doi:10.1016/S0370-2693(03)00036-4}}.

\bibitem{Burkardt:2000za}
M.~Burkardt, {Impact parameter dependent parton distributions and off forward
  parton distributions for $\zeta \to 0$}, Phys. Rev. D62 (2000) 071503,
  [Erratum: Phys. Rev.D66,119903(2002)].
\newblock \href {http://arxiv.org/abs/hep-ph/0005108}
  {\path{arXiv:hep-ph/0005108}}, \href
  {http://dx.doi.org/10.1103/PhysRevD.62.071503, 10.1103/PhysRevD.66.119903}
  {\path{doi:10.1103/PhysRevD.62.071503, 10.1103/PhysRevD.66.119903}}.

\bibitem{Ralston:2001xs}
J.~P. Ralston, B.~Pire, {Femtophotography of protons to nuclei with deeply
  virtual Compton scattering}, Phys. Rev. D66 (2002) 111501.
\newblock \href {http://arxiv.org/abs/hep-ph/0110075}
  {\path{arXiv:hep-ph/0110075}}, \href
  {http://dx.doi.org/10.1103/PhysRevD.66.111501}
  {\path{doi:10.1103/PhysRevD.66.111501}}.

\bibitem{Diehl:2002he}
M.~Diehl, {Generalized parton distributions in impact parameter space}, Eur.
  Phys. J. C25 (2002) 223--232, [Erratum: Eur. Phys. J.C31,277(2003)].
\newblock \href {http://arxiv.org/abs/hep-ph/0205208}
  {\path{arXiv:hep-ph/0205208}}, \href
  {http://dx.doi.org/10.1007/s10052-002-1016-9}
  {\path{doi:10.1007/s10052-002-1016-9}}.

\bibitem{Goeke:2001tz}
K.~Goeke, M.~V. Polyakov, M.~Vanderhaeghen, {Hard exclusive reactions and the
  structure of hadrons}, Prog. Part. Nucl. Phys. 47 (2001) 401--515.
\newblock \href {http://arxiv.org/abs/hep-ph/0106012}
  {\path{arXiv:hep-ph/0106012}}, \href
  {http://dx.doi.org/10.1016/S0146-6410(01)00158-2}
  {\path{doi:10.1016/S0146-6410(01)00158-2}}.

\bibitem{Diehl:2003ny}
M.~Diehl, {Generalized parton distributions}, Phys. Rept. 388 (2003) 41--277.
\newblock \href {http://arxiv.org/abs/hep-ph/0307382}
  {\path{arXiv:hep-ph/0307382}}, \href
  {http://dx.doi.org/10.1016/j.physrep.2003.08.002}
  {\path{doi:10.1016/j.physrep.2003.08.002}}.

\bibitem{Belitsky:2005qn}
A.~V. Belitsky, A.~V. Radyushkin, {Unraveling hadron structure with generalized
  parton distributions}, Phys. Rept. 418 (2005) 1--387.
\newblock \href {http://arxiv.org/abs/hep-ph/0504030}
  {\path{arXiv:hep-ph/0504030}}, \href
  {http://dx.doi.org/10.1016/j.physrep.2005.06.002}
  {\path{doi:10.1016/j.physrep.2005.06.002}}.

\bibitem{Boffi:2007yc}
S.~Boffi, B.~Pasquini, {Generalized parton distributions and the structure of
  the nucleon}, Riv. Nuovo Cim. 30~(9) (2007) 387--448.
\newblock \href {http://arxiv.org/abs/0711.2625} {\path{arXiv:0711.2625}},
  \href {http://dx.doi.org/10.1393/ncr/i2007-10025-7}
  {\path{doi:10.1393/ncr/i2007-10025-7}}.

\bibitem{Mueller:2014hsa}
D.~M{\"{u}}ller, {Generalized Parton Distributions -- visions, basics, and
  realities --}, Few Body Syst. 55 (2014) 317--337.
\newblock \href {http://arxiv.org/abs/1405.2817} {\path{arXiv:1405.2817}},
  \href {http://dx.doi.org/10.1007/s00601-014-0894-3}
  {\path{doi:10.1007/s00601-014-0894-3}}.

\bibitem{Frankfurt:1999fp}
L.~L. Frankfurt, P.~V. Pobylitsa, M.~V. Polyakov, M.~Strikman, {Hard exclusive
  pseudoscalar meson electroproduction and spin structure of a nucleon}, Phys.
  Rev. D60 (1999) 014010.
\newblock \href {http://arxiv.org/abs/hep-ph/9901429}
  {\path{arXiv:hep-ph/9901429}}, \href
  {http://dx.doi.org/10.1103/PhysRevD.60.014010}
  {\path{doi:10.1103/PhysRevD.60.014010}}.

\bibitem{Pire:2004ie}
B.~Pire, L.~Szymanowski, {Hadron annihilation into two photons and backward VCS
  in the scaling regime of QCD}, Phys. Rev. D71 (2005) 111501.
\newblock \href {http://arxiv.org/abs/hep-ph/0411387}
  {\path{arXiv:hep-ph/0411387}}, \href
  {http://dx.doi.org/10.1103/PhysRevD.71.111501}
  {\path{doi:10.1103/PhysRevD.71.111501}}.

\bibitem{Pire:2005ax}
B.~Pire, L.~Szymanowski, {QCD analysis of {$ \bar{p} N \to \gamma^* \pi$} in
  the scaling limit}, Phys. Lett. B622 (2005) 83--92.
\newblock \href {http://arxiv.org/abs/hep-ph/0504255}
  {\path{arXiv:hep-ph/0504255}}, \href
  {http://dx.doi.org/10.1016/j.physletb.2005.06.080}
  {\path{doi:10.1016/j.physletb.2005.06.080}}.

\bibitem{Lansberg:2011aa}
J.~P. Lansberg, B.~Pire, K.~Semenov-Tian-Shansky, L.~Szymanowski, {A consistent
  model for {$\pi N$} transition distribution amplitudes and backward pion
  electroproduction}, Phys. Rev. D85 (2012) 054021.
\newblock \href {http://arxiv.org/abs/1112.3570} {\path{arXiv:1112.3570}},
  \href {http://dx.doi.org/10.1103/PhysRevD.85.054021}
  {\path{doi:10.1103/PhysRevD.85.054021}}.

\bibitem{Pire:2015kxa}
B.~Pire, K.~Semenov-Tian-Shansky, L.~Szymanowski, {QCD description of backward
  vector meson hard electroproduction}, Phys. Rev. D91~(9) (2015) 094006.
\newblock \href {http://arxiv.org/abs/1503.02012} {\path{arXiv:1503.02012}},
  \href {http://dx.doi.org/10.1103/PhysRevD.91.094006}
  {\path{doi:10.1103/PhysRevD.91.094006}}.

\bibitem{Lansberg:2012ha}
J.~P. Lansberg, B.~Pire, K.~Semenov-Tian-Shansky, L.~Szymanowski, {Accessing
  baryon to meson transition distribution amplitudes in meson production in
  association with a high invariant mass lepton pair at GSI-FAIR with
  $\overline {P}ANDA$}, Phys. Rev. D86 (2012) 114033, [Erratum: Phys.
  Rev.D87,no.5,059902(2013)].
\newblock \href {http://arxiv.org/abs/1210.0126} {\path{arXiv:1210.0126}},
  \href {http://dx.doi.org/10.1103/PhysRevD.87.059902,
  10.1103/PhysRevD.86.114033} {\path{doi:10.1103/PhysRevD.87.059902,
  10.1103/PhysRevD.86.114033}}.

\bibitem{Pire:2013jva}
B.~Pire, K.~Semenov-Tian-Shansky, L.~Szymanowski, {QCD description of
  charmonium plus light meson production in $\bar{p}-N$ annihilation}, Phys.
  Lett. B724 (2013) 99--107, [Erratum: Phys. Lett.B764,335(2017)].
\newblock \href {http://arxiv.org/abs/1304.6298} {\path{arXiv:1304.6298}},
  \href {http://dx.doi.org/10.1016/j.physletb.2013.06.015,
  10.1016/j.physletb.2016.11.049} {\path{doi:10.1016/j.physletb.2013.06.015,
  10.1016/j.physletb.2016.11.049}}.

\bibitem{Lepage:1980fj}
G.~P. Lepage, S.~J. Brodsky, {Exclusive Processes in Perturbative Quantum
  Chromodynamics}, Phys. Rev. D22 (1980) 2157.
\newblock \href {http://dx.doi.org/10.1103/PhysRevD.22.2157}
  {\path{doi:10.1103/PhysRevD.22.2157}}.

\bibitem{Chernyak:1983ej}
V.~L. Chernyak, A.~R. Zhitnitsky, {Asymptotic Behavior of Exclusive Processes
  in QCD}, Phys. Rept. 112 (1984) 173.
\newblock \href {http://dx.doi.org/10.1016/0370-1573(84)90126-1}
  {\path{doi:10.1016/0370-1573(84)90126-1}}.

\bibitem{Chernyak:1987nv}
V.~L. Chernyak, A.~A. Ogloblin, I.~R. Zhitnitsky, {Calculation of Exclusive
  Processes With Baryons}, Z. Phys. C42 (1989) 583, [Yad. Fiz.48,1398(1988);
  Sov. J. Nucl. Phys.48,889(1988)].
\newblock \href {http://dx.doi.org/10.1007/BF01557664}
  {\path{doi:10.1007/BF01557664}}.

\bibitem{Stefanis:1999wy}
N.~G. Stefanis, {The Physics of exclusive reactions in QCD: Theory and
  phenomenology}, Eur. Phys. J direct C 7 (1999) 1.
\newblock \href {http://arxiv.org/abs/hep-ph/9911375}
  {\path{arXiv:hep-ph/9911375}}, \href
  {http://dx.doi.org/10.1007/s1010599c0007} {\path{doi:10.1007/s1010599c0007}}.

\bibitem{Braun:1999te}
V.~M. Braun, S.~E. Derkachov, G.~P. Korchemsky, A.~N. Manashov, {Baryon
  distribution amplitudes in QCD}, Nucl. Phys. B553 (1999) 355--426.
\newblock \href {http://arxiv.org/abs/hep-ph/9902375}
  {\path{arXiv:hep-ph/9902375}}, \href
  {http://dx.doi.org/10.1016/S0550-3213(99)00265-5}
  {\path{doi:10.1016/S0550-3213(99)00265-5}}.

\bibitem{Park:2017irz}
K.~Park, et~al., {Hard exclusive pion electroproduction at backward angles with
  CLAS}, Phys. Lett. B780 (2018) 340--345.
\newblock \href {http://arxiv.org/abs/1711.08486} {\path{arXiv:1711.08486}},
  \href {http://dx.doi.org/10.1016/j.physletb.2018.03.026}
  {\path{doi:10.1016/j.physletb.2018.03.026}}.

\bibitem{Li:2019xyp}
W.~B. Li, et~al., {Unique Access to $u$-Channel Physics: Exclusive
  Backward-Angle Omega Meson Electroproduction}, Phys. Rev. Lett. 123~(18)
  (2019) 182501.
\newblock \href {http://arxiv.org/abs/1910.00464} {\path{arXiv:1910.00464}},
  \href {http://dx.doi.org/10.1103/PhysRevLett.123.182501}
  {\path{doi:10.1103/PhysRevLett.123.182501}}.

\bibitem{Diehl:2020uja}
S.~Diehl, et~al., {Extraction of Beam-Spin Asymmetries from the Hard Exclusive
  $\pi^+$ Channel off Protons in a Wide Range of Kinematics}, Phys. Rev. Lett.
  125~(18) (2020) 182001.
\newblock \href {http://arxiv.org/abs/2007.15677} {\path{arXiv:2007.15677}},
  \href {http://dx.doi.org/10.1103/PhysRevLett.125.182001}
  {\path{doi:10.1103/PhysRevLett.125.182001}}.

\bibitem{Lutz:2009ff}
M.~F.~M. Lutz, et~al., {Physics Performance Report for PANDA: Strong
  Interaction Studies with Antiprotons} (2009).
\newblock \href {http://arxiv.org/abs/0903.3905} {\path{arXiv:0903.3905}}.

\bibitem{Li:2017xcf}
W.~Li,
  \href{https://misportal.jlab.org/ul/publications/view_pub.cfm?pub_id=15234}{{Exclusive
  Backward-Angle Omega Meson Electroproduction}}, Ph.D. thesis, Regina U.
  (2017-10).
\newblock \href {http://arxiv.org/abs/1712.03214} {\path{arXiv:1712.03214}}.
\newline\urlprefix\url{https://misportal.jlab.org/ul/publications/view_pub.cfm?pub_id=15234}

\bibitem{Li:2020nsk}
W.~B. Li, et~al., {Backward-angle Exclusive $\pi^0$ Production above the
  Resonance Region} (8 2020).
\newblock \href {http://arxiv.org/abs/2008.10768} {\path{arXiv:2008.10768}}.

\bibitem{WP:21}
C.~A. Gayoso, et~al., {Progress and opportunities in backward angle (u-channel)
  physics}, Eur. Phys. J. A 57~(12) (2021) 342.
\newblock \href {http://arxiv.org/abs/2107.06748} {\path{arXiv:2107.06748}},
  \href {http://dx.doi.org/10.1140/epja/s10050-021-00625-2}
  {\path{doi:10.1140/epja/s10050-021-00625-2}}.

\bibitem{Singh:2014pfv}
B.~P. Singh, et~al., {Experimental access to Transition Distribution Amplitudes
  with the PANDA experiment at FAIR}, Eur. Phys. J. A51~(8) (2015) 107.
\newblock \href {http://arxiv.org/abs/1409.0865} {\path{arXiv:1409.0865}},
  \href {http://dx.doi.org/10.1140/epja/i2015-15107-y}
  {\path{doi:10.1140/epja/i2015-15107-y}}.

\bibitem{Singh:2016qjg}
B.~Singh, et~al., {Feasibility study for the measurement of $\pi N$ transition
  distribution amplitudes at $\overline P$ANDA in $\bar{p}p\to J/\psi\pi^0$},
  Phys. Rev. D95~(3) (2017) 032003.
\newblock \href {http://arxiv.org/abs/1610.02149} {\path{arXiv:1610.02149}},
  \href {http://dx.doi.org/10.1103/PhysRevD.95.032003}
  {\path{doi:10.1103/PhysRevD.95.032003}}.

\bibitem{Lansberg:2007ec}
J.~P. Lansberg, B.~Pire, L.~Szymanowski, {Hard exclusive electroproduction of a
  pion in the backward region}, Phys. Rev. D75 (2007) 074004, [Erratum: Phys.
  Rev.D77,019902(2008)].
\newblock \href {http://arxiv.org/abs/hep-ph/0701125}
  {\path{arXiv:hep-ph/0701125}}, \href
  {http://dx.doi.org/10.1103/PhysRevD.75.074004, 10.1103/PhysRevD.77.019902}
  {\path{doi:10.1103/PhysRevD.75.074004, 10.1103/PhysRevD.77.019902}}.

\bibitem{Lansberg:2007se}
J.~P. Lansberg, B.~Pire, L.~Szymanowski, {Production of a pion in association
  with a high-$Q^2$ dilepton pair in antiproton-proton annihilation at
  GSI-FAIR}, Phys. Rev. D76 (2007) 111502.
\newblock \href {http://arxiv.org/abs/0710.1267} {\path{arXiv:0710.1267}},
  \href {http://dx.doi.org/10.1103/PhysRevD.76.111502}
  {\path{doi:10.1103/PhysRevD.76.111502}}.

\bibitem{Pire:2010if}
B.~Pire, K.~Semenov-Tian-Shansky, L.~Szymanowski, {A Spectral representation
  for baryon to meson transition distribution amplitudes}, Phys. Rev. D82
  (2010) 094030.
\newblock \href {http://arxiv.org/abs/1008.0721} {\path{arXiv:1008.0721}},
  \href {http://dx.doi.org/10.1103/PhysRevD.82.094030}
  {\path{doi:10.1103/PhysRevD.82.094030}}.

\bibitem{Pire:2011xv}
B.~Pire, K.~Semenov-Tian-Shansky, L.~Szymanowski, {{$\pi N$} transition
  distribution amplitudes: their symmetries and constraints from chiral
  dynamics}, Phys. Rev. D84 (2011) 074014.
\newblock \href {http://arxiv.org/abs/1106.1851} {\path{arXiv:1106.1851}},
  \href {http://dx.doi.org/10.1103/PhysRevD.84.074014}
  {\path{doi:10.1103/PhysRevD.84.074014}}.

\bibitem{Pire:2016gut}
B.~Pire, K.~Semenov-Tian-Shansky, L.~Szymanowski, {Backward charmonium
  production in $\pi N$ collisions}, Phys. Rev. D95~(3) (2017) 034021.
\newblock \href {http://arxiv.org/abs/1611.07234} {\path{arXiv:1611.07234}},
  \href {http://dx.doi.org/10.1103/PhysRevD.95.034021}
  {\path{doi:10.1103/PhysRevD.95.034021}}.

\bibitem{Gribov:1972ri}
V.~N. Gribov, L.~N. Lipatov, {Deep inelastic $e p$ scattering in perturbation
  theory}, Sov. J. Nucl. Phys. 15 (1972) 438--450.

\bibitem{Altarelli:1977zs}
G.~Altarelli, G.~Parisi, {Asymptotic Freedom in Parton Language}, Nucl. Phys. B
  126 (1977) 298--318.
\newblock \href {http://dx.doi.org/10.1016/0550-3213(77)90384-4}
  {\path{doi:10.1016/0550-3213(77)90384-4}}.

\bibitem{Dokshitzer:1977sg}
Y.~L. Dokshitzer, {Calculation of the Structure Functions for Deep Inelastic
  Scattering and $e^+ e^-$ Annihilation by Perturbation Theory in Quantum
  Chromodynamics.}, Sov. Phys. JETP 46 (1977) 641--653.

\bibitem{Efremov:1979qk}
A.~V. Efremov, A.~V. Radyushkin, {Factorization and Asymptotical Behavior of
  Pion Form-Factor in QCD}, Phys. Lett. 94B (1980) 245--250.
\newblock \href {http://dx.doi.org/10.1016/0370-2693(80)90869-2}
  {\path{doi:10.1016/0370-2693(80)90869-2}}.

\bibitem{Lepage:1979zb}
G.~P. Lepage, S.~J. Brodsky, {Exclusive Processes in Quantum Chromodynamics:
  Evolution Equations for Hadronic Wave Functions and the Form-Factors of
  Mesons}, Phys. Lett. 87B (1979) 359--365.
\newblock \href {http://dx.doi.org/10.1016/0370-2693(79)90554-9}
  {\path{doi:10.1016/0370-2693(79)90554-9}}.

\bibitem{Efremov:1978rn}
A.~V. Efremov, A.~V. Radyushkin, {Asymptotical Behavior of Pion Electromagnetic
  Form-Factor in QCD}, Theor. Math. Phys. 42 (1980) 97--110.
\newblock \href {http://dx.doi.org/10.1007/BF01032111}
  {\path{doi:10.1007/BF01032111}}.

\bibitem{Lepage:1979za}
G.~Lepage, S.~J. Brodsky, {Exclusive Processes in Quantum Chromodynamics: The
  Form-Factors of Baryons at Large Momentum Transfer}, Phys. Rev. Lett. 43
  (1979) 545--549, [Erratum: Phys.Rev.Lett. 43, 1625--1626 (1979)].
\newblock \href {http://dx.doi.org/10.1103/PhysRevLett.43.545}
  {\path{doi:10.1103/PhysRevLett.43.545}}.

\bibitem{Matveev:1972gb}
V.~A. Matveev, R.~M. Muradyan, A.~N. Tavkhelidze, {Automodelity in strong
  interactions}, Lett. Nuovo Cim. 5 (1972) 907--912.
\newblock \href {http://dx.doi.org/10.1007/BF02777988}
  {\path{doi:10.1007/BF02777988}}.

\bibitem{Brodsky:1973kr}
S.~J. Brodsky, G.~R. Farrar, {Scaling laws at large transverse momentum}, Phys.
  Rev. Lett. 31 (1973) 1153--1156.
\newblock \href {http://dx.doi.org/10.1103/PhysRevLett.31.1153}
  {\path{doi:10.1103/PhysRevLett.31.1153}}.

\bibitem{Amaryan:2021cnj}
M.~J. Amaryan, W.~J. Briscoe, M.~G. Ryskin, I.~I. Strakovsky, {High Energy
  Behaviour of the Light Meson Photoproduction and the ''Quark Counting
  Rules''} (2021).
\newblock \href {http://arxiv.org/abs/2102.03633} {\path{arXiv:2102.03633}}.

\bibitem{Botts:1989nd}
J.~Botts, G.~F. Sterman, {Sudakov Effects in Hadron Hadron Elastic Scattering},
  Phys. Lett. B 224 (1989) 201, [Erratum: Phys.Lett.B 227, 501 (1989)].
\newblock \href {http://dx.doi.org/10.1016/0370-2693(89)91074-5}
  {\path{doi:10.1016/0370-2693(89)91074-5}}.

\bibitem{Li:1992nu}
H.-n. Li, G.~F. Sterman, {The Perturbative pion form-factor with Sudakov
  suppression}, Nucl. Phys. B 381 (1992) 129--140.
\newblock \href {http://dx.doi.org/10.1016/0550-3213(92)90643-P}
  {\path{doi:10.1016/0550-3213(92)90643-P}}.

\bibitem{Li:1996gi}
H.-n. Li, {Resummation in hard QCD processes}, Phys. Rev. D 55 (1997) 105--119.
\newblock \href {http://arxiv.org/abs/hep-ph/9604267}
  {\path{arXiv:hep-ph/9604267}}, \href
  {http://dx.doi.org/10.1103/PhysRevD.55.105}
  {\path{doi:10.1103/PhysRevD.55.105}}.

\bibitem{Dagaonkar:2014yea}
S.~K. Dagaonkar, P.~Jain, J.~P. Ralston, {Uncovering the Scaling Laws of Hard
  Exclusive Hadronic Processes in a Comprehensive Endpoint Model}, Eur. Phys.
  J. C 74~(8) (2014) 3000.
\newblock \href {http://arxiv.org/abs/1404.5798} {\path{arXiv:1404.5798}},
  \href {http://dx.doi.org/10.1140/epjc/s10052-014-3000-6}
  {\path{doi:10.1140/epjc/s10052-014-3000-6}}.

\bibitem{Dagaonkar:2015laa}
S.~Dagaonkar, P.~Jain, J.~P. Ralston, {The Dirac Form Factor Predicts the Pauli
  Form Factor in the Endpoint Model}, Eur. Phys. J. C 76~(7) (2016) 368.
\newblock \href {http://arxiv.org/abs/1503.06938} {\path{arXiv:1503.06938}},
  \href {http://dx.doi.org/10.1140/epjc/s10052-016-4224-4}
  {\path{doi:10.1140/epjc/s10052-016-4224-4}}.

\bibitem{Farrar:1989wb}
G.~R. Farrar, G.~F. Sterman, H.-y. Zhang, {Absence of Sudakov Factors in Large
  Angle Photoproduction and Compton Scattering}, Phys. Rev. Lett. 62 (1989)
  2229.
\newblock \href {http://dx.doi.org/10.1103/PhysRevLett.62.2229}
  {\path{doi:10.1103/PhysRevLett.62.2229}}.

\bibitem{Colangelo:2000dp}
P.~Colangelo, A.~Khodjamirian, {QCD sum rules, a modern perspective} (2000)
  1495--1576\href {http://arxiv.org/abs/hep-ph/0010175}
  {\path{arXiv:hep-ph/0010175}}, \href
  {http://dx.doi.org/10.1142/9789812810458_0033}
  {\path{doi:10.1142/9789812810458_0033}}.

\bibitem{Lenz:2003tq}
A.~Lenz, M.~Wittmann, E.~Stein, {Improved light cone sum rules for the
  electromagnetic form-factors of the nucleon}, Phys. Lett. B 581 (2004)
  199--206.
\newblock \href {http://arxiv.org/abs/hep-ph/0311082}
  {\path{arXiv:hep-ph/0311082}}, \href
  {http://dx.doi.org/10.1016/j.physletb.2003.12.009}
  {\path{doi:10.1016/j.physletb.2003.12.009}}.

\bibitem{Braun:2001tj}
V.~M. Braun, A.~Lenz, N.~Mahnke, E.~Stein, {Light cone sum rules for the
  nucleon form-factors}, Phys. Rev. D 65 (2002) 074011.
\newblock \href {http://arxiv.org/abs/hep-ph/0112085}
  {\path{arXiv:hep-ph/0112085}}, \href
  {http://dx.doi.org/10.1103/PhysRevD.65.074011}
  {\path{doi:10.1103/PhysRevD.65.074011}}.

\bibitem{Lenz:2009ar}
A.~Lenz, M.~Gockeler, T.~Kaltenbrunner, N.~Warkentin, {The Nucleon Distribution
  Amplitudes and their application to nucleon form factors and the $N \to
  \Delta$ transition at intermediate values of $Q^2$}, Phys. Rev. D 79 (2009)
  093007.
\newblock \href {http://arxiv.org/abs/0903.1723} {\path{arXiv:0903.1723}},
  \href {http://dx.doi.org/10.1103/PhysRevD.79.093007}
  {\path{doi:10.1103/PhysRevD.79.093007}}.

\bibitem{Mueller:1998fv}
D.~M{\"{u}}ller, D.~Robaschik, B.~Geyer, F.~M. Dittes, J.~Ho{\v{r}}ej{\v{s}}i,
  {Wave functions, evolution equations and evolution kernels from light ray
  operators of QCD}, Fortsch. Phys. 42 (1994) 101--141.
\newblock \href {http://arxiv.org/abs/hep-ph/9812448}
  {\path{arXiv:hep-ph/9812448}}, \href
  {http://dx.doi.org/10.1002/prop.2190420202}
  {\path{doi:10.1002/prop.2190420202}}.

\bibitem{Radyushkin:1996nd}
A.~V. Radyushkin, {Scaling limit of deeply virtual Compton scattering}, Phys.
  Lett. B 380 (1996) 417--425.
\newblock \href {http://arxiv.org/abs/hep-ph/9604317}
  {\path{arXiv:hep-ph/9604317}}, \href
  {http://dx.doi.org/10.1016/0370-2693(96)00528-X}
  {\path{doi:10.1016/0370-2693(96)00528-X}}.

\bibitem{Ji:1996nm}
X.-D. Ji, {Deeply virtual Compton scattering}, Phys. Rev. D55 (1997)
  7114--7125.
\newblock \href {http://arxiv.org/abs/hep-ph/9609381}
  {\path{arXiv:hep-ph/9609381}}, \href
  {http://dx.doi.org/10.1103/PhysRevD.55.7114}
  {\path{doi:10.1103/PhysRevD.55.7114}}.

\bibitem{Collins:1996fb}
J.~C. Collins, L.~Frankfurt, M.~Strikman, {Factorization for hard exclusive
  electroproduction of mesons in QCD}, Phys. Rev. D 56 (1997) 2982--3006.
\newblock \href {http://arxiv.org/abs/hep-ph/9611433}
  {\path{arXiv:hep-ph/9611433}}, \href
  {http://dx.doi.org/10.1103/PhysRevD.56.2982}
  {\path{doi:10.1103/PhysRevD.56.2982}}.

\bibitem{Radyushkin:1997ki}
A.~V. Radyushkin, {Nonforward parton distributions}, Phys. Rev. D56 (1997)
  5524--5557.
\newblock \href {http://arxiv.org/abs/hep-ph/9704207}
  {\path{arXiv:hep-ph/9704207}}, \href
  {http://dx.doi.org/10.1103/PhysRevD.56.5524}
  {\path{doi:10.1103/PhysRevD.56.5524}}.

\bibitem{Diehl:1998dk}
M.~Diehl, T.~Gousset, B.~Pire, O.~Teryaev, {Probing partonic structure in
  $\gamma^* \gamma \to \pi \pi$ near threshold}, Phys. Rev. Lett. 81 (1998)
  1782--1785.
\newblock \href {http://arxiv.org/abs/hep-ph/9805380}
  {\path{arXiv:hep-ph/9805380}}, \href
  {http://dx.doi.org/10.1103/PhysRevLett.81.1782}
  {\path{doi:10.1103/PhysRevLett.81.1782}}.

\bibitem{Dittes:1988xz}
F.~M. Dittes, D.~M{\"{u}}ller, D.~Robaschik, B.~Geyer, J.~Horejsi, {The
  Altarelli-Parisi Kernel as Asymptotic Limit of an Extended Brodsky-Lepage
  Kernel}, Phys. Lett. B 209 (1988) 325--329.
\newblock \href {http://dx.doi.org/10.1016/0370-2693(88)90955-0}
  {\path{doi:10.1016/0370-2693(88)90955-0}}.

\bibitem{Cosyn:2019aio}
W.~Cosyn, S.~Cotogno, A.~Freese, C.~Lorc\'e, {The energy-momentum tensor of
  spin-1 hadrons: formalism}, Eur. Phys. J. C 79~(6) (2019) 476.
\newblock \href {http://arxiv.org/abs/1903.00408} {\path{arXiv:1903.00408}},
  \href {http://dx.doi.org/10.1140/epjc/s10052-019-6981-3}
  {\path{doi:10.1140/epjc/s10052-019-6981-3}}.

\bibitem{Polyakov:2018zvc}
M.~V. Polyakov, P.~Schweitzer, {Forces inside hadrons: pressure, surface
  tension, mechanical radius, and all that}, Int. J. Mod. Phys. A 33~(26)
  (2018) 1830025.
\newblock \href {http://arxiv.org/abs/1805.06596} {\path{arXiv:1805.06596}},
  \href {http://dx.doi.org/10.1142/S0217751X18300259}
  {\path{doi:10.1142/S0217751X18300259}}.

\bibitem{Lorce:2018egm}
C.~Lorc\'e, H.~Moutarde, A.~P. Trawi\'nski, {Revisiting the mechanical
  properties of the nucleon}, Eur. Phys. J. C 79~(1) (2019) 89.
\newblock \href {http://arxiv.org/abs/1810.09837} {\path{arXiv:1810.09837}},
  \href {http://dx.doi.org/10.1140/epjc/s10052-019-6572-3}
  {\path{doi:10.1140/epjc/s10052-019-6572-3}}.

\bibitem{Radyushkin:1996ru}
A.~V. Radyushkin, {Asymmetric gluon distributions and hard diffractive
  electroproduction}, Phys. Lett. B 385 (1996) 333--342.
\newblock \href {http://arxiv.org/abs/hep-ph/9605431}
  {\path{arXiv:hep-ph/9605431}}, \href
  {http://dx.doi.org/10.1016/0370-2693(96)00844-1}
  {\path{doi:10.1016/0370-2693(96)00844-1}}.

\bibitem{Musatov:1999xp}
I.~V. Musatov, A.~V. Radyushkin, {Evolution and models for skewed parton
  distributions}, Phys. Rev. D61 (2000) 074027.
\newblock \href {http://arxiv.org/abs/hep-ph/9905376}
  {\path{arXiv:hep-ph/9905376}}, \href
  {http://dx.doi.org/10.1103/PhysRevD.61.074027}
  {\path{doi:10.1103/PhysRevD.61.074027}}.

\bibitem{Polyakov:1999gs}
M.~V. Polyakov, C.~Weiss, {Skewed and double distributions in pion and
  nucleon}, Phys. Rev. D60 (1999) 114017.
\newblock \href {http://arxiv.org/abs/hep-ph/9902451}
  {\path{arXiv:hep-ph/9902451}}, \href
  {http://dx.doi.org/10.1103/PhysRevD.60.114017}
  {\path{doi:10.1103/PhysRevD.60.114017}}.

\bibitem{Teryaev:2001qm}
O.~V. Teryaev, {Crossing and radon tomography for generalized parton
  distributions}, Phys. Lett. B510 (2001) 125--132.
\newblock \href {http://arxiv.org/abs/hep-ph/0102303}
  {\path{arXiv:hep-ph/0102303}}, \href
  {http://dx.doi.org/10.1016/S0370-2693(01)00564-0}
  {\path{doi:10.1016/S0370-2693(01)00564-0}}.

\bibitem{Radyushkin:2011dh}
A.~V. Radyushkin, {Generalized Parton Distributions and Their Singularities},
  Phys. Rev. D 83 (2011) 076006.
\newblock \href {http://arxiv.org/abs/1101.2165} {\path{arXiv:1101.2165}},
  \href {http://dx.doi.org/10.1103/PhysRevD.83.076006}
  {\path{doi:10.1103/PhysRevD.83.076006}}.

\bibitem{Belitsky:2000vk}
A.~V. Belitsky, D.~M{\"{u}}ller, A.~Kirchner, A.~Schafer, {Twist three analysis
  of photon electroproduction off pion}, Phys. Rev. D 64 (2001) 116002.
\newblock \href {http://arxiv.org/abs/hep-ph/0011314}
  {\path{arXiv:hep-ph/0011314}}, \href
  {http://dx.doi.org/10.1103/PhysRevD.64.116002}
  {\path{doi:10.1103/PhysRevD.64.116002}}.

\bibitem{Pire:2002ut}
B.~Pire, L.~Szymanowski, {Impact representation of generalized distribution
  amplitudes}, Phys. Lett. B 556 (2003) 129--134.
\newblock \href {http://arxiv.org/abs/hep-ph/0212296}
  {\path{arXiv:hep-ph/0212296}}, \href
  {http://dx.doi.org/10.1016/S0370-2693(03)00134-5}
  {\path{doi:10.1016/S0370-2693(03)00134-5}}.

\bibitem{Collins:1998be}
J.~C. Collins, A.~Freund, {Proof of factorization for deeply virtual Compton
  scattering in QCD}, Phys. Rev. D 59 (1999) 074009.
\newblock \href {http://arxiv.org/abs/hep-ph/9801262}
  {\path{arXiv:hep-ph/9801262}}, \href
  {http://dx.doi.org/10.1103/PhysRevD.59.074009}
  {\path{doi:10.1103/PhysRevD.59.074009}}.

\bibitem{Frankfurt:2002kz}
L.~Frankfurt, M.~Polyakov, M.~Strikman, D.~Zhalov, M.~Zhalov, {Novel hard
  semiexclusive processes and color singlet clusters in hadrons}, in:
  {Exclusive Processes at High Momentum Transfer}, 2002, pp. 361--368.
\newblock \href {http://arxiv.org/abs/hep-ph/0211263}
  {\path{arXiv:hep-ph/0211263}}, \href
  {http://dx.doi.org/10.1142/9789812776211_0049}
  {\path{doi:10.1142/9789812776211_0049}}.

\bibitem{Wallon:2011zx}
S.~Wallon, {A short review of the theory of hard exclusive processes}, Few Body
  Syst. 53 (2012) 59--80.
\newblock \href {http://arxiv.org/abs/1109.6187} {\path{arXiv:1109.6187}},
  \href {http://dx.doi.org/10.1007/s00601-012-0308-3}
  {\path{doi:10.1007/s00601-012-0308-3}}.

\bibitem{Ji:1998xh}
X.-D. Ji, J.~Osborne, {One loop corrections and all order factorization in
  deeply virtual Compton scattering}, Phys. Rev. D 58 (1998) 094018.
\newblock \href {http://arxiv.org/abs/hep-ph/9801260}
  {\path{arXiv:hep-ph/9801260}}, \href
  {http://dx.doi.org/10.1103/PhysRevD.58.094018}
  {\path{doi:10.1103/PhysRevD.58.094018}}.

\bibitem{Sterman:1978bi}
G.~F. Sterman, {Mass divergences in annihilation processes. 1. Origin and
  nature of divergences in cut vacuum polarization diagrams}, Phys. Rev. D 17
  (1978) 2773.
\newblock \href {http://dx.doi.org/10.1103/PhysRevD.17.2773}
  {\path{doi:10.1103/PhysRevD.17.2773}}.

\bibitem{Sterman:1978bj}
G.~F. Sterman, {Mass divergences in annihilation processes. 2. Cancellation of
  divergences in cut vacuum polarization diagrams}, Phys. Rev. D 17 (1978)
  2789.
\newblock \href {http://dx.doi.org/10.1103/PhysRevD.17.2789}
  {\path{doi:10.1103/PhysRevD.17.2789}}.

\bibitem{Libby:1978bx}
S.~B. Libby, G.~F. Sterman, {Mass divergences in two particle inelastic
  scattering}, Phys. Rev. D 18 (1978) 4737.
\newblock \href {http://dx.doi.org/10.1103/PhysRevD.18.4737}
  {\path{doi:10.1103/PhysRevD.18.4737}}.

\bibitem{Collins:1981ta}
J.~C. Collins, G.~F. Sterman, {Soft partons in QCD}, Nucl. Phys. B 185 (1981)
  172--188.
\newblock \href {http://dx.doi.org/10.1016/0550-3213(81)90370-9}
  {\path{doi:10.1016/0550-3213(81)90370-9}}.

\bibitem{CSS}
J.~C. Collins, D.~E. Soper, G.~F. Sterman, {in Perturbative QCD, ed. A. H.
  Mueller}, World Scientific, Singapore, 1989.

\bibitem{Coleman:1965xm}
S.~Coleman, R.~Norton, {Singularities in the physical region}, Nuovo Cim. 38
  (1965) 438--442.
\newblock \href {http://dx.doi.org/10.1007/BF02750472}
  {\path{doi:10.1007/BF02750472}}.

\bibitem{Radyushkin:1983wh}
A.~V. Radyushkin, {On Spectral Properties of Parton Correlation Functions and
  Multiparton Wave Functions}, Phys. Lett. 131B (1983) 179--182.
\newblock \href {http://dx.doi.org/10.1016/0370-2693(83)91116-4}
  {\path{doi:10.1016/0370-2693(83)91116-4}}.

\bibitem{Radyushkin:1983ea}
A.~V. Radyushkin, {Alpha Representation and Spectral Properties of Multiparton
  Functions}, Theor. Math. Phys. 61 (1985) 1144, [Teor. Mat. Fiz.61,284(1984)].
\newblock \href {http://dx.doi.org/10.1007/BF01029116}
  {\path{doi:10.1007/BF01029116}}.

\bibitem{Bauer:2002nz}
C.~W. Bauer, S.~Fleming, D.~Pirjol, I.~Z. Rothstein, I.~W. Stewart, {Hard
  scattering factorization from effective field theory}, Phys. Rev. D 66 (2002)
  014017.
\newblock \href {http://arxiv.org/abs/hep-ph/0202088}
  {\path{arXiv:hep-ph/0202088}}, \href
  {http://dx.doi.org/10.1103/PhysRevD.66.014017}
  {\path{doi:10.1103/PhysRevD.66.014017}}.

\bibitem{Kivel:2010ns}
N.~Kivel, M.~Vanderhaeghen, {Soft spectator scattering in the nucleon form
  factors at large $Q^2$ within the SCET approach}, Phys. Rev. D 83 (2011)
  093005.
\newblock \href {http://arxiv.org/abs/1010.5314} {\path{arXiv:1010.5314}},
  \href {http://dx.doi.org/10.1103/PhysRevD.83.093005}
  {\path{doi:10.1103/PhysRevD.83.093005}}.

\bibitem{Kivel:2013sya}
N.~Kivel, M.~Vanderhaeghen, {QCD radiative corrections to the soft spectator
  contribution in the wide angle Compton scattering}, Nucl. Phys. B 883 (2014)
  224--255.
\newblock \href {http://arxiv.org/abs/1312.5456} {\path{arXiv:1312.5456}},
  \href {http://dx.doi.org/10.1016/j.nuclphysb.2014.03.019}
  {\path{doi:10.1016/j.nuclphysb.2014.03.019}}.

\bibitem{Collins:1999yw}
J.~C. Collins, {Factorization for hard exclusive electroproduction}, in: {6th
  INT / Jlab Workshop on Exclusive and Semiexclusive Processes at High Momentum
  Transfer}, 1999, pp. 159--166.
\newblock \href {http://arxiv.org/abs/hep-ph/9907513}
  {\path{arXiv:hep-ph/9907513}}.

\bibitem{Mankiewicz:1997uy}
L.~Mankiewicz, G.~Piller, T.~Weigl, {Hard exclusive meson production and
  nonforward parton distributions}, Eur. Phys. J. C5 (1998) 119--128.
\newblock \href {http://arxiv.org/abs/hep-ph/9711227}
  {\path{arXiv:hep-ph/9711227}}, \href
  {http://dx.doi.org/10.1007/s100529800829, 10.1007/s100520050253}
  {\path{doi:10.1007/s100529800829, 10.1007/s100520050253}}.

\bibitem{Diehl:1998pd}
M.~Diehl, T.~Gousset, B.~Pire, {Exclusive electroproduction of vector mesons
  and transversity distributions}, Phys. Rev. D 59 (1999) 034023.
\newblock \href {http://arxiv.org/abs/hep-ph/9808479}
  {\path{arXiv:hep-ph/9808479}}, \href
  {http://dx.doi.org/10.1103/PhysRevD.59.034023}
  {\path{doi:10.1103/PhysRevD.59.034023}}.

\bibitem{Collins:1999un}
J.~C. Collins, M.~Diehl, {Transversity distribution does not contribute to hard
  exclusive electroproduction of mesons}, Phys. Rev. D 61 (2000) 114015.
\newblock \href {http://arxiv.org/abs/hep-ph/9907498}
  {\path{arXiv:hep-ph/9907498}}, \href
  {http://dx.doi.org/10.1103/PhysRevD.61.114015}
  {\path{doi:10.1103/PhysRevD.61.114015}}.

\bibitem{Duncan:1979hi}
A.~Duncan, A.~H. Mueller, {Asymptotic Behavior of Composite Particle
  Form-Factors and the Renormalization Group}, Phys. Rev. D 21 (1980) 1636.
\newblock \href {http://dx.doi.org/10.1103/PhysRevD.21.1636}
  {\path{doi:10.1103/PhysRevD.21.1636}}.

\bibitem{Milshtein:1981cy}
A.~Milshtein, V.~S. Fadin, {On Applicability of the Renormalization Group to
  Calculation of Baryonic Form-factors. (In Russian)}, Yad. Fiz. 33 (1981)
  1391--1395.

\bibitem{Milshtein:1982js}
A.~Milshtein, V.~S. Fadin, {On Double Logarithmic Suppression of
  Nonrenormalization Group Contribution to Baryon Form-factor. (in Russian)},
  Yad. Fiz. 35 (1982) 1603--1609.

\bibitem{Li:1992ce}
H.-n. Li, {Sudakov suppression and the proton form-factor in QCD}, Phys. Rev. D
  48 (1993) 4243--4254.
\newblock \href {http://dx.doi.org/10.1103/PhysRevD.48.4243}
  {\path{doi:10.1103/PhysRevD.48.4243}}.

\bibitem{Bolz:1994hb}
J.~Bolz, R.~Jakob, P.~Kroll, M.~Bergmann, N.~Stefanis, {A Critical analysis of
  the proton form-factor with Sudakov suppression and intrinsic transverse
  momentum}, Z. Phys. C 66 (1995) 267--278.
\newblock \href {http://arxiv.org/abs/hep-ph/9405340}
  {\path{arXiv:hep-ph/9405340}}, \href {http://dx.doi.org/10.1007/BF01496601}
  {\path{doi:10.1007/BF01496601}}.

\bibitem{Jones:1999rz}
M.~Jones, et~al., {$G(E(p)) / G(M(p))$ ratio by polarization transfer in
  polarized $e p \to e $ polarized $p$}, Phys. Rev. Lett. 84 (2000) 1398--1402.
\newblock \href {http://arxiv.org/abs/nucl-ex/9910005}
  {\path{arXiv:nucl-ex/9910005}}, \href
  {http://dx.doi.org/10.1103/PhysRevLett.84.1398}
  {\path{doi:10.1103/PhysRevLett.84.1398}}.

\bibitem{Puckett:2010ac}
A.~Puckett, et~al., {Recoil Polarization Measurements of the Proton
  Electromagnetic Form Factor Ratio to $Q^2$ = 8.5 GeV$^2$}, Phys. Rev. Lett.
  104 (2010) 242301.
\newblock \href {http://arxiv.org/abs/1005.3419} {\path{arXiv:1005.3419}},
  \href {http://dx.doi.org/10.1103/PhysRevLett.104.242301}
  {\path{doi:10.1103/PhysRevLett.104.242301}}.

\bibitem{Riordan:2010id}
S.~Riordan, et~al., {Measurements of the Electric Form Factor of the Neutron up
  to $Q^2=3.4$ GeV$^2$ using the Reaction ${ }^{3}
  \overrightarrow{\mathrm{He}}\left(\vec{e}, e^{\prime} n\right) p p$}, Phys.
  Rev. Lett. 105 (2010) 262302.
\newblock \href {http://arxiv.org/abs/1008.1738} {\path{arXiv:1008.1738}},
  \href {http://dx.doi.org/10.1103/PhysRevLett.105.262302}
  {\path{doi:10.1103/PhysRevLett.105.262302}}.

\bibitem{Kivel:2012mf}
N.~Kivel, {Factorizing the hard and soft spectator scattering contributions for
  the nucleon form factor $F_1$ at large $Q^2$}, Eur. Phys. J. A 48 (2012) 156.
\newblock \href {http://arxiv.org/abs/1202.4944} {\path{arXiv:1202.4944}},
  \href {http://dx.doi.org/10.1140/epja/i2012-12156-8}
  {\path{doi:10.1140/epja/i2012-12156-8}}.

\bibitem{Isgur:1984jm}
N.~Isgur, C.~Llewellyn~Smith, {Asymptotic {$Q^2$} for Exclusive Processes in
  Quantum Chromodynamics}, Phys. Rev. Lett. 52 (1984) 1080.
\newblock \href {http://dx.doi.org/10.1103/PhysRevLett.52.1080}
  {\path{doi:10.1103/PhysRevLett.52.1080}}.

\bibitem{Isgur:1988iw}
N.~Isgur, C.~Llewellyn~Smith, {The Applicability of Perturbative QCD to
  Exclusive Processes}, Nucl. Phys. B 317 (1989) 526--572.
\newblock \href {http://dx.doi.org/10.1016/0550-3213(89)90532-4}
  {\path{doi:10.1016/0550-3213(89)90532-4}}.

\bibitem{Isgur:1989cy}
N.~Isgur, C.~Llewellyn~Smith, {Perturbative QCD in exclusive processes}, Phys.
  Lett. B 217 (1989) 535--538.
\newblock \href {http://dx.doi.org/10.1016/0370-2693(89)90092-0}
  {\path{doi:10.1016/0370-2693(89)90092-0}}.

\bibitem{Ioffe:1982qb}
B.~Ioffe, A.~V. Smilga, {Meson Widths and Form-Factors at Intermediate Momentum
  Transfer in Nonperturbative QCD}, Nucl. Phys. B 216 (1983) 373--407.
\newblock \href {http://dx.doi.org/10.1016/0550-3213(83)90291-2}
  {\path{doi:10.1016/0550-3213(83)90291-2}}.

\bibitem{Nesterenko:1982gc}
V.~Nesterenko, A.~Radyushkin, {Sum Rules and Pion Form-Factor in QCD}, Phys.
  Lett. B 115 (1982) 410.
\newblock \href {http://dx.doi.org/10.1016/0370-2693(82)90528-7}
  {\path{doi:10.1016/0370-2693(82)90528-7}}.

\bibitem{Braun:2006hz}
V.~Braun, A.~Lenz, M.~Wittmann, {Nucleon Form Factors in QCD}, Phys. Rev. D 73
  (2006) 094019.
\newblock \href {http://arxiv.org/abs/hep-ph/0604050}
  {\path{arXiv:hep-ph/0604050}}, \href
  {http://dx.doi.org/10.1103/PhysRevD.73.094019}
  {\path{doi:10.1103/PhysRevD.73.094019}}.

\bibitem{Musatov:1997pu}
I.~Musatov, A.~Radyushkin, {Transverse momentum and Sudakov effects in
  exclusive QCD processes: $\gamma^{*} \gamma \pi^{0}$ form-factor}, Phys. Rev.
  D 56 (1997) 2713--2735.
\newblock \href {http://arxiv.org/abs/hep-ph/9702443}
  {\path{arXiv:hep-ph/9702443}}, \href
  {http://dx.doi.org/10.1103/PhysRevD.56.2713}
  {\path{doi:10.1103/PhysRevD.56.2713}}.

\bibitem{Berger:2001zn}
E.~R. Berger, M.~Diehl, B.~Pire, {Probing generalized parton distributions in
  $\pi N \to l^+ l^- N$}, Phys. Lett. B 523 (2001) 265--272.
\newblock \href {http://arxiv.org/abs/hep-ph/0110080}
  {\path{arXiv:hep-ph/0110080}}, \href
  {http://dx.doi.org/10.1016/S0370-2693(01)01345-4}
  {\path{doi:10.1016/S0370-2693(01)01345-4}}.

\bibitem{Berger:2001xd}
E.~R. Berger, M.~Diehl, B.~Pire, {Time - like Compton scattering: Exclusive
  photoproduction of lepton pairs}, Eur. Phys. J. C 23 (2002) 675--689.
\newblock \href {http://arxiv.org/abs/hep-ph/0110062}
  {\path{arXiv:hep-ph/0110062}}, \href
  {http://dx.doi.org/10.1007/s100520200917} {\path{doi:10.1007/s100520200917}}.

\bibitem{Yu:2018ydp}
B.-G. Yu, K.-J. Kong, {Features of $\omega$ photoproduction off proton target
  at backward angles : Role of nucleon Reggeon in $u$-channel with parton
  contributions}, Phys. Rev. D 99~(1) (2019) 014031.
\newblock \href {http://arxiv.org/abs/1810.11645} {\path{arXiv:1810.11645}},
  \href {http://dx.doi.org/10.1103/PhysRevD.99.014031}
  {\path{doi:10.1103/PhysRevD.99.014031}}.

\bibitem{Laget:2019tou}
J.~Laget, {Exclusive Meson Photo- and Electro-production, a Window on the
  Structure of Hadronic Matter}, Prog. Part. Nucl. Phys. 111 (2020) 103737.
\newblock \href {http://arxiv.org/abs/1911.04825} {\path{arXiv:1911.04825}},
  \href {http://dx.doi.org/10.1016/j.ppnp.2019.103737}
  {\path{doi:10.1016/j.ppnp.2019.103737}}.

\bibitem{Mueller:1982bq}
A.~H. Mueller, {Topics in High-Energy Perturbative QCD Including Interactions
  with Nuclear Matter}, in: {17th Rencontres de Moriond on Elementary Particle
  Physics: I. Electroweak Interactions and Grand Unified Theories}, 1982, pp.
  13--43.

\bibitem{Brodsky:1982kg}
S.~J. Brodsky, {Testing Quantum Chromodynamics}, in: {XIII International
  Symposium on Multiparticle Dynamics}, 1982, pp. 963--1002.

\bibitem{Frankfurt:1991rk}
L.~Frankfurt, M.~Strikman, {Color screening and color transparency in hard
  nuclear processes}, Prog. Part. Nucl. Phys. 27 (1991) 135--193.
\newblock \href {http://dx.doi.org/10.1016/0146-6410(91)90004-8}
  {\path{doi:10.1016/0146-6410(91)90004-8}}.

\bibitem{Kopeliovich:1991pu}
B.~Z. Kopeliovich, B.~G. Zakharov, {Quantum effects and color transparency in
  charmonium photoproduction on nuclei}, Phys. Rev. D 44 (1991) 3466--3472.
\newblock \href {http://dx.doi.org/10.1103/PhysRevD.44.3466}
  {\path{doi:10.1103/PhysRevD.44.3466}}.

\bibitem{Frankfurt:1992dx}
L.~Frankfurt, G.~A. Miller, M.~Strikman, {Color transparency phenomenon and
  nuclear physics}, Comments Nucl. Part. Phys. 21~(1) (1992) 1--39.

\bibitem{Nikolaev:1992si}
N.~N. Nikolaev, {Quantum mechanics of color transparency}, Comments Nucl. Part.
  Phys. 21~(1) (1992) 41--70.

\bibitem{Jain:1995dd}
P.~Jain, B.~Pire, J.~P. Ralston, {Quantum color transparency and nuclear
  filtering}, Phys. Rept. 271 (1996) 67--179.
\newblock \href {http://arxiv.org/abs/hep-ph/9511333}
  {\path{arXiv:hep-ph/9511333}}, \href
  {http://dx.doi.org/10.1016/0370-1573(95)00071-2}
  {\path{doi:10.1016/0370-1573(95)00071-2}}.

\bibitem{Dutta:2012ii}
D.~Dutta, K.~Hafidi, M.~Strikman, {Color Transparency: past, present and
  future}, Prog. Part. Nucl. Phys. 69 (2013) 1--27.
\newblock \href {http://arxiv.org/abs/1211.2826} {\path{arXiv:1211.2826}},
  \href {http://dx.doi.org/10.1016/j.ppnp.2012.11.001}
  {\path{doi:10.1016/j.ppnp.2012.11.001}}.

\bibitem{Bhetuwal:2020jes}
D.~Bhetuwal, et~al., {Ruling out color transparency in quasi-elastic
  $^{12}$C(e,e'p) up to $Q^2$ of 14.2 (GeV/c)$^2$}, Phys. Rev. Lett. 126~(8)
  (2021) 082301.
\newblock \href {http://arxiv.org/abs/2011.00703} {\path{arXiv:2011.00703}},
  \href {http://dx.doi.org/10.1103/PhysRevLett.126.082301}
  {\path{doi:10.1103/PhysRevLett.126.082301}}.

\bibitem{Carroll:1988rp}
A.~S. Carroll, et~al., {Nuclear transparency to large angle $p p$ elastic
  scattering}, Phys. Rev. Lett. 61 (1988) 1698--1701.
\newblock \href {http://dx.doi.org/10.1103/PhysRevLett.61.1698}
  {\path{doi:10.1103/PhysRevLett.61.1698}}.

\bibitem{Ralston:1988rb}
J.~P. Ralston, B.~Pire, {Fluctuating proton size and oscillating nuclear
  transparency}, Phys. Rev. Lett. 61 (1988) 1823.
\newblock \href {http://dx.doi.org/10.1103/PhysRevLett.61.1823}
  {\path{doi:10.1103/PhysRevLett.61.1823}}.

\bibitem{Pasquini:2009ki}
B.~Pasquini, M.~Pincetti, S.~Boffi, {Parton content of the nucleon from
  distribution amplitudes and transition distribution amplitudes}, Phys. Rev.
  D80 (2009) 014017.
\newblock \href {http://arxiv.org/abs/0905.4018} {\path{arXiv:0905.4018}},
  \href {http://dx.doi.org/10.1103/PhysRevD.80.014017}
  {\path{doi:10.1103/PhysRevD.80.014017}}.

\bibitem{Lansberg:2006uh}
J.~P. Lansberg, B.~Pire, L.~Szymanowski, {Backward DVCS and Proton to Photon
  Transition Distribution Amplitudes}, Nucl. Phys. A782 (2007) 16--23.
\newblock \href {http://arxiv.org/abs/hep-ph/0607130}
  {\path{arXiv:hep-ph/0607130}}, \href
  {http://dx.doi.org/10.1016/j.nuclphysa.2006.10.014}
  {\path{doi:10.1016/j.nuclphysa.2006.10.014}}.

\bibitem{Berger:2001zb}
E.~R. Berger, F.~Cano, M.~Diehl, B.~Pire, {Generalized parton distributions in
  the deuteron}, Phys. Rev. Lett. 87 (2001) 142302.
\newblock \href {http://arxiv.org/abs/hep-ph/0106192}
  {\path{arXiv:hep-ph/0106192}}, \href
  {http://dx.doi.org/10.1103/PhysRevLett.87.142302}
  {\path{doi:10.1103/PhysRevLett.87.142302}}.

\bibitem{Kirchner:2003wt}
A.~Kirchner, D.~M{\"{u}}ller, {Deeply virtual Compton scattering off nuclei},
  Eur. Phys. J. C 32 (2003) 347--375.
\newblock \href {http://arxiv.org/abs/hep-ph/0302007}
  {\path{arXiv:hep-ph/0302007}}, \href
  {http://dx.doi.org/10.1140/epjc/s2003-01415-x}
  {\path{doi:10.1140/epjc/s2003-01415-x}}.

\bibitem{Guzey:2003jh}
V.~Guzey, M.~Strikman, {DVCS on spinless nuclear targets in impulse
  approximation}, Phys. Rev. C 68 (2003) 015204.
\newblock \href {http://arxiv.org/abs/hep-ph/0301216}
  {\path{arXiv:hep-ph/0301216}}, \href
  {http://dx.doi.org/10.1103/PhysRevC.68.015204}
  {\path{doi:10.1103/PhysRevC.68.015204}}.

\bibitem{Cano:2003ju}
F.~Cano, B.~Pire, {Deep electroproduction of photons and mesons on the
  deuteron}, Eur. Phys. J. A 19 (2004) 423--438.
\newblock \href {http://arxiv.org/abs/hep-ph/0307231}
  {\path{arXiv:hep-ph/0307231}}, \href
  {http://dx.doi.org/10.1140/epja/i2003-10127-x}
  {\path{doi:10.1140/epja/i2003-10127-x}}.

\bibitem{Polyakov:1998ze}
M.~V. Polyakov, {Hard exclusive electroproduction of two pions and their
  resonances}, Nucl. Phys. B555 (1999) 231.
\newblock \href {http://arxiv.org/abs/hep-ph/9809483}
  {\path{arXiv:hep-ph/9809483}}, \href
  {http://dx.doi.org/10.1016/S0550-3213(99)00314-4}
  {\path{doi:10.1016/S0550-3213(99)00314-4}}.

\bibitem{Diehl:2000uv}
M.~Diehl, T.~Gousset, B.~Pire, {Exclusive production of pion pairs in $\gamma^*
  \gamma$ collisions at large $Q^2$}, Phys. Rev. D 62 (2000) 073014.
\newblock \href {http://arxiv.org/abs/hep-ph/0003233}
  {\path{arXiv:hep-ph/0003233}}, \href
  {http://dx.doi.org/10.1103/PhysRevD.62.073014}
  {\path{doi:10.1103/PhysRevD.62.073014}}.

\bibitem{Pobylitsa:2001cz}
P.~V. Pobylitsa, M.~V. Polyakov, M.~Strikman, {Soft pion theorems for hard near
  threshold pion production}, Phys. Rev. Lett. 87 (2001) 022001.
\newblock \href {http://arxiv.org/abs/hep-ph/0101279}
  {\path{arXiv:hep-ph/0101279}}, \href
  {http://dx.doi.org/10.1103/PhysRevLett.87.022001}
  {\path{doi:10.1103/PhysRevLett.87.022001}}.

\bibitem{Braun:2006td}
V.~M. Braun, D.~{\relax Yu}. Ivanov, A.~Lenz, A.~Peters, {Deep inelastic pion
  electroproduction at threshold}, Phys. Rev. D75 (2007) 014021.
\newblock \href {http://arxiv.org/abs/hep-ph/0611386}
  {\path{arXiv:hep-ph/0611386}}, \href
  {http://dx.doi.org/10.1103/PhysRevD.75.014021}
  {\path{doi:10.1103/PhysRevD.75.014021}}.

\bibitem{Jaffe:1983hp}
R.~L. Jaffe, {Parton Distribution Functions for Twist Four}, Nucl. Phys. B 229
  (1983) 205--230.
\newblock \href {http://dx.doi.org/10.1016/0550-3213(83)90361-9}
  {\path{doi:10.1016/0550-3213(83)90361-9}}.

\bibitem{Alfaro_red_book}
V.~De~Alfaro, S.~Fubini, G.~Furlan, C.~Rossetti,
  \href{https://cds.cern.ch/record/1559182}{{Currents in hadron physics}},
  North-Holland Publ, Amsterdam, 1973.
\newline\urlprefix\url{https://cds.cern.ch/record/1559182}

\bibitem{Radyushkin:1998bz}
A.~V. Radyushkin, {Symmetries and structure of skewed and double
  distributions}, Phys. Lett. B449 (1999) 81--88.
\newblock \href {http://arxiv.org/abs/hep-ph/9810466}
  {\path{arXiv:hep-ph/9810466}}, \href
  {http://dx.doi.org/10.1016/S0370-2693(98)01584-6}
  {\path{doi:10.1016/S0370-2693(98)01584-6}}.

\bibitem{Radyushkin:1998es}
A.~V. Radyushkin, {Double distributions and evolution equations}, Phys. Rev.
  D59 (1999) 014030.
\newblock \href {http://arxiv.org/abs/hep-ph/9805342}
  {\path{arXiv:hep-ph/9805342}}, \href
  {http://dx.doi.org/10.1103/PhysRevD.59.014030}
  {\path{doi:10.1103/PhysRevD.59.014030}}.

\bibitem{Gelfand_Graev}
I.~Gelfand, M.~Graev, N.~Vilenkin, {Generalized functions}, Academic Press,
  N.Y.-London, 1966.

\bibitem{Kumericki:2008di}
K.~Kumericki, D.~M{\"{u}}ller, K.~Passek-Kumericki, {Sum rules and dualities
  for generalized parton distributions: Is there a holographic principle?},
  Eur. Phys. J. C58 (2008) 193--215.
\newblock \href {http://arxiv.org/abs/0805.0152} {\path{arXiv:0805.0152}},
  \href {http://dx.doi.org/10.1140/epjc/s10052-008-0741-0}
  {\path{doi:10.1140/epjc/s10052-008-0741-0}}.

\bibitem{Avdeenko:1981twg}
V.~A. Avdeenko, S.~E. Korenblit, V.~L. Chernyak, {Asymptotic behavior of
  electromagnetic and weak nucleon formfactors in QCD}, Sov. J. Nucl. Phys.
  33~(2) (1981) 252--256, [Yad. Fiz.33,481(1981)].

\bibitem{Chernyak:1984bm}
V.~L. Chernyak, I.~R. Zhitnitsky, {Nucleon Wave Function and Nucleon
  Form-Factors in QCD}, Nucl. Phys. B246 (1984) 52--74.
\newblock \href {http://dx.doi.org/10.1016/0550-3213(84)90114-7}
  {\path{doi:10.1016/0550-3213(84)90114-7}}.

\bibitem{SemenovTianShansky:2007hv}
K.~M. Semenov-Tian-Shansky, A.~V. Vereshagin, V.~V. Vereshagin, {Bootstrap and
  the physical values of $\pi N$ resonance parameters}, Phys. Rev. D77 (2008)
  025028.
\newblock \href {http://arxiv.org/abs/0706.3672} {\path{arXiv:0706.3672}},
  \href {http://dx.doi.org/10.1103/PhysRevD.77.025028}
  {\path{doi:10.1103/PhysRevD.77.025028}}.

\bibitem{Farrar:1988vz}
G.~R. Farrar, H.~Zhang, A.~A. Ogloblin, I.~R. Zhitnitsky, {Baryon Wave
  Functions and Cross-sections for Photon Annihilation to Baryon Pairs}, Nucl.
  Phys. B311 (1989) 585--612.
\newblock \href {http://dx.doi.org/10.1016/0550-3213(89)90169-7}
  {\path{doi:10.1016/0550-3213(89)90169-7}}.

\bibitem{Itzykson}
C.~Itzykson, J.~Zuber, {Quantum Field Theory}, McGraw-Hill, New York, USA,
  1980.

\bibitem{Ericson_Weise}
T.~Ericson, W.~Weise, {Pions and Nuclei}, Clarendon Press, Oxford, 1988.

\bibitem{Brodsky:1981rp}
S.~J. Brodsky, G.~P. Lepage, {Large Angle Two Photon Exclusive Channels in
  Quantum Chromodynamics}, Phys. Rev. D24 (1981) 1808.
\newblock \href {http://dx.doi.org/10.1103/PhysRevD.24.1808}
  {\path{doi:10.1103/PhysRevD.24.1808}}.

\bibitem{Blumlein:1997pi}
J.~Blumlein, B.~Geyer, D.~Robaschik, {On the evolution kernels of twist-2 light
  ray operators for unpolarized and polarized deep inelastic scattering}, Phys.
  Lett. B406 (1997) 161--170.
\newblock \href {http://arxiv.org/abs/hep-ph/9705264}
  {\path{arXiv:hep-ph/9705264}}, \href
  {http://dx.doi.org/10.1016/S0370-2693(97)00680-1}
  {\path{doi:10.1016/S0370-2693(97)00680-1}}.

\bibitem{Balitsky:1987bk}
I.~Balitsky, V.~M. Braun, {Evolution Equations for QCD String Operators}, Nucl.
  Phys. B 311 (1989) 541--584.
\newblock \href {http://dx.doi.org/10.1016/0550-3213(89)90168-5}
  {\path{doi:10.1016/0550-3213(89)90168-5}}.

\bibitem{Bukhvostov:1985rn}
A.~Bukhvostov, G.~Frolov, L.~Lipatov, E.~Kuraev, {Evolution Equations for
  Quasi-Partonic Operators}, Nucl. Phys. B 258 (1985) 601--646.
\newblock \href {http://dx.doi.org/10.1016/0550-3213(85)90628-5}
  {\path{doi:10.1016/0550-3213(85)90628-5}}.

\bibitem{Peskin:1979mn}
M.~E. Peskin, {Anomalous Dimensions of Three Quark Operators}, Phys. Lett. B 88
  (1979) 128--132.
\newblock \href {http://dx.doi.org/10.1016/0370-2693(79)90129-1}
  {\path{doi:10.1016/0370-2693(79)90129-1}}.

\bibitem{Braun:2003rp}
V.~Braun, G.~Korchemsky, D.~M\"uller, {The Uses of conformal symmetry in QCD},
  Prog. Part. Nucl. Phys. 51 (2003) 311--398.
\newblock \href {http://arxiv.org/abs/hep-ph/0306057}
  {\path{arXiv:hep-ph/0306057}}, \href
  {http://dx.doi.org/10.1016/S0146-6410(03)90004-4}
  {\path{doi:10.1016/S0146-6410(03)90004-4}}.

\bibitem{Bateman:100233}
H.~Bateman, A.~Erd\'{e}lyi, {Higher transcendental functions}, Vol.~II,
  McGraw-Hill, New York, USA, 1955.

\bibitem{Braun:2001qx}
V.~M. Braun, G.~Korchemsky, A.~Manashov, {Evolution equation for the structure
  function $g_2 (x, Q^2)$}, Nucl. Phys. B 603 (2001) 69--124.
\newblock \href {http://arxiv.org/abs/hep-ph/0102313}
  {\path{arXiv:hep-ph/0102313}}, \href
  {http://dx.doi.org/10.1016/S0550-3213(01)00165-1}
  {\path{doi:10.1016/S0550-3213(01)00165-1}}.

\bibitem{Braun:1998id}
V.~M. Braun, S.~E. Derkachov, A.~Manashov, {Integrability of three particle
  evolution equations in QCD}, Phys. Rev. Lett. 81 (1998) 2020--2023.
\newblock \href {http://arxiv.org/abs/hep-ph/9805225}
  {\path{arXiv:hep-ph/9805225}}, \href
  {http://dx.doi.org/10.1103/PhysRevLett.81.2020}
  {\path{doi:10.1103/PhysRevLett.81.2020}}.

\bibitem{Mueller:2005ed}
D.~M{\"{u}}ller, A.~Schafer, {Complex conformal spin partial wave expansion of
  generalized parton distributions and distribution amplitudes}, Nucl. Phys. B
  739 (2006) 1--59.
\newblock \href {http://arxiv.org/abs/hep-ph/0509204}
  {\path{arXiv:hep-ph/0509204}}, \href
  {http://dx.doi.org/10.1016/j.nuclphysb.2006.01.019}
  {\path{doi:10.1016/j.nuclphysb.2006.01.019}}.

\bibitem{Kirch:2005tt}
M.~Kirch, A.~Manashov, A.~Schafer, {Evolution equation for generalized parton
  distributions}, Phys. Rev. D 72 (2005) 114006.
\newblock \href {http://arxiv.org/abs/hep-ph/0509330}
  {\path{arXiv:hep-ph/0509330}}, \href
  {http://dx.doi.org/10.1103/PhysRevD.72.114006}
  {\path{doi:10.1103/PhysRevD.72.114006}}.

\bibitem{Manashov:2005xp}
A.~Manashov, M.~Kirch, A.~Schafer, {Solving the leading order evolution
  equation for GPDs}, Phys. Rev. Lett. 95 (2005) 012002.
\newblock \href {http://arxiv.org/abs/hep-ph/0503109}
  {\path{arXiv:hep-ph/0503109}}, \href
  {http://dx.doi.org/10.1103/PhysRevLett.95.012002}
  {\path{doi:10.1103/PhysRevLett.95.012002}}.

\bibitem{Shuvaev:1999fm}
A.~Shuvaev, {Solution of the off forward leading logarithmic evolution equation
  based on the Gegenbauer moments inversion}, Phys. Rev. D 60 (1999) 116005.
\newblock \href {http://arxiv.org/abs/hep-ph/9902318}
  {\path{arXiv:hep-ph/9902318}}, \href
  {http://dx.doi.org/10.1103/PhysRevD.60.116005}
  {\path{doi:10.1103/PhysRevD.60.116005}}.

\bibitem{Noritzsch:2000pr}
J.~D. Noritzsch, {Off forward parton distributions and Shuvaev's
  transformations}, Phys. Rev. D 62 (2000) 054015.
\newblock \href {http://arxiv.org/abs/hep-ph/0004012}
  {\path{arXiv:hep-ph/0004012}}, \href
  {http://dx.doi.org/10.1103/PhysRevD.62.054015}
  {\path{doi:10.1103/PhysRevD.62.054015}}.

\bibitem{Polyakov:2002wz}
M.~Polyakov, A.~Shuvaev, {On 'dual' parametrizations of generalized parton
  distributions} (7 2002).
\newblock \href {http://arxiv.org/abs/hep-ph/0207153}
  {\path{arXiv:hep-ph/0207153}}.

\bibitem{Polyakov:2008aa}
M.~V. Polyakov, K.~M. Semenov-Tian-Shansky, {Dual parametrization of GPDs
  versus double distribution Ansatz}, Eur. Phys. J. A 40 (2009) 181--198.
\newblock \href {http://arxiv.org/abs/0811.2901} {\path{arXiv:0811.2901}},
  \href {http://dx.doi.org/10.1140/epja/i2008-10759-2}
  {\path{doi:10.1140/epja/i2008-10759-2}}.

\bibitem{Muller:2014wxa}
D.~M\"uller, M.~V. Polyakov, K.~M. Semenov-Tian-Shansky, {Dual parametrization
  of generalized parton distributions in two equivalent representations}, JHEP
  03 (2015) 052.
\newblock \href {http://arxiv.org/abs/1412.4165} {\path{arXiv:1412.4165}},
  \href {http://dx.doi.org/10.1007/JHEP03(2015)052}
  {\path{doi:10.1007/JHEP03(2015)052}}.

\bibitem{szeg1939orthogonal}
G.~Szeg{\"{o}}, \href{https://books.google.ru/books?id=RemVAwAAQBAJ}{Orthogonal
  Polynomials}, no. v. 23 in American Math. Soc: Colloquium publ, American
  Mathematical Society, 1939.
\newline\urlprefix\url{https://books.google.ru/books?id=RemVAwAAQBAJ}

\bibitem{CarlsonTh}
F.~Carlson, {Sur une classe de s\'{e}ries de Taylor}, Ph.D. thesis, Uppsala
  University, Sweden (1914).

\bibitem{Anikin:2007yh}
I.~Anikin, O.~Teryaev, {Dispersion relations and subtractions in hard exclusive
  processes}, Phys. Rev. D 76 (2007) 056007.
\newblock \href {http://arxiv.org/abs/0704.2185} {\path{arXiv:0704.2185}},
  \href {http://dx.doi.org/10.1103/PhysRevD.76.056007}
  {\path{doi:10.1103/PhysRevD.76.056007}}.

\bibitem{Polyakov:2007rv}
M.~Polyakov, {Tomography for amplitudes of hard exclusive processes}, Phys.
  Lett. B 659 (2008) 542--550.
\newblock \href {http://arxiv.org/abs/0707.2509} {\path{arXiv:0707.2509}},
  \href {http://dx.doi.org/10.1016/j.physletb.2007.11.012}
  {\path{doi:10.1016/j.physletb.2007.11.012}}.

\bibitem{Muller:2015vha}
D.~M\"uller, K.~M. Semenov-Tian-Shansky, {$J=0$ fixed pole and $D$-term form
  factor in deeply virtual Compton scattering}, Phys. Rev. D 92~(7) (2015)
  074025.
\newblock \href {http://arxiv.org/abs/1507.02164} {\path{arXiv:1507.02164}},
  \href {http://dx.doi.org/10.1103/PhysRevD.92.074025}
  {\path{doi:10.1103/PhysRevD.92.074025}}.

\bibitem{Kumericki:2009uq}
K.~Kumeri\v{c}ki, D.~M{\"{u}}ller, {Deeply virtual Compton scattering at small
  $x_B$ and the access to the GPD H}, Nucl. Phys. B 841 (2010) 1--58.
\newblock \href {http://arxiv.org/abs/0904.0458} {\path{arXiv:0904.0458}},
  \href {http://dx.doi.org/10.1016/j.nuclphysb.2010.07.015}
  {\path{doi:10.1016/j.nuclphysb.2010.07.015}}.

\bibitem{Burkardt:2002hr}
M.~Burkardt, {Impact parameter space interpretation for generalized parton
  distributions}, Int. J. Mod. Phys. A18 (2003) 173--208.
\newblock \href {http://arxiv.org/abs/hep-ph/0207047}
  {\path{arXiv:hep-ph/0207047}}, \href
  {http://dx.doi.org/10.1142/S0217751X03012370}
  {\path{doi:10.1142/S0217751X03012370}}.

\bibitem{Lichtenberg:1968zz}
D.~B. Lichtenberg, L.~J. Tassie, P.~J. Keleman, {Quark-Diquark Model of Baryons
  and SU (6)}, Phys. Rev. 167 (1968) 1535--1542.
\newblock \href {http://dx.doi.org/10.1103/PhysRev.167.1535}
  {\path{doi:10.1103/PhysRev.167.1535}}.

\bibitem{Carroll:1969ty}
J.~Carroll, D.~B. Lichtenberg, J.~Franklin, {Electromagnetic properties of
  baryons in a quark-diquark model with broken su(6)}, Phys. Rev. 174 (1968)
  1681--1688.
\newblock \href {http://dx.doi.org/10.1103/PhysRev.174.1681}
  {\path{doi:10.1103/PhysRev.174.1681}}.

\bibitem{Cutkosky:1977kd}
R.~E. Cutkosky, R.~E. Hendrick, {Does the Baryon Spectrum Reveal a Diquark
  Structure?}, Phys. Rev. D 16 (1977) 2902.
\newblock \href {http://dx.doi.org/10.1103/PhysRevD.16.2902}
  {\path{doi:10.1103/PhysRevD.16.2902}}.

\bibitem{Anselmino:1987vk}
M.~Anselmino, P.~Kroll, B.~Pire, {Diquarks in Exclusive Reactions at Large
  Momentum Transfer}, Z. Phys. C 36 (1987) 89.
\newblock \href {http://dx.doi.org/10.1007/BF01556169}
  {\path{doi:10.1007/BF01556169}}.

\bibitem{Dziembowski:1990md}
Z.~Dziembowski, J.~Franklin, {The nucleon distribution amplitude and diquark
  clustering}, Phys. Rev. D42 (1990) 905--910.
\newblock \href {http://dx.doi.org/10.1103/PhysRevD.42.905}
  {\path{doi:10.1103/PhysRevD.42.905}}.

\bibitem{Pire:2019nwa}
B.~Pire, K.~Semenov-Tian-Shansky, L.~Szymanowski, {Nucleon-to-meson transition
  distribution amplitudes in impact parameter space}, PoS LC2019 (2019) 012.
\newblock \href {http://arxiv.org/abs/1912.05165} {\path{arXiv:1912.05165}},
  \href {http://dx.doi.org/10.22323/1.374.0012}
  {\path{doi:10.22323/1.374.0012}}.

\bibitem{Frankfurt:1998et}
L.~L. Frankfurt, A.~Freund, M.~Strikman, {Deeply virtual Compton scattering at
  HERA: A Probe of asymptotia}, Phys. Lett. B460 (1999) 417--424.
\newblock \href {http://arxiv.org/abs/hep-ph/9806535}
  {\path{arXiv:hep-ph/9806535}}, \href
  {http://dx.doi.org/10.1016/S0370-2693(99)00803-5}
  {\path{doi:10.1016/S0370-2693(99)00803-5}}.

\bibitem{Penttinen:1999th}
M.~Penttinen, M.~V. Polyakov, K.~Goeke, {Helicity skewed quark distributions of
  the nucleon and chiral symmetry}, Phys. Rev. D62 (2000) 014024.
\newblock \href {http://arxiv.org/abs/hep-ph/9909489}
  {\path{arXiv:hep-ph/9909489}}, \href
  {http://dx.doi.org/10.1103/PhysRevD.62.014024}
  {\path{doi:10.1103/PhysRevD.62.014024}}.

\bibitem{Matsinos:2019kqi}
E.~Matsinos, {A brief history of the pion-nucleon coupling constant} (2019).
\newblock \href {http://arxiv.org/abs/1901.01204} {\path{arXiv:1901.01204}}.

\bibitem{Dumbrajs:1983jd}
O.~Dumbrajs, R.~Koch, H.~Pilkuhn, G.~Oades, H.~Behrens, J.~De~Swart, P.~Kroll,
  {Compilation of Coupling Constants and Low-Energy Parameters. 1982 Edition},
  Nucl. Phys. B 216 (1983) 277--335.
\newblock \href {http://dx.doi.org/10.1016/0550-3213(83)90288-2}
  {\path{doi:10.1016/0550-3213(83)90288-2}}.

\bibitem{PDG2020}
P.~Z. et~al. (Particle Data~Group), {The Review of Particle Physics (2020)},
  Prog. Theor. Exp. Phys. 2020 (2020) 083C01.
\newblock \href {http://dx.doi.org/https://doi.org/10.1093/ptep/ptaa104}
  {\path{doi:https://doi.org/10.1093/ptep/ptaa104}}.

\bibitem{Grein:1979nw}
W.~Grein, P.~Kroll, {Two Pion and Three Pion Cut Contributions to
  Nucleon-Nucleon Scattering}, Nucl. Phys. A 338 (1980) 332--348.
\newblock \href {http://dx.doi.org/10.1016/0375-9474(80)90036-6}
  {\path{doi:10.1016/0375-9474(80)90036-6}}.

\bibitem{Mergell:1995bf}
P.~Mergell, U.~G. Meissner, D.~Drechsel, {Dispersion theoretical analysis of
  the nucleon electromagnetic form-factors}, Nucl. Phys. A 596 (1996) 367--396.
\newblock \href {http://arxiv.org/abs/hep-ph/9506375}
  {\path{arXiv:hep-ph/9506375}}, \href
  {http://dx.doi.org/10.1016/0375-9474(95)00339-8}
  {\path{doi:10.1016/0375-9474(95)00339-8}}.

\bibitem{Meissner:1997qt}
U.-G. Meissner, V.~Mull, J.~Speth, J.~W. van Orden, {Strange vector currents
  and the OZI rule}, Phys. Lett. B 408 (1997) 381--386.
\newblock \href {http://arxiv.org/abs/hep-ph/9701296}
  {\path{arXiv:hep-ph/9701296}}, \href
  {http://dx.doi.org/10.1016/S0370-2693(97)00828-9}
  {\path{doi:10.1016/S0370-2693(97)00828-9}}.

\bibitem{Kivel:2002ia}
N.~Kivel, M.~V. Polyakov, {One loop chiral corrections to hard exclusive
  processes: 1. Pion case} (2002).
\newblock \href {http://arxiv.org/abs/hep-ph/0203264}
  {\path{arXiv:hep-ph/0203264}}.

\bibitem{Brodsky:1997de}
S.~J. Brodsky, H.-C. Pauli, S.~S. Pinsky, {Quantum chromodynamics and other
  field theories on the light cone}, Phys. Rept. 301 (1998) 299--486.
\newblock \href {http://arxiv.org/abs/hep-ph/9705477}
  {\path{arXiv:hep-ph/9705477}}, \href
  {http://dx.doi.org/10.1016/S0370-1573(97)00089-6}
  {\path{doi:10.1016/S0370-1573(97)00089-6}}.

\bibitem{Pincetti:2008fh}
M.~Pincetti, B.~Pasquini, S.~Boffi, {Nucleon to Pion Transition Distribution
  Amplitudes in a Light-Cone Quark Model}, in: {2nd International Workshop on
  Transverse Polarization Phenomena in Hard Processes}, 2008.
\newblock \href {http://arxiv.org/abs/0807.4861} {\path{arXiv:0807.4861}},
  \href {http://dx.doi.org/10.1142/9789814277785\_0033}
  {\path{doi:10.1142/9789814277785\_0033}}.

\bibitem{PincettiPhD}
M.~Pincetti, {Exploring the partonic structure of the nucleon on the
  light-cone}, Ph.D. thesis, University of Pavia (2008).

\bibitem{Dziembowski:1997vh}
Z.~Dziembowski, H.~Holtmann, A.~Szczurek, J.~Speth, {Pionic corrections to
  nucleon electromagnetic properties in a light cone framework}, Annals Phys.
  258 (1997) 1--36.
\newblock \href {http://dx.doi.org/10.1006/aphy.1997.5693}
  {\path{doi:10.1006/aphy.1997.5693}}.

\bibitem{Speth:1996pz}
J.~Speth, A.~W. Thomas, {Mesonic contributions to the spin and flavor structure
  of the nucleon}, Adv. Nucl. Phys. 24 (1997) 83--149.
\newblock \href {http://dx.doi.org/10.1007/0-306-47073-X\_2}
  {\path{doi:10.1007/0-306-47073-X\_2}}.

\bibitem{Boffi:2002yy}
S.~Boffi, B.~Pasquini, M.~Traini, {Linking generalized parton distributions to
  constituent quark models}, Nucl. Phys. B 649 (2003) 243--262.
\newblock \href {http://arxiv.org/abs/hep-ph/0207340}
  {\path{arXiv:hep-ph/0207340}}, \href
  {http://dx.doi.org/10.1016/S0550-3213(02)01016-7}
  {\path{doi:10.1016/S0550-3213(02)01016-7}}.

\bibitem{Pasquini:2008ax}
B.~Pasquini, S.~Cazzaniga, S.~Boffi, {Transverse momentum dependent parton
  distributions in a light-cone quark model}, Phys. Rev. D 78 (2008) 034025.
\newblock \href {http://arxiv.org/abs/0806.2298} {\path{arXiv:0806.2298}},
  \href {http://dx.doi.org/10.1103/PhysRevD.78.034025}
  {\path{doi:10.1103/PhysRevD.78.034025}}.

\bibitem{Kofler:2017uzq}
S.~Kofler, B.~Pasquini, {Collinear parton distributions and the structure of
  the nucleon sea in a light-front meson-cloud model}, Phys. Rev. D 95~(9)
  (2017) 094015.
\newblock \href {http://arxiv.org/abs/1701.07839} {\path{arXiv:1701.07839}},
  \href {http://dx.doi.org/10.1103/PhysRevD.95.094015}
  {\path{doi:10.1103/PhysRevD.95.094015}}.

\bibitem{Pasquini:2006dv}
B.~Pasquini, S.~Boffi, {Virtual meson cloud of the nucleon and generalized
  parton distributions}, Phys. Rev. D 73 (2006) 094001.
\newblock \href {http://arxiv.org/abs/hep-ph/0601177}
  {\path{arXiv:hep-ph/0601177}}, \href
  {http://dx.doi.org/10.1103/PhysRevD.73.094001}
  {\path{doi:10.1103/PhysRevD.73.094001}}.

\bibitem{Pasquini:2007iz}
B.~Pasquini, S.~Boffi, {Electroweak structure of the nucleon, meson cloud and
  light-cone wavefunctions}, Phys. Rev. D 76 (2007) 074011.
\newblock \href {http://arxiv.org/abs/0707.2897} {\path{arXiv:0707.2897}},
  \href {http://dx.doi.org/10.1103/PhysRevD.76.074011}
  {\path{doi:10.1103/PhysRevD.76.074011}}.

\bibitem{Knodlseder:2015vmu}
M.~Knodlseder, {Nucleon electromagnetic form factors in perturbative QCD},
  Ph.D. thesis, Regensburg U. (2015).

\bibitem{Ji:1986uh}
C.-R. Ji, A.~Sill, R.~Lombard, {Leading order perturbative QCD calculation of
  nucleon Dirac form-factors}, Phys. Rev. D 36 (1987) 165.
\newblock \href {http://dx.doi.org/10.1103/PhysRevD.36.165}
  {\path{doi:10.1103/PhysRevD.36.165}}.

\bibitem{Brooks:2000nb}
T.~C. Brooks, L.~J. Dixon, {Recalculation of proton Compton scattering in
  perturbative QCD}, Phys. Rev. D 62 (2000) 114021.
\newblock \href {http://arxiv.org/abs/hep-ph/0004143}
  {\path{arXiv:hep-ph/0004143}}, \href
  {http://dx.doi.org/10.1103/PhysRevD.62.114021}
  {\path{doi:10.1103/PhysRevD.62.114021}}.

\bibitem{Thomson:2006ny}
R.~Thomson, A.~Pang, C.-R. Ji, {Real and virtual nucleon Compton scattering in
  the perturbative limit}, Phys. Rev. D 73 (2006) 054023.
\newblock \href {http://arxiv.org/abs/hep-ph/0602164}
  {\path{arXiv:hep-ph/0602164}}, \href
  {http://dx.doi.org/10.1103/PhysRevD.73.054023}
  {\path{doi:10.1103/PhysRevD.73.054023}}.

\bibitem{Stefanis:1987vr}
N.~G. Stefanis, {Quark content of the nucleon in QCD: Perturbative and
  nonperturbative aspects}, Phys. Rev. D 40 (1989) 2305, [Erratum: Phys.Rev.D
  44, 1616 (1991)].
\newblock \href {http://dx.doi.org/10.1103/PhysRevD.40.2305}
  {\path{doi:10.1103/PhysRevD.40.2305}}.

\bibitem{Gaillard:1982zm}
M.~K. Gaillard, L.~Maiani, R.~Petronzio, {Soft Pion Emission in $p \bar{p}$
  Resonance Formation}, Phys. Lett. B 110 (1982) 489--492.
\newblock \href {http://dx.doi.org/10.1016/0370-2693(82)91044-9}
  {\path{doi:10.1016/0370-2693(82)91044-9}}.

\bibitem{Brodsky:1981kj}
S.~J. Brodsky, G.~P. Lepage, {Helicity Selection Rules and Tests of Gluon Spin
  in Exclusive QCD Processes}, Phys. Rev. D24 (1981) 2848.
\newblock \href {http://dx.doi.org/10.1103/PhysRevD.24.2848}
  {\path{doi:10.1103/PhysRevD.24.2848}}.

\bibitem{Goloskokov:2015zsa}
S.~V. Goloskokov, P.~Kroll, {The exclusive limit of the pion-induced
  Drell\textendash{}Yan process}, Phys. Lett. B 748 (2015) 323--327.
\newblock \href {http://arxiv.org/abs/1506.04619} {\path{arXiv:1506.04619}},
  \href {http://dx.doi.org/10.1016/j.physletb.2015.07.016}
  {\path{doi:10.1016/j.physletb.2015.07.016}}.

\bibitem{Sawada:2016mao}
T.~Sawada, W.-C. Chang, S.~Kumano, J.-C. Peng, S.~Sawada, K.~Tanaka, {Accessing
  proton generalized parton distributions and pion distribution amplitudes with
  the exclusive pion-induced Drell-Yan process at J-PARC}, Phys. Rev. D 93~(11)
  (2016) 114034.
\newblock \href {http://arxiv.org/abs/1605.00364} {\path{arXiv:1605.00364}},
  \href {http://dx.doi.org/10.1103/PhysRevD.93.114034}
  {\path{doi:10.1103/PhysRevD.93.114034}}.

\bibitem{Mulders:1990xw}
P.~Mulders, {Modifications of Nucleons in Nuclei and Other Consequences of the
  Quark Substructure}, Phys. Rept. 185 (1990) 83--169.
\newblock \href {http://dx.doi.org/10.1016/0370-1573(90)90086-H}
  {\path{doi:10.1016/0370-1573(90)90086-H}}.

\bibitem{Kroll:1995pv}
P.~Kroll, M.~Schurmann, P.~A. Guichon, {Virtual Compton scattering off protons
  at moderately large momentum transfer}, Nucl. Phys. A 598 (1996) 435--461.
\newblock \href {http://arxiv.org/abs/hep-ph/9507298}
  {\path{arXiv:hep-ph/9507298}}, \href
  {http://dx.doi.org/10.1016/0375-9474(96)00002-4}
  {\path{doi:10.1016/0375-9474(96)00002-4}}.

\bibitem{Arens:1996xw}
T.~Arens, O.~Nachtmann, M.~Diehl, P.~V. Landshoff, {Some tests for the helicity
  structure of the pomeron in e p collisions}, Z. Phys. C 74 (1997) 651--669.
\newblock \href {http://arxiv.org/abs/hep-ph/9605376}
  {\path{arXiv:hep-ph/9605376}}, \href
  {http://dx.doi.org/10.1007/s002880050430} {\path{doi:10.1007/s002880050430}}.

\bibitem{Huang:2000kd}
H.~W. Huang, P.~Kroll, {Large momentum transfer electroproduction of mesons},
  Eur. Phys. J. C 17 (2000) 423--435.
\newblock \href {http://arxiv.org/abs/hep-ph/0005318}
  {\path{arXiv:hep-ph/0005318}}, \href
  {http://dx.doi.org/10.1007/s100520000500} {\path{doi:10.1007/s100520000500}}.

\bibitem{Hand:1963bb}
L.~Hand, {Experimental investigation of pion electroproduction}, Phys. Rev. 129
  (1963) 1834--1846.
\newblock \href {http://dx.doi.org/10.1103/PhysRev.129.1834}
  {\path{doi:10.1103/PhysRev.129.1834}}.

\bibitem{Burkert:2004sk}
V.~Burkert, T.~Lee, {Electromagnetic meson production in the nucleon resonance
  region}, Int. J. Mod. Phys. E 13 (2004) 1035--1112.
\newblock \href {http://arxiv.org/abs/nucl-ex/0407020}
  {\path{arXiv:nucl-ex/0407020}}, \href
  {http://dx.doi.org/10.1142/S0218301304002545}
  {\path{doi:10.1142/S0218301304002545}}.

\bibitem{Lansberg:2010mf}
J.~Lansberg, B.~Pire, L.~Szymanowski, {Spin Observables in
  Transition-Distribution-Amplitude Studies}, J. Phys. Conf. Ser. 295 (2011)
  012090.
\newblock \href {http://arxiv.org/abs/1011.6635} {\path{arXiv:1011.6635}},
  \href {http://dx.doi.org/10.1088/1742-6596/295/1/012090}
  {\path{doi:10.1088/1742-6596/295/1/012090}}.

\bibitem{Amaldi:1979vh}
E.~Amaldi, S.~Fubini, G.~Furlan, {Pion electroproduction. Electroproduction at
  low-energy and hadron form-factors}, Vol.~83, 1979.
\newblock \href {http://dx.doi.org/10.1007/BFb0048209}
  {\path{doi:10.1007/BFb0048209}}.

\bibitem{Drechsel:1992pn}
D.~Drechsel, L.~Tiator, {Threshold pion photoproduction on nucleons}, J. Phys.
  G 18 (1992) 449--497.
\newblock \href {http://dx.doi.org/10.1088/0954-3899/18/3/004}
  {\path{doi:10.1088/0954-3899/18/3/004}}.

\bibitem{Ahmad:2008hp}
S.~Ahmad, G.~R. Goldstein, S.~Liuti, {Nucleon Tensor Charge from Exclusive
  $\pi^0$ Electroproduction}, Phys. Rev. D 79 (2009) 054014.
\newblock \href {http://arxiv.org/abs/0805.3568} {\path{arXiv:0805.3568}},
  \href {http://dx.doi.org/10.1103/PhysRevD.79.054014}
  {\path{doi:10.1103/PhysRevD.79.054014}}.

\bibitem{Goloskokov:2011rd}
S.~V. Goloskokov, P.~Kroll, {Transversity in hard exclusive electroproduction
  of pseudoscalar mesons}, Eur. Phys. J. A 47 (2011) 112.
\newblock \href {http://arxiv.org/abs/1106.4897} {\path{arXiv:1106.4897}},
  \href {http://dx.doi.org/10.1140/epja/i2011-11112-6}
  {\path{doi:10.1140/epja/i2011-11112-6}}.

\bibitem{Braun:2000kw}
V.~Braun, R.~J. Fries, N.~Mahnke, E.~Stein, {Higher twist distribution
  amplitudes of the nucleon in QCD}, Nucl. Phys. B589 (2000) 381--409,
  [Erratum: Nucl. Phys.B607,433(2001)].
\newblock \href {http://arxiv.org/abs/hep-ph/0007279}
  {\path{arXiv:hep-ph/0007279}}, \href
  {http://dx.doi.org/10.1016/S0550-3213(00)00516-2,
  10.1016/S0550-3213(01)00254-1} {\path{doi:10.1016/S0550-3213(00)00516-2,
  10.1016/S0550-3213(01)00254-1}}.

\bibitem{Kivel:2019wjh}
N.~Kivel, {A study of power suppressed contributions in $J/\psi \rightarrow
  p\bar{p}$ decay}, Eur. Phys. J. A 56~(2) (2020) 64.
\newblock \href {http://arxiv.org/abs/1910.02850} {\path{arXiv:1910.02850}},
  \href {http://dx.doi.org/10.1140/epja/s10050-020-00064-5}
  {\path{doi:10.1140/epja/s10050-020-00064-5}}.

\bibitem{Borodulin:2017pwh}
V.~I. Borodulin, R.~N. Rogalyov, S.~R. Slabospitskii, {CORE 3.1 (COmpendium of
  RElations, Version 3.1) } (2017).
\newblock \href {http://arxiv.org/abs/1702.08246} {\path{arXiv:1702.08246}}.

\bibitem{Kroll:2012sm}
P.~Kroll, H.~Moutarde, F.~Sabatie, {From hard exclusive meson electroproduction
  to deeply virtual Compton scattering}, Eur. Phys. J. C 73~(1) (2013) 2278.
\newblock \href {http://arxiv.org/abs/1210.6975} {\path{arXiv:1210.6975}},
  \href {http://dx.doi.org/10.1140/epjc/s10052-013-2278-0}
  {\path{doi:10.1140/epjc/s10052-013-2278-0}}.

\bibitem{Goldstein:2013gra}
G.~R. Goldstein, J.~O. Gonzalez~Hernandez, S.~Liuti, {Flexible parametrization
  of generalized parton distributions: the chiral-odd sector}, Phys. Rev. D
  91~(11) (2015) 114013.
\newblock \href {http://arxiv.org/abs/1311.0483} {\path{arXiv:1311.0483}},
  \href {http://dx.doi.org/10.1103/PhysRevD.91.114013}
  {\path{doi:10.1103/PhysRevD.91.114013}}.

\bibitem{Favart:2015umi}
L.~Favart, M.~Guidal, T.~Horn, P.~Kroll, {Deeply Virtual Meson Production on
  the nucleon}, Eur. Phys. J. A 52~(6) (2016) 158.
\newblock \href {http://arxiv.org/abs/1511.04535} {\path{arXiv:1511.04535}},
  \href {http://dx.doi.org/10.1140/epja/i2016-16158-2}
  {\path{doi:10.1140/epja/i2016-16158-2}}.

\bibitem{Bedlinskiy:2012be}
I.~Bedlinskiy, et~al., {Measurement of Exclusive $\pi^0$ Electroproduction
  Structure Functions and their Relationship to Transversity GPDs}, Phys. Rev.
  Lett. 109 (2012) 112001.
\newblock \href {http://arxiv.org/abs/1206.6355} {\path{arXiv:1206.6355}},
  \href {http://dx.doi.org/10.1103/PhysRevLett.109.112001}
  {\path{doi:10.1103/PhysRevLett.109.112001}}.

\bibitem{Dlamini:2020ulg}
M.~Dlamini, et~al., {Deep exclusive electroproduction of $\pi^0$ at high $Q^2$
  in the quark valence regime} (2020).
\newblock \href {http://arxiv.org/abs/2011.11125} {\path{arXiv:2011.11125}}.

\bibitem{Bolz:1996sw}
J.~Bolz, P.~Kroll, {Modeling the nucleon wave function from soft and hard
  processes}, Z. Phys. A 356 (1996) 327.
\newblock \href {http://arxiv.org/abs/hep-ph/9603289}
  {\path{arXiv:hep-ph/9603289}}, \href
  {http://dx.doi.org/10.1007/s002180050186} {\path{doi:10.1007/s002180050186}}.

\bibitem{Gockeler:2008xv}
M.~Gockeler, et~al., {Nucleon distribution amplitudes from lattice QCD}, Phys.
  Rev. Lett. 101 (2008) 112002.
\newblock \href {http://arxiv.org/abs/0804.1877} {\path{arXiv:0804.1877}},
  \href {http://dx.doi.org/10.1103/PhysRevLett.101.112002}
  {\path{doi:10.1103/PhysRevLett.101.112002}}.

\bibitem{Braun:2014wpa}
V.~M. Braun, S.~Collins, B.~Gl\"a\ss{}le, M.~G\"ockeler, A.~Sch\"afer, R.~W.
  Schiel, W.~S\"oldner, A.~Sternbeck, P.~Wein, {Light-cone Distribution
  Amplitudes of the Nucleon and Negative Parity Nucleon Resonances from Lattice
  QCD}, Phys. Rev. D 89 (2014) 094511.
\newblock \href {http://arxiv.org/abs/1403.4189} {\path{arXiv:1403.4189}},
  \href {http://dx.doi.org/10.1103/PhysRevD.89.094511}
  {\path{doi:10.1103/PhysRevD.89.094511}}.

\bibitem{Bali:2019ecy}
G.~S. Bali, et~al., {Light-cone distribution amplitudes of octet baryons from
  lattice QCD}, Eur. Phys. J. A 55~(7) (2019) 116.
\newblock \href {http://arxiv.org/abs/1903.12590} {\path{arXiv:1903.12590}},
  \href {http://dx.doi.org/10.1140/epja/i2019-12803-6}
  {\path{doi:10.1140/epja/i2019-12803-6}}.

\bibitem{King:1986wi}
I.~King, C.~T. Sachrajda, {Nucleon Wave Functions and QCD Sum Rules}, Nucl.
  Phys. B 279 (1987) 785--803.
\newblock \href {http://dx.doi.org/10.1016/0550-3213(87)90019-8}
  {\path{doi:10.1016/0550-3213(87)90019-8}}.

\bibitem{Gari:1986ue}
M.~Gari, N.~Stefanis, {Electromagnetic Form-factors of the Nucleon From
  Perturbative QCD and QCD Sum Rules}, Phys. Lett. B 175 (1986) 462--466.
\newblock \href {http://dx.doi.org/10.1016/0370-2693(86)90624-6}
  {\path{doi:10.1016/0370-2693(86)90624-6}}.

\bibitem{Radyushkin:1990te}
A.~V. Radyushkin, {Hadronic form-factors: Perturbative QCD versus QCD sum
  rules}, Nucl. Phys. A 532 (1991) 141--154.
\newblock \href {http://dx.doi.org/10.1016/0375-9474(91)90691-X}
  {\path{doi:10.1016/0375-9474(91)90691-X}}.

\bibitem{Anikin:2013aka}
I.~V. Anikin, V.~M. Braun, N.~Offen, {Nucleon Form Factors and Distribution
  Amplitudes in QCD}, Phys. Rev. D 88 (2013) 114021.
\newblock \href {http://arxiv.org/abs/1310.1375} {\path{arXiv:1310.1375}},
  \href {http://dx.doi.org/10.1103/PhysRevD.88.114021}
  {\path{doi:10.1103/PhysRevD.88.114021}}.

\bibitem{Mecking:2003zu}
B.~A. Mecking, et~al., {The CEBAF Large Acceptance Spectrometer (CLAS)}, Nucl.
  Instrum. Meth. A 503 (2003) 513--553.
\newblock \href {http://dx.doi.org/10.1016/S0168-9002(03)01001-5}
  {\path{doi:10.1016/S0168-9002(03)01001-5}}.

\bibitem{Kubarovskiy:2012yz}
A.~Kubarovskiy, {Electroproduction of $\pi^0$ at high momentum transfers in
  non-resonant region with CLAS}, AIP Conf. Proc. 1560~(1) (2013) 576--578.
\newblock \href {http://dx.doi.org/10.1063/1.4826845}
  {\path{doi:10.1063/1.4826845}}.

\bibitem{Park:2012rn}
K.~Park, et~al., {Deep exclusive $\pi^+$ electroproduction off the proton at
  CLAS}, Eur. Phys. J. A 49 (2013) 16.
\newblock \href {http://arxiv.org/abs/1206.2326} {\path{arXiv:1206.2326}},
  \href {http://dx.doi.org/10.1140/epja/i2013-13016-9}
  {\path{doi:10.1140/epja/i2013-13016-9}}.

\bibitem{Laget:2009hs}
J.~M. Laget, {Unitarity constraints on charged pion photoproduction at large
  $p_\bot$}, Phys. Lett. B 685 (2010) 146--150.
\newblock \href {http://arxiv.org/abs/0912.1942} {\path{arXiv:0912.1942}},
  \href {http://dx.doi.org/10.1016/j.physletb.2010.01.052}
  {\path{doi:10.1016/j.physletb.2010.01.052}}.

\bibitem{Laget:2004qu}
J.~M. Laget, {The Space-time structure of hard scattering processes}, Phys.
  Rev. D 70 (2004) 054023.
\newblock \href {http://arxiv.org/abs/hep-ph/0406153}
  {\path{arXiv:hep-ph/0406153}}, \href
  {http://dx.doi.org/10.1103/PhysRevD.70.054023}
  {\path{doi:10.1103/PhysRevD.70.054023}}.

\bibitem{Laget_TBP}
J.~M. Laget, {Unitarity constraints on meson electroproduction at backward
  angles}, Phys. Rev. C 104~(2) (2021) 025202.
\newblock \href {http://arxiv.org/abs/2104.13078} {\path{arXiv:2104.13078}},
  \href {http://dx.doi.org/10.1103/PhysRevC.104.025202}
  {\path{doi:10.1103/PhysRevC.104.025202}}.

\bibitem{Morand:2005ex}
L.~Morand, et~al., {Deeply virtual and exclusive electroproduction of omega
  mesons}, Eur. Phys. J. A 24 (2005) 445--458.
\newblock \href {http://arxiv.org/abs/hep-ex/0504057}
  {\path{arXiv:hep-ex/0504057}}, \href
  {http://dx.doi.org/10.1140/epja/i2005-10032-4}
  {\path{doi:10.1140/epja/i2005-10032-4}}.

\bibitem{Anderson:1969jw}
R.~L. Anderson, D.~Gustavson, J.~R. Johnson, I.~Overman, D.~Ritson, B.~H. Wiik,
  {High-energy photoproduction of charged pions at backward angles}, Phys. Rev.
  Lett. 23 (1969) 721--724.
\newblock \href {http://dx.doi.org/10.1103/PhysRevLett.23.721}
  {\path{doi:10.1103/PhysRevLett.23.721}}.

\bibitem{Anderson:1969dv}
J.~L. Anderson, J.~W. Ryon, {Electromagnetic radiation in accelerated systems},
  Phys. Rev. 181 (1969) 1765--1775.
\newblock \href {http://dx.doi.org/10.1103/PhysRev.181.1765}
  {\path{doi:10.1103/PhysRev.181.1765}}.

\bibitem{Boyarski:1967sp}
A.~Boyarski, F.~Bulos, W.~Busza, R.~E. Diebold, S.~D. Ecklund, G.~E. Fischer,
  J.~R. Rees, B.~Richter, {5-GeV - 16-GeV single {$\pi^+$} photoproduction from
  hydrogen}, Phys. Rev. Lett. 20 (1968) 300--303.
\newblock \href {http://dx.doi.org/10.1103/PhysRevLett.20.300}
  {\path{doi:10.1103/PhysRevLett.20.300}}.

\bibitem{Fischer:2021kcr}
C.~S. Fischer, J.~Haidenbauer, C.~Hanhart, M.~F.~M. Lutz, S.~M. Ryan, {PANDA
  Phase One} (1 2021).
\newblock \href {http://arxiv.org/abs/2101.11877} {\path{arXiv:2101.11877}}.

\bibitem{Wiedner:2011mf}
U.~Wiedner, {Future Prospects for Hadron Physics at PANDA}, Prog. Part. Nucl.
  Phys. 66 (2011) 477--518.
\newblock \href {http://arxiv.org/abs/1104.3961} {\path{arXiv:1104.3961}},
  \href {http://dx.doi.org/10.1016/j.ppnp.2011.04.001}
  {\path{doi:10.1016/j.ppnp.2011.04.001}}.

\bibitem{Sudol:2009vc}
M.~Sudol, et~al., {Feasibility studies of the time-like proton electromagnetic
  form factor measurements with PANDA at FAIR}, Eur. Phys. J. A 44 (2010)
  373--384.
\newblock \href {http://arxiv.org/abs/0907.4478} {\path{arXiv:0907.4478}},
  \href {http://dx.doi.org/10.1140/epja/i2010-10960-8}
  {\path{doi:10.1140/epja/i2010-10960-8}}.

\bibitem{Mora:2012}
M.~C. Mora~Espi,
  \href{{https://panda.gsi.de/system/files/user_uploads/m.c.moraespi/TH-PHD-2012-002.pdf}}{{Feasibility
  studies for accessing nucleon structure observables with the PANDA experiment
  at the future FAIR facility}}, Ph.D. thesis, {Johannes Gutenberg Universitat,
  Mainz} ({2012}).
\newline\urlprefix\url{{https://panda.gsi.de/system/files/user_uploads/m.c.moraespi/TH-PHD-2012-002.pdf}}

\bibitem{Lundborg:2005am}
A.~Lundborg, T.~Barnes, U.~Wiedner, {Charmonium production in p anti-p
  annihilation: Estimating cross sections from decay widths}, Phys. Rev. D 73
  (2006) 096003.
\newblock \href {http://arxiv.org/abs/hep-ph/0507166}
  {\path{arXiv:hep-ph/0507166}}, \href
  {http://dx.doi.org/10.1103/PhysRevD.73.096003}
  {\path{doi:10.1103/PhysRevD.73.096003}}.

\bibitem{Lin:2012ru}
Q.-Y. Lin, H.-S. Xu, X.~Liu, {Revisiting the production of charmonium plus a
  light meson at PANDA}, Phys. Rev. D 86 (2012) 034007.
\newblock \href {http://arxiv.org/abs/1203.4029} {\path{arXiv:1203.4029}},
  \href {http://dx.doi.org/10.1103/PhysRevD.86.034007}
  {\path{doi:10.1103/PhysRevD.86.034007}}.

\bibitem{Ma:2014pka}
B.~Ma, B.~Pire, K.~Semenov-Tian-Shansky, L.~Szymanowski, {${\pi}N$ TDAs from
  charmonium production in association with a forward pion at
  $\overline{P}ANDA$}, EPJ Web Conf. 73 (2014) 05006.
\newblock \href {http://arxiv.org/abs/1402.0413} {\path{arXiv:1402.0413}},
  \href {http://dx.doi.org/10.1051/epjconf/20147305006}
  {\path{doi:10.1051/epjconf/20147305006}}.

\bibitem{MA:2014scq}
B.~Ma, \href{{https://tel.archives-ouvertes.fr/tel-01131186}}{{Simulation of
  electromagnetic channels for PANDA@FAIR}}, Ph.D. thesis, {Universit{\'{e}}
  Paris-Sud, Orsay} ({2014}).
\newline\urlprefix\url{{https://tel.archives-ouvertes.fr/tel-01131186}}

\bibitem{Boer:2011fh}
D.~Boer, et~al., {Gluons and the quark sea at high energies: Distributions,
  polarization, tomography}\href {http://arxiv.org/abs/1108.1713}
  {\path{arXiv:1108.1713}}.

\bibitem{AbelleiraFernandez:2012cc}
J.~Abelleira~Fernandez, et~al., {A Large Hadron Electron Collider at CERN:
  Report on the Physics and Design Concepts for Machine and Detector}, J. Phys.
  G 39 (2012) 075001.
\newblock \href {http://arxiv.org/abs/1206.2913} {\path{arXiv:1206.2913}},
  \href {http://dx.doi.org/10.1088/0954-3899/39/7/075001}
  {\path{doi:10.1088/0954-3899/39/7/075001}}.

\bibitem{Anderle:2021wcy}
D.~P. Anderle, et~al., {Electron-ion collider in China}, Front. Phys. (Beijing)
  16~(6) (2021) 64701.
\newblock \href {http://arxiv.org/abs/2102.09222} {\path{arXiv:2102.09222}},
  \href {http://dx.doi.org/10.1007/s11467-021-1062-0}
  {\path{doi:10.1007/s11467-021-1062-0}}.

\bibitem{AbdulKhalek:2021gbh}
R.~Abdul~Khalek, et~al., {Science requirements and detector concepts for the
  Electron-Ion Collider: EIC yellow report. }\href
  {http://arxiv.org/abs/2103.05419} {\path{arXiv:2103.05419}}.

\bibitem{EIC:RDHandbook}
E.-C. Aschenauer, et~al.,
  \href{http://eicug.org/web/sites/default/files/EIC_HANDBOOK_v1.2.pdf}{Electron-ion
  collider detector requirements and {R\&D} handbook.} (2020).
\newline\urlprefix\url{http://eicug.org/web/sites/default/files/EIC_HANDBOOK_v1.2.pdf}

\bibitem{Hagler:2009ni}
P.~Hagler, {Hadron structure from lattice quantum chromodynamics}, Phys. Rept.
  490 (2010) 49--175.
\newblock \href {http://arxiv.org/abs/0912.5483} {\path{arXiv:0912.5483}},
  \href {http://dx.doi.org/10.1016/j.physrep.2009.12.008}
  {\path{doi:10.1016/j.physrep.2009.12.008}}.

\bibitem{Lin:2017snn}
H.-W. Lin, et~al., {Parton distributions and lattice QCD calculations: a
  community white paper}, Prog. Part. Nucl. Phys. 100 (2018) 107--160.
\newblock \href {http://arxiv.org/abs/1711.07916} {\path{arXiv:1711.07916}},
  \href {http://dx.doi.org/10.1016/j.ppnp.2018.01.007}
  {\path{doi:10.1016/j.ppnp.2018.01.007}}.

\bibitem{Aoki:2017puj}
Y.~Aoki, T.~Izubuchi, E.~Shintani, A.~Soni, {Improved lattice computation of
  proton decay matrix elements}, Phys. Rev. D 96~(1) (2017) 014506.
\newblock \href {http://arxiv.org/abs/1705.01338} {\path{arXiv:1705.01338}},
  \href {http://dx.doi.org/10.1103/PhysRevD.96.014506}
  {\path{doi:10.1103/PhysRevD.96.014506}}.

\bibitem{Ji:2013dva}
X.~Ji, {Parton Physics on a Euclidean Lattice}, Phys. Rev. Lett. 110 (2013)
  262002.
\newblock \href {http://arxiv.org/abs/1305.1539} {\path{arXiv:1305.1539}},
  \href {http://dx.doi.org/10.1103/PhysRevLett.110.262002}
  {\path{doi:10.1103/PhysRevLett.110.262002}}.

\bibitem{Radyushkin:2017cyf}
A.~V. Radyushkin, {Quasi-parton distribution functions, momentum distributions,
  and pseudo-parton distribution functions}, Phys. Rev. D 96~(3) (2017) 034025.
\newblock \href {http://arxiv.org/abs/1705.01488} {\path{arXiv:1705.01488}},
  \href {http://dx.doi.org/10.1103/PhysRevD.96.034025}
  {\path{doi:10.1103/PhysRevD.96.034025}}.

\bibitem{Lin:2014zya}
H.-W. Lin, J.-W. Chen, S.~D. Cohen, X.~Ji, {Flavor Structure of the Nucleon Sea
  from Lattice QCD}, Phys. Rev. D 91 (2015) 054510.
\newblock \href {http://arxiv.org/abs/1402.1462} {\path{arXiv:1402.1462}},
  \href {http://dx.doi.org/10.1103/PhysRevD.91.054510}
  {\path{doi:10.1103/PhysRevD.91.054510}}.

\bibitem{Orginos:2017kos}
K.~Orginos, A.~Radyushkin, J.~Karpie, S.~Zafeiropoulos, {Lattice QCD
  exploration of parton pseudo-distribution functions}, Phys. Rev. D 96~(9)
  (2017) 094503.
\newblock \href {http://arxiv.org/abs/1706.05373} {\path{arXiv:1706.05373}},
  \href {http://dx.doi.org/10.1103/PhysRevD.96.094503}
  {\path{doi:10.1103/PhysRevD.96.094503}}.

\bibitem{Roberts:1994dr}
C.~D. Roberts, A.~G. Williams, {Dyson-Schwinger equations and their application
  to hadronic physics}, Prog. Part. Nucl. Phys. 33 (1994) 477--575.
\newblock \href {http://arxiv.org/abs/hep-ph/9403224}
  {\path{arXiv:hep-ph/9403224}}, \href
  {http://dx.doi.org/10.1016/0146-6410(94)90049-3}
  {\path{doi:10.1016/0146-6410(94)90049-3}}.

\bibitem{Cloet:2007pi}
I.~C. Cloet, A.~Krassnigg, C.~D. Roberts, {Dynamics, symmetries and hadron
  properties}, eConf C070910 (2007) 125.
\newblock \href {http://arxiv.org/abs/0710.5746} {\path{arXiv:0710.5746}}.

\bibitem{Chang:2013pq}
L.~Chang, I.~C. Cloet, J.~J. Cobos-Martinez, C.~D. Roberts, S.~M. Schmidt,
  P.~C. Tandy, {Imaging dynamical chiral symmetry breaking: pion wave function
  on the light front}, Phys. Rev. Lett. 110~(13) (2013) 132001.
\newblock \href {http://arxiv.org/abs/1301.0324} {\path{arXiv:1301.0324}},
  \href {http://dx.doi.org/10.1103/PhysRevLett.110.132001}
  {\path{doi:10.1103/PhysRevLett.110.132001}}.

\bibitem{Shi:2018mcb}
C.~Shi, C.~Mezrag, H.-s. Zong, {Pion and kaon valence quark distribution
  functions from Dyson-Schwinger equations}, Phys. Rev. D 98~(5) (2018) 054029.
\newblock \href {http://arxiv.org/abs/1806.10232} {\path{arXiv:1806.10232}},
  \href {http://dx.doi.org/10.1103/PhysRevD.98.054029}
  {\path{doi:10.1103/PhysRevD.98.054029}}.

\bibitem{Mezrag:2014jka}
C.~Mezrag, L.~Chang, H.~Moutarde, C.~D. Roberts, J.~Rodr\'\i{}guez-Quintero,
  F.~Sabati\'e, S.~M. Schmidt, {Sketching the pion's valence-quark generalised
  parton distribution}, Phys. Lett. B 741 (2015) 190--196.
\newblock \href {http://arxiv.org/abs/1411.6634} {\path{arXiv:1411.6634}},
  \href {http://dx.doi.org/10.1016/j.physletb.2014.12.027}
  {\path{doi:10.1016/j.physletb.2014.12.027}}.

\bibitem{Bednar:2018htv}
K.~D. Bednar, I.~C. Clo\"et, P.~C. Tandy, {Nucleon quark distribution functions
  from the Dyson\textendash{}Schwinger equations}, Phys. Lett. B 782 (2018)
  675--681.
\newblock \href {http://arxiv.org/abs/1803.03656} {\path{arXiv:1803.03656}},
  \href {http://dx.doi.org/10.1016/j.physletb.2018.06.020}
  {\path{doi:10.1016/j.physletb.2018.06.020}}.

\bibitem{Cloet:2008re}
I.~C. Cloet, G.~Eichmann, B.~El-Bennich, T.~Klahn, C.~D. Roberts, {Survey of
  nucleon electromagnetic form factors}, Few Body Syst. 46 (2009) 1--36.
\newblock \href {http://arxiv.org/abs/0812.0416} {\path{arXiv:0812.0416}},
  \href {http://dx.doi.org/10.1007/s00601-009-0015-x}
  {\path{doi:10.1007/s00601-009-0015-x}}.

\bibitem{Muller:2012yq}
D.~{M\"{u}}ller, B.~Pire, L.~Szymanowski, J.~Wagner, {On timelike and spacelike
  hard exclusive reactions}, Phys. Rev. D 86 (2012) 031502.
\newblock \href {http://arxiv.org/abs/1203.4392} {\path{arXiv:1203.4392}},
  \href {http://dx.doi.org/10.1103/PhysRevD.86.031502}
  {\path{doi:10.1103/PhysRevD.86.031502}}.

\bibitem{Hakioglu:1991pn}
T.~Hakioglu, M.~D. Scadron, {Vector meson dominance, one loop order quark
  graphs, and the chiral limit}, Phys. Rev. D 43 (1991) 2439--2442.
\newblock \href {http://dx.doi.org/10.1103/PhysRevD.43.2439}
  {\path{doi:10.1103/PhysRevD.43.2439}}.

\bibitem{Schildknecht:2005xr}
D.~Schildknecht, {Vector meson dominance}, Acta Phys. Polon. B 37 (2006)
  595--608.
\newblock \href {http://arxiv.org/abs/hep-ph/0511090}
  {\path{arXiv:hep-ph/0511090}}.

\bibitem{Brodsky:2008qu}
S.~J. Brodsky, F.~J. Llanes-Estrada, A.~P. Szczepaniak, {Local Two-Photon
  Couplings and the {$J=0$} Fixed Pole in Real and Virtual Compton Scattering},
  Phys. Rev. D 79 (2009) 033012.
\newblock \href {http://arxiv.org/abs/0812.0395} {\path{arXiv:0812.0395}},
  \href {http://dx.doi.org/10.1103/PhysRevD.79.033012}
  {\path{doi:10.1103/PhysRevD.79.033012}}.

\bibitem{Fucini:2020vpr}
S.~Fucini, M.~Rinaldi, S.~Scopetta, {Generalized parton distributions of light
  nuclei}, Few Body Syst. 62~(1) (2021) 3.
\newblock \href {http://arxiv.org/abs/2010.12212} {\path{arXiv:2010.12212}},
  \href {http://dx.doi.org/10.1007/s00601-020-01590-0}
  {\path{doi:10.1007/s00601-020-01590-0}}.

\bibitem{Hattawy:2017woc}
M.~Hattawy, et~al., {First Exclusive Measurement of Deeply Virtual Compton
  Scattering off $^4$He: Toward the 3D Tomography of Nuclei}, Phys. Rev. Lett.
  119~(20) (2017) 202004.
\newblock \href {http://arxiv.org/abs/1707.03361} {\path{arXiv:1707.03361}},
  \href {http://dx.doi.org/10.1103/PhysRevLett.119.202004}
  {\path{doi:10.1103/PhysRevLett.119.202004}}.

\bibitem{Brodsky:1983vf}
S.~J. Brodsky, C.-R. Ji, G.~Lepage, {Quantum Chromodynamic Predictions for the
  Deuteron Form-Factor}, Phys. Rev. Lett. 51 (1983) 83.
\newblock \href {http://dx.doi.org/10.1103/PhysRevLett.51.83}
  {\path{doi:10.1103/PhysRevLett.51.83}}.

\bibitem{Yero:2020cbq}
C.~Yero, et~al., {Probing the Deuteron at Very Large Internal Momenta}, Phys.
  Rev. Lett. 125~(26) (2020) 262501.
\newblock \href {http://arxiv.org/abs/2008.08058} {\path{arXiv:2008.08058}},
  \href {http://dx.doi.org/10.1103/PhysRevLett.125.262501}
  {\path{doi:10.1103/PhysRevLett.125.262501}}.

\bibitem{Yero:2020urm}
C.~Yero, {Cross Section Measurements of Deuteron Electro-Disintegration at Very
  High Recoil Momenta and Large 4-Momentum Transfers $(Q^2)$}, Ph.D. thesis,
  Florida Intl. U. (2020).
\newblock \href {http://arxiv.org/abs/2009.11343} {\path{arXiv:2009.11343}},
  \href {http://dx.doi.org/10.2172/1644044} {\path{doi:10.2172/1644044}}.

\bibitem{WeissTalk20}
C.~Weiss,
  \href{{https://indico.jlab.org/event/375/contributions/5998/attachments/5051/6298/weiss20_backward20.pdf}}{Exploring
  the soft-hard transition in forward and backward meson production},
  {B}ackward-angle ($u$-channel) Physics Workshop, 21 September 2020- 23
  September 2020, online event, USA (2020).
\newline\urlprefix\url{{https://indico.jlab.org/event/375/contributions/5998/attachments/5051/6298/weiss20_backward20.pdf}}

\bibitem{Vinnikov:2006xw}
A.~V. Vinnikov, {Code for prompt numerical computation of the leading order GPD
  evolution } (2006).
\newblock \href {http://arxiv.org/abs/hep-ph/0604248}
  {\path{arXiv:hep-ph/0604248}}.

\bibitem{Pire:1998nw}
B.~Pire, J.~Soffer, O.~Teryaev, {Positivity constraints for off - forward
  parton distributions}, Eur. Phys. J. C 8 (1999) 103--106.
\newblock \href {http://arxiv.org/abs/hep-ph/9804284}
  {\path{arXiv:hep-ph/9804284}}, \href
  {http://dx.doi.org/10.1007/s100529901063} {\path{doi:10.1007/s100529901063}}.

\bibitem{Pobylitsa:2002iu}
P.~V. Pobylitsa, {Positivity bounds on generalized parton distributions in
  impact parameter representation}, Phys. Rev. D 66 (2002) 094002.
\newblock \href {http://arxiv.org/abs/hep-ph/0204337}
  {\path{arXiv:hep-ph/0204337}}, \href
  {http://dx.doi.org/10.1103/PhysRevD.66.094002}
  {\path{doi:10.1103/PhysRevD.66.094002}}.

\bibitem{Diehl:2017wew}
M.~Diehl, J.~R. Gaunt, {Double parton scattering theory overview}, Adv. Ser.
  Direct. High Energy Phys. 29 (2018) 7--28.
\newblock \href {http://arxiv.org/abs/1710.04408} {\path{arXiv:1710.04408}},
  \href {http://dx.doi.org/10.1142/9789813227767_0002}
  {\path{doi:10.1142/9789813227767_0002}}.

\bibitem{Lorce:2011dv}
C.~Lorce, B.~Pasquini, M.~Vanderhaeghen, {Unified framework for generalized and
  transverse-momentum dependent parton distributions within a 3Q light-cone
  picture of the nucleon}, JHEP 05 (2011) 041.
\newblock \href {http://arxiv.org/abs/1102.4704} {\path{arXiv:1102.4704}},
  \href {http://dx.doi.org/10.1007/JHEP05(2011)041}
  {\path{doi:10.1007/JHEP05(2011)041}}.

\bibitem{Goritschnig:2012vs}
A.~T. Goritschnig, B.~Pire, W.~Schweiger, {Double handbag description of
  proton-antiproton annihilation into a heavy meson pair}, Phys. Rev. D 87~(1)
  (2013) 014017, [Erratum: Phys.Rev.D 88, 079903 (2013)].
\newblock \href {http://arxiv.org/abs/1210.8095} {\path{arXiv:1210.8095}},
  \href {http://dx.doi.org/10.1103/PhysRevD.87.014017}
  {\path{doi:10.1103/PhysRevD.87.014017}}.

\bibitem{Haisch:2021hvj}
U.~Haisch, A.~Hala, {Light-cone sum rules for proton decay}, JHEP 05 (2021)
  258.
\newblock \href {http://arxiv.org/abs/2103.13928} {\path{arXiv:2103.13928}},
  \href {http://dx.doi.org/10.1007/JHEP05(2021)258}
  {\path{doi:10.1007/JHEP05(2021)258}}.

\bibitem{Vladimirov}
V.~S. Vladimirov, { Equations of mathematical physics}, Mir, Moscow, 1984.

\end{thebibliography}

\end{document}